\documentclass[twoside]{book}

\usepackage{quotchap}

\usepackage{minitoc}

\usepackage{fancyhdr}
\usepackage{amssymb}
\usepackage{amsmath}
\usepackage{bm}
\usepackage{graphicx}
\usepackage{wrapfig}
\usepackage{subfig} 
\usepackage{fancybox}
\usepackage{mathdots}
\usepackage{rotating}
\usepackage{makeidx}
\usepackage{enumitem} 
\usepackage{color}
\usepackage{framed} 
\usepackage{wasysym}
\usepackage{gravity} 
\usepackage{pdfpages}

\usepackage{tikz}

\usetikzlibrary{
  calc,
  decorations.markings,
  decorations.pathmorphing,
  decorations.pathreplacing,
  patterns.meta,
  shapes.misc
}
\colorlet{collapsingmatter}{blue!35}

\usepackage{graphicx,epsf,epsfig,amssymb} 
\usepackage{cite} 
\usepackage{aas_macros}

\usepackage[
colorlinks=true,
linkcolor=blue,
urlcolor=magenta,
filecolor=blue,
citecolor=red,
pdfstartview=FitV,
pdftitle={},
pdfauthor={},
pdfsubject={},
pdfkeywords={},
pdfpagemode={},
bookmarksopen=true
]{hyperref} 

\usepackage{float}

\usepackage{geometry}
\newcommand{\textcite}[1]{\texttt{#1}}

\usepackage{nameref}
\usepackage{xparse}
\usepackage{etoolbox} 

\makeatletter
\NewDocumentCommand{\codefile}{O{} m}{%
  \item \texttt{#2}%
  \phantomsection%
  \def\@currentlabelname{#2}%
  \label{#1}%
}
\makeatother

\newcommand{\coderef}[1]{%
  \begingroup
  \hypersetup{hidelinks}%
  \textcolor{red}{\hyperref[#1]{\texttt{\nameref{#1}}}}%
  \endgroup
}
\definecolor{ss}{rgb}{0.54, 0.17, 0.89}%

\newcounter{mnotecount}[section]

\usepackage{epigraph} 
\usepackage{lipsum}
\usepackage[T1]{fontenc} 
\usepackage{lmodern} 

\usepackage{textcomp, gensymb}
\usepackage{bm} 

\usepackage{mathrsfs} 

\def\be{\begin{equation}}
\def\ee{\end{equation}}
\def\beq{\begin{eqnarray}}
\def\eeq{\end{eqnarray}}

\let\dotlessitext\i

\newcommand{\beqn}{\begin{eqnarray*}}
\newcommand{\eeqn}{\end{eqnarray*}}
\newcommand{\p}{\partial}

\def\op{ \ $ }
\def\cl{$ \ }

\def\l2{\op \bf L^2\cl}

\def\l{{\cal L}}

\renewcommand{\vec}[1]{\boldsymbol{#1}}

\makeatletter

\let\oldchaptermark\chaptermark
\renewcommand{\chaptermark}[1]{\oldchaptermark{\@chaptermark}}
\let\old@chapter\@chapter
\def\@chapter[#1]#2{%
  \def\@chaptermark{#1}
  \old@chapter[#2]{#2}%
}

\let\oldsectionmark\sectionmark
\renewcommand{\sectionmark}[1]{\oldsectionmark{\@sectionmark}}
\let\oldsubsectionmark\subsectionmark
\renewcommand{\subsectionmark}[1]{\oldsubsectionmark{\@subsectionmark}}
\let\old@sect\@sect
\def\@sect#1#2#3#4#5#6[#7]#8{%
  \@namedef{@#1mark}{#7}
  \old@sect{#1}{#2}{#3}{#4}{#5}{#6}[#8]{#8}%
}

\usepackage{xcolor}
\usepackage{colortbl}

\newcommand{\Msun}{\,\mathrm{M}_\odot}
\newcommand{\Rsun}{\,\mathrm{R}_\odot}
\newcommand{\Lsun}{\,\mathrm{L}_\odot}

\newcommand{\blankpage}{\newpage\null\thispagestyle{empty}\newpage}
\usepackage{lmodern} \normalfont 
\DeclareFontShape{T1}{lmr}{bx}{sc} { <-> ssub * cmr/bx/sc }{}

\newcommand{\chapterauthor}[1]{%
  {\parindent0pt\vspace*{-20pt}%
   \linespread{1.1}\large\scshape#1%
   \par\nobreak\vspace*{30pt}}%
  \@afterheading%
}

\renewcommand\chapter{\clearpage
\thispagestyle{plain}%
\global\@topnum\z@
\@afterindentfalse
\secdef\@chapter\@schapter}
\makeatother
\makeindex
\begin{document}

\begin{titlepage}
\includepdf{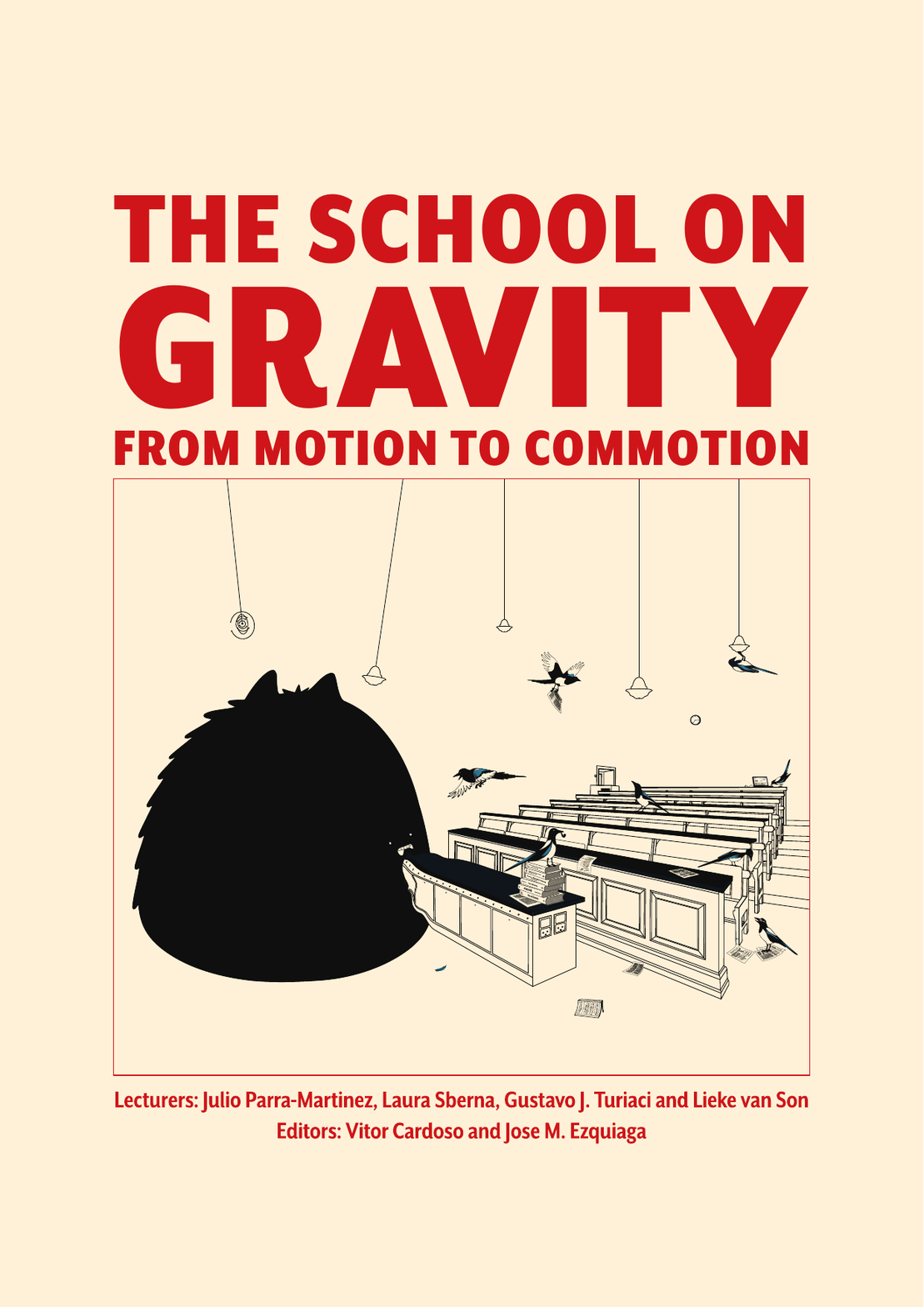} 
\end{titlepage}

\blankpage
\thispagestyle{empty}

\vspace{2em}
\begin{center} 
\Huge \sc The School on Gravity  
\\[0.2cm]
\huge \sc Lecture Notes\\[1.2in]

\Large {\scshape \textsl{The Center of Gravity @ The Niels Bohr Institute (2026)}}\\[0.6in]

\Large {\scshape \textsl{Editors:}}\\[0.2in]
\large \sc Vitor Cardoso \textsuperscript{a} and Jose M. Ezquiaga \textsuperscript{a}\\[0.2in]
\vspace{0.3em}
\Large {\scshape \textsl{Lecturers:}}\\[0.2in]
\large \sc Julio Parra-Martinez \textsuperscript{1}, \, Laura Sberna \textsuperscript{2}, \\
Gustavo J. Turiaci \textsuperscript{3},\, Lieke van Son \textsuperscript{4} \\[0.2in]

\large \sc (Dated: \today)\\[1in]
\begin{minipage}{0.9\textwidth}
\hrule height 0.5pt
\vspace{0.3em}
\small
\raggedright
\textsuperscript{A} Center of Gravity, Niels Bohr Institute, Blegdamsvej 17, 2100 Copenhagen, Denmark\\

\textsuperscript{1} Institut des Hautes \'Etudes Scientifiques, 91440 Bures-sur-Yvette, France\\

\textsuperscript{2} School of Mathematical Sciences, University of Nottingham,
University Park, Nottingham NG7 2RD, United Kingdom\\

\textsuperscript{3} Department of Physics, University of Washington, Seattle, WA 98195, USA\\

\textsuperscript{4} Department of Astrophysics/IMAPP, Radboud University, P.O. Box 9010, NL-6500 GL Nijmegen,
The Netherlands\\[0.3em]
\end{minipage}\\[0.8in]

\end{center}

\blankpage
\newpage

\thispagestyle{empty}
\vspace*{1.5in}

\begin{center}
    {\Huge \sc Abstract}
\end{center}

\vspace{0.5in}

\begin{center}
\begin{minipage}{0.85\textwidth}
\normalsize

Understanding the gravitational interaction is one of the great scientific endeavors of the 21st century. The Nobel Prize-winning discovery of gravitational waves in 2015, and subsequent detections by the LIGO-Virgo-KAGRA Collaboration, have marked a revolutionary step forward in physics and astrophysics by opening an entirely new window for exploring the Universe. Future upgrades to these detectors, along with upcoming facilities such as LISA, the Einstein Telescope, and radio and optical instruments, promise other breakthrough discoveries that the next generation of scientists will need to interpret.
\\

The School on Gravity (22–26 June 2026, Copenhagen) aimed to introduce junior scientists—graduate students and advanced undergraduates with a background in general relativity—to a wide range of foundational topics in the exciting new field of gravitational-wave physics. 
Topics spanned from the theory of motion, to new challenges introduced by quantum physics, to the astrophysics of black holes and source modeling for gravitational-wave detectors.
\\

The School on Gravity took place at the legendary Auditorium A of the Niels Bohr Institute, inviting students and lecturers to challenge current paradigms in gravitational physics in the spirit of the quantum revolution that unfolded in this very room. 
\\

These lecture notes aim at opening up those new ideas presented in Auditorium A to the rest of the scientific community. 

\end{minipage}
\end{center}

\newpage
\thispagestyle{empty}
\paragraph{Scientific Organizing Committee:}
Vitor Cardoso (Director of CoG), Emil Bjerrum-Bohr, Jose M. Ezquiaga, Troels Harmark, Niels Obers, Marta Orselli, Alessia Platania, Johan Samsing, Maarten van de Meent, Ziqi Yan

\paragraph{Local Organizing Committee:}
Vitor Cardoso, Giada Caneva Santoro, Jose M. Ezquiaga, Simon Maenaut, Emily Kate Mansfield, Fanny Töpper, Luka Vujeva, and Guangzi Xu

\paragraph{CoG internal reviewers:} 
Giada Caneva Santoro, 
Francesco del Porro, 
Brian Seymour,
Nabha Shah

\paragraph{Design:}
Ana Carvalho

\paragraph{Participants:} \textcolor{white}{.}

\noindent Alejandro Svyatkovskyy Kholyavka (University of the Balearic Islands) \\
Alessandro Mollicone (University of Naples Federico II) \\
April Cheng (Princeton University) \\
Beatrice Lazzati (University of Milano-Bicocca) \\
Bert Depoorter (KU Leuven) \\
Brian Seymour (Niels Bohr Institute) \\
Cecilia Maria Fabbri (University of Nottingham) \\
Charlotte Rizo (ENS Paris-Saclay) \\
Dario Rossi (University of Pisa) \\
Davide Panella (Niels Bohr Institute \& University of Perugia) \\
Davina Nwachukwu (King’s College London) \\
Devesh Giri (IISER \& TIFR) \\
Disha Hegde (UCLouvain) \\ 
Dylan Hebrail (Lund University) \\
Eloisa Margherita Calafiore (Università degli Studi di Milano-Bicocca) \\
Emil Bjerrum-Bohr (Niels Bohr Institute) \\
Filipe Santos (Técnico Lisboa) \\
Francesco Mattia Carone (La Sapienza, University of Rome) \\
Giulia Cuomo (University of Milano-Bicocca) \\
Giulia Ortame (University of Cambridge) \\
Guangzi Xu (Niels Bohr Institute) \\
Irmeli Nix (Ruprecht-Karls-Universität Heidelberg) \\
Irvin Martinez-Rodriguez (Carnegie Mellon University) \\
Jessica Murphy (Trinity College Dublin) \\
João Joaquim (Técnico Lisboa) \\
João Sieiro dos Santos (Técnico Lisboa) \\
Johan Samsing (Niels Bohr Institute) \\
Jonas Broe Bendtsen (University of Southern Denmark) \\
Jorge Expósito Patiño (Técnico Lisboa) \\
Jupiter Stevenson (University of Nottingham) \\
Karen Kang (University of Cambridge) \\
Konstantinos Topaloglou (University of Bologna) \\
Laura Michelutti (Aix-Marseille University) \\
Leart Sabani (Niels Bohr Institute) \\
Luca Sebastiani (MPI for Gravitational Physics, Albert Einstein Institute) \\
Lucía Vélez Tartajo (Niels Bohr Institute) \\
Luka Vujeva (Niels Bohr Institute) \\
Lukas Schnabel (University of Hamburg) \\
Maria Gonçalves Salgado (Leiden University) \\
Marie Sommerfeld (Heidelberg University) \\
Mariia Murga (TU Darmstadt) \\
Marta Cocco (University of Perugia \& Niels Bohr Institute) \\
Mathilde Menu (Aix-Marseille University) \\
Matilde Garcia (University of Milano-Bicocca) \\
Milos Ertola Urtubey (Instituto Argentino de Radioastronomía) \\
Nathan Castet (IHES, IPhT) \\
Nao Nakamoto (Kyoto University) \\
Niels Obers (Niels Bohr Institute) \\
Oscar Combaluzier-Szteinsznaider (Université Paris-Cité) \\
Polyxeni Rafaela Tsigdinou (Niels Bohr Institute) \\
Qixuan Xu (Técnico Lisboa) \\
Rachel Wang (Institut des Hautes Études Scientifiques) \\
Robrecht Keijzer (KU Leuven) \\
Sofia Dossena (University of Milano-Bicocca) \\
Sofía Bussières López (University of the Balearic Islands) \\
Swasti Jain (Cardiff University) \\
Theophanes Karydas (University of Amsterdam) \\
Thippayawis (Tong) Cheunchitra (University of Melbourne) \\
Thomas Apostolidis (APC Paris, Université Paris Cité) \\
Tom Colin (LUX, Observatoire de Paris) \\
Tom van der Steen (KU Leuven) \\
Tommaso Corbetta (University of Padova) \\
Utkarsh Basu (Radboud University) \\
Vincent He (Institut des Hautes Études Scientifiques) \\
Yuhe Zeng (Shanghai Jiao Tong University) \\
Yuvraj Sharma (Rhode Island University) \\

\thispagestyle{empty} 

\thispagestyle{empty}
\tableofcontents 
\thispagestyle{empty}

\blankpage
\frontmatter
\chapter*{Why a School on Gravity?}
\chapterauthor{Vitor Cardoso and Jose M. Ezquiaga}
\label{ch:introduction} 
\phantomsection
\addcontentsline{toc}{chapter}{Why a School on Gravity?}
\markboth{Why a School on Gravity?}{Why a School on Gravity?}
\allowdisplaybreaks
\minitoc
Gravity is everywhere, now more than ever. 
We have wondered for millennia about how best to throw rocks into a pond, reach the highest fruit on a tree or understand planetary motion. 
Technological and theoretical developments have led us to elucidate the trajectories of falling objects, the universality of gravitational forces and the deflection of light by massive objects. 
The last decade has, however, brought a true paradigm shift: we can now directly image black holes and measure the ripples of spacetime itself. 
In other words, we now have observational access to explore gravity in its strongest and most dynamical regime.

It is fair to say that, despite the tremendous recent progress, we have more questions than answers. 
The present limitations are clear. 
We know that general relativity is incomplete. 
We know that determining the astrophysical origin of the observed colliding black holes is hard. 
We know that a single perturbative method cannot describe the whole spectrum of moving compact objects. 
And there are so many other things we know we do not know. 
We definitely need new ideas and approaches to understand gravity.

Understanding gravity in the 21st century requires a holistic perspective: 
deviations from general relativity cannot be established without properly modeling detector noise;
the photon ring cannot be mapped without informed priors on the accretion disks; 
gravitational-wave waveform developments require new ideas from different communities; 
and 
new observational facilities will not be funded without theoretical motivations and support from the general public. 
We believe this reality should be reflected in the training of the next generation of gravity researchers. 

The ``School on Gravity: from motion to commotion'' is the first in a series of annual PhD schools held at the historic Auditorium A of the Niels Bohr Institute. 
This series aims to cover the extensive landscape of gravity research in a way that challenges current paradigms and standard assumptions, to nurture a community that exploits synergies across fields. 
We hope you join us on this journey! 

\mainmatter

\pagenumbering{arabic}

\chapter{Black Holes and Neutron Stars}
\chapterauthor{Lieke van Son}
\label{ch:part1} 
\allowdisplaybreaks
\minitoc
%
%


These lecture notes are an introduction to the astrophysics of stellar-mass black holes and neutron stars. They address four questions: 
\begin{itemize}[nolistsep]
    \item How do massive stars form black holes and neutron stars?
    \item How do we observe and study black holes and neutron stars?
    \item How do we form the {\it merging} pairs of compact objects seen by gravitational-wave detectors?
    \item What are the open problems that gravitational-wave populations pose today?
\end{itemize}
Broadly, we will move from stars and electromagnetic observations towards black holes and gravitational waves.

\section{Stellar black hole formation}\label{sec:day1}
All stellar-mass black holes (and neutron stars) were once massive stars\footnote{notwithstanding primordial black holes, whose existence remains speculative}. The learning goal of this first section is therefore: how do we get from a massive star to a stellar-mass black hole?
We begin with some of the most consequential uncertainties in massive (binary) stellar evolution, and how they propagate into compact object formation.
We then follow the core of a massive star to collapse, and briefly discuss supernova explosion theory and the resulting supernova zoo. Throughout, we focus on the mechanisms that set the boundaries of black hole formation, so that by the end of this section we can answer three questions:
I) What is the least massive black hole you can form from stars?
II) Are stellar black holes born with low spin?
III) What is the most massive black hole you can form from stars in our Galaxy? Is this different from the maximum black hole mass in the early Universe? 

\subsection{The uncertain (late) evolution of massive stars \label{ss:massive_stars}}

\subsubsection{Why care about massive stars? \label{sss: why_care}}

Massive binary stars impact every part of modern astrophysics. They helped reionise the early Universe \cite{Stanway2016,Gotberg2020}, they shape their host galaxies through the mechanical feedback of stellar winds and their explosive deaths \cite{Hopkins2012,Agertz2013,Sarbadhicary2022}, and they drive the chemical evolution of the Universe \cite{Maiolino2019}, in particular producing the elements that ultimately form the building blocks of life \cite{Kitadai2017}. Finally, and most relevant here, massive stars are the progenitors of the most compact objects in our Universe: neutron stars and black holes.
Unfortunately, massive stars are difficult to catch while alive: they are intrinsically rare, and they live fast and die young.


\subsubsection{Massive stars live fast, die young, and are rare \label{sss: live fast die young}}

By ``massive star'' we mean a star that is at least $8-11$ times more massive than the Sun, as these are the stars that reach the core temperatures required for non-degenerate carbon ignition ($T_c \gtrsim 5\times10^{8}$\,K).
Non-degenerate carbon ignition requires a CO-core mass above approximately $\sim 1.06\Msun$ after helium burning, corresponding to a zero-age main-sequence (ZAMS) mass of roughly $8\Msun$. Stars with masses above $11\Msun$ are thought to subsequently ignite heavier elements until an Fe core is formed which will collapse towards a black hole or neutron star. The exact boundary depends on uncertainties in e.g., convective overshooting, and stellar winds shifting it between $\sim 8$ and $11\Msun$.

How can we estimate how long stars live? 
An appropriate estimate for a star's lifetime is the nuclear timescale: how long a star can maintain hydrostatic equilibrium by producing as much nuclear energy as it radiates away. Nuclear fusion converts a small fraction $\phi \approx 0.007$ (for hydrogen fusion) of the rest mass of the reacting nuclei into energy, and only a fraction $f_{\rm nuc}$ of the stellar mass serves as fuel (we take $f_{\rm nuc} \approx 0.1$, a bit low for massive stars but fine to order of magnitude):
\be
\tau_{\rm nuc} = \frac{E_{nuc}}{L} =\phi\, f_{\rm nuc}\, \frac{M c^2}{L}
\approx 10^{10}\, \left(\frac{M}{\Msun}\right) \left(\frac{\Lsun}{L}\right)\,\mathrm{yr}.
\ee
Using the (main-sequence) mass--luminosity relation $L \propto M^{3.5}$,
\be
\tau_{\rm nuc} \propto 10^{10}\, \left(\frac{M}{\Msun}\right)^{-2.5}\,\mathrm{yr},
\ee
so a $10\Msun$ star lives only $\tau_{\rm nuc} \approx 10^{10-2.5}\,\mathrm{yr} \approx 30$\,Myr: massive stars live fast and die young.

Massive stars are also intrinsically rare. The initial mass function (IMF), i.e., how often stars of a given mass are born, declines steeply above one solar mass, $\xi(m) \propto m^{-2.35}$: for every 1000 stars formed, only approximately 2 will be massive stars.


\subsubsection{The Hertzsprung--Russell diagram \label{sss: HRD}}

The Hertzsprung--Russell (HR) diagram is the most common tool astronomers use to judge stars by their appearance. Since 1913, following Ejnar Hertzsprung and Henry Norris Russell, we sort stars by colour (temperature) on the horizontal axis and luminosity on the vertical axis. This pattern links the main properties of stars: more massive stars are brighter; and using the Stefan--Boltzmann law,
\be
L = 4\pi R^2 \sigma T_{\rm eff}^4,
\ee
stars in the bottom left have small radii while those in the top right are large. The HR diagram also encodes evolutionary state: as stars age they become bloated and puffy, moving to the cool, red side. Where and how long a star spends at different locations is primarily set by its mass; we catch most stars on the main sequence, where they spend $\sim$90\% of their lives, versus $\lesssim$10\% fusing helium and beyond.

Over the years, a lot of jargon has emerged based on where stars are found in this diagram. Some are self-explanatory, but it is useful to have seen those related to massive stars at least once: B stars (hot, $10{,}000\,\mathrm{K} \lesssim T \lesssim 28{,}000$\,K) and O stars ($T \gtrsim 28{,}000$\,K) are hot massive stars; red supergiants (RSG, $T \lesssim 4800$\,K), yellow supergiants (YSG, $4800\,\mathrm{K} \lesssim T \lesssim 7500$\,K) and blue supergiants (BSG, $T \gtrsim 10{,}000$\,K) are very luminous evolved massive stars; luminous blue variables (LBVs) are hot, extremely luminous stars undergoing episodic outbursts and luminosity variations; and Wolf--Rayet (WR) stars are hot stars whose spectra are dominated by broad emission lines of He/N/C/O with almost no hydrogen.


\subsubsection{Stellar winds  $\rightarrow$ \normalfont{\textit{setting the masses of black holes throughout cosmic history}}} 

Mass loss is important during all phases of massive-star evolution for $M \gtrsim 15\Msun$. This is partially an empirical fact (we observe massive stars to have very strong outflows) and partially understood through the theory of radiatively driven winds. The timescale for mass loss can be comparable to or smaller than $\tau_{\rm nuc}$, but the driving mechanisms are unfortunately often not well understood and mass-loss rates are correspondingly uncertain. 
Crucially, the mass of a black hole is largely set long before the star collapses as it is set by how much mass the star loses.  

Different types of winds dominate in different regions of the HR diagram, for a review, see e.g., \cite{Vink2022}:

{\bf Hot, line-driven winds.} On the hot side we find \textit{radiation}- or line-driven winds (the ``CAK'' theory) \cite{LucySolomon1970,CAK1975}: photons are captured or scattered in the absorption lines of metal ions in the outflow, transferring momentum outward. Iron-group lines in particular dominate the radiative acceleration. Because this only works if partially ionised metals are available, line-driven winds are strongly metallicity dependent, roughly
\be
\dot{M}_{\rm wind} \propto Z^{\alpha}, \qquad \alpha \sim 0.85,
\ee
so winds are expected to be much weaker at low metallicity. At $Z \lesssim Z_\odot/30$ intermediate-mass elements (mainly CNO) gain importance in the driving, flattening the $\dot M$--$Z$ relation \cite{Vink2001}. Within the hot winds we distinguish main-sequence OB winds (with a possible separate regime for very massive stars) and Wolf--Rayet-like winds.

{\bf Wolf--Rayet winds.} Wolf--Rayet stars are essentially very massive stars that have been stripped of their hydrogen envelopes. They are thought to be the direct progenitors of black holes. Their observational classification, characterized by strong emission lines, \textit{stems directly} from their strong, optically thick and radiatively-driven winds ($\dot M \sim 10^{-5}$--$10^{-4}\Msun\,\mathrm{yr}^{-1}$)\cite{2026enap....2..569S}. 
As such, their wind strength is very sensitive to the iron opacity bumps \cite{Vink2017,SanderVink2020}, but the exact strength of WR winds remains highly uncertain, with prescriptions differing by orders of magnitude. This has important consequences for black hole and neutron star formation, especially at higher metallicities.

{\bf Cool (RSG) winds.} On the cool ($T\lesssim 10^4$ K) side the gas becomes mostly neutral and line driving is inefficient. The driving mechanism of cool stars, such as RSG winds is unclear. Potentially, they are a combination of stellar pulsations and radiation pressure on dust grains that form in the cool outer atmosphere \cite{2021ARA&A..59..337D}. 
The lack of theory makes this an empirical story, and here too the mass-loss estimates vary by multiple orders of magnitude \cite{2018MNRAS.475...55B,2020MNRAS.492.5994B,2022ApJ...933...41B}.

{\bf LBV outbursts and the Eddington limit.} Luminous Blue Variable (LBV) stars do not actually experience a wind but rather strong outbursts with $\dot M \gtrsim 10^{-3}\Msun\,\mathrm{yr}^{-1}$. LBVs are thought to sit close to the Eddington luminosity. 
The Eddington limit reappears repeatedly throughout these lectures (the same limit caps black-hole accretion), so it is worth deriving once here.

Consider a thin spherical shell of mass $\mathrm{d}m$ at radius $r$ around a central object of mass $M$, assuming both accretion and the radiation field are spherically symmetric. 
The gravitational force on the shell is 
\be
F_{\rm grav} = \frac{GM\,\mathrm{d}m}{r^2}. 
\ee
The momentum flux carried by radiation through the shell is $L/c$. 
The fraction of this momentum that is intercepted by the shell is its total cross-section, $\kappa\,\mathrm{d}m$ (the opacity $\kappa$ has units of area per mass), divided by the shell area $4\pi r^2$. Hence
\be
F_{\rm rad} = \frac{L}{c} \frac{\kappa }{4\pi r^2}\,\mathrm{d}m .
\ee
Setting $F_{\rm rad} = F_{\rm grav}$ and solving for $L$ gives the Eddington luminosity
\be
L_{\rm Edd} = \frac{4\pi G M c}{\kappa}.
\ee
The value of $\kappa$ depends on the dominant opacity source; for ionised hydrogen, electrons scatter with the Thomson cross-section while protons carry the mass, $\kappa = \sigma_{\rm T}/m_{\rm p}$. This results in 
\be
L_{\rm Edd} \approx 1.3\times10^{38}\,\frac{M}{\Msun}\,\mathrm{erg\,s^{-1}} \approx 3\times10^{4}\,\frac{M}{\Msun} \Lsun
\ee
We define the Eddington factor
\be
\Gamma \equiv \frac{L}{L_{\rm Edd}}.
\ee
Very massive stars and WR stars can approach or exceed $\Gamma \sim 1$. 
This is indeed what is observed: very few stars sit above the empirical Humphreys--Davidson limit in the HR diagram \cite{HumphreysDavidson1979}. 
Stars that luminous are essentially living at their Eddington limit, making them very unstable, and easily blow their outer layers away. 
They never become big fluffy red supergiants but instead move towards the compact very-massive-star and WR regime.
Over a dozen LBVs are identified in the Milky Way and Magellanic Clouds, with a similar number in other Local Group galaxies e.g., \cite{Smith2025LBVreview}.


{\bf The maximum black-hole mass as a function of metallicity.} If the most massive stars blow their outer layers away, those layers can of course never contribute to a later formed compact object. 
High metallicity means strong winds means low black-hole masses; the maximum stellar black-hole mass one can form is set by winds at high (e.g., galactic) metallicities. At very low metallicities, a different mechanism called `pair instability' comes into play (Section~\ref{sec:PISN}), placing an upper limit on the black-hole mass that can be formed from stars, even for weak winds \cite{Spera2015,SperaMapelli2017,Vink2021,Merritt2026}.
Because cosmic star formation is also heavily metallicity dependent, the maximum black-hole mass we can form from stars is a function of redshift: at high redshift, when the Universe was more metal poor, we expect to be able to form much more massive black holes than today. This is one of the reasons we suspect many of the merging binary black holes detected by LIGO/Virgo/KAGRA were born in the early Universe, and not in the Milky Way today e.g., \cite{FishbachvanSon2023}.
%

\subsubsection{Rotation and angular-momentum transport $\rightarrow$ \normalfont{\textit{setting the spins of black holes and neutron stars}}}

The second big uncertainty is rotation, and specifically how angular momentum (AM) is transported inside the star. 
This matters because it sets the final spin of the neutron star or black hole, and because rapid rotation can unlock an entirely different evolutionary pathway (chemically homogeneous evolution). Broadly, four classes of mechanisms transport AM: hydrodynamical instabilities, rotationally induced circulations, magnetic torques, and internal gravity waves e.g., \cite{Zahn1975,ChaboyerZahn1992,MeynetMaeder1997,Rieutord2025}. We focus on the middle two.

{\bf Rotationally induced (meridional) circulations.} 
A rotating star is deformed into an oblate shape (think of a flattened ball), so its equipotential surfaces are packed more tightly at the poles than at the equator. 
If the rotation is conservative (angular velocity constant on cylinders), these equipotential surfaces are also surfaces of constant pressure, density and temperature. This means that the radiative flux then scales with the local effective gravity: $F \propto g_{\rm eff}$. 
This is known as \textit{von Zeipel's theorem}. 
The observable consequence is gravity darkening: a rapid rotator has hot, bright poles and a cooler, darker equator. 
Real stars are thought to rotate closer to shellular ($\Omega \approx \Omega(r)$), which breaks the assumption, and the true dependence is somewhat shallower.    
A star with this flux imbalance cannot simultaneously satisfy hydrostatic and thermal equilibrium, and the resulting mismatch drives large-scale meridional (Eddington--Sweet) circulation \cite{1925Obs....48...73E,1950MNRAS.110..548S} that transports energy from where there is a surplus to where there is a deficit.
How these circulations actually develop under differential rotation is considerably more complicated than the 1970s picture and remains an active research area: they are suppressed by horizontal turbulence and by composition gradients \cite{2006A&A...451.1025R,2009A&A...495..271D,Rieutord2025}.
The same flows induce chemical mixing, transporting fresh hydrogen from the envelope into the hot burning core.
If the mixing is efficient enough, the star never develops the classical core-envelope structure. 
Rather than burning only its innermost material, it can burn through effectively its entire hydrogen supply and remains \textbf{(quasi-)chemically homogeneous (CHE)}.
Such stars do not evolve towards the cool red side of the HR diagram: they stay compact and simply become hotter and brighter, moving up and to the left \cite{Brott2011}.

Because CHE stars stay compact \textit{and} massive, they have been proposed as a particularly efficient route to gravitational-wave sources \cite{deMinkMandel2016,MandeldeMink2016,Marchant2016}, and as the most likely progenitors of long gamma-ray bursts \cite{2005A&A...443..643Y,2006ApJ...637..914W,2006A&A...460..199Y}.
Having spent their lives rotating rapidly, they plausibly leave rapidly rotating black holes behind; we return to this in Section~\ref{sec:day3che}.
Observationally, however, a CHE star is hard to tell apart from an ordinary WR star that has simply lost its envelope, which is why there are to date only \textit{candidates} for CHE stars and no confirmed examples \cite{2019A&A...623A...8K}.


{\bf Magnetic torques.} As a star evolves and develops a core--envelope structure, the contracting core spins up while the expanding envelope spins down, creating strong differential rotation. 
This shear winds up the poloidal field into a strong toroidal field; the toroidal field is in turn unstable to the Tayler instability, whose displacements regenerate a poloidal component, which the shear winds up again. 
Even from a weak seed field, this loop can sustain a strong toroidal field: the Tayler--Spruit dynamo \cite{Spruit2002}. 
The field lines act as lever arms coupling core to envelope. This can transport angular momentum outward: the core spins down as the envelope expands (and slows down). The eventual remnant is left rotating (more) slowly.

{\bf Comparing predicted to observed rotation rates.}
The spins of young neutron stars and white dwarfs let us work backwards: given the angular momentum a main-sequence progenitor's core starts with, how much must have been removed before collapse?
Observed remnant angular momenta fall far below what simple angular-momentum conservation would give \cite{Suijs2008}, so efficient core-to-envelope transport is required. 
Asteroseismology of red giants points the same way, measuring core rotation rates well below the predictions of purely hydrodynamic transport \cite{2012Natur.481...55B,2012A&A...548A..10M,2012ApJ...756...19D,2014A&A...564A..27D}.
This has motivated revised prescriptions \cite{Fuller2019} giving considerably stronger torques than the original, and correspondingly slower compact-object spins \cite{FullerMa2019}.

On the gravitational-wave side, individual component spins are poorly measured; what is best constrained is the effective spin $\chi_{\rm eff}$, the mass-weighted projection of the two component spins onto the orbital angular momentum (defined in Section~\ref{ss: day2gw}).
With hundreds of events we can nonetheless constrain the spin distribution at the population level, and although the uncertainties are large, the emerging picture is that merging binary black holes are mostly slowly spinning \cite{GWTC5_pop}.

All this seems to point towards black holes being born with slow spins.
However, black holes in X-ray binaries tell a different story; they are significantly more rapidly spinning than the merging binary black holes \cite{FishbachKalogera2022}.
We do not understand why. Are we comparing apples and oranges? For the low-mass X-ray binaries, accretion spin-up is at least arithmetically available, though it demands an average spin gain of $\simeq 0.47$ if these black holes were born with binary-black-hole-like spins, which is a significant amount of accreted angular momentum. High-mass systems are harder to explain still: they are young and wind-fed, leaving little opportunity to spin up after birth, so their near-maximal spins are likely to be natal. 
Another possibility is that the X-ray spin measurements are systematically biased, since they rely on assumptions about the inner accretion-disk structure; this remains actively debated \cite{FishbachKalogera2022,2024A&A...690A..21B,2024ApJ...962..101Z,2024ApJ...967L...9Z}.
In general, the origin of black hole spins remains an open question.


\subsubsection{Massive stars are not alone: binaries}

The third large uncertainty: massive stars do not live alone. The multiplicity fraction rises with mass; already $\sim$50\% of solar-mass stars have a companion, and essentially \textit{all massive stars are born in binaries}, or even higher-order multiples \cite{Sana2012,Offner2022}. 
Binarity changes stellar lives completely. The main interactions are i) mass transfer, altering their appearance, the orbital evolution, and the final masses of both stars, possibly causing stellar mergers, and ii) tides, which exchange spin and orbital angular momentum, pushing systems towards aligned, synchronised, circularised orbits. 
Mass transfer is the biggest and most impactful uncertainty, while tides are thought to be particularly important as a discriminator between isolated binary evolution and dynamical formation of compact binaries, where the objects need not be tidally synchronised. Since binary interactions are at the core of forming merging double compact objects, we defer the details to Section~\ref{sec:day3}.

\subsection{Core-collapse \label{ss:core collapse}}

\subsubsection{Advanced burning and the pre-supernova star}
Stars maintain hydrostatic equilibrium by balancing their luminosity against nuclear energy generation.\footnote{For much more in depth treatment of stellar structure and evolution than we can give here, see the lecture notes by Onno Pols \url{https://www.astro.ru.nl/~onnop/education/stev_utrecht_notes/}.} 
When a star exhausts hydrogen in its core, the core falls out of equilibrium, contracts and heats until it is hot enough to ignite helium; the same repeats for subsequent fuels. 
Only stars with $M_{\rm CO,core}\gtrsim 1.4\Msun$ will keep contracting through the full sequence
\be
\mathrm{H} \rightarrow \mathrm{He} \rightarrow \mathrm{C} \rightarrow \mathrm{Ne} \rightarrow \mathrm{O} \rightarrow \mathrm{Si} \rightarrow \mathrm{Fe},
\ee
each occurring at ever higher central temperature and at ever increasing speed.
Above $T \gtrsim 5\times10^{8}$\,K thermal neutrinos dominate the energy losses. 
At this point $L_\nu \gg L_{\rm photon}$, see e.g., the neutrino-equivalent HR diagram of \cite{Farag2020}. 
Hydrostatic and thermal equilibrium in the core are then set by neutrino emission, and the timescales on which the core burns, shrinks and expands are set by the neutrino luminosity. 
The nuclear burning times of these late stages become much shorter than the thermal timescale of the envelope. The outer layers of the star have no time to respond and the star looks unchanged from carbon burning onward in the HR diagram. 
On the inside the interior freezes into the final ``onion-skin'' structure of compositional layers separated by burning shells (e.g., onion figures in \cite{Laplace2021}).

\subsubsection{Core collapse}
Iron-56 sits at the peak of the binding-energy-per-nucleon curve: fusing it costs energy rather than releasing it. Once an iron core has formed, hydrostatic equilibrium can no longer be maintained by further fusion. The Fe core is supported mostly by relativistic electron degeneracy pressure, 
\be
P_e = K_{\rm ER}\, \rho^{4/3}.
\ee
Two processes then conspire to remove the remaining support at high density and temperature: electron captures on heavy nuclei (inverse $\beta$-decay)
\be
p + e^- \rightarrow n + \nu,
\ee
which make the core neutron rich (neutronization) and remove the electrons that supplied the $P_e$ pressure; and photodisintegration of heavy nuclei at $T \gtrsim 10^{10}$\,K,
\be
^{56}\mathrm{Fe} + \gamma \leftrightarrow 13\,^4\mathrm{He} + 4n,
\ee
which absorbs $\sim$2\,MeV per nucleon from the internal energy of the gas, decreasing the gas pressure. The result is the dynamical collapse of the iron core on almost the free-fall time $t_{\rm dyn} \approx (G\rho)^{-1/2} \sim 10$\,ms.

The collapse proceeds until the central density reaches $\sim 3\times10^{14}\,\mathrm{g\,cm^{-3}}$, about 1.5 times the density of an atomic nucleus. At this point neutrinos become degenerate and trapped, matter stiffens as the strong force becomes relevant, the pressure rises drastically, and the inner core bounces back like a spring: we have created a proto-neutron star. The still-infalling outer core now slams supersonically into it, launching an outward shock. 

For many years it was unclear whether this shock could unbind the envelope (and thus produce a supernova) at all. 
Photodisintegration of the infalling iron-group nuclei drains the energy, while electron captures behind the shock radiate the energy away as neutrinos that escape easily. 
Models predicted that the shock stalls within tens of milliseconds: naively, supernovae should never happen.

The solution is that the proto-neutron star is now so dense that it is partially opaque to neutrinos; and a fraction of those diffusing out are reabsorbed in the ``gain layer'' behind the stalled shock. 
This heating drives convective overturn, and the shock is further perturbed by what is known as the `standing accretion-shock instability' (SASI). 
The resulting asymmetries hold matter in the gain region for longer, letting the shock harvest more neutrino energy. These processes are intrinsically multi-dimensional, which is why it took 2D and 3D models to show how to revive the shock. 
These simulations have only become computationally feasible over the past decade or so (see \cite{Janka2012,Muller2016,BurrowsVartanyan2021}, and references therein for a more in depth review).

\subsection{Explosive endpoints (supernovae) \label{ss: supernovae}}

\subsubsection{The supernova zoo}
A successful explosion looks like a supernova. 
With peak photon luminosities reaching $\sim10^{42}\,\mathrm{erg\,s^{-1}}$ they are comparable to or brighter than their host galaxy ($\sim10^{41}$--$10^{44}$),  
Which is why they are among the oldest recorded astronomical observations: Han-dynasty astronomers in 185\,CE noted a new star that took months to fade. 
Modern transient surveys have transformed the field, and today, of the order $10^5$ events are known.\footnote{See also this great visualization of all known supernovae by Isaac Shivvers 
 \url{https://ishivvers.github.io/maps/sne.html}}

As samples grew, a classification emerged, based initially on the question: is there hydrogen in the spectrum? Yes (type II) or no (type I).
However, as the sample size increased, this classification branched into the complex decision tree still in use today \cite{Filippenko1997}. 
For our purposes, two things are worth highlighting. First, almost every supernova class is an exploding massive star; only Type Ia are white dwarfs. 
Second, hydrogen essentially tracks the envelope. Its presence means the envelope survived (Type II, exploding supergiants). Instead, Type Ib (stripped of H) and Type Ic (stripped of both H and He) supernovae imply the star was stripped of its envelope before explosion (WR-like progenitors), either by stellar winds or binary interaction. 
%

\subsubsection{To explode or not to explode: black hole formation and the lower mass gap}
Very crudely it is often assumed that successful explosion $\rightarrow$ neutron star; while  unsuccessful explosion (where the shock was never revived) $\rightarrow$ black hole. 
But deciding which stars explode or not is complicated. 
Explodability depends on the core and envelope structure at collapse, which depends on the star's entire prior evolution. 
Simulating that collapse in 3D is computationally expensive drastically limiting the number of progenitors that have been modelled in detail. 
Even within a single code, modest parameter variations such as the efficiency of neutrino-driven convection change which progenitors succeed \cite{Boccioli2023}.
The explodability behaviour is furthermore not monotonic: summaries of 3D models show complex correlations emerging between exploding and non-exploding progenitors e.g., \cite{Burrows2024}. 

What would black-hole formation look like? A failed supernova should look like a star that 
simply disappears. Kochanek et al.\ \cite{2008ApJ...684.1336K} 
proposed ``a survey about nothing'' to hunt for them, and finding nothing has itself proved publishable: 
the first compelling candidate for such a disappearing star was NGC\,6946-BH1 \cite{2015MNRAS.450.3289G,2017MNRAS.468.4968A,2021MNRAS.508.1156B}; 
the best-observed to date is M31-2014-DS1 \cite{2026Sci...391..689D}. 
Whether such stars are \textit{truly} gone or merely clouded by the dust they shed remains contested, with intermediate-luminosity transients and stellar mergers both offering alternative readings \cite{2026MNRAS.546ag052B}.

The `explode or not' picture is in any case likely too simple. 
A star can explode without ejecting everything: fallback accretion onto the proto-neutron star can still produce a black hole. 
Unfortunately, we cannot easily tell which happened from the supernova itself, because while the core sets the fate of the star, it is the envelope and circumstellar material that determine most of the appearance of the supernova.
Fallback matters enormously for the \textit{minimum black-hole mass}.
In the simplified picture, stars that explode leave neutron stars ($\lesssim 2\Msun$), while failed explosions retain their C, O and He layers on top of the iron core, adding several solar masses. In the latter case the lightest black holes would be $\sim$6--9$\Msun$. This implies a \textit{`lower mass gap'} between the heaviest neutron stars and the lightest black holes (see Section~\ref{sec:day4lowgap}).
However, fallback smooths this discontinuity back into a continuous remnant-mass function. 
The minimum black-hole mass is therefore either just above the maximum neutron-star mass, or at $\sim6\Msun$. Gravitational-wave detections of low-mass compact objects have begun to populate the gap \cite{GW230529}, so it is at most a dearth rather than a truly empty gap, but the theory of fallback and of the minimum black-hole mass remains an active area of research.


\subsubsection{Natal kicks}\label{day1:natal kicks}
Radio pulsars are observed moving through the Galaxy at $100$--$1000\,\mathrm{km\,s^{-1}}$, much faster than the stellar populations they were born from. The explanation for this is that neutron stars must be kicked at birth. The shape of the kick-velocity distribution remains actively debated \cite{DisbergMandel2025,2005MNRAS.360..974H,2017A&A...608A..57V,2020MNRAS.494.3663I,2021ApJ...920L..37W}.
So does the mechanism. Asymmetry is already built into the explosion, since the hydrodynamic instabilities that revive the shock produce aspherical flows. In the ``tug-boat'' picture, a dense clump of slow ejecta gravitationally drags the newborn compact object towards it, which predicts a stronger, faster shock in the direction opposite the kick \cite{Janka2013}. 
Other simulations find that momentum conservation in asymmetric ejecta suffices to produce sizeable kicks on its own \cite{2023ApJ...957...68B,2025ApJ...987..164B}.

Whether black holes are kicked at birth is closely tied to the theoretical questions around fallback. Observationally the question is much harder for black holes. Large kicks tend to unbind the binary, and an isolated black hole is nearly impossible to detect (Section~\ref{sec:day2}). The sample is thus small and selected against precisely the systems that would have received substantial kicks. 
A handful of systems have inferred 3D velocities, but translating a present-day velocity back to a natal kick is harder still, requiring the binary's full orbital and Galactic history.
Only two things seem clear: some black holes received a substantial kick ($\sim200\,\mathrm{km\,s^{-1}}$), and some received essentially none ($\lesssim 5\,\mathrm{km\,s^{-1}}$).

\subsubsection{Pair-instability supernovae and the upper mass gap}\label{sec:PISN}
At the other end of the mass spectrum, the cores of very massive stars reach temperatures ($T \gtrsim 7\times10^{8}$--$10^{9}$\,K) at which photons convert into electron--positron pairs, before oxygen ignites. 
This is a problem, because those photons were supplying the radiation pressure holding the core up: pair creation drains pressure support and drives the adiabatic index $\Gamma_1 = (\partial \ln P/\partial \ln \rho)_{\rm ad}$ towards the dynamical instability threshold $4/3$. 
The core contracts, oxygen ignites explosively, and the thermonuclear energy released can unbind the star entirely, leaving no remnant at all: a pair-instability supernova (PISN) \cite{FowlerHoyle1964,RakavyShaviv1967,Barkat1967,Fraley1968}. 
This sets an upper limit to the black-hole mass that stars can form.

Onset of the instability does not require full disruption. For CO cores above $\sim$38$\Msun$  the pulse ejects mass rather than unbinding the star. 
The star may undergo one or several such pulses before either disrupting or restabilising and collapsing to a black hole: \textit{pulsational} pair-instability supernovae (PPISN) \cite{Renzo2020a}. 
Full disruption sets in only above about 60$\Msun$ in CO core, though the exact threshold is debated and heavily dependent on the uncertain $^{12}\mathrm{C}(\alpha,\gamma)^{16}\mathrm{O}$ reaction rate \cite{2017ApJ...836..244W,Farmer2019,Farmer2020,Mehta2022,Farag2022}.

For the most massive stars (helium cores $\gtrsim 130\Msun$) a second instability takes over; the energy released by explosive burning is spent photodisintegrating nuclei rather than unbinding the star, cancelling the dynamical effect of the pair instability. The star collapses to a black hole after all.

The resulting remnant-mass landscape, as a function of core mass, has four regimes: I) core-collapse black holes at low masses; II) a PPISN-driven turnover, where increasing core mass gives a \textit{decreasing} remnant mass; III) an empty pair-instability gap; and IV) above it, black holes again with masses of roughly 130$\Msun$ and beyond. 
We return to the most recent observational constraints on this picture in Section~\ref{sec:highmassgap}. 


\subsubsection{Recap: The boundaries of stellar black hole formation}
\begin{figure}[!ht]
    \centering
    \includegraphics[width=0.55\linewidth]{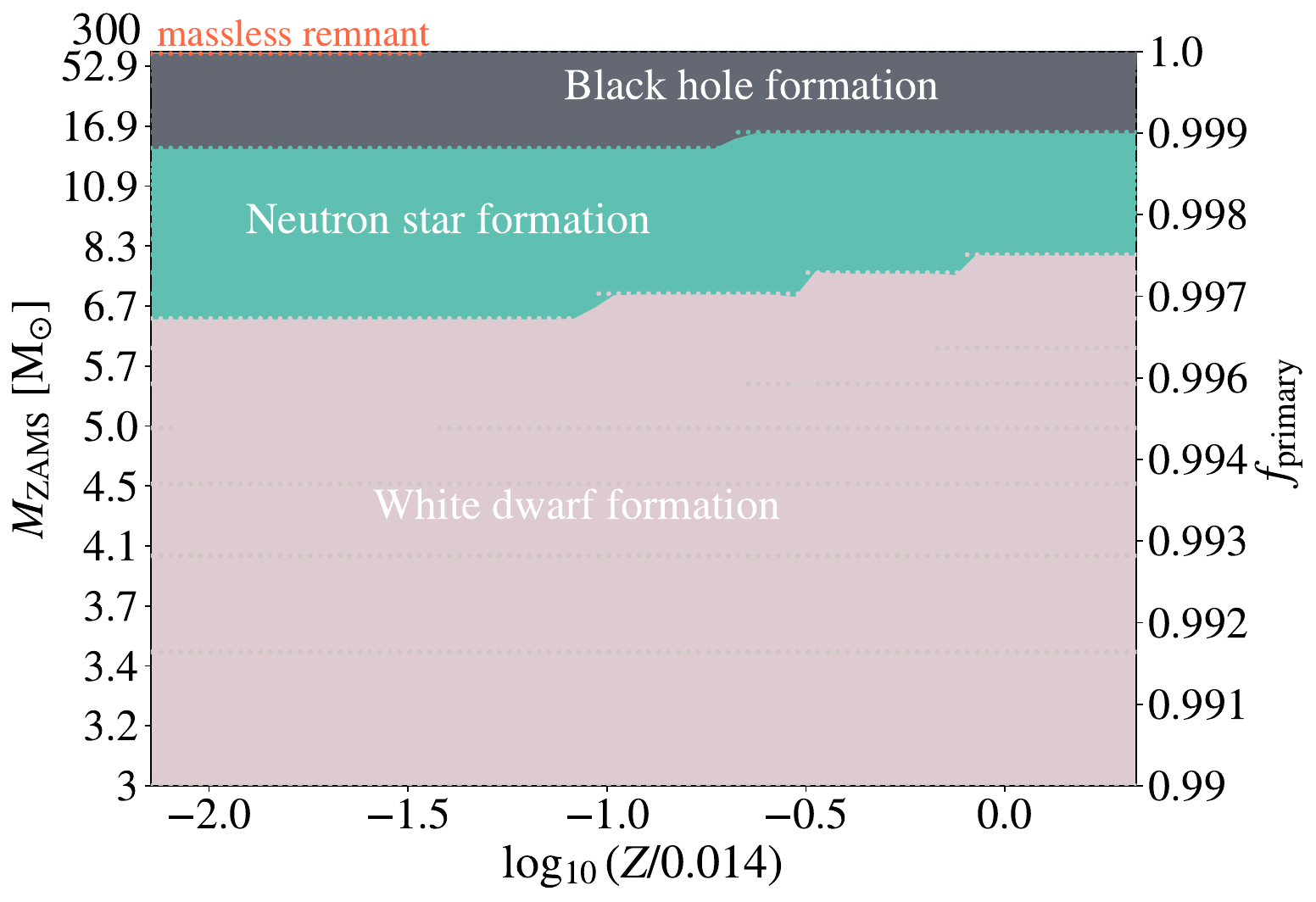}
    \caption{\footnotesize Approximate regions of WD, NS and BH formation from a star of a given ZAMS mass, at different metallicities. The $y$-axis is scaled to follow a Salpeter IMF so that the surface corresponds to the fraction of primary masses that contribute (note this is only the upper per cent). Figure from \cite{vanSon2025}.}
    \label{fig:remnant_type_per_Z}
\end{figure}

Figure~\ref{fig:remnant_type_per_Z} summarises where the boundaries between the different remnant types fall, and how they shift with metallicity.
Returning to the questions with which we opened this section: 
%
What is the least massive black hole you can form from stars? 
About 2--3$\Msun$: with fallback, black holes can form just above the maximum neutron-star mass. 
%
Are stellar black holes born with low spin? 
Nobody knows whether all are, but at least some must be: we observe merging BBHs to be slowly spinning while XRB black holes appear to spin quickly. 
%
What is the most massive black hole you can form from stars in our Galaxy (i.e., at roughly solar metallicity)? 
About $20\Msun$: set by stellar winds. 
And the most massive black hole from stars in the early Universe? 
Roughly 150--350$\Msun$, from above the PISN gap; the answer depends on how massive you believe stars can be born.

\section{Hunting for neutron stars and stellar black holes}\label{sec:day2}

Having formed neutron stars and black holes, how do we find them?
To answer this question, let us first ask ourselves what a neutron star looks like.

Neutron stars pack 1--2$\Msun$ into a radius of $R \approx 13$\,km (roughly the diameter of Manhattan), implying a mean density $\rho \approx 3\times 10^{14}$\,g\,cm$^{-3}$: one to two times the density of an atomic nucleus, with central densities several times higher still. 
We do not really know what happens in the deeper layers of a neutron star, beyond expecting it to be very neutron rich (for reviews, see \cite{LattimerPrakash2004, OzelFreire2016, Lattimer2021}).
Typical surface temperatures are around $T \sim 10^{6}$\,K (see e.g., \cite{Potekhin2020}). 
Neutron stars are thus far hotter than ordinary stars: typical HR diagrams are not even drawn out that far, stopping before $\log(T/{\rm K}) \approx 5.3$ at the hot end of the Wolf--Rayet stars.

Where would a neutron star be found on the HR diagram?
Given its radius and surface temperature, we can naively apply the Stefan--Boltzmann law $L = 4\pi R^2 \sigma T_{\rm eff}^4$ (ignoring that neutron stars are not really perfect black bodies) to get a value of the luminosity $\log(L/\Lsun) \approx -0.5$, comparable to that of a $0.8\Msun$ main-sequence star. 
However, the HR diagram is drawn for observations in the optical and infrared. 
At $T \sim 10^{6}$\,K nearly all of the thermal flux emerges in soft X-rays and the extreme UV. 
This makes it incredibly hard to detect neutron stars directly, raising the question: how did astronomers manage to detect them? The same question is even more pressing for stellar black holes. 

\subsection{(Radio) Pulsars}

\paragraph{The first detection of a neutron star}
In 1967, Jocelyn Bell, a PhD student at Cambridge, was studying distant radio sources with a detector built by her advisor Antony Hewish. 
While sifting the data for interesting phenomena she found a strange signal unlike anything seen before: pulses arriving precisely every 1.3373\,s.
The discovery of this first so-called radio pulsar was announced in 1968 \cite{Hewish1968}.\footnote{Since the pulses were so regular, one of the first explanations considered was a signal from an extraterrestrial intelligence (which is why the source was dubbed ``LGM-1'' for ``Little Green Men'').} 
Not long after, a pulsar was discovered at the centre of the Crab Nebula \cite{StaelinReifenstein1968}, whose nebula is linked to the supernova of 1054 recorded by Chinese astronomers at the time. 
The Crab pulsar's pulses have also been detected in the optical \cite{CockeDisneyTaylor1969} and in X-rays \cite{Fritz1969, Bradt1969}.
This confirmed the connection between pulsars and dying massive stars, which had already been predicted decades earlier: in 1934 Baade and Zwicky proposed that ``super-novae represent the transitions from ordinary stars into neutron stars'' \cite{BaadeZwicky1934}, and \cite{OppenheimerVolkoff1939} computed the first models of their structure. 

Pulsars (short for ``pulsating radio source'') are now known to be rapidly rotating, strongly magnetised neutron stars: the Crab spins on its axis roughly 30 times per second, and inferred surface fields are typically $\sim$$10^{12}$\,G, ranging from $\sim$$10^{8}$\,G for recycled millisecond pulsars up to $\sim$$10^{15}$\,G for magnetars.
Together, the rotation and the field channel a small fraction of the star's rotational energy into a narrow, coherent radio beam along the magnetic axis. 
Because that axis is generally misaligned with the spin axis, the beam sweeps around like a lighthouse and we see a pulse each time it crosses our line of sight.
These are the characteristic pulses to which they owe their name, and which are regular enough to enable pulsar timing (see Section~\ref{ss: day2gw}).
%

{\bf Pulsar evolution.}
Thousands of pulsars are now known. They are born rapidly rotating and strongly magnetised. 
As the pulsar ages, it loses energy in the form of a relativistic particle wind as well as some electromagnetic radiation (the radio pulses make up only a tiny fraction of the latter).
The energy they radiate is extracted from their rotational energy; the star thus spins down.

Two properties of a pulsar can be measured with exquisite precision: the spin period $P$, and its rate of change $\dot P$. 
Most other parameters we would like to know follow from these two. 
If the spin-down is driven by magnetic dipole braking then $\dot P \propto B^{2}/P$, which inverts to give the surface field
\begin{equation}
    B \simeq 3.2\times10^{19}\,\sqrt{P\dot P}~{\rm G} \qquad (P \textrm{ in s}),
\end{equation}
together with a characteristic age $\tau = P/2\dot P$ and a spin-down luminosity $\dot E = 4\pi^{2} I \dot P/P^{3}$. 
The radio emission depends on $B$ and $P$ through the polar-cap potential drop $\Delta\phi \propto B/P^{2}$. 

This is why pulsars are studied in the $P$--$\dot P$ diagram. 
Young pulsars start at the top left of the diagram, where they are born at short $P$ and high $\dot P$.
As the star spins down (longer $P$, moving to the right), $\dot P$ will also decrease.
This is because $\dot P \propto B^{2}/P$: at fixed field, a longer period also decreases the rate of change ($\dot P$). 
It therefore lingers longer at long periods, which is why most pulsars are observed in the ``traffic jam'' towards the centre-right of the diagram. 
Eventually it crosses the ``death line'', below which $\Delta\phi$ is too small to sustain the pair production that feeds the radio emission, and the pulsar switches off. 
A typical radio lifetime is $\sim10^{7}$\,yr.

This explains classical radio pulsars. 
However, there is another class of observed pulsars with very rapid rotation yet slowly changing periods (low $\dot P$ at short $P$, hence low $B$).
If pulsars only ever evolve towards longer periods, how can these exist? 
The solution is binaries: in the binary situation, a dying pulsar can accrete matter from its companion. 
The accreted material carries angular momentum, spinning the star back up to millisecond periods, while at the same time reducing its magnetic field.
No new physics is needed to switch the radio back on: since $\Delta\phi \propto B/P^{2}$, the fall in $P$ more than compensates for the fall in $B$, and the star is lifted back above the death line.
These ``recycled'' pulsars are what we observe as \textit{millisecond pulsars} \cite{BhattacharyaVanDenHeuvel1991}.


\subsection{Accretion and X-ray binaries \label{ss: XRB}}
The second way to find neutron stars and black holes is through the light they emit when they accrete matter from a companion star.
This is how the first stellar-mass black hole was found: Cygnus X-1.
It was picked up as a bright X-ray source in 1964 \cite{Bowyer1965}, but not recognised as a black hole until 1972\cite{WebsterMurdin1972, Bolton1972}.\footnote{Stephen Hawking famously bet Kip Thorne \textit{against} Cyg X-1 being a black hole, and conceded the bet in 1990.}
Accretion remained the only way to find stellar-mass black holes for the next fifty years; 
\textit{all} of our knowledge of the stellar-BH mass distribution from before gravitational waves came from X-ray binaries. 
Only in the last decade or so have we added black hole detections through the other methods discussed in this chapter (radial velocities, astrometry, microlensing, and gravitational waves).

Matter falling onto a compact object is an extraordinarily efficient energy source: an accretion disk converts of order 10\% of the rest-mass energy of infalling material into radiation (compare to hydrogen fusion, which liberates less than 1\%). The reason is the depth of the potential well for these compact objects:
\begin{equation}
    E_{\rm acc} \approx \frac{G M m}{2 R} \approx 0.1\, m c^2 .
\end{equation}
With $M$ the mass of the compact object, $m$ the mass of the accreted material, and $R$ the radius at which the accretion flow stops and releases its energy (this is either the neutron star surface or roughly the innermost stable circular orbit $R_{\rm ISCO}$ for a black hole). 
Filling in the numbers for an electron falling onto a neutron star ($M = 1.4\Msun$, $R = 10$\,km, electron rest mass $\sim$500\,keV), we get $\sim$50\,keV per electron, indeed $10\%$ of the electron rest-mass energy. 
Since the X-ray band spans energies of $\sim$0.5--100\,keV, we can also immediately see that accretion onto compact objects is X-ray bright. 
The compact object itself is invisible; it is the shock-heated inner disk (and bremsstrahlung from electron--proton collisions, and the heated NS surface) that shines.
Systems in which a NS or BH accretes gas from an ``alive'' companion star are called X-ray binaries (XRBs).
We distinguish two types based on the \textit{companion} mass (the labels thus, confusingly, say nothing about whether the accretor is a NS or BH). 

\paragraph{High-mass X-ray binaries (HMXB):} donor masses $\gtrsim 8\Msun$ (OB star + NS/BH).
These are mostly powered by wind accretion, so no Roche-lobe overflow is required.
They come in two sub-groups: \textit{wind-fed HMXBs} ($P \lesssim 10$\,d, $e \lesssim 0.1$, O star often close to RLOF, causing persistent X-rays), and \textit{Be/X-ray binaries}, in which the compact object accretes from the decretion disk of a Be star on an eccentric orbit ($P \sim 15$--300\,d, $e \lesssim 0.5$, far from RLOF, transient X-rays).
HMXBs are a natural stage of massive binary evolution, expected right before the moment of reverse mass transfer (i.e., when the initially less massive star evolves to fill its Roche lobe and dumps mass onto the compact object).
As such, they also represent a progenitor stage of GW sources.
If both the compact object (via radial velocity of the X-ray pulses) and the donating star are seen, the system is effectively a double-lined binary, and both masses can be measured. 
A single-lined system (i.e., only the accreting object is visible) yields only a lower limit via the mass function. 
Spins come for free if the accretor is a pulsar, since the (orbit-corrected) pulse period \textit{is} the spin period.
For black holes the spin has to be inferred by fitting a model to the disk emission.

\paragraph{Low-mass X-ray binaries (LMXB):} donor's masses $\lesssim 1$--$2\Msun$ (G/K star + NS/BH).
LMXBs are powered by long-term Roche-lobe overflow, which builds genuine accretion discs.
The optical light is usually dominated by X-rays reprocessed in the disc, drowning out the faint donor.
Many LMXBs are transient sources, fluctuating between a quiescent and an outburst state. In quiescence the donor may become visible, allowing an RV measurement and providing a lower limit on the mass of the compact object.
NS LMXBs show Type I X-ray bursts: thermonuclear flashes of accreted H/He on the NS surface (BHs have no surface to burst on).
LMXBs are the ancestors of millisecond pulsars, but their formation is harder to explain than for HMXBs.
This is because long-term RLOF requires dynamically stable mass transfer, while the extreme mass ratios involved are expected to lead to unstable mass transfer instead. 
Nonetheless, several standard formation scenarios exist for LMXBs: (i) (inefficient) common-envelope evolution, since a low-mass star is expected to lack the orbital energy required to unbind a massive envelope, (ii) accretion-induced collapse of a massive white dwarf, and (iii) a triple-star origin (e.g., V404 Cygni) \cite{1998ApJ...493..351K,2003MNRAS.341..385P,2025ApJ...983..115S}.

\subsection{Indirect detections through a companion star \label{ss: indirect detect}}

Beyond accretion, a black hole or neutron star may betray itself through the \textit{motion} of a luminous companion.
There are several observational techniques to find such `dormant' compact objects. 

\paragraph{Radial velocities.} By observing the spectrum of the luminous companion, we can discover the periodic blue- and redshifting of its spectral lines caused by the orbital motion of an unseen companion. From this, we can infer the orbit. 
The radial-velocity amplitude combined with Kepler's laws yields the binary mass function, and hence a lower limit on the mass of the unseen companion.  
Several tens of NS systems are known in this way, but only a handful of BH candidates, some of which remain debated, e.g.\ VFTS 243 \cite{Shenar2022}, a system in the globular cluster  NGC~3201 \cite{Giesers2019}, and HD~130298 \cite{Mahy2022}.

\paragraph{Astrometry.} By tracking the position of a star on the sky with very high precision, the presence of an unseen companion can be inferred from the periodic `wobble' of its photocentre. 
Accounting for the motion on the sky due to parallax and the proper motion of the source, any residual periodic motion betrays the presence of a companion. This technique has been transformed by \textit{Gaia}.

To date, \textit{Gaia} astrometry has revealed a sample of about 20 neutron stars \cite{ElBadry2024NS}, and three dormant stellar-mass black holes \cite{ElBadry2023BH1,ElBadry2023BH2,Panuzzo2024}.
Gaia BH1 and BH2 both have a black hole of $\sim 9\Msun$ in a wide orbit ($\sim200$ d and $\sim 1300$ d respectively) with a solar-type star, while Gaia BH3 has a $\sim 33\Msun$ black hole in an $\sim 11$-yr orbit with an old, metal-poor giant companion.
Gaia~BH3 is the most massive stellar black hole known in the Milky Way. Its low-metallicity companion fits well with the expectation that the most massive black holes form in metal-poor environments (recall the winds discussed in Section~\ref{ss:massive_stars}).

The orbits and mass ratios of the \textit{Gaia} black holes pose an evolutionary puzzle: they are too unequal in mass for stable mass transfer, yet too wide to be the product of a common envelope. 
Suggested solutions include modified isolated-binary evolution
\cite{2024A&A...692A.141K,2024MNRAS.535L..44G,2024MNRAS.535.3577K,2024A&A...690A.144I,2025arXiv251110728O},
hierarchical triple evolution \cite{2024PASP..136a4202N,2024ApJ...964...83G,2024ApJ...975L...8L,2025OJAp....8E..79T},
and dynamical assembly in star clusters \cite{2023MNRAS.526..740R,2024MNRAS.527.4031T,2024ApJ...965...22D,2024A&A...688L...2M}, as invoked for oMEGACat~BH-2, or, for Gaia~BH3, formation in a now-disrupted star cluster, e.g.\ \cite{Balbinot2024}. 
To date, their origin remains an open question, making it especially valuable to study their properties in the context of other post-mass-transfer systems \cite{vanson2026_post_mt}.

Since it is not clear how exactly the \textit{Gaia} black holes and neutron stars form, it is equally unclear how many more such systems we should expect to find.
Predictions for the number of black holes detectable by \textit{Gaia} have in fact grown steadily more pessimistic over time, from early estimates of $\sim10^3--10^{5}$ \cite{MashianLoeb2017,Breivik+2017ApJ...850L..13B,Yalinewich+2018MNRAS.481..930Y,Yamaguchi+2018ApJ...861...21Y} down to hundreds \cite{ShaoLi2019ApJ...885..151S,Shikauchi+2020PASJ...72...45S,Chawla+2022ApJ...931..107C,Andrews+2022arXiv220700680A}, and currently tens to a hundred for DR4 (scheduled for release in December 2026) \cite{Nagarajan+2025PASP..137d4202N,Chawla+2026ApJ..1007..170C}.

Although \textit{Gaia} has clearly revolutionised the field of astrometry, it is not the only source of astrometric data. 
Very recently, oMEGACat~BH-2 was identified in $\omega$~Cen from the extensive HST and JWST astrometric dataset \cite{Whitaker2026}.
This marked the first \textit{astrometric} discovery of a stellar-mass black hole in a star cluster (compared to NGC~3201 above, which was found spectroscopically). 
At $\sim4.5\Msun$ it falls in the putative `lower mass gap' (see Section~\ref{sec:day4lowgap}), while it moreover has the longest orbital period of any known black hole + luminous companion system. Lastly, it is comparatively metal rich, and has been argued to favour a dynamical origin.

\paragraph{Ellipsoidal variations.} A star that is tidally deformed by a close companion has a projected surface area that varies over the course of an orbit. This results in a coherent photometric modulation with two maxima and two minima per orbital period. 
This effect is only measurable if the star fills a \textit{substantial} fraction of its Roche lobe, so such systems are necessarily close, interaction-shaped binaries.
Candidates are typically identified photometrically and then confirmed spectroscopically. This has yielded large samples of genuine close binaries, most of them red giants with low-mass, often white dwarf, companions, as well as bloated stripped stars caught after a phase of mass transfer (e.g., \cite{El-Badry2022_unicorns}).
While there have been active searches \cite{Green2025}, no black hole or neutron star has been unambiguously discovered through this channel to date.

\subsection{Gravitational microlensing}
When a mass passes almost exactly in front of a more distant background star, it acts as a gravitational lens and temporarily magnifies the source. 
The observable is the Einstein crossing time $t_{\rm E}$, that is, the duration of the event:
\begin{equation}
    t_{\rm E} = \frac{\theta_{\rm E}}{\mu_{\rm rel}}.
    \label{eq:einstein_crossing_time}
\end{equation}
Here, $\mu_{\rm rel}$ is the relative proper motion of lens and source, and $\theta_{\rm E}$ is the Einstein radius:
\begin{equation}
    \theta_{\rm E} = \sqrt{\frac{4GM_{\rm L}}{c^{2}}\,\frac{D_{\rm S}-D_{\rm L}}{D_{\rm L}D_{\rm S}}},
    \label{eq:einstein_radius}
\end{equation}
where $M_{\rm L}$ is the lens mass, $D_{\rm L}$ and $D_{\rm S}$ are the distances from the observer to the lens and to the source. 
For a lens of a few solar masses at a few kpc, $\theta_{\rm E}$ is of order a milliarcsecond, far too small to resolve, so the event is observed as a smooth, achromatic brightening of the source lasting $t_{\rm E}$.
Since $t_{\rm E}\propto\sqrt{M_{\rm L}}$, black hole lenses produce events lasting hundreds of days, compared to weeks for typical stellar lenses. Long-duration events are therefore the more natural place to search, and the large surveys monitoring the Galactic bulge (OGLE, MOA and KMTNet) provide the required cadence and source density.
There is a degeneracy between the lens mass, $\mu_{\rm rel}$ and the distances; a slow, nearby, low-mass lens can appear like a fast, distant, massive one. Breaking this degeneracy requires a second observable: the microlens parallax $\pi_{\rm E}$ (measurable when the orbital motion of the Earth modulates the light curve), or the astrometric shift of the source centroid, which measures $\theta_{\rm E}$.

To date, only a single isolated stellar-mass black hole has been found through microlensing: OGLE-2011-BLG-0462 / MOA-2011-BLG-191, observed in 2011 and confirmed as a black hole in 2022. 
Its mass determination has a rocky history. HST astrometry initially gave $7.1\pm1.3\Msun$ \cite{Sahu2022}, while an independent analysis of the same event found $1.6$ to $4.4\Msun$ \cite{Lam2022}, which would have placed the lens in the lower mass gap. 
A re-analysis of the OGLE photometry identified systematic errors in the earlier reductions and recovered $7.88\pm0.82\Msun$ \cite{Mroz2022}, and with additional HST epochs and re-reduced photometry the low value was revised upward to $\sim6.0^{+1.2}_{-1.0}\Msun$, consistent with a black hole \cite{LamLu2023}. 

Yet, the Galaxy is expected to contain a large population of isolated black holes.
Population synthesis predicts that the Galaxy hosts $\sim 1.2\times10^{8}$ isolated black holes, against only $\sim 9.3\times10^{6}$ residing in binaries \cite{Olejak2020_synthetic_catalog}.
In other words, single black holes outnumber those in binaries by more than ten to one!
For these single black holes, gravitational microlensing is essentially the only available detection method.

\subsection{Gravitational waves \label{ss: day2gw}}

\paragraph{The Hulse--Taylor binary.} 
The first evidence for gravitational waves came from radio observations: In 1974 Russell Hulse and his PhD supervisor Joseph Taylor discovered PSR~B1913+16: a pulsar in a $7.75$\,hr, highly eccentric ($e \simeq 0.62$) orbit around another neutron star \cite{HulseTaylor1975}. 
The pulsar acts as a clock carried around the orbit, so timing its pulses over years yields the orbital parameters, and their evolution, with extraordinary precision. 
The orbital period was found to be shrinking by $\sim$76\,$\mu$s per year, in perfect agreement with the decay predicted by the quadrupole formula for gravitational-wave emission \cite{TaylorWeisberg1982}.
This was the first strong-field test of general relativity and the first demonstration that gravitational radiation carries energy away from a real astrophysical system, recognised with the 1993 Nobel Prize. 
It also established that double neutron stars exist and merge within a Hubble time: B1913+16 will merge in roughly 300\,Myr. 
Pulsar timing has since been promoted from indirect evidence to a gravitational-wave detector in its own right.

\paragraph{The first direct detection of gravitational waves.} 
About a decade before the writing of this text, on 14 September 2015, the first direct detection of gravitational waves was made by the LIGO Scientific and Virgo Collaborations: GW150914, the merger of two black holes of roughly $36$ and $29\Msun$ \cite{GW150914}. 
This monumental discovery opened a new window on the Universe, and has since revolutionised the study of stellar-mass compact objects. 
In this short decade we have gone from a few tens of known stellar-mass black holes (the \textit{total} yield of all the aforementioned methods)\footnote{This fun plot shows how gravitational observations have dwarfed the number of known stellar-mass black holes from other methods. \url{https://media.ligo.northwestern.edu/gallery/mass-plot}}, to several hundred: at the time of writing the counter stands at 267 confident reported detections of merging black holes and neutron stars \cite{GWTC5,GWTC5_pop}, corresponding to more than 500 individual compact objects, the vast majority of them black holes.

\paragraph{What a gravitational wave measures.} 
The total mass of a merging binary sets the frequency at which its signal ends. 
For merging black holes, we can use the innermost stable circular orbit of a Schwarzschild black hole as a rough guide,
\begin{equation}
    f_{\rm ISCO} \simeq 4.4\,\mathrm{kHz}\; \frac{\Msun}{M_{\rm tot}},
\label{eq:fisco}
\end{equation}
where $M_{\rm tot} = m_1 + m_2$ is the total mass and $f_{\rm ISCO}$ the \textit{gravitational-wave} frequency at that orbit, twice the orbital frequency. 
A detector's frequency band therefore determines which masses it can observe. 
For stellar-mass compact objects this means $\sim$10\,Hz to a few kHz, which is reached with kilometre-scale laser interferometers: LIGO Hanford, LIGO Livingston, Virgo and KAGRA (jointly the LVK network). 
They measure the strain $h = \Delta L / L$, the fractional change in arm length induced by the passing wave, of order $10^{-21}$ for the loudest sources. 
Signals are extracted by matched filtering: banks of template waveforms, computed for given masses and spins, are cross-correlated with the data, and the source properties follow from those that fit best. 
The masses are imprinted directly on the phase evolution, and the leading-order combination, the chirp mass,
\begin{equation}
    \mathcal{M}_{\rm chirp} = \frac{(m_1 m_2)^{3/5}}{(m_1 + m_2)^{1/5}},
\end{equation}
where $m_1$ and $m_2$ are the component masses, is typically recovered to sub-percent precision. 

Beyond the masses, the picture becomes blurrier \cite{ChristensenMeyer2022}. 
The spin information is compressed into the effective spin,
\begin{equation}
    \chi_{\rm eff} = \frac{ \left( \frac{\vec S_1}{m_1} + \frac{\vec S_2}{m_2} \right) \cdot \hat{L} }{m_1 + m_2},
\end{equation}
where $m_i$ is the mass, and $\vec S_i$ the spin angular momentum of component $i$. $\hat{L}$ is the unit vector along the orbital angular momentum.
$\chi_{\rm eff}$ is degenerate with the mass ratio (a large aligned spin resembles a more equal-mass binary).
This degeneracy is hard to break without the detection of higher-order multipoles, which requires asymmetric component masses and a favourable orientation (see e.g., \cite{GW190412, GW241011_GW241110}).
Individual spin magnitudes and tilts imprint only weakly on the waveform, and are therefore more difficult to constrain \cite{Purrer2016}.
Eccentricity is effectively inaccessible for field binaries, which circularise during the inspiral and are expected to be circular by the time they enter the relevant frequency band. 
The luminosity distance is degenerate with the inclination, since both scale the amplitude. 
The main dynamical effect of the spins is on the duration of the inspiral: components anti-aligned with the orbital angular momentum plunge sooner, while aligned spins hang the orbit up and add cycles in band.

\begin{table}[!ht]
\centering
\small
\setlength{\tabcolsep}{5pt}
\renewcommand{\arraystretch}{1.15}
%
\definecolor{cardbg}{HTML}{EAF3F7}
\newcommand{\rowbg}{\rowcolor{cardbg}[\tabcolsep][\tabcolsep]}
\newcommand{\rowbgflat}{\rowcolor{cardbg}[0pt][0pt]}
\newcommand{\phcard}[2][7mm]{\raisebox{-0.35\height}{\includegraphics[height=#1]{Plots/bhns_cards/#2}}}
\newcommand{\detects}[3]{%
  \ifx\rowbg#1\rowbgflat\fi
  \multicolumn{3}{@{}l@{}}{\rule[-1.1ex]{0pt}{4.2ex}%
    \hbox to \dimexpr2.5cm+6cm+5.5cm+6\tabcolsep\relax
    {\leaders\hbox{\rule{2.5pt}{0.5pt}\hskip3pt}\hfil}}\tabularnewline
  #1\multicolumn{3}{l}{\rule{0pt}{5.5ex}\textit{Astrophysical phenomena:}\hspace{2.5mm}#2\hspace{3mm}\textbf{#3}\rule[-5ex]{0pt}{0pt}}\tabularnewline
  \hline}
\begin{tabular}{p{2.5cm}p{6cm}p{5,5cm}}
\textbf{Technique} & \textbf{Can constrain} & \textbf{Selection effects} \tabularnewline
\hline
\raggedright (Radio) pulsars & \raggedright Spin period and spin-down, magnetic field, characteristic age; both masses to $\sim$0.1\% in relativistic binaries & \raggedright Only neutron stars whose beam is directed towards us \tabularnewline
\detects{}{\phcard[10mm]{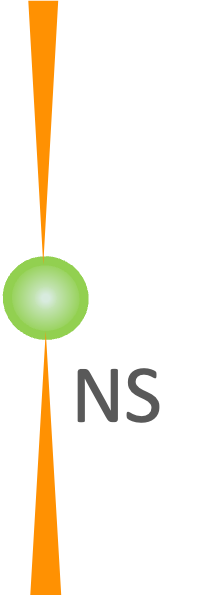}\hspace{2mm}\phcard{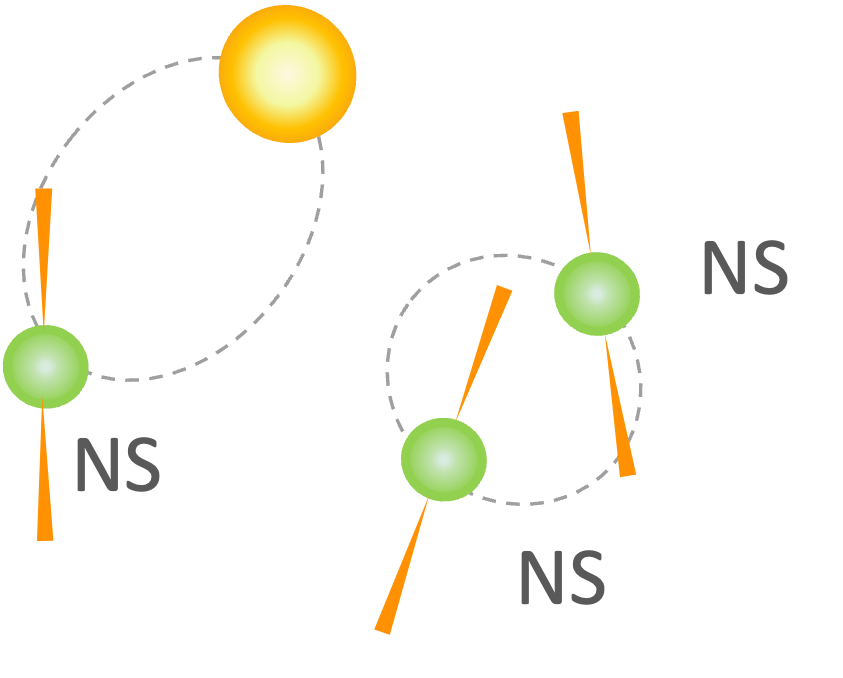}}{Single NS, and pulsars in binaries}
\rowbg \raggedright X-ray binaries & \raggedright Accretion rate, spin from the disc, and a dynamical mass given an inclination & \raggedright Only while accreting, so transients are missed between outbursts; inclination stays degenerate \tabularnewline
\detects{\rowbg}{\phcard{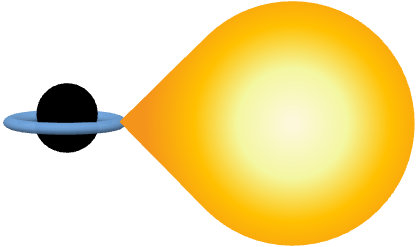}\hspace{2mm}\phcard{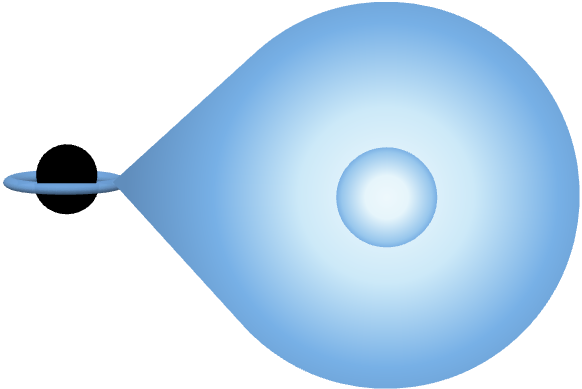}}{Accreting BH or NS with a live donor}
\raggedright Radial velocities & \raggedright Orbit and mass function, a lower limit on the unseen mass & \raggedright Faint or rapidly rotating companions drop out, no face-on orbits \tabularnewline
\raggedright Astrometry & \raggedright Full orbit, a mass and distance measurement & \raggedright Only nearby systems (within a few kpc), and only for periods shorter than the mission time \tabularnewline
\raggedright Ellipsoidal variations & \raggedright Orbit and Roche-filling factor; a candidate mass function & \raggedright Only near-Roche-filling systems, and those are swamped by contact binaries \tabularnewline
\detects{}{\phcard[10mm]{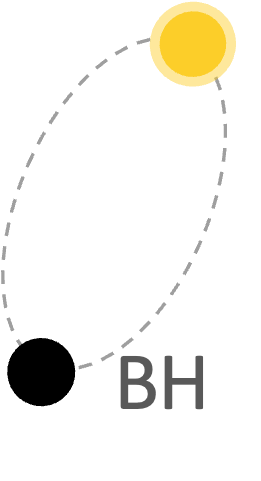}}{Dormant BH or NS with a luminous companion}
\rowbg \raggedright Microlensing & \raggedright Lens mass, distance and proper motion & \raggedright Chance alignment; needs a crowded background field \tabularnewline
\detects{\rowbg}{\phcard{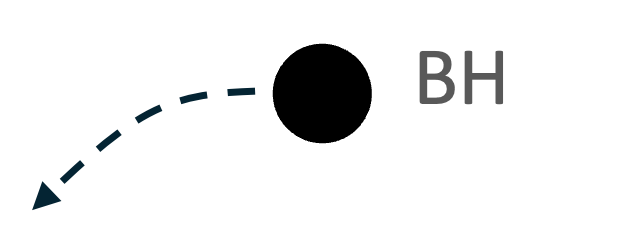}}{Single BH}
\raggedright Gravitational waves & \raggedright $\mathcal{M}_{\rm chirp}$, $\chi_{\rm eff}$, luminosity distance & \raggedright Only binaries that merge within a Hubble time \tabularnewline
\detects{}{\phcard{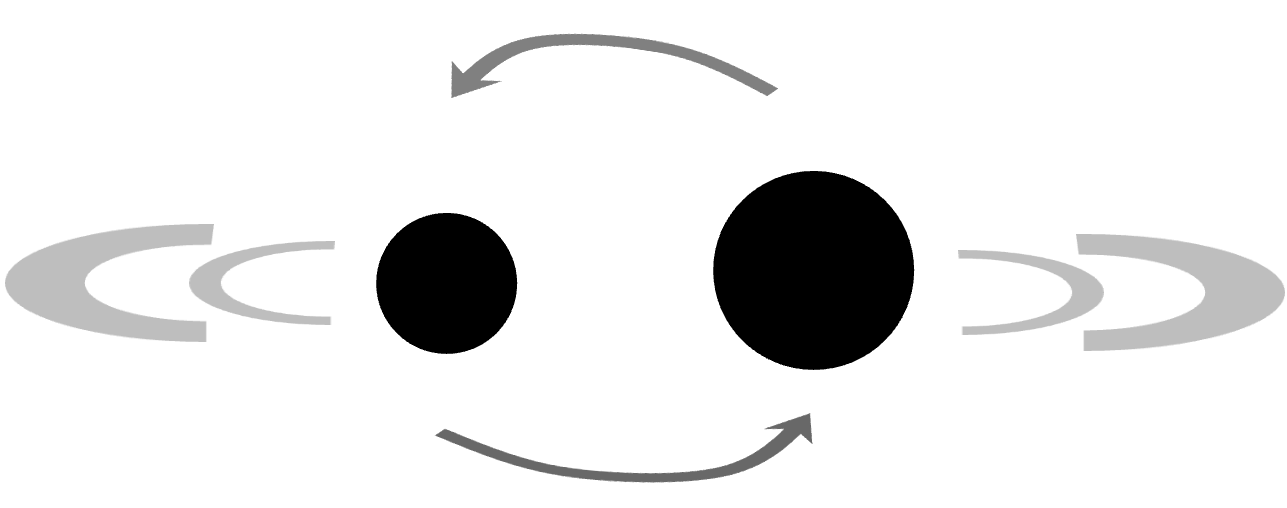}\hspace{2mm}\phcard{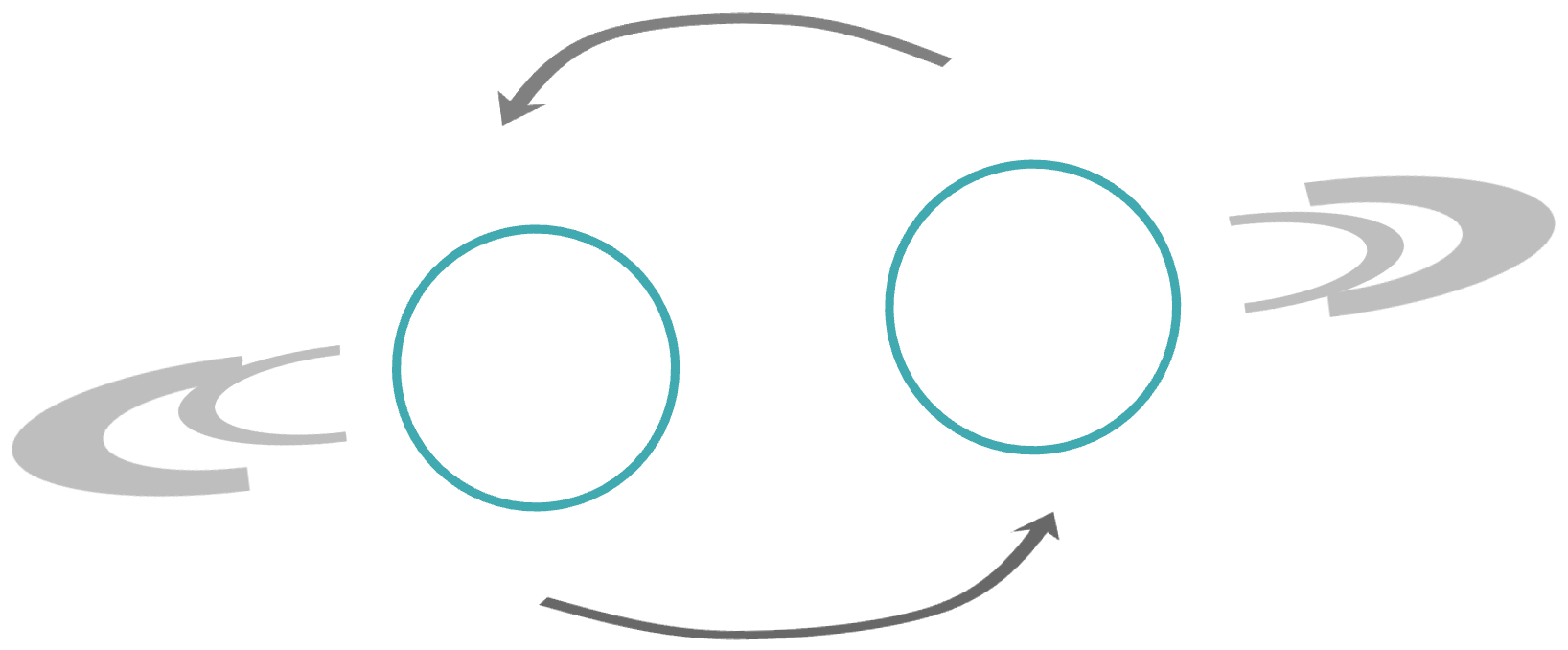}\hspace{2mm}\phcard{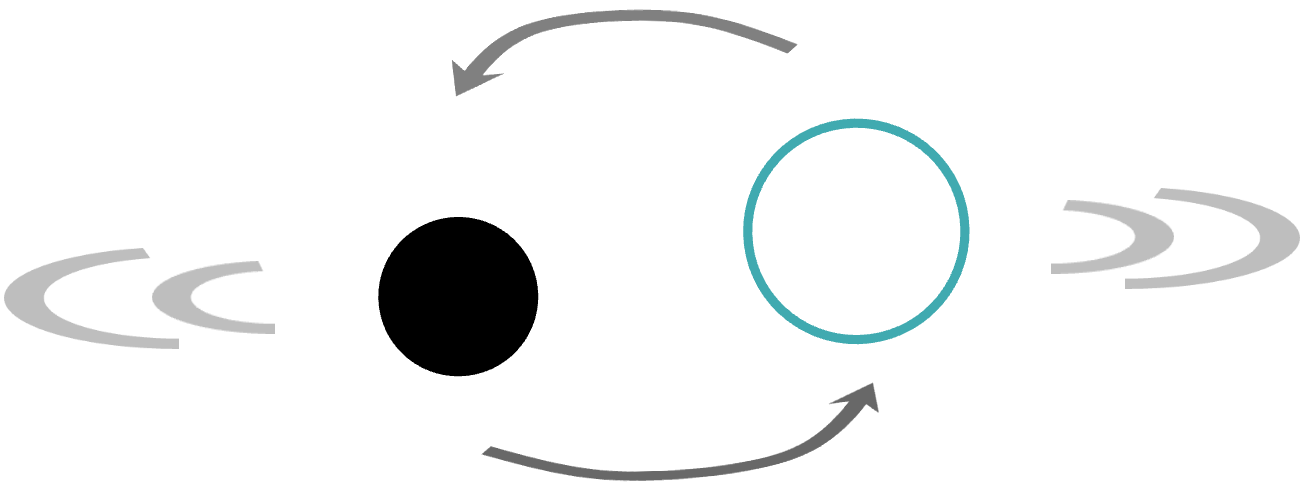}}{Merging compact binary}
\rowbg \raggedright Pulsar timing arrays & \raggedright Amplitude of the nHz stochastic background & \raggedright No individual sources resolved yet; needs decades of timing \tabularnewline
\detects{\rowbg}{\phcard{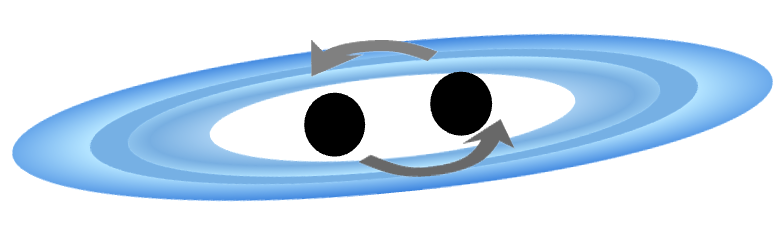}}{Supermassive BH binary}
\end{tabular}
\caption{The different techniques to observe black holes and neutron stars.}
\label{tab:catching}
\end{table}

\paragraph{The gravitational-wave spectrum.} 
Just as for the electromagnetic spectrum, different frequencies reveal different astrophysical phenomena. 
For a circular binary the dominant harmonic is emitted at $f_{\rm GW} = 2 f_{\rm orb}$, hence at lower frequencies we observe wider orbits, while the terminal frequency scales inversely with total mass (Eq.~\ref{eq:fisco}). 
A detector's band thus fixes both the size of the orbit and the mass of the system it can observe.
We distinguish three main frequency bands, each with its own detector technology and astrophysical targets.

\begin{itemize}
    \item \textit{Hz to kHz, kilometre-scale interferometers.} 
    These detectors are most sensitive to the merging black hole and neutron star binaries discussed above.
    Because what we measure with interferometers is a strain \textit{amplitude}, which falls off as $1/d_{\rm L}$ (rather than a flux, which falls off as $1/d_{\rm L}^{2}$), increasing our sensitivity by a certain factor buys the same factor of increase in distance, (or its cube in accessible volume). 
    Therefore building larger and more sensitive detectors pays off much more steeply than it does in the electromagnetic case.
    The next generation of GW detectors, such as the Einstein Telescope and Cosmic Explorer, with arms of $10$ to $40$\,km, will
    cover a similar mass range but reach $z\sim20$ for black hole binaries and $z\simeq2$--$3$ for neutron star binaries, against $z\simeq1$ and $z\simeq0.2$ today \cite{ETBlueBook,CosmicExplorer2021}. Stellar-mass mergers will thus be observed out to the epoch of the very first stars and perhaps even primordial black holes!
    \item \textit{mHz, LISA} \cite{LISA2024}. 
    With arms of $2.5$ million km trailing the Earth, LISA will reach orbits that are far wider or central masses that are much larger. Specific astrophysical targets include (i) Galactic double white dwarfs, (ii) the early inspiral of stellar-mass NS and BH binaries, years to decades before they enter the LVK band, (iii) extreme mass-ratio inspirals, since the ISCO of a $\sim$$10^{6}\Msun$ black hole already lies at mHz and the inspiralling body completes an enormous number of cycles in band, and (iv) mergers of $10^{5-6}\Msun$ massive black holes.
    \item \textit{nHz, pulsar timing arrays.} 
    In this case, the detector is galaxy-sized, with ``arms'' of tens of thousands of light years long. 
    The astrophysical targets are $\sim$$10^{9}\Msun$ supermassive black hole binaries with orbital periods of years. 
    Because we know precisely when each pulse should arrive, a passing wave reveals itself as \textit{correlated} deviations in arrival times across
    pulsar pairs: those in similar directions arrive early together, while pairs separated by $\sim$90$^\circ$ anti-correlate. 
    This dependence on angular separation is the Hellings--Downs curve \cite{HellingsDowns1983}.
    The first tentative evidence for such a background (plausibly the hum of unresolved supermassive black hole mergers), was reported in 2023 \cite{NANOGrav2023}
\end{itemize}

\subsubsection{Summary: how to catch black holes and neutron stars}
No single method sees the whole population of stellar-mass black holes and neutron stars. 
Each is sensitive to a particular corner of parameter space, and each suffers from its own selection effects: radio pulsars require a beam pointed at us; X-ray binaries require ongoing accretion; the dormant-binary methods require a luminous companion whose light we can track; microlensing requires a chance alignment; and gravitational waves require the binary to merge within the age of the Universe. 
Table~\ref{tab:catching} provides an overview of the different techniques discussed in this section. 
This material can also be explored through the ``Hunting for black holes \& neutron stars'' card game: \url{https://liekevanson.github.io/bhns_cards.html}, that was used during the actual course.

\newpage
\section{Gravitational-wave formation channels}\label{sec:day3}

As discussed in section ~\ref{ss: day2gw}, the number of known stellar-mass black holes has grown rapidly over the last decade, dwarfing the number of previously known stellar-mass black holes from all electromagnetic methods combined.
And this is only the beginning, as next-generation detectors promise $10^{5}$--$10^{6}$ detections per year, cf. \cite{BorhanianSathyaprakash2024,Gupta2024}.
All these observations naturally raise the question: how did they form?


\subsection{How to make a merging binary}
\paragraph{The separation challenge} Gravity is quite a weak force when compared to other fundamental forces, and gravitational-wave emission depends very strongly on the separation between the two compact objects. 
The leading-order, orbit-averaged coalescence time for two point masses $m_1$ and $m_2$ spiralling in from an initial separation $a_0$ and initial eccentricity $e_0$ is \cite{Peters1964}
\begin{equation}
t_{\rm coal} = \frac{5}{256}\,\frac{c^{5}\, a_0^{4}}{G^{3}\, m_1 m_2 (m_1 + m_2)}\; g(e_0)
\;\lesssim\; t_{\rm H} \approx 14\ \mathrm{Gyr},
\label{eq:t_col}
\end{equation}
where $c$ is the speed of light, $G$ the gravitational constant, and $t_{\rm H}$ the Hubble time. 
The factor $g(e_0)$ collects the enhancement from eccentricity,
\begin{equation}
\begin{split}
g(e_0) = {} & \frac{48}{19}
\left( \frac{1 - e_0^{2}}
            {e_0^{12/19}\left(1 + \frac{121}{304}\,e_0^{2}\right)^{870/2299}} \right)^{\!4} \\[4pt]
& \times \int_{0}^{e_0}
\frac{e^{29/19}\left(1 + \frac{121}{304}\,e^{2}\right)^{1181/2299}}
     {\left(1 - e^{2}\right)^{3/2}}\;\mathrm{d}e ,
\end{split}
\end{equation}
with $e$ the eccentricity, running from $e_0$ down to zero as the orbit circularises.
A binary only enters the observed (merging) population if its coalescence time is less than the age of the universe, i.e., if its coalescence time is less than $t_{\rm H}$. 

The steep $a_0^{4}$ dependence means the compact objects must be brought \textit{very} close together before gravitational-wave emission starts to dominate the orbital evolution: two $10\Msun$ black holes must sit within $\lesssim 20\Rsun$, while their progenitor stars were of order $1000\Rsun$.
At such separations the progenitors would not have fitted; you would get a stellar merger instead.
So how do you bring the cores of stars, or their remnants, close enough that they can merge through gravitational-wave emission?
This is often referred to as the \textit{separation challenge}, and it is what all formation channels aim to solve.

\begin{figure}[!ht]
\centering
\includegraphics[width=0.8\textwidth]{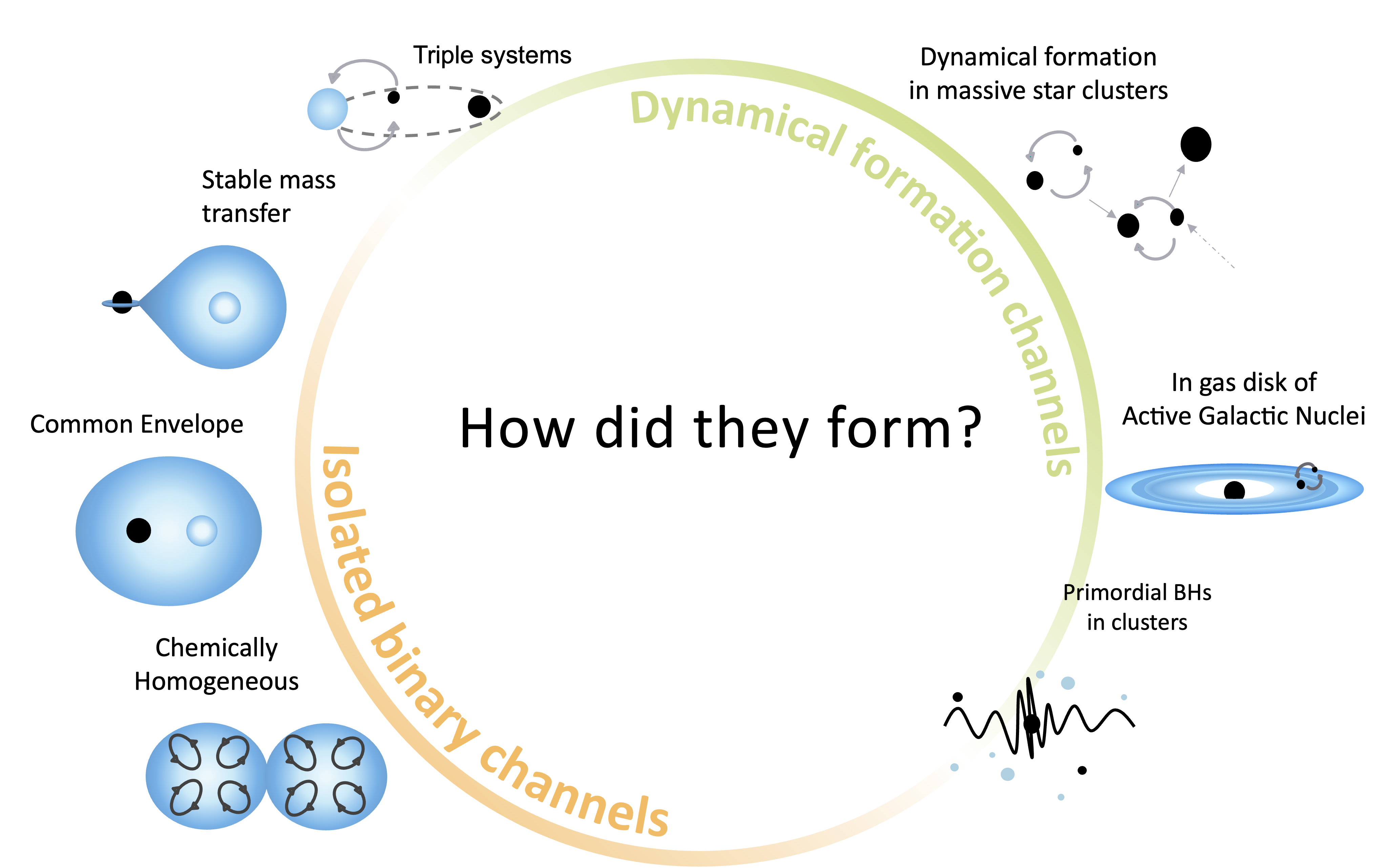}
\caption{Cartoon depictions of the main formation channels for merging compact binaries.}
\label{fig:how_did_they_form}
\end{figure}

\paragraph{Formation channels} Many formation channels have been proposed. 
They are often broadly divided into two families: 
(I) isolated binary channels, in which stars born together shrink their orbit through binary interactions such as mass transfer and tides; and 
(II) dynamical channels, which bring compact objects together through interactions in dynamically dense environments. 
Finding which channel dominates is a central question in gravitational-wave astronomy: if we know how an observed merger formed, we can turn the question around and use gravitational waves to learn about the underlying physics of that channel.
This is why gravitational waves hold the promise of advancing our understanding of things ranging from binary evolution to the discs of active galactic nuclei.

Many gravitational-wave astronomers started with a ``one channel to rule them all'' mindset: the game was to find which channel \textit{dominates} the merger rate \cite{Zevin2021}.
However, as more detections came in over the years, more and more formation channels were proposed, and the field drifted towards ``one channel for every observation''. 
At present, all proposed channels have enough flexibility to predict merger rates of the order of the observed local BBH merger rate \cite{MandelBroekgaarden2022}.
The huge error bars in merger-rate predictions are driven by uncertainties in the underlying binary stellar physics as well as in the cosmic initial conditions (i.e.\ the birth metallicity distribution and star formation history). 

Luckily, all is not lost: with more detections also came better constraints on the \textit{properties} of the observed systems, rather than just on the merger rate.
There are now enough detections to measure the distributions of chirp masses, mass ratios and effective spins, and there are tentative claims of residual eccentricity and of redshift evolution in those distributions~\cite{GWTC5_pop,FishbachvanSon2023,2022ApJ...940..171R}~.

This section summarizes the main formation channels, focusing on what each channel predicts for the GW observables. 
Because this is a rapidly evolving field, we stick as much as possible to intuitive relations rather than quantitative results.

\subsection{Isolated binary evolution channels}
For isolated binary evolution, the most important physics is mass transfer and tides.
Together they set the final masses, separations and spins of the merging compact objects, and hence the gravitational-wave observables.

\textbf{Mass transfer} is one of the defining features of what sets binary stars apart from single stars. 
In the context of GW formation, it determines how the binary loses angular momentum, and hence the final masses and separations.
More specifically, by moving mass, between stars or out of the system entirely, it changes the stellar evolution and the final compact-object masses in a very direct way.  
Since the mass carries angular momentum, it also changes the binary separation, which affects any potential future interactions and sets the coalescence time (eq. \ref{eq:t_col}).

For our purposes, the second most important binary interaction is \textbf{tides}.
Tidal effects depend steeply on the ratio of stellar radius to orbital separation ($R/a$), and are therefore only relevant for close binaries.
In general, tides work to circularise and synchronise the orbit.
Circularisation depends more steeply on $R/a$ than synchronisation, $\tau_{\rm circ} \sim (R/a)^{-8}$ against $\tau_{\rm sync} \sim (R/a)^{-6}$ \cite{1977A&A....57..383Z,1981A&A....99..126H}. 
Since $R < a$, the steeper power is also the \textit{longer} timescale, so $\tau_{\rm circ} > \tau_{\rm sync}$: systems synchronise well before they circularise.
This means that even though many observed post-mass-transfer systems still have non-zero eccentricity \cite{vanson2026_post_mt}, they are still likely to be spin--orbit aligned. 
Because gravitational-wave sources must be short-period, interacting binaries, tides are generally assumed to be efficient enough in the isolated formation channels that we expect \textbf{low eccentricity and orbit-aligned spins} for merging BBHs (ignoring any complications caused by natal kicks).
Tides are moreover essential for the Chemically Homogeneous channel (discussed below) and can set the spin of the first-born black hole.

The type of mass transfer is so decisive that the isolated channels are classified by it. 
Since it is the stability of the mass transfer that separates the two, we first look at Roche-lobe overflow and what sets its stability, and only then turn to the channels themselves.

\subsubsection{Roche-lobe overflow: when is mass transfer (un)stable?}

{\bf The Roche potential} approximates each star as a point mass, assumes a circular orbit of constant separation $a$, and works in the co-rotating frame, where a test particle feels the gravity of both stars plus the centrifugal force. 
The equipotential surfaces of this Roche potential tell us whether a parcel of mass will move towards the donor or the accretor. 
The Lagrangian points L1, L2 and L3, are saddle points of this potential. The volume-equivalent radius of the donor's Roche lobe can be approximated as a function of the mass ratio $q$\cite{Eggleton1983} 
\begin{equation}
    \frac{R_{\rm L,1}}{a}  \approx \frac{0.49q^{2/3}}{0.6q^{2/3} + \ln (1+q^{1/3})}.
\end{equation}
%
{\bf When is mass transfer (un)stable?}
To first order, the dynamical stability of mass transfer can be estimated by comparing how the donor’s radius and its Roche-lobe radius respond to mass loss. 
In equation form we compare logarithmic derivatives ($\zeta$-parameters): the Roche-lobe response
\begin{equation}
    \zeta_{\rm L} \equiv \frac{\mathrm{d}\ln R_{\rm L}}{\mathrm{d}\ln M_{\rm d}},
\end{equation}
against the donor's radius ($R_\star$) response, which comes on two timescales: 
\begin{equation}
    \underbrace{\zeta_{\rm ad} \equiv (\partial \ln R_\star / \partial \ln M_\star)_{\rm ad}}_{\text{adiabatic}} \quad
\underbrace{\zeta_{\rm eq} \equiv (\partial \ln R_\star / \partial \ln M_\star)_{\rm eq}}_{\text{thermal}},
\end{equation}
where the adiabatic response works on a dynamical timescale, restoring hydrostatic equilibrium, while the thermal timescale response restores thermal equilibrium. 
If both $\zeta_{\rm ad} \geq \zeta_{\rm L}$ and $\zeta_{\rm eq} \geq \zeta_{\rm L}$, the new equilibrium radius fits within the Roche lobe: secularly stable mass transfer. 
In the intermediate case $\zeta_{\rm ad} \geq \zeta_{\rm L} > \zeta_{\rm eq}$, mass transfer is dynamically stable and proceeds on the donor's thermal timescale. 
If $\zeta_{\rm L} > \zeta_{\rm ad}$, the donor cannot shrink inside its lobe on a dynamical timescale and cannot recover hydrostatic equilibrium: likely runaway (unstable) mass transfer. 
Given a particular mass transfer efficiency (i.e., the fraction of the donated mass that is accreted by the companion, as opposed to being lost from the system), the $\zeta$-parameters map onto a critical mass ratio \cite{Soberman1997}.
I.e., mass transfer is dynamically stable (proceeding on the nuclear/thermal timescale) when $M_{\rm a}/M_{\rm d} > q_{\rm crit}$.

Mass transfer stability thus depends heavily on the donor's envelope structure:
donors with radiative envelopes remain stable down to more extreme mass ratios ($q_{\rm crit} \sim 0.25$) than donors with convective envelopes ($q_{\rm crit} \sim 0.8$--1) e.g., \cite{HjellmingWebbink1987,Ge2015,Ge2020,Temmink2023}, since convective envelopes \textit{expand} upon mass loss (this is an effect of the entropy profile).
At extreme mass ratios mass transfer is destabilised by the Darwin instability, a tidal synchronisation instability that sets in once the donor's rotational angular momentum exceeds one third of the orbital angular momentum \cite{Darwin1879RSPS...29..168D,Hut1980A&A....92..167H}.
The exact values of $q_{\rm crit}$ are debated as mass-transfer stability remains an active area of research. The aforementioned simple parametrisation does not include effects such as delayed dynamical instability and the response of the {\it accretor} e.g., \cite{HjellmingWebbink1987,Ge2020,Temmink2023,2025A&A...703A.121T,2026A&A...709L...6T,2021ApJ...923..277R}.
Some mass-ratio dependence is unavoidable, however, through the dependence of $R_{\rm L}$ on $q$.


{\bf Angular-momentum budget of non-conservative mass transfer.}
To follow the orbital evolution of a mass-transferring binary we need to know how much angular momentum leaves with the matter that leaves the system, i.e.\ the \textit{specific angular momentum of the ejected matter}.
This is set by \textit{where} the mass leaves the system, and is commonly assumed to be one of three modes \cite{Soberman1997}:
(1) a fast isotropic wind from the vicinity of the donor (the ``Jeans mode'');
(2) isotropic re-emission from the vicinity of the accretor (e.g.\ accretion-driven winds or jets from a compact accretor);
or (3) outflow through the outer Lagrangian point L2, which, lacking the energy to escape the binary potential, settles into a circumbinary ring at radius $a_{\rm ring}$ (the ``intermediate mode'', relevant at high transfer rates and for (near) contact binaries).
Writing the specific angular momentum of the ejecta as $\gamma$ times the mean specific orbital angular momentum $h_{\rm orb} = J_{\rm orb}/(M_{\rm d} + M_{\rm a})$, the three modes correspond to
\be
\gamma_{\rm Jeans} = \frac{M_{\rm a}}{M_{\rm d}}, \qquad
\gamma_{\rm iso} = \frac{M_{\rm d}}{M_{\rm a}}, \qquad
\gamma_{\rm ring} = \frac{(M_{\rm d} + M_{\rm a})^{2}}{M_{\rm d} M_{\rm a}} \sqrt{\frac{a_{\rm ring}}{a}},
\ee
with $M_{\rm d}$ and $M_{\rm a}$ the donor and accretor mass and $a$ the orbital separation.
Winds (Jeans mode) mass loss will always widen the orbit, isotropic re-emission shrinks can both widen or shrink the orbit depending on the mass ratio. Mass lost from a circumbinary ring always drains angular momentum from the orbit: since $(M_{\rm d} + M_{\rm a})^{2}/(M_{\rm d} M_{\rm a}) \geq 4$ and $a_{\rm ring} > a$, we have $\gamma_{\rm ring} \gtrsim 5$, and the orbit shrinks rapidly.
As we will see below, the common envelope shrinks orbits more efficiently still.
How the mass leaves the system also sets the delay times of the stable channel.


\subsubsection{The stable mass-transfer channel}\label{sec:day3smt}
In the stable mass-transfer channel the binary never goes through a common envelope.
In most cases it experiences two phases of stable mass transfer: one from the initially more massive star onto its companion, and a second one in the opposite direction once the mass ratio has been reversed \cite{Inayoshi2017,vandenHeuvel2017,Bavera2021,GallegosGarcia2021,Marchant2021,vanSon2022_no_peaks,Picco2024}.
Requiring both phases to remain stable constrains the masses and mass ratios at each interaction, so this channel makes comparatively sharp predictions for the masses, mass ratios, separations and delay times of the binaries it produces.
It predicts, for example, both a minimum and a maximum \textit{final} mass ratio \cite{vanSon2022_no_peaks,ChenBreivikvanSon2026,2026arXiv260523083G}, while the limit on how tight stable mass transfer can make the orbit sets a minimum separation, and hence a minimum delay time \cite{Klencki2026}.

Because more massive stars have more radiative envelopes, \textbf{more massive stars are more likely to undergo stable mass transfer}.
Requiring two phases of stable mass transfer also sets a minimum remnant mass \cite{vanSon2022_no_peaks}, and it is why the stable channel does not produce NSNS mergers, and the progenitors of double neutron stars cannot avoid a common envelope \cite{VignaGomez2018,Broekgaarden2026} (see a more extended discussion in Section~
\ref{sec:day4nsns}).
\subsubsection{The common-envelope channel}
In the `classical' common-envelope (CE) channel the first mass-transfer phase is stable, while the second one is unstable \cite{Belczynski2007,PostnovYungelson2014,Belczynski2016,VignaGomez2018}.
Once mass transfer becomes unstable, the donor dumps ever more mass onto its companion while the orbit shrinks, until the companion effectively plunges into the donor's envelope. 
Both stars then share a single envelope, hence a `common envelope' (see \cite{IvanovaJusthamRicker2020} for a review).
The inspiralling companion deposits orbital energy into the envelope, which may lift and eventually unbind the gas.
The outcome is either a close detached binary (envelope successfully ejected) or, if the ejection fails, a stellar merger, which for a neutron-star companion produces the exotic Thorne--\.Zytkow objects \cite{ThorneZytkow1975,ThorneZytkow1977}.
For gravitational waves we need the first outcome.

{\bf The energy formalism.}
Once the companion is inside the envelope the problem is no longer a mass-transfer episode, but a star or compact object moving through dense gas, which is inherently 3D and hydrodynamical.
Such simulations are computationally expensive and have only recently become feasible \cite{Ohlmann2016,RoepkeDeMarco2023}. 
Hence, population studies instead resort to the energy formalism \cite{vandenHeuvel1976,TutukovYungelson1979,IbenTutukov1984,Webbink1984,LivioSoker1988}.
Here we assume that a fraction $\alpha_{\rm CE}$ of the orbital energy released during the inspiral is available to unbind the envelope,
\begin{equation}
    \alpha_{\rm CE} = \frac{E_{\rm bind}}{\Delta E_{\rm orb}}, \qquad
\Delta E_{\rm orb} = \frac{G M_{\rm a} M_{\rm core}}{2 a_{\rm f}} - \frac{G M_{\rm a} M_{\rm d}}{2 a_{\rm i}},
\end{equation}
where $M_{\rm d}$ and $M_{\rm a}$ are the donor and companion mass, $M_{\rm core}$ the mass of the donor's core, and $a_{\rm i}$ and $a_{\rm f}$ the separation before and after the inspiral.
The energy needed to unbind the envelope follows from integrating over the donor's stellar structure,
\be
E_{\rm bind} = \int_{M_{\rm core}}^{M_{\rm d}} \left( \frac{Gm}{r} - u \right) \mathrm{d}m,
\ee
with $m$ the mass coordinate, $r(m)$ the radius at that coordinate and $u$ the specific internal energy (including corrections such as the dissociation energy of molecular hydrogen).
Some algebra later, the post-CE separation is
\be
a_{\rm f} = M_{\rm core} M_{\rm a} \left( \frac{2 E_{\rm bind}}{\alpha_{\rm CE} G} + \frac{M_{\rm d} M_{\rm a}}{a_{\rm i}} \right)^{-1},
\ee
where $M_{\rm core} = M_{\rm d} - M_{\rm env}$, with $M_{\rm env}$ the envelope mass, and where we assumed the companion accretes negligibly during the inspiral.
Note that $\alpha_{\rm CE} \leq 1$ if only orbital energy is available.
Some studies nonetheless adopt values of $\alpha_{\rm CE} > 1$, which requires additional energy sources, such as recombination.
%
The key characteristic: {\bf the separations that come out of a CE are very small}, much smaller than what stable mass transfer can produce.

\subsubsection{Comparing the stable mass transfer and common envelope channels}
The steep separation dependence of the coalescence time (eq.~\ref{eq:t_col}) means that shorter post-mass-transfer periods translate into shorter delay times.
Since a common envelope shrinks the orbit far more efficiently than stable mass transfer, the stable channel is left with wider orbits and hence longer delay times.
The two channels also differ in mass: more massive stars are more likely to undergo stable mass transfer, and on top of that two phases of stable mass transfer cannot produce the lowest-mass double compact objects (although where exactly this mass boundary falls is model dependent).
In summary, {\bf the stable channel produces more massive systems with longer delay times, while the CE channel produces lower-mass systems with shorter delay times} \cite{vanSon2022_rate_redshift}, a separation that turns out to be quite robust against most model uncertainties e.g., \cite{Bavera2021,GallegosGarcia2021,Marchant2021,Olejak2021,Picco2024,Klencki2026}.
This correlation between mass and delay time predicts a mass-dependent redshift evolution of the merger rate: high-mass systems, with their long delay times, should evolve more shallowly, while low-mass systems, with short delays, should track the star-formation rate more closely e.g., \cite{vanSon2022_rate_redshift}.

The short post-CE separations are also important for the spins of the merging black holes.
In the `classical' CE channel, we are left with a compact object in a very tight orbit around a stripped helium core after the successful envelope ejection.
If angular-momentum transport is efficient, the He core is expected to be spinning slowly. 
However, the tides in short orbits can be strong enough to spin it back up before it collapses, hence producing a spinning second-born black hole \cite{FullerMa2019,FullerLu2022}.
In the stable channel the post-mass-transfer orbits are too wide for tides to spin up the helium star, so its black holes are expected to be born slowly spinning \cite{Bavera2020}.

\subsubsection{Chemically homogeneous evolution}\label{sec:day3che}
We have seen in Section~\ref{sec:day1} that rapid rotation can completely alter the life of a star.
If the star rotates rapidly enough, meridional (Eddington--Sweet) circulations mix fresh hydrogen into the burning core, and the star never develops a core--envelope structure.
This matters for GW sources because such a star stays compact: it never expands, never fills its Roche lobe, and so the binary does not need mass transfer to solve the separation challenge.
The two stars simply collapse into black holes in the tight orbit they were born in \cite{deMinkMandel2016,MandeldeMink2016,Marchant2016}.
It is perhaps the simplest BBH channel: start close, stay compact and merge.

Rotational mixing is most efficient in the most massive stars, which are more radiation-pressure dominated.
In a radiative gas the adiabatic temperature gradient $\nabla_{\rm ad} = \left( \mathrm{d} \ln T / \mathrm{d} \ln P \right)_{\rm ad}$, with $T$ the temperature and $P$ the pressure, drops from 0.4 for an ideal gas to 0.25 when radiation pressure dominates.
Since the meridional currents depend on the sub-adiabaticity $(\nabla_{\rm ad} - \nabla)$, where $\nabla = \mathrm{d}\ln T/\mathrm{d}\ln P$ is the actual gradient, a lower $\nabla_{\rm ad}$ makes the star more susceptible to them \cite{Kippenhahn1974,Zahn1983,Maeder1987,Hastings2020}.
Hence {\bf higher-mass stars go CHE more easily}, and the channel only produces black holes above roughly $20\Msun$.
Working against the mixing is the gradient in mean molecular weight ($\mu$-gradient) that the burning itself builds up, and which suppresses the circulations once it becomes too steep.
Mixing therefore has to be strong from the start: a star can only evolve chemically homogeneously if it is born rapidly rotating and compact, before any core--envelope structure (and corresponding $\mu$-gradient) has had time to develop.

CHE can thus only occur in a narrow window of initial conditions.
At the lowest masses the adiabatic temperature gradient is too high for the circulations to develop; at long periods tides are too weak to keep the stars spinning fast enough; and at the shortest periods the stars already overflow L2 on the ZAMS and most likely merge.
Together these limits carve out a banana-shaped region in the plane of initial period versus initial mass \cite{deMinkMandel2016,MandeldeMink2016,Marchant2016,Hastings2020,Sharpe2024}.
The window is further restricted to low metallicity, since strong winds spin down the stars and widen the orbit, working against the tidal synchronisation needed to maintain the rapid rotation \cite{MandeldeMink2016,Marchant2016,Riley2021}.
CHE also favours near-equal masses: both stars have to satisfy the CHE condition at the same orbital period, and since the rotation rate required for homogeneous mixing depends on mass, the two windows only overlap when the components are similar.
\cite{Hastings2020} indeed find that the window shrinks significantly for unequal-mass ($q = 0.7$) binaries.

What about the \textbf{spins}?
One would expect CHE to be a prime way to form rapidly rotating black holes, since the stars are tidally locked in a very tight orbit right up to collapse.
However, the big uncertainty remains the angular-momentum transport: the Tayler--Spruit prescription of \cite{Spruit2002} leaves very high spins at carbon depletion, whereas the more efficient transport of \cite{FullerLu2022} yields slowly or non-spinning cores at collapse. So even this channel can form slowly spinning black holes.\\

In summary, the CHE channel is expected to produce massive ($\gtrsim 20\Msun$), near-equal-mass black holes with short delay times (depending on metallicity and stellar winds), with spins that hinge entirely on the assumed angular-momentum transport.

\subsubsection{Triples}\label{day3:triples}
Most massive stars do not just live in binaries but in triples or higher-order multiples \cite{Offner2022}.
Triples are dynamically stable when they occur in a hierarchical configuration, meaning that the outer orbit is much larger than the inner orbit, $a_{\rm out} \gg a_{\rm in}$, with $a_{\rm in}$ and $a_{\rm out}$ the semi-major axes of the inner and outer orbit.
In this case the secular approximation applies: we can use phase-averaged, long-term evolution in which the orbits torque each other and exchange angular momentum but not energy, so their shapes and orientations change on timescales much longer than the orbital periods, while the semi-major axes do not (see the review by \cite{Naoz2016}).

The hallmark effects are the von Zeipel-Lidov-Kozai (ZLK) cycles \cite{vonZeipel1910,Lidov1962,Kozai1962}: mutual inclination is exchanged for inner-orbit eccentricity and vice versa.
Even systems starting from zero eccentricity can be pumped to extremely high eccentricities (e.g., $e_{1,\rm max} = 0.689$ for initial mutual inclination $i_{\rm tot} = 90^\circ$, and $e_{1,\rm max} = 0.99$ for $i_{\rm tot} = 158^\circ$ in the classic examples of \cite{Naoz2016}), which, given the steep eccentricity dependence of the coalescence time (eq.~\ref{eq:t_col}), strongly affects which systems can merge within a Hubble time.
The quadrupole-order approximation of ZLK assumes a circular outer orbit and a test-particle inner orbit. Relaxing either of these assumptions requires the octupole level, at which the inner orbit reaches even higher eccentricities and can flip from prograde to retrograde (the eccentric Kozai--Lidov mechanism).
Octupole-order effects can thus further boost BBH mergers for low-$q$ inner binaries \cite{Su2021b}. 
If the ZLK timescale is short enough to act while the stars are still alive, triples cause more, and earlier, interactions: mergers of the inner binary (turning the triple into a binary), or mass transfer from the tertiary onto the inner binary, which is even more complicated than the binary mass transfer we already struggle to model \cite{Toonen2020A&A...640A..16T}.

Triples might be especially important for BBH formation: even when the three stars never interact, a tertiary can drive inner binaries to merge that would otherwise be too wide, so \textbf{the presence of triples is generally expected to increase the BBH merger rate} e.g., \cite{FragioneLoeb2019a}.
For NSNS and NSBH on the other hand, NS natal kicks might easily unbind the necessarily hierarchical, and thus also fragile triple system.

The expected GW signatures of triple-induced mergers are \textbf{high eccentricity}, possibly retained all the way into the LIGO band e.g., \cite{Antonini2017,Liu2019}, and a \textbf{flat(ter) mass-ratio distribution} than for either isolated binaries or dense clusters\cite{Martinez2022,Stegmann2022,SuLai2021,Dorozsmai2024}.

\subsection{Dynamical formation channels}\label{sec:day3dyn}
Everything so far assumed that the binary evolves undisturbed.
This assumption breaks down in the cores of star clusters, where the stellar density is high enough that the timescale for a close encounter drops below the age of the cluster.
There, the orbit of a binary is no longer purely set by its own evolution but by its encounter history.
Note that this matters particularly \textit{after} the compact objects have formed: their progenitors live only a few Myr, but the black holes they leave behind are essentially permanent, and so have ample time to be hardened, exchanged or ejected.

Dynamical channels can operate wherever the density of objects is high enough for the processes discussed below to act within the lifetime of the environment.
In practice the focus is on the densest environments in the Universe.
The density sets the interaction rate, while the escape velocity sets what stays behind after an interaction.
Ordered by increasing mass and density, these are:
\textbf{open and young massive clusters} ($M \sim 10^{2}$--$10^{5}\Msun$, $v_{\rm esc}$ of order a few $\mathrm{km\,s^{-1}}$, possibly still gas-rich);
\textbf{globular clusters} ($M \sim 10^{5}$--$10^{7}\Msun$, $v_{\rm esc} \sim 10$--$100\,\mathrm{km\,s^{-1}}$, old and gas-poor, commonly approximated as single-metallicity although the multiple-population problem has long been known);
and \textbf{nuclear star clusters} ($M \sim 10^{6}$--$10^{8}\Msun$, $v_{\rm esc}$ up to several hundred $\mathrm{km\,s^{-1}}$, the densest of the three, found at galactic centres and possibly influenced by a central massive black hole) e.g., \cite{PortegiesZwart2010,Harris2010,Neumayer2020}.
Separate from these, the gas discs of \textbf{active galactic nuclei} (AGN) provide a dense environment of their own, in which gas adds processes that are absent in gas-poor clusters.

\subsubsection{Dynamical processes}

Dynamics adds a rich zoo of processes on top of the single and binary stellar evolution discussed so far e.g., \cite{SigurdssonHernquist1993,PortegiesZwartMcMillan2000,MillerLauburg2009,Rodriguez2015,Rodriguez2016a,Antonini2016}.
Broadly, they act in three ways:\\
\textbf{I) which objects will interact}
Which objects interact at all is set by two-body relaxation, which drives the cluster towards energy equipartition.
Since the dynamical-friction timescale scales inversely with the mass of the infalling object, the heaviest objects sink to the centre first, so mass segregation determines which objects populate the dense core (we return to the consequences below).
In AGN discs the same role is played by gas dynamical friction and by migration traps.\\
\textbf{II) assemble new binaries}
Dynamics can assemble binaries from compact objects that were never born together.
Three initially unbound objects can form a binary if a close interaction leaves two of them bound while the third carries away the excess energy (\textit{three-body binary formation}). Similarly, two unbound compact objects can become bound if they pass close enough at pericentre that gravitational-wave emission removes enough energy (\textit{a GW capture}).
The latter requires a very small pericentre, so these binaries are born highly eccentric and merge on very short timescales.\\
\textbf{III) modify existing binaries}
Once a binary exists, dynamical encounters can keep reshaping it.
In a binary-single resonant encounter the three bodies enter a chaotic, temporarily bound phase in which they repeatedly exchange partners, until the least massive body is preferentially ejected and the remaining two are left more tightly bound.
This is \textit{Heggie's law}: hard binaries (their binding energy exceeds the typical kinetic energy of the surrounding stars) tend to harden, while soft binaries soften.
Weaker perturbations do the same job less dramatically: stellar flybys and close encounters change the orbit incrementally, and the secular ZLK mechanism introduced above applies whenever the perturber stays bound.

\subsubsection{What dynamical channels predict for the GW observables}

\textbf{Isotropic spins.} Dynamically assembled binaries are expected to be paired in random orientations, so their spin orientations are isotropic, in contrast to the aligned spins expected from isolated evolution. 
Spins leave only a weak imprint on the waveform and are correspondingly hard to measure; but the isotropy prediction nevertheless plays a central role in channel-fraction inference e.g., \cite{Rodriguez2016b,Vitale2017,Farr2017,Zevin2021,Bouffanais2021,Baibhav2023}.\\

\noindent\textbf{Eccentricity.} Dynamical channels also generically allow higher eccentricities, but most such binaries circularise before they enter the band of ground-based detectors. Only a small fraction, formed at very small pericentre separations (GW captures in particular), retain a measurable eccentricity at $\sim$10 Hz. Space-based detectors such as LISA (and perhaps DECIGO) probe an earlier stage of the inspiral, where the memory of an eccentric formation survives \cite{Breivik2016}. \\

\noindent\textbf{Hierarchical mergers.}\label{ss: hierarchical mergers} 
Dynamical environments uniquely allow the product of a double compact object merger to merge again.
For such a `hierarchical' merger to occur, the remnant must be retained in the cluster after the first merger.
Whether it is retained is decided by the gravitational recoil imparted at merger, compared against the escape velocity of the environment.

The recoil can reach hundreds to thousands of $\mathrm{km\,s^{-1}}$, and its amplitude depends sensitively on the spins of the merging black holes.
Low spins, and spins aligned or anti-aligned with the orbital angular momentum, give small recoils, while large spins with substantial components \textit{in the orbital plane} produce the largest ones \cite{2007PhRvL..98w1101G,2007PhRvL..98w1102C,2011PhRvL.107w1102L}.
Compared against the escape velocities listed above, such large recoils translate into low retention of second-generation (2G) black holes, and higher-generation mergers are therefore expected to be far more common in nuclear star clusters than in young clusters \cite{Rodriguez2019}.

The spin of the merger product, in contrast, is almost independent of what went in: \textbf{2G black holes peak at a spin of $\sim0.7$}, a robust numerical-relativity result e.g., \cite{Pretorius2005,Buonanno2008,Hofmann2016}.
Two things contribute to this final spin: the spins of the progenitors and the orbital angular momentum at merger. 
The orbital term dominates for anything but extreme mass ratios, causing the final spin to be largely independent of the properties of the incoming black holes, and peak strongly around $\sim0.7$\cite{FishbachHolzFarr2017}. 
This $\sim0.7$ spin feeds back into larger recoils for the mergers of 2G black holes themselves, drastically suppressing anything beyond the second generation.

Together, the $\sim0.7$ spin magnitude and the isotropic spin orientations lead to a \textbf{strong prediction for the effective inspiral spin of hierarchical mergers.}
For a 2G primary with $\chi_1 \approx 0.7$, which is about twice as massive as its non-spinning 1G companion, the maximum is
\begin{equation}
    \max(\chi_{\rm eff}) \approx \frac{2 m \cdot 0.7 + m \cdot 0}{3m} = \frac{1.4}{3} \approx 0.47 .
\end{equation}
For isotropic orientations $\mathrm{E}[\cos\theta] = 0$ and $\mathrm{Var}(\cos\theta) = \int_{-1}^{1} u^2\,\tfrac{1}{2}\,\mathrm{d}u = 1/3$. In other words, the distribution is centred on zero and broad: plugging in these values yields a $\chi_{\rm eff}$ distribution roughly uniform between $-0.5$ and $+0.5$ e.g., \cite{Antonini2025a}. This broad, flat distribution has played an important role in interpreting recent observations, i.e.\ in the hunt for hierarchical mergers in GWTC-5 data; we return to this in Section~\ref{sec:highmassgap}. \\


{\bf Dynamical channels strongly favour BBH mergers over BHNS and NSNS mergers.} 
The reason is mass segregation: the dynamical-friction (relaxation) timescale scales inversely with the mass of the infalling object, so black holes sink to the centre first. 
Once there they dominate the cluster core, preventing NSs and their progenitors from segregating and suppressing their interaction rate. 
The binding energy released as black-hole binaries harden is injected into the core, expanding it and lowering the central density, so the black holes are depleted only slowly in the ``BH burning'' process \cite{Spitzer1967,BreenHeggie2013}. 
Only once that population is depleted can NSs begin to interact, which in most globular clusters never happens: black holes dominate the core over essentially the entire cluster lifetime \cite{Morscher2015,Ye2020}.
On top of that, natal kicks eject the tightest NS progenitors at birth, so only low-kick channels such as electron-capture supernova products are likely retained in the cluster environments. 
The net effect: {\bf the cluster contribution to the NSNS merger rate falls orders of magnitude below the field contribution} e.g., \cite{Clausen2013,Ye2020,Rastello2020,ArcaSedda2020}.

\subsubsection{Summary of the dynamical channels}
Dynamical formation is expected to produce isotropic spin orientations, and hence a $\chi_{\rm eff}$ distribution centred on zero; near-equal-mass mergers for the first-generation population; and on average higher eccentricities than isolated binaries.
In environments with escape velocities high enough to retain the remnants, it additionally produces hierarchical mergers, recognisable by a next-generation spin of $\sim0.7$, a broad $\chi_{\rm eff}$ distribution, and $q \sim 0.5$ for the 2G$+$1G pairs.
Finally, dynamical channels strongly favour BBH mergers over BHNS and NSNS.

\section{Open problems in gravitational-wave astronomy}\label{sec:day4}

Armed with the formation channels of Section~\ref{sec:day3}, we now turn to what the observed population shows, what this has already taught us about the likely formation channels for different types of compact object binaries, and some of the open problems that remain.
We focus \textit{mostly} on black holes, simply because they dominate the catalogue: of the 267 compact binary mergers in the GWTC-5.0 population sample, only (about) three are NSBH and two are NSNS \cite{GWTC4pop,GWTC5_pop}.
We moreover concentrate most of our discussion on masses, because mass is the most solidly constrained property of a GW event, and the features in the mass distribution are correspondingly the most robust ones we have today.

\subsection{Why NSNS mergers only form through the common envelope channel}\label{sec:day4nsns}
Before turning to the black holes, it is worth a moment on the neutron stars.
Most channels  (as discussed in Section~\ref{sec:day3}) are simply not very efficient at producing low-mass merging systems.
Let's go through the main channels one by one, and see why they are not expected to contribute significantly to the NSNS merger rate.

Most channels simply do not operate at low mass.
Cluster environments disfavour NSNS mergers as opposed to BBHs (as discussed in Section~\ref{sec:day3}): mass segregation places black holes in the core first, and they dominate it over essentially the entire cluster lifetime, so neutron stars do not segregate and their interaction rate remains suppressed \cite{Morscher2015,Ye2020}.
In addition, natal kicks eject the tightest NS progenitors at birth, so only low-kick products such as electron-capture supernovae are retained.
The cluster contribution to the NSNS merger rate therefore falls orders of magnitude below the field contribution e.g., \cite{Clausen2013,Ye2020,Rastello2020,ArcaSedda2020}.
Secular triples are easily unbound by NS natal kicks, so they contribute only if kicks are low e.g., \cite{HamersThompson2019,SamsingHotokezaka2021,Lu2021,Hamers2021}.
CHE needs the radiation-pressure-dominated envelopes of high-mass stars, and thus typically yields only black holes with masses $\gtrsim 20\Msun$ e.g., \cite{deMink2009,Song2013,MandeldeMink2016,Marchant2016,Riley2021}.
Population III binaries peak around $\sim20\Msun$ e.g., \cite{Marigo2001,Belczynski2004,Kinugawa2014,Inayoshi2017}.
That leaves the AGN discs, where the rates are highly uncertain, the smoking gun of AGN-disk assisted double compact object formation are rather very high-mass N-th generation hierarchical merger (due to the very deep potential wells) \cite{Baruteau2011,Bellovary2016,Leigh2018,Yang2019,Secunda2019,McKernan2020,Cantiello2021}, which is not relevant for NSNS mergers.
At the lowest masses we are therefore left with only two contenders to contribute to the NSNS merger rate: the two `classical' isolated channels: stable mass transfer and the common envelope.

However, as we hinted at in Section~\ref{sec:day3smt}, the stable mass transfer channel also fails to make a significant contribution to the NSNS merger rate. Mass transfer is more stable for comparable masses and less stable at extreme mass ratios, and since the core-mass fraction grows with stellar mass, this restriction on the mass ratio translates into a restriction on the mass itself. Requiring two phases of stable mass transfer imposes a minimum remnant mass which (for default physics assumptions) lies at about $\sim 5\Msun$, i.e. above the maximum NS mass \cite{vanSon2022_no_peaks}.
The channel loses contribution towards lower primary masses and disappears before reaching neutron-star masses.
Whether the simulations represent the real world is a separate question, but the mass-ratio and stability argument behind this limit should hold at least qualitatively.

The conclusion is that {\bf NSNS mergers require a common envelope}, and essentially all population-synthesis codes agree on this \cite{VignaGomez2018,2023MNRAS.524..426I,2022ApJ...937..118W,Broekgaarden2026}
\textbf{the higher we go in mass, the more channels can contribute}.

\begin{figure}[h!]
    \centering
    \includegraphics[width=0.9\textwidth]{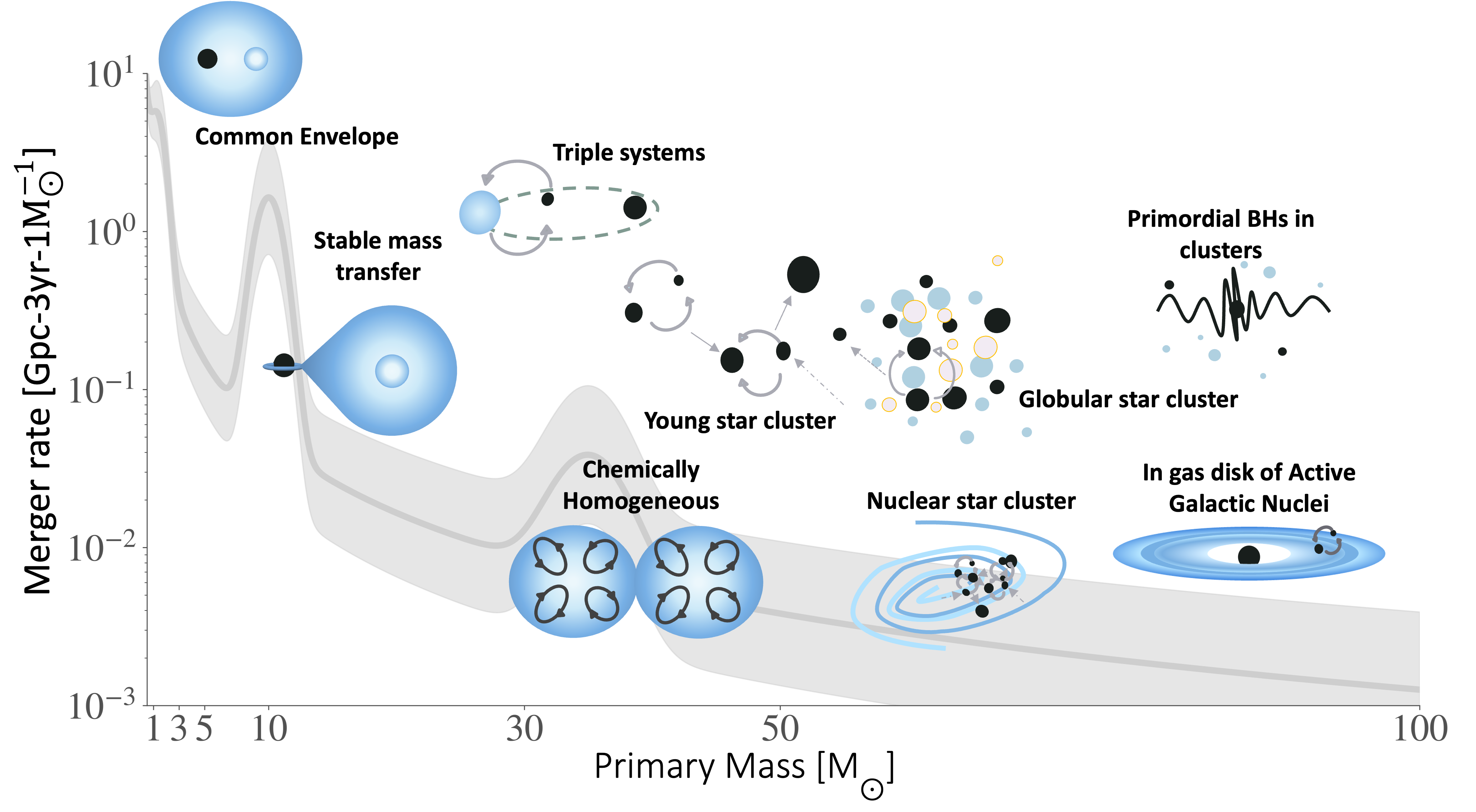}
    \caption{The higher we go in mass, the more channels can contribute. The cartoon shows the main channels discussed in Section~\ref{sec:day3}. The grey background represents the rough location of the gaps and peaks in the mass function as discussed in this Section. }
    \label{fig:day4_channel_cartoon}
\end{figure}


\subsection{Features in the BBH mass function}
With hundreds of detections we no longer have just a merger rate of BBHs, but a full mass function, and multiple features have now been detected \cite{GWTC5_pop}.
In the rest of this section we walk up the mass function, from the lower mass gap to the highest masses.

The two most robustly identified are peaks, at $\sim 9 \Msun$ (Section~\ref{sec:peak10Msun}) and $\sim 35 \Msun$ (Section~\ref{sec:35peak}), recovered consistently with both strongly parametrised \cite{gwtc-3pop,Toubiana2023,MaganaHernandez2024,Gennari2025} and flexible `non-parametric' methods \cite{Fishbach_2020,Veske2021,TiwariFairhurst2021,Ray_2023,PhysRevD.111.L061305,Sadiq_2023,Edelman2023,Rinaldi_2024,CallisterFarr2024}.
Despite this strong observational evidence, the physical origin of both features remains uncertain:
several formation scenarios have been proposed for each\cite{vanSon2022_no_peaks,2023MNRAS.526.4130H,Antonini_2023,2025A&A...695A..71L,2025arXiv251007573W}, but mass alone appears insufficient to distinguish among them \cite{Roy2025}.

Alongside the peaks sit two extensively studied `gaps', which come at the problem from the opposite direction: both were predicted before there was a BBH mass function.
The (putative) `lower mass gap' between the heaviest neutron stars and the lightest black holes (Section~\ref{sec:day4lowgap}) follows from how, and whether, massive stars explode, while the `upper mass gap', or pair-instability supernova (PISN) mass gap (Section~\ref{sec:highmassgap}), follows from the pair instability itself.

Since the release of GWTC-5.0 \cite{GWTC5_pop} there are enough events to zoom in on an individual mass feature and ask whether the systems there share other properties, such as mass ratios and spins, and first indications of such sub-populations have emerged \cite{2025arXiv250915646B}.
Linking such sub-populations to formation channels is like playing BINGO with the observables: each channel fills in different squares for mass, mass ratio, $\chi_{\rm eff}$, and redshift evolution.
Multi-dimensional approaches of this kind, combining mass, spin and eccentricity, are exactly what is now beginning to locate the PISN gap edge and to identify mergers of dynamical or even hierarchical origin \cite{2025arXiv251022698W,2025arXiv251018867R,2025arXiv250904637A,2025arXiv251105316T,2025ApJ...994L..47S}.

\subsection{The (putative) lower mass gap}\label{sec:day4lowgap}
A gap between the heaviest neutron stars and the lightest black holes was first proposed more than twenty years ago.
It was suggested on the basis of (seven) black-hole mass measurements in low-mass X-ray binaries, which indicated a lack of black holes below about $6\Msun$ \cite{Bailyn1998}, and revisited a decade later with sixteen systems, reaching the same conclusion below about $5\Msun$ \cite{Ozel2010}.
The statistical significance of the gap has been debated ever since \cite{Kreidberg2012,WyrzykowskiMandel2020,Jonker2021,Siegel2023}.

If the data do support a gap (or dearth) of objects at these masses, there are three possible explanations.
First, this could be caused by an observational bias against detecting such systems.
Lower-mass black holes produce weaker X-ray emission, and the disk modelling used to infer their masses can bias those masses upwards; \cite{Kreidberg2012} showed that this accounts for part of the gap in XRBs, but not for all of it.
Gravitational-wave catalogues carry a well-understood bias against the lowest masses \cite{Fishbach2017}, but this that can (and is) corrected by LVK when they report rates.

Second, black holes in this mass range may not form at all.
This is the supernova picture from Section~\ref{sec:day1}: successful explosions leave low-mass neutron stars, while failed explosions retain their outer core layers and produce black holes of $\gtrsim 6\Msun$ \cite{SchneiderPodsiadlowskiLaplace2023, Kochanek2014}
However, uncertainties in fallback can smooth the gap away again \cite{Fryer2012}.

Third, the gap may be an evolutionary selection effect: BHs at these masses do form, but do not end up in the systems we observe.
Such a bias is again expected for both XRBs and double compact object's:
LMXBs are difficult to form to begin with, and merging double compact objects comprise only the very tail of the full binary population (roughly $1\Msun$ per $10^{6}\Msun$ of star formation ends up in a merging double compact object system \cite{vanSon2025}).

The GW data have since shown that any such gap is at least not empty.
GW230529 has a primary mass in the middle of the proposed gap \cite{GW230529}. 
Earlier events had approached the gap from the secondary-mass side (the secondary of GW190814, GW190917 and GW200210), but none was conclusive on its own.
The GWTC-5 rate estimates now rule out a completely empty lower mass gap \cite{GWTC5_pop}.
%
Similar claims have emerged on the EM side. 
Although many of the earlier candidate black holes turned out to be stripped stars, so-called ``BH impostors'' \cite{2020A&A...639L...6S,2022MNRAS.511L..24E,El-Badry2022_unicorns}, multiple candidates have nonetheless survived scrutiny, including the millisecond-pulsar companion in PSR J0514$-$4002E \cite{Barr2024}, the $\sim 3.3\Msun$ companion of 2MASS J05215658$+$4359220, and, very recently, oMEGACat BH-2 \cite{Whitaker2026}.

The EM samples nevertheless continue to show a paucity \cite{ElBadry2024_GaiaNSs}. 
The question has shifted from whether the gap is empty to how sparsely it is populated, and whether that sparsity reflects an evolutionary bias, a genuine preference of the supernova mechanism, or both.
That GWs find gap objects while Gaia binaries and XRBs do not is itself informative, since the two probe different formation histories.
The lower mass gap is thus a good example of a long-standing problem that is being resolved by combining constraints from both EM and GW.

\subsection{The global peak of the BBH mass function }\label{sec:peak10Msun}
The global maximum of the BBH mass function sits at a primary mass of $\sim9-10\Msun$ \cite{GWTC5_pop}.
The exclusion of many of the formation channels at the lowest masses as we discussed in Section~\ref{sec:day4nsns} is blurrier here than at neutron-star masses.
It is unclear whether black holes receive natal kicks \cite{Janka2013,2020MNRAS.499.3214M}, which brings triples back into play \cite{Antonini2017,Stegmann2022,Martinez2022}, and cluster models have been shown to be able to produce peaks at low BBH masses as well \cite{ClaireYe2026,2026arXiv260614846O}.
Explosion physics and isolated binary evolution are nonetheless likely to dominate in this mass range, and the two leading explanations for the peak come from exactly those two directions.

The first is that the peak is a supernova feature.
The compactness of a star at the onset of core collapse is not a monotonic function of its mass \cite{2011ApJ...730...70O,2012ApJ...757...69U,2014ApJ...783...10S,2018ApJ...860...93S,2020ApJ...890...43C,Farag2022}: models of both single and binary-stripped stars show peaks in compactness \cite{SchneiderPodsiadlowskiLaplace2023,2025A&A...695A..71L,2021A&A...645A...5S,2024A&A...686A..45S}.
Where those peaks sit, and how sharp they are, depends on the treatment of convective boundary mixing, carbon burning and the size of the nuclear reaction network \cite{2016ApJS..227...22F,2024A&A...682A.123T,2024RNAAS...8..152R}.
%
Since high compactness disfavours a successful explosion (likely) leading to black-hole formation, while low compactness leads to an explosion and (likely) a neutron star \cite{2011ApJ...730...70O,2016MNRAS.460..742M,2016ApJ...818..124E,2025A&A...700A..20M}, structure in compactness imprints structure on the remnant-mass function \cite{Kochanek2014,2020MNRAS.499.2803P,2020MNRAS.499.3214M}.
Several studies have since propagated this into full populations and tested it against the GW data \cite{Disberg2023,2025A&A...694A.186G,2026arXiv260525994G,2025arXiv251007573W,2025arXiv250820787W,2026arXiv260401420L}.

The second proposed theory is that the peak is an evolutionary selection effect of the stable mass-transfer channel, following the same low-mass exclusion argument used in Sections \ref{sec:day3smt} and \ref{sec:day4nsns}.
The minimum black-hole mass of the channel acts as a truncation, and a truncated power law naturally produces a peak just above the cut, at $\sim 9\Msun$ \cite{vanSon2022_no_peaks,vanSon2022_rate_redshift}, close to the rise seen in the GW data \cite{Farah2021}.
Even if mass-transfer stability is treated differently, the channel still imposes boundaries in mass and separation \cite{Klencki2026}, making this a more robust prediction.

The same argument predicts \textit{two} sub-populations rather than one, caused by what is known as mass-ratio reversal (MRR) \cite{2022ApJ...938...45B,2022MNRAS.517.2738M}.  This means that the black hole from the initially less massive star can become the more massive black hole in the binary. In the stable mass transfer scenario, the minimum mass differs for black holes originating from the initially more massive star as opposed to those originating from the initially less massive star, thus creating two peaks in the mass and mass ratio functions.
The non-reversed population then contributes to the peak near $10\Msun$ (with mass ratios extending to lower values) and the reversed one near $15$--$20\Msun$ (with more equal masses) \cite{ChenBreivikvanSon2026,Maclean2026arXiv260628506M,2026arXiv260901784G,2026arXiv260521580S}.
Because the two differ in which black hole forms first, they should separate in spin and redshift as well as in mass \cite{2022ApJ...938...45B,2024A&A...689A.305O}, which is what makes the prediction falsifiable and what comparisons with the GW data have begun to test \cite{2023ApJ...958...13A,2024ApJ...964L...6A,2025arXiv251105316T,2026arXiv260700565F,2026arXiv260523083G}.
How much weight such spin-based inferences can carry does, however, still heavily depend on the population model assumed \cite{2026ApJ..1007L..45W}.

Most likely both mechanisms contribute, and their relative importance remains an open question: compactness alone does not account for the full drop in rate between $\sim10\Msun$ and $\sim15\Msun$ \cite{2026arXiv260401420L}, and the most recent works combine compactness peaks with formation-channel components \cite{2026arXiv260525994G}.


\subsection{The $\sim$35$\Msun$ feature}\label{sec:35peak}
The $\sim 35\Msun$ feature is arguably the most-studied of the BBH features, in part because it was among the first to emerge.
In the first two GW transient catalogues the favoured explanation was a pulsational pair-instability pile-up (see also Section~\ref{sec:PISN}): PPISN mass loss produces a turnover in the relation between core mass and remnant mass, and the resulting degeneracy piles black holes up just below the pair-instability gap e.g., \cite{Stevenson2019,Woosley2020,Farmer2019}.
The shape of that turnover matters: a flat turnover produces a strong pile-up, a sharp one only a knee or a drop.

However, stellar models never placed the onset of the pair-instability gap at $35\Msun$.
The dominant uncertainty setting the location of the PISN gap is the uncertain $^{12}\mathrm{C}(\alpha,\gamma)^{16}\mathrm{O}$ rate \cite{deBoer2017}. 
More carbon, at the expense of oxygen, shields the core during explosive oxygen burning and pushes the gap edge up, while more oxygen brings it down, but only to $\sim 45\Msun$ \cite{Farmer2019,Farmer2020,Shen2023}.
With updated rates and converged models the edge has moved up rather than down, towards $\sim 50$--$60\Msun$ \cite{Farag2022,Mehta2022,Winch2024}, and explaining the $35\Msun$ feature with PPISN appears to be inconsistent with the nuclear reaction rates \cite{2024ApJ...976..121G}.
Shifting the rate far enough to engineer a $35\Msun$ pile-up creates a second problem: the implied PISN and PPISN rates come into tension with the observed rate of hydrogen-poor superluminous supernovae \cite{Hendriks2023}.

Multiple alternative explanations have been proposed for the $\sim 35\Msun$ feature, including chemically homogeneous evolution \cite{2023MNRAS.520.5724B}, hierarchical mergers \cite{TiwariFairhurst2021,Mahapatra2025}, and formation in globular clusters \cite{AntoniniGieles2020,Antonini2023}.
A further possibility is that the relevant mass scale is not the maximum black-hole mass but the \textit{onset} of pulsations, which sits closer to $35\Msun$ \cite{vanSon2022_rate_redshift}: the radial expansion driven by the pulsations can trigger mass transfer in a binary, and would act exactly where the feature is observed.

Some explanations do appear to be preferred over others. 
The hierarchical-merger explanation is for example disfavoured by the mass ratio and effective spin distribution \cite{Roy2025}.
Some subpopulation studies have found that the events making up the $35\Msun$ feature prefer equal masses and carry a $\chi_{\rm eff}$ distribution that is narrow and symmetric about zero \cite{2026ApJ..1005L..54S,2026ApJ..1005L..55R,Banagiri2026}, which could be an indication of a dynamical origin (e.g., clusters).
\cite{2026ApJ..1005L..54S} find that this subpopulation can be made up \textit{exclusively} of globular-cluster binaries provided black hole birth spins lie in the range $0.1$--$0.2$. 
However, the statistical significance is still low, and the distribution could still be consistent with a $\chi_{\rm eff}$ centered at small but positive values \cite{Roy2025,2026arXiv260722011G}.
%
How much of this structure is robust to the population model at all is itself contested \cite{2026arXiv260516510D,2025arXiv250819208M,2026arXiv260306566A,Maclean2026arXiv260628506M}, and cleanly separating clusters from e.g., field triples will likely require eccentricity measurements from next-generation detectors \cite{2026MNRAS.545f1938D}.
Though some channels seem more likely than others, for now the origin of binary black holes in their mid-thirties remains unsolved \cite{Roy2025}.


\subsection{The pair-instability gap and the highest masses}\label{sec:highmassgap}
The pair-instability mechanism is expected to produce a gap in the black-hole mass function, between roughly 45 and 120$\Msun$ in black-hole  masses.
Although we are not sure \textit{where} exactly the gap begins (see previous section), it is quite certain that pair pulsations \textit{should} occur at all (see the review by \cite{RenzoSmith2024}, and references therein).

Yet, the GW data rule out a completely \textit{empty} gap between $\sim 40-100\Msun$ \cite{GWTC5_pop}.
This is not completely surprising: since the first detections in or above the presumed location of the PISN gap, many mechanisms have been proposed to populate it, ranging from super-Eddington accretion \cite{vanSon2020}, to hierarchical mergers in either globular clusters or AGN disks \cite{Rodriguez2019,Yang2019} or oversized envelopes combined with fallback \cite{DiCarlo2019,DiCarlo2020,Renzo2020}.
Finding the edges of the PISN gap has thus changed from finding very sharp cut-offs in the mass function to finding the transition between populations with different formation channels.

Hierarchical mergers have received the most attention, as they are a natural consequence of the dynamical channels discussed in Section~\ref{sec:day3dyn}.
Moreover, as discussed in Section~\ref{ss: hierarchical mergers}, hierarchical mergers give very specific predictions for the effective spin distribution due to the hierarchical mergers being about twice as massive as the first generation, having component spins of $\sim 0.7$, and being isotropically oriented. 
Recent searches for such a distinct spin population at high mass have indeed found evidence for a transition: above roughly $45$--$50\Msun$ the $\chi_{\rm eff}$ distribution becomes broad and symmetric about zero, as expected for dynamically assembled mergers, and the transition mass lies much closer to the predicted PISN edge than to $35\Msun$ \cite{Antonini2025a}.
The same transition has been recovered independently, together with a steep drop in the secondary-mass distribution setting in at a comparable mass \cite{Tong2026}.

There are reasons for caution, however.
A model constructed to contain a gap will return one, a shallower decline in the secondary mass fits the current data just as well, and the mass ratios of the high-mass systems are not peaked near $q \sim 0.5$, as would be expected for second-generation black holes paired with first-generation companions (2G$+$1G) \cite{RayKalogera2026}.
The latest catalogue analyses likewise recover a broad mass-ratio distribution for the high-mass subpopulation \cite{GWTC5_pop}.
The mass-ratio distribution alone therefore neither confirms nor excludes such mergers.

Many studies have tried to use similar arguments to `measure' the location of the PISN gap from GW observations.
Flexible fits to the mass spectrum place it at $\sim 44\Msun$, with uncertainties of several solar masses, and some analyses go as far as returning the $^{12}\mathrm{C}(\alpha,\gamma)^{16}\mathrm{O}$ $S$-factor itself as a fit parameter \cite{2025arXiv250904637A,Tong2026,2025PhRvD.112f3040A}.
How much to trust any single number is another matter: inferred edge locations spread from roughly $45$ to $85\Msun$ depending on the model adopted \cite{AfrozMukherjee2025,Edelman2021,Ulrich2024}. 
With sufficiently flexible models the evidence for a sharp drop-off anywhere in the forties weakens considerably \cite{RayKalogera2026}.
However, the prospects for a clean measurement are slowly improving with catalogue size \cite{2026PhRvD.113j3021M}.

All of this concerns the \textit{lower} edge of the gap, simply because that is where most of the detections are.
A few systems have nonetheless been argued to lie (partially) \textit{above} the PISN gap, including GW190521 \cite{GW190521}, and GW231123 \cite{GW231123_2025}. 
%
The latter is the most compelling example: with component masses of $\sim137$ and $\sim103\Msun$ and a total mass of $190$--$265\Msun$, both components may sit in or above the gap.
It is unusual in another respect as well, since both components are rapidly spinning.
Whether it represents regime IV of Section~\ref{sec:PISN}, repeated hierarchical growth, or something else entirely is unresolved.

\subsection{Closing}
Gravitational-wave astronomy has moved from detection to population science in a single decade.
The BBH mass function has revealed multiple features, and those features are beginning to acquire (tentative) sub-population identities in mass ratio, spin, eccentricity and delay time.
At the time of writing, each of these is a battleground between formation channels, whose physics we reviewed in these lectures: mass-transfer stability, common envelopes, pair instability and dynamical assembly.

That is also what makes this an exciting field to enter, since we can expect a great deal of progress on these questions over the next decade.
The open questions we have walked through are exactly the problems that the next generation of astrophysicists, i.e.\ you, will help solve.

\chapter{Motion}
\chapterauthor{Julio Parra-Martinez}
\label{ch:part2}
\allowdisplaybreaks
\minitoc
\usetikzlibrary{arrows.meta,fadings,patterns,positioning,intersections}
\tikzfading[name=blob fade,inner color=transparent!0,outer color=transparent!100]
\definecolor{mapPN}{RGB}{60,105,175}
\definecolor{mapSF}{RGB}{55,135,90}
\definecolor{mapNR}{RGB}{215,140,45}
\providecommand{\cA}{{\mathcal A}}
\providecommand{\cF}{{\mathcal F}}
\providecommand{\cG}{{\mathcal G}}
\providecommand{\cM}{{\mathcal M}}
\providecommand{\cO}{{\mathcal O}}
\providecommand{\es}[2]{\begin{equation} \label{#1} \begin{split} #2 \end{split} \end{equation}}
\providecommand{\ta}{\widetilde{a}}
\providecommand{\vect}{\bm}

\section{Introduction}
\label{sec:motion:introduction}
Physics is supposed to be cumulative. We dig deeper, our theories grow
more sophisticated, and the problems that once defeated us become textbook
exercises. The problem of motion of $N$ bodies somehow never got the message.
Off to a good start, Newton handed us the two-body problem already solved: two
masses, one inverse-square
force, a relative coordinate tracing a Kepler ellipse that closes on itself. The
$N\geq3$-body problem proved far less obliging. Then, after many years of
struggle, quantum mechanics arrived to
explain that we may not, after all, say where a body is and how fast it is
moving at the same time. Quantum field theory went one better and revealed
that it is no longer clear how many bodies there are at all (particles being
created and annihilated at the slightest provocation), nor even that the empty
space between them is empty. Three centuries of deepening our most fundamental
account of motion have carried the problem of motion from a closed-form
solution to a state of affairs in which we are unsure where the bodies
are, what they are made of, or how many of them there are. So much for
progress.

Relativity offers a way back up, since for the large and heavy objects of
astrophysics, such as black holes and neutron stars, quantum effects are
negligible. The question therefore returns to the classical two-body problem,
now made more interesting by relativity and, in particular, by the presence of
radiation. That
sharpened problem is hard, but it is well posed, and in the last decades we have
learned to solve it to great accuracy. We
know this because we have begun to listen.

On 14 September 2015, the two LIGO detectors recorded a signal lasting only a
few tenths of a second: a gravitational wave whose frequency and amplitude rose
together, reached a peak, and rapidly faded~\cite{LIGOScientific:2016aoc}.
 The shape of this \emph{chirp} encoded the final orbits and merger of two black
 holes, each with a mass of roughly thirty Suns, moving at a sizable fraction of
 the speed of light. To interpret such a signal, we must be able to predict its
 \emph{waveform}. That is, to answer the question: given two compact bodies
 bound by their mutual gravity, how do they move, and how do they radiate?

In Newtonian gravity, the motion reduces to the integrable Kepler problem. In
General Relativity, the orbit precesses, and gravitational radiation carries
away energy and angular momentum and drives the binary toward merger. There is
no general closed-form solution, and no single approximation covers the entire
problem. Rather, the appropriate method to solve for the motion depends on the
binary's separation, velocity, and mass ratio. At large separations, the field
is weak and the motion can be treated perturbatively. Similarly, when one body is much lighter than the other, the extreme mass ratio of the two masses provides a different perturbative
parameter. Near merger, the field becomes strong and we must turn to computers
to solve Einstein's equations.

These four lectures introduce the principal ideas and methods behind our modern
description of this relativistic motion and radiation. The lectures
are aimed at master's and beginning Ph.D. students, and assume only a first
course in general relativity and field theory. We take a somewhat unorthodox
route: rather than organizing the problem around solving Einstein's equations
for the spacetime metric and the bodies' trajectories, we will focus as directly
as possible on gauge-invariant observables and the symmetries governing the
problem. This approach, combining effective field theory with ideas from
scattering theory, has been very productive in recent years. In Lecture~1, we
will introduce the parameters of the problem, examine the basic anatomy of a
gravitational-wave signal, and map the parameter space of the problem and the
methods used to solve it in each corner. We will then focus on
effective-field-theory methods for the analytically tractable regions of this
space. In Lecture~2, we will approximate compact bodies by point particles and
study the validity of this approximation. In Lecture~3, we will use scattering of particles to understand the conservative two-body dynamics and bound-orbit observables at
leading order in the weak field. Finally, in Lecture~4, we will incorporate
radiation.

\section{Mapping the two-body problem}
Our goal in this first lecture is to build a map of the two-body problem and
understand the basics of the various methods used to solve it. First of all we
need to find coordinates on the parameter space of the problem.
The most obvious are the masses of the two bodies, $m_1$
and $m_2$. It is convenient instead to define the total and reduced masses,
\be
M = m_1 + m_2\,, \qquad \mu = \frac{m_1 m_2}{m_1+m_2}\,,
\ee
and the \emph{symmetric mass ratio}
\be
  \nu \;\equiv\; \frac{\mu}{M} \;=\; \frac{m_1 m_2}{(m_1 + m_2)^2}\,,
   \label{eq:lec1:nu}
\ee
which runs from $\nu = \tfrac14$ for equal masses down to $\nu = 0$ in the
test-mass limit, where one body is infinitely lighter than the other.

The motion itself is characterized by the relative orbital separation $r = |\bm
r_1 - \bm r_2|$ and velocity $v = |\bm v_1 - \bm v_2|$. From these and
the speed of light, $c$, we can build two dimensionless numbers,
\be
  \frac{v}{c}
  \qquad\text{and}\qquad
  \frac{GM}{r\, c^2} ,
  \label{eq:lec1:axes}
\ee
the first measuring how relativistic the motion is, the second how strong
(equivalently, how curved) the gravitational field is at the orbital scale.

These would be independent expansion parameters for general motion. For a
bound orbit, however, they are tied together. Balancing the centripetal
acceleration against the Newtonian attraction, $v^2 / r \sim GM/r^2$, gives
the virial relation,
\be
  \frac{v^2}{c^2} \;\sim\; \frac{GM}{r\, c^2} ,
  \label{eq:lec1:virial}
\ee
so that weak field and slow motion are, on a bound orbit, governed by a single
small parameter.

Consider instead an unbound encounter: two bodies that
fall in from infinity, deflect through some angle, and recede back to
infinity. Now the velocity at infinity $v$ and the impact parameter $b$ are
independent pieces of initial data, and nothing forbids an ultrarelativistic
flyby, $v \to c$, in a weak field, $GM/(b\,c^2) \ll 1$. The two axes are
therefore independent for scattering.

\subsection{The Keplerian limit and its breakdown}
\label{sec:lec1:kepler}

In the weak-field and non-relativistic limit the two-body problem reduces to its Newtonian counterpart; let us recall why that problem was solvable.

Pass to the relative coordinate $\bm r = \bm r_1 - \bm r_2$, so that the
two bodies become a single fictitious particle of reduced mass, $\mu$, moving in
the central potential $V(r) = -G M \mu / r$. The motion has three degrees of
freedom and hence a six-dimensional phase space $(\bm r, \bm v)$. Fortunately,
the Kepler problem comes
with a large number of conserved quantities. The force does not depend on time,
so the energy,
\be 
E = \tfrac12 \mu\, v^2 - \frac{G M \mu}{r}\,,
\ee
is of course conserved. As is the angular momentum,
\be
\bm L = \mu\, \bm r \times \bm v\,,
\ee
because the force is central. Its direction pins the orbit to a fixed plane, and
its magnitude sets the centrifugal barrier that
keeps the particle off the center. These two are generic: \emph{any} central
force conserves them, and any central force is therefore integrable in the
sense of Liouville, with three integrals ($E$, $|\bm L|$, and one component
of $\bm L$) in involution. 

Generic integrability, however, does not suffice to
close an orbit. A generic central potential gives a rosette-like trajectory: the
particle
oscillates between periastron and apoastron while the line of apsides slowly
turns, densely filling an annulus. That Kepler's orbits nonetheless close is
peculiar to the $1/r$
potential, and it follows from one more conserved quantity: the
Laplace--Runge--Lenz vector,
\be
  \bm A \;=\; \mu \bm v \times \bm L \;-\; G M \mu^2\, \hat{\bm r}\,,
  \label{eq:lec1:lrl}
\ee
which is conserved only for the exact inverse-square force. It points from the
center toward periastron and its length is proportional to the eccentricity,
so conserving it freezes both the shape of the orbit and its orientation. The
seven components of $(E, \bm L, \bm A)$ obey
two relations, $\bm A \cdot \bm L = 0$ and
$A^2 = (G M \mu^2)^2 + 2\mu E\, L^2$, leaving five independent constants on
a six-dimensional phase space. Five constraints restrict the motion to a
closed curve, leaving no freedom for the orbit to precess. The associated
hidden $SO(4)$ symmetry makes the Kepler problem superintegrable.

Relativistic effects break this structure in a definite order. The
Laplace--Runge--Lenz vector is conserved only for the exact $1/r$ law,
so the first relativistic correction to the potential, which we will meet
later, sets $\bm A$ turning. The correction is conservative: it drains
no energy from the orbit but merely rotates $\bm A$, by
\be
  \frac{|\Delta \bm A|}{|\bm A|} \;\sim\; \left(\frac{v}{c}\right)^{2}
  \qquad \text{per orbit} ,
  \label{eq:lec1:dA}
\ee
so the periastron advances and the ellipse no longer closes, an effect
Einstein famously computed for Mercury.

Energy and angular momentum are conserved by every conservative
correction, and only radiation, carrying them off to infinity, changes
them. The rate follows from a parametric pass through Einstein's quadrupole
formula~\cite{Einstein:1918btx}, which gives the power radiated in the form of
gravitational waves as
\[
  P \sim \frac{G}{c^5}\,\langle\dddot{Q}_{ij}\,\dddot{Q}_{ij}\rangle ,
\]
in terms of the binary's mass quadrupole, $Q_{ij}$, with the precise prefactor
left to Lecture~4. The quadrupole scales as $Q_{ij}\sim\mu\,r^2$ and each time
derivative brings down a factor of the orbital frequency $\Omega\sim v/r$, so
$\dddot{Q}\sim\mu v^3/r$ and hence
\be
P\sim \frac{G\mu^2 v^6}{r^2c^5}\,.
\ee
The energy radiated in one orbit is this power times the period,
$\Delta E = P\, t_{\text{orbital}}$ with $t_{\text{orbital}}\sim r/v$; weighed
against the binding energy $|E|\sim GM\mu/r\sim\mu v^2$, the factors of $G$
and $r$ cancel, leaving
\be
\frac{\Delta E}{|E|}   \;\sim\;
  \nu \left(\frac{v}{c}\right)^{5}
  \qquad \text{per orbit} .
  \label{eq:lec1:adiabatic}
\ee
A similar exercise shows that angular momentum changes at the same rate, as
Peters and Mathews worked out in
detail~\cite{Peters:1963ux,Peters:1964zz}. The
odd powers of $v/c$ in \eqref{eq:lec1:adiabatic} are the hallmark of radiation
reaction, and the factor of $\nu$ marks the loss as a two-body effect: a test
particle ($\nu \to 0$) radiates negligibly and would orbit forever.

\subsection{The parameter space and methods}
\label{sec:lec1:param-space}

\begin{figure}[t]
  \centering
\begin{tikzpicture}[x=1cm,y=1cm,
    lab/.style={font=\small,align=center,inner sep=1.5pt},
    ax/.style={font=\footnotesize,inner sep=1.5pt}]

  \fill[mapPN,fill opacity=0.30] (0,3.8) rectangle (11,5);
  \fill[mapPN,fill opacity=0.30,path fading=south] (0,2.6) rectangle (11,3.8);

  \fill[mapSF,fill opacity=0.32] (9.5,0) rectangle (11,5);
  \fill[mapSF,fill opacity=0.32,path fading=west] (5.8,0) rectangle (9.5,5);

  \begin{scope}
    \clip (0,0) rectangle (11,5);
    \fill[mapNR,fill opacity=0.52,path fading=blob fade] (-3.8,-1.8) rectangle
    (3.8,1.8);
  \end{scope}

  \node[lab] at (4.3,4.4) {post-Newtonian / post-Minkowskian};
  \node[lab,rotate=90] at (10.35,2.5) {self-force};
  \node[lab] at (1.9,0.75) {numerical relativity};

  \draw[thin] (0,0) rectangle (11,5);
  \node[ax,anchor=north west] at (0.05,-0.1) {$\nu=\tfrac14$};
  \node[ax,anchor=north east] at (10.95,-0.1) {$\nu\to0$};
  \node[ax,anchor=north] at (5.5,-0.62) {$\nu$};
  \node[ax,rotate=90] at (-0.3,4.2) {weak};
  \node[ax,rotate=90] at (-0.3,0.8) {strong};
  \node[ax,rotate=90] at (-0.85,2.5) {$rc^2/GM$};
\end{tikzpicture}
  \caption{A map of methods in the plane of mass ratio and field strength. No
  domain ends at a line: each fades out as its expansion parameter stops being
  small, and where two overlap either method applies.}
  \label{fig:lec1:methods}
\end{figure}
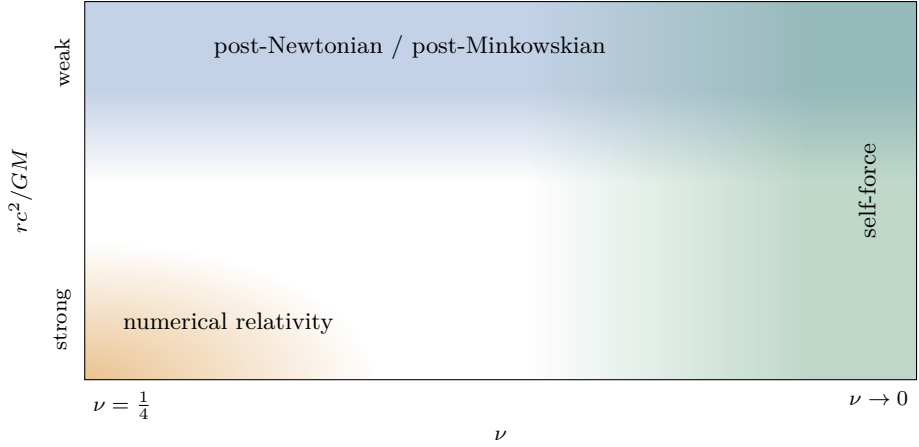

We are finally in a position to draw a map of the two-body problem and explain
which methods are used to solve it in each region. Each available method expands
about a different solvable limit. Near such a
limit, the conserved quantities are perturbed only slightly and their drift
can be computed order by order. The relevant limits occupy different
boundaries of parameter space. Figure~\ref{fig:lec1:methods} displays them
in the plane of mass ratio and field strength.

The most symmetric limit is the one with no relativity at all: at
$c \to \infty$ the problem reduces to the Kepler problem of
Sec.~\ref{sec:lec1:kepler}, superintegrable, with $E$, $\bm L$, and $\bm A$ all
conserved and the orbit closed.
\emph{Post-Newtonian} (PN) theory expands about this point, writing the
metric as
\be
  g_{\mu\nu} = \eta_{\mu\nu} + h_{\mu\nu} ,
  \qquad h \sim \frac{GM}{rc^2} \sim \frac{v^2}{c^2} ,
  \label{eq:lec1:pn}
\ee
and solving Einstein's equations iteratively~\cite{Blanchet:2013haa}, order by
order in $(v/c)^2$; by
convention an $n$PN correction is one suppressed by
$(v/c)^{2n} \sim (GM/rc^2)^n$ relative to the Newtonian result. Since the
unperturbed motion conserves everything, the corrections break the
conservation laws only slightly, in the order established in
Sec.~\ref{sec:lec1:kepler}: the periastron precession \eqref{eq:lec1:dA} is the
leading, first post-Newtonian (1PN) correction, and radiation reaction
\eqref{eq:lec1:adiabatic}, at $(v/c)^5$, enters at order 2.5PN. PN theory
covers the weak-field band at any mass
ratio (large separations, the top of Figure~\ref{fig:lec1:methods}) and
breaks down when $v\to c$ near merger.

\emph{Post-Minkowskian} (PM) theory occupies the same weak-field band, organized
along a different direction. For unbound encounters the velocity is an
independent piece of initial data, not something one may expand in, so the
expansion is organized about free trajectories at $G \to 0$
instead, in powers of $G$ at fixed velocity; an $n$PM correction carries $n$
powers of $G$. 

The two expansions truncate the same double series in $G$ and $v/c$ along
different lines (Table~\ref{tab:lec1:pnpm}). A correction
$(GM/rc^2)^{k-1}(v/c)^{2j}$ to the Newtonian interaction carries $k$ powers of
$G$ and sits at PN order $n=k+j-1$, so a fixed PN order collects contributions
from
$n+1$ successive PM orders, while a fixed PM order contributes one term to every
PN order at once.

A different integrable limit sits at the opposite edge of the map where the mass ratio is sent to zero. The light body becomes a test mass on a geodesic of the
\emph{exact} spacetime of its heavy companion, at no cost in field strength or
velocity: the background is the full nonlinear black-hole geometry. This limit
is integrable too: a geodesic conserves its energy and angular momentum from
the time-translation and rotational symmetries of the metric. It is not
superintegrable (a Schwarzschild geodesic precesses, so $\bm A$ is not
conserved), but it stays on a torus and is solvable.

The expansion about this geodesic is controlled not by $\nu$ but by the plain
mass ratio,
\be
  q \;\equiv\; \frac{m_2}{m_1} \;\leq\; 1 ,
  \qquad \nu = \frac{q}{(1+q)^2} ,
  \label{eq:lec1:q}
\ee
with $m_2$ the lighter body, the two coinciding, $\nu \simeq q$, in exactly the
corner where the expansion applies. \emph{Gravitational self-force} theory
expands in powers of $q$,
\be
  g_{\mu\nu} = g^{(0)}_{\mu\nu} + q\, h^{(1)}_{\mu\nu}
  + q^2 h^{(2)}_{\mu\nu} + \dots ,
  \label{eq:lec1:sf}
\ee
with $g^{(0)}$ the Schwarzschild metric of the heavy body; at $O(q)$ the light
body modifies the metric, and the perturbation, $h^{(1)}$, can be computed using
techniques from \emph{black-hole perturbation theory}.  In turn the light body
feels the \emph{self-force} of its own field
$h^{(1)}$~\cite{Mino:1996nk,Quinn:1996am,Detweiler:2002mi,Poisson:2011nh,Barack:2018yvs}.
Valid at any field strength, it is the method for studying the
extreme-mass-ratio inspirals LISA will observe~\cite{LISA:2017pwj}.

\begin{table}[t]
  \centering
{\small\setlength{\tabcolsep}{12pt}%
\arrayrulecolor{black!25}%
\newcommand{\pnpmw}[1]{\makebox[48pt]{#1}}%
\newcommand{\pnpmstrut}{\rule[-12.5pt]{0pt}{34pt}}%
\newcommand{\pnpmhstrut}{\rule[-5pt]{0pt}{17.5pt}}%
\begin{tabular}{c|c|c|c|c|c}
\pnpmw{\pnpmhstrut} & \pnpmw{0PN} & \pnpmw{1PN} & \pnpmw{2PN} & \pnpmw{3PN}
    & \pnpmw{$\cdots$}\\
\hline
\pnpmw{\pnpmstrut 1PM} & $1$ & $\displaystyle\frac{v^{2}}{c^{2}}$
    & $\displaystyle\frac{v^{4}}{c^{4}}$
    & $\displaystyle\frac{v^{6}}{c^{6}}$ & $\cdots$\\
\hline
\pnpmw{\pnpmstrut 2PM} & & $\displaystyle\frac{GM}{rc^{2}}$
    & $\displaystyle\frac{GM}{rc^{2}}\,\frac{v^{2}}{c^{2}}$
    & $\displaystyle\frac{GM}{rc^{2}}\,\frac{v^{4}}{c^{4}}$ & $\cdots$\\
\hline
\pnpmw{\pnpmstrut 3PM} & & & $\displaystyle\frac{G^{2}M^{2}}{r^{2}c^{4}}$
    & $\displaystyle\frac{G^{2}M^{2}}{r^{2}c^{4}}\,\frac{v^{2}}{c^{2}}$ &
    $\cdots$\\
\hline
\pnpmw{\pnpmstrut 4PM} & & & & $\displaystyle\frac{G^{3}M^{3}}{r^{3}c^{6}}$ &
$\cdots$\\
\hline
\pnpmw{\pnpmhstrut $\vdots$} & & & & & $\ddots$\\
\end{tabular}}
  \caption{Corrections to the Newtonian interaction $-Gm_1m_2/r$, arranged by
  post-Minkowskian order (rows) and post-Newtonian order (columns). Only
  integer PN orders are displayed; radiation reaction gives also half-integer PN orders, starting at 4PN.}
  \label{tab:lec1:pnpm}
\end{table}

The weak-field expansions and self-force theory expand about different limits
in different parameters: PN and PM in field strength at any mass ratio,
self-force in mass ratio at any field strength. On the map the first is
therefore a horizontal band and the second a vertical one, overlapping in the
weak-field, extreme-mass-ratio corner.

Near \emph{neither} limit (comparable masses and strong field at once, the
merger and plunge) there is no integrable point to expand about and no small
parameter, and every conserved quantity changes by order one each orbit. The
only option is to solve the Einstein equations
numerically~\cite{Pretorius:2005gq,Campanelli:2005dd,Baker:2005vv}.
\emph{Numerical relativity} makes no approximation in the equations, but it has
a small
parameter of its own, the grid spacing $\Delta x$, and stability ties the time
step to the spacing through the
Courant condition
\be
  c\,\Delta t \lesssim \Delta x ,
  \label{eq:lec1:cfl}
\ee
so no signal crosses more than one cell per step. Because the grid must
resolve the smaller hole, the time step is driven down to a fraction of that
hole's light-crossing time, the shortest scale in the problem, while the
evolution must span the far longer inspiral. That gap makes numerical
relativity the only first-principles method for the comparable-mass binary
coalescence but
an expensive one; the cost climbs rapidly
as the mass ratio grows extreme. 

Where the domains overlap, results obtained in one expansion check those of
another: a weak-field, comparable-mass binary can be treated by PN theory or
by PM results continued to bound motion, and a weak-field, extreme-mass-ratio
inspiral by PN or self-force theory. One recent example is the
agreement between PM scattering angles and numerical-relativity simulations
of the same encounters~\cite{Damour:2014afa,Damour:2022ybd}.

No one method makes first-principles predictions over the whole parameter
space. Covering it is the job of \emph{waveform models}, which are often
phenomenological and tie together the predictions from all different corners.
Perhaps the most famous is the
\emph{effective-one-body} (EOB) framework: an analytic
resummation that maps the two-body dynamics to one particle in an effective
metric, calibrated against numerical
relativity~\cite{Buonanno:1998gg,Buonanno:2000ef}.

\subsection{The stages of a coalescence}
\label{sec:lec1:stages}

The parameters of the problem are not static: a bound binary
radiates, loses energy, and spirals inward, so that over its lifetime it
sweeps across the parameter space of Figure~\ref{fig:lec1:methods} towards the
strong-field/relativistic region. Its evolution passes through a long
\emph{inspiral} in which the bodies are far from each other, an often violent
\emph{merger}, when they coalesce, and a brief \emph{ringdown} in which the
final object relaxes to a stationary state. To understand these stages
qualitatively we analyze the relevant timescales, following the conserved
quantities of the Kepler problem as they cease to be conserved.

We first need a reference timescale, and the natural
one is the orbital period, which in the Keplerian limit is
\be
  t_{\text{orbital}} = 2\pi\sqrt{\frac{r^3}{GM}}
  = 2\pi\,\frac{GM}{c^3}\left(\frac{rc^2}{GM}\right)^{3/2} ,
  \label{eq:lec1:torb}
\ee
the dynamical time $GM/c^3$ stretched by the separation measured in
gravitational radii.

Dividing the period by the per-orbit drift \eqref{eq:lec1:dA} of the
Laplace--Runge--Lenz vector gives the time for the periastron to swing once
around,
\be
  t_{\text{prec}} \sim t_{\text{orbital}}\left(\frac{v}{c}\right)^{-2} .
  \label{eq:lec1:tprec}
\ee
Since $v/c \ll 1$ on the inspiral, the ellipse turns slowly: $t_{\text{prec}}$
is already long compared with the period.

Dividing instead by the per-orbit loss \eqref{eq:lec1:adiabatic} gives the
inspiral time,
\be
  t_{\text{inspiral}} \sim \frac{t_{\text{orbital}}}{\nu}
  \left(\frac{v}{c}\right)^{-5} ,
  \label{eq:lec1:tinsp}
\ee
three inverse powers of $v/c$ longer than $t_{\text{prec}}$, and longer still
by $1/\nu$. The orbit therefore shrinks much more slowly than it precesses:
$t_{\text{orbital}}\ll t_{\text{prec}}\ll t_{\text{inspiral}}$. This separation
allows us to treat the orbit as approximately conservative over many cycles
while following the slow radiative evolution of $E$ and $L$.

This separation of timescales closes as the orbit tightens. At separations of
a few $GM/c^2$, the velocity becomes relativistic and the field becomes
strong during the resulting merger and
plunge, so comparable-mass systems must be treated with numerical
relativity.

What remains is a single distorted black hole, which sheds its deformations as
a superposition of damped oscillations, the ringdown, and settles to a
stationary Kerr geometry. These are the remnant's \emph{quasinormal
modes}~\cite{Regge:1957td,Zerilli:1970se,Vishveshwara:1970zz,Teukolsky:1973ha,Berti:2009kk},
with frequency and decay rate fixed by the only scale left in the problem, the
light-crossing time of the final hole of mass $M_f$,
\be
  t_{\text{ringdown}} \sim \frac{GM_f}{c^3} .
  \label{eq:lec1:tring}
\ee
The deformations are small, so once more the dynamics is perturbative (the
regime of black-hole perturbation theory), and they damp away after just a few
cycles.

Ordered by duration, the four timescales are
\be
  t_{\text{ringdown}} \;\lesssim\; t_{\text{orbital}} \;\ll\; t_{\text{prec}}
  \;\ll\; t_{\text{inspiral}} .
  \label{eq:lec1:timescales}
\ee
This chain orders the timescales by magnitude, not by when they
occur. The three inspiral timescales $t_{\text{orbital}} \ll t_{\text{prec}} \ll
t_{\text{inspiral}}$ refer to the long early epoch, where $v/c \ll 1$ makes
the gaps wide. The ringdown comes last yet has the shortest timescale of all,
because by merger the separation has shrunk to $r \sim GM/c^2$ and
$t_{\text{orbital}}$ with it, down to the bare dynamical time $GM/c^3$; there
$t_{\text{ringdown}} \lesssim t_{\text{orbital}}$ is a near-equality rather
than a strong inequality.

\begin{figure}[t]
  \centering
\subfloat[\label{fig:lec1:waveforms-a}]{%
\begin{tikzpicture}[
    every node/.style={inner sep=1pt},
    ax/.style={-{Stealth[length=4pt]},thin},
    wave/.style={black,line width=0.4pt},
    env/.style={thin,black!45},
    dim/.style={thin,{Bar[width=4.5pt]}-{Bar[width=4.5pt]}},
    tlab/.style={font=\footnotesize,text height=1.6ex,text depth=0.7ex},
    guide/.style={thin,dotted,black!65}]
\newcommand{\wfAturns}[1]{5.039*(2.542-pow(4.45-#1,0.625))}%
\newcommand{\wfAamp}[1]{(0.23+0.63*pow(#1/4.0,2.2))}%
\newcommand{\wfAring}[1]{0.86*exp(-(#1-4.0)/0.26)*sin(360*mod(0.75+4.2*(#1-4.0),1))}%
\begin{scope}
  \draw[ax] (0,-1.45) -- (5.60,-1.45) node[below=3pt,font=\small] {$t$};
  \draw[ax] (0,-1.45) -- (0,1.15) node[left=3pt,font=\small] {$h$};

  \draw[env,samples=200,domain=0:4.0,variable=\x] plot (\x,{\wfAamp{\x}});
  \draw[env,samples=200,domain=0:4.0,variable=\x] plot (\x,{-(\wfAamp{\x})});

  \draw[wave,samples=600,domain=0:4.0,variable=\x]
        plot (\x,{\wfAamp{\x}*sin(360*mod(\wfAturns{\x},1))});
  \draw[wave,samples=300,domain=4.0:5.05,variable=\x] plot (\x,{\wfAring{\x}});

  \draw[guide] (4.0,-1.08) -- (4.0,0.95);
  \node[font=\footnotesize,anchor=base] at (4.0,-1.30) {merger};

  \draw[dim] (0.08,1.00) -- (3.50,1.00);
  \node[tlab,above=3pt] at (1.79,1.00) {$t_{\text{inspiral}}$};
  \draw[dim] (4.0,1.00) -- (4.26,1.00);
  \node[tlab,above=3pt] at (4.13,1.00) {$t_{\text{ringdown}}$};

  \draw[guide] (0.5433,0.0) -- (0.5433,-0.90);
  \draw[guide] (1.5469,0.0) -- (1.5469,-0.90);
  \draw[dim] (0.5433,-0.82) -- (1.5469,-0.82);
  \node[tlab,below=3pt] at (1.045,-0.82) {$t_{\text{orbital}}$};

\end{scope}
\end{tikzpicture}}
  \hspace{0.7cm}
  \subfloat[\label{fig:lec1:waveforms-b}]{%
\begin{tikzpicture}[
    every node/.style={inner sep=1pt},
    ax/.style={-{Stealth[length=4pt]},thin},
    wave/.style={black,line width=0.4pt},
    env/.style={thin,black!45},
    dim/.style={thin,{Bar[width=4.5pt]}-{Bar[width=4.5pt]}},
    tlab/.style={font=\footnotesize,text height=1.6ex,text depth=0.7ex},
    guide/.style={thin,dotted,black!65}]
\newcommand{\wfBenv}[1]{((0.34+0.51*#1/4.3)*(0.60+0.40*sin(360*3.25*#1/4.3)))}%
\newcommand{\wfBturns}[1]{13*#1/4.3+3*(#1/4.3)*(#1/4.3)}%
\newcommand{\wfBring}[1]{0.80*exp(-(#1-4.3)/0.26)*sin(360*mod(6.5*(#1-4.3),1))}%

\begin{scope}
  \draw[ax] (0,-1.45) -- (5.60,-1.45) node[below=3pt,font=\small] {$t$};
  \draw[ax] (0,-1.45) -- (0,1.15) node[left=3pt,font=\small] {$h$};

  \draw[env,samples=240,domain=0:4.3,variable=\x] plot (\x,{\wfBenv{\x}});
  \draw[env,samples=240,domain=0:4.3,variable=\x] plot (\x,{-(\wfBenv{\x})});

  \draw[wave,samples=620,domain=0:4.3,variable=\x]
        plot (\x,{\wfBenv{\x}*sin(360*mod(\wfBturns{\x},1))});
  \draw[wave,samples=240,domain=4.3:5.35,variable=\x] plot (\x,{\wfBring{\x}});

  \draw[guide] (4.3,-1.08) -- (4.3,0.90);
  \node[font=\footnotesize,anchor=base] at (4.3,-1.30) {plunge};

  \draw[guide] (0.3251,0.10) -- (0.3251,0.95);
  \draw[guide] (0.9444,0.10) -- (0.9444,0.95);
  \draw[dim] (0.3251,0.95) -- (0.9444,0.95);
  \node[tlab,above=3pt] at (0.635,0.95) {$t_{\text{orbital}}$};

  \draw[guide] (1.654,0.55) -- (1.654,0.95);
  \draw[guide] (2.977,0.71) -- (2.977,0.95);
  \draw[dim] (1.654,0.95) -- (2.977,0.95);
  \node[tlab,above=3pt] at (2.32,0.95) {$t_{\text{prec}}$};

\end{scope}
\end{tikzpicture}}
  \caption{Schematic strain $h(t)$ for a comparable-mass binary (a) and an
  extreme-mass-ratio inspiral (b), with the timescales of
  \eqref{eq:lec1:timescales} marked on the features that carry them. Neither
  panel is to scale, in time or in number of cycles.}
  \label{fig:lec1:waveforms}
\end{figure}
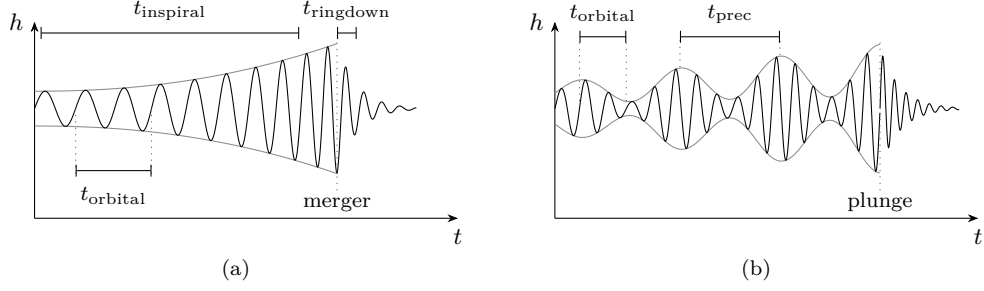

This hierarchy also explains the qualitative shape of the full
\emph{inspiral--merger--ringdown} signal (Figure~\ref{fig:lec1:waveforms}).
Write the waveform as $h(t)\cos\varphi(t)$. The phase $\varphi$ winds on the
orbital scale (at twice the orbital frequency, since the mass quadrupole
returns to itself after only half an orbit), while the amplitude $h(t)$ and the
frequency drift only on the far longer $t_{\text{inspiral}}$. The number of wave
cycles spent in a detector's band is therefore the ratio of the two,
$t_{\text{inspiral}}/t_{\text{orbital}} \sim \nu^{-1}(c/v)^5$: from tens for a
heavy black-hole binary to tens of thousands for a neutron-star binary, and as
large as $\sim\!10^5$ for an extreme-mass-ratio inspiral. A comparable-mass
binary (Figure~\ref{fig:lec1:waveforms-a}) has, moreover, circularized by the
time it
enters the band, so there is a single carrier, which sweeps smoothly upward
beneath a monotonically swelling envelope closed by a
ringdown of just a few cycles.

In an extreme-mass-ratio inspiral
(Figure~\ref{fig:lec1:waveforms-b}) the inspiral time is stretched by $1/\nu$
while the per-orbit precession is unchanged (it carried no factor of $\nu$), so
a light
body wheels through a great many precession cycles before it spirals in. On an
eccentric or inclined orbit the turning of $\bm A$ splits the carrier into a
comb of nearby lines, whose beats modulate the amplitude on the intermediate
scale $t_{\text{prec}}$; the long inspiral then carries an oscillating
envelope which persists until the smaller object plunges and is swallowed in a
blip.

To make this picture quantitative, we split the dynamics into a
\emph{conservative} part (time-symmetric, derivable from a potential or, for
scattering, summarized by the scattering angle) and a \emph{dissipative} part,
the radiation reaction that drains energy and angular momentum into the
gravitational waves measured at infinity. The timescale hierarchy
\eqref{eq:lec1:timescales} justifies the split: a \emph{fast} orbital phase,
winding once per period, is carried along by a \emph{slowly} drifting ellipse
whose energy $E$ and angular momentum $L$ creep along only on the inspiral
clock. Over a single orbit those labels barely move
($\Delta E/|E|\sim\nu(v/c)^5\ll1$ from \eqref{eq:lec1:adiabatic}), so to
leading order the motion is a fixed, conservative orbit of definite $E$ and
$L$, with $E$ and $L$ acting as adiabatic invariants that drift across a
sequence of such orbits. The conservative and dissipative sectors are not
decoupled, but this separation of timescales lets us compute them as
though they are.

The slow drift is then fixed by balance: whatever the waves carry to infinity
is lost by the orbit,
\be
  \frac{dE}{dt} = -\cF , \qquad \frac{dL}{dt} = -\cG ,
  \label{eq:lec1:balance}
\ee
with $\cF$ and $\cG$ the orbit-averaged fluxes of energy and angular momentum.
The balance itself is exact conservation; the \emph{adiabatic approximation}
consists of evaluating the fluxes on a conservative orbit held fixed during
that calculation. Both PN and self-force treatments of the inspiral organize
their
leading radiative evolution in this way.

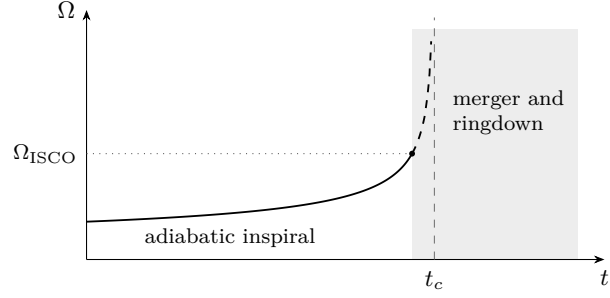
\begin{figure}[t]
  \centering
\begin{tikzpicture}[x=1cm,y=1cm,
    every node/.style={inner sep=1pt},
    ax/.style={-{Stealth[length=4pt]},thin},
    guide/.style={thin,dotted,black!65}]
  \fill[black!7] (4.305,0) rectangle (6.5,3.05);
  \draw[ax] (0,0) -- (6.85,0) node[below=3pt,font=\small] {$t$};
  \draw[ax] (0,0) -- (0,3.3) node[left=3pt,font=\small] {$\Omega$};
  \draw[thin,dashed,black!55] (4.6,0) -- (4.6,3.2);
  \node[font=\small,anchor=north] at (4.6,-0.12) {$t_c$};
  \draw[guide] (0,1.4) -- (4.305,1.4);
  \node[font=\footnotesize,anchor=east] at (-0.08,1.4) {$\Omega_{\text{ISCO}}$};
  \draw[line width=0.7pt,samples=200,smooth,domain=0:4.305,variable=\x]
    plot (\x,{0.8861*pow(4.6-\x,-0.375)});
  \draw[line
  width=0.7pt,dashed,samples=160,smooth,domain=4.305:4.5576,variable=\x]
    plot (\x,{0.8861*pow(4.6-\x,-0.375)});
  \fill (4.305,1.4) circle (1.1pt);
  \node[font=\footnotesize,anchor=north] at (1.9,0.46) {adiabatic inspiral};
  \node[font=\footnotesize,anchor=west,align=left]
    at (4.80,1.95) {merger and\\ ringdown};
\end{tikzpicture}
  \caption{The leading-order frequency sweep \eqref{eq:lec1:chirp},
  $\Omega\propto(t_c-t)^{-3/8}$, against time. Both the expansion in
  $v/c\sim(GM\Omega/c^3)^{1/3}$ and the two-timescale argument behind
  \eqref{eq:lec1:Omegadot} fail near the ISCO, while $\Omega$ is still finite:
  the curve is dashed beyond it, and cut off at the top of the frame rather
  than at $t_c$. The shaded region is where numerical relativity takes over.}
  \label{fig:lec1:chirp}
\end{figure}

The two ingredients of the split, a conservative orbit and a flux
computed on it, already fix the waveform, as we now show at the level of
scalings. Take the orbit to be quasi-circular, as a comparable-mass
binary has become by the time it radiates in band, so
that a single number labels it; we use the orbital frequency $\Omega$. The
conservative dynamics supplies the orbit's energy as a function of that label,
at leading (Newtonian) order,
\be
  E(\Omega) \sim -\,\mu\,(GM\Omega)^{2/3} ,
  \label{eq:lec1:EOmega}
\ee
and the quadrupole emission of Sec.~\ref{sec:lec1:kepler} supplies the flux.
Energy lost equals energy radiated, $dE/dt=-\cF$, which the chain rule turns
into an equation for how fast the frequency sweeps,
\be
  \frac{d\Omega}{dt} = -\,\frac{\cF(\Omega)}{dE/d\Omega} .
  \label{eq:lec1:Omegadot}
\ee
Putting in the leading scalings ($\cF\sim(G/c^5)\,\mu^2(GM)^{4/3}\Omega^{10/3}$
from the quadrupole formula, together with \eqref{eq:lec1:EOmega}), the two
masses enter only through the combination called {\it chirp mass}, $\cM$,
\be
  \frac{d\Omega}{dt} \;\sim\; \left(\frac{G\cM}{c^3}\right)^{5/3}\Omega^{11/3} ,
  \qquad
  \cM \equiv \nu^{3/5} M = \mu^{3/5} M^{2/5}\,,
  \label{eq:lec1:chirpmass}
\ee
which is therefore the best-measured parameter
of a low-mass signal. Integrating once gives the frequency sweep,
\be
  \Omega(t) \;\sim\; \frac{c^3}{G\cM}
  \left(\frac{c^3\,(t_c-t)}{G\cM}\right)^{-3/8} ,
  \label{eq:lec1:chirp}
\ee
the \emph{chirp}. The leading expression formally diverges at the
coalescence time $t_c$, outside the domain in which the approximation is
valid (Figure~\ref{fig:lec1:chirp}). A quasi-circular inspiral ends before
then, at the \emph{innermost stable circular orbit} (ISCO), which for a test
mass in Schwarzschild is the last stable circular geodesic, at
$r=6GM/c^2$~\cite{Wald:1984rg};
inside it the bodies plunge. Finally, one further integration gives the actual phase. Radiating at
twice the
orbital frequency, the wave carries $\varphi(t)=2\int\Omega\,dt$, so that,
measured back from coalescence,
\be
  \varphi_c - \varphi(t) \;\sim\;
  \left(\frac{c^3\,(t_c-t)}{G\cM}\right)^{5/8} ,
  \label{eq:lec1:phase}
\ee
and the cycles remaining from a given frequency to merger number
$N\sim(\varphi_c-\varphi)/2\pi\sim(G\cM\Omega/c^3)^{-5/3}\sim\nu^{-1}(v/c)^{-5}$,
matching the cycle count above.

Accurate conservative dynamics is valuable because, once combined with the
fluxes, it determines the inspiral waveform through
\eqref{eq:lec1:Omegadot}. We develop the conservative input in Lectures~2
and~3, and the radiation and fluxes in Lecture~4. This construction ceases to be
adiabatic near merger, where numerical relativity takes over.

\subsection{Exercises}
\label{sec:lec1:exercises}

\begin{enumerate}

\item \textbf{The Laplace--Runge--Lenz vector and the onset of precession.}
\begin{enumerate}
  \item[(a)] For the Newtonian Kepler problem, show that the
  Laplace--Runge--Lenz vector is conserved. 
  \item[(b)] Dot $\bm A$ into $\bm r$ to derive the orbit equation
  $r(\phi)=p/(1+e\cos\phi)$, with $k=GM\mu$, semi-latus rectum $p=L^2/(k\mu)$ and
  eccentricity $e=A/(k\mu)$. Conclude that $\bm A$ points from the focus to
  periastron and has length $\propto e$.
  \item[(c)] Recall the timelike geodesics of Schwarzschild. In the equatorial
  plane, writing $u\equiv1/r$ and taking $l$ to be the angular momentum
  \emph{per unit mass}, the orbit obeys
  the relativistic Binet equation,
  \[
    \frac{d^2u}{d\phi^2}+u=\frac{GM}{l^2}+\frac{3GM}{c^2}\,u^2 ,
  \]
  Verify that dropping the  term $3GM u^2/c^2$ returns the
  Newtonian ellipse of part~(b).
  \item[(d)] Treating $3GM u^2/c^2$ as a perturbation, show that its secular
  (resonant) part advances the periastron by
  \[
    \Delta\phi_{\rm prec}=\frac{6\pi G^2M^2}{c^2l^2}=\frac{6\pi GM}{c^2 p}
  \]
  per orbit. To leading post-Newtonian order the orbit is a precessing ellipse
  of fixed shape, so $\bm A$ keeps its length and merely co-rotates with the
  periastron; its fractional rotation per orbit is then
  $|\Delta\bm A|/|\bm A|\approx\Delta\phi_{\rm prec}$. Explain why this implies
  the scaling
  $|\Delta\bm A|/|\bm A|\sim(v/c)^2$ claimed in the main lecture.
\end{enumerate}

\item \textbf{Why numerical relativity struggles for EMRIs.}
\begin{enumerate}
  \item[(a)] Estimate that the number of time steps needed to numerically evolve
  a
          coalescence of mass ratio $q=m/M$ through merger grows as $N\sim
          q^{-2}$. \emph{Hint:} one power of $q^{-1}$
  from the number of orbits to merger and
  another from the steps per orbit.
  \item[(b)] A comparable-mass run ($q\sim 1$) takes about a day on a
  supercomputer. How long would the same code need for an extreme-mass-ratio
  inspiral with $m_2/m_1\sim10^{-6}$?
\end{enumerate}

\item \textbf{How accurately must a neutron-star inspiral be measured to feel
its tides?} The tidal deformability of a neutron star imprints on the
gravitational-wave phase only at $5$PN order~\cite{Flanagan:2007ix}. Take a
quasi-circular neutron-star
binary that sweeps through the band over $N_{\rm orb}\sim10^4$ orbits.
\begin{enumerate}
  \item[(a)] Estimate the leading accumulated phase $\phi$, and the tidal
  dephasing $\delta\phi$ of a body with tides $\lambda\sim(Gm/c^2)^5/G$, the
  deformability scale of a maximally compact body.
  \item[(b)] To be sensitive to such an imprint, how accurately must the phase
  be measured and modelled? Give the answer both in radians and as a fractional
  accuracy in $\phi$.
  \item[(c)] Now use the real neutron-star value
  $\lambda\sim(10^2\text{--}10^3)\,(Gm/c^2)^5/G$. At fixed masses and frequency
  interval, how does the leading tidal dephasing scale with $\lambda$?
  Explain why the $5$PN weighting concentrates the imprint in the late inspiral. This
  required accuracy has a theory-side counterpart: to what post-Newtonian order
  must the point-mass phasing be known so its truncation does not masquerade as
  a tidal signal?
\end{enumerate}

\end{enumerate}

\section{The one-body problem and effective field theory}
Thus far we have been treating each body, whether a black hole or a neutron
star, as a structureless point, although a black hole hides a region of
unbounded curvature behind a horizon and a neutron star has internal dynamics
described by complicated astrophysics. In this lecture we explain why this
approximation is warranted and where it breaks.

The reason we can approximate a messy compact object by a point particle is,
once again, thanks to a large separation of scales:
A compact object of
mass $m$ has a size
$R$ of order its Schwarzschild radius, $R \sim Gm/c^2$ (a neutron star is a few
times larger, but parametrically the same), while orbits before merger stay well
outside it, $r \gg Gm/c^2$. From the vantage of the orbit each body is
effectively a point, and its
internal makeup enters only through how it responds to the slowly varying field
its companion supplies.

Using again the virial relation we find that the relevant ratio of scales is
\be
  \frac{R}{r} \;\sim\; \frac{Gm}{r\,c^2} \;\sim\; \left(\frac{v}{c}\right)^2 ,
  \label{eq:lec2:Rr}
\ee
so the pointlike description is valid in
the same regime as the weak-field expansion. In the rest of this lecture we
write down the gauge-invariant
observables that correspond to the various ways a gravitational wave couples to
the point particle. We then interpret them in terms of an effective action, one
that describes an arbitrary compact object as an expansion in this ratio of
scales. We work throughout in the language of scattering amplitudes, which is
a direct way to access gauge-invariant observables. From here on we mostly
set $c=1$, restoring it where post-Newtonian counting requires it.

\subsection{Bootstrapping gravitational interactions}
\label{sec:lec2:bootstrap}

Let us see how far we can get by using only the symmetries of the problem,
without ever writing down an action. We will be working in the weak-field limit,
so the symmetries are those of the (flat) background spacetime, the Poincar\'e group,
together with gauge invariance and dimensional analysis. We will see that they
are enough to fix the one- and two-graviton amplitudes of a point particle.

The simplest subgroup of Poincar\'e is the translations, whose invariance we can
trivialize by going to momentum space.\footnote{Our Fourier convention is
$\phi(x)=\int\!\frac{d^4k}{(2\pi)^4}\,e^{ik\cdot x}\tilde\phi(k)$.}
Lorentz invariance then implies that every amplitude is a scalar function of the
momenta and polarizations at hand. For a
spinless point mass these rules leave very little freedom, because the body
brings so little to build with: its mass $m$ and its velocity $u^\mu$
($u^2=-1$), while each on-shell graviton brings its light-like momentum $k^\mu$
($k^2=0$) and its
transverse-traceless polarization, which we can write, without loss of
generality, as a product of polarization vectors
$\varepsilon_{\mu\nu}(k)=\varepsilon_\mu(k)\,\varepsilon_\nu(k)$, satisfying the
on-shell conditions
\be
k\cdot\varepsilon=0, \qquad \varepsilon\cdot\varepsilon=0.
\ee
Furthermore, linearized gauge invariance requires that the amplitude be inert
under the replacement
$\varepsilon_\mu\to\varepsilon_\mu+k_\mu$. We will see that these constraints
and dimensional analysis go rather far in fixing various amplitudes by pure
consistency, an approach sometimes called
\emph{bootstrapping}~\cite{Arkani-Hamed:2017jhn}.

Start with a single graviton with momentum $k$, sourced by a body of momentum
$p=m\,u$. After emitting the graviton the body carries $p-k$, and
keeping it on shell
requires
\be
(p-k)^2+m^2=-2m\,u\cdot k=0,
\ee
i.e.,\ $u\cdot k=0$ so the graviton carries no
energy in the body's rest frame. This is consistent, as a uniformly moving mass
cannot emit
or absorb a graviton on shell, and its one-point amplitude carries a factor
$2\pi\,\delta(u\cdot k)$ enforcing this. Furthermore, under a translation
$x\to x+x_0$ the momentum-space amplitude picks up a phase, $e^{-ik\cdot x_0}$,
so a line through the point $x_0$ rather than the origin carries that phase.
What multiplies these two factors is a Lorentz scalar quadratic in
$\varepsilon$, and the only non-zero such scalar is $(u\cdot\varepsilon)^2$.
Hence
\be
  \mathcal{A}(h)=\sqrt{8\pi G}\,m\,(u\cdot\varepsilon)^2\,
  e^{-ik\cdot x_0}\,2\pi\,\delta(u\cdot k).
  \label{eq:lec2:1pt}
\ee
The coefficient is the body's gravitational charge; that it equals the
inertial $m$ of $p=mu$ is the equivalence principle, which in this on-shell
language is forced rather than assumed~\cite{Weinberg:1964ew}, and $\sqrt{8\pi
G}$ is the
normalization that will later reproduce Newton's potential
(Sec.~\ref{sec:lec2:worldline}).
As a check, gauge invariance holds on the support of the delta function: under
$\varepsilon\to\varepsilon+k$ the change in $(u\cdot\varepsilon)^2$ is
proportional to $u\cdot k=0$.

Let us now bootstrap the leading coupling the body would have were it not
structureless. If the body carried a dipole moment, then we would have an
additional vector $D^\mu$ transverse to $u$, and write an amplitude
$(u\cdot\varepsilon)^2\,(k\cdot D)$. This coupling is not physical though, as
it can be eliminated by a translation of the worldline, $x_0\to x_0+D/m$,
which shifts the phase $e^{-ik\cdot x_0}$ and cancels the dipole:
\be
  \sqrt{8\pi G}\,(u\cdot\varepsilon)^2\,\big[m-i\,k\cdot D\big]\,
  e^{-ik\cdot x_0}\,2\pi\,\delta(u\cdot k)
  =\sqrt{8\pi G}\,m\,(u\cdot\varepsilon)^2\,e^{-ik\cdot(x_0+D/m)}\,
  2\pi\,\delta(u\cdot k)+O(D^2).
  \label{eq:lec2:dipole-shift}
\ee
Indeed, a mass dipole is the offset of the body's centroid from the worldline,
which we remove by a change of coordinates in which the worldline sits at the
centroid. In other words, the mass dipole is a gauge artifact. The only physical
dipole
is the current dipole, the spin, which we will not have time to cover in these
lectures.

A body with a permanent shape can, however, carry a quadrupole moment,
$Q^{\mu\nu}$, orthogonal to $u$ so it carries the correct number of degrees of
freedom. Its amplitude is
most conveniently written in terms of the graviton's linearized field strength,
\be
  F^{\mu\nu}=k^\mu\varepsilon^\nu-k^\nu\varepsilon^\mu ,
  \label{eq:lec2:F}
\ee
which is inert under $\varepsilon\to\varepsilon+k$ identically. Contracting the
velocity into it gives the gauge-invariant vector
\be
  W^\mu=u_\nu F^{\nu\mu}=(u\cdot k)\,\varepsilon^\mu-(u\cdot\varepsilon)\,k^\mu,
  \label{eq:lec2:W}
\ee
which vanishes outright when $\varepsilon\to k$, and the quadrupolar analog of
\eqref{eq:lec2:1pt} for a permanent moment is
\be
  \mathcal{A}_Q(h)=-\tfrac12\sqrt{8\pi G}\,Q^{\mu\nu}\,W_\mu W_\nu\,
  e^{-ik\cdot x_0}\,2\pi\,\delta(u\cdot k).
  \label{eq:lec2:1ptQ}
\ee
Gauge invariance does not fix the overall factor, which is a choice of
normalization for $Q^{\mu\nu}$; we take it so that $Q^{\mu\nu}$ is the usual
mass
quadrupole (Sec.~\ref{sec:lec2:worldline}).
Note that $W^2=0$ due to the graviton on-shell conditions, so we can take the
quadrupole to be traceless and symmetric without losing any information.

Now a more interesting process: the scattering of a gravitational wave off the
point-like object. This is sometimes called gravitational Compton
scattering~\cite{Bautista:2021wfy,Bautista:2022wjf,Bautista:2023sdf,Ivanov:2024sds,Saketh:2024juq,Caron-Huot:2025tlq,Bjerrum-Bohr:2025bqg,Bautista:2026qse,Bjerrum-Bohr:2026fhs,Ivanov:2026icp,Bautista:2026fcp,Brunello:2026rdk,Brunello:2026lzf,Correia:2026utp},
in analogy with the electromagnetic process. This can be described as a process
with an incoming graviton with momentum $k_1$ and an outgoing one with momentum
$-k_2$, so that both momenta are counted as incoming. Away from $u\cdot k_i=0$
gauge invariance is no longer automatic, so each external leg
enters through its field strength $F_i$ of \eqref{eq:lec2:F} or, contracted
with the velocity, its $W_i$ of \eqref{eq:lec2:W}. As usual, each polarization
must enter
quadratically, since
$\varepsilon_i^{\mu\nu}=\varepsilon_i^\mu\varepsilon_i^\nu$,
so out of $F_1$, $F_2$, the velocity, and the momenta there are just two
independent gauge-invariant scalars of the right type:\footnote{The remaining
product one might form, $(F_1\cdot F_2)\big[(F_1\cdot F_2)+4\,W_1\cdot
W_2\big]$,
is proportional to the Gram determinant of
$\{u,k_1,k_2,\varepsilon_1,\varepsilon_2\}$ and vanishes in four
dimensions, so it is not an independent structure.}
\be
  (W_1\cdot W_2)^2=(u\cdot F_1\cdot F_2\cdot u)^2 \quad \text{and} \quad
  (F_1\cdot F_2)^2\equiv(F_1^{\mu\nu}F_{2\,\mu\nu})^2 .
  \label{eq:lec2:structures}
\ee
Each structure can be weighted
by a scalar function of the momentum invariants $k_1\cdot k_2$, $u\cdot k_1$ and
$u\cdot k_2$ (with $k_i^2=0$). By energy conservation and translation
covariance, the amplitude thus carries the
factor $e^{-i(k_1+k_2)\cdot x_0}\,2\pi\,\delta\big(u\cdot(k_1+k_2)\big)$ of a
static source, which we leave implicit; on its support $u\cdot k_2=-u\cdot k_1$,
two invariants remain, so the most general gauge-invariant amplitude of this
type is
\be
  \mathcal{M}(1,2)=f(k_1\!\cdot\!k_2,\,u\!\cdot\!k_1)\,(W_1\cdot W_2)^2
        +g(k_1\!\cdot\!k_2,\,u\!\cdot\!k_1)\,\big(F_1\cdot F_2\big)^2 ,
  \label{eq:lec2:compton-general}
\ee
with the scalar coefficients $f,g$ still undetermined.

\begin{figure}[t]
  \centering
\subfloat[graviton pole]{%
\begin{tikzpicture}[
  x=1cm, y=1cm,
  worldline/.style={line width=0.8pt, dash pattern=on 4pt off 2.5pt},
  fluct/.style={line width=0.8pt},
  graviton/.style={line width=0.65pt, decorate,
                   decoration={snake, amplitude=2.4pt, segment length=6pt,
                               pre length=2pt, post length=2pt}},
  blob/.style={draw=black, line width=0.6pt, fill=black!15},
  flow/.style={-{Latex[length=4.5pt,width=3.5pt]}, line width=0.5pt},
  lbl/.style={font=\footnotesize, inner sep=1.5pt},
  panelcap/.style={font=\small},
]
\def\blobr{0.32}

\begin{scope}
  \coordinate (aL) at (0,0);       \coordinate (aR) at (5.4,0);
  \coordinate (bA) at (2.7,1.75);  \coordinate (bW) at (2.7,0);
  \coordinate (e1) at (1.35,2.75); \coordinate (e2) at (4.05,2.75);

  \draw[worldline, -{Latex[length=5pt,width=4pt]}] (aL) -- (aR);
  \node[lbl, anchor=west] at (0.05,0.34) {$m,\ u^{\mu}$};

  \draw[graviton] (e1) -- (bA);
  \draw[graviton] (e2) -- (bA);
  \draw[flow] ($(e1)+(-0.36,0.14)$) -- ($(e1)+(0.14,-0.31)$);
  \draw[flow] ($(e2)+(0.36,0.14)$)  -- ($(e2)+(-0.14,-0.31)$);
  \node[lbl, anchor=east] at ($(e1)+(-0.34,0.26)$) {$k_1$};
  \node[lbl, anchor=west] at ($(e2)+(0.34,0.26)$)  {$k_2$};

  \draw[graviton] (bA) -- (bW);
  \filldraw[blob] (bA) circle[radius=\blobr];
  \filldraw[blob] (bW) circle[radius=\blobr];
  \node[lbl, anchor=east] at (2.36,1.14) {$k_1+k_2$};

  \node[lbl, anchor=west] at (3.04,0.88) {$k_1\cdot k_2$};

\end{scope}
\end{tikzpicture}}
  \hspace{1cm}
  \subfloat[worldline pole]{%
\begin{tikzpicture}[
  x=1cm, y=1cm,
  worldline/.style={line width=0.8pt, dash pattern=on 4pt off 2.5pt},
  fluct/.style={line width=0.8pt},
  graviton/.style={line width=0.65pt, decorate,
                   decoration={snake, amplitude=2.4pt, segment length=6pt,
                               pre length=2pt, post length=2pt}},
  blob/.style={draw=black, line width=0.6pt, fill=black!15},
  flow/.style={-{Latex[length=4.5pt,width=3.5pt]}, line width=0.5pt},
  lbl/.style={font=\footnotesize, inner sep=1.5pt},
  panelcap/.style={font=\small},
]
\def\blobr{0.32}

\begin{scope}
  \coordinate (bL) at (0,0);       \coordinate (bR) at (5.4,0);
  \coordinate (p1) at (1.7,0);     \coordinate (p2) at (4.1,0);
  \coordinate (q1) at (1.7,2.75);  \coordinate (q2) at (4.1,2.75);

  \draw[worldline] (bL) -- (p1);
  \draw[fluct] (p1) -- (p2);
  \draw[worldline, -{Latex[length=5pt,width=4pt]}] (p2) -- (bR);

  \draw[graviton] (q1) -- (p1);
  \draw[graviton] (q2) -- (p2);
  \draw[flow] ($(q1)+(-0.42,0.14)$) -- ($(q1)+(-0.42,-0.41)$);
  \draw[flow] ($(q2)+(0.42,0.14)$)  -- ($(q2)+(0.42,-0.41)$);
  \node[lbl, anchor=east] at ($(q1)+(-0.34,0.26)$) {$k_1$};
  \node[lbl, anchor=west] at ($(q2)+(0.34,0.26)$)  {$k_2$};
  \filldraw[blob] (p1) circle[radius=\blobr];
  \filldraw[blob] (p2) circle[radius=\blobr];

  \node[lbl] at (2.9,-0.52) {$u\cdot k_1$};

\end{scope}

\end{tikzpicture}}
  \caption{The two channels in which the amplitude \eqref{eq:lec2:compton}
  factorizes: the exchanged graviton, on shell at $k_1\cdot k_2\to0$ (a), and
  the recoiling worldline, a double pole at $u\cdot k_1\to0$ (b).}
  \label{fig:lec2:factorization}
\end{figure}
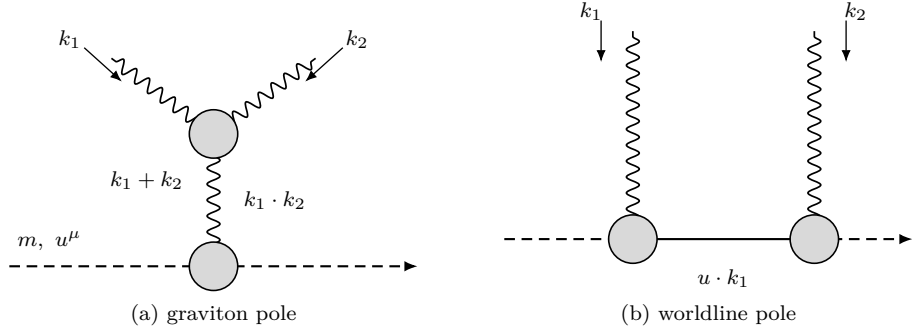

To go further let us invoke dimensional analysis. A classical amplitude contains
no $\hbar$, so the body can enter only through the length $Gm$, and at order
$G$ the functions $f$ and $g$ are $Gm$ times functions of the momenta of
dimension (length)$^4$, both tensor structures being quartic in momenta. Poles are
the only non-polynomial dependence a tree amplitude can have, and they sit where
an intermediate state goes on shell, which happens in two places
(Figure~\ref{fig:lec2:factorization}). Either the two waves fuse into a single
graviton, which propagates across and is absorbed by the source, on shell when
$k_1+k_2$ becomes null, i.e. at $k_1\cdot k_2\to0$; or the body absorbs one
wave, recoils, and re-emits the other. A free body has no restoring force, so
its displacement responds to a periodic force of frequency $\omega=u\cdot k_1$
with the $1/(m\omega^2)$ of Newton's second law, a \emph{double} pole
$1/(u\cdot k_1)^2$. These four powers of momentum exhaust the dimension, so $f$
and $g$ are constants divided by $(k_1\cdot k_2)(u\cdot k_1)^2$, and the answer
is
\be
  \mathcal{M}(1,2)=- i 8\pi G\,m \,
  \frac{(W_1\cdot W_2)^2}{(k_1\cdot k_2)\,(u\cdot k_1)^2},
  \label{eq:lec2:compton}
\ee
with the second structure absent. In other words, to this order the entire
response of a point mass to a passing gravitational wave is one number, the
mass,
times a fixed kinematic factor. The interior of the body has not appeared
anywhere.

Both the constant and the absence of the second term follow from factorization: at each pole the
residue is a product of lower-point amplitudes summed over intermediate on-shell
states and polarizations. At the worldline pole both
gravitons attach to the line through the one-point coupling
\eqref{eq:lec2:1pt}, two powers of velocity each, so the residue is the square
of that coupling divided by the inertia $m$ and carries four powers of $u$. At
the graviton pole the residue is the one-point coupling of the fused graviton
times the graviton's self-coupling, the three-graviton amplitude,\footnote{The
bracket is the three-gluon amplitude of Yang--Mills. The fact that the graviton
amplitude is its square
is a first glimpse of the \emph{double
copy}~\cite{Kawai:1985xq,Bern:2008qj,Bern:2022wqg}, which we do not have time to
develop
further.}
\be
  {\cal M}_3(1,2,3)=-\sqrt{32\pi G}\,
  \Big[\,(\varepsilon_1\!\cdot\!\varepsilon_2)(\varepsilon_3\!\cdot\!k_1)
        +(\varepsilon_2\!\cdot\!\varepsilon_3)(\varepsilon_1\!\cdot\!k_2)
+(\varepsilon_3\!\cdot\!\varepsilon_1)(\varepsilon_2\!\cdot\!k_3)\,\Big]^2,
  \label{eq:lec2:M3}
\ee
with $k_1+k_2+k_3=0$. Explicitly, with $K=k_1+k_2$ the momentum of the
exchanged graviton,
\be
  \lim_{(k_1+k_2)^2\to0}\,(k_1+k_2)^2\,\mathcal{M}(1,2)
  \;=\;i\sum_\lambda{\cal M}_3(1,2,-K)\,\mathcal{A}(K),
  \label{eq:lec2:graviton-pole}
\ee
where the third leg of ${\cal M}_3$ carries polarization $\varepsilon^\lambda$,
the graviton absorbed in $\mathcal{A}$ carries $\varepsilon^{\lambda*}$, and the
phase and delta function of $\mathcal{A}$ are the overall factor left implicit
in $\mathcal{M}(1,2)$. The sum runs over the two physical polarizations
$\varepsilon^\lambda_{\mu\nu}=\varepsilon^\lambda_\mu\varepsilon^\lambda_\nu$
of the intermediate graviton, squares of polarization four-vectors transverse
to its momentum, $K\cdot\varepsilon^\lambda=0$. They are complete on the
transverse plane and sum to its projector,
\be
  \sum_\lambda\varepsilon^\lambda_\mu\,\varepsilon^{\lambda*}_\nu
  =\eta_{\mu\nu}-\frac{k_\mu\bar k_\nu+k_\nu\bar k_\mu}{k\cdot\bar k}
  \equiv\Pi_{\mu\nu},
  \label{eq:lec2:vecsum}
\ee
which is orthogonal to the momentum $k$ and to a reference null vector
$\bar k$ that drops out of any gauge-invariant quantity. The graviton
polarization sum is the symmetric, trace-free square of $\Pi_{\mu\nu}$,
\be
  \sum_\lambda\varepsilon^{\lambda*}_{\mu\nu}\,\varepsilon^\lambda_{\alpha\beta}
=\tfrac12\big(\Pi_{\mu\alpha}\Pi_{\nu\beta}+\Pi_{\mu\beta}\Pi_{\nu\alpha}\big)
    -\tfrac12\,\Pi_{\mu\nu}\Pi_{\alpha\beta}
   \equiv\Lambda_{\mu\nu,\alpha\beta}.
  \label{eq:lec2:polsum}
\ee
Between ${\cal M}_3$, transverse in its third leg, and the coupling
\eqref{eq:lec2:1pt}, the $k$, $\bar k$ and trace terms of $\Lambda$ drop, so
the sum replaces $\varepsilon_3^\mu\varepsilon_3^\nu$ by $u^\mu u^\nu$ in the
squared bracket of \eqref{eq:lec2:M3}, with $k_3=-k_1-k_2$:
\begin{align}
  i\sum_\lambda{\cal M}_3(1,2,-K)\,\mathcal{A}(K)
  &=-16\pi i\,Gm\,\big[(\varepsilon_1\cdot\varepsilon_2)(u\cdot k_1)
   +(u\cdot\varepsilon_2)(\varepsilon_1\cdot k_2)
   -(u\cdot \varepsilon_1)(\varepsilon_2 \cdot k_1)\big]^2 \nonumber \\
  &=-16\pi i\,Gm\,\frac{(W_1\cdot W_2)^2}{(u\cdot k_1)^2} 
   \bigg|_{k_1\cdot k_2=0}.
  \label{eq:lec2:fact}
\end{align}
The last step is an identity at $k_1\cdot k_2=0$, where $W_1\cdot W_2$
carries an explicit factor $u\cdot k_1$, and it fixes the result in
\eqref{eq:lec2:compton}.

One loophole remains. Poles are the only singularities allowed, but nothing
forbids adding to $f$ and $g$ polynomials in the invariants, which factorization
does not constrain. Since a classical amplitude has no scale other than $Gm$, the
lowest such terms are
\be
  \mathcal{M}(1,2)\;\supset\;(Gm)^5\Big[\alpha_1\,(W_1\cdot W_2)^2
  +\alpha_2\,\big(F_1\cdot F_2\big)^2\Big]+\dots,
  \label{eq:lec2:contact}
\ee
with dimensionless constants $\alpha_1$, $\alpha_2$ that symmetry alone cannot
fix, and
the dots standing for higher powers of the invariants, each accompanied by two
more powers of $Gm$. We will come back to the interpretation of these contact
terms later in this lecture.

\subsection{Worldline effective field theory}
\label{sec:lec2:worldline}

Let us interpret the results of the bootstrap using the more familiar language
of action principles. More concretely, we follow the approach of
\cite{Goldberger:2004jt,Goldberger:2007hy,Porto:2016pyg} (see also
\cite{Damour:1992we}, where compact bodies were first \emph{skeletonized} into
point sources carrying field-dependent couplings) where an effective action is
constructed for the long-distance degrees of freedom of a compact object,
systematically incorporating the effects of its finite size and internal
structure.
The EFT construction has three steps:

\begin{enumerate}
  \item \textbf{Degrees of freedom and symmetries.} Identify what survives at
  long distances together with the symmetries it
  must respect.
  \item \textbf{The most general action.} Write down every term those
  symmetries allow, each with an unknown coefficient.
  \item \textbf{Power counting.} Use the small ratio to order this infinite
  list, so that only finitely many terms contribute to a given accuracy.
\end{enumerate}

These steps might leave some coefficients unfixed. Such \emph{Wilson
coefficients} carry the short-distance physics the effective theory does not
resolve, here the
interior of a neutron star or black hole, and must be fixed from outside, by
comparing an observable with the same one computed in a more complete theory,
or by measuring it. Once fixed they can be reused in any long-distance
calculation. In Sec.~\ref{sec:lec2:bootstrap} the mass $m$ played this role for
the one- and two-graviton amplitudes, and $\alpha_1$, $\alpha_2$ in
\eqref{eq:lec2:contact} are the next ones.

At distances much larger than its size $R$, the retained degree of freedom of
a compact body is a worldline $x^\mu(\sigma)$; because spin is excluded, it
carries no further internal data. Two symmetries constrain its action: general
covariance requires a spacetime scalar built from the metric and the worldline,
and reparametrization invariance requires independence of how we label points
along it, $\sigma \to \sigma'(\sigma)$. The only such object with no
derivatives of the metric is the elapsed proper time, so the leading action is
\be
  S_{\rm pp} \;=\; -\,m \int d\tau,
  \qquad
  d\tau \;=\; \sqrt{-\,g_{\mu\nu}\,dx^\mu dx^\nu},
  \label{eq:lec2:pp-action}
\ee
with $c=1$. Varying it gives the geodesic equation: at leading order the body
falls freely, as the equivalence principle demands.

Now compare with Sec.~\ref{sec:lec2:bootstrap}. On the flat-space expansion
$g_{\mu\nu}=\eta_{\mu\nu}+\sqrt{32\pi G}\,h_{\mu\nu}$ the term linear in $h$ is
the minimal coupling $\sqrt{8\pi G}\int d^4x\,h_{\mu\nu}T^{\mu\nu}$ to the
stress tensor of a point particle,
\be
  T^{\mu\nu}(x)=m\!\int\! d\tau\,u^\mu u^\nu\,\delta^4\!\big(x-x(\tau)\big),
  \label{eq:lec2:T}
\ee
the only symmetric tensor a structureless body can build from its velocity. On
a straight worldline $x^\mu=x_0^\mu+u^\mu\tau$ its Fourier transform is
\be
  T^{\mu\nu}(k)=m\,u^\mu u^\nu\,e^{-ik\cdot x_0}\,2\pi\,\delta(k\cdot u),
  \label{eq:lec2:Tk}
\ee
and contracting it with $\varepsilon_{\mu\nu}$ gives the one-point amplitude
\eqref{eq:lec2:1pt}, translation phase and delta function included.
 
Expanding instead about the straight line,
$x^\mu=x_0^\mu+u^\mu\tau+z^\mu(\tau)$, gives the kinetic term
$\tfrac{m}{2}\int d\tau\,\dot z^2$, whose propagator $1/(m\omega^2)$ is the
double pole of the worldline channel in
Figure~\ref{fig:lec2:factorization}(b), together with a vertex linear in $h$
and in $z$ that attaches a graviton to the displaced line. Expanding
\eqref{eq:lec2:pp-action} to second order in $h$ and adding the bulk
three-graviton vertex, obtained by expanding the Einstein--Hilbert action to
non-linear order, for the other channel would reproduce the
Compton amplitude \eqref{eq:lec2:compton}. Note that this would require
significant effort, whereas in the effective action a single coefficient, the
mass $m$, controls the leading two-graviton amplitude.

Finite size means the body feels not just the field but its variation across
its own extent, that is, the curvature. Step 2 is therefore to list the
reparametrization- and diffeomorphism-invariant operators built from the
curvature on the worldline, the velocity $u^\mu=dx^\mu/d\tau$, and whatever
moments the body itself carries. Step 3
orders them by the number of derivatives they carry, each costing a power of
$R/r$,
\be
  S \;=\; -\,m\int d\tau \;+\; \int d\tau\,\big(\text{curvature operators}\big).
  \label{eq:lec2:tower}
\ee
Which operators actually appear depends on what the body is.

A body with \emph{intrinsic} structure, a permanent shape fixed in its own
rest frame, carries multipole moments. The first genuine moment is the
mass quadrupole $Q^{\mu\nu}$, a symmetric, trace-free tensor orthogonal to $u$.
It couples linearly to the electric tidal tensor, the velocity projection of the
Weyl tensor $E_{\mu\nu}=C_{\mu\alpha\nu\beta}u^\alpha u^\beta$ (equal to the
Riemann curvature
$R_{\mu\alpha\nu\beta}u^\alpha u^\beta$ in vacuum),
\be
  S_{\rm mult}\;=\;-\tfrac12\int d\tau\,Q^{\mu\nu}E_{\mu\nu}\;+\;\cdots,
  \label{eq:lec2:multipole}
\ee
the dots standing for higher mass moments, which couple to derivatives of $E$,
and current (mass-flow) moments, which couple to the magnetic Weyl tensor
$B_{\mu\nu}$. On an on-shell graviton the linearized Riemann tensor is
$\sqrt{8\pi G}\,F_{\mu\alpha}F_{\nu\beta}$ in terms of the field strength
\eqref{eq:lec2:F}, so $E_{\mu\nu}=\sqrt{8\pi G}\,W_\mu W_\nu$ and
\eqref{eq:lec2:multipole} is the coupling behind the quadrupole amplitude
\eqref{eq:lec2:1ptQ}, normalization included: the tensor $Q^{\mu\nu}W_\mu W_\nu$
found there by gauge invariance alone is the energy of the quadrupole in the
tidal field.

The curvature \emph{scalars}, those with no body moment contracted into them,
do not contribute. The lowest, $\int d\tau\,R$ and $\int d\tau\,
R_{\mu\nu}u^\mu u^\nu$, are proportional to the Ricci tensor, hence to the
vacuum field equations, and any such operator can be removed by a
redefinition of the metric supported on the worldline, so it cannot affect an
observable. We can show this explicitly in an example. Consider the two scalar
operators,
\be
  S_R \;=\; c_1\int d\tau\,R \;+\; c_2\int d\tau\,R_{\mu\nu}u^\mu u^\nu,
  \label{eq:lec2:redundant-ops}
\ee
and restore the bulk Einstein--Hilbert action $\tfrac{1}{16\pi G}\int d^4x\,
\sqrt{-g}\,R$, whose variation under $g_{\mu\nu}\to g_{\mu\nu}+\delta
g_{\mu\nu}$
is 
\be
\delta S = -\tfrac{1}{16\pi G}\int d^4x\,\sqrt{-g}\,(R^{\mu\nu}-\tfrac12
g^{\mu\nu}R)\,
\delta g_{\mu\nu}.
\ee 
A shift supported on the worldline $x^\mu(\tau)$,
\be
  \delta g_{\mu\nu}(x) \;=\; 16\pi G\left[\big(\tfrac12 c_2 - c_1\big)g_{\mu\nu}
  \;+\; c_2\,u_\mu u_\nu\right]\int d\tau\,
  \frac{\delta^4\big(x-x(\tau)\big)}{\sqrt{-g}},
  \label{eq:lec2:field-redef}
\ee
turns the variation into exactly $-S_R$, cancelling the two
operators. The same procedure, but with a different redefinition, works for any
operator containing a Ricci tensor or Ricci scalar.

\subsection{Tidal operators and Love numbers}
\label{sec:lec2:finite-size}

Beyond the intrinsic multipole moments, the leading finite-size terms in the
worldline action of a structureless body
are the two quadratic Weyl invariants of the
tower \eqref{eq:lec2:tower} introduced in Sec.~\ref{sec:lec2:worldline},
\be
  S_{\rm FS} \;=\; \int d\tau\,
  \Big[\, \frac14 \lambda\,E_{\mu\nu}E^{\mu\nu}
        + \frac14\lambda_B\,B_{\mu\nu}B^{\mu\nu} \,\Big],
  \label{eq:lec2:SFS}
\ee
with $E_{\mu\nu}$ the electric tidal tensor of \eqref{eq:lec2:multipole} and
$B_{\mu\nu}$ the magnetic (gravitomagnetic) one, built from the dual of the
Riemann tensor. The two coefficients, the electric- and magnetic-type tidal
Wilson coefficients~\cite{Bini:2012gu}, are free parameters of the worldline
theory, of dimension (mass)$\times$(length)$^4$ (or (length)$^5$ once $G=c=1$).

Power counting identifies these coefficients with the constants left free in
Sec.~\ref{sec:lec2:bootstrap}. Each term of \eqref{eq:lec2:SFS} is quadratic in
the curvature, so for two gravitons it yields a contact vertex interaction, with
nothing propagating between the two legs and hence no poles. With
$E_{\mu\nu}=\sqrt{8\pi G}\,W_\mu
W_\nu$ on a single graviton, $E_{\mu\nu}E^{\mu\nu}$ produces $(W_1\cdot W_2)^2$
(you will work this out in Exercise~2). The magnetic operator produces both
structures in \eqref{eq:lec2:structures}, because on a single on-shell graviton
the linearized curvature is self-dual or anti-self-dual according to the
helicity, so its magnetic and electric parts are proportional,
\be
  B_{\mu\nu}=\mp\,i\,E_{\mu\nu}
  \qquad\text{for helicity }\pm ,
  \label{eq:lec2:BE}
\ee
and $B_{\mu\nu}B^{\mu\nu}$ therefore contributes $-E_1\!\cdot\!E_2$ between two
gravitons of equal helicity and $+E_1\!\cdot\!E_2$ between two of opposite
helicity. The same duality relates the structures of
\eqref{eq:lec2:structures}: in four dimensions $F_1\!\cdot\!F_2=-4\,W_1\cdot
W_2$
for equal helicities, while $F_1\!\cdot\!F_2$ vanishes for opposite ones.
Matching
the two helicity configurations separately therefore fixes both constants of
\eqref{eq:lec2:contact} outright,
\be
  \alpha_2\,(Gm)^5 \;=\; -\tfrac{\pi}{2}\, G\lambda_B ,
  \qquad
  \alpha_1\,(Gm)^5 \;=\; 4\pi G\,(\lambda+\lambda_B) ,
  \label{eq:lec2:alpha-lambda}
\ee
so $\alpha_2$ fixes the magnetic coefficient and $\alpha_1$, once $\lambda_B$
is subtracted, the electric one. 

\begin{figure}[t]
  \centering
\begin{tikzpicture}[
  x=1cm, y=1cm,
  body/.style={draw=black, line width=0.6pt, inner color=white, outer
  color=black!22},
  ghost/.style={draw=black!50, line width=0.4pt, dash pattern=on 2pt off 1.6pt},
  field/.style={draw=black!60, line width=0.45pt},
  tip/.style={field, -{Stealth[length=3.6pt,width=3pt]}},
  place/.style={-{Stealth[length=6pt,width=5pt]}, line width=0.8pt},
  lbl/.style={font=\footnotesize, inner sep=1.5pt},
]
\def\bodyR{0.85}

\shade[body] (0,0) circle[radius=\bodyR];

\draw[place] (1.45,0) -- (2.75,0);

\begin{scope}[shift={(5.7,0)}]
  \def\winW{2.5}   
  \def\winH{1.75}  
  \begin{scope}
    \clip (-\winW,-\winH) rectangle (\winW,\winH);
    \foreach \xo/\pos in {0.06/0.30, 0.22/0.32, 0.55/0.36, 1.15/0.45} {
      \pgfmathsetmacro{\tend}{ln(\winW/\xo)/2}
      \pgfmathsetmacro{\ta}{\pos*\tend}
      \pgfmathsetmacro{\tb}{\ta-0.02}
      \foreach \sx in {1,-1} { \foreach \sy in {1,-1} {
        \draw[field, domain=0:\tend, samples=50, smooth, variable=\t]
          plot ({\sx*\xo*exp(2*\t)}, {\sy*\winH*exp(-\t)});
        \draw[tip] ({\sx*\xo*exp(2*\tb)}, {\sy*\winH*exp(-\tb)})
                -- ({\sx*\xo*exp(2*\ta)}, {\sy*\winH*exp(-\ta)});
      }}
    }
    \foreach \s in {1,-1} {
      \draw[field] (0,\s*\winH) -- (0,\s*\bodyR);
      \draw[tip]   (0,\s*1.50) -- (0,\s*1.45);
      \draw[field] (\s*\bodyR,0) -- (\s*\winW,0);
      \draw[tip]   (\s*1.95,0) -- (\s*2.00,0);
    }
  \end{scope}
  \shade[body] (0,0) ellipse[x radius=1.28*\bodyR, y radius=0.72*\bodyR];
  \draw[ghost] (0,0) circle[radius=\bodyR];
  \node[lbl, anchor=south west] at (1.55,1.0) {$E_{ij}$};
  \node[lbl] at (0,0) {$Q_{ij}$};
\end{scope}
\end{tikzpicture}
  \caption{A spherical body placed in the tidal field of a distant companion
  lying along the horizontal axis. The lines are field lines of the tidal
  acceleration $-E_{ij}x^j$, directed inward across the axis and outward along
  it; the body stretches along the axis and acquires the quadrupole
  $Q_{ij}=-\lambda E_{ij}$ of \eqref{eq:lec2:Q-response}. The dashed circle is
  the undeformed shape.}
  \label{fig:lec2:tidal}
\end{figure}
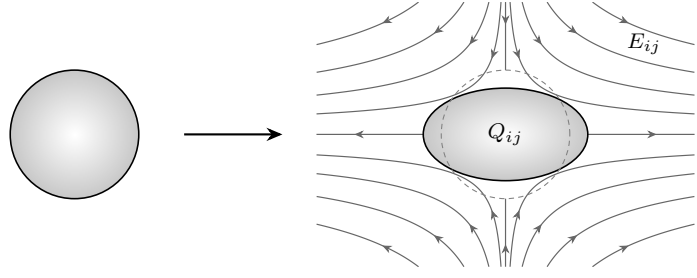

In Newtonian language, $\lambda$ measures the response every extended body
shows in an external field. Concretely, a body in an external potential
$\Phi_{\rm ext}$ feels across its extent the tidal field,
\be
  E_{ij} \;=\; \partial_i \partial_j \Phi_{\rm ext}\,.
  \label{eq:lec2:tidal-newt}
\ee
The body stretches along the tidal axis
(Figure~\ref{fig:lec2:tidal}) and develops an induced mass quadrupole, linear
in the field for a weak one,
\be
  Q_{ij} \;=\; -\,\lambda\, E_{ij}.
  \label{eq:lec2:Q-response}
\ee
The constant $\lambda$, the \emph{tidal deformability},  or Love number,
measures how easily the
body distorts, and it can be shown, with some work, to be the same $\lambda$
that
multiplies the electric operator in \eqref{eq:lec2:SFS} by separating the
gravitational field into a tidal and response field and reading the resulting
induced quadrupole moment by comparing to the coupling in
\eqref{eq:lec2:multipole}.

In practice, the coefficient of the electric operator is fixed, like every
Wilson
coefficient, by matching: place the body in a static external tidal field and
compute the quadrupole it develops, once in the full theory and once in the
effective one, and set them equal. The full theory in our case would be a full
hydrodynamic description of a neutron star, depending e.g., on its equation of
state; or a detailed description of a black hole in full General Relativity,
including the zone near the horizon.

Now ask the same of a black hole. There is no fluid and no equation of state,
but the question is still well posed: solve the static quadrupolar
($\ell=2$) perturbation of the Schwarzschild geometry and read off the induced
quadrupole. There is none. Regularity at the horizon forces the static tidal
response to vanish identically,
\be
  \lambda^{\rm BH} \;=\; 0 .
  \label{eq:lec2:bh-love}
\ee
A Schwarzschild black hole has no induced static quadrupole response: the most compact object
there is has
exactly zero conservative static tidal response (in the standard language, its Love
numbers vanish). This was established by direct calculation in
black-hole perturbation
theory~\cite{Binnington:2009bb,Damour:2009wj,Kol:2011vg,Hui:2020xxx},
and the vanishing has since been traced to a
hidden symmetry of the perturbation equations, a ladder structure relating
solutions of different multipole order, still under active
study~\cite{Hui:2021vcv,Charalambous:2021mea,Charalambous:2022rre,Parra-Martinez:2025bcu}.
Nor is it an artifact of linear response: the quadratic Love numbers of a
Schwarzschild black hole vanish as
well~\cite{Riva:2023rcm,Iteanu:2024dvx}.

Equation \eqref{eq:lec2:bh-love} is a statement about the conservative, static
response only. A black hole still absorbs:
tidal forces do work on the horizon and energy flows down it, an effect no real
value of $\lambda$ can capture. Dissipation lives instead in the imaginary part
of the response, or equivalently in a separate worldline operator with its own
coefficient, fixed by matching not to a static deformation but to the flux the
horizon absorbs~\cite{Hawking:1972hy,Hartle:1973zz,Goldberger:2005cd}; we set it
aside in these lectures.

Once again, dimensional analysis
gives the most important property of these tidal operators: their coefficients
have
dimensions of (mass)$\times$(length)$^4$ with $c=1$, so their natural scaling is
\be
  \lambda \;\sim\; \frac{R^5}{G} \;\sim\; G^4m^5 \qquad (c=1),
  \label{eq:lec2:love}
\ee
with a dimensionless prefactor set by the internal structure.
Using this we can estimate the size of the contribution of the electric term to
the binding energy of the binary. The companion of mass $m_2$ at separation $r$
sets up a tidal field $E \sim Gm_2/r^3$ at the body, which through
\eqref{eq:lec2:Q-response} induces a quadrupole $Q \sim \lambda\,Gm_2/r^3$. The
energy of that quadrupole in the companion's field is of order
$\lambda E^2 \sim \lambda\,G^2 m_2^2/r^6$. Comparing with the Newtonian binding
energy, $U_N \sim Gm_1 m_2/r$,
\be
  \frac{U_{\rm tidal}}{U_N}
  \;\sim\; \frac{\lambda\, G m_2}{m_1\, r^5}
  \;\sim\; \left(\frac{R_1}{r}\right)^{\!5}
  \;\sim\; \left(\frac{v}{c}\right)^{\!10},
  \label{eq:lec2:5PN}
\ee
using $\lambda \sim R_1^5/G$, $R_1 \sim Gm_1/c^2$, and the virial relation; the
mass-ratio factors are of order one and do not change the counting. So these
tidal corrections first enter at $(v/c)^{10}$ relative to Newtonian, that is at
5PN. They are detectable all the same, because a neutron star is several times
larger than $Gm/c^2$,
which enhances $(R_1/r)^5$ by a large factor.

\subsection{Exercises}
\label{sec:lec2:exercises}

\begin{enumerate}

\item \textbf{The gravitational Rutherford cross section.} Take the
gravitational Compton amplitude \eqref{eq:lec2:compton} for a wave scattering
off a structureless mass $m$. Going to the rest frame of the mass and summing
over the in- and outgoing graviton helicities, show that the cross section
takes the Rutherford-like form,
\[
  \frac{d\sigma}{d\Omega}\;\propto\;(Gm)^2\,
  \frac{\cos^8(\theta/2)+\sin^8(\theta/2)}{\sin^4(\theta/2)},
\]
with $\theta$ the scattering angle.

\item \textbf{The leading tidal amplitude.} The finite-size operator
$\tfrac{\lambda}{4}\!\int\!d\tau\,E_{\mu\nu}E^{\mu\nu}$ of
\eqref{eq:lec2:SFS} is quadratic in the curvature: a contact vertex attaching
two gravitons to the worldline.
\begin{enumerate}
  \item[(a)] For $g_{\mu\nu}=\eta_{\mu\nu}+\sqrt{32\pi G}\,h_{\mu\nu}$ the
  linearized Riemann tensor is
  \[
    R_{\mu\alpha\nu\beta}=\sqrt{8\pi G}\,\big(
        \partial_\alpha\partial_\nu h_{\mu\beta}
      + \partial_\mu\partial_\beta h_{\alpha\nu}
      - \partial_\alpha\partial_\beta h_{\mu\nu}
      - \partial_\mu\partial_\nu h_{\alpha\beta}\big).
  \]
  Compute the electric tidal tensor
  $E^i_{\mu\nu}=R^i_{\mu\alpha\nu\beta}u^\alpha u^\beta$ of a single on-shell
  graviton $(k_i,\varepsilon_i)$ in terms of its field strength
  \eqref{eq:lec2:F}, confirming $E_{\mu\nu}=\sqrt{8\pi G}\,W_\mu W_\nu$, and
  show that the two-graviton contraction takes the form,
  \[
    E^1_{\mu\nu}\,E^{2\,\mu\nu}
    =8\pi G\,(W_1\cdot W_2)^2 ,
  \]
  so the leading tidal amplitude is $\mathcal{M}^{\rm tidal}\propto
  G\lambda\,(W_1\cdot W_2)^2$, the identification \eqref{eq:lec2:alpha-lambda}
  with its pure number: the same gauge-invariant structure as the Compton
  amplitude \eqref{eq:lec2:compton}, now with no poles, the signature of a
  contact term.
  \item[(b)] \emph{An EMRI's smaller body: does its tide matter for LISA?} A
  neutron star of mass $m$ spirals into a massive black hole $M\gg m$,
  deforming in the hole's tidal field $E\sim GM/r^3$ with deformability
  $\lambda\sim (Gm/c^2)^5/G$, cf.~\eqref{eq:lec2:love}. Estimate the star's
  tidal contribution to the binding energy relative to the leading dynamics in
  the strong-field regime that dominates the LISA signal ($r\sim GM/c^2$,
  $v/c\sim1$). At what order in the mass ratio $q=m/M$ of \eqref{eq:lec1:q}
  does it enter? LISA-quality EMRI waveforms require the dynamics through
  second order in the self-force expansion~\cite{Wardell:2021fyy} (successive
  powers of $q$); is the
  smaller body's tide relevant?
\end{enumerate}

\item \textbf{Multipole expansion revisited.} The tidal operator above is
quadratic in the curvature. Its linear-in-curvature parent is the multipole
coupling \eqref{eq:lec2:multipole} of a quadrupole moment $Q^{\mu\nu}(\tau)$ to
the tidal field,
\[
  S_Q=-\tfrac12\int d\tau\;Q^{\mu\nu}\,E_{\mu\nu},
\]
with $E_{\mu\nu}$ linearized as in Exercise~2. Recall the de~Donder
propagator $iP_{\mu\nu,\alpha\beta}/q^2$, with $P_{\mu\nu,\alpha\beta}=\tfrac12(
\eta_{\mu\alpha}\eta_{\nu\beta}+\eta_{\mu\beta}\eta_{\nu\alpha}
-\eta_{\mu\nu}\eta_{\alpha\beta})$, which returns in Lecture~3.
\begin{enumerate}
  \item[(a)] Vary $S_Q$ with respect to $h_{\mu\nu}$ to read off the effective
  source $T_Q^{\mu\nu}(x)$ that the quadrupole adds to the linearized Einstein
  equations. How does its derivative structure differ from the point-mass
  monopole source \eqref{eq:lec2:T}, and why does that difference identify
  a quadrupole?
  \item[(b)] For a static quadrupole at rest ($u^\mu=(1,\vect 0)$, $Q^{ij}$
  constant), solve the linearized field equation with the de~Donder propagator
  and find the Newtonian potential $\Phi(\vect x)$ it sources. How does it fall
  off with distance compared with the monopole? \emph{Hint:} in momentum space
  the source is $\propto q_i q_j Q^{ij}$; Fourier-transform against the $1/q^2$
  propagator.
  \item[(c)] \emph{An intrinsic quadrupole: at what PN order?} Suppose the body
  carries a permanent quadrupole $Q$ rather than an induced one, so its
  coupling to the companion's tidal field $E\sim Gm_2/r^3$ is now \emph{linear}
  in $E$, $U_Q\sim Q\,Gm_2/r^3$. Taking the natural compact scaling
  $Q\sim m_1 R_1^2$ with $R_1\sim Gm_1/c^2$, estimate the PN order at which it
  enters the dynamics. How does this compare with the induced tide
  \eqref{eq:lec2:5PN}, and what accounts for the difference?
\end{enumerate}

\end{enumerate}

\section{Conservative dynamics}
Now that we understand the effective long-wavelength description of a compact
body,
we are ready to tackle the problem of motion of two bodies. For the moment, we
will ignore radiation and focus on the conservative motion, which is justified
by the hierarchy of scales we explained in Lecture~1.
Instead of solving for the detailed motion of each body (which depends on a
choice of coordinates),
we will focus on a gauge-invariant quantity, the Hamilton--Jacobi on-shell
action, which encodes a large number of observables as a function of
scattering data. Using this function, we will be able to calculate the
observables relevant for all four classical tests of General Relativity, as well
as the leading corrections to the Newtonian potential and the energy of a
circular orbit.

\subsection{Scattering and the on-shell action}
\label{sec:lec3:setup}

\begin{figure}[t]
  \centering
\begin{tikzpicture}[
  x=1.2cm, y=1.2cm,
  worldline/.style={line width=0.9pt},
  free/.style={line width=0.5pt, dash pattern=on 3pt off 2.2pt},
  vel/.style={line width=1.1pt,-{Stealth[length=2.4mm]}},
  vec/.style={line width=0.8pt,-{Stealth[length=2.2mm]}},
  lbl/.style={font=\footnotesize}
]


\draw[free] (-3.45,0.95) -- (3.0,0.95);
\draw[free] (-3.45,-0.95) -- (3.0,-0.95);
\draw[free] (1.0,0.625) -- (2.95,-0.009);

\draw[worldline] plot[domain=-2.6:2.6, samples=121, smooth]
  (\x, {0.95 - 0.16245*(\x + sqrt(\x*\x + 3.24))});
\draw[worldline] plot[domain=-2.6:2.6, samples=121, smooth]
  (\x, {-0.95 + 0.16245*(sqrt(\x*\x + 3.24) - \x)});

\draw[vel] (-2.2,0.66) -- (-1.35,0.66);
\node[lbl] at (-1.78,0.49) {$u_1$};
\draw[vel] (2.2,-0.66) -- (1.35,-0.66);
\node[lbl] at (1.78,-0.49) {$u_2$};

\draw[vec] (-3.1,-0.95) -- (-3.1,0.95);
\node[lbl,anchor=east] at (-3.2,0) {$b^\mu$};

\draw[line width=0.5pt] (1.9,0.95) arc[start angle=0, end angle=-18,
radius=1.9];
\node[lbl] at (2.14,0.67) {$\chi$};

\end{tikzpicture}
  \caption{A scattering encounter in the center-of-mass frame.  $\chi$ is the
  scattering angle.}
  \label{fig:lec3:geometry}
\end{figure}
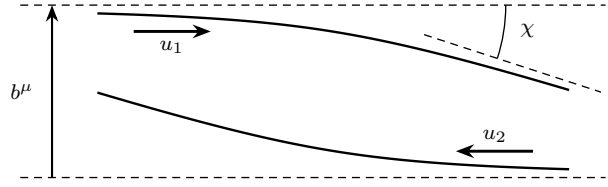

We focus our attention first on unbound encounters, in which the two bodies come
in from infinity, interact, and then separate again and travel along free
trajectories (see Figure~\ref{fig:lec3:geometry}). With radiation neglected,
energy and angular momentum are conserved, so the state long after the
encounter differs from the initial state only by a rotation in
the plane of the motion through the scattering angle $\chi$, a time delay
$\Delta t$ and also a proper-time delay $\Delta\tau_i$ for each body. These are
the observables we will extract from the on-shell action.

The kinematics of this process is fixed by data at infinity, where each body
travels on a straight worldline
\be
  x_i^\mu(\tau)=b_i^\mu+u_i^\mu\tau,
  \label{eq:lec3:worldlines}
\ee
with constant time-like four-velocity $u_i^\mu$, with $u_i^2=-1$. Two unit
timelike vectors form a single invariant, the relative Lorentz factor
\be
  \gamma\;\equiv\;-\,u_1\!\cdot\!u_2\;=\;\frac{1}{\sqrt{1-v_{\rm
  rel}^2}}\;\geq\;1,
  \label{eq:lec3:gamma}
\ee
with $v_{\rm rel}$ the relative velocity measured in either body's rest frame.
The offsets $b_i^\mu$ are each defined only up to a shift along $u_i^\mu$, which
merely resets the origin of proper time, so the physical information they carry
is the impact
parameter $b^\mu=b_1^\mu-b_2^\mu$ projected orthogonally to both velocities: a
spacelike vector in the two-plane transverse to $u_1$ and $u_2$, of length
$b=\sqrt{b\cdot b}$. Together with the masses, $(\gamma,b)$ fixes the incoming
state completely.

The same data have a center-of-mass description. Squaring the total momentum
gives the center-of-mass energy,
\be
  E^2=-(m_1u_1+m_2 u_2)^2=m_1^2+m_2^2+2 m_1 m_2\,\gamma,
  \label{eq:lec3:Ecm}
\ee
and in that frame the two spatial momenta are equal and
opposite with common magnitude $p_\infty$,
\be
  p_\infty=\frac{m_1 m_2\sqrt{\gamma^2-1}}{E},
  \label{eq:lec3:pinf}
\ee
so the angular momentum is $J=p_\infty b$. The pairs $(\gamma,b)$ and $(E,J)$
are
interchangeable, and we use whichever is more convenient.

In the center-of-mass frame the conservative relative motion is a one-body
problem, a
separation $\vect r$ and its conjugate momentum $\vect p$. Whatever the
dynamics, conservation of the energy $E$ and of the angular momentum
$J=|\vect r\times\vect p|$ reduces it to a single function of the separation,
the squared momentum $\vect p^2(r,E)$ carried at radius $r=|\vect r|$, which
tends to $p_\infty^2$ as $r\to\infty$. Splitting off the angular part leaves
the radial momentum,
\be
  p_r^2=\vect p^2(r,E)-\frac{J^2}{r^2},
  \label{eq:lec3:pr-def}
\ee
which vanishes at the turning points of the orbit, the periastron $r_{\min}$
for an unbound encounter.

The \emph{on-shell action}, $\tilde S$, is the action evaluated on the solution
of the equations of motion, the phase the interaction accumulates along the two
trajectories. For the relative motion it is the phase integral of the radial
momentum, with the free propagation subtracted,
\be
  \tilde S(E,J)
  =2\int_{r_{\min}}^{\infty}\!dr\;p_r(r,E,J)
  \;-\;2\int_{J/p_\infty}^{\infty}\!dr\;p_r^{\rm free}(r,E,J),
  \label{eq:lec3:S-radial}
\ee
the factor of $2$ counting the incoming and outgoing branches, and the
subtraction being
the same quantity for free motion at the same $E$ and $J$, with
$p_r^{\rm free}=\sqrt{p_\infty^2-J^2/r^2}$ and turning point $J/p_\infty$. 

The on-shell action is a Hamilton--Jacobi function, so its first derivatives
are the displacements conjugate to the variables it depends on, as the
derivatives of a free energy are the entropy and the pressure. The boundary
data of the scattering form the conjugate pairs $(\phi,J)$ for angle and
angular momentum, $(t,E)$ for time and energy, and $(\tau_i,m_i)$ for each
body's proper time and mass. Differentiating with respect to $J$, $E$ and $m_i$
gives the three observables listed above,
\be
  \chi=-\frac{\p\tilde S}{\p J},
  \qquad
  \Delta t=\frac{\p\tilde S}{\p E},
  \qquad
  \Delta\tau_i=-\frac{\p\tilde S}{\p m_i}\bigg|_{E,J}.
  \label{eq:lec3:generating}
\ee

These relations can be checked directly against \eqref{eq:lec3:S-radial}.
Differentiating under the integral sign with respect to $J$ at
fixed $E$ (the boundary terms are absent because $p_r$ vanishes at the turning
points) yields the rate at which the
azimuth advances along the orbit,
\be
  -\frac{\p p_r}{\p J}\bigg|_{E,r}=\frac{J}{r^2 p_r}=\frac{d\phi}{dr},
  \label{eq:lec3:azimuth}
\ee
so each integral is the azimuth swept along its orbit, and the difference of the
two is the scattering angle $\chi$. The second relation follows in the same way.
At fixed $(r,J)$ the energy derivative
of the radial momentum is the rate at which coordinate time elapses along the
orbit,
\be
  \frac{\p p_r}{\p E}\bigg|_{r,J}=\frac{dt}{dr},
  \label{eq:lec3:elapsed}
\ee
the analogue of \eqref{eq:lec3:azimuth} for the time, so
$\p\tilde S/\p E$ is the time the true orbit takes minus the time the straight
line takes at the same $E$ and $J$. 

The third relation follows because each worldline enters the action as
$-m_i\!\int d\tau_i$, so the mass derivative gives the proper time that body
accumulates, again in excess of free flight.

\subsection{The simplest gravitational interaction}
\label{sec:lec3:glue}

Now that we understand our target, the on-shell action, let us actually compute
it. We will do so at leading order in the post-Minkowskian expansion.

Take the straight worldlines \eqref{eq:lec3:worldlines} as the motion at zeroth
order in $G$; the leading interaction is the first correction to that free
motion. Each body contributes a stress-tensor source \eqref{eq:lec2:T} of
Lecture~2, and at order $G$ they interact by exchanging a single graviton,
sourced by either body.
Substituting the leading-order solution for the gravitational field back into
the action gives
$\tilde S$ at order $G$. Equivalently, we can compute the action by gluing the
two
momentum-space sources, the Fourier transforms \eqref{eq:lec2:Tk} of the
straight-worldline stress tensors at momentum $q$, to the ends of the
graviton propagator $iP_{\mu\nu,\alpha\beta}/q^2$:
\be
  \tilde S
  =8\pi G\int\!\frac{d^4q}{(2\pi)^4}\;
   \tilde T_1^{\mu\nu}(q)\,\frac{P_{\mu\nu,\alpha\beta}}{q^2}\,
   \tilde T_2^{\alpha\beta}(-q),
  \qquad
  \tilde T_i^{\mu\nu}(q)=m_i\,u_i^\mu u_i^\nu\,e^{-iq\cdot
  b_i}\,2\pi\,\delta(q\cdot u_i),
  \label{eq:lec3:glue}
\ee
where we used de Donder gauge with
\be
  P_{\mu\nu,\alpha\beta}
  =\tfrac12\big(\eta_{\mu\alpha}\eta_{\nu\beta}
              +\eta_{\mu\beta}\eta_{\nu\alpha}
              -\eta_{\mu\nu}\eta_{\alpha\beta}\big).
  \label{eq:lec3:prop}
\ee
The pole describes the long-range gravitational force, and the numerator can
also be obtained in a gauge-invariant way, as the polarization sum
$\Lambda_{\mu\nu,\alpha\beta}$ of \eqref{eq:lec2:polsum} over the exchanged
graviton.

Each source carries $u_i^\mu u_i^\nu$, so the entire kinematics reduces to
the contraction of two such tensors through
\eqref{eq:lec3:prop},
\be
  u_1^\mu u_1^\nu\,\Lambda_{\mu\nu,\alpha\beta}\,u_2^\alpha u_2^\beta = u_1^\mu
  u_1^\nu\,P_{\mu\nu,\alpha\beta}\,u_2^\alpha u_2^\beta
  =(u_1\!\cdot\!u_2)^2-\tfrac12\,u_1^2\,u_2^2
  =\tfrac12\big(2\gamma^2-1\big),
  \label{eq:lec3:contraction}
\ee
where we used $u_i^2=-1$ and the relative Lorentz factor
\eqref{eq:lec3:gamma}. Note that between conserved
sources most of the structure of $\Lambda_{\mu\nu,\alpha\beta}$ drops out: every
term
beyond the plain metrics carries at least one factor of $k$, which contracts
to zero against a conserved source, and $\Lambda$ reduces to the de Donder
numerator \eqref{eq:lec3:prop}.

Putting everything together, the on-shell action is
\be
  \tilde S
  =4\pi G\,m_1 m_2\,(2\gamma^2-1)
   \int\!\frac{d^4q}{(2\pi)^2}\,
   \delta(q\cdot u_1)\,\delta(q\cdot u_2)\,\frac{e^{-iq\cdot b}}{q^2}.
  \label{eq:lec3:glue-reduced}
\ee
The delta functions force $q$ into the
two-plane orthogonal to both velocities, the transverse plane containing
$b^\mu$,
\be
  \int\!
  \frac{d^4q}{(2\pi)^2}\,\delta(q\!\cdot\!u_1)\,\delta(q\!\cdot\!u_2)\,(\cdots)
  =\frac{1}{\sqrt{\gamma^2-1}}\int\! \frac{d^2q_\perp}{(2\pi)^2}\,(\cdots).
  \label{eq:lec3:gram}
\ee
The Jacobian from the two delta functions is the square root of (minus) the Gram
determinant of the velocities, $-\det(u_i\cdot u_j)=\gamma^2-1$. The remaining
integral is a two-dimensional Fourier transform of
the massless propagator, which is logarithmically divergent at small $q_\perp$.
This divergence is simply the infinite range of the
$1/r$ force seen in momentum space. We can regulate it by discarding the
exchange of transverse wavelengths longer
than $b_c\gg b$, i.e., by cutting off the integral at
$|q_\perp|=1/b_c$, which gives
\be
  \int_{|q_\perp|>1/b_c}\!\frac{d^2q_\perp}{(2\pi)^2}\,
  \frac{e^{-iq_\perp\cdot b}}{q_\perp^2}
  =-\frac{1}{2\pi}\,\Big[\ln\frac{b}{2b_c}+\gamma_E\Big],
  \label{eq:lec3:2dft}
\ee
with $\gamma_E$ the Euler-Mascheroni constant. Neither it nor the particular
value of $b_c$ is physical: both can be absorbed into an additive constant in
the action, which does not affect any observables.
Assembling \eqref{eq:lec3:glue-reduced}--\eqref{eq:lec3:2dft} and dropping
$\gamma_E$ gives the leading-order on-shell action
\be
  \tilde S(\gamma,b)
  =-\frac{2G m_1 m_2\,(2\gamma^2-1)}{\sqrt{\gamma^2-1}}\,\ln\frac{b}{2b_c},
  \label{eq:lec3:eikonal}
\ee
or, equivalently, in terms of the center-of-mass variables $(E,J)$,
\be
  \tilde S(E,J)
  =-\frac{2G m_1 m_2\,(2\gamma^2-1)}{\sqrt{\gamma^2-1}}\,
    \ln\frac{J}{2p_\infty b_c}.
  \label{eq:lec3:eikonal-J}
\ee

\subsection{Tests of General Relativity from the on-shell action}
\label{sec:lec3:observables}

As explained above, each observable is now simply obtained by taking a
derivative of \eqref{eq:lec3:eikonal}. We would like to see how the four
classical tests of General Relativity emerge from this simple calculation. Some
of them,
however, are properties of bound orbits, while \eqref{eq:lec3:eikonal} describes
a scattering encounter. The counterpart of \eqref{eq:lec3:S-radial} for bound
orbits is the phase
integral over one radial period, taken between the two turning points
$r_{\min}$ and $r_{\max}$,
\be
  \tilde S^{\rm bound}(E,J)
  =2\int_{r_{\min}}^{r_{\max}}\!dr\;p_r(r,E,J)\;+\;2\pi J,
  \label{eq:lec3:S-bound-def}
\ee
where the $2\pi J$ takes
the place of the free propagation: the orbit sweeps $2\pi$ per radial period.
Its derivatives are the bound observables, taken in
the same variables as \eqref{eq:lec3:generating},
\be
  \Delta\phi=-\frac{\p\tilde S^{\rm bound}}{\p J},
  \qquad
  T_r=\frac{\p\tilde S^{\rm bound}}{\p E},
  \qquad
  \tau_i=-\frac{\p\tilde S^{\rm bound}}{\p m_i}\bigg|_{E,J},
  \label{eq:lec3:bound-generating}
\ee
the periastron advance $\Delta\phi$, the radial period $T_r$, and the proper
time each
body accumulates over an orbital period, $\tau_i$.

Fortunately, we do not need to compute the bound action directly: in the
conservative sector the scattering data already determine the bound dynamics,
through the \emph{boundary-to-bound} map~\cite{Kalin:2019rwq,Kalin:2019inp},
extended to the radiative sector in~\cite{Cho:2021arx}.
The existence of such a map is already hinted at by \eqref{eq:lec3:pr-def},
which
makes no reference to any particular orbit: the function $\vect p^2(r,E)$ is the
same for scattering and for bound orbits. The two cases
differ only in the range of $r$ the motion covers, set by the
turning points of the integral, where $p_r$ vanishes. The phase integrals
\eqref{eq:lec3:S-radial} and \eqref{eq:lec3:S-bound-def} therefore share the
integrand but differ in their endpoints, so it is enough to follow the turning
points, which are roots of the same function, as the energy is lowered through
$E=M$.

Expanding $\vect p^2$ in the weak-field limit gives a post-Minkowskian series in
$G$,
\be
\vect p^2(r,E)=p_\infty^2+\sum_{k\geq1}P_k(E)\frac{G^k}{r^k},
\label{eq:lec3:p2-expansion}
\ee
and keeping the
leading term gives the radial momentum the familiar Newtonian shape,
\be
  p_r(r)=\sqrt{\,p_\infty^2+\frac{P_1 G}{r}-\frac{J^2}{r^2}\,},
  \label{eq:lec3:pr}
\ee
which vanishes where the quadratic $p_\infty^2r^2+P_1Gr-J^2$ does. For
scattering orbits the energy exceeds the total rest mass, $E>M$, so
$p_\infty^2>0$ and only one of the two roots is positive,
\be
  r_{\min}=\frac{-P_1 G+\sqrt{P_1^2 G^2+4 p_\infty^2 J^2}}{2 p_\infty^2} =
  \frac{2J^2}{P_1G+\sqrt{P_1^2G^2+4p_\infty^2J^2}},
  \label{eq:lec3:rmin}
\ee
the periastron of a hyperbola of eccentricity
$e=\sqrt{1+4p_\infty^2J^2/P_1^2G^2}>1$, while the other root is negative and
unphysical.

\begin{figure}[t]
  \centering
\subfloat[]{%
\begin{tikzpicture}[>={Stealth[length=4pt,width=3pt]},line
join=round,font=\footnotesize]

\tikzset{
  hyp/.style   ={black,line width=0.7pt},
  par/.style   ={black,line width=0.7pt,dash pattern=on 3pt off 2pt},
  ell/.style   ={black,line width=1.0pt,dash pattern=on 0.4pt off 2.2pt,line
  cap=round},
  guide/.style ={black!55,line width=0.35pt,densely dotted},
  ax/.style    ={black,line width=0.5pt},
  cut/.style   ={red!70!black,line width=0.6pt,decorate,
                 decoration={snake,amplitude=0.7pt,segment length=3.5pt}},  
  root/.style  ={fill=black,draw=black,circle,inner sep=0pt,minimum size=3pt},
  xroot/.style ={fill=white,draw=black,circle,inner sep=0pt,minimum size=3pt},
}

\def\pp{0.85}
\def\eh{2.4}    
\def\eell{0.60}   

\begin{scope}

  \draw[guide] (-2.20,0) -- (0.70,0);

  \draw[hyp] plot[variable=\t,domain=-105.5:105.5,samples=201,smooth]
    ({\pp/(1+\eh*cos(\t))*cos(\t)},{\pp/(1+\eh*cos(\t))*sin(\t)});
  \draw[par] plot[variable=\t,domain=-139.5:139.5,samples=201,smooth]
    ({\pp/(1+cos(\t))*cos(\t)},{\pp/(1+cos(\t))*sin(\t)});
  \draw[ell] plot[variable=\t,domain=0:360,samples=201,smooth]
    ({\pp/(1+\eell*cos(\t))*cos(\t)},{\pp/(1+\eell*cos(\t))*sin(\t)});

  \node[anchor=west,align=left,inner sep=1pt]        at (0.02,1.85)
  {hyperbola\\$e>1$};
  \node[anchor=north east,align=right,inner sep=1pt] at (-2.10,1.98)
  {parabola\\$e=1$};
  \node[anchor=north east,align=right,inner sep=1pt] at (-2.25,-0.06)
  {ellipse\\$e<1$};

  \node[root] at (0,0) {};
  \node[anchor=south east,inner sep=1.5pt] at (-0.05,0.04) {focus};

  \foreach \rmin in {0.250,0.425,0.531}{\node[root,minimum size=2.4pt] at
  (\rmin,0) {};}
  \draw[guide] (0.70,0.22) -- (0.55,0.05);
  \node[anchor=west,inner sep=1.5pt] at (0.68,0.24) {$r_{\min}$};
  \node[root,minimum size=2.4pt] at (-2.125,0) {};
  \node[anchor=south east,inner sep=1.5pt] at (-2.16,0.06) {$r_{\max}$};

  \draw[ax,->] (-0.05,-0.60) -- (0.80,-0.60);
  \node[anchor=west,align=left,inner sep=2pt] at (0.83,-0.60) {decreasing\\$E$};

\end{scope}

\end{tikzpicture}}
  \hspace{1.5em}
  \subfloat[]{%
\begin{tikzpicture}[>={Stealth[length=4pt,width=3pt]},line
join=round,font=\footnotesize]

\tikzset{
  hyp/.style   ={black,line width=0.7pt},
  par/.style   ={black,line width=0.7pt,dash pattern=on 3pt off 2pt},
  ell/.style   ={black,line width=1.0pt,dash pattern=on 0.4pt off 2.2pt,line
  cap=round},
  guide/.style ={black!55,line width=0.35pt,densely dotted},
  ax/.style    ={black,line width=0.5pt},
  cut/.style   ={red!70!black,line width=0.6pt,decorate,
                 decoration={snake,amplitude=0.7pt,segment length=3.5pt}},  
  root/.style  ={fill=black,draw=black,circle,inner sep=0pt,minimum size=3pt},
  xroot/.style ={fill=white,draw=black,circle,inner sep=0pt,minimum size=3pt},
}

\def\pp{0.85}
\def\eh{2.4}    
\def\eell{0.60}   

\begin{scope}

  \def\xl{-2.40}   
  \def\xr{4.40}    
  \def\ya{2.15}    
  \def\yb{0.00}    
  \def\yc{-2.15}   

  \foreach \y in {\ya,\yb,\yc}{
    \fill[black!8] (\xl,\y-0.52) rectangle (0,\y+0.52);
    \draw[ax,->] (\xl,\y) -- (\xr,\y) node[anchor=west,inner sep=2pt] {$r$};
    \draw[ax] (0,\y-0.08) -- (0,\y+0.08);
    \node[anchor=north,inner sep=2pt] at (0,\y-0.07) {$0$};
  }
  \node[anchor=south west,inner sep=1.5pt,black!60] at (\xl+0.06,\yc-0.43)
  {$r<0$};

  \node[anchor=south west,inner sep=1pt] at (\xl,\ya+0.72) {$p_\infty^2>0$
  ($E>M$): scattering};
  \draw[cut] (1.00,\ya) -- (\xr-0.12,\ya);
  \draw[cut] (\xl+0.10,\ya) -- (-1.35,\ya);
  \node[root] at (1.00,\ya) {};
  \node[anchor=south,inner sep=2.5pt] at (1.00,\ya+0.03) {$r_{\min}$};
  \node[anchor=south,inner sep=2.5pt] at (2.95,\ya+0.03) {$p_r^2\geq0$};
  \draw[cut] (\xl,\ya) -- (-1.35,\ya);
  \node[xroot] at (-1.35,\ya) {};
  \node[anchor=north,inner sep=3pt] at (-1.30,\ya-0.10) {unphysical};

  \node[anchor=south west,inner sep=1pt] at (\xl,\yb+0.72) {$p_\infty^2=0$
  ($E=M$): threshold};
  \draw[cut] (1.45,\yb) -- (\xr-0.12,\yb);
  \node[root] at (1.45,\yb) {};
  \node[anchor=south,inner sep=2.5pt] at (1.45,\yb+0.03) {$r_{\min}$};
  \draw[ax,->] (-1.35,\yb-0.24) -- (\xl+0.04,\yb-0.24);
  \node[anchor=north west,inner sep=2.5pt] at (0.40,\yb-0.44) {second root at
  $r=\pm\infty$};

  \node[anchor=south west,inner sep=1pt] at (\xl,\yc+0.72) {$p_\infty^2<0$
  ($E<M$): bound};
  \draw[cut] (1.85,\yc) -- (3.55,\yc);
  \node[root] at (1.85,\yc) {};
  \node[root] at (3.55,\yc) {};
  \node[anchor=south,inner sep=2.5pt] at (1.85,\yc+0.03) {$r_{\min}$};
  \node[anchor=south,inner sep=2.5pt] at (3.55,\yc+0.03) {$r_{\max}$};
  \draw[ax,->] (\xr-0.12,\yc-0.24) -- (3.68,\yc-0.24);       
  \node[anchor=north east,inner sep=2.5pt] at (\xr-0.10,\yc-0.30) {back from
  $+\infty$};

  \draw[guide,->] (1.02,\ya-0.18) -- (1.40,\yb+0.48);        
  \draw[guide,->] (1.62,\yb-0.88) -- (1.83,\yc+0.48);
  \draw[guide,->] (-1.38,\ya-0.62) -- (-1.88,\yb+0.48);      
  \node[anchor=west,inner sep=1pt,black!60] at (1.55,\ya-0.48) {as $E$
  decreases};

\end{scope}

\end{tikzpicture}}
  \caption{Lowering the energy through $E=M$ closes the orbit (a) as the second
  turning point moves through infinity (b). The wiggly line in (b) is the branch
  cut of $p_r$, the stretch of the axis with $p_r^2\geq0$.}
  \label{fig:lec3:turning}
\end{figure}
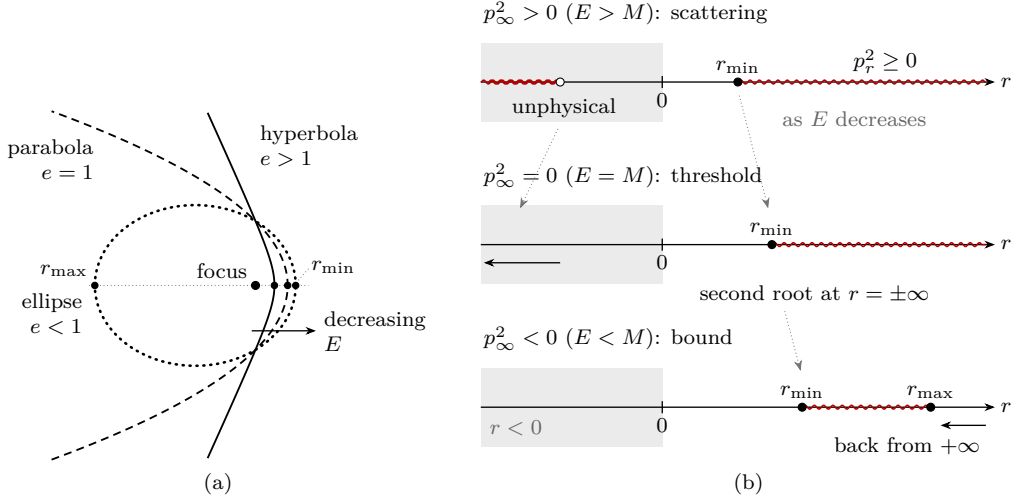

As the energy is lowered the periastron moves smoothly, so it is analytic in the
energy across
$E=M$ and the scattering periastron simply becomes the bound one. The negative
root, on the other hand, runs off to $-\infty$ as $p_\infty^2\to0$ and
returns from $+\infty$ once $p_\infty^2<0$. At threshold itself $e=1$ and the
orbit
is a parabola whose second turning point sits at infinity; below it both roots
are positive, the second is the apoastron $r_{\max}$, and the orbit closes into
an ellipse with $e<1$ (Figure~\ref{fig:lec3:turning}). The higher $P_k$
displace the turning points but do not change this picture, order by order
in $G$.

Both \eqref{eq:lec3:S-radial} and \eqref{eq:lec3:S-bound-def} integrate the
same $p_r$, and they differ only in the limits of integration: the scattering
integral runs outward from the periastron, the bound one from the periastron to
the apoastron. Since
\[
  \int_{r_{\min}}^{\infty}-\int_{r_{\max}}^{\infty}
  =\int_{r_{\min}}^{r_{\max}},
\]
the bound integral is the difference of two scattering integrals, one started at
the periastron and one at the apoastron, and it remains to identify the latter.
Under $J\to-J$ the radial momentum, which contains $J$ only as
$J^2$, is unchanged. The turning points are not: in the post-Minkowskian
expansion the square root in \eqref{eq:lec3:rmin} is expanded about its free
value $2p_\infty J$, which is odd in $J$, so the two roots differ by the sign
of $2p_\infty J$ and periastron and apoastron trade places as $J\to -J$. That
makes $\tilde S(E,-J)$ the scattering integral taken from the apoastron, and the
difference of the two the bound phase integral
\be
  \tilde S^{\rm bound}(E,J)=\tilde S(E<M,J)-\tilde S(E<M,-J),
  \label{eq:lec3:b2b}
\ee
where $E<M$ is understood as the analytic continuation of the scattering action
below threshold.
This relation allows us to compute the bound action from the scattering one.

Let us apply the map at 1PM, to the $\tilde S$ of \eqref{eq:lec3:eikonal}.
Antisymmetrizing, the cutoff drops out because $\ln(p_\infty b_c)$ is
independent of
$J$, and the logarithm contributes $\ln J-\ln(-J)=-i\pi$;
continuing $p_\infty\to i\sqrt{-p_\infty^2}$ onto the bound branch then makes
the result real,
\be
  \tilde S^{\rm bound}(E,J)
  =\frac{2\pi G\,m_1 m_2\,(2\gamma^2-1)}{\sqrt{1-\gamma^2}}
  =\frac{2\pi G\,m_1^2 m_2^2\,(2\gamma^2-1)}{E\,\sqrt{-p_\infty^2}}.
  \label{eq:lec3:S-bound}
\ee
Direct integration of $p_r$ between the turning points
gives the same result (see Exercise~2). With this in hand, we can now compute
the four classical tests of General Relativity.

\paragraph{The scattering angle and the bending of light.} Using our dictionary
for extracting observables from the on-shell action, the
angular-momentum derivative of the unbound action gives the deflection angle,
\be
  \chi\;=\;-\frac{\partial\tilde S}{\partial J}
  \;=\;\frac{2G m_1 m_2\,(2\gamma^2-1)}{J\,\sqrt{\gamma^2-1}}
  \;+\;\mathcal{O}(G^2).
  \label{eq:lec3:angle}
\ee
It is exact in velocity at order $G$, and is the leading term of the
post-Minkowskian expansion of Lecture~1, first obtained by
Westpfahl~\cite{Westpfahl:1979gu,Westpfahl:1985tsl}.

Now consider the following limit: send body~1 past a stationary mass at
nearly the speed of light. Take the probe limit first, $E_1\ll m_2=M$, with
$E_1=\gamma m_1$ the energy of body~1 in the rest frame of body~2: the recoil
of body~2 is then negligible, so it stays at rest, the center-of-mass frame is
its rest frame with $E\simeq M$, and the relative Lorentz factor of
\eqref{eq:lec3:gamma} is body~1's own. Now let body~1 approach the speed of
light, $\gamma\to\infty$ at fixed $E_1$, so that $m_1\to0$. In this limit
$p_\infty=m_1 M\sqrt{\gamma^2-1}/E\to E_1$, so $J\to E_1 b$, and
$2\gamma^2-1\to2\gamma^2$.
Substituting into \eqref{eq:lec3:angle}, the probe's energy cancels, and the
angle reduces to
\be
  \chi\;\longrightarrow\;\frac{4GM}{b}.
  \label{eq:lec3:bending}
\ee
This is the famous general-relativistic light-bending angle.

\paragraph{The Shapiro time delay.} The $E$-derivative in
\eqref{eq:lec3:generating} is the time delay of the trajectory. Here $t$ is the
time of a clock at rest far from the binary, in the frame where
the total momentum has no spatial part, so $\Delta t$ is the delay as read by a
distant observer at rest with respect to the center of mass.  Taken at fixed
$J$, with $\gamma$ and
$p_\infty$ written out through $E$ and the masses, that derivative gives
\be
  \Delta t=-\frac{2G\,\gamma\,(2\gamma^2-3)\,E}{(\gamma^2-1)^{3/2}}
  \,\ln\frac{b}{2b_c},
  \label{eq:lec3:timedelay}
\ee
up to a $b$-independent term, the energy derivative of the $\ln p_\infty$ in
\eqref{eq:lec3:eikonal-J}, which drops out of any difference between impact
parameters. Now
send the same fast particle past $M$. In the massless probe limit,
$\gamma\to\infty$ at fixed $E\to M$, the kinematic factor reduces to $2M$ and
\be
  \Delta t=-4GM\,\ln\frac{b}{2b_c}
  =2GM\,\ln\!\frac{4b_c^2}{b^2},
  \label{eq:lec3:shapiro}
\ee
the Shapiro time delay~\cite{Shapiro:1964uw}, the excess light-travel time a
signal accrues skimming a mass at impact parameter $b$. The standard form of the
result is $2GM\ln(4r_1r_2/b^2)$, written between endpoints at distances
$r_1,r_2$ on either side of the mass, and comparing the two identifies
$b_c=\sqrt{r_1r_2}$: the infrared cutoff is the baseline over which the delay
accumulates. Note that the time delay diverges for an infinite baseline, so the
physically measurable quantity is the difference in time delay between two
different impact parameters, which is finite and independent of the cutoff.

\paragraph{The gravitational redshift.} The mass derivative in
\eqref{eq:lec3:generating} is the excess proper time accumulated by body~$i$
during the encounter, taken at fixed $E$ and $J$, the arguments of the
Hamilton--Jacobi function. 
At 1PM,
\be
  \Delta\tau_1=-\frac{2G\,\big[\gamma\,(2\gamma^2-3)\,m_1-m_2\big]}
  {(\gamma^2-1)^{3/2}}\,\ln\frac{b}{2b_c},
  \label{eq:lec3:redshift}
\ee
and $\Delta\tau_2$ with $1\leftrightarrow2$; only the $\ln b$ is physical, as
for the time delay. Comparing with \eqref{eq:lec3:timedelay},
\be
  \Delta\tau_i=\frac{m_i}{E}\,\Delta t
  +\frac{2G\,m_j}{(\gamma^2-1)^{3/2}}\,\ln\frac{b}{2b_c},
  \qquad j\neq i,
  \label{eq:lec3:tau-vs-t}
\ee
exactly at this order. Body~$i$ takes its rest-mass share of the
coordinate-time delay common to both worldlines, and what is left over is fixed
by the companion's mass alone. As a check, in the ultrarelativistic probe limit
taken for the
bending \eqref{eq:lec3:bending}, $\Delta\tau_1$ vanishes, as it must: a massless
probe accrues
no proper time.

On the bound branch the same derivative, now of
\eqref{eq:lec3:bound-generating}, gives the proper time body~1 accumulates over
one radial period,
\be
  \tau_1=-\frac{\p\tilde S^{\rm bound}}{\p m_1}\bigg|_{E,J}
  =-\frac{2\pi G\,\big[\,\gamma\,(2\gamma^2-3)\,m_1-m_2\,\big]}
          {(1-\gamma^2)^{3/2}}.
  \label{eq:lec3:tau-bound}
\ee
The coordinate time over the same period is the
$E$-derivative,
\be
  T_r=\frac{\p\tilde S^{\rm bound}}{\p E}
  =-\frac{2\pi G\,E\,\gamma\,(2\gamma^2-3)}{(1-\gamma^2)^{3/2}},
  \label{eq:lec3:Tr}
\ee
the relativistic Kepler period (Exercise~3), and the ratio of the two is the
orbit average of $d\tau_1/dt$, often called the \emph{Detweiler
redshift}~\cite{Detweiler:2008ft,LeTiec:2011ab,Barack:2009ey},
\be
  z_1=\frac{\tau_1}{T_r}=\frac{m_1}{E}-\frac{m_2}{E\,\gamma\,(2\gamma^2-3)}.
  \label{eq:lec3:z1}
\ee
To read \eqref{eq:lec3:z1} as something
familiar, make body~1 a probe of $m_2\simeq M$ on a circular orbit of radius $a$ and
let the motion be slow, so that 
$\gamma\simeq 1+(E-M)/\mu$ with binding energy $E-M=-G\mu M/(2a)$,
retaining $M=m_1+m_2$. Then
\be
  z_1\;\longrightarrow\;1-\frac{3GM}{2a}.
  \label{eq:lec3:zschw}
\ee
This is the redshift for an observer travelling along a circular orbit. A clock
held static at
the same radius runs instead at $1-GM/a$, which is what Pound and Rebka
measured~\cite{Pound:1960zz}; the difference between the two is the time
dilation of the orbital motion.

\paragraph{The periastron advance.} Finally, let us briefly discuss the
periastron advance. The angular-momentum derivative of
\eqref{eq:lec3:S-bound} vanishes, since the 1PM action carries no $J$ at all.
This means that precession is a 2PM
effect, and to see it we would have to work harder and compute the next order
correction to the on-shell action. So in this lecture we will not get the $43''$
per century that the turning
Laplace--Runge--Lenz vector of Mercury traces out, but you will get there in
Exercise~2. The
modern literature goes very much further, and corrections to the on-shell action
(or more generally to the conservative dynamics) have been computed up to the
fifth post-Minkowskian order,
$\cO(G^5)$~\cite{Cheung:2018wkq,Bern:2019nnu,Kalin:2020mvi,Kalin:2020fhe,Bern:2021dqo,Bern:2021yeh,Dlapa:2021npj,Driesse:2024xad,Bern:2024adl,Dlapa:2025biy,Dlapa:2026oyq,Bern:2025zno,Driesse:2026qiz,Bern:2025wyd}.

The classical tests of general relativity (the perihelion advance of Mercury,
the bending of starlight, the gravitational redshift, and Shapiro's later
fourth, the time delay) were each, historically, a separate calculation in its
own setting. Here they all follow from the on-shell action, without reference to
a gauge or coordinate choice and without solving the equations of motion
explicitly.

\subsection{Corrections to Newton's potential and the binding energy}
\label{sec:lec3:matching}

We have had some fun computing observables in a gauge-invariant way and relating
unbound and bound orbits.
An alternative route to connect the orbits is through coordinate-dependent
potentials. Since these are somewhat more familiar let us describe how we can
extract them from the on-shell action.

Let us assume, in the center-of-mass frame, that the conservative relative
dynamics is
governed by a Hamiltonian of the isotropic form,
\be
  H(\vect r,\vect p)
  =\sqrt{\vect p^2+m_1^2}+\sqrt{\vect p^2+m_2^2}+V(r,\vect p^2),
  \qquad
  V=\sum_{k\geq1}\frac{c_k(\vect p^2)}{r^k}\,G^k,
  \label{eq:lec3:hamiltonian}
\ee
with the unknown coefficients $c_k$, one per order in $G$. Solving $H=E$ gives
the
$\vect p^2(r,E)$ of \eqref{eq:lec3:pr-def}, whose leading coefficient in the
expansion \eqref{eq:lec3:p2-expansion} is
\be
  P_1=-\frac{2E_1 E_2}{E}\,c_1 ,
  \label{eq:lec3:P1c1}
\ee
with $E_i\equiv\sqrt{p_\infty^2+m_i^2}$ the asymptotic energies of the two
bodies, $E_1+E_2=E$. The leading correction to the on-shell action, written as
an integral of the radial momentum, is
\be
  \tilde S
  =P_1 G\int_{J/p_\infty}^{b_c}\!\frac{dr}{r\,\sqrt{p_\infty^2-J^2/r^2}}
  +\mathcal{O}(G^2)
  =-\frac{P_1 G}{p_\infty}\,\ln\frac{J}{2p_\infty b_c}+\dots,
  \label{eq:lec3:S-pert}
\ee
where again we had to cut off the integral at large $r$ by $b_c$. Equating this
to \eqref{eq:lec3:eikonal-J} fixes $P_1$,
\be
  \vect p^2(r,E)=p_\infty^2
  +\frac{2G\,m_1^2 m_2^2\,(2\gamma^2-1)}{E\,r}+\mathcal{O}(G^2),
  \label{eq:lec3:p2r}
\ee
and inverting \eqref{eq:lec3:P1c1} gives the potential itself,
$c_1=-m_1^2 m_2^2(2\gamma^2-1)/E_1 E_2$, so the matched 1PM interaction is
\be
  V(r)=-\frac{G\,m_1^2 m_2^2\,(2\gamma^2-1)}{E_1
  E_2}\,\frac{1}{r}+\mathcal{O}(G^2),
  \label{eq:lec3:V1PM}
\ee
with the velocity dependence carried by $2\gamma^2-1$ and $E_1E_2$.

Since \eqref{eq:lec3:V1PM} is exact in velocity at
order $G$, expanding it in $v/c$ gives the leading relativistic correction to
Newton,
\be
  V(r)=-\frac{G m_1
  m_2}{r}\left[\,1+\frac{3+2\nu}{2}\,\frac{v^2}{c^2}+\dots\right]
  +\mathcal{O}(G^2),
  \label{eq:lec3:pn}
\ee
with $v$ the relative velocity and $\nu=\mu/M$. This is not, however, the
full first post-Newtonian potential. Recall from Lecture~1 that a fixed PN order
draws on several PM orders. The virial relation $GM/r\sim v^2$ lifts the
2PM ($G^2/r^2$) term to the same 1PN order, so \eqref{eq:lec3:pn} is only
its $G$-linear piece. 

It is important to stress that while the on-shell action is unique, the
potential
is not, since a coordinate change shifts $V$ while leaving every observable
unchanged. On the other hand, the Hamiltonian describes both bound and unbound
orbits.
Let us use it to compute the binding energy of a quasi-circular inspiral.
Circular means that
the radial momentum vanishes and stays zero, so $p_r^2$ and its radial
derivative vanish together. At 1PM the two conditions
read
\be
  p_\infty^2+\frac{P_1 G}{r}-\frac{J^2}{r^2}=0,
  \qquad
  -\frac{P_1 G}{r^2}+\frac{2J^2}{r^3}=0,
  \label{eq:lec3:circ-cond}
\ee
and they fix the radius and the energy of the orbit carrying angular momentum
$J$,
\be
  r=\frac{2J^2}{P_1 G},
  \qquad
  p_\infty^2(E)=-\frac{G^2 P_1(E)^2}{4J^2}.
  \label{eq:lec3:circ}
\ee
The second equation is the Kepler relation between energy and angular momentum on a
circular orbit, with the Newtonian quantities replaced by the relativistic
$p_\infty^2(E)$ of \eqref{eq:lec3:pinf} and $P_1(E)$ of \eqref{eq:lec3:p2r},
and it is exact in velocity at order $G$. Its right-hand side is negative, so
$E<M$ and the orbit is bound. The radius has dropped out of it, so solving for
$E$ gives the energy of the circular orbit carrying angular momentum $J$, a
relation $E(J)$ along the sequence of circular orbits.

To evolve a waveform we would like to express the energy as a function of the
orbital frequency rather than of $J$. Along that sequence the radius varies as
well, $E(J)=H(r_c(J),0,J)$ with $r_c(J)$ the circular radius of
\eqref{eq:lec3:circ} and the radial momentum held at zero, so differentiating
gives two terms,
\be
  \frac{dE}{dJ}=\frac{\p H}{\p r}\,\frac{dr_c}{dJ}+\frac{\p H}{\p J}
  =\frac{\p H}{\p J}=\Omega ,
  \label{eq:lec3:dEdJ}
\ee
of which the first vanishes because there is no radial force on a circular
orbit, $\p H/\p r=0$, while the second is $\dot\phi$, the orbital frequency, by
Hamilton's equations. Differentiating the $E(J)$ of \eqref{eq:lec3:circ}
therefore gives $\Omega(J)$, and eliminating $J$ between the two leaves
$E(\Omega)$. In the slow-motion limit, this procedure gives the quasicircular
binding energy
\be
  E(\Omega) -M =-\frac{\mu}{2}\,(GM\Omega)^{2/3}
  \left[\,1+\frac{15-\nu}{12}\,(GM\Omega)^{2/3}+\dots\right].
  \label{eq:lec3:EOmega}
\ee
The leading term is the Newtonian binding energy of a circular orbit
\eqref{eq:lec1:EOmega} of Lecture~1 now with its coefficient.
The correction has the same status as \eqref{eq:lec3:pn}: it is the 1PM piece
of a 1PN coefficient. The $G^2/r^2$ term of the potential contributes at
the same PN order. Including it shifts the bracket to the complete 1PN coefficient, $-(9+\nu)/12$~\cite{Blanchet:2013haa}.

Let us finish the lecture by noting that using the Hamiltonian was a convenience
here rather than a necessity. The same binding energy of the binary, $E(\Omega)$, is a gauge-invariant quantity that can be computed directly from the
bound on-shell action. You will explore this in more detail in Exercises~3
and~4.

\subsection{Exercises}
\label{sec:lec3:exercises}

\begin{enumerate}

\item \textbf{The impulse from the angle.} 
\begin{enumerate}
  \item[(a)] Elastic scattering rotates the
  momentum of each particle by an amount known as the impulse, $\Delta p_1^\mu$.
  Show from kinematics that  in the center of mass frame,
  \[
    \frac{|\Delta p_1|}{p_\infty}=2\sin\frac{\chi}{2}\,.
  \]
  \item[(b)] At 1PM the angle is small, $2\sin(\chi/2)\approx\chi$, and the
  deflection is attractive: the impulse points from body~1 toward body~2,
  along $-b^\mu$ with $b^\mu\equiv b_1^\mu-b_2^\mu$ pointing from body~2 to
  body~1. Combining with \eqref{eq:lec3:angle} and $J=p_\infty b$, show that
  \[
    \Delta p_1^\mu
    =-\frac{2G m_1 m_2\,(2\gamma^2-1)}{\sqrt{\gamma^2-1}}\,\frac{b^\mu}{b^2}
    +\cO(G^2).
  \]
  
\end{enumerate}

\item \textbf{Boundary-to-bound, and why precession starts at 2PM.} The bound
on-shell action \eqref{eq:lec3:S-bound} was obtained from scattering data
alone, with the remark that one could also get it by integrating $p_r$ between
the turning points. Do so.
\begin{enumerate}
  \item[(a)] For a bound orbit ($p_\infty^2<0$), take the radial momentum
  \eqref{eq:lec3:pr} with the matched 1PM coefficient \eqref{eq:lec3:p2r} and
  evaluate the phase integral of \eqref{eq:lec3:S-bound-def} between the two
  turning points:
  \[
    \tilde S^{\rm bound}(E,J)=2\int_{r_{\min}}^{r_{\max}}\!p_r\,dr+2\pi J
    =\frac{2\pi G m_1^2 m_2^2\,(2\gamma^2-1)}{E\sqrt{-p_\infty^2}}\,,
  \]
  in agreement with \eqref{eq:lec3:S-bound}. 
  \item[(b)] Expand the scattering angle in inverse angular momentum,
  \[
    \chi(E,J)=\sum_{n\geq1}\chi_n(E)\,J^{-n},
  \]
  and argue that $\chi_n\propto G^n$.
 
  \item[(c)]  The boundary-to-bound map keeps only the even powers,
  \[
    \Delta\phi(E,J)=\chi(E,J)+\chi(E,-J)=2\!\!\sum_{n\
    \mathrm{even}}\!\!\chi_n(E)\,J^{-n}.
  \]
  so the 1PM term ($n=1$) cancels and the leading periastron advance is 2PM.
  How is this consistent with the claim that precession
  is a 1PN effect? 
  \item[(d)] Now compute it. The next term of the radial momentum does nothing
  but shift the centrifugal barrier,
  \[
    p_r^2=p_\infty^2+\frac{P_1G}{r}-\frac{J^2-P_2G^2}{r^2},
  \]
  so the phase integral of (a) applies with $J$ traded for
  $\sqrt{J^2-P_2G^2}$ in the angular part. Show that
  \[
    \tilde S^{\rm bound}
    =2\pi\bigg[\frac{P_1G}{2\sqrt{-p_\infty^2}}-\sqrt{J^2-P_2G^2}+J\bigg],
  \]
  which returns \eqref{eq:lec3:S-bound} at $P_2=0$, and so
  \[
  \Delta\phi=\frac{\pi P_2 G^2}{J^2}+\cO(G^4).
  \]
  The coefficient $P_2(E)$
  would come from two-graviton exchange; in the
  non-relativistic limit it is $P_2\to6\mu^2M^2/c^2$. Use this to recover the
  1PN advance $\Delta\phi\to6\pi GM/[c^2a(1-e^2)]$ for an elliptical orbit of
  semi-major axis $a$ and eccentricity
  $e$, and specialize to
  Mercury's orbit to obtain the $43''$ per century.
\end{enumerate}

\item \textbf{From the time delay to the orbital period.} The energy derivative
of the on-shell action is a time: on the unbound branch the interaction time
delay, continued to the bound branch the radial period,
\[
  T_r=\frac{\partial \tilde S^{\rm bound}}{\partial E}\,.
\]
(Hamilton's equation $dt=(\partial p_r/\partial E)\,dr$ gives the elapsed
radial time as $\partial_E\!\oint p_r\,dr$.) Take the bound action of the
previous exercise, with $p_\infty$ and $E$ as in \eqref{eq:lec3:pinf}.
\begin{enumerate}
  \item[(a)] Show that $T_r$ takes the form,
  \[
    T_r=-2\pi G E\,\frac{\gamma\,(2\gamma^2-3)}{(1-\gamma^2)^{3/2}}\,.
  \]
  Note that the result depends on the energy but not on $J$. This is the
  relativistic
  Kepler degeneracy at 1PM.
  \item[(b)] Take the non-relativistic limit $E\to M$, $\gamma\to1$, with
  binding energy $\varepsilon=E-M$ and $M=m_1+m_2$, $\mu=m_1 m_2/M$. Show that
  $-p_\infty^2\to2\mu|\varepsilon|$ and reproduce Kepler's third law,
  \[
    T_r\to\pi\,G m_1 m_2\sqrt{\tfrac{\mu}{2}}\;|\varepsilon|^{-3/2}
    =2\pi\sqrt{\frac{a^3}{GM}},
  \]
  \item[(c)] Keeping one more order in the non-relativistic expansion gives
  its first relativistic correction. Show that it takes the form,
  \[
    T_r=\pi\,G m_1 m_2\sqrt{\tfrac{\mu}{2}}\;|\varepsilon|^{-3/2}
    \left[\,1+\frac{15-\nu}{4}\,\frac{|\varepsilon|}{\mu}+\cdots\right],
    \qquad \nu=\frac{\mu}{M}.
  \]
  This happens to be the complete 1PN correction to $T_r(E)$: the leading
  2PM term in the bound action is independent of $E$.
\end{enumerate}

\item \textbf{The binding energy without a potential.} The circular-orbit
energy \eqref{eq:lec3:EOmega} was obtained from the matched Hamiltonian. Get it
from the bound on-shell action instead.
\begin{enumerate}
  \item[(a)] A circular orbit does not oscillate radially. Show that the phase
  integral in \eqref{eq:lec3:S-bound-def} then vanishes, so that
  $\tilde S^{\rm bound}=2\pi J$, which is Sommerfeld's $n_r=0$, and that with
  \eqref{eq:lec3:S-bound} this fixes the angular momentum of the circular orbit
  of energy $E$,
  \[
    J=\frac{G m_1 m_2\,(2\gamma^2-1)}{\sqrt{1-\gamma^2}}\,.
  \]
  \item[(b)] Differentiate that condition along the sequence of circular
  orbits. Using $T_r=\p\tilde S^{\rm bound}/\p E$ and
  $\Delta\phi=-\p\tilde S^{\rm bound}/\p J$, show that
  $T_r\,dE/dJ=2\pi+\Delta\phi$. The right-hand side is the azimuth swept in one
  radial period, so $dE/dJ$ is the orbital frequency $\Omega$.
  \item[(c)] Take the non-relativistic limit as in the previous exercise and
  recover \eqref{eq:lec3:EOmega}. At 1PM the advance vanishes, so
  $\Omega=2\pi/T_r$ and the radial and azimuthal frequencies coincide.
\end{enumerate}

\end{enumerate}

\section{Radiation}
We are finally ready to say something about the gravitational radiation produced
by the motion of two bodies.

At leading order in $G$ the field is linear in the source, and far from the
source it is free, a superposition of on-shell plane waves,
\be
  h_{\mu\nu}(x)=\sqrt{32\pi
  G}\,\sum_\lambda\int\!\frac{d^3k}{(2\pi)^3\,2\omega}\,
  \Big[\varepsilon^\lambda_{\mu\nu}(k)\,a_\lambda(k)\,e^{ik\cdot x}+{\rm
  c.c.}\Big],
  \qquad \omega=|\vect k|,
  \label{eq:lec4:onshell}
\ee
where $h_{\mu\nu}$ is the physical perturbation,
$g_{\mu\nu}=\eta_{\mu\nu}+h_{\mu\nu}$,
hence the $\sqrt{32\pi G}$ relative to the canonically normalized field of
Lecture~2. The detector sits at $\vect x=r\hat{\vect n}$, and we take
$r\to\infty$ at fixed retarded time $u=t-r$, following the same wavefront
outwards. The phase of each plane wave then reads
$e^{ik\cdot x}=e^{-i\omega u}\,e^{-i\omega r(1-\hat{\vect k}\cdot\hat{\vect
n})}$,
finite at $\hat{\vect k}=\hat{\vect n}$ and rapidly oscillating elsewhere, so
for $\omega r\gg1$ the angular integral is fixed by its stationary points
$\hat{\vect k}=\pm\hat{\vect n}$:
\be
  \int\! d\Omega_k\,e^{i\omega r\,\hat{\vect k}\cdot\hat{\vect n}}\,f(\hat{\vect
  k})
  =\frac{2\pi}{i\omega r}\Big[f(\hat{\vect n})\,e^{i\omega r}
   -f(-\hat{\vect n})\,e^{-i\omega r}\Big]+\mathcal O\big((\omega r)^{-2}\big).
  \label{eq:lec4:saddle}
\ee
The first term gives $e^{-i\omega u}$, the outgoing wave; the second,
$e^{-i\omega(u+2r)}$, oscillates without bound as $r\to\infty$ and averages
away in the $\omega$ integral. The $1/r$ of the wave is the
solid angle $\sim(\omega r)^{-1}$ of the plane waves that still interfere
constructively at distance $r$. Fourier transforming in time,
$\tilde\phi(\omega)=\int dt\,e^{i\omega t}\phi(t)$, and projecting on a
polarization, $h_\lambda=\varepsilon^{\lambda*}_{\mu\nu}h^{\mu\nu}$,
\be
  h_\lambda(\omega,\vect x)=-\,\frac{i\sqrt{32\pi G}}{4\pi r}\,e^{i\omega r}\,
  a_\lambda(k),
  \qquad k^\mu=\omega(1,\hat{\vect n}).
  \label{eq:lec4:wave-coeff}
\ee
The coefficient is fixed by the source. To first order in the coupling the
worldlines produce the emission amplitude, 
\be
a_\lambda(k)=i\mathcal A_\lambda(k)= i\sqrt{8\pi
G}\,\varepsilon^{\lambda*}_{\mu\nu}\tilde T^{\mu\nu}(k),
\ee
which is the one-point amplitude
\eqref{eq:lec2:1pt} evaluated on the two worldlines, so
\be
  h_\lambda(\omega,\vect x)= \frac{\sqrt{32\pi G}}{4\pi r}\,e^{i\omega r}\,
  \mathcal A_\lambda(k) = \frac{4G}{r}\,e^{i\omega r}\,
  \varepsilon^{\lambda*}_{\mu\nu}\,\tilde T^{\mu\nu}(k),
  \label{eq:lec4:memory-strain}
\ee
where the particle stress tensor
\be
\tilde T^{\mu\nu}(k)=\sum_a m_a\!\int\! d\tau\;u_a^\mu u_a^\nu\,
  e^{-ik\cdot x_a(\tau)},
\ee
is the generalization of \eqref{eq:lec2:Tk} to
worldlines that are not straight. This particle contribution suffices for the
leading soft pole computed below.

\subsection{The gravitational memory}
\label{sec:lec4:memory}

\begin{figure}[t]
  \centering
\begin{tikzpicture}[
  x=1cm, y=1cm,
  body1/.style={line width=0.9pt, mapPN},
  body2/.style={line width=0.9pt, mapNR},
  arr/.style={postaction={decorate,decoration={markings,
      mark=at position 0.25 with {\arrow{Stealth[length=2.2mm]}},
      mark=at position 0.75 with {\arrow{Stealth[length=2.2mm]}}}}},
  lens/.style={line width=0.5pt},
  zoom/.style={line width=0.4pt, black!55, dotted},
  lbl/.style={font=\footnotesize}
]


\draw[body1,arr] (-3.6,0) -- (0,0) -- (3.118,-1.800);
\draw[body2,arr] (3.6,0) -- (0,0) -- (-3.118,1.800);
\node[lbl,anchor=north] at (-2.3,-0.12) {$p_1^{\rm in}$};
\node[lbl,anchor=north east] at (2.3,-1.42) {$p_1^{\rm out}$};
\node[lbl,anchor=south] at (2.3,0.12) {$p_2^{\rm in}$};
\node[lbl,anchor=south west] at (-2.3,1.42) {$p_2^{\rm out}$};

\draw[lens] (0,0) circle[radius=0.18];
\draw[zoom] (-0.075,0.163) -- (1.397,3.090);
\draw[zoom] (0.167,-0.067) -- (3.014,1.554);
\begin{scope}[shift={(1.9,2.0)}]
  \clip (0,0) circle[radius=1.2];
  \draw[body1] plot[domain=-1.25:1.25, samples=81, smooth]
    (\x, {0.6 - 0.28868*(\x + sqrt(\x*\x + 0.16))});
  \draw[body2] plot[domain=-1.25:1.25, samples=81, smooth]
    (\x, {-0.6 + 0.28868*(sqrt(\x*\x + 0.16) - \x)});
\end{scope}
\draw[lens] (1.9,2.0) circle[radius=1.2];

\end{tikzpicture}
  \caption{A scattering encounter seen from far away: each body moves on a
  straight line with a kink.}
  \label{fig:lec4:kink}
\end{figure}
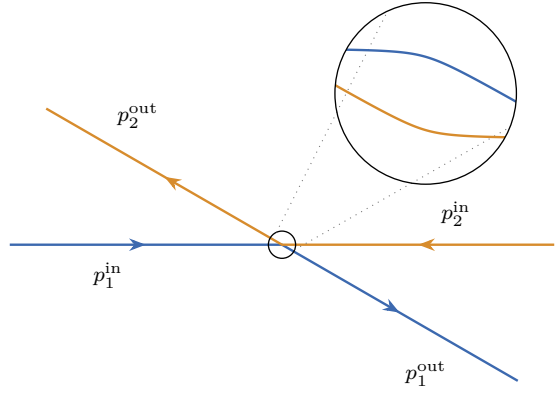

We begin at low frequency, where the waveform can be computed without
solving for the motion in detail. Consider again a scattering encounter. As the
bodies sweep past each other, a
burst of radiation goes out. After it has passed, the gravitational wave strain
does not relax to
zero but settles at a new constant, leaving a pair of free test masses
straddling the wave permanently displaced. The offset is the so-called (linear)
gravitational-wave \emph{memory}. This effect is a consequence of the soft
graviton theorem, which says that the low-frequency
radiation cannot depend on the encounter's history, only on its asymptotic
states, so the memory is computable from the free worldlines before and after
the collision.

Model the encounter as a kick (Figure~\ref{fig:lec4:kink}), $u_a=u_a^{\rm in}$
for $\tau<0$ and $u_a^{\rm out}$ for $\tau>0$, continuous at $x_a(0)=b_a$, with
arbitrary in and out velocities/momenta. The proper-time integral in
\eqref{eq:lec4:memory-strain}
splits at the kick into two half-line exponentials, each regulated by its
convergence at large $|\tau|$,
\[
  \int_0^\infty\! d\tau\,e^{-ia\tau}=\frac{-i}{a-i\epsilon},
  \qquad
  \int_{-\infty}^0\! d\tau\,e^{-ia\tau}=\frac{i}{a+i\epsilon},
\]
so that, with $p_a=m_a u_a$,
\be
  \varepsilon^*_{\mu\nu}\tilde T^{\mu\nu}
  =-\,i\sum_a\left[\frac{(\varepsilon^*\!\cdot p_a^{\rm out})^2}{k\cdot p_a^{\rm
  out}-i\epsilon}
                 -\frac{(\varepsilon^*\!\cdot p_a^{\rm in})^2}{k\cdot p_a^{\rm
                 in}+i\epsilon}\right]
    e^{-ik\cdot b_a}.
  \label{eq:lec4:memory-poles}
\ee
Every factor $k\cdot p_a=\omega\,(\hat k\cdot p_a)$ (we write
$\hat k^\mu=(1,\hat{\vect n})$) vanishes linearly with the frequency, and the
phases $e^{-ik\cdot b_a}\to1$, so at low frequency the strain is dominated by a
pole,
\be
  h\;\xrightarrow{\;\omega\to0\;}\;
  -\,\frac{4Gi}{r}\,\frac1\omega
  \sum_a\left[\frac{(\varepsilon^*\!\cdot p_a^{\rm out})^2}{\hat k\cdot p_a^{\rm
  out}}
             -\frac{(\varepsilon^*\!\cdot p_a^{\rm in})^2}{\hat k\cdot p_a^{\rm
             in}}\right].
  \label{eq:lec4:soft-pole}
\ee
Note that it is independent of the
interaction that turned the in-momenta into the out-momenta: the same factor
multiplies the emission from any process with these asymptotic states,
gravitational or not. This universality in the amplitudes language of
Lecture~3 is known as the leading soft graviton theorem by
Weinberg~\cite{Weinberg:1965nx}.

Back in the time domain the two poles are steps, the half-line integrals
that produced them read in reverse,
$\int\frac{d\omega}{2\pi}\,\frac{e^{-i\omega t}}{\omega\pm i\epsilon}
=\mp i\,\theta(\pm t)$: the out-state term of \eqref{eq:lec4:memory-poles}
switches on after the encounter and the in-state term switches off. The
strain sits at a constant built from the incoming momenta before, at the
corresponding outgoing constant after, and the transient between them is the
burst, carried by the finite-frequency terms dropped in
\eqref{eq:lec4:soft-pole}. A constant strain alone is unobservable, flat space
in shifted coordinates; the observable is the offset,
\be
  \Delta h\;\equiv\;h(+\infty)-h(-\infty)
  =-\,\frac{4G}{r}\sum_a\left[
    \frac{(\varepsilon^*\!\cdot p_a^{\rm out})^2}{\hat k\cdot p_a^{\rm out}}
   -\frac{(\varepsilon^*\!\cdot p_a^{\rm in})^2}{\hat k\cdot p_a^{\rm
   in}}\right]\,.
  \label{eq:lec4:memory}
\ee

Momentum conservation ties the two kicks together, $\Delta p_2=-\Delta p_1$,
up to the momentum carried off by the radiation itself, which enters only at
$\cO(G^3)$~\cite{Herrmann:2021lqe,Jakobsen:2021smu,Damour:2020tta,Dlapa:2022lmu}.
Writing $p_a^{\rm out}=p_a+\Delta p_a$ with $p_a\equiv p_a^{\rm in}$ and
expanding \eqref{eq:lec4:memory} to first order in the impulse,
\be
  \Delta h=-\,\frac{4G}{r}\sum_a\left[
    \frac{2\,(\varepsilon^*\!\cdot p_a)}{\hat k\cdot p_a}\,\varepsilon^{*\mu}
   -\frac{(\varepsilon^*\!\cdot p_a)^2}{(\hat k\cdot p_a)^2}\,\hat k^\mu\right]
  \Delta p_{a\mu}\,,
  \label{eq:lec4:memory-impulse}
\ee
the derivative of the soft factor in \eqref{eq:lec4:soft-pole} with respect
to the momenta, contracted with the impulse. The impulse is one more
derivative of the on-shell action \eqref{eq:lec3:eikonal} of Lecture~3: the
angle was its derivative in the angular momentum, \eqref{eq:lec3:angle}, and
differentiating instead in body~1's transverse position $b_1$ gives the
momentum transferred to body~1,
\be
  \Delta p_1^\mu
  =\frac{\partial\tilde S}{\partial b_{1\mu}}
  =-\frac{2G m_1 m_2\,(2\gamma^2-1)}{\sqrt{\gamma^2-1}}\,\frac{b^\mu}{b^2}
  \;+\;\cO(G^2),
  \label{eq:lec4:impulse}
\ee
with $b^\mu\equiv b_1^\mu-b_2^\mu$ pointing from body~2 to body~1, so the
minus sign is an attractive kick. Because $\tilde S$ depends on the two
endpoints only through the difference $b^\mu$, the derivative in $b_2$ is
opposite, $\Delta p_2=-\Delta p_1$;
dividing the magnitude by $p_\infty$ reproduces the angle, the
elastic-scattering kinematics of Lecture~3's first exercise. Inserting this
into \eqref{eq:lec4:memory-impulse} gives the leading-order memory of the
encounter,
\es{eq:lec4:memory-1pm}{
  \Delta h=\frac{8G^2 m_1 m_2\,(2\gamma^2-1)}{r\,b^2\,\sqrt{\gamma^2-1}}
  \bigg\{&\,2\,(\varepsilon^*\!\cdot b)\left[\frac{\varepsilon^*\!\cdot
  p_1}{\hat k\cdot p_1}
        -\frac{\varepsilon^*\!\cdot p_2}{\hat k\cdot p_2}\right]\\
   &-(\hat k\cdot b)\left[\frac{(\varepsilon^*\!\cdot p_1)^2}{(\hat k\cdot
   p_1)^2}
        -\frac{(\varepsilon^*\!\cdot p_2)^2}{(\hat k\cdot
        p_2)^2}\right]\bigg\}\,.
}
Neither of the two terms in braces in \eqref{eq:lec4:memory-1pm} is separately invariant
under $\varepsilon^*\to\varepsilon^*+\beta\hat k$; only their combination is.
The
invariance is manifest in terms of the linearized field strength
\eqref{eq:lec2:F} of the outgoing graviton, built here on the null direction
$\hat k$ rather than on $k$,
\be
  F^{\mu\nu}=\hat k^\mu\varepsilon^{*\nu}-\hat k^\nu\varepsilon^{*\mu},
  \label{eq:lec4:F}
\ee
in which the same memory reads
\be
  \Delta h=\frac{8G^2m_1m_2\,(2\gamma^2-1)}{r\,b^2\,\sqrt{\gamma^2-1}}\;
  \frac{p_1\!\cdot\!F\!\cdot\!p_2}{(\hat k\cdot p_1)(\hat k\cdot p_2)}\;
  \sum_a\frac{b\!\cdot\!F\!\cdot\!p_a}{\hat k\cdot p_a}\,,
  \label{eq:lec4:memory-1pm-F}
\ee
with $p\!\cdot\!F\!\cdot\!p'\equiv p_\mu F^{\mu\nu}p'_\nu$ the contraction that
appeared in the Compton amplitude \eqref{eq:lec2:compton-general}.

At leading order in the velocities $\hat k\cdot p_a\simeq-m_a$, and the
bracket in \eqref{eq:lec4:memory} reduces to
$-\,\varepsilon^*_{\mu\nu}\,\Delta\big(\sum_a m_a u_a^\mu u_a^\nu\big)$. On a
free
worldline the second mass moment $Q^{\mu\nu}=\sum_a m_a x_a^\mu x_a^\nu$
grows quadratically with time, $\ddot Q^{\mu\nu}=2\sum_a m_a u_a^\mu
u_a^\nu$, constant but not zero, so
\be
  \Delta h\;\simeq\;\frac{2G}{r}\,\varepsilon^*_{\mu\nu}\,\Delta\ddot Q^{\mu\nu}
  \label{eq:lec4:memory-quad}
\ee
(only the trace-free part of $Q^{\mu\nu}$ survives the traceless
polarization): the memory is the change, across the encounter, in the second
time derivative of the system's mass quadrupole.

What we have computed is the so-called \emph{linear} memory, sourced by the
change in the momenta of the massive bodies. There is also a \emph{nonlinear}
contribution, in which the radiated gravitons themselves carry energy to
infinity and source a low-frequency strain of their
own~\cite{Zeldovich:1974gvh,Braginsky:1987kwo,Christodoulou:1991cr,Wiseman:1991ss,Blanchet:1992br,Favata:2010zu,
Georgoudis:2025vkk}.
Neither has yet been measured: a single event's offset lies below current
sensitivity, and the ongoing searches stack many
detections~\cite{Lasky:2016knh}.
The subleading terms of the expansion \eqref{eq:lec4:soft-pole} are universal
as well, including a $\ln\omega$ from the wave scattering off the binary's
own long-range field, the \emph{tail}; soft theorems thus fix the
lowest-frequency end of every gravitational waveform, whatever the process
that emits it~\cite{Blanchet:1987wq,Sahoo:2018lxl,Saha:2019tub}.

\subsection{The binary as a radiating point source}
\label{sec:lec4:multipoles}

The memory needed only the endpoints of the motion, but a detector's signal
sits at finite frequency, at the orbital scale $\omega\sim\Omega$, where the
source integral in \eqref{eq:lec4:memory-strain} depends on the whole
trajectory. The
orbit has size $r$ and turns at a frequency $\Omega\sim v/r$, and it radiates
at roughly that frequency, so its gravitational waves have wavelength
$\lambda_{\rm GW}\sim c/\Omega\sim r\,(c/v)$: the radiating system is small
compared
with the wavelength it produces,
\be
  \frac{r}{\lambda_{\rm GW}}\;\sim\;\frac{v}{c}\;\ll\;1
  \label{eq:lec4:rlambda}
\ee
(cf.\ $R/r$ in \eqref{eq:lec2:Rr}), which is what a multipole expansion
exploits. Seen from the wave zone, where the waves are measured, the orbit is
therefore a
point, and we apply the worldline construction of Sec.~\ref{sec:lec2:worldline}
to the binary's center of mass.

What survives at wavelengths $\lambda_{\rm GW}\gg r$ is the center-of-mass
worldline
together with the
long-wavelength field it radiates, and the only imprint the unresolved interior
can leave on the far field is a set of multipole moments, organized by
rotational
invariance about the source into two families, mass moments and current
(mass-flow) moments~\cite{Thorne:1980ru}. We
write the moments covariantly as symmetric trace-free tensors orthogonal to the
worldline four-velocity $u^\mu$; in the binary's rest frame $u^\mu=(1,\vect0)$
they
are ordinary spatial STF tensors $Q^{ij},\,O^{ijk},\dots$ of rank $2,3,\dots$

The worldline couples to the radiation field
through these moments, each term a scalar built from the field's curvature on
the
worldline contracted with a moment of the source, as in
Sec.~\ref{sec:lec2:worldline}. After the monopole, the leading mass couplings
are
\be
  S \;=\; \int d\tau\,\Big[\,-\,M
        \;-\;\tfrac12\,Q^{\mu\nu}(\tau)\,E_{\mu\nu}
        \;-\;\tfrac16\,O^{\mu\nu\rho}(\tau)\,\nabla_\rho E_{\mu\nu}
        \;+\;\cdots\Big],
  \label{eq:lec4:multipole-action}
\ee
with $E_{\mu\nu}=C_{\mu\alpha\nu\beta}u^\alpha u^\beta$ the electric tidal
tensor
of the radiation field at the center of mass (the same object as in
Sec.~\ref{sec:lec2:worldline}), and $\nabla_\rho E_{\mu\nu}$ its gradient, to
which
the octupole couples one derivative higher. The dots stand for still higher mass
moments and for the current moments, which couple to the magnetic tidal tensor
$B_{\mu\nu}$.
The quadrupole term is the coupling \eqref{eq:lec2:multipole} of
Lecture~2, where a body's induced quadrupole responded to its companion's tidal
field; here the binary's quadrupole is instead the source of the radiation
field.

Each additional multipole costs a power of $v/c$: successive
moments resolve successively finer features of the source, each suppressed by a
further power of $r/\lambda_{\rm GW}\sim v/c$.

The monopole $M$ is the total mass-energy, which is conserved, and a constant
radiates nothing. The mass dipole is the center of mass, whose rate of change is
the total momentum, also conserved; gravity thus has no mass-dipole radiation,
for the same reason a system with a uniform charge-to-mass ratio has no
electric-dipole radiation. The current dipole is the total angular momentum,
conserved as well. The first moment free to vary is therefore the mass
quadrupole
$Q^{ij}$, and its radiation is the leading gravitational wave.

The moment is fixed by matching, computing it in the full theory, here
from the orbital motion of Lecture~3. For the mass quadrupole the match is the
classical, familiar one: $Q^{ij}$ is the trace-free second mass moment of the
source,
\be
  Q^{ij}(t)=\sum_a m_a\Big(x_a^i x_a^j-\tfrac13\delta^{ij}\,\vect x_a^2\Big)
  \;\xrightarrow{\;\rm CM\;}\;
  \mu\Big(x^i x^j-\tfrac13\delta^{ij}\,\vect x^2\Big),
  \label{eq:lec4:quad-moment}
\ee
the last form in the center-of-mass frame, with $\vect x=\vect x_1-\vect x_2$
the
relative orbit and $\mu$ the reduced mass.

Because the quadrupole term in
\eqref{eq:lec4:multipole-action} is linear in the radiation field, it acts as
a graviton source. Using the quadrupolar vertex \eqref{eq:lec2:1ptQ}, a
time-dependent moment $Q^{\mu\nu}(\tau)$ emits a graviton of momentum $k$ and
polarization $\varepsilon$ with amplitude,
\be
  \cA_Q=-\tfrac12\sqrt{8\pi G}\;\tilde
  Q^{\mu\nu}(\omega)\,(u\!\cdot\!F)_\mu\,(u\!\cdot\!F)_\nu,
  \label{eq:lec4:quad-emission}
\ee
gauge invariant because it depends on the graviton only through the field
strength
$F^{\mu\nu}=k^\mu\varepsilon^\nu-k^\nu\varepsilon^\mu$, in the combination
$(u\!\cdot\!F)^\mu$ of \eqref{eq:lec2:W}. Evaluate it in
the binary's rest frame $u^\mu=(1,\vect0)$ on a transverse-traceless mode: then
$u\!\cdot\!\varepsilon=0$ and $u\!\cdot\!k=-\omega$, so each factor becomes
$(u\!\cdot\!F)_\mu\to-\omega\,\varepsilon_\mu$ and
\be
  \cA_Q=-\tfrac12\sqrt{8\pi G}\;\omega^2\,\varepsilon_{\mu\nu}\,\tilde
  Q^{\mu\nu}(\omega),
  \qquad \varepsilon_{\mu\nu}=\varepsilon_\mu\varepsilon_\nu .
  \label{eq:lec4:quad-emission-rest}
\ee
The two powers of $\omega$ are the two derivatives the coupling carries: the
quadrupole couples to the tidal tensor $E_{\mu\nu}$, two gradients of the field,
and
each gradient of an outgoing wave brings down a power of its momentum. Carried
out
to the detector as at the start of this lecture, the wave is this amplitude
spread
over the wave zone as $e^{i\omega r}/r$, and its normalization is fixed
by the leading-order strain \eqref{eq:lec4:memory-strain}: conservation of the stress tensor gives
$\int d^3x\,T^{ij}=\tfrac12\,\partial_t^2\!\int
d^3x\,T^{00}x^ix^j$,
so at long wavelength
$\varepsilon^*_{\mu\nu}\tilde
T^{\mu\nu}=-\tfrac12\,\omega^2\,\varepsilon^*_{\mu\nu}\tilde Q^{\mu\nu}$
against a traceless polarization, which reproduces
\eqref{eq:lec4:quad-emission-rest}. The radiated strain is then
\be
  h(\omega,\vect r)=-\frac{2G\omega^2}{r}\,e^{i\omega r}\,
  \varepsilon^*_{\mu\nu}\,\tilde Q^{\mu\nu}(\omega)\,.
  \label{eq:lec4:waveform-contracted}
\ee

Undoing the rest-frame,
transverse-traceless choice for a general line of sight $\hat n$ reinstates the
projector $\Lambda_{ij,kl}(\hat n)=\Pi_{ik}\Pi_{jl}-\tfrac12\,\Pi_{ij}\Pi_{kl}$
($\Pi_{ij}=\delta_{ij}-\hat n_i\hat n_j$, the spatial form of the polarization
sum \eqref{eq:lec2:polsum} of Lecture~2), and transforming back to the time
domain
trades $-\omega^2$ for two time derivatives,
\be
  h_{ij}^{\rm TT}(t,\vect r)=\frac{2G}{r}\,\Lambda_{ij,kl}(\hat n)\,
  \ddot Q_{kl}(t-r),
  \label{eq:lec4:quadrupole}
\ee
Einstein's quadrupole formula for the waveform~\cite{Einstein:1918btx}.
Radiation requires the two time derivatives: a
uniformly varying quadrupole radiates nothing, just as a uniformly moving mass
did
not in \eqref{eq:lec2:1pt}. The formula's permanent piece is the memory: the
offset \eqref{eq:lec4:memory-quad} is \eqref{eq:lec4:quadrupole} evaluated on
the
change in $\ddot Q$ across an encounter. For a circular orbit at frequency
$\Omega$ the moment
\eqref{eq:lec4:quad-moment} goes as $Q_{ij}\sim\mu r^2 e^{2i\Omega t}$, so
$\tilde Q^{\mu\nu}(\omega)$ peaks at $\omega=2\Omega$: the dominant
gravitational wave is emitted at twice the orbital frequency.

\subsection{The radiated power}
\label{sec:lec4:power}

A gravitational wave carries energy because the
graviton is a propagating field, and a propagating field stores energy
quadratically in its amplitude. The rate at which the wave
\eqref{eq:lec4:waveform-contracted} drains the orbit is therefore its squared
strain, summed over the graviton's two polarizations and over every direction
$\hat n$ of propagation. In Lecture~2 we normalized it canonically,
$g_{\mu\nu}=\eta_{\mu\nu}+\sqrt{32\pi G}\,h_{\mu\nu}$; expanding the
Einstein--Hilbert action to second order in the transverse-traceless fluctuation
gives each polarization the action of a free massless field,
$-\tfrac12(\partial h^{\rm can})^2$, with the canonically normalized field
$h^{\rm can}=h^{\rm TT}/\sqrt{32\pi G}$. A massless field radiates energy
outward at a rate $\langle\dot\phi^2\rangle$ per unit area, exactly as the
electromagnetic field does. Summing the two polarizations and
undoing the normalization, the power crossing a sphere of radius $r$ per unit
solid
angle is
\be
  \frac{dE}{dt\,d\Omega}=\frac{r^2}{32\pi G}\,
   \sum_{\lambda}\big\langle\,|\dot h_\lambda|^2\,\big\rangle ,
  \label{eq:lec4:flux}
\ee
with $\langle\cdot\rangle$ an average over many wave cycles, which turns an
oscillating flux into the steady rate at which the orbit loses energy.

The strain in polarization $\lambda$ is the contraction
$h_\lambda=\varepsilon^{\lambda*}_{\mu\nu}h^{\mu\nu}$ of
\eqref{eq:lec4:waveform-contracted},
with
$\varepsilon^\lambda_{\mu\nu}=\varepsilon^\lambda_\mu\varepsilon^\lambda_\nu$
built from the
polarization four-vector of Lecture~2 (transverse,
$k\cdot\varepsilon^\lambda=0$). The
sum over $\lambda$ is the polarization sum \eqref{eq:lec2:polsum} of Lecture~2,
$\sum_\lambda\varepsilon^{\lambda*}_{\mu\nu}\,\varepsilon^\lambda_{\alpha\beta}
=\Lambda_{\mu\nu,\alpha\beta}$, the covariant transverse-traceless projector;
between the conserved sources of Lecture~3 it could be replaced by the
propagator numerator; here it cannot. For a wave travelling along $\hat n$,
$k^\mu=\omega(1,\hat n)$, the reference vector $\bar k^\mu=\omega(1,-\hat n)$
fixes the frame and reduces the vector projector \eqref{eq:lec2:vecsum} to
$\Pi_{ij}=\delta_{ij}-\hat n_i\hat n_j$ with $\Pi_{0\mu}=0$, making $\Lambda$
the same projector as in the waveform \eqref{eq:lec4:quadrupole}.

Read \eqref{eq:lec4:waveform-contracted} in the time domain,
$h_\lambda=(2G/r)\,\varepsilon^{\lambda*}_{\mu\nu}\,\ddot Q^{\mu\nu}(t-r)$; each
time
derivative brings down a $\dddot Q$, and \eqref{eq:lec2:polsum} folds the sum
over
$\lambda$ into the projector:
\be
  \frac{dE}{dt\,d\Omega}=\frac{G}{8\pi}\,\Lambda_{\mu\nu,\alpha\beta}(\hat n)\,
   \big\langle\dddot Q^{\mu\nu}\,\dddot Q^{\alpha\beta}\big\rangle .
  \label{eq:lec4:dPdOmega}
\ee
With $Q^{\mu\nu}$ spatial and trace-free in the rest frame the projector reduces
to
its spatial part $\Lambda_{ij,kl}$, and the average over directions is
elementary,
\be
  \int d\Omega\,\Lambda_{ij,kl}(\hat n)
   =\frac{2\pi}{15}\big(11\,\delta_{ik}\delta_{jl}
     -4\,\delta_{ij}\delta_{kl}+\delta_{il}\delta_{jk}\big),
\ee
giving the quadrupole formula for the radiated power,
\be
  \frac{dE}{dt}=\frac{G}{5}\,\big\langle\dddot Q_{ij}\,\dddot Q^{ij}\big\rangle
  .
  \label{eq:lec4:luminosity}
\ee
Restoring units multiplies the right-hand side by $1/c^5$.

Evaluated on the moment \eqref{eq:lec4:quad-moment}, this is the rate at which
the binary loses energy, and the
balance $\dot E_{\rm orbit}=-\,dE/dt$ tightens the orbit. For a circular orbit
of
separation $r$ and frequency $\Omega$ the moment turns over at $2\Omega$ and
$\dddot Q_{ij}\sim\mu r^2\Omega^3$, so
\be
  \frac{dE}{dt}=\frac{32}{5}\,G\,\mu^2 r^4\Omega^6 ,
  \label{eq:lec4:circular-power}
\ee
the $32/5$ from the circular-orbit average. Set against the binding energy
$|E|\sim Gm_1 m_2/r$ over one period, this is the per-orbit loss
$\Delta E/|E|\sim\nu\,(v/c)^5$ of \eqref{eq:lec1:adiabatic}. The quadrupole formula,
applied to the eccentric orbit of the binary pulsar PSR~B1913+16~\cite{Hulse:1974eb},
predicts an orbital period that shrinks as the system radiates. Taylor and
Weisberg measured that drift and found it to agree with the
prediction~\cite{Taylor:1982zz}, which is how gravitational waves were first
shown to carry energy, some three decades before one was detected
directly~\cite{LIGOScientific:2016aoc}.

The wave carries angular momentum as well as energy, and the balance laws
\eqref{eq:lec1:balance} of Lecture~1 involve both fluxes. The angular-momentum
flux
$\cG$ follows from the same average as the power. We do not have time to derive
it, but the result is
\be
  \cG^i=\frac{2G}{5c^5}\,\epsilon^{ijk}\,
  \big\langle\,\ddot Q_{jl}\,\dddot Q_{kl}\,\big\rangle\,.
  \label{eq:lec4:Lflux}
\ee
 On a circular orbit, evaluating both fluxes on the
moment \eqref{eq:lec4:quad-moment} gives $\cG$ along the orbital axis with
\be
  \cF=\Omega\,\cG .
  \label{eq:lec4:FG}
\ee
Thus energy and angular-momentum balance are equivalent for a
quasi-circular orbit. The relation also follows mode by mode: an $m=2$
harmonic with $\omega=2\Omega$ carries energy and angular momentum in the
ratio $\omega/m=\Omega$. For eccentric motion the two fluxes contain
independent information and together determine the evolution of the
eccentricity~\cite{Peters:1963ux,Peters:1964zz}.

\subsection{Concluding remarks}

Reaching the radian-level accuracy the detectors require means computing the
two inputs $E(\Omega)$ and $\cF(\Omega)$ far past the leading order in $v/c$
and $Gm/rc^2$ we have worked to here. The
post-Minkowskian expansion now reaches the fifth power of $G$ in the
conservative
sector~\cite{Bern:2021dqo,Dlapa:2021npj,Driesse:2024xad,Driesse:2026qiz} and
the third in the radiative
waveform~\cite{Goldberger:2009qd,Brandhuber:2023hhy,Herderschee:2023fxh,Georgoudis:2023lgf},
and the post-Newtonian waveform is comparably
deep~\cite{Blanchet:2013haa,Blanchet:2023bwj,Blanchet:2023sbv}. We have left out
spin: astrophysical objects rotate, and spin corrections
enter the waveform at 1.5PN order~\cite{Levi:2018nxp}. And beyond the
weak-field, slow-velocity regime lie numerical relativity and
the self-force expansion, which divide the strong-field corner of the
parameter space of Lecture~1 between them. The templates the detectors actually
compare
against are a synthesis of all of these
methods~\cite{Buonanno:1998gg,Buonanno:2000ef,Pretorius:2005gq,Barack:2018yvs}.

We set out to describe how two compact bodies move and what they radiate,
and, truth be told, these lectures are only an appetizer to that
problem. Coming detectors (LISA~\cite{LISA:2017pwj}, the Einstein
Telescope~\cite{ET:2019dnz}, Cosmic Explorer~\cite{Reitze:2019iox}) will
require waveform accuracies beyond every current model, for sources (extreme
mass
ratios, hour-long neutron-star signals) that stress the boundaries between
different methods. Three centuries on, and we are still not done with the
problem of motion of two bodies, nor is it done with us.

\subsection{Exercises}
\label{sec:lec4:exercises}

\begin{enumerate}

\item \textbf{Revisiting the chirp.} In Lecture~1 we estimated the scaling of
the inspiral phase at leading order in PN. Use the luminosity
\eqref{eq:lec4:circular-power} and the circular-orbit energy to fix the
coefficients: eliminate the separation in favor of $\Omega$, integrate the
balance law $dE/dt=-\cF$ for the frequency sweep, and integrate once more for
the phase $\varphi=2\!\int\!\Omega\,dt$ of the wave, emitted at twice the
orbital
frequency. Show that
  \[
       \varphi_c-\varphi(t)=2\left[\frac{c^3\,(t_c-t)}{5\,G\cM}\right]^{5/8},
  \]
with $\cM$ the chirp mass \eqref{eq:lec1:chirpmass}, so that at leading order
the phase measures the two masses only through that one combination.

\item \textbf{Octupolar radiation.} The next term in the multipole action
\eqref{eq:lec4:multipole-action} is the mass octupole,
\[
-\tfrac16\int d\tau\;O^{\mu\nu\rho}(\tau)\,\nabla_\rho E_{\mu\nu}\,,\] 
with
$O^{\mu\nu\rho}$ the symmetric-traceless, $u$-orthogonal octupole moment and
$E_{\mu\nu}$ the electric tidal tensor.
\begin{enumerate}
  \item[(a)] Following the quadrupole derivation of
  Sec.~\ref{sec:lec4:multipoles}, show that the octupole radiates as
  \[
    h_O=\frac{2G}{3r}\,\varepsilon^*_{\mu\nu}\,\hat n_\rho\,
    \dddot{O}^{\mu\nu\rho}(t-r).
  \]
  \item[(b)] Find the formula for the power it radiates. How is it suppressed
  relative to the
  quadrupolar power?
\end{enumerate}

\item \textbf{Angular-momentum balance on a circular orbit.}
        \begin{enumerate}
  \item[(a)] Evaluate the flux of angular momentum \eqref{eq:lec4:Lflux} on the
  circular-orbit moment
  \eqref{eq:lec4:quad-moment}, $\vect x=r\,(\cos\Omega t,\sin\Omega t,0)$, and
  show that $\cG$ points along the orbital axis with magnitude
  \[
    \cG=\frac{32}{5}\,G\,\mu^2 r^4\Omega^5=\frac{\cF}{\Omega}\,,
  \]
  with $\cF$ the luminosity \eqref{eq:lec4:circular-power}.
  \item[(b)] Using the Newtonian circular-orbit relations $L=\mu\sqrt{GMr}$
  and $E=-GM\mu/2r$, show that $dE/dL=\Omega$ along the sequence of circular
  orbits, so that the two balance laws \eqref{eq:lec1:balance} predict the
  same shrinkage rate $\dot r$. This shows that a quasi-circular inspiral
  can be evolved with either flux, as claimed in the lecture.
\end{enumerate}

\end{enumerate}
 

\chapter{Environmental effects in gravitational waves}
\chapterauthor{Laura Sberna}
\label{ch:part3}
\allowdisplaybreaks
\minitoc

These notes are partly based on an upcoming review on environmental effects by Francisco Duque, Laura Sberna and Rodrigo Vicente. 

\section{Introduction} %

\subsection{To vacuum or not to vacuum: a zoology of environments}

The standard theory of gravitational waves assumes that its sources are isolated objects. However, gravitational wave sources can form in a variety of astrophysical environments, where the density of gas, dark matter, or other compact objects and stars can be high and where this approximation will be less accurate. To give a sense of the scales involved, I compare in Table~\ref{tab:scales} the densities of a variety of realistic environments. In particular, I compare: a region of our Galaxy like the one we live in (galactic field, closer to the edge than to the center);  inner regions around active and quiet supermassive black holes. 
For the density of dark matter, I also compare vanilla with more exotic scenarios. Besides the quantities listed in the table, the electromagnetic field could also make the environment ``less empty" than commonly assumed. And lucky for us, the most interesting (in the sense of promising, or problematic!) environments are also the hardest ones to model: galactic nuclei are filled with ionised, dense gas, intense electromagnetic fields, instabilities and turbulent gas.

The goal of these lectures is to explore the effect that the environment can have on a gravitational wave signal. 

\begin{table}
\caption{Typical scales of a variety of realistic environments. 
The density of the dark matter spike is reported at $100$ gravitational radii. Having milli-Herz gravitational wave sources in mind, we focus on low-mass supermassive black holes.
}\label{tab:scales}
\begin{tabular}{c|c|c|c}
Environment &  matter & dark matter [$\rm GeV /cm^3$] & stars \\ \hline 
Galactic field & $0.1$ part./$\rm cm^3$  & $0.01 - 1$ \cite{Read:2014qva}  & 0.1 star/$\rm pc^3$ \\ \hline
Globular cluster & // &  $0.1 - 10^3$ \cite{Garani:2023esk} & $\lesssim 10^{6}M_\odot$/$\rm pc^3$ \cite{Benacquista:2011kv}\\ \hline
Quiet galactic nucleus & $\lesssim 10^{-9} \, \rm g/cm^3$ & $10^3$ & $10^{6-7}M_\odot$/$\rm pc^3$ \cite{Neumayer2020}\\ \hline
Active galactic nucleus & $10^{-9}-10^{-4} \, \rm g/cm^3$ \cite{Mallick:2022lfr} & $10^3$ & // \\ \hline
Dark matter ``spike''  & & $10^{15}$  ($10^{13} M_\odot/{\rm pc}^3$) \cite{Gondolo:1999ef} & \\
\end{tabular}
\end{table}

\subsection{Motivation and prospects}

Spoiler alert: we have not detected any environmental effects in gravitational wave signals. An analysis of LIGO-Virgo events from the first gravitational wave catalog, GWTC-1, found no evidence for them \cite{CanevaSantoro:2023aol}. The analysis was able to put constraints -- under some favourable assumptions -- on the local density, 
\begin{equation}
    \rho \lesssim 20 \, \rm g / cm^3 \, ,
\end{equation}
but as you can see this is far from realistic (compare to Table~\ref{tab:scales}). As the reference points out, the current constraint is roughly the density of gold at room temperature on Earth.

While we can neglect the environment of LIGO-Virgo sources, we believe it could be important for some sources at lower GW frequencies. For future detectors, such as LISA, environmental effects will be both a blessing and a curse. If left unmodelled when we are actually sensitive to them, this could bias our inference of the source's properties, or we could mistake them for a deviation from general relativity. If we model correctly and detected environmental effects, we could learn how the source formed -- binary formation is very hard to model, from stellar binaries to massive black hole binaries. 
We could also learn about the environment itself. For instance, accretion disks are very complicated, only partially understood systems, and environmental effects could directly probe their density and properties. Gravitational waves could also probe regions and distances inaccessible to standard electromagnetic observations, and allow for interesting multimessenger synergies. Finally, environmental effects could be our only probe of matter that does not interact or interacts too weakly with ordinary matter, such as dark matter or other exotic fields/particles.

\subsection{Classes of environmental effects}
Environmental effects relevant to gravitational wave observations can be broadly categorised as follows:
\begin{itemize}
    \item \textbf{Accretion} modifies the mass of the compact objects over time, indirectly affecting their dynamics. If the masses change slowly compared to the typical timescale of the system, $\dot{m}_i/m\ll \Omega$ (adiabatic approximation), the angular momentum is conserved by the process (i.e., it is an adiabatic invariant). Therefore, we can write
    \begin{equation}
        \frac{dL}{dt} = \frac{\partial L}{\partial m_i} \frac{dm_i}{dt} + \frac{\partial L}{\partial a} \frac{da}{dt} =0 \, ,
    \end{equation}
    with $L=\mu a^2 \Omega$.
    Therefore, even in the absence of an external interaction taking away/donating angular momentum to the binary, (adiabatic) mass evolution implies evolution of the binary separation, and will impact their gravitational wave signal. 
    For a compact object embedded in a gas disk, the accretion rate is typically small compared to other environmental effects and is usually treated as a subdominant correction (but this is up for debate!). We will not cover accretion further in these lectures, but if you are curious you can have a look at e.g.~\cite{Caputo:2020irr} (2020);
    \item \textbf{Dynamical friction} is the dominant interaction between a compact object and a surrounding medium -- gas, stars, or a bosonic field -- through gravitational wakes. It will be the central tool in Examples 1 and 2.
    \item \textbf{Propagation effects} arise when the gravitational wave itself travels through an external gravitational field or medium. The leading example in these lectures is the Rømer delay and relativistic Doppler effect induced by a third body (Example 3). Other propagation effects include, for example, the Shapiro delay and gravitational lensing.
    \item \textbf{Three-body dynamical effects}: a third body can directly perturb the inner binary's orbital evolution through tidal forces, secular (long-timescale) interactions such as Kozai–Lidov oscillations (see e.g.~\cite{Naoz2016} (2016)) which can drive large oscillations in eccentricity and inclination, or even chaotic dynamics. 
\end{itemize}

\subsection{An environmental classic: dynamical friction}\label{sec:DF}

Dynamical friction refers to a force (typically a drag, hence friction) of purely gravitational origin experienced by objects (black holes, stars, or even clusters or galaxies) moving through a medium (gas, stars, a bosonic field). The physical mechanism is simple: the gravitational field of the object perturbs the motion or state of the components of the medium. In response, the perturbed medium forms a wake around the body (typically behind with respect to its direction of travel), which exerts a force (typically a drag) on the object. 

The theory of dynamical friction was originally developed by Chandrasekhar in 1943 for a collisionless medium composed of particles, and a point mass moving at constant non-relativistic velocity and in a straight line through it. We will briefly review his result, following \cite{Binney1987} (1987).

Let us consider a body of mass $\mu$ and a medium composed of individual masses $\mu_m\ll \mu$. Let us also assume that the medium has a total mass much larger than $\mu$ and that it is extended and homogeneous. Then, the collection of particles in the medium induces a drag
\begin{equation}
    \vec{F}_{\rm DF} = - 4 \pi G^2 \mu^2 \mu_m \ln\Lambda \int d^3\vec{v}_m f(\vec{v}_m) \frac{\vec{v}_p - \vec{v}_m}{|\vec{v}_p - \vec{v}_m|^3} \, , \text{(collisionless particle medium, non-relativistic)}
\end{equation}
where $\vec{v}_p$ is the velocity of the object, $\vec{v}_m$ the velocity field of the particles, $f$ is the phase-space density of particles, normalised to the local number density $\int d^3\vec{v}_m f(\vec{v}_m) = n$. The logarithmic factor is known as the Coulomb logarithm (due to the Coulomb nature of the gravitational potential and an analogous effect in plasma physics) and its argument
\begin{equation}
    \Lambda = \frac{b_{\rm max}}{b_{\rm min}}
\end{equation}
is the ratio between the minimum and maximum impact parameter with which particles can encounter the object. The minimum impact parameter is set by the maximum between the radius of the object (if non-zero) and the impact parameter causing a deflection of $90^\circ$, $b_{\rm min}=G \mu/\langle v_m\rangle^2$. With $b_{\rm min}$, we are separating the effect of dynamical friction from the effect of accretion.
The maximum impact parameter $b_{\rm max}$ is set by the spatial extension of the medium. 

If the particles have an isotropic velocity distribution, the problem simplifies. The force experienced by the particle is
\begin{equation} \label{eq:DF_iso}
    \vec{F}_{\rm DF} = - 16 \pi^2 G^2 \mu^2 \rho \ln\Lambda \left[ \int_0^{v_p} dv_m \tilde{f}(v_m) v_m^2 \right] \, \frac{\hat{v}_p}{v_p^2} \, , \quad \text{(collisionless, isotropic medium, non-relativistic)}
\end{equation}
where we have taken into account the fact that particles with $v_m\gg v_p$ induce a negligible force. This is due to the fact that the object induces an asymmetry in slow particles depending on their direction of travel, but fails to do the same for fast particles: slow particles are either deflected by the object behind it, or fail to overtake it; fast particles always manage to fly past, and their effect cancels when integrated over the sphere. 
We have further assumed that the phase-space density factorizes $f \sim \tilde{f} n$ (normalised so that $\int_0^{\infty} dv_m \tilde{f}(v_m) v_m^2=1$) and introduced the density $\rho=\mu_m n$. 
The scaling with the object's velocity is easy to understand: $v_\mu$ sets the radius within which particles are gravitationally affected by the object, $R\sim G \mu/v_p^2$, so that the force is approximately given by the gravitational force of the particles within the wake, $F \sim G \mu \rho R^3/R^2$.

Let us look at two interesting limits of Eq.~\eqref{eq:DF_iso}. When $v_p$ is small, we can approximate $\tilde{f}(v_m)\simeq \tilde{f}(0)$ and integrate over the particles' velocities,
\begin{equation} \label{eq:DF_iso_slow}
    \vec{F}_{\rm DF} \simeq - 16/3 \pi^2 G^2 \mu^2 \rho \ln\Lambda \tilde{f}(0) v_p \, \hat{v}_p \, , \quad \text{(collisionless, isotropic medium, non-relativistic, slow)}
\end{equation}
In this limit, gravitational drag is proportional to the velocity, just like viscous drag.
On the other hand, if the perturber is fast the integral reduces to unity (up to $4 \pi$),
\begin{equation} \label{eq:DF_iso_fast}
    \vec{F}_{\rm DF} \simeq - 4 \pi G^2 \mu^2 \rho \ln\Lambda \, \frac{\hat{v}_p}{v_p^2} \, , \quad \text{(collisionless, isotropic medium, non-relativistic but fast)}
\end{equation}
This is very different compared to viscous drag: fast perturbers experience very little gravitational drag. 

{\bf Questions:} Try to derive Chandrasekhar's formula following \cite{Binney1987}.

{\bf Further reading:} Dynamical friction in the (relativistic) effective field theory language \cite{Katz:2019rgf}.

\section{Example 1: asymmetric binary in an accretion disk}

In our first example of environmental effects in gravitational waves, we will consider a stellar-mass black hole spiralling towards a massive black hole (an extreme-mass-ratio inspiral) through an accretion disk. 
We denote the massive black hole mass by $M$, the secondary mass by $\mu$, with $\mu/M\simeq 10^{-6}-10^{-5}$. The orbit of compact objects inspiralling in a disk tend to align with the disk and circularise, due to disk interactions. This typically happens on short timescales, when the compact object is still outside the band of gravitational wave detectors. Therefore, we will assume the compact object is on circular orbit in the disk plane. 

\subsection{Ingredient 1: the environment}



The first environment we will consider is the accretion disk of a massive black hole. In an accretion disk, gas circularizes and progressively loses angular momentum,  eventually falling into the central black hole (or being expelled, in minor part, through jets and winds). Accretion disks are ubiquitous in astrophysics: they form around young stars during the early stages of planet formation and appear around compact objects of all masses, from stellar-mass black holes in X-ray binaries to the active galactic nuclei (AGN) powered by accreting massive black holes. 

We will model the disk under the following assumptions:
\begin{enumerate}
\item \textit{Non-relativistic}: We will assume the disk is non-relativistic, i.e. use Newtonian physics. 

\item \textit{Central mass}: in the absence of other forces, gas elements admit Keplerian orbits with angular velocity $\Omega_K$,
\begin{equation}\label{eq:omega_k}
\Omega \approx \Omega_\text{K} = \sqrt{G M / r^3}.
\end{equation} 
\item \textit{Steady state/stationary disk}:
all disk quantities vary slowly with respect to the orbital time: the surface density $\Sigma \simeq$ const., etc.
\item  \textit{Thin disk}: the vertical scale of the disk is much smaller than its radial extension, $z\ll r$. This will allow us to integrate the equations of the gas over the vertical direction, and define vertically-integrated quantities such as the integrated surface density $\Sigma=\int \rho dz \simeq \rho 2 H$.
\item  \textit{Light disk}: we neglect the disk's self-gravity compared to the gravitational pull of the central black hole.
\vspace{2 mm}

The disk structure follows from assumptions 1-4 and the physical principles of conservation of mass, energy and momentum. Neglecting viscosity, the disk is governed by the (compressible) Euler equations and the continuity equation, 
\begin{align}
    &\frac{d\vec{v}}{dt} =  -\frac{1}{\rho}\vec{\nabla}P -G\frac{M}{r^2}\hat{r} \, , \\
    &\frac{d}{d t} \rho= \frac{\partial}{\partial t}\rho +\vec{\nabla}\cdot(\rho \vec{v}) = 0 \, .
\end{align}
To exploit the geometry of the disk, we will use cylindrical coordinates. 
We start by studying the ``thin aspect" of the disk, i.e.~its vertical structure, projecting the Euler equation in the $z$ direction. To have vertical hydrostatic equilibrium, the gas pressure must compensate the gravity of the central object (using the fact that the disk’s self-gravity is negligible),
\begin{equation}
    \frac{\partial}{\partial z} P = c_s^2\frac{\partial}{\partial z} \rho = - \rho \frac{G M}{d^2} \frac{z}{d} \simeq - \rho \Omega_K^2 z
\end{equation}
where we used an equation of state with a constant speed of sound. This tells us that the density profile has a typical height (the disk's \emph{scale height}) of 
\begin{equation}\label{eq:speed_sound}
    H = c_s /\Omega_K \, .
\end{equation}
To demand that the disk is thin is to demand that $ H/r \ll 1$, or $c_s /v_K \ll 1$, i.e., that the disk is supersonic.

Integrating the continuity equation over $z$, one obtains 
\begin{equation}
    \frac{d}{d t} \Sigma r = r\frac{\partial \Sigma}{\partial t} + \frac{\partial}{\partial r} (r \Sigma v_r) =0 \xrightarrow{\rm stationary}  \frac{\partial}{\partial r} (r \Sigma v_r) = 0 \, .
\end{equation}
This tells us that, to have a stationary disk, we need a mass sink (the accreting black hole!), and that the radial mass flux is approximately constant: 
\begin{equation}
\dot{M} = -2 \pi r \Sigma v_r = \rm{const}  \, .  
\end{equation}
One of the defining properties of AGN is their high luminosity (in a compact region), which largely surpasses the host galaxy luminosity,  and can be close to the respective \textit{Eddington luminosity} of the central black hole
\begin{equation}
L_\text{Edd} = \frac{4\pi G M m_p c}{\sigma_T} \simeq 3.2 \times 10^4 \frac{M}{M_\odot} L_\odot \, ,     
\end{equation}
with $m_p$ the mass of the proton and $\sigma_T$ the Thomson scattering cross section. This luminosity is the one at which radiation pressure exerted by photons scattering on the electrons of the ionized disk gas can balance the gravitational pull of the black hole on the gas itself, which is dominated by the gravitational interaction with the protons. Because this limit strictly applies only under spherical symmetry, it is not directly applicable to geometrically thin accretion disks. Nonetheless, it provides a useful benchmark, and AGN bolometric luminosities (integrated over all frequencies) are typically written $L_\text{bol} = f_\text{Edd} \, L_\text{Edd}$. For AGN, $f_\text{Edd} \gtrsim 0.01$, including values above unity.

The connection between the AGN luminosity and the black hole accretion rate ($\dot{M}$) is parametrised by the radiative efficiency $\eta$,
\begin{equation}
\dot{M} = L_\text{bol}/(\eta c^2)    \, . 
\end{equation}
Typically, $\eta \sim 0.1$ for AGN. The (fractional) Eddington mass accretion is then
\begin{equation}
\dot{M}_\text{Edd} / M = 2.2 \times 10^{-8} \left(0.1 / \eta \right) \, \text{yr}^{-1} \, ,    
\end{equation}
Accretion rates are often expressed in terms of the dimensionless quantity $f_\text{Edd}$ defined by $\dot{M} = f_\text{Edd}\dot{M}_\text{Edd}$.

Finally, we use the projection of momentum conservation (or the Navier-Stokes equation) in the radial direction, involving the centrifugal force, the radial pressure gradient and the gravitational force of the central body,
\begin{equation}
    \frac{v_\phi^2}{r}= \frac{G M}{r^2} + \frac{1}{\rho} \frac{\partial P}{\partial r} \, .
\end{equation}
The pressure term, $\frac{1}{\rho} \frac{\partial P}{\partial r} \simeq \frac{1}{\rho} c_s^2 \frac{\rho}{r} = v_K^2 \frac{H^2}{r^3} $, which gives us
\begin{equation}
    v_\phi^2\simeq v_K^2 + v_K^2 \frac{H^2}{r^2} \simeq v_K^2
\end{equation}
so we learn that  pressure gradients are negligible in thin disks, and that a thin disk is quasi-Keplerian. This also tells us that the angular momentum per unit mass
\begin{equation}
    l = r^2 \Omega_K = \sqrt{ G M r} \, ,
\end{equation}
increases as a function of radius. As a consequence, for material to actually be accreted (i.e., move to smaller radii), it requires a mechanism to \emph{lose} angular momentum. 

There are two classes of mechanisms that can achieve angular momentum loss: ``viscosity" of some kind, i.e., a redistribution of angular momentum within the disk; or a ``wind" of some kind, that takes angular momentum away from the disk.\footnote{This is the reason why accretion is deeply connected with energy release, and why accretion is really the power house of our Universe. The most luminous phenomena in the Universe are connected to accretion.} Disks at the center of galaxies are typically viscous disks. We therefore introduce an additional assumption on the nature of the disk viscosity to determine the disk structure.

\item \textit{$\alpha$-viscosity}
The structure of an accretion disk is determined by the interplay between gravity, pressure (e.g. hydrodynamical or, in some regimes, radiative), and the physical mechanism that transports angular momentum outward. This is the ingredient that ``destabilizes'' the gas flow from (stable) circular motion, allowing gas to fall inward and be accreted. Naively, one would expect that this would come from typical viscous stresses due to molecular collisions between neighboring fluid rings, giving rise to the stress energy tensor
\begin{equation}
    T_{ij} = \nu \rho \left( \partial v^i \partial x^j  + \partial v^j \partial x^i  \right) \, ,
\end{equation}
where $\nu$ is the \textit{kinematic viscosity}. However, the temperatures in accretion disks are sufficiently high ($T \sim 10^5 - 10^7 \, K$), that the gas is fully ionized and the Coulomb mean free path becomes very large:$\sim 3 \times 10^4 \,  \text{cm}$, which should be contrasted with $\sim 1$ cm in the interiors of stars. This makes the resulting microscopic viscosity \emph{too small} to account for the observed accretion rates. 
The shear stresses in AGN are actually believed to originate from \emph{magneto-rotational instability} (MRI). MRI arises because the disk supports a magnetic field that amplifies small radial perturbations between neighbouring fluid elements. These generate turbulent flows, whose fluctuations in velocity and magnetic field transport angular momentum outwards and act as an effective viscosity. Detailed modelling of this turbulent flow requires magnetohydrodynamics simulations. This goes beyond the scope of this lecture; we will instead focus on a phenomenological disk model. 


By far the most commonly adopted disk model follows the seminal prescription of Shakura and Sunyaev from 1973 \cite{Shakura:1972te}. Their approach relies on an educated “guess’’ for the form of the effective viscosity, motivated purely by dimensional arguments. Kinematic viscosity has units
\begin{equation}
[ \nu ] = [\rm{length}^2 ]/ [\rm{time} ]= [velocity] \times [length] \, .
\end{equation}
The maximum speed for propagation of information in the gas is the speed of sound $c_s$, while the maximum length scale for correlations in the turbulent flow, which determine the effective viscosity, is the characteristic scale height of the disk. Shakura and Sunyaev therefore postulated that the effective viscosity is a fraction $\alpha$ of the product of these two quantities ($\alpha$ viscosity)
\begin{equation}\label{eq:alpha_viscosity}
 \nu  = \alpha c_s H \, ,
\end{equation}

Quite remarkably, using this prescription and a series of simplifying assumptions, they were able to construct a fully analytic accretion disk model. 
The assumptions we made so far, including the $\alpha$-viscosity prescription, define algebraic relations between the surface density $\Sigma(r)$, pressure $P(r)$, scale height $H(r)$ and the accretion rate $\dot{M}$. The full solution is determined by the physics of energy dissipation and radiative transport. For these, Shakura and Sunyaev made the following additional assumptions: 

\item \textit{Radiatively efficient}: All heat generated by viscous stresses is assumed to be radiated away locally and instantaneously, rather than advected inward. This “radiative cooling’’ condition relates the local energy generation rate (and the accretion rate) to the disk temperature. 
\item \textit{Optically thick}: the surface density is sufficiently high so that the mean free path of photons is much smaller than the disk scale height, so that they are easily scattered and absorbed. Radiation is then trapped in the disk and can only escape through diffusion, which determines a relation between the effective ``black-body'' temperature and the much larger temperature at the disk midplane. 

To complete the system, one must also specify an opacity law 
$\kappa(\rho,T)$ and an equation of state $P(\rho,T)$. 
Different choices describe different radial regions of the disk 
depending on whether electron scattering or bound-free absorption 
dominates, or whether gas or radiation pressure provides vertical 
support. 
\end{enumerate}

\begin{figure}
    \centering
    \includegraphics[trim={.cm .cm .cm .cm}, width=0.95\linewidth]{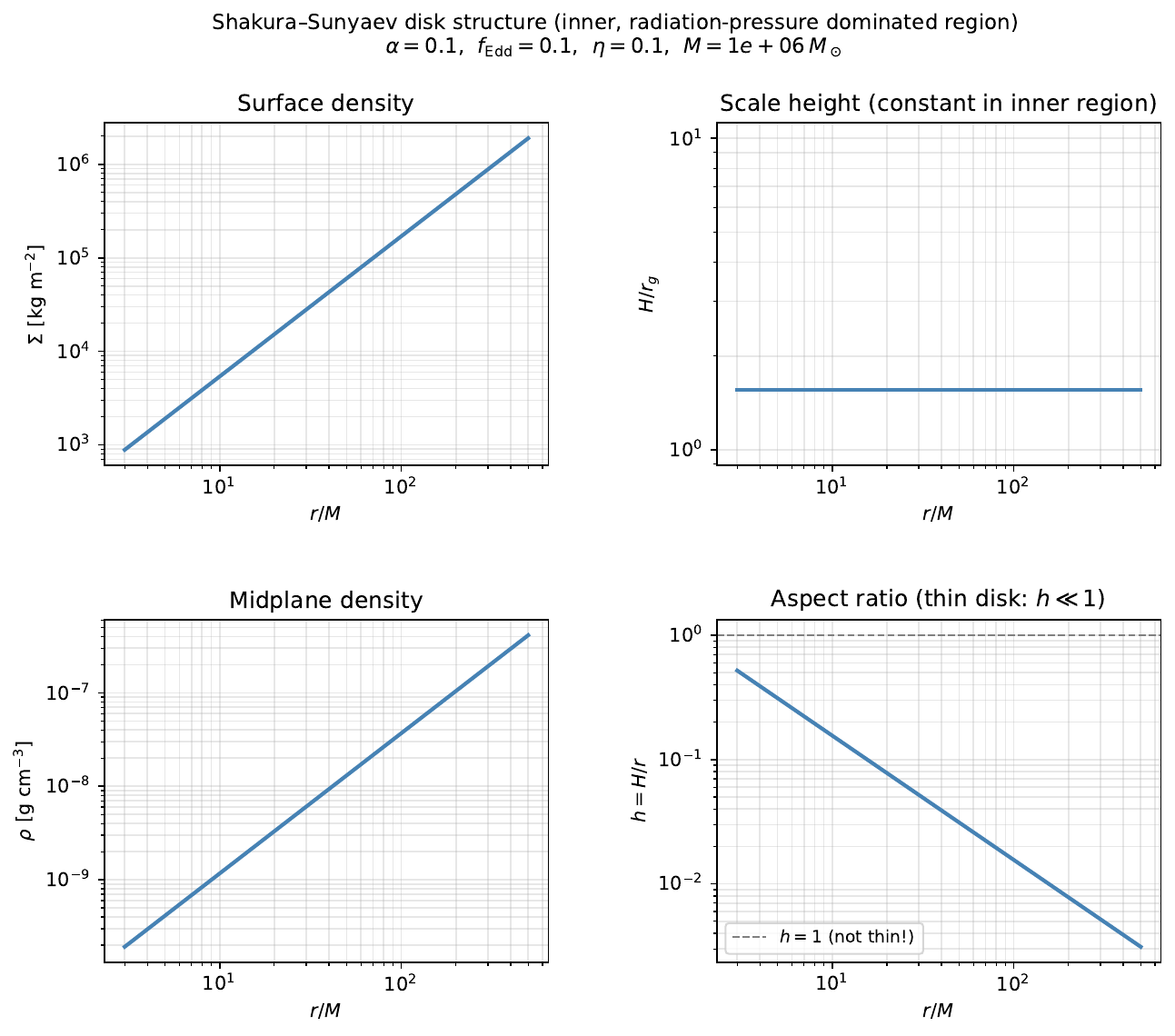}
    \caption{The Shakura-Sunyaev disk in the inner regions.}
    \label{fig:SS_disk}
\end{figure}
For a given accretion rate ratio $f_\text{Edd}$ and a given viscosity parameter 
$\alpha$, the Shakura-Sunyaev model provides closed-form power-law type expressions for 
the full radial disk structure (midplane temperature, density, surface 
density, scale height, effective temperature, etc.). In Fig.~\ref{fig:SS_disk} 
we illustrate some of these quantities for representative values of 
$f_{\rm Edd}$ and $\alpha$.

In the inner regions relevant to our example, which are \emph{radiation-pressure} and \emph{electron scattering} dominated, the disk structure is given by
\begin{align}
    \Sigma(r) &= \Sigma_0 \, (r/M)^{3/2}, \quad {\rm with} \quad  \Sigma_0 = 1.7 \times 10^2 \, {\rm kg/m^2} \left(\frac{\alpha}{0.1}\right)^{-1} \left(\frac{f_{\rm Edd}}{0.1}\frac{0.1}{\eta}\right)^{-1} \, , \label{eq:Sigma_alpha} \\
    H(r) &= H_0 , \quad {\rm with} \quad H_0 = 2.3 \times 10^9 {\rm m} \left(\frac{f_{\rm Edd}}{0.1}\frac{0.1}{\eta}\right) \left(\frac{M}{10^6 M_\odot}\right) \, .
\end{align}
Other useful quantities are the midplane density, which may be approximated as
\begin{equation}
    \rho \simeq \frac{\Sigma}{2H} \, .
\end{equation}
and the disk aspect ratio, 
\begin{equation}
    h \equiv H/r \, .
\end{equation}

{\bf Question}: Do you think these assumptions are justified? Try to make some back-of-the-envelope estimates to assess this. For instance: is the disk really non-relativistic in the regions probed by extreme mass ratio inspirals?\\
{\bf Further reading}: Review on accretion disk theory \cite{Abramowicz:2011xu}.

\subsection{Ingredient 2: the interaction}\label{sec:migration}

Next, we need to assess how the lighter compact object interacts with the accretion disk. There are two complementary ways of characterizing this interaction. The interaction can be seen as a form of \emph{dynamical friction} (reviewed in Sec.~\ref{sec:DF}), due to \emph{a differentially rotating} medium (the disk). 
In the simplest interpretation, the perturber focuses background particles into an overdense wake trailing behind it; this results in a gravitational drag force that opposes its motion relative to the medium. 
Alternatively, the interaction between the perturber and the disk  can be seen as the cumulative effect (\emph{Lindblad or resonant torque}) of all resonances (known as \emph{Lindblad resonances}) between the perturber's oscillatory gravitational potential and the rotating medium.  
Resonant torques and Lindblad resonances play a role in, among other systems, the migration of planets in proto-planetary disks, and in the orbital decay of binaries surrounded by circumbinary disks. 

In this lecture, we will exploit the first formulation (dynamical friction) to find an approximation for the perturber-disk interaction. Consider a compact object of mass $\mu$ on a 
circular orbit of radius $r_p$ around a central mass $M$, with Keplerian angular frequency $\Omega_p=\Omega_K(r_p)$ [Eq.~\eqref{eq:omega_k}]. 
The surrounding gas forms a disk with surface density $\Sigma(r)$ and vertical scale height 
$H(r)$, related to the speed of sound in the gas by $H = c_s /\Omega_K$ [Eq.~\eqref{eq:speed_sound}].
Gas that co-orbits the compact object has no relative velocity with respect to it, so it does not exert any dynamical friction. If we consider a radial offset $r= r_p + \delta r$, the relative (azimuthal) velocity between the gas and the perturber is 
\begin{equation}\label{eq:vrel_disk}
    |v_{\rm rel}| = \left| \Omega_{K}(r) r -\Omega_p r_p \right| \simeq
     \left| \frac{\partial \Omega }{ \partial r } (r_p) \, r_p \, \delta r +\Omega_p \delta r \right|
    = \frac{1}{2} \, \Omega_p \, |\delta r| \, ,
\end{equation}

To first approximation, we will neglect the disk's viscosity and treat the disk as a gaseous (collisional) but inviscid medium. For such a medium, dynamical friction was derived by Ostriker in 1999 \cite{Ostriker:1998fa}. Compared to dynamical friction in a collisionless medium reviewed in Sec.~\ref{sec:DF}, drag in a collisional gas is affected by pressure waves, which reduce the drag in the subsonic limit. As a result, regions of the disk with subsonic relative speeds compared to the perturber will contribute less.
When the relative speed is supersonic, the dynamical friction force is given by\footnote{Since our goal is to get a rough estimate, we neglect the logarithmic factor in the dynamical friction formula.} 
\begin{equation}\label{eq:df_ostriker}
    F_{\rm DF,\,  out} \sim 
   - 4\pi G^2 \mu^2 \frac{\rho}{v^2} 
  \quad ({\rm collisional, nonrelativistic, supersonic})\, .
\end{equation}
From this formula we see that the dominant contribution to dynamical friction comes actually from the region where the flow transitions from subsonic to supersonic,
\begin{equation}
    v_{\rm rel} \sim c_s 
    \quad \Rightarrow \quad |\delta r| \sim H \, .
\end{equation}
where we have used Eq.~\eqref{eq:vrel_disk}.
Therefore, the torque is generated primarily by gas located within a radial distance $\sim H$ of the perturber's orbit, exerting a force
\begin{equation}\label{eq:df_ostriker_cs}
    F_{\rm DF,\,  out} \sim 
   - 4\pi G^2 \mu^2 \frac{\rho}{c_s^2} 
    \simeq - 8\pi G^2 \mu^2 \frac{\Sigma}{H^3 \Omega_p^2}  \quad ({\rm collisional, nonrelativistic, supersonic})\, .
\end{equation}

Eq.~\eqref{eq:df_ostriker_cs} is the contribution just from the \emph{outer} disk, where particles move slower than the compact body. The inner disk region instead rotates faster and exerts a forward push, which naively we could assume to have the same magnitude and opposite side as $F_{\rm DF,\,  out}$. The net effect would then cancel. However, the disk ($\Sigma(r)$, $H(r)$, $c_s(r)$) varies over the scale $\delta r \sim H$, leading to a fractional asymmetry of order $H/r_p$. Thus, the net dynamical friction force is of the order of 
\begin{equation}
    F_{\rm DF, \, net} \sim \frac{H}{r_p} F_{\rm DF,\, out}
    \sim -8\pi G^2 \mu^2 \frac{\Sigma}{H^2 \Omega_p^2 r_p} \, .
\end{equation}
The resulting torque on the perturber is
\begin{equation}\label{eq:torque_mig}
    T_{\rm disk} \simeq r_p \, F_{\rm DF, \, net}
    = - 8\pi G^2 \mu^2 \frac{\Sigma}{H^2 \Omega_p^2} \, \simeq - C_{\rm L}\mu G \frac{\mu}{M} \Sigma \frac{r^3}{H^2} .
\end{equation}
where we have introduced $C_{\rm L} \sim \mathcal{O}(1)$ to absorb all numerical factors. 

This back-of-the-envelope estimate captures the main features of the perturber-disk interaction: the proportionality to the mass ratio $\mu/M$ and the disk surface density $\Sigma$, the inverse proportionality to the disk scale height. In a more careful calculation, the numerical prefactor $C_{\rm L}$ would depend on the density and temperature gradients, on the tri-dimensional nature of the disk, on thermal and radiative processes, etc. 
The sign of the torque is determined by the relative strength of the inner and outer component. In a Shakura-Sunyaev, isothermal disk, the torque is negative, i.e. the perturber loses angular momentum to the disk and inspirals.

As anticipated, the scaling \eqref{eq:torque_mig} was also derived by Goldreich and Tremaine in their seminal work in 1979-1980 \cite{Goldreich:1980wa,GoldreichTremaine1},  by looking at the linear response of a fluid disk to the time-varying gravitational potential of an embedded perturber.
They expanded the planet's potential in azimuthal Fourier modes with frequency $m \Omega_p$
\begin{equation}
   \Phi_p (r, \phi, t)
   = \sum_{m \ge 0} \Phi_m(r)\, e^{i m (\phi - \Omega_p t)} \, ,
\end{equation}
and directly solved a wave equation obtained from linearizing the mass continuity and the Euler equation that govern the disk response. At a radius $r$, the gas experiences perturbations at a Doppler-shifted frequency $m\!\left[\Omega_p - \Omega(r)\right]$ in its rest frame.  
There will be radii at which this forcing frequency matches the \emph{epicyclic frequency} $\kappa(r)$ of the disk, i.e., the frequency at which a fluid element oscillates about its circular orbit when slightly displaced, acquiring a small eccentricity. For a Keplerian disk the epicyclic frequency reduces to $\kappa = \Omega$. These resonances, known as \emph{Lindblad resonances}, occur where
\begin{equation}
    D(r) = \pm m\!\left[\Omega_p - \Omega(r)\right] - \kappa  = 0 \, .
\end{equation}
At each of these resonances spiral density waves are launched: inner Lindblad waves propagate inward and remove angular momentum from the disk, while outer Lindblad resonances propagate outward and add angular momentum.  
By conservation of angular momentum, the perturber experiences equal and opposite torques: the inner resonances push it outward and the outer resonances pull it inward. 

The total torque is obtained by summing over all azimuthal modes $m$. 
However, modes with very large $m$ (short azimuthal wavelength) lie so close together in radius that they overlap and interfere destructively. This occurs for $m \gtrsim r_p / H$, i.e.~perturbations generated at distances $\sim H$ from the perturber's orbit, which is the same region excluded by sub-sonic motion in the dynamical friction estimate above.  
Summing over all resonances outside this region reproduces the scaling of Eq.~\eqref{eq:torque_mig}, confirming that the dynamical-friction and linear-wave pictures are equivalent representations of the same underlying physics. This equivalence is not a coincidence: both pictures are linear responses of a fluid to a gravitational perturbation, just described in different ways.

{\bf Questions}: 1) Do you know, or can you think of other physical settings where this interaction would play a role? 2) What is the radial dependence of the torque in the Shakura-Sunyaev disk of the previous section? \\
{\bf Further reading}: The original papers on Lindblad torques \cite{GoldreichTremaine1, Goldreich:1980wa}.
\subsection{Ingredient 3: gravitational waves}\label{sec:accdisk}
We are finally ready to assess how the interaction with the disk affects the orbit and gravitational wave signal of the binary.
Extreme mass-ratio inspirals (EMRIs) can be modelled perturbatively exploiting the mass ratio $\epsilon\equiv q$ as a small parameter. The metric of the primary black hole is expanded perturbatively as $g_{\mu\nu} =g^{\rm Kerr}_{\mu\nu} + \epsilon h^{(1)}_{\mu\nu} + \mathcal{O}(\epsilon^2)$. During the inspiral, an additional approximation (the ``two-timescale" expansion) can be made by exploiting the disparate times scales of the problem, also related to the small mass ratio $q$. In the following, we first describe the construction of vacuum waveforms implemented in Fast EMRI Waveforms (FEW),  a highly modular framework for the rapid generation of accurate EMRI waveforms.

At zeroth order in the mass ratio, the EMRI travels on a geodesic of fixed radial, azimuthal and polar frequencies $\Omega_r$, $\Omega_\phi$, $\Omega_\theta$. The EMRI separation (semi-latus rectum) $p$, eccentricty $e$ and inclination $x_I=\cos(\theta)$ are all constant at leading order, while the phases of orbital motion oscillate on the fast (orbital) timescale.
At first order in $\epsilon$, the EMRI emits gravitational waves that carry away energy and angular momentum on a timescale longer $\sim q^{-1}$ than the orbital one. This affects the EMRI orbit, causing the orbital constants to evolve slowly with time. This is captured by the system of equations
\begin{align}
    \frac{d}{dt}p &= \epsilon \, f_p(a, p, e, x_I) \label{eq:p} \\
    \frac{d}{dt}e &= \epsilon \, f_e(a, p, e, x_I) \label{eq:e} \\
    \frac{d}{dt}x_I &= \epsilon \, f_{x_I}(a, p, e, x_I) \label{eq:x} \\
     \frac{d}{dt}\Phi_{\varphi,\theta,r} &= \Omega_{\varphi, \theta, r}(a, p, e, x_I)/M 
\end{align}
with initial condition $\{\Phi_{\varphi0},\Phi_{\theta0},\Phi_{r0},
p_0,e_0,x_{I0}\}$. The right-hand sides $f_i$ are determined by the fluxes of energy and angular momentum (and a quantity related to the Kerr Carter constant in non-equatorial orbits). For example the first equation is derived from $\frac{dL}{dt}= \frac{dp}{dt} \frac{dL}{dp}= \dot{L}$ where $\dot{L}$ is the flux of angular momentum in vacuum. {\bf In the presence of the disk, the disk torque contributes to the right hand side of Eqs.~\eqref{eq:p}, \eqref{eq:e} and \eqref{eq:x}.}

At a large distance from the source, the waveform $h$ is given by
\begin{equation}\label{eq:main_wave}
    h = \frac{\mu}{d_L}\sum_{lmkn} A_{lmkn}(t) \, S_{lmkn}(\theta,\phi) \, \exp[-i\Phi_{mkn}(t)],
\end{equation}
where $t$ is the time of arrival of the gravitational wave at the solar system baricenter, $d_L$ is the luminosity distance of the source, and $(l,m,k,n)$ are the indices describing the frequency-domain harmonic mode decomposition (or simply ``harmonic'' or ``mode''): $l$ denotes the orbital angular momentum mode index and can take the value $l=2,3,...$; $m$ denotes the azimuthal mode index and can take the values from $-l\le m \le l$; $k$ and $n$ denote the polar and radial mode indices and can take values from $-\infty$ up to $+\infty$. 
The phases $\Phi_{mkn}$ are related to the orbital phases by $\Phi_{mkn} =m\Phi_{\varphi} + k\Phi_{\theta} + n\Phi_{r}$. These phases will be affected by the environmental interaction through the correction to the EMRI trajectory.

The amplitudes $A_{lmkn}(t)$ quantify the size of each harmonic $(l,m,k,n)$, and are determined by solving for the metric perturbation sourced by an EMRI on geodesic motion at fixed orbital parameters $A_{lmkn}(p(t), e(t), x_I(t))$. For example, in sources with comparable masses and no precession or eccentricity, the strongest mode is the $(l=2,m=2,k=0,n=0)$. Finally, the angular function $S_{lmkn}(\theta,\phi)$ describes how the amplitude is modulated depending on the source-frame polar viewing angle $\theta$, the source-frame azimuthal viewing angle $\phi$. For Schwarzchild black holes these functions reduce to the spin-weighted spherical harmonics ${}_{-2}Y_{lm}$.
Because waveform amplitudes are already proportional to the small mass $\mu$, and because we are considering disks with negligible self-gravity, the effect of the environment on the amplitudes is negligible.

At leading order in the velocity of the secondary object (i.e., at leading order in the post-Newtonian expansion), the binary emits gravitational waves that carry away angular momentum at the rate
\begin{equation}\label{eq:Ldot_gw}
    \dot{L}_{\rm gw} = -\frac{32}{5} \frac{G^{7/2}}{c^5} \mu^2 M^{5/2} r^{-7/2} \, .
\end{equation}
To get a sense of how this compares with the angular momentum exchanged with a Shakura-Sunyaev disk, we can look at the ratio
\begin{equation}
    \frac{T_{\rm disk}}{\dot{L}_{\rm gw}} \simeq  \frac{5}{32} \frac{c^5}{G^{5/2}} C_L \Sigma(r) \,  H(r)^{-2} M^{-7/2} r^{13/2} 
    \, ,
\end{equation}
where we used Eq.~\eqref{eq:torque_mig} for the disk torque. If we use Eq.~\eqref{eq:Sigma_alpha} for the disk surface density profile and express the ratio in terms of the gravitational wave frequency $f_{\rm gw} = \Omega(r)/\pi$, we find
\begin{align}
    \frac{T_{\rm disk}}{\dot{L}_{\rm gw}} \simeq  & \frac{5}{32} \frac{c^8}{G^{4/3}} C_L \Sigma_0 H_0^{-2} M^{-7/3} (\pi f_{\rm gw})^{-16/3}  \nonumber \\
    \simeq & 1 \times 10^{-9} C_L \left(\frac{\alpha}{0.1}\right)^{-1} \left(\frac{f_{\rm Edd}}{0.1}\frac{0.1}{\eta}\right)^{-3} \left(\frac{M}{10^6 M_\odot}\right) \left(\frac{r}{10 M}\right)^{8}
    \, .
\end{align}
This is a very small number! However, the EMRIs detectable by milli-Herz gravitational wave detectors are extremely sensitive probes. These systems have tens of thousands of cycles detectable in band ($n_{\rm cycle} \sim (\mu/M)^{-1} \sim 10^5 -10^6$). Small angular momentum losses accumulate over each cycle, adding up to a measurable dephasing.
We can also gain a couple of orders of magnitude by looking at disks with more favorable (smaller, but not unrealistic) accretion rates and viscosities. Finally, the coefficient $C_L$ depends on the details of the disk interaction, the disk thermodynamics and even relativistic corrections, and can be larger than 1.

Orders in the post-Newtonian expansion are typically counted in powers of the velocity squared $(v/c)^2 \sim f_{\rm gw}^{2/3}$. In this sense, the disk interaction (obtained with Newtonian physics) comes at post-Newtonian order $n=-8$. Negative post-Newtonian effects are relatively more important in the orbital evolution of the binary at larger distances compared to gravitational-wave emission. This is the case for most environmental effects.  For example, disk interactions affect the migration of the secondary compact object from larger radii in the disk, and should be taken into account when studying the formation of these systems in gas-rich environments, such as active galactic nuclei. The catch is that, at larger separations, the overall amplitude of the gravitational wave signal is also smaller, and the orbital evolution slower, so that the source accumulates less signal-to-noise ratio in this region.

In practice, we might want to be more agnostic when analysing gravitational wave data, and use a parametrisation of the type
\begin{equation}
    T_{\rm disk} = A_{\rm disk} (r/10 M)^{n_{\rm disk}} \dot{L}_{\rm gw} \, ,
\end{equation}
inspired by our model so far. We can then use observations to constrain both $A_{\rm disk}$ and $n_{\rm disk}$, or fix the latter to the scaling predicted by a given disk and torque model. 

You can find out more on the tutorial notebook at \href{https://github.com/stephengreen/DirtyEMRI-tutorial}{https://github.com/stephengreen/DirtyEMRI-tutorial}.

{\bf Further reading}: More on the construction of EMRI waveforms behind the FEW package \cite{Drasco:2005kz,Hughes:2021exa}.
\section{Example 2: asymmetric binary in a boson cloud}
In this example, we will continue to work with the same gravitational wave source (an extreme mass ratio inspiral around a massive black hole) but place it in a different environment: a boson cloud.

\subsection{Ingredient 1: the environment}

Ultralight bosons or axions (with masses $\mu_b\sim 10^{-22}-10^{-10} $ eV) arise in many extensions of the Standard Model and general relativity. For example, the Peccei-Quinn axion arises in the resolution of the strong charge-parity problem in quantum chromo-dynamics; ultralight bosons arise from compactification in string theory; scalar fields nonminimally couple to the metric in many alternative theories of gravity. 

Ultralight bosons can naturally accumulate in over-densities around spinning black holes through the \emph{superradiant instability}. When the superradiant condition is met 
\begin{equation}
    \omega < m \Omega_{\rm H} \, ,
\end{equation}
a wave of frequency $\omega$ and azimuthal number $m$ can extract energy and angular moment from a black hole with horizon angular frequency $\Omega_{\rm H}$. In addition, the mass of the boson acts as a confining mechanism for waves with $\omega < \mu_b$, allowing the scalar field to grow on a sufficiently short timescale when the parameter 
\begin{equation}
    \alpha = \mu_b M \lesssim 1 \, ,
\end{equation}
where $\alpha$ is the ratio between the gravitational radius of the black hole and the Compton wavelength of the boson.
The parameter $\alpha$ can be seen as the gravitational coupling between the boson and the black hole.
As the scalar field grows, the black hole spins down until the superradiant condition is no longer met. The result is a long-lived overdensity (up to $\sim 10\%$ of the black hole mass) known as a boson or superradiant cloud, or gravitational atom. 

A simple example of an ultralight boson is a massive scalar field of mass $\mu_b$. On a curved spacetime with metric $g_{\alpha\beta}$, the field satisfies the Klein-Gordon equation  
\begin{equation}\label{eq:KG}
    \left(g^{\alpha\beta} \nabla_{\alpha}\nabla_{\beta} - \mu_b^2\right) \Phi = 0 \, .
\end{equation}
On an axi-symmetric, stationary black hole background, this equation admits \textbf{quasi-bound} mode solutions of the form
\begin{equation}
    \Phi(t,r,\theta,\varphi) = e^{-i\omega t} e^{i m \phi} S_{\ell m}(\theta) \frac{1}{r} \phi_{\ell \omega} (r) \, ,
\end{equation}
where $\ell, m$ are the angular numbers, $S$ and $R$ solve separated angular and radial equations and the frequency $\omega=\omega_{R} + i \omega_I$ is complex and takes value in a spectrum ${\omega_{n \ell}}$. Quasi-bound states are confined in the vicinity of the black hole, within approximately a Bohr radius
\begin{equation}\label{eq:boson_radius}
    r_b \simeq M \alpha^{-2} = 100 M \left(\frac{0.1}{\alpha}\right)^{2} \, ,
\end{equation}
(inside the gravitational wave-emitting region of an EMRI!)
and typically decay over time ($\omega_I<0$), unless superradiantly unstable ($\omega_I>0$) or unless the superradiant condition is saturated $\omega_I=0$. 

For small $\alpha \ll 1$ and at large distances from the black hole $r/M\gg1$, the Klein Gordon equation reduces to the Shr\"odinger equation for the hydrogen  atom, with the gravitational coupling $\alpha$ playing the role of the fine structure constant. The eigenfrequencies are set by the energy levels $\omega_{n\ell m} \simeq \mu_b[1 - \alpha^2/(2n^2)]$, and the spatial profiles are hydrogen-like wavefunctions with a characteristic Bohr radius $r_b = n^2/(\alpha \mu_b)$. Explicitly, with the ansatz
\begin{equation}
    \Phi = \frac{1}{\sqrt{2 \mu_b}} \left(\psi(t,r)e^{-i\mu_b t} + {\rm c.c.} \right)
\end{equation}
and in the limits $\alpha\ll 1$ and $r/M\gg1$,
the Klein-Gordon equation reduces to a time-dependent Shr\"odinger equation with a Coulomb potential,
\begin{equation}
    i \partial_t \psi(t,r) = \left( -\frac{1}{2\mu_b} \nabla^2 -\frac{\alpha}{r} \right) \psi(t,r)
\end{equation}
The eigenfunctions are given by, 
\begin{align}
    \psi(t,r) =& e^{-i(\omega_{n\ell}-\mu_b)t} R_{n\ell}(r) Y_{\ell m}(\theta,\phi) \, , \\
    R_{n\ell}(r) =& \sqrt{\left(\frac{2\mu_b \alpha}{n}\right)^3\frac{(n-\ell -1)!}{2n(n+\ell)!}} \left(\frac{2\mu_b \alpha r}{n}\right)^\ell e^{-\frac{\mu_b \alpha r}{n}}
    L_{n-\ell-1}^{2\ell+1} \left(\frac{2\mu_b \alpha r}{n}\right) \label{eq:atom_radial}
\end{align}

Superradiance is one natural production mechanism, but dense bosonic environments can also arise from gravitational cooling (wave condensation around compact objects), from direct dark matter accretion, or from a pre-existing ultralight dark matter background. 
Ultralight bosonic fields are indeed a potential candidate for dark matter (known as \emph{fuzzy dark matter}), as they behave as cold dark matter on large scales, while suppressing the growth of structure at small scales due to their large de Broglie wavelength,
\begin{equation}
    \lambda_{\rm dB} = \frac{\hbar}{p} \simeq \frac{\hbar}{\mu_b \sigma} \simeq 2. \times 10^{-2} {\rm pc} \left(\frac{\mu}{10^{-18} {\rm eV}} \right)^{-1} \left(\frac{\sigma}{100 {\rm km/s}} \right)^{-1}\, ,
\end{equation}
where $\sigma$ is the velocity dispersion of the bosons.
As dark matter, ultralight bosons must be produced (as quantum particles) and condensate into a Bose-Einstein condensate in the early Universe, after which their distribution is characherised by a classical wave function satisfying a nonlinear Schr\"{o}dinger equation.
Boson stars -- gravitationally bound solitons of a self-interacting scalar field -- are another realisation of bosonic overdensities, though they are not directly connected to the central black hole.

{\bf Further reading:} Review on black hole superradiance \cite{Brito:2015oca}. 

\subsection{Ingredient 2: the interaction} 
The simplest way to describe the interaction between the secondary and the boson cloud is again through (surprise!) \emph{dynamical friction}. 
Let us compare the set up originally reviewed in Sec.~\ref{sec:DF} with the case at hand. Besides the intrinsic limitations of Chandrasekhar's formula, we have neglected the fact that our medium is composed of waves, rather than particles. In the system under consideration, the de-Broglie wavelength is comparable with other scales of the problem, and we therefore cannot neglect wave effects. The dynamical friction formula for a bosonic field medium is more involved; We will use here an approximation valid when the system is non-relativistic and when the velocity of the perturber is $v\gg \mu_b \mu$,
\begin{equation}\label{eq:DF_boson}
    \vec{F}_{\rm DF} \simeq - 4 \pi G^2 \mu^2 \rho  \left[ \ln (v_p \mu_b b_{\rm max}) +\gamma_E \right] \, \frac{\hat{v}_p}{v_p^2} \, , \quad \text{(scalar medium, non-relativistic),}
\end{equation}
where $\gamma_E\simeq 0.577$ is the Euler-Mascheroni constant. 

This formula was derived assuming the environment is uniform and the perturber moves with constant velocity.
For our perturber on a quasi-geodesic orbit through a boson cloud, this is still an approximation: the motion of the perturber is not uniform and straight; the boson cloud is not a uniform medium; there is an ambiguity in the choice of $b_{\rm max}$. For instance, even in the non-relativistic approximation, the density of the superradiant cloud is given by,
\begin{equation}
    \rho_\phi \simeq \mu_b^2 |\Phi|^2 \, .
\end{equation}
whose radial dependence is given by Eq.~\eqref{eq:atom_radial}, and the cloud is never spherically symmetric, as the superradiant instability extracts angular momentum from the host black hole. 

Despite these limitations, the formula \eqref{eq:DF_boson} gives the right order of magnitude for the perturber-cloud interaction. The resulting torque on the perturber is
\begin{equation}
    T_\text{boson cloud} \simeq r_p \, F_{\rm DF} = - 4 \pi G^2 \mu^2 \rho  \left[ \ln (v_p \mu_b b_{\rm max}) +\gamma_E \right] \, \frac{r_p}{v_p^2} \, .
\end{equation}
We will now use this formula to estimate the effect of the cloud on gravitational wave observations. 

In the non-relativistic approximation, dynamical friction between the perturber and the cloud is more naturally described as \emph{ionization}. As in quantum mechanics, the perturber's potential induces transitions between the cloud's unperturbed state and the continuum (or unbound) spectrum. The torque is related to the power exchanged over all allowed jumps (which must satisfy the angular selection rules and energy conservation). 


{\bf Further reading:} Relativistic dynamical friction in ultralight dark matter \cite{Vicente:2022ivh}; the perturber-cloud interaction in the hydrogen-atom approximation \cite{Baumann:2021fkf}.

\subsection{Ingredient 3: gravitational waves} Similarly to the accretion disk case (Sec.~\ref{sec:accdisk}), we can compare the torque exerted by the cloud with the loss of angular momentum due to gravitational wave emission, Eq.~\eqref{eq:Ldot_gw},
\begin{align}
    T_\text{boson cloud}/ \dot{L}_{\rm gw} =  &
    \frac{5}{8} \pi C_b \frac{c^5}{G^{5/2}} \rho_{\rm b} \frac{r^{11/2}}{M^{7/2}} \nonumber \\
    \simeq& 5 \times 10^{-3} C_b 
    \left(\frac{M_b}{0.05 M}\right)
    \left(\frac{\mu_b}{10^{-17} {\rm eV}}\right)^6 
    \left(\frac{M}{10^{6} M_{\odot}}\right)^{6}
    \left(\frac{r}{10M}\right)^{11/2}
    \, ,
\end{align}
where in the second step we took $\rho_b \simeq M_b/r_b^3$ with the radius $r_b$ given in Eq.~\eqref{eq:boson_radius} and with $M_b$ the mass of the boson cloud.
To simplify this formula further and get an order of magnitude estimate, we have neglected the dependence of the logarithmic piece on the radius and velocity and absorbed it into a coefficient $C_b$. 
Note that numerically the coefficient $C_{\rm b} \simeq \mathcal{O}(1)$. If we take $b_{\rm max}\simeq r_b$ to be the approximate extension of the cloud and the boson mass so that 
\begin{equation}
    \alpha = 0.075 \left(\frac{\mu_b}{10^{-17} {\rm eV}}\right) \left(\frac{M}{10^{6} M_{\odot}}\right) \, .
\end{equation}
we have $C_b\simeq 2.0$.

You can find out more on the tutorial notebook at \href{https://github.com/stephengreen/DirtyEMRI-tutorial}{https://github.com/stephengreen/DirtyEMRI-tutorial}.

\section{Example 3: comparable-mass binary around a third body}
In this example, we will talk about the effect of a third body. The third object could be
\begin{itemize}
    \item a third compact object of stellar origin, with the three objects forming a triple system. Triples are very common star systems, especially for massive stars;
    \item a supermassive black hole, typically at the center of a galactic nucleus;
    \item a diffuse source of gravity, such as a stellar cluster; 
\end{itemize}

As our source of gravitational waves, we will depart from previous examples and consider a comparable-mass compact-object binary of stellar origin. Comparable-mass binaries composed of black holes and neutron stars are currently detected LIGO and Virgo, and could be detected by future milli-Herz missions such as LISA. Our binary will have source-frame masses $M_1>M_2$, total mass $M=M_1+M_2$ and chirp mass $\mathcal{M}=(M_1M_2)^{3/5}M^{-1/5}$.

\subsection{Ingredient 1: the environment}
For concreteness, we will focus on the case in which the third object is a supermassive black hole, with mass $M_3$ between $10^5 M_\odot$ and $10^{11} M_\odot$ and spin $a_3$. We denote the radial separation from the binary by $R_3$.

Comparable-mass, stellar-origin black hole binaries can form in the accretion disk around the supermassive black hole, if the latter is actively accreting, or in the surrounding nuclear cluster. When formed or captured by the accretion disk, the binary is likely to orbit around the supermassive black hole on a circular outer orbit (due to the migration torques reviewed in Sec.~\ref{sec:migration}!), aligned with the plane of the disk (itself aligned with the plane perpendicular to the supermassive black hole's spin). The orbit is also likely to take place in a so-called migration trap, where outward and inward disk torques cancel and black holes can meet and merge efficiently. The location of migration traps is sensitive to disk physics and to the value of $M_3$, typically ranging between $R_3/M_{3}\simeq 10 - 10^3$ for a central supermassive black hole of mass $M_{3}$. 

{\bf Further reading:} The original proposal for the AGN formation channel for stellar-mass binaries \cite{Bartos:2016dgn}.

\begin{figure}
\vspace{2cm}
    \centering
    \includegraphics[trim={.cm 1.cm .cm 8.5cm}, width=0.5\linewidth]{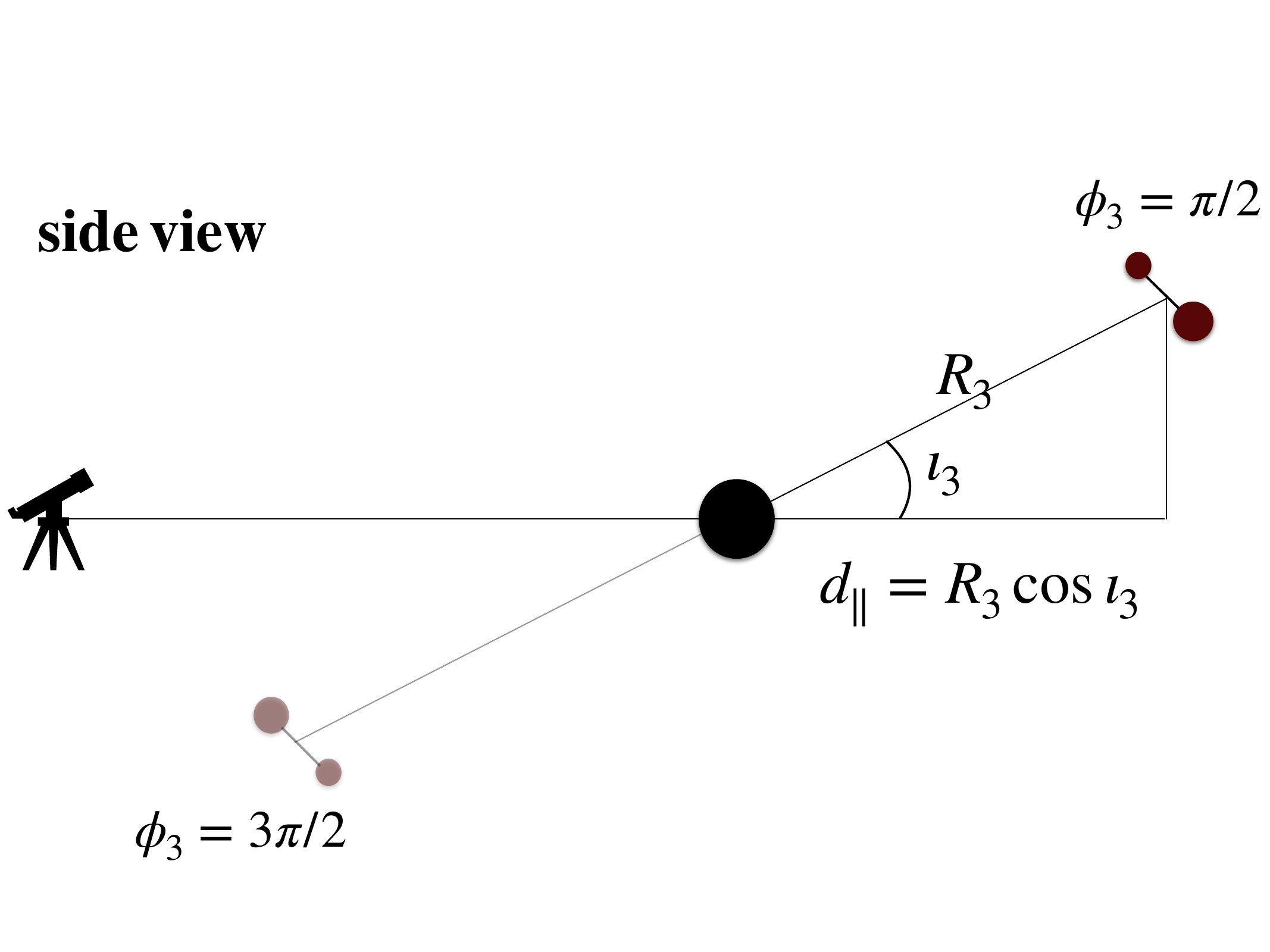}
    \caption{Parameters of the hierarchical triple system.}
    \label{fig:triple_configuration}
\end{figure}


\subsection{Ingredient 2: the interaction}

The biggest effect of the presence of the supermassive black hole is the \emph{motion of the binary} around the third body. Even in the Newtonian limit, the component of the motion along the line of sight induces a time-dependent propagation delay between the emission and observation of the signal. If the source has a position $d_{\parallel}(t)$ projected along the direction of the observer (pointing from the observer to the source), and in the non-relativistic approximation, the observer time is related to the source-frame time by
\begin{equation}\label{eq:time_delay}
    t_{\rm o} = t_s + d_{\parallel}(t_s)/c \, .
\end{equation}
Overall, this effect is known as the \emph{Rømer delay}, and plays an important role in pulsar timing and other astronomical observations.

Expanding the motion around $t=0$, we can write
\begin{equation}\label{eq:time_expansion}
    d_{\parallel}(t_s) =d_{\parallel,0} + v_{\parallel,0} t_s + a_{\parallel,0} t_s^2/2 + \mathcal{O}(t^3) \, ,
\end{equation}
where $v_{\parallel,0}$, $a_{\parallel,0}$ are the velocity and acceleration projected along the line of sight, respectively.  
A constant line-of-sight velocity $v_\parallel$ produces a \emph{Doppler} effect on the waveform, inducing a redshift or blueshift,
\begin{equation}
    1+z = \frac{dt_o}{dt_s} = 1 + v_{\parallel,0} + \mathcal{O}(t) \, .
\end{equation}
The redshift is not a gravitational-wave observable: it is completely degenerate with the intrinsic source-frame masses. It is also indistinguishable from a redshift of cosmological origin. 
We can therefore set $v_\parallel=0$ or, equivalently, absorb it in the definition of the observed masses, distance, etc. In the following, we will therefore focus on the effect of the line-of-sight acceleration $a_{\parallel,0}$. 

If the source is on a circular orbit of radius $R_3$ around the third body, we can write the delay exactly as
\begin{equation}
    d_\parallel(t) = R_3 \cos{\iota_3} \sin(\Omega_{K,3} t + \phi_3) \, ,
\end{equation}
If we expand this in $t$ around $t=0$, and define the angle between the acceleration vector and the line of sight at $t=0$, $\cos\theta_{\parallel,0} = -\cos(\iota_3)\sin(\phi_3)$, we find that the initial line-of-sight acceleration is given by 
\begin{align}
    a_{\parallel,0}= \cos\theta_{\parallel,0} M_3/R_3^2  \simeq 2 \times 10^{-5} \cos\theta_{\parallel,0} \left(\frac{M_{\rm SMBH}}{10^6 M_\odot}\right)^{-1} \left(\frac{R}{100 \, M_{\rm SMBH}}\right)^{-2}  c/s\, .
\end{align}
Along the full outer orbit, the acceleration along the line of sight varies on the Keplerian timescale
\begin{equation}
    T_3 = 2\pi \frac{R_3^{3/2}}{M_3^{1/2}} \simeq 0.001 \, {\rm yr} \left(\frac{M_{\rm SMBH}}{10^6  M_{\odot}}\right) \left(\frac{R}{100 \,  M_{\rm SMBH}}\right)^{3/2} \, .
\end{equation}
For this system, this timescale is much longer than the observation time in current gravitational-wave detectors. Therefore, it is typically appropriate to use a small $t$ expansion and neglect higher derivatives beyond the acceleration. However, future detectors such as LISA could observe the same source for years, and be sensitive to higher orders of the expansion, or even to the full motion. 


The effect of the third body, of course, is not restricted to a Newtonian time delay. In the next section, we will briefly review the relativistic extension of the Doppler effect. The third body affects the propagation of the gravitational wave signals also through its gravitational field (e.g.~the \emph{Shapiro} effect, lensing). Finally, the third body can also directly affect the dynamics of the source binary. Three-body gravitating systems are notoriously difficult to model, with some part of the parameter space exhibiting chaos and instabilities. To be dynamically stable (i.e., live long enough to be observed!) triples need to be hierarchical, consisting of an inner binary (with masses $M_{1,2}$ and separation $a_{12}$) with a tertiary companion (with mass $M_3$ and separation $a_{3}$) on a much wider orbit around the binary’s centre of mass. These systems are typically modeled perturbatively in the small ratio of the semi-major axes $a_{12}/a_{3}$ of the inner and outer binary.


{\bf Further reading:} An interesting approach to hierarchical triple systems: the effective two-body approach of \cite{Kuntz:2021ohi}.

\subsection{Ingredient 3: gravitational waves}
So far we have used the language of Newtonian physics, and focused on time delays. We will now review the effect of relativistic motion on the full gravitational wave signal.  

Let us begin with the simplest example: a boosted source, i.e., a source moving with constant velocity with respect to a reference source. We need to understand how the gravitational wave emitted in the source frame (i.e.~a vacuum waveform model) propagates to the observer, and how it is seen in the observer's reference frame. This requires the definition of three reference frames: the boosted source's four-velocity $\tilde{u}_s$, the reference (unboosted) source's four-velocity $u_s$ and the observer's four-velocity $u_o$.
We then write the gravitationa wave in the WKB or eikonal approximation, an expansion in the rapidly oscillating phase,
\begin{equation}\label{eq:boost_WKB}
    h_{\mu\nu} = {\rm Re}(H_{\mu\nu}e^{-i \Phi}) \, .
\end{equation}
where $H_{\mu\nu}$ is the (slowly evolving) tensor amplitude and $\Phi$ is the rapidly evolving phase. 
The Doppler factor is defined as the ratio between the energy in the boosted source frame and in the reference source frame,
\begin{equation}\label{eq:doppler_source}
    \mathcal{D}_s \equiv \frac{\tilde{E}_s}{E_s} = \frac{\tilde{u}_s^\mu k_\mu}{u_s^\mu k_\mu}  = \gamma (1+n_\mu v^\mu)\, ,
\end{equation}
where $k_\mu = \partial_\mu \Phi$ is the wave's eikonal momentum, $n$ is the direction of observation and $v$ is the boost velocity, both measured with respect to the reference source, and $\gamma = 1/\sqrt{1-v_iv^i}$. 
The redshift induced by the boost is given by 
\begin{equation}\label{eq:redshift}
   1+z \equiv \frac{\tilde{u}_s^\mu k_\mu}{u_o^\mu k_\mu} = \frac{\tilde{u}_s^\mu k_\mu}{u_s^\mu k_\mu} \frac{u_s^\mu k_\mu}{u_o^\mu k_\mu}  = \mathcal{D}_s (1+ z_{\rm ref})\, .
\end{equation} 
where $z_{\rm ref}$ is the reference source's redshift, e.g. its cosmological redshift.
The observer and reference source proper times are then related by 
\begin{equation}
    d\tau_o = (1+z) d\tilde{\tau}_s. 
\end{equation}
This is the relativistic version of the time delay \eqref{eq:time_delay}. Indeed, both the Doppler shift factor \eqref{eq:doppler_source} and the redshift \eqref{eq:redshift} include the relativistic gamma (or Lorentz) factor. The latter contains both the the longitudinal and transverse components of the boost with respect to the direction of observation, and gives rise to the purely relativistic \emph{longitudinal} Doppler effect.

In addition to rescaling the time with respect to the source frame, the boost has also an effect on the gravitational wave amplitude.
Far from the source, the gravitational wave amplitude scales as $h\sim  1/d_L$ where $d_L$ is the luminosity distance. The luminosity distance is defined from the relation between the energy flux and the intrinsic luminosity, so that $F=L/(4\pi d_L^2)$. When the source is boosted ($v_\parallel>0$), two effects suppress the observed flux relative to the unboosted case. First, each graviton arrives with energy redshifted by $\mathcal{D}_s$, reducing the energy per graviton. Second, the rate at which gravitons arrive is also reduced by $\mathcal{D}_s$, because the emission rate is Doppler-shifted along with the frequency. Together these give $F \propto \mathcal{D}_s^{-2}$, resulting in  $h\propto  \mathcal{D}_s^{-1}$. Equivalently, the boost makes the source appear to lie at a luminosity distance larger by a factor $\mathcal{D}_s$. A similar argument applies to the case $v_\parallel<0$.

The boost has also an effect on the \emph{observed direction of propagation} of the gravitational wave, an analog of the \emph{aberration} effect experienced by electromagnetic waves. We will neglect this effect here.
Putting all these ingredients together, the gravitational wave seen by the observer is
\begin{equation}\label{eq:boosted_gw}
    h^o_{ij}(\tau_o) \simeq \mathcal{D}_s^{-1} h^s_{ij} (\tau_s(\tau_o)) \, .
\end{equation}
This relation applies also to the case of our interest (a constant acceleration), as long as the the boost evolves on a time-scale that is much longer than the period of the gravitational wave, so that it can be treated as an instantaneous boost. 

Expanding the parallel component of the velocity in $\tau_s$ around $0$ and keeping the leading, observable order, we have
\begin{equation}
    \tau_o \simeq \gamma(\tau_s + a_\parallel \tau_s^2/2) \simeq \tau_s + a_\parallel \tau_s^2/2 \, , 
\end{equation}
and 
\begin{equation}
    \mathcal{D}_s \simeq \gamma (1+ a_\parallel \tau_s) \simeq 1+ a_\parallel \tau_s \, .
\end{equation}
We have also taken $\gamma\simeq1$, as corrections will scale quadratically in the acceleration and $\tau_s$. At leading order, we can also invert
\begin{equation}\label{eq:ts_to}
    \tau_s(\tau_o) \simeq \tau_o - a_\parallel \tau_o^2/2 \, , 
\end{equation}
and find
\begin{equation}\label{eq:D_to}
    \mathcal{D}_s(\tau_o) \simeq 1+ a_\parallel \tau_o \, .
\end{equation}

Equations \eqref{eq:boosted_gw}, \eqref{eq:ts_to} and \eqref{eq:D_to} are all you need when the reference waveform model is defined in the time domain. You can find out more on the tutorial notebook at this \href{https://github.com/stephengreen/DirtyEMRI-tutorial}{link}.
However, they do not give us a lot of intuition on this effect, and they cannot be readily implemented in waveform models native to the frequency domain. In the following, we will use the stationary phase approximation to see how peculiar acceleration affects the waveform in the frequency domain. For this, we will focus on the transformation of the argument of the waveform, and neglect the Doppler factor in the amplitude. This is a good approximation, as amplitude effects are typically subdominant in gravitational wave observations.

We start by writing the frequency-domain waveform,
\begin{align}
    h_o(f) = &\int dt_o \,  h_s(t_o - a_\parallel \tau_o^2/2) e^{2\pi i f t_o} \\
    \simeq &  \int dt  A_s(t) e^{i \varphi_s(t)} e^{2\pi i f (t+a_\parallel t^2/2)} \, .
\end{align}
In the second line we have defined the new integration variable $t=t_o - a_\parallel \tau_o^2/2$, introduced the waveform amplitude $A$ and phase $\varphi$ and neglected an amplitude factor containing $a_\parallel$. At leading order in $a_\parallel$, we can approximate the Fourier transform by applying the stationary phase approximation as we would in vacuum, identifying the stationary point $t(f)$ where the derivative of the phase vanishes as the solution of
\begin{equation}
    2 \pi f = \frac{d \varphi_s}{dt}(t)\, .
\end{equation}
At leading order in the post-Newtonian expansion, the stationary point is given by
\begin{equation}
    t_f = \frac{5}{256} \mathcal{M} (\pi \mathcal{M} f)^{-8/3} \, ,
\end{equation}
where $\mathcal{M}$ is the inner binary's chirp mass.
Focusing on the phase, we find that in the frequency domain
\begin{equation}
    \tilde{\varphi_o}(f) = \tilde{\varphi_s}(f) + 2\pi f a_\parallel t_f^2/2 \, ,
\end{equation}
so that the correction to the phase due to line-of-sight acceleration, at leading post-Newtonian order and leading order in $a_\parallel$, is
\begin{equation}
    \delta \varphi(f) = + 2\pi f a_\parallel t_f^2/2 = \frac{25}{65536}a_\parallel \mathcal{M} (\pi \mathcal{M} f)^{-13/3}\, .
\end{equation}
This tells us that acceleration comes in at negative post-Netwonian order.

{\bf Question}: What is the leading post-Newtonian order of line-of-sight acceleration? Compare it with the first example we considered, the interaction with an accretion disk in the Shakura-Sunyaev model. Do these two effects come in a the same order?

{\bf Further reading}: Boosted gravitational waves \cite{Cusin:2024git}. First constraints on line-of-sight acceleration in gravitational wave sources \cite{Vijaykumar:2023tjg}

\vspace{0.5cm}

\chapter{Completing Einstein:\\ Quantum Black Holes}
\chapterauthor{Gustavo J. Turiaci}
\label{ch:part4}
\allowdisplaybreaks
\minitoc
\def\d{\mathrm{d}}
\def\i{\mathrm{i}}
\usetikzlibrary{
  calc,
  decorations.markings,
  decorations.pathmorphing,
  decorations.pathreplacing,
  patterns.meta,
  shapes.misc
}
\colorlet{collapsingmatter}{blue!35}

\section{Introduction to quantum gravity and black holes}

The purpose of this section is to identify the regimes when quantum field theory is applicable, when
quantum field theory has to be considered in curved spacetime, and when full quantum gravity is necessary. This will be useful later when we discuss
exceptions to these generic expectations.

\subsection{Quantum fields, gravity, and the Planck scale}

The Planck scale plays such a prominent role that it will be useful to define it from the onset. We will see that it is the scale at which gravitational and quantum effects become
simultaneously important. In four spacetime dimensions, the Planck mass,
length, and time are those constructed out of Newton's constant $G_N$ together with Planck's constant $\hbar$ and the speed of light $c$, namely
\begin{equation}
    m_{\rm P}=\sqrt{\frac{\hbar c}{G_N}}
    \sim 10^{-5}\,\mathrm{g},
    \qquad
    \ell_{\rm P}=\sqrt{\frac{\hbar G_N}{c^3}}
    \sim 10^{-35}\,\mathrm{m},
    \qquad
    t_{\rm P}=\frac{\ell_{\rm P}}{c}
    \sim 10^{-44}\,\mathrm{s}.
\end{equation}
To simplify the discussion we will work in units where $c=k_{\rm B}=1$, while
keeping $G_N$ and $\hbar$ explicit.

\smallskip

Suppose we have an object with energy $E$.  When is classical gravity
important?  Gravity should be weak far from the object and strong close to it.
We can compare the magnitude of the
gravitational energy of a probe mass $m$ with its rest energy.  The former is
$E_{\rm grav}\sim G_N E m/L$, while the latter is $E_{\rm rest}=m$.
Gravity becomes strong when their ratio is of order one,
\begin{equation}
\frac{E_{\rm grav}}{E_{\rm rest}}\sim \frac{G_N E}{L} \gtrsim 1,
~~~\Rightarrow~~~
L\lesssim L_{\rm grav}=2G_N E .
\end{equation}
The scaling argument fixes this length only up to a numerical factor; we have
arbitrarily chosen it so that $L_{\rm grav}$ is the Schwarzschild radius.
For the Sun it is $3\,\mathrm{km}$, while for the Earth it is about
$9\,\mathrm{mm}$.  These lengths are much smaller than the actual sizes
of those objects, so their gravitational fields are weak at their surfaces.  If
an object is compact enough that one can probe distances $r\sim L_{\rm grav}$,
then gravity becomes strong. This is the regime near a black hole horizon, where general relativity is clearly important.

\smallskip

Independently, when are quantum effects are relevant?  For a system of
energy $E$, the associated Compton wavelength is of the order $L_{\rm quant}=\hbar/E$.
Quantum mechanics becomes relevant at length scales smaller than the Compton
length.  If the size of the system is kept fixed, quantum effects become strong
at low energies. The average energy of a single quantum is of order $\hbar/L$,
so the classical limit $L\gg L_{\rm quant}$ is equivalent to a large-occupation-number limit.  We can
now combine the two scales:

\begin{itemize}

\item Consider first a regime in which classical gravity is strong while quantum
effects remain weak. To reach this regime, we must be able to approach
$L_{\rm grav}$ while staying far from $L_{\rm quant}$. This requires
$L_{\rm grav} \gg L_{\rm quant}$, or, equivalently, energies much larger than the
Planck mass,
$E \gg \sqrt{\frac{\hbar}{G_N}} = m_{\rm P}$.
Since $m_{\rm P} \sim 10^{-5}\,\mathrm{g}$, astrophysical black holes lie comfortably
within this regime. More generally, if we are given a theory of quantum gravity such as string theory, strong classical-gravity effects can be
isolated by keeping $G_N$ fixed while taking the limit $\hbar \to 0$.

\smallskip

\item Another regime arises when gravity is weak but quantum effects are
important. This corresponds to the opposite limit,
$L_{\rm grav} \ll L_{\rm quant}$, or equivalently $E \ll m_{\rm P}$.
In this case gravity remains weak even well within the scales where quantum
effects become significant. This is the regime described by quantum field
theory, including quantum field theory on a prescribed curved background.
For example, an electron at rest has
a Compton wavelength of
$\sim 10^{-13}\,{\rm m}$, enormously
larger than its gravitational radius 
$\sim 10^{-58}\,{\rm m}$.
Given a theory of quantum gravity, this regime is
obtained by keeping $\hbar$ fixed while taking $G_N\to0$. It describes most
of the physics we see around us.

\smallskip

\item Finally, both quantum and gravitational effects are
important when the two scales are comparable,
$L_{\rm grav} \sim L_{\rm quant}$, or $E \sim m_{\rm P}$. Of course energies above the Planck scale are common in everyday life but the length scales necessary to observe quantum gravity effects would be ridiculously tiny.
In this case both gravity and quantum mechanics are important, and we need a
theory of quantum gravity. This is the regime in which neither $G_N$ nor
$\hbar$ can be neglected. An example is provided by the early universe.

\end{itemize}

These elementary considerations provide the most naive approach to understanding the interplay between gravity and quantum mechanics.  Later we will see examples in black hole
physics where quantum gravity effects appear in unexpected regimes.

\subsection{Black hole solutions}

Next we review some basic properties of black hole solutions that will be necessary later on. 

\subsubsection*{Schwarzschild black hole}

The simplest black hole solution of Einstein gravity is the Schwarzschild
solution.  It describes the exterior geometry of a static, spherically
symmetric object of mass $M$.  In four
spacetime dimensions, the metric is
\begin{equation}
    \mathrm{d}s^2
    =
    -f(r)\,\mathrm{d}t^2
    +
   \frac{\mathrm{d}r^2}{f(r)}
    +
    r^2 \mathrm{d}\Omega^2 ,~~~~f(r) = 1-\frac{2G_N M}{r},
\end{equation}
where $\mathrm{d}\Omega^2=\mathrm{d}\theta^2
+\sin^2\theta\,\mathrm{d}\phi^2$ is the line
element on the unit two-sphere.  This is a vacuum solution of
pure Einstein gravity with no cosmological constant, described by an action
\begin{equation}
    S_{\rm EH}
    =
    \frac{1}{16\pi G_N}
    \int \mathrm{d}^4x\,\sqrt{-g}\,R .
\end{equation}
Boundary terms must be added to ensure a well-defined variational problem.  We will return to this issue later.  Varying this action gives the vacuum Einstein equations
\begin{equation}
R_{\mu\nu}-\frac{1}{2}R g_{\mu\nu}=0,
\end{equation}
away from matter sources.  In the
Schwarzschild solution the mass $M$ is
encoded in the asymptotic boundary data rather than in a local stress tensor in
the exterior region.
The Schwarzschild radius is $r_s = 2G_N M$.  At $r=r_s$, the
coefficient of $\mathrm{d}t^2$ vanishes and the coefficient of
$\mathrm{d}r^2$ diverges.  This is not a physical singularity, but rather a
coordinate singularity.  Curvature invariants such as
$
    R_{\mu\nu\rho\sigma}R^{\mu\nu\rho\sigma}
    =
    48 G_N^2 M^2/r^6
$
diverge only at $r=0$, suggesting that this is the only true singularity in the spacetime.

\smallskip 

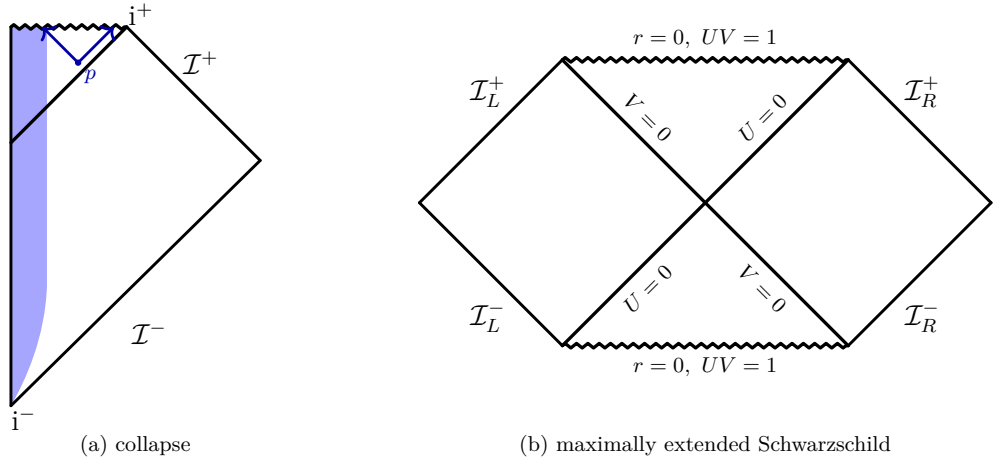
\begin{figure}[t!]
    \centering
    \begin{tikzpicture}[
        x=0.42cm,
        y=0.42cm,
        scale=0.9,
        transform shape,
        line cap=round,
        line join=round,
        every node/.style={font=\small}
    ]
    \hspace{-1cm}
        \begin{scope}
            \coordinate (iminus) at (0,0);
            \coordinate (lefttop) at (0,13.25);
            \coordinate (iplus) at (4.05,13.25);
            \coordinate (right) at (8.72,8.58);
            \coordinate (horizonstart) at (0,9.20);
            \coordinate (insideevent) at (2.35,12.00);

            \path[fill=collapsingmatter]
                (iminus) -- (lefttop) -- (1.26,13.25) -- (1.26,4.20)
                .. controls (1.20,2.40) and (0.52,0.90) .. (iminus)
                -- cycle;

            \draw[black, line width=1.1pt] (iminus) -- (lefttop);
            \draw[line width=1.25pt,decoration={zigzag,segment length=1.5mm, amplitude=0.3mm},decorate]
                (lefttop) -- (iplus);
            \draw[black, line width=1.1pt] (iminus) -- (right);
            \draw[black, line width=1.1pt] (iplus) -- (right);
            \draw[black, line width=1.25pt] (horizonstart) -- (iplus);

            \draw[blue!65!black, line width=1.05pt, ->]
                (insideevent) -- (1.10,13.25);
            \draw[blue!65!black, line width=1.05pt, ->]
                (insideevent) -- (3.60,13.25);
            \fill[blue!65!black] (insideevent) circle[radius=0.11];
            \node[blue!65!black, below right=-1pt] at (insideevent) {$p$};

            \node[font=\large] at (4.55,13.70) {$\mathrm{i}^+$};
            \node[font=\large] at (0.45,-0.45) {$\mathrm{i}^-$};
            \node[font=\large] at (6.65,12) {$\mathcal I^+$};
            \node[font=\large] at (4.85,2.55) {$\mathcal I^-$};
            \node[font=\small] at (4.35,-1.40) {(a) collapse};
        \end{scope}
\hspace{2cm}
        \begin{scope}[shift={(14.0,0)}]
            \coordinate (left) at (-5,7.10);
            \coordinate (right) at (15,7.10);
            \coordinate (topl) at (0,12.10);
            \coordinate (topr) at (10,12.10);
            \coordinate (botl) at (0,2.10);
            \coordinate (botr) at (10,2.10);
            \coordinate (bif) at (5,7.10);

            \draw[black, line width=1.1pt] (left) -- (topl);
            \draw[black, line width=1.1pt] (left) -- (botl);
            \draw[black, line width=1.1pt] (right) -- (topr);
            \draw[black, line width=1.1pt] (right) -- (botr);
            \draw[line width=1.25pt,decoration={zigzag,segment length=1.5mm, amplitude=0.3mm},decorate]
                (topl) -- (topr);
            \draw[line width=1.25pt,decoration={zigzag,segment length=1.5mm, amplitude=0.3mm},decorate]
                (botl) -- (botr);

            \draw[black, line width=1.25pt] (bif) -- (topl);
            \draw[black, line width=1.25pt] (bif) -- (topr);
            \draw[black, line width=1.25pt] (bif) -- (botl);
            \draw[black, line width=1.25pt] (bif) -- (botr);
            \node[font=\small, rotate=45, above] at (7.40,9.70) {$U=0$};
            \node[font=\small, rotate=-45, above] at (2.60,9.70) {$V=0$};
            \node[font=\small, rotate=45, below] at (2.60,4.50) {$U=0$};
            \node[font=\small, rotate=-45, below] at (7.40,4.50) {$V=0$};

            \node[font=\large] at (-2.60,11) {$\mathcal I^+_L$};
            \node[font=\large] at (-2.60,3.10) {$\mathcal I^-_L$};
            \node[font=\large] at (12.60,11) {$\mathcal I^+_R$};
            \node[font=\large] at (12.60,3.10) {$\mathcal I^-_R$};
            \node[font=\small] at (5,12.85) {$r=0,\ UV=1$};
            \node[font=\small] at (5,1.35) {$r=0,\ UV=1$};
            \node[font=\small] at (5,-1.40)
                {(b) maximally extended Schwarzschild};
        \end{scope}
    \end{tikzpicture}
    \caption{(a) Penrose diagrams for a black hole formed by collapse and (b) for the
    maximally extended Schwarzschild solution.  The collapse geometry has one
    asymptotic exterior, while the eternal solution contains two asymptotic
    exterior regions joined through the bifurcation surface.  In panel (a),
    the blue lines show the future light cone of an event $p$ inside the
    horizon.}
    \label{fig:collapse-vs-eternal}
\end{figure}

Expressing the geometry in Kruskal coordinates makes evident that the horizon is smooth.
First, it is useful to introduce the tortoise coordinate and null coordinates
\begin{equation}
    r_\ast
    = \int \frac{\mathrm{d}r}{f(r)}
    =
    r+r_s\log\left(\frac{r}{r_s}-1\right),
    \qquad
    u=t-r_\ast,\qquad v=t+r_\ast .
\end{equation}
The temporal/radial part of the metric becomes proportional to $-f(r)\,\mathrm{d}u\,\mathrm{d}v$, which is still singular. Kruskal coordinates exponentiate these null
coordinates and keep the metric smooth along the horizon. Take
$U=-e^{-u/(2r_s)}$ and $V=e^{v/(2r_s)}$.  They obey
$U V=-(\frac{r}{r_s}-1)e^{r/r_s}$, which can be used to solve for $r$ in terms of $U$ and $V$. The horizons are located at $U=0$ and $V=0$, while the singularity is at $UV=1$. The metric becomes
\begin{equation}
    \mathrm{d}s^2
    =
    -\frac{4r_s^3}{r}e^{-r/r_s}\,\mathrm{d}U\,\mathrm{d}V
    +
    r^2\mathrm{d}\Omega^2 .
\end{equation}
Now $g_{UV}$ is finite and
nonzero at $r=r_s$.  Thus the future and past horizons are regular null surfaces in these coordinates. 

\smallskip

The surface $r=r_s$ is an event horizon, as illustrated in figure~\ref{fig:collapse-vs-eternal}.  Once an observer crosses this
surface at an event such as $p$ in figure~\ref{fig:collapse-vs-eternal}, future-directed timelike and null trajectories necessarily move toward
smaller $r$ and eventually reach the singularity at $r=0$.  Thus the horizon separates the region from which signals can escape to infinity
from the region from which they cannot.

\smallskip 

Classically, the Schwarzschild black hole is completely characterized by its
mass. Below we identify some other geometric properties of the spacetime that will play an important role in whats next:

\begin{itemize}

\item The horizon has the geometry of a two-sphere with $r=r_s$, so its area is
\begin{equation}
    A
    =
    \int_{S^2}\sqrt{g_{\theta\theta}g_{\phi\phi}}\,
    \mathrm{d}\theta\,\mathrm{d}\phi
    =
    4\pi r_s^2
    =
    16\pi G_N^2 M^2 .
\end{equation}
This quantity plays a central role in black hole thermodynamics.

\item The horizon is a surface of infinite redshift, which can be expressed in a
coordinate-independent way. For a stationary black hole, Hawking's rigidity
theorem \cite{Hawking:1972BlackHoles} implies that the horizon is generated
by a Killing vector field $\chi^\mu$ that becomes null on the horizon. A
particle with four-momentum $p^\mu$ has a conserved Killing energy
$
E=-p_\mu\chi^\mu.
$
Outside the horizon, where $\chi^\mu$ is timelike, an observer moving along
the orbits of $\chi^\mu$ has four-velocity
$
u_{\rm obs}^\mu=\chi^\mu/\sqrt{-\chi^2}
$
and measures the particle's local energy to be
$
E_{\rm loc}=-p_\mu u_{\rm obs}^\mu
=E/\sqrt{-\chi^2}.
$
As the emission point approaches the horizon, the denominator vanishes $\chi^2\to0$. Consequently, a
family of particles emitted with the same finite local energy has
$E=\sqrt{-\chi^2}\,E_{\rm loc}\to0$. This is the sense in which the
horizon is a surface of infinite redshift. For a Schwarzschild black hole,
$\chi=\partial_t$ and $\chi^2=-f(r)$, while $E$ is the energy measured
at infinity. The general relation therefore reduces to
$E_\infty=\sqrt{f(r)}\,E_{\rm loc}\to0$ as $r\to r_s$.

\item Another important geometric notion is that of the horizon surface gravity.  Let
$\chi^\mu$ be the Killing vector field that becomes null at the horizon, normalized so that
$\chi^2\to -1$ at infinity. The
surface gravity $\kappa$ is defined through\footnote{For any Killing horizon generated by the normal
$\chi^\mu$, the quantity $\chi_\mu\chi^\mu$ vanishes on the horizon.  Its gradient is therefore normal to the horizon and proportional to $\chi_\mu$.  Using the Killing equation, one finds
$\nabla_\mu(\chi_\nu\chi^\nu)
=2\chi^\nu\nabla_\mu\chi_\nu
=-2\chi^\nu\nabla_\nu\chi_\mu$.}
\begin{equation}
    \chi^\nu\nabla_\nu\chi^\mu
    =
    \kappa\,\chi^\mu ,
    \qquad \Rightarrow \qquad
    \kappa
    =
    \frac{f'(r_s)}{2}
    =
    \frac{1}{2r_s}
    =
    \frac{1}{4G_N M}.
\end{equation}
On the right we indicated the value it takes for a Schwarszchild black hole. The surface gravity has a simple interpretation. Consider a non-inertial observer with four-velocity along the Killing vector field,
$u^\mu=\chi^\mu/\sqrt{-\chi^2}$.  The proper acceleration is $a^\mu=u^\nu\nabla_\nu u^\mu$, and it can be rewritten as
$$
   a^\mu
    =
   \frac{\chi^\nu \nabla_\nu \chi^\mu}{-\chi^\rho \chi_\rho} ,\qquad \Rightarrow\qquad \kappa =  \sqrt{-\chi^2}\, a\Big|_{\text{horizon}}.
$$
The quantity on the right-hand side is also equal to the force needed to hold a unit mass at the horizon as measured at infinity; see \cite{wald1984general}.  The surface gravity is a measure of the gravitational field strength at the horizon.

\end{itemize}

\begin{figure}[t!]
    \centering
    \begin{tikzpicture}[
        scale=0.8,x=0.52cm,
        y=0.52cm,
        line cap=round,
        line join=round,
        every node/.style={font=\small}
    ]
        \coordinate (iminus) at (0,0);
        \coordinate (lefttop) at (0,13.25);
        \coordinate (iplus) at (4.05,13.25);
        \coordinate (right) at (8.72,8.58);
        \coordinate (horizonstart) at (0,9.20);
        \coordinate (observerevent) at (4.5673,12.4855);
        \coordinate (raymatter) at (1.26,9.1782);

        \draw[purple!75!black, line width=1.15pt,
            postaction={decorate},
            decoration={markings,mark=at position 0.5 with {\arrow{>}}}]
            (raymatter) -- (observerevent);

        \path[fill=collapsingmatter]
            (iminus) -- (lefttop) -- (1.26,13.25) -- (1.26,4.20)
            .. controls (1.20,2.40) and (0.52,0.90) .. (iminus)
            -- cycle;

        \draw[blue!70!black, line width=1.25pt]
            (iminus)
            .. controls (2.80,3.50) and (7.00,9.80) .. (iplus);

        \draw[black, line width=1.1pt] (iminus) -- (lefttop);
        \draw[line width=1.25pt,decoration={zigzag,segment length=1.5mm, amplitude=0.3mm},decorate]
            (lefttop) -- (iplus);
        \draw[black, line width=1.1pt] (iminus) -- (right);
        \draw[black, line width=1.1pt] (iplus) -- (right);

        \draw[black, line width=1.25pt] (horizonstart) -- (iplus);

        \node[blue!70!black, anchor=west] at (5.30,9.55) {obs};
        \node[font=\large] at (6.65,12) {$\mathcal I^+$};
    \end{tikzpicture}
    \caption{A distant observer follows the blue timelike worldline.  The
    outgoing light ray leaves the surface of the collapsing matter and is
    received by the observer.}
    \label{fig:outgoing-ray-collapse}
\end{figure}
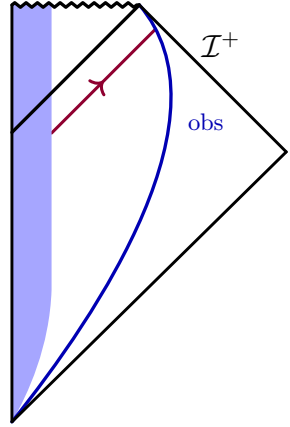

What does a distant observer see during gravitational collapse? Once the
collapsing matter crosses the event horizon, no signal emitted by it can
reach the exterior. Nevertheless, in Schwarzschild coordinates the matter
appears to approach the horizon only asymptotically. This can be understood
using the relationship
$
U=-e^{-\kappa u},
$
where $\kappa=1/2r_s$ is the surface gravity of the Schwarzschild black hole. Let
the trajectory of a radially infalling particle be
$\bigl(U(\tau),V(\tau)\bigr)$, where $\tau$ is its proper time, and suppose
that it crosses the future horizon at the finite proper time $\tau_0$, so
that $U(\tau_0)=0$. See figure~\ref{fig:outgoing-ray-collapse}. Because the trajectory is smooth and timelike in the
regular Kruskal coordinates,
$
U(\tau)\approx-A(\tau_0-\tau),
$
with $A$ finite and non-zero. An outgoing radial light ray follows a line of constant $U$. Therefore, a
ray emitted by the particle at proper time $\tau$ reaches
$\mathcal{I}^+$ at the retarded time
$$
u(\tau)
=-\frac{1}{\kappa}\log\!\bigl(-U(\tau)\bigr)
=-\frac{1}{\kappa}\log\!\bigl(\tau_0-\tau\bigr)
+\cdots .
$$
Thus as the particle approaches the horizon $\tau\to\tau_0$ and $u$ diverges. For an observer at a fixed large
radius $r_{\rm obs}$ the same logarithmic divergence occurs in
Schwarzschild time. Equivalently,
$\tau_0-\tau\propto e^{-\kappa u}$, and light emitted with a finite proper
frequency is observed with a frequency that decays as
$
\omega_{\rm obs}\propto e^{-\kappa u}.
$
The observer never sees the matter cross the horizon. Instead, its image rapidly becomes redshifted and faint. Notice that the timescale is very short, for a solar-mass black hole $1/\kappa \sim 10^{-5}\,{\rm s}$. This exponential
relation between the near-horizon and asymptotic coordinates also underlies
the thermal Hawking radiation discussed in the next section.

\subsubsection*{Reissner-Nordstrom black hole}

In the last lecture the charged generalization will play a central role. This is the Reissner-Nordstrom spacetime.  It is a
static, spherically symmetric solution of the Einstein-Maxwell equations. The metric takes the form
\begin{equation}
    \mathrm{d}s^2
    =
    -f(r)\mathrm{d}t^2
    +
    \frac{\mathrm{d}r^2}{f(r)}
    +
    r^2 \mathrm{d}\Omega^2 ,
    \qquad
    A
    =
    -\frac{Q}{4\pi r}\,\mathrm{d}t + \mathrm{d} \alpha ,
\end{equation}
with
\begin{equation}
    f(r)
    =
    1-\frac{2G_N M}{r}
    +
    \frac{G_N Q^2}{4\pi r^2} ,
\end{equation}
where $Q$ is the electric charge, in conventions where the Maxwell action is
\begin{equation}
    S_{\rm Maxwell}
    =
    -\frac{1}{4}\int \mathrm{d}^4x \sqrt{-g}\,
    F_{\mu\nu}F^{\mu\nu}.
\end{equation}
The
chemical potential
is defined as a potential difference between infinity and the horizon.  With
$\chi=\partial_t$, the electrostatic potential is $\chi_\mu A^\mu$. The chemical potential is then
\beq
\Phi&=\left.\chi\mathbin{\cdot}A\right|_{\infty}
-\left.\chi\mathbin{\cdot}A\right|_{\mathcal H^+}= \frac{Q}{4 \pi r_+} .
\eeq
In the gauge with $\alpha=0$, $A_t\to 0$ at infinity and
$\left.\chi\mathbin{\cdot}A\right|_{\mathcal H^+}=-\Phi$. 
Equivalently, one may choose a gauge regular on the future horizon, where
$\chi\mathbin{\cdot}A=0$ at $r=r_+$; then the same chemical potential appears as the
asymptotic value at infinity.

\smallskip

There is also a magnetic version of the solution.  For a purely magnetic charge
$P$, the metric has the same form with $Q^2$ replaced by $P^2$, while the
field strength is
$F=\frac{P}{4\pi}\sin\theta\,\mathrm{d}\theta\wedge\mathrm{d}\phi$.
The corresponding gauge potential is only locally defined, as usual for a
magnetic monopole.  More generally, for a dyonic black hole with electric charge
$Q$ and magnetic charge $P$, the metric depends on $Q^2+P^2$.

\smallskip

The contribution from the electric charge to the redshift factor $f(r)$ has a simple Newtonian interpretation.  In the weak-field
limit one writes $g_{tt}=-(1+2\Phi)$, so $f(r)=1+2\Phi(r)$, and for small $\Phi(r)$ it has the interpretation of a Newtonian potential which at least at large $r$ behaves as $\Phi(r) = - G_N M(r)/r$ with $M(r)$ being the total energy inside the sphere of radius $r$. The effective Newtonian potential for the Reissner-Nordstrom solution is
$\Phi(r)=-\frac{G_N M}{r}+\frac{G_N Q^2}{8\pi r^2}$.
The positive term can be interpreted as a gravitational effect due to the energy of the electric field. With the normalization above, $E_r=Q/(4\pi r^2)$, since we can reproduce
\begin{equation}
    M(r)=M-
    \frac{1}{2}\int_r^\infty
    \,
    \underbrace{\left(\frac{Q}{4\pi r'^2}\right)^2}_{\text{electric field square}} \, 4\pi r'^2
    \mathrm{d}r'
    =
   M - \frac{Q^2}{8\pi r}.
\end{equation}
This illustrate the fact that the quantity $M$ is not only the mass of the constituents of the collapsing matter, but the total energy supporting the black hole spacetime.

\smallskip

The horizons are the zeros of $f(r)$.  They are located at
\begin{equation}
    r_\pm
    =
    G_N M
    \pm
    \sqrt{G_N^2 M^2 - \frac{G_N Q^2}{4\pi}} .
\end{equation}
The outer horizon $r_+$ is the event horizon, while $r_-$ is an inner
Cauchy horizon.
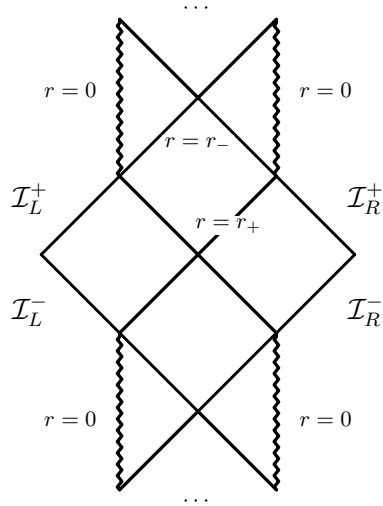
\begin{figure}[t!]
    \centering
    \begin{tikzpicture}[
        x=0.46cm,
        y=0.46cm,
        scale=0.92,
        transform shape,
        line cap=round,
        line join=round,
        every node/.style={font=\small}
    ]
        \coordinate (bif) at (0,0);
        \coordinate (tl) at (-2.45,2.45);
        \coordinate (tr) at (2.45,2.45);
        \coordinate (bl) at (-2.45,-2.45);
        \coordinate (br) at (2.45,-2.45);
        \coordinate (future) at (0,4.90);
        \coordinate (past) at (0,-4.90);
        \coordinate (left) at (-4.90,0);
        \coordinate (right) at (4.90,0);
        \coordinate (topl) at (-2.45,7.35);
        \coordinate (topr) at (2.45,7.35);
        \coordinate (botl) at (-2.45,-7.35);
        \coordinate (botr) at (2.45,-7.35);

        \draw[black, line width=1.1pt] (left) -- (tl);
        \draw[black, line width=1.1pt] (left) -- (bl);
        \draw[black, line width=1.1pt] (right) -- (tr);
        \draw[black, line width=1.1pt] (right) -- (br);

        \draw[black, line width=1.25pt] (bif) -- (tl);
        \draw[black, line width=1.25pt] (bif) -- (tr);
        \draw[black, line width=1.25pt] (bif) -- (bl);
        \draw[black, line width=1.25pt] (bif) -- (br);

        \draw[black, line width=1.1pt] (tl) -- (future);
        \draw[black, line width=1.1pt] (tr) -- (future);
        \draw[black, line width=1.1pt] (bl) -- (past);
        \draw[black, line width=1.1pt] (br) -- (past);

        \draw[black, line width=1.25pt] (future) -- (topl);
        \draw[black, line width=1.25pt] (future) -- (topr);
        \draw[black, line width=1.25pt] (past) -- (botl);
        \draw[black, line width=1.25pt] (past) -- (botr);
        \node at (0,7.70) {$\cdots$};
        \node at (0,-7.70) {$\cdots$};

        \draw[line width=1.25pt,decoration={zigzag,segment length=1.5mm, amplitude=0.3mm},decorate]
            (tl) -- (topl);
        \draw[line width=1.25pt,decoration={zigzag,segment length=1.5mm, amplitude=0.3mm},decorate]
            (tr) -- (topr);
        \draw[line width=1.25pt,decoration={zigzag,segment length=1.5mm, amplitude=0.3mm},decorate]
            (botl) -- (bl);
        \draw[line width=1.25pt,decoration={zigzag,segment length=1.5mm, amplitude=0.3mm},decorate]
            (botr) -- (br);

        \node[font=\large] at (-5.25,1.75) {$\mathcal I^+_L$};
        \node[font=\large] at (-5.25,-1.75) {$\mathcal I^-_L$};
        \node[font=\large] at (5.25,1.75) {$\mathcal I^+_R$};
        \node[font=\large] at (5.25,-1.75) {$\mathcal I^-_R$};
        \node[font=\small, fill=white, inner sep=1pt] at (0.95,0.90)
            {$r=r_+$};
        \node[font=\small, fill=white, inner sep=1pt] at (0,3.45)
            {$r=r_-$};
        \node[font=\small, anchor=east] at (-2.90,5.15) {$r=0$};
        \node[font=\small, anchor=west] at (2.90,5.15) {$r=0$};
        \node[font=\small, anchor=east] at (-2.90,-5.15) {$r=0$};
        \node[font=\small, anchor=west] at (2.90,-5.15) {$r=0$};
    \end{tikzpicture}
    \caption{Schematic Penrose diagram for the nonextremal
    Reissner-Nordstrom black hole.  The outer horizons $r=r_+$ are event
    horizons, while the inner horizons $r=r_-$ are Cauchy horizons.  The
    wavy vertical lines are the timelike singularities at $r=0$; only one
    repeating portion of the maximal extension is shown.}
    \label{fig:rn-penrose}
\end{figure}
The black hole exists provided $G_N^2 M^2 \geq \frac{G_N Q^2}{4\pi}$.
Equivalently, the mass must be large enough compared to the charge.  When the
inequality is saturated, the two horizons coincide, $r_+ = r_-$.
This is the extremal Reissner-Nordstrom black hole.

It is useful to rewrite
$f(r)=\left(1-\frac{r_+}{r}\right)\left(1-\frac{r_-}{r}\right)$.
The outer horizon is a Killing horizon generated by
$\chi=\partial_t$.  Its surface gravity is
\begin{equation}
    \kappa
    =
    \frac{1}{2} f'(r_+)
    =
    \frac{r_+-r_-}{2 r_+^2}.
\end{equation}
Thus the extremal limit is a degenerate-horizon limit:
$r_+ \to r_-$. This implies that the surface gravity vanishes $\kappa \to 0$ which can be taken as a definition of extremality.

\smallskip 

The area of the outer horizon is $A_+=4\pi r_+^2$ which is a function of $M$ and $Q$.
This remains finite in the extremal limit:
\begin{equation}
    A_{\rm ext}
    =
    4\pi r_{\rm ext}^2
    =
    G_N Q^2 .
\end{equation}
Equivalently, $r_{\rm ext}^2=G_NQ^2/(4\pi)$.  Restoring constants and
using the rationalized convention $\alpha=e^2/(4\pi\hbar c)$, this gives
$$
    \frac{Q_{\rm ext}}{e}
    =
    \frac{r_{\rm ext}}{\sqrt{\alpha}\,\ell_{\rm P}}
    \simeq
    2.2\times 10^{39}
    \left(\frac{r_{\rm ext}}{3\,\mathrm{km}}\right).
$$
We see that achieving a macroscopic extremal black hole requires a very large charge. The fact that the extremal area is finite will play a central role later.

\subsubsection*{Near-extremal geometry}

Near extremality, the region close to the horizon develops a universal geometry
\cite{Kunduri:2007vf}. Instead of reproducing the general proof, we will illustrate it for the Reissner-Nordstrom black hole. In
the exactly extremal limit, call $r_+ = r_- = r_{\rm ext}$, and
$f(r)=\left(1-\frac{r_{\rm ext}}{r}\right)^2$.
To zoom in on the horizon, set $r=r_{\rm ext}+\rho$ with $\rho\ll r_{\rm ext}$.  Then
$$
    f(r)
    =
    \frac{\rho^2}{(r_{\rm ext}+\rho)^2}
    \simeq
    \frac{\rho^2}{r_{\rm ext}^2},
    \qquad
    r^2\mathrm{d}\Omega^2\simeq r_{\rm ext}^2\mathrm{d}\Omega^2 .
$$
The near-horizon metric is therefore
\begin{equation}
    \mathrm{d}s^2
    \simeq
    -\frac{\rho^2}{r_{\rm ext}^2}\mathrm{d}t^2
    +
    \frac{r_{\rm ext}^2}{\rho^2}\mathrm{d}\rho^2
    +
    r_{\rm ext}^2\mathrm{d}\Omega^2 .
\end{equation}
With $z=r_{\rm ext}^2/\rho$, this becomes
\begin{equation}
    \mathrm{d}s^2
    =
    r_{\rm ext}^2\underbrace{\left(\frac{-\mathrm{d}t^2+\mathrm{d}z^2}{z^2}\right)}_{\text{AdS$_2$ with unit radius}}
    +
    r_{\rm ext}^2\mathrm{d}\Omega^2 ,
\end{equation}
which is the product of two dimensional Anti-de Sitter (AdS) geometry and a two-sphere $\mathrm{AdS}_2\times S^2$, with both factors having radius $r_{\rm ext}$.
For a near-extremal black hole the length of the AdS throat can be measured by the proper
distance from the horizon to a fixed reference radius.  Let $r=r_+ + \rho$,
$\Delta = r_+-r_-$, and $r_{\rm ref}=r_+ + \rho_{\rm ref}$, with
$\Delta\ll \rho_{\rm ref}\ll r_+$.  Denoting the common extremal radius by
$r_{\rm ext}\simeq r_+\simeq r_-$, one finds in this region
$$
    f(r)
    =
    \frac{(r-r_+)(r-r_-)}{r^2}
    \simeq
    \frac{\rho(\rho+\Delta)}{r_{\rm ext}^2},
$$
so the radial proper length is
\begin{equation}
    \ell_{\rm throat}
    =
    \int_{r_+}^{r_{\rm ref}}\frac{\mathrm{d}r}{\sqrt{f(r)}}
    \simeq
    r_{\rm ext}\int_0^{\rho_{\rm ref}}
    \frac{\mathrm{d}\rho}{\sqrt{\rho(\rho+\Delta)}}
    =
    2r_{\rm ext}\,\operatorname{arcsinh}\sqrt{\frac{\rho_{\rm ref}}{\Delta}} .
\end{equation}
Since $\kappa=(r_+-r_-)/(2r_+^2)\simeq \Delta/(2r_{\rm ext}^2)$, this becomes
\begin{equation}
    \ell_{\rm throat}
    =
    r_{\rm ext}\log\!\left(\frac{4\rho_{\rm ref}}{\Delta}\right)+\cdots
    =
    r_{\rm ext}\log\!\left(\frac{2\rho_{\rm ref}}{\kappa r_{\rm ext}^2}\right)+\cdots .
\end{equation}
Thus, at fixed reference radius, $\ell_{\rm throat}$ diverges logarithmically
as $r_{\rm ext}\log(1/(\kappa r_{\rm ext}))$ when $\kappa\to0$.
This is important because many questions about near-extremal charged black
holes reduce, at low energies, to questions about the dynamics of nearly
$\mathrm{AdS}_2$ gravity.

\begin{figure}[t!]
    \centering
    \begin{tikzpicture}[scale=1.1, baseline={([yshift=0cm]current bounding box.center)}]
        \draw[thick] (0,-2) ellipse (0.5 and 0.2);
        \draw[thick] (-1,1) to [bend left=20] (-0.5,-0.5) -- (-0.5,-2);
        \draw[thick] (1,1) to [bend right=20] (0.5,-0.5) -- (0.5,-2);
        \draw[thick] (-3,0) -- (3,0) -- (3.5,2) -- (-2.5,2) -- (-3,0);
        \draw (2,2.2) node {\small Asymptotically flat};
        \draw (1.2,-2) node {\small Horizon};
        \draw[->,thick] (-1,1.3) -- (1,1.3);
        \draw[->,thick] (1,1.3) -- (-1,1.3);
        \draw (0,1.6) node {\small $Q$};
        \draw (1.3,-1) node {\small AdS${}_2\times$S${}^2$};
        \fill[blue!50,nearly transparent] (-.74,.6) to [bend left=15] (-0.5,-0.5) -- (-0.5,-2) to [bend right=50] (.5,-2) -- (.5,-.5) to [bend left=15] (.74,.6) to [bend right=22] (-.74,.6);
        \draw[thick,blue] (0,.6) ellipse (0.72 and 0.15);
        \draw (0,-3) node {(a)};
    \end{tikzpicture}
    \hspace{2cm}
    \begin{tikzpicture}[scale=2,baseline={([yshift=0cm]current bounding box.center)}]
        \draw[thick,red] (.5,-.5) to[bend right=10] (.4,-1.2);
        \draw[thick,red] (-.5,-.5) to[bend left=10] (-.4,-1.2);
        \draw[thick,red] (.5,.5) to[bend left=10] (.4,1.2);
        \draw[thick,red] (-.5,.5) to[bend right=10] (-.4,1.2);
        \draw[thick,blue] (-.5,-.5) to[bend right=15] (-.5,.5);
        \draw[thick,blue] (.5,-.5) to[bend left=15] (.5,.5);
        \fill[blue!50,nearly transparent] (-.4,-1.2) to[bend right=10] (-.5,-.5) to[bend right=15] (-.5,.5) to[bend right=10] (-.4,1.2) -- (.4,1.2) to[bend right=10] (.5,.5) to[bend right=15] (.5,-.5) to[bend right=10] (.4,-1.2) -- (-.5,-1.2);
        \draw[thick] (-1,0) -- (0,1) -- (1,0) -- (0,-1) -- (-1,0);
        \draw[thick] (-.5,.5)--(.5,-.5);
        \draw[thick] (.5,.5)--(-.5,-.5);
        \draw[thick,decoration={zigzag,segment length=1.5mm, amplitude=0.3mm},decorate] (-.5,.5)--(-.5,1.2);
        \draw[thick,decoration={zigzag,segment length=1.5mm, amplitude=0.3mm},decorate] (.5,.5)--(.5,1.2);
        \draw[thick,decoration={zigzag,segment length=1.5mm, amplitude=0.3mm},decorate] (-.5,-.5)--(-.5,-1.2);
        \draw[thick,decoration={zigzag,segment length=1.5mm, amplitude=0.3mm},decorate] (.5,-.5)--(.5,-1.2);
        \draw (0,-1.5) node {(b)};
    \end{tikzpicture}
    \caption{Two schematic views of the near-extremal Reissner-Nordstrom
    geometry. (a) On a fixed time slice, the asymptotically flat exterior opens
    into a long $\mathrm{AdS}_2\times S^2$ throat, whose transverse size is
    set by the charge $Q$, ending at the horizon. (b) The same throat region
    appears as the blue surface in the Penrose diagram.}
    \label{fig:rn-ads2-throat}
\end{figure}
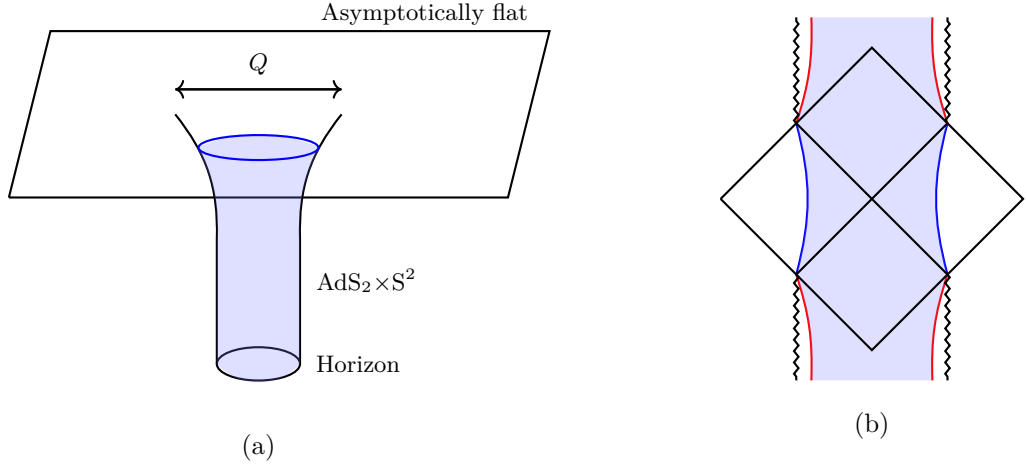

\subsubsection*{Kerr black holes}

The rotating, asymptotically flat black hole in vacuum is the Kerr solution
\cite{Kerr:1963ud}.  We will not need the full metric here.  It is enough to
introduce $m\equiv G_N M$ and $a\equiv J/M$, where $J$ is the
angular momentum.  The horizons are located at the zeros of
the function $\Delta(r)=r^2-2mr+a^2$.
Thus
\begin{equation}
    r_\pm
    =
    G_NM
    \pm
    \sqrt{G_N^2M^2-\frac{J^2}{M^2}} .
\end{equation}
The outer horizon is $r_+$, while $r_-$ is the inner horizon.  The horizon
exists provided $G_NM^2\geq |J|$.  Since Kerr is stationary and axisymmetric,
it has two commuting Killing vectors, the time translation $\partial_t$ and
the rotation $\partial_\phi$.  The horizon is not generated by $\partial_t$
alone.  Its null generators co-rotate with the black hole, so the Killing
horizon is generated by
\begin{equation}
\chi=\partial_t+\Omega\partial_\phi,~~~~~\Omega=\frac{a}{r_+^2+a^2} .
\end{equation}
This is the Killing vector that becomes null $\chi^2=0$
on the horizon $r=r_+$. $\Omega$ can be interpreted as the angular velocity of the black hole. The surface gravity is defined using this $\chi$,
by $\chi^\nu\nabla_\nu\chi^\mu=\kappa\chi^\mu$ on the horizon, giving
\begin{equation}
    \kappa
    =
    \frac{r_+-r_-}{2(r_+^2+a^2)} .
\end{equation}
As for Reissner-Nordstrom, the extremal limit is the limit in which the two
horizons coincide and $\kappa\to0$. The near-horizon geometry of extremal Kerr is also universal, and is known as the near-horizon extremal Kerr (NHEK) geometry \cite{Bardeen:1999px}. It is a warped and fibered geometry with an $\mathrm{AdS}_2$ base, with the warp factors depending on the polar angle $\theta$.

\smallskip

At extremality the total energy and horizon radius is given by
\begin{equation}
    M_{\rm ext}
    =
    \sqrt{\frac{|J|}{G_N}},
    \qquad
    r_{\rm ext}
    =
    G_NM_{\rm ext}
    =
    \sqrt{G_N|J|}.
\end{equation}
The area of a Kerr horizon is $A_+=4\pi(r_+^2+a^2)$, and therefore
\begin{equation}
    A_{\rm ext}
    =
    8\pi r_{\rm ext}^2
    =
    8\pi G_N|J| .
\end{equation}
Restoring constants, the angular momentum needed to make an extremal Kerr black
hole of radius $r_{\rm ext}$ is
\begin{equation}
    \frac{J_{\rm ext}}{\hbar}
    =
    \frac{r_{\rm ext}^2}{\ell_{\rm P}^2}
    \simeq
    3.4\times10^{76}
    \left(\frac{r_{\rm ext}}{3\,\mathrm{km}}\right)^2 .
\end{equation}
For $r_{\rm ext}=3\,\mathrm{km}$, this is
$J_{\rm ext}\simeq3.6\times10^{42}\,\mathrm{kg\,m^2/s}$.  The spin angular
momentum of the Earth is approximately
$J_\oplus\simeq5.8\times10^{33}\,\mathrm{kg\,m^2/s}$, so this Kerr black hole
would carry about $6\times10^8$ times the Earth's spin angular momentum. Among well-measured astrophysical black holes, Cygnus X-1 is often quoted as a
candidate close to extremality.  Its mass is estimated to be
$M\sim 20\,M_\odot$ \cite{Miller-Jones:2021cyg}, and continuum-fitting
analyses \cite{Zhao:2021cyg} give the dimensionless spin
$a_\ast\equiv\frac{J}{G_N M^2}>0.9985$.

\section{Black hole thermodynamics}

In Lecture~1, we described black holes as classical spacetimes characterized
by their mass, electric charge, and angular momentum. We now turn to the
first indications that black holes should also be regarded as quantum
systems with thermodynamic properties. The classical laws of black hole
mechanics \cite{Bardeen:1973gs} already exhibit a striking parallel with the laws of
thermodynamics: the surface gravity plays the role of temperature, while the
horizon area plays the role of entropy. Hawking's analysis of quantum fields in a black hole
background transformed this formal analogy into a physical statement:
black holes emit radiation characterized by a definite temperature
\cite{Hawking:1975vcx}. We will first derive the Hawking temperature and
discuss black hole evaporation and greybody factors. We will then introduce
the Bekenstein-Hawking entropy and formulate the generalized laws of black
hole thermodynamics.

\subsection{The black hole temperature}\label{subsec:hawking-radiation}

We begin with a discussion of the black hole quantity that plays the role of temperature. The answer is suggested by the zeroth law of black-hole mechanics. Combining
Einstein's equations with the dominant energy condition, one can show that the
surface gravity $\kappa$, as defined in the previous section, is constant along
the black hole horizon. Thermodynamic equilibrium is likewise characterized
by a uniform temperature throughout the system. It is therefore natural to
identify the black hole temperature with the surface gravity.

\subsubsection*{The Unruh effect}\label{subsec:unruh-effect}

Before explaining Hawking's derivation that the temperature is determined by the surface gravity, we review a simpler effect that captures almost the same physics. This is the Unruh effect which states that a uniformly accelerated observer describes the
Minkowski vacuum as a thermal state
\cite{Bisognano:1975ih,Unruh:1976db} and see \cite{Witten:2024bht, Harlow:2014yka} for other reviews. 

\smallskip

\begin{figure}[t!]
    \centering
    \begin{minipage}[t]{0.53\textwidth}
        \centering
        \begin{tikzpicture}[
            scale=0.72,
            x=1.25cm,
            y=1.25cm,
            line cap=round,
            line join=round,
            every node/.style={font=\small}
        ]
            \begin{scope}
                \clip (-3.45,-2.85) rectangle (3.45,2.85);

                \path[fill=blue!7] (0,0) -- (3.45,3.45) --
                    (3.45,-3.45) -- cycle;

                \foreach \yy in {-1.80,0,1.80} {
                    \draw[orange!75!black, line width=0.65pt]
                        (0,0) -- (3.45,\yy);
                }

                \foreach \rr in {1.30,2.05} {
                    \draw[blue!70!black, line width=0.9pt]
                        plot[domain=-1.55:1.55,samples=100]
                        ({\rr*cosh(\x)},{\rr*sinh(\x)});
                }
            \end{scope}

            \draw[gray!70, dashed, line width=0.9pt]
                (-2.85,-2.85) -- (2.85,2.85);
            \draw[gray!70, dashed, line width=0.9pt]
                (-2.85,2.85) -- (2.85,-2.85);
            \draw[->, line width=0.9pt] (-3.45,0) -- (3.55,0)
                node[below right] {$x$};
            \draw[->, line width=0.9pt] (0,-2.85) -- (0,2.95)
                node[above left] {$t$};

            \fill (0,0) circle (1.1pt);
        \end{tikzpicture}

        \smallskip
        {\small\textbf{(a)} Lorentzian}
    \end{minipage}%
    \hfill
    \begin{minipage}[t]{0.45\textwidth}
        \centering
        \begin{tikzpicture}[
            x=1.05cm,
            y=1.05cm,
            line cap=round,
            line join=round,
            every node/.style={font=\small}
        ]
            \path[fill=orange!9] (-2.75,0) rectangle (2.75,-2.55);
            \foreach \angle in {225,270,315} {
                \draw[gray!35, line width=0.55pt]
                    (0,0) -- ({2.35*cos(\angle)},{2.35*sin(\angle)});
            }

            \draw[->, line width=0.9pt] (-2.75,0) -- (2.85,0)
                node[below right] {$x$};
            \draw[->, line width=0.9pt] (0,-2.55) -- (0,2.45)
                node[above left] {$t_E$};

            \draw[orange!75!black, ->, line width=1.0pt]
                plot[domain=180:360,samples=100]
                ({1.75*cos(\x)},{1.75*sin(\x)});
            \node[fill=orange!9, inner sep=1.5pt] at (0,-1.72)
                {$e^{-\pi K_R}$};

            \node[above] at (-1.85,0.08) {$|n^\ast\rangle_L$};
            \node[above] at (1.85,0.08) {$|n\rangle_R$};
            \fill (0,0) circle (1.1pt);
        \end{tikzpicture}

        \smallskip
        {\small\textbf{(b)} Euclidean}
    \end{minipage}
    \caption{Lorentzian and Euclidean descriptions of the Rindler
    construction.  (a) The dashed lines $x=\pm t$ are the Rindler horizons;
    the shaded right wedge is covered by constant-$\rho$ hyperbolae and
    constant-$\eta$ rays.  (b) The lower-half-plane Euclidean path integral
    evolves from the left half-line to the right through an angle $\pi$.
    Acting on a boost eigenstate, $e^{-\pi K_R}$ produces the weight
    $e^{-\pi\omega_n}$ in equation~\eqref{eq:rindler-vacuum}.}
    \label{fig:minkowski-rindler}
\end{figure}
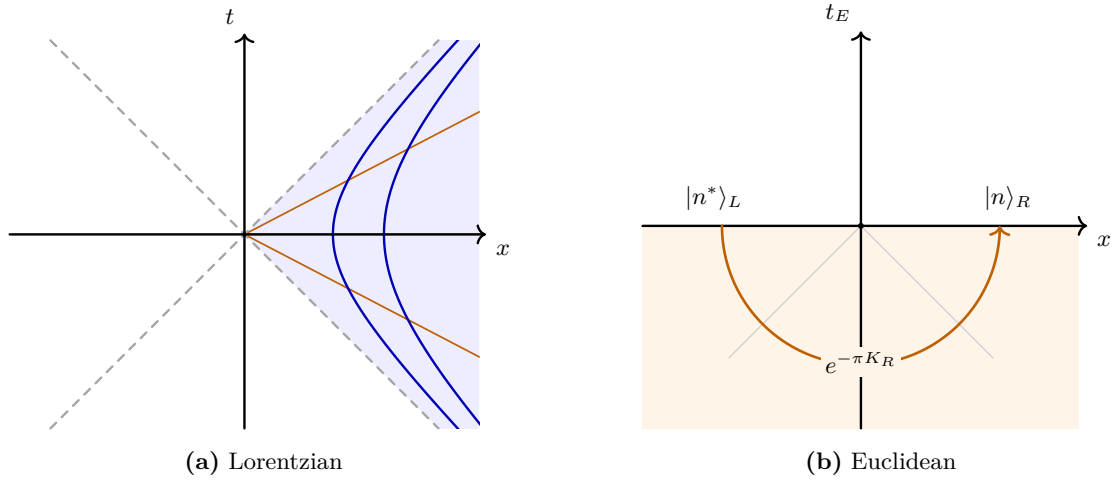

Consider Minkowski space with coordinates $(t,x,\vec y)$ and metric
$\mathrm{d}s^2=-\mathrm{d}t^2+\mathrm{d}x^2+\mathrm{d}\vec y^{\,2}$. We study
a family of accelerated observers moving along the $x$-direction, while
$\vec y$ parametrizes the transverse space. We introduce Rindler coordinates
\begin{equation}
    t=\rho\sinh(a\eta),
    \qquad
    x=\rho\cosh(a\eta),
    \qquad
    \rho>0.
    \label{eq:rindler-coordinates}
\end{equation}
The parameter $a$ is an arbitrary normalization factor with units of inverse
length. At fixed $\rho$, these equations describe the trajectory of a
uniformly accelerated observer. The metric becomes
\begin{equation}
    \mathrm{d}s^2
    =
    -a^2\rho^2\,\mathrm{d}\eta^2
    +
    \mathrm{d}\rho^2
    +
    \mathrm{d}\vec y^{\,2}.
\end{equation}
The relation between the inertial and accelerated coordinate systems is shown
in figure~\ref{fig:minkowski-rindler}(a). We see that these coordinates cover only the right Rindler wedge $x>|t|$.  The left wedge is covered by a similar set of coordinates.

\smallskip

Consider an accelerated observer at fixed $\rho$. It follows from the metric
that the proper time along the observer's trajectory satisfies
$\mathrm{d}\tau=a\rho\,\mathrm{d}\eta$, implying that the four-velocity is
$u^\mu=(\sinh a\eta,\cosh a\eta,\vec 0)$. Differentiating once more with
respect to proper time shows that the acceleration satisfies
$a_\mu a^\mu=1/\rho^2$. Equivalently, the proper acceleration is
$a_{\text{prop}}=1/\rho$. As $\rho\to0$, the hyperbolae approach the null
surfaces $x=\pm t$, the
Rindler horizons. An observer confined to the right wedge cannot receive
signals from the left wedge. 

\smallskip

How an accelerated observer describes a quantum field in the
Minkowski vacuum? An inertial observer evolves according to Minkowski time
$t$, while the accelerated observer evolves according to Rindler time $\eta$.
Observers can probe the state of the field by measuring correlation functions
of local operators as functions of time. For example, consider a free scalar
field $\Phi$ and the two-point function
$\langle\Phi(t,x,\vec y)\Phi(t',x',\vec y')\rangle$. When this correlator is
expressed in the accelerated coordinates, the relation
$t=\rho\sinh(a\eta)$ immediately implies periodicity in imaginary Rindler time:
\begin{equation}
\langle \Phi(\eta,\rho,\vec y)\Phi(\eta',\rho',\vec y')\rangle,~~~~~\eta\sim\eta+\frac{2\pi\mathrm{i}}{a}.
\end{equation}
This suggests that the accelerated observer sees a thermal state with
temperature
\begin{equation}
    T = \hbar \frac{a}{2\pi},\label{eq:unruh-temperature}
\end{equation}
known as the Unruh temperature. The imaginary-time periodicity of correlators
measured by an accelerated observer can be checked explicitly in specific
field theories, but it is a general property of the map between inertial and
accelerated coordinates. One can show that the only stationary state for
which all correlators are periodic in imaginary time is a thermal state. This
is the Kubo-Martin-Schwinger (KMS) condition.

\smallskip 

The temperature in equation~\eqref{eq:unruh-temperature} is conjugate to the Hamiltonian generating
$\partial_\eta$. We can instead compute the temperature measured with
respect to the generator of proper-time translations
$\partial_\tau$ associated with the accelerated observer.
Since $\mathrm{d}\tau=a\rho\,\mathrm{d}\eta$, the proper temperature is
$T_{\text{prop}}=\frac{\hbar}{2\pi\rho}$. This is the temperature measured by
a thermometer carried by the accelerated observer, and it increases as the
observer's trajectory approaches the Rindler horizon.

\smallskip

There is another perspective that will be useful in the next lecture. Consider
the Euclidean continuation of the Rindler metric, obtained by taking
$\eta=-\mathrm{i}\hat{\eta}$.
The metric is then
\begin{equation}
    \mathrm{d}s^2
    =
    a^2\rho^2\,\mathrm{d}\hat{\eta}^2
    +
    \mathrm{d}\rho^2+\mathrm{d}\vec y^{\,2}.
\end{equation}
For fixed $\vec y$, this is the flat plane in polar coordinates, with angle
$a\hat{\eta}$. Smoothness at $\rho=0$ requires the identification
$\hat{\eta}\sim\hat{\eta}+2\pi/a$. All correlators in a state that is smooth
at $\rho=0$, such as the Minkowski vacuum, are periodic in the Euclidean time
$\hat{\eta}$. By the KMS condition, that state is thermal, and an observer
whose evolution is generated by $\eta$-translations sees
$T=\hbar a/(2\pi)$. This is the same result as before, now derived from the
smoothness of the Euclidean geometry.

\smallskip 

We can construct the Minkowski
vacuum in a basis of fixed Rindler-energy states for an accelerated observer.
Its path-integral preparation is shown in figure~\ref{fig:minkowski-rindler}(b).
Infinite Euclidean evolution projects any initial state onto the Minkowski
vacuum. The Hilbert space at a fixed time slice decomposes as $\mathcal{H}=\mathcal{H}_L \times \mathcal{H}_R$. Let
$K_R=H_\eta/a$ be the dimensionless boost
generator restricted to the right wedge, normalized so that it translates
the dimensionless Rindler angle $a\eta$. The path integral for $t<0$ can be thought of as evolution over $a\eta$ by a Euclidean angle $\pi$.  If
$K_R|n\rangle_R=\omega_n|n\rangle_R$, the Euclidean path integral on the
lower half-plane prepares the Minkowski vacuum in the form
\begin{equation}
    |0_M\rangle
    =
    \frac{1}{\sqrt{Z}}
    \sum_n
    e^{-\pi\omega_n}
    |n^\ast\rangle_L |n\rangle_R.
    \label{eq:rindler-vacuum}
\end{equation}
Here $|n^\ast\rangle_L$ is the CPT-conjugate state.  Euclidean evolution
through an angle $\pi$ produces the factor $e^{-\pi\omega_n}$.  The full
state is pure but entangled between the wedges. Tracing over the left wedge
gives
\begin{equation}
    \rho_R
    =
    \operatorname{Tr}_{\mathcal{H}_L} |0_M\rangle\langle 0_M|
    =
    \frac{e^{-2\pi K_R}}
    {Z},\qquad Z = \operatorname{Tr}_{\mathcal H_R}e^{-2\pi K_R},
\end{equation}
where $Z$ is the partition function. The Minkowski vacuum is therefore exactly thermal with respect to the boost
Hamiltonian.

\smallskip 

For a free bosonic field the thermal state can be expressed mode by mode. The right-wedge occupation number
obeys the Planck distribution
\begin{equation*}
    \langle N_{\omega\mathbf{k}}\rangle_R
    =
    \frac{1}{e^{2\pi\omega}-1}.
\end{equation*}
The same structure will soon reappear at a black hole horizon for weakly
coupled matter and will be the origin of Hawking radiation. Tracing over the partner
behind the horizon makes the exterior mode thermal.

\smallskip

We can also estimate the entropy of the right Rindler wedge directly from this
thermal description. Let us denote by $\Sigma$ the $x=0$ surface separating the two wedges. In four spacetime dimensions the thermal entropy density
of a relativistic field scales as
$s_{\rm th}\sim T_{\rm prop}^3/\hbar^3\sim1/\rho^3$.  On a constant-$\eta$
slice, $\mathrm{d}V=\mathrm{d}\rho\,\mathrm{d}^2y$, and hence, introducing a
proper-distance cutoff $\rho\geq\epsilon$ and denoting the transverse area by
$A_\Sigma=\int\mathrm{d}^2y$, we get
\begin{equation}
    S_R\sim
A_\Sigma\int_\epsilon^\infty\mathrm{d}\rho/\rho^3
\sim A_\Sigma/\epsilon^2.
\end{equation}
Thus the Rindler entropy has a UV divergence whose
leading term is proportional to the area of the horizon in units of the cutoff scale.

\subsubsection*{Hawking radiation}
\label{sec:black-hole-radiation}

In this section, we explain the conceptual basis of Hawking radiation. We treat matter fields quantum mechanically on the
fixed geometry of a collapsing star
\cite{Hawking:1975vcx,Davies:1976ei,Christensen:1977jc} and \cite{Witten:2024bht} for a recent review.
In general, this treatment is reliable when the horizon is much larger than
the Planck length. Later, we will present a counterexample in the context of
near-extremal black holes, but for now we focus on the conventional analysis.

\smallskip

Consider an observer held at a fixed radius far from the black hole. In the
limit $r\to\infty$, the radiation measured by this observer can be described
in terms of outgoing modes at future null infinity $\mathcal{I}^+$. The
quantum field is initially placed in the vacuum at past null infinity
$\mathcal{I}^-$, with no incoming particles, and the resulting state is
required to be smooth across the future horizon. At late times, this state
is known as the Unruh vacuum. Its regularity at the horizon is analogous to
the regularity of the Minkowski vacuum across a Rindler horizon. Nevertheless,
it is not empty when decomposed into positive-frequency modes defined with
respect to the asymptotic Schwarzschild time $t$. A distant stationary
observer therefore detects an outgoing flux with a thermal spectrum, up to
greybody factors, rather than the vacuum associated with Schwarzschild time
translations. The latter is the Boulware vacuum, which is analogous to the
Rindler vacuum and is singular at the horizon.

\smallskip

\begin{figure}[t!]
    \vspace*{0.7cm}
    \centering
    \begin{minipage}[t]{0.56\textwidth}
    \centering
    \begin{tikzpicture}[
        scale=0.9,x=0.52cm,
        y=0.52cm,
        line cap=round,
        line join=round,
        every node/.style={font=\small}
    ]
        \coordinate (iminus) at (0,0);
        \coordinate (lefttop) at (0,13.25);
        \coordinate (iplus) at (4.05,13.25);
        \coordinate (right) at (8.72,8.58);
        \coordinate (horizonstart) at (0,9.20);

        \path[fill=collapsingmatter]
            (iminus) -- (lefttop) -- (1.26,13.25) -- (1.26,4.20)
            .. controls (1.20,2.40) and (0.52,0.90) .. (iminus)
            -- cycle;

        \draw[black, line width=1.1pt] (iminus) -- (lefttop);
        \draw[line width=1.25pt,decoration={zigzag,segment length=1.5mm, amplitude=0.3mm},decorate]
            (lefttop) -- (iplus);
        \draw[black, line width=1.1pt] (iminus) -- (right);
        \draw[black, line width=1.1pt] (iplus) -- (right);

        \draw[black, line width=1.25pt] (horizonstart) -- (iplus);

        \coordinate (rayone) at (2.471,11.289);
        \coordinate (raytwo) at (2.719,11.232);
        \coordinate (raythree) at (2.974,11.168);
        \coordinate (rayfour) at (3.325,11.073);
        \coordinate (rayfive) at (3.782,10.937);
        \coordinate (raysix) at (4.353,10.746);

        \draw[purple!75!black, line width=1.15pt] (rayone) -- (4.241,13.059);
        \draw[purple!75!black, line width=1.15pt] (raytwo) -- (4.393,12.907);
        \draw[purple!75!black, line width=1.15pt] (raythree) -- (4.553,12.747);
        \draw[purple!75!black, line width=1.15pt] (rayfour) -- (4.776,12.524);
        \draw[purple!75!black, line width=1.15pt] (rayfive) -- (5.072,12.228);

        \draw[blue!70!black, line width=1.25pt]
            (0,11.55)
            .. controls (1.45,11.63) and (4.75,11.12) .. (right);

        \node[blue!70!black, font=\large] at (5.85,9.35) {$\mathcal S$};
    \end{tikzpicture}
    
    \smallskip
    {\small\textbf{(a)} Collapse geometry}
    \end{minipage}%
    \hfill
    \begin{minipage}[t]{0.40\textwidth}
    \centering
    \begin{tikzpicture}[
        scale=0.9,
        line cap=round,
        line join=round,
        every node/.style={font=\small}
    ]
        \coordinate (bifurcation) at (0,0);
        \coordinate (intersection) at (1.05,-1.05);
        \coordinate (raytop) at (3.15,1.05);

        \draw[black, line width=1.2pt] (-2.75,-2.75) -- (2.75,2.75);
        \draw[black, line width=1.2pt] (-2.75,2.75) -- (2.75,-2.75);
        \node[above left=1pt] at (2.35,2.35) {$\mathcal H^+$};
        \node[below left=1pt] at (2.35,-2.35) {$\mathcal H^-$};

        \draw[purple!75!black, line width=1.2pt,
            postaction={decorate},
            decoration={markings,
                mark=at position 0.58 with {\arrow{>}}}]
            (intersection) -- (raytop);
        \fill[purple!75!black] (intersection) circle (1.5pt);
        \node[purple!75!black, above left=1pt] at (2.55,0.55) {ray};

        \draw[blue!70!black, <->, line width=0.8pt]
            ($(bifurcation)+(0.17,0.17)$) --
            ($(intersection)+(0.17,0.17)$)
            node[midway, above right=-1pt] {$\lambda$};
        \fill (bifurcation) circle (1.5pt);
    \end{tikzpicture}

    \smallskip
    {\small\textbf{(b)} Stationary extension}
    \end{minipage}
    \caption{Schematic Penrose diagrams for Hawking's calculation, following
    the setup emphasized in \cite{Witten:2024bht}.  (a) A Cauchy slice
    $\mathcal S$ crosses the future horizon outside the collapsing matter.
    Outgoing null rays received at later Killing times $t$ cross
    $\mathcal S$ at exponentially smaller values of a smooth transverse
    coordinate $u$.  (b) In the maximally extended stationary geometry, a ray
    propagated backwards intersects the past horizon $\mathcal H^-$ at affine
    separation $\lambda$ from the bifurcation point.}
    \label{fig:hawking-rays}
\end{figure}
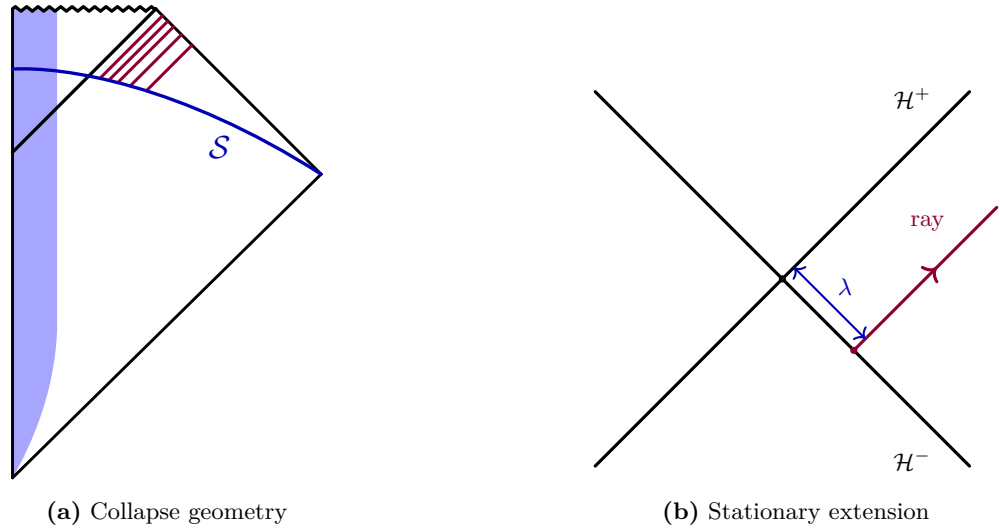

The mechanism is encoded in the causal geometry of the collapse. Consider a
family of outgoing wave packets that reach $\mathcal{I}^+$ at successively
later retarded times. When propagated backward along their null
characteristics to an earlier Cauchy surface $\mathcal{S}$, these wave
packets intersect $\mathcal{S}$ increasingly close to the event horizon, as
illustrated in figure~\ref{fig:hawking-rays}. Consequently, the dependence of
a mode on retarded time at $\mathcal{I}^+$ is mapped to a dependence on its
distance from the horizon on $\mathcal{S}$, where the precise notion of
distance will be defined below.

\smallskip

To make this relation more precise, following Hawking, we replace the
collapse geometry by the maximally extended stationary black hole spacetime
that describes the late-time exterior.  In this extended geometry, an
outgoing null ray reaching $\mathcal{I}^+$ at retarded time $u$ can be
continued backward until it intersects the past horizon $\mathcal{H}^-$. Let
$\lambda$ be an affine parameter along $\mathcal{H}^-$, chosen to vanish at
the bifurcation surface. Then $\lambda$ measures the affine-parameter
separation between the bifurcation surface and the point at which the ray
intersects $\mathcal{H}^-$. Time translations are generated by the
stationary Killing vector $\chi=\partial_t$. At $\mathcal{I}^+$, this Killing
flow acts as a translation of the retarded time $u$, whereas on
$\mathcal{H}^-$ it acts as a rescaling of $\lambda$. The main geometric
input is therefore the relation between $u$ and $\lambda$, which we now
derive.

\smallskip

Let 
$
    k=\partial_\lambda
$
be the affine tangent vector along the past horizon, such that
$k^\mu\nabla_\mu k^\nu=0$. The stationary Killing vector
$\chi=\partial_t$ is tangent to $\mathcal{H}^-$ but points toward the
bifurcation surface, so on $\mathcal{H}^-$ it can be written as
$\chi=f(\lambda)k$, with $f(\lambda)<0$. We then have
\begin{equation}
    \chi^\mu\nabla_\mu\chi^\nu
    =
    f k^\mu\nabla_\mu(fk^\nu)
    =
    f\frac{\mathrm{d}f}{\mathrm{d}\lambda}k^\nu
    =
    \frac{\mathrm{d}f}{\mathrm{d}\lambda}\chi^\nu.
\end{equation}
From the definition of surface gravity it follows that
$\mathrm{d}f/\mathrm{d}\lambda=-\kappa$. Since $\chi$ vanishes at the
bifurcation surface, $f(0)=0$, and hence
$
    \chi
    =
    -\kappa\lambda\,\partial_\lambda.
$ along $\mathcal{H}^-$.
At $\mathcal{I}^+$, the Killing flow acts as a translation of the retarded
time $u$, so that $\chi=\partial_u$. Therefore,
$
    \partial_u=-\kappa\lambda\,\partial_\lambda,
$
which leads to a differential equation for $\lambda(u)$, namely
$
   \mathrm{d}\lambda/\mathrm{d}u
    =
    -\kappa\lambda.
$
Integrating gives
\begin{equation}
    \lambda=\lambda_0e^{-\kappa u},
    \qquad\text{or equivalently}\qquad
    u=-\frac{1}{\kappa}
    \log\, \lambda + \text{constant},
\end{equation}
where $\lambda_0$ reflects the freedom to rescale the affine parameter and
to shift the origin of $u$. 

\smallskip

Let us emphasize that the maximally extended black hole geometry is only an
auxiliary construction. Once the exterior geometry has settled down to a
stationary black hole, however, this construction provides a convenient way
to define $\lambda$ as a measure of the separation on $\mathcal{S}$ close to the horizon. This
separation decreases exponentially at late retarded times,
$\lambda=\lambda_0 e^{-\kappa u}$. If the quantum state is regular across the horizon, its short-distance
correlation functions have the universal vacuum form when expressed in
terms of the affine coordinate $\lambda$. After the exponential change of
variables from $\lambda$ to $u$, these correlation functions satisfy the
KMS condition with imaginary-time period
$
    u\longrightarrow u+\mathrm{i}\,\frac{2\pi}{\kappa},
$
with periodicity for bosonic fields and antiperiodicity for fermionic fields.
At infinity, translations of $u$ coincide with translations of the
observer's proper time. The outgoing modes therefore have thermal
occupation numbers at the Hawking temperature
\begin{equation}
    T
    =
    \frac{\hbar\kappa}{2\pi},
\end{equation}
up to the greybody factors associated with propagation through the exterior
geometry, which we discuss below. This is the essence of Hawking radiation. Importantly, the argument relies only on the universal
short-distance behavior of quantum-field correlation functions near the
horizon, at separations that can remain much larger than the Planck length
while being much smaller than the horizon scale, see the review \cite{Witten:2024bht} for a more detailed discussion on this point.

\smallskip

For a non-rotating black hole, Killing time $u$ coincides with Schwarzschild
time $t$ at fixed $r$, so its physical Hamiltonian generates propagation in $t$. The radiation density matrix is
\begin{equation}
\rho \sim e^{-\beta H},
\end{equation}
where $\beta=1/T$ and $H$ is the energy operator in the QFT. Since we only
used the definition of a Killing horizon in this argument, the exact same
logic applies to a rotating black hole. The main modification in that case is
the following. Killing time is not identified with $t$ at fixed $r$,
since instead the Killing vector field is
$\chi=\partial_t+\Omega\partial_\phi$, with $\Omega$ the black hole angular
velocity we computed in Lecture 1. The
Hamiltonian conjugate to Killing time is therefore
$H-\Omega J$. In that case
\begin{equation}
\rho \sim e^{-\beta(H-\Omega J)},
\end{equation}
where $H$ and $J$ are the physical energy and angular-momentum operators in
the QFT. This implies that Hawking radiation is in a grand canonical state of
temperature $T=\hbar\kappa/(2\pi)$ and angular velocity $\Omega$. This
generalizes in a straightforward way for neutral matter around a charged
black hole. For charged matter the argument has to be slightly adjusted to
account for the Lorentz force when tracking rays, see
\cite{Hawking:1975vcx}. The upshot is that the matter state is in the grand
canonical ensemble
\begin{equation}
\rho \sim e^{-\beta(H - \Omega J - \Phi Q)}
\end{equation}
with $\Phi$ the chemical potential of the black hole, which we computed in Lecture 1, and $Q$ the generator of $\mathrm{U}(1)$ transformations on the matter field. 

\smallskip

Given the thermal nature of Hawking radiation we can quickly estimate the evaporation time for a black hole. For a neutral non-rotating black hole, a typical emitted quantum has wavelength of order $r_s$ and
energy of order $T\sim\hbar/r_s$.  The luminosity is therefore
parametrically
\begin{equation}
    \frac{\mathrm{d}M}{\mathrm{d}t}
    \sim
    -\frac{A T^4}{\hbar^3}
    \sim
    -\frac{\hbar}{G_N^2 M^2},
    \label{eq:schwarzschild-mass-loss}
\end{equation}
up to numerical factors and the number of light species. Thus evaporation is
extremely slow for astrophysical black holes since the temperature, being originated by a quantum effect, is down by a power of $\hbar$. Within this estimation, integration gives
$t_{\rm evap}\simeq5120\pi G_N^2M^3/\hbar
\simeq2.1\times10^{67}(M/M_{\odot})^3\,\mathrm{yr}$, so a black hole with
$r_s\simeq3\,\mathrm{km}$ has $t_{\rm evap}\sim2\times10^{67}$ years.

\subsubsection*{Emission rates and greybody factors}\label{subsec:greybody-factors}

Although the horizon emits radiation at temperature $T$, not all of that
radiation reaches infinity. A potential barrier outside the black hole
reflects some of it back into the hole. The fraction transmitted to infinity
is called the greybody factor. This effect can be described explicitly for
free quantum fields on the black hole background.

\smallskip

\begin{samepage}
Let us begin by neglecting this effect. For a mode of angular frequency
$\omega$, the emission spectrum has the Planck or Fermi-Dirac form,
depending on the quantum statistics:
\begin{equation}
    \frac{\mathrm{d}E}{\mathrm{d}t\,\mathrm{d}\omega}
    =
   \sum_{\ell,m} \frac{\hbar\omega}{2\pi}
   \frac{1}{e^{\beta\hbar\omega}\mp 1},
\end{equation}
where we take $+1$ for fermions and $-1$ for bosons. The sum runs over the
angular-momentum quantum numbers $\ell$ and $m$. We next outline how to include
the effect of the potential barrier.
\end{samepage}

\smallskip

\begin{figure}[t!]
    \centering
    \begin{tikzpicture}[
        x=0.62cm,
        y=1.75cm,
        line cap=round,
        line join=round,
        every node/.style={font=\small}
    ]
        \def\xmin{-6}
        \def\xmax{8}
        \def\ymax{2.05}

        \draw[->, line width=0.8pt] (-6.25,0) -- (8.65,0)
            node[below right=2pt] {$r_\ast/r_s$};
        \draw[->, line width=0.8pt] (-6,-0.02) -- (-6,2.20)
            node[above] {$r_s^2 V_\ell$};

        \foreach \xt in {-6,-4,-2,0,2,4,6,8} {
            \draw (\xt,0.025) -- (\xt,-0.025)
                node[below=3pt] {$\xt$};
        }
        \foreach \yt in {0.5,1.0,1.5,2.0} {
            \draw (-5.92,\yt) -- (-6.08,\yt)
                node[left=3pt] {$\yt$};
        }

        \begin{scope}
            \clip (-6,0) rectangle (8,2.05);
            \draw[blue!70!black, line width=1.15pt]
                plot[domain=-7:1.8, samples=320, smooth]
                ({1+exp(\x)+\x},
                 {(1-1/(1+exp(\x)))*
                 (1/((1+exp(\x))*(1+exp(\x))*(1+exp(\x))))});
            \draw[red!75!black, line width=1.15pt]
                plot[domain=-7:1.8, samples=320, smooth]
                ({1+exp(\x)+\x},
                 {(1-1/(1+exp(\x)))*
                 (2/((1+exp(\x))*(1+exp(\x)))+
                 1/((1+exp(\x))*(1+exp(\x))*(1+exp(\x))))});
            \draw[green!45!black, line width=1.15pt]
                plot[domain=-7:1.8, samples=320, smooth]
                ({1+exp(\x)+\x},
                 {(1-1/(1+exp(\x)))*
                 (6/((1+exp(\x))*(1+exp(\x)))+
                 1/((1+exp(\x))*(1+exp(\x))*(1+exp(\x))))});
            \draw[purple!75!black, line width=1.15pt]
                plot[domain=-7:1.8, samples=320, smooth]
                ({1+exp(\x)+\x},
                 {(1-1/(1+exp(\x)))*
                 (12/((1+exp(\x))*(1+exp(\x)))+
                 1/((1+exp(\x))*(1+exp(\x))*(1+exp(\x))))});
        \end{scope}

        \begin{scope}[shift={(4.00,1.38)}]
            \draw[fill=white, draw=gray!40, rounded corners=1pt]
                (-0.18,-0.18) rectangle (2.30,0.95);
            \draw[blue!70!black, line width=1.15pt] (0,0.75) -- (0.38,0.75);
            \node[anchor=west] at (0.55,0.75) {$\ell=0$};
            \draw[red!75!black, line width=1.15pt] (0,0.50) -- (0.38,0.50);
            \node[anchor=west] at (0.55,0.50) {$\ell=1$};
            \draw[green!45!black, line width=1.15pt] (0,0.25) -- (0.38,0.25);
            \node[anchor=west] at (0.55,0.25) {$\ell=2$};
            \draw[purple!75!black, line width=1.15pt] (0,0.00) -- (0.38,0.00);
            \node[anchor=west] at (0.55,0.00) {$\ell=3$};
        \end{scope}
    \end{tikzpicture}
    \caption{Dimensionless Schwarzschild effective potential for scalar
    partial waves, plotted against the tortoise coordinate with $r_s=1$.
    The horizon is at $r_\ast/r_s\to -\infty$, spatial infinity is at
    $r_\ast/r_s\to +\infty$, and the barrier becomes higher for larger
    angular momentum.}
    \label{fig:greybody-potential}
\end{figure}
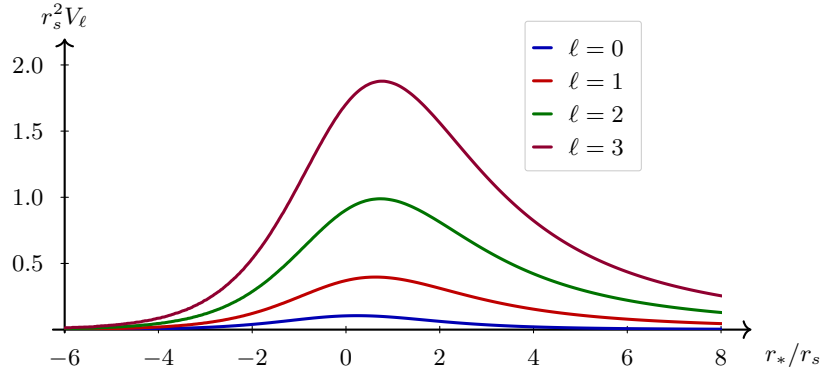

Consider a massless scalar field for simplicity. Expand
$\phi=\sum_{\ell m}\chi_{\ell m}(t,r)Y_{\ell m}(\Omega)/r$ and use the
tortoise coordinate $r_\ast$ introduced in Lecture 1.
Each partial wave obeys a one-dimensional wave equation
\begin{equation}
    \left[
        -\partial_t^2+\partial_{r_\ast}^2
        -V_\ell(r)
    \right]\chi_{\ell m}=0,
\end{equation}
with effective potential
\begin{equation}
    V_\ell(r)
    =
    \left(1-\frac{r_s}{r}\right)
    \left[
        \frac{\ell(\ell+1)}{r^2}
        +
        \frac{r_s}{r^3}
    \right].
\end{equation}
The potential vanishes at both the horizon and infinity but has a barrier in
between, as anticipated. The potential entering the greybody factors is shown
in figure~\ref{fig:greybody-potential}. It is most naturally plotted as a
function of the tortoise coordinate. Setting $x=r/r_s$,
\begin{equation}
    \frac{r_\ast}{r_s}
    =
    x+\log(x-1),
    \qquad
    r_s^2 V_\ell
    =
    \left(1-\frac{1}{x}\right)
    \left[
        \frac{\ell(\ell+1)}{x^2}
        +
        \frac{1}{x^3}
    \right],
    \qquad x\geq 1.
\end{equation}

Let $T_{\omega\ell}$ be the transmission amplitude through this barrier and
define the greybody factor
$\Gamma_{\omega\ell}=|T_{\omega\ell}|^2$. 
The emission spectrum measured at infinity is the thermal spectrum multiplied
by this transmission probability. For a bosonic field,
\begin{equation}
    \frac{\mathrm{d}E}{\mathrm{d}t\,\mathrm{d}\omega}
    =
    \sum_{\ell,m}
    \frac{\hbar\omega}{2\pi}
    \frac{\Gamma_{\omega\ell}}
    {e^{\beta\hbar\omega}\mp 1}.
\end{equation}
For the $s$-wave, the transmission problem can be solved analytically in the
low-frequency regime. Matching the near-horizon and asymptotic solutions for
$\ell=0$ and $\omega r_s\ll1$ gives
$\Gamma_{\omega 0}\simeq4(\omega r_s)^2$. Equivalently, the corresponding
absorption cross section for plane waves approaches
$\sigma_{\rm abs}^{(\ell=0)}=\pi\Gamma_{\omega 0}/\omega^2
\simeq4\pi r_s^2=A$, the area of the horizon. Page
analyzed this effect carefully for Schwarzschild and Kerr black holes,
obtaining precise estimates of their evaporation rates
\cite{Page:1976df,Page:1976ki}.

For a stationary charged and rotating black hole, the thermal factor is shifted
by the chemical potentials already introduced in Lecture 1. A field mode
proportional to $e^{-\mathrm{i}\omega t+\mathrm{i}m\phi}$, with electric
charge $q$, has horizon-frame energy
$
    E
    =
    \hbar(\omega-m\Omega)-q\Phi.
$
The Hawking energy spectrum at infinity takes the general form
\begin{equation}
    \frac{\mathrm{d}E}{\mathrm{d}t\,\mathrm{d}\omega}
    =
    \sum_{\ell,m}
    \frac{\hbar\omega}{2\pi}\,
    \frac{\Gamma_{\omega\ell m q}}
    {\exp\!\left[\beta
    \left(\hbar(\omega-m\Omega)-q\Phi\right)\right]
    \mp 1},
    \label{eq:charged-rotating-hawking-spectrum}
\end{equation}
where the upper sign is for bosons and the lower sign for fermions. Here
$\Gamma_{\omega\ell m q}$ is the flux-normalized absorption coefficient. In
the Schwarzschild case it reduces to the nonnegative transmission probability
defined above, but for a rotating or charged hole it can have either sign.

\smallskip

For bosons, the horizon-frame energy can be negative. In the superradiant
range
$0<\hbar\omega<\hbar m\Omega+q\Phi$, the absorption
coefficient $\Gamma_{\omega\ell m q}$ is negative: an incident wave is
reflected with greater amplitude while extracting energy, angular momentum, or
charge from the black hole
\cite{Zeldovich:1971mw,Starobinsky:1973aij}. The denominator in
equation~\eqref{eq:charged-rotating-hawking-spectrum} is negative in the same range,
so the spontaneous outgoing flux remains positive. Fermionic fields have the
Fermi-Dirac shift shown above but do not exhibit classical superradiant
amplification.

\smallskip

A charged black hole has a second discharge channel. Its near-horizon electric field can create charged
pairs through the Schwinger effect, even in the zero-temperature extremal
limit. For a particle of mass $\mu$ and charge $q$, a locally constant
field $E$ gives the characteristic suppression
$\exp[-\pi\mu^2/(\hbar|qE|)]$
\cite{Schwinger:1951nm}. One member of the pair can fall into the black hole
while the other escapes, reducing its charge. Schwinger production is driven directly by the
electric field and can dominate the discharge of a sufficiently cold charged black hole.

\subsection{The Bekenstein-Hawking entropy}

The preceding discussion shows that black holes possess a genuine
thermodynamic temperature. Under the standard assumptions of the no-hair
theorems, a black hole formed by collapse radiates away its transient
multipole structure and approaches a stationary geometry
characterized by only its mass $M$, angular momentum $J$, and electric charge
$Q$ \cite{Israel:1967wqNoHair,Carter:1971zcNoHair,Robinson:1975bv}. This
relaxation is analogous to thermalization: the details of the initial state
cease to appear as independent macroscopic parameters, and the equilibrium
configuration is specified by a small number of conserved charges. The
zeroth law of black hole mechanics ensures that the surface gravity
$\kappa$, and hence the Hawking temperature
$T=\hbar\kappa/(2\pi)$, is constant over the horizon of a stationary
black hole. This naturally raises the question of what entropy should be
assigned to the black hole. As first anticipated by Bekenstein
\cite{Bekenstein:1973ur}, the answer is proportional to the horizon area. In the rest of the section we review the argument.

\subsubsection*{First and second laws}

We begin with the classical motivation for associating black hole entropy
with the horizon area. The mass of a black hole need not increase: energy
can be extracted from a rotating or charged black hole together with angular
momentum or electric charge. Nevertheless, Hawking's area theorem states
that, assuming the null energy condition and appropriate global regularity
conditions, the area of the future event horizon cannot decrease in
classical general relativity,
\begin{equation}
    \delta A\geq 0
\end{equation}
\cite{Hawking:1971tu}. This monotonicity is the gravitational analogue of
the increase of entropy required by the second law of thermodynamics and
motivates the identification of $A$ as an entropy-like
quantity. The precise proportionality between entropy and area, including
its normalization, will follow from the first law and the Hawking
temperature.

\smallskip

To determine the proportionality constant between entropy and area, we turn to the first law of black hole mechanics. For infinitesimal variations
between nearby stationary black holes, the first law
relates the change in mass to the variations of the horizon area, angular
momentum, and electric charge
\begin{equation}
    \delta M
    =
    \frac{\kappa}{8\pi G_N}\,\delta A
    +
    \Omega\,\delta J
    +
    \Phi\, \delta Q,
\end{equation}
see \cite{Bardeen:1973gs}. The area term in the first law can be written in terms of Hawking temperature.
The first law of black hole mechanics therefore takes precisely the
thermodynamic form
$
    \delta M
    =
    T\,\delta S_{\rm BH}
    +
    \Omega\,\delta J
    +
    \Phi \,\delta Q,
$
provided that
\begin{equation}
    S_{\rm BH}
    =
    \frac{A}{4G_N\hbar},
\end{equation}
up to an additive constant. This is the Bekenstein-Hawking entropy. Wald generalized the first-law construction to a
broad class of diffeomorphism-invariant theories of gravity
\cite{Wald:1993nt}. In such theories, the entropy is given by a Noether
charge and need not be proportional to the horizon area.

\smallskip

For a Schwarzschild black hole of one solar mass, the Bekenstein-Hawking
formula gives $S_{\rm BH}\sim 10^{77}$. By comparison, the Sun
contains approximately
$N_{\rm baryon}\sim M_\odot/m_p\sim10^{57}$ baryons. Assigning an entropy of
order one to each constituent therefore gives the rough estimate
$S_{\rm Sun}\sim10^{57}$. A solar-mass black hole thus
has roughly twenty orders of magnitude more entropy than an ordinary star of the same mass. Although substantial progress has been made in recent decades, a general microscopic derivation of black hole entropy and a complete description of the underlying microstates remain outstanding problems.

\subsubsection*{Generalized entropy}

The finiteness of the Bekenstein-Hawking entropy may seem unremarkable at
first, but it is striking in light of the Unruh effect. There, the fine-grained von Neumann entropy
across the Rindler horizon is UV divergent. This implies that the entropy of
quantum matter fields across a black-hole event horizon is also infinite:
\begin{equation}
    S_{\rm out} \sim \frac{A}{\epsilon^2}.
\end{equation}
This suggests that the black-hole entropy should include these infinite
contributions from unobservable vacuum modes behind the horizon.

A related problem concerns the status of the second law in the presence of
matter configurations that themselves carry entropy. For example, suppose we
throw a cup of tea into a black hole. As it falls through the horizon, the
total entropy of the spacetime must remain nondecreasing. The correct
second-law statement therefore cannot concern the ordinary entropy outside the
black hole alone. This motivates the generalized entropy
\begin{equation}
    S_{\rm gen}
    =
    \frac{A}{4G_N\hbar}
    +
    S_{\rm out},
\end{equation}
where $S_{\rm out}$ is the entropy of matter and radiation outside the
horizon. The generalized second law states that the generalized entropy cannot
decrease between successive cuts of a future causal horizon,
$\delta S_{\rm gen}\geq0$. Subject to appropriate semiclassical assumptions,
Wall proved this statement for rapidly changing quantum fields and arbitrary
horizon slices \cite{Wall:2011hj}.

\smallskip

We can now return to the opening comments about the finiteness of the entropy.
The preceding argument naively implies that both $S_{\rm out}$ and
$S_{\rm gen}$ are always divergent: regardless of the state outside the black
hole, the matter fields approach the vacuum at very short distances from the
horizon. Susskind and Uglum realized that, although $S_{\text{out}}$ is divergent, the generalized entropy is actually finite
\cite{Susskind:1994sm}. The Newton constant appearing in the
Bekenstein-Hawking term also receives divergent corrections from UV physics.
They
showed that these divergences cancel against those in $S_{\rm out}$, so the
generalized entropy of a black hole remains finite. More recently, \cite{Chandrasekaran:2022eqq} gave an
operator-algebraic derivation in which the generalized entropy is identified
with the von Neumann entropy of a type~$\mathrm{II}_\infty$ algebra of
semiclassical observables.

\section{The gravitational path integral}

In the previous lecture, black-hole mechanics led us to a striking
thermodynamic analogy, culminating in the Bekenstein-Hawking entropy.  Modern
holography turns this analogy into a precise quantum duality: the fundamental
degrees of freedom of quantum gravity can be encoded on a lower-dimensional
boundary \cite{tHooft:1993dmi,Susskind:1994vu}.  In its sharpest realization,
AdS/CFT, quantum gravity in an asymptotically $\mathrm{AdS}$ spacetime is
equivalent to an ordinary nongravitational conformal field theory (CFT) on its
boundary \cite{Maldacena:1997re,Gubser:1998bc,Witten:1998qj}. In this duality, AdS arises as the geometry near the horizon of extremal black holes and black branes.

This motivates the \emph{central dogma} of black hole physics: as viewed from
the exterior, a black hole behaves as an
ordinary quantum system with unitary microscopic evolution and approximately
$\exp{\left(S_{\text{BH}}\right)}$ states.  The Bekenstein-Hawking entropy thus counts actual
microstates, while the apparent dissipation due to the horizon found in the geometric description
arises from coarse graining rather than fundamental information loss.  AdS/CFT
provides the best-controlled setting for this statement, while it also serves
as an organizing principle for asymptotically flat black holes. In this duality AdS is the near-horizon black hole spacetime while the CFT is the quantum system. Near-extremal black
holes are especially useful because their low-temperature expansion, specially in supergravity when the extremal black hole preserves some supercharges, gives a
controlled arena in which to compare the geometric and microscopic pictures.

The gravitational path integral provides a direct calculational
bridge between the geometric and quantum descriptions.  The thermal trace of
the quantum system is represented in the bulk by a sum over Euclidean
geometries, and the black-hole saddle computes its free energy and hence its
entropy.  This method, pioneered by Gibbons and Hawking \cite{Gibbons:1976ue},
has since become central to holographic entanglement entropy
\cite{Ryu:2006ef,Lewkowycz:2013nqa}, the Page curve
\cite{Almheiri:2019qdq,Penington:2019kki}, near-extremal black holes
\cite{Iliesiu:2020qvm}, and protected comparisons with microscopic string
theory \cite{Cabo-Bizet:2018ehj,Iliesiu:2021}.  We begin by reviewing the
finite-temperature path integral in ordinary quantum mechanics, where its
equivalence to a trace over states is explicit, before promoting the same
construction to dynamical spacetime.

\subsection{Path integral at finite temperature}

In the canonical formalism, the free energy can be computed from the partition function by counting states. In the path integral formulation of quantum mechanics, the partition function is expressed as a sum over all possible paths with periodic Euclidean time. The
canonical partition function is
\begin{equation}
    Z(\beta)
    =
    \operatorname{Tr}_{\mathcal H} e^{-\beta H}.
\end{equation}
The evolution operator in quantum mechanics over Lorentzian time $t$ is $e^{- \i H t/\hbar}$. The operator $e^{-\beta H}$ can be interpreted as evolution in Euclidean time $\hbar \beta$. In general Euclidean time $\tau$ is defined as 
$t=-\mathrm{i}\tau$.  Its matrix elements are Euclidean path
integrals on an interval of length $\hbar\beta$,
\begin{equation}
    \langle \phi_f|e^{-\beta H}|\phi_i\rangle
    =
    \int_{\phi(0)=\phi_i}^{\phi(\hbar \beta)=\phi_f}
    \mathcal{D}\phi\, e^{-I_E[\phi]/\hbar}.
\end{equation}
Taking the trace identifies the final and initial states, making effectively Euclidean time a circle of length $\hbar \beta$, such that:
\begin{equation}
    Z(\beta)
    =
    \int_{\phi(\hbar \beta)=\phi(0)}
    \mathcal{D}\phi\, e^{-I_E[\phi]/\hbar}\,.
\end{equation}
 Bosons are periodic and fermions antiperiodic
around this thermal circle. The path integral computes the trace without
displaying the energy eigenstates explicitly. In quantum mechanics this is not a problem since the path integral and canonical quantization are equivalent. In quantum gravity, we find that taking the path integral as more fundamental to the spacetime description leads to surprising results.

\subsubsection*{The partition function of a compact oscillator}

As an example of the path integral formulation of the partition function, we will compute the partition function of a compact oscillator in quantum mechanics. Consider a degree of freedom $\phi(t)$ that takes value on a circle of radius $R$,
\begin{equation}
    \phi(t)\sim\phi(t)+2\pi R.
\end{equation}
This could be the angular coordinate of a particle moving on a circle. For simplicity we consider a free particle with Euclidean action
\begin{equation}
    I_E[\phi]
    =
    \int_0^{\beta\hbar}\mathrm{d}\tau\,
        \frac{m}{2}\dot\phi^{\,2}.
\end{equation}
The path integral with such an action is quadratic, and therefore can be written exactly as a sum over saddle points with their one-loop determinant associated to the fluctuations around them. The saddle points are the classical solutions of the Euclidean equations of motion, which are simply $\partial_\tau^2\phi=0$. The general solution is
\begin{equation}
    \phi_w(\tau)
    =
    \bar\phi
    +2\pi Rw \frac{\tau}{\hbar \beta},\qquad 0\leq\bar\phi<2\pi R,\qquad w\in\mathbb{Z}.
\end{equation}
If the field $\phi(\tau)$ was not defined on a circle, periodicity along thermal circle would require $w=0$. For finite $R$, however, the field can wind around the circle $w$ times. The winding number $w$ is a topological invariant of the path and cannot be changed by continuous deformations. The Euclidean action evaluated on this solution is
\begin{equation}
    I_E[\phi]
    =
    \frac{2\pi^2mR^2w^2}{\hbar \beta}.
\end{equation}
The path integral is therefore a sum over winding $w$, as well as an integral over $\bar{\phi}$. We have computed the classical action on each saddle point which depends on $w$ but is independent of $\bar{\phi}$. The next step is to compute the one-loop determinant of the fluctuations around each saddle point. 

In order to do this we expand the field as
\begin{equation}
    \phi(\tau)
    =
   \phi_w(\tau)
    +\eta(\tau),
    \qquad
    \eta(\hbar\beta)=\eta(0).
\end{equation}
The periodicity condition guarantees that $\eta(\tau)$ keeps us in the same winding sector. We can absorb the mode $\bar\phi$ as a constant mode in $\eta(\tau)$. The action for the fluctuation is
\begin{equation}
    I_E[\eta]
    =
    \frac{2\pi^2mR^2w^2}{\hbar\beta}
    +
    \frac{m}{2}\int_0^{\hbar\beta}\mathrm{d}\tau\,\dot\eta^{\,2}.
\end{equation}
The partition function is given, after integration by parts in the fluctuation term, by
\begin{equation}
    Z(\beta)
    =
    \sum_{w\in\mathbb{Z}}
    \exp\left(-\frac{2\pi^2mR^2w^2}{\beta\hbar^2}\right)
    \int \mathcal{D}\eta\,
    \exp\left[
        -\frac{m}{2\hbar}
        \int_0^{\hbar\beta}\mathrm{d}\tau\,
        \eta(-\partial_\tau^2)\eta
    \right]\,.
\end{equation}
In the next of this section we evaluate this path integral and the sum over windings.

To compute the integral over the fluctuations we need to diagonalize the quadratic operator in the action. For this expression to be well-defined we need to give a prescription to determine the measure. We define the measure implicitly through the Gaussian integral
\begin{equation}
    \int \mathcal{D}\eta \, e^{- || \eta||^2} = 1, \qquad || \eta||^2 = \mu \int_0^{\hbar \beta} \mathrm{d}\tau \eta^2.
\end{equation} 
Here $\mu$ is a normalization constant with appropriate units giving a one-parameter family of measures. We refer to this choice as the ultra-local measure.

Next we shall evaluate the Gaussian integrals involved in the definition of the measure, as well as in the partition function. To diagonalize the operator $-\partial_\tau^2$ we can expand the fluctuations in Fourier modes
\begin{equation}
    \eta(\tau)=\sum_{\substack{n\in\mathbb{Z}\\ n\neq0}}\eta_n e^{\frac{2\pi \i n }{\hbar \beta}\tau},
    \qquad
    \eta_{-n}=\eta_n^*.
\end{equation}  
The action in terms of $\eta_n$ is given by
\begin{equation}
I_E[\phi]
=
\frac{2\pi^2 m R^2 w^2}{\hbar \beta}
+
\frac{m\hbar\beta}{2}
\sum_{\substack{n\in\mathbb{Z}\\ n\neq0}}
\left( \frac{2\pi n}{\hbar \beta} \right)^2 |\eta_n|^2.
\end{equation}
We can take care of the measure in the following simple way. We rescale $\eta_n$ such that the inner product is normalized. The inner product is given by
\begin{equation}
    ||\eta||^2 = \mu \int_0^{\hbar \beta} \mathrm{d}\tau \eta^2 = \mu \sum_{\substack{n\in\mathbb{Z}\\ n\neq0}} |\eta_n|^2 \hbar \beta.
\end{equation}
We can define the normalized modes $\hat{\eta}_n = \sqrt{\mu \hbar \beta} \eta_n$ such that $||\eta||^2 = \sum_{n\neq0} |\hat{\eta}_n|^2$. The action in terms of the normalized modes is
\begin{equation}
\frac{I_E[\phi]}{\hbar}
=
\frac{2\pi^2 m R^2 w^2}{\beta\hbar^2}
+
\frac{m}{2 \mu \hbar}
\sum_{\substack{n\in\mathbb{Z}\\ n\neq0}}
\left( \frac{2\pi n}{\hbar \beta} \right)^2 |\hat{\eta}_n|^2.
\end{equation}
Since the field is real, the modes with labels $n$ and $-n$ form a complex-conjugate pair, or equivalently two real Gaussian variables. Excluding the constant mode for the moment, the Gaussian integral over the nonzero modes is
\begin{equation}
    Z_{\rm nz}
    =
    \prod_{n=1}^{\infty}
    \left[
        \frac{m}{2\mu\hbar}
        \left(\frac{2\pi n}{\hbar\beta}\right)^2
    \right]^{-1}.
\end{equation}

This infinite product is defined using zeta-function regularization. Introducing
$
    A
    =
    \frac{m}{2\mu\hbar}
   (\frac{2\pi}{\hbar\beta})^2,
$
we use $\zeta(0)=-1/2$ and $\zeta'(0)=-\frac{1}{2}\log(2\pi)$ to obtain\footnote{Zeta-function regularization replaces a divergent sum by a convergent function of a complex parameter, analytically continues it, and evaluates it at the value corresponding to the original expression. Infinite products are defined by applying this prescription to their logarithms.}
\begin{equation}
    \log Z_{\rm nz}
    =-\sum_{n=1}^{\infty}\log(An^2)=-\zeta(0)\log A+2\zeta'(0)
    =\frac{1}{2}\log A-\log(2\pi).
\end{equation}
%
%
%
Therefore the nonzero-mode determinant is
\begin{equation}
    Z_{\rm nz}
    =
    \frac{1}{\hbar\beta}
    \sqrt{\frac{m}{2\mu\hbar}}.
\end{equation}

The omitted $n=0$ mode is annihilated by $-\partial_\tau^2$, so it cannot be included in the determinant. It is precisely the collective coordinate $\bar\phi$ that was separated from $\eta$ above, and it must be integrated over independently. The ultralocal norm of a constant variation is
\begin{equation}
    ||\delta\bar\phi||^2
    =
    \mu\int_0^{\hbar\beta}\d\tau\,(\delta\bar\phi)^2
    =
    \mu\hbar\beta\,(\delta\bar\phi)^2.
\end{equation}
The normalized coordinate is $\widehat{\bar\phi}=\sqrt{\mu\hbar\beta}\,\bar\phi$. The defining Gaussian integral for the ultralocal measure therefore implies
\begin{equation}
    \frac{\d\widehat{\bar\phi}}{\sqrt{\pi}}
    =
    \sqrt{\frac{\mu\hbar\beta}{\pi}}\,\d\bar\phi.
\end{equation}
Since the action is independent of $\bar\phi$, integration over its compact range gives
\begin{equation}
    Z_0
    =
    \sqrt{\frac{\mu\hbar\beta}{\pi}}
    \int_0^{2\pi R}\d\bar\phi
    =
    2\pi R\sqrt{\frac{\mu\hbar\beta}{\pi}}.
\end{equation}

Combining the zero mode, the nonzero-mode determinant, and the sum over winding sectors, the arbitrary normalization $\mu$ cancels and we find
\begin{equation}
    Z(\beta)
    =
    \sqrt{\frac{2\pi mR^2}{\beta\hbar^2}}
    \sum_{w\in\mathbb{Z}}
    \exp\left(-\frac{2\pi^2mR^2w^2}{\beta\hbar^2}\right).
\end{equation}

There is a useful mnemonic for the zero-mode factor. In dimensional regularization the regulated number of modes of a local field vanishes, $\sum_{n\in\mathbb{Z}}1=1+2\zeta(0)=0$, so the exponent in the product over all modes vanishes. Removing the $n=0$ mode and integrating it separately supplies the factor $\beta^{1/2}$ in $Z_0$; removing $N$ modes similarly gives $\beta^{N/2}$. Although $Z_0$ and $Z_{\rm nz}$ separately depend on $\mu$, their product, which is the complete one-loop factor, is independent of $\mu$ in dimensional regularization. Thus every ultralocal choice of measure gives the same $Z(\beta)$. Moreover, the absolute normalization of the one-loop determinant agrees exactly with canonical quantization, as the following Poisson resummation makes manifest.

To make contact with the canonical description, use the Poisson resummation formula
\begin{equation}
    \sum_{w\in\mathbb{Z}}e^{-\pi a w^2}
    =
    \frac{1}{\sqrt{a}}
    \sum_{k\in\mathbb{Z}}e^{-\pi k^2/a},\qquad  a=\frac{2\pi mR^2}{\beta\hbar^2}.
\end{equation}
The prefactor in the winding representation is exactly $\sqrt{a}$, and hence
\begin{equation}
    Z(\beta)
    =
    \sum_{k\in\mathbb{Z}}
    \exp\left(-\beta\frac{\hbar^2k^2}{2mR^2}\right).
\end{equation}
Indeed, the normalized momentum eigenstates of a particle on a circle are
\begin{equation}
    \langle\phi|k\rangle
    =
    \frac{1}{\sqrt{2\pi R}}e^{\i k\phi/R},
    \qquad
    p_k=\frac{\hbar k}{R},
    \qquad
    E_k=\frac{p_k^2}{2m}=\frac{\hbar^2k^2}{2mR^2}.
\end{equation}
Thus the Poisson-resummed path integral is precisely
\begin{equation}
    Z(\beta)
    =
    \sum_{k\in\mathbb{Z}}e^{-\beta E_k}
    =
    \operatorname{Tr}_{\mathcal H}e^{-\beta H}.
\end{equation}
Poisson resummation exchanges the winding sectors of the Euclidean path integral with the quantized momentum eigenstates of the canonical description. This example shows that for quantum mechanics the path integral is equivalent to the canonical formulation, and also that the thermal trace is naturally a Euclidean object in the path integral approach.

\smallskip

This example contains one further lesson that will be important in quantum
gravity. Complex solutions of the classical equations of motion frequently
arise in both quantum mechanics and gravity. Although such solutions do not
represent physical Lorentzian histories, they can contribute as essential
saddles of the path integral when they lie on the appropriate integration
contour. In the present example, the sum over winding sectors is necessary
to reproduce the Hilbert-space result, even though the corresponding
saddles are complex in Lorentzian signature. We will encounter a
gravitational example of the same phenomenon at the end of this section.
Thus, obtaining the correct physics may require complexified configurations
that have no direct Lorentzian interpretation. While this is sometimes
presented as a difficulty peculiar to gravitational path integrals, the
present example shows that it is a \textit{feature} already familiar from
the path-integral quantization of ordinary quantum systems.

\subsection{The gravitational path integral}

We have just seen how the microstates of a free particle moving on a loop emerge from path-integral quantization. In quantum mechanics, however, one usually proceeds in the opposite direction: canonical quantization is taken as the starting point and used to illuminate the path-integral formulation.

\smallskip

In quantum gravity, by contrast, the situation is reversed. The path integral provides more direct access to coarse-grained properties of the underlying microstates. For example, a naive canonical quantization of matter and gravity yields a divergent black hole entropy because of ultraviolet divergences at the horizon, rather than the finite Bekenstein-Hawking entropy. We will now see how the path integral reproduces this finite result.

\smallskip

It is important to emphasize, however, that black holes can indeed admit a canonical description in which their microstates are states in a Hilbert space governed by a Hamiltonian. A central lesson of string theory is that this canonical description may bear little resemblance to the original geometric formulation of the black hole. The AdS/CFT correspondence provides an explicit realization of this idea.

\subsubsection*{Neutral black holes}

The central idea of Gibbons and Hawking, which laid the foundation for the gravitational path integral, is the following. In quantum field theory, the thermal partition function, and hence the free energy or entropy at inverse temperature $\beta$, is obtained by compactifying Euclidean time on a circle of circumference $\hbar\beta$. In gravity, however, the spacetime geometry is itself dynamical, so this prescription cannot be imposed throughout the bulk \emph{a priori}. Gibbons and Hawking instead interpreted the field-theory construction as an asymptotic boundary condition: one fixes the thermal circle at the asymptotic boundary and then sums over all smooth bulk geometries compatible with it. Thus, to compute the free energy, one fixes the circumference $\hbar\beta$ of the asymptotic Euclidean time circle and integrates over all of its smooth bulk fillings.

One such filling is thermal flat space,
\begin{equation}\label{eq:thermal-flat-space}
    \mathrm{d}s_E^2
    =
    \mathrm{d}\tau^2
    +
    \mathrm{d}r^2
    +
    r^2\mathrm{d}\Omega^2,
    \qquad
    \tau \sim \tau + \hbar\beta.
\end{equation}
For this saddle, the inverse temperature $\beta$ is arbitrary. Although the only modification of the classical geometry is the compactification of Euclidean time, its effect on the quantum matter fields is non-trivial: rather than preparing the vacuum, the Euclidean path integral now describes a thermal ensemble at inverse temperature $\beta$.

Wick rotating the Lorentzian Schwarzschild solution according to
$t=-\mathrm{i}\tau$ yields another candidate filling, the Euclidean
Schwarzschild geometry,
\begin{equation}
    \mathrm{d}s_E^2
    =
    f(r)\,\mathrm{d}\tau^2
    +
    \frac{\mathrm{d}r^2}{f(r)}
    +
    r^2\mathrm{d}\Omega^2,
    \qquad
    f(r)=1-\frac{r_s}{r},
    \qquad
    r_s=2G_NM .
\end{equation}
At large $r$, we have $f(r)\to 1$. Thus, after imposing the identification
$\tau\sim\tau+\hbar\beta$, this geometry satisfies the same asymptotic
boundary condition as thermal flat space. It therefore represents a black
hole of mass $M$ immersed in a thermal bath at inverse temperature $\beta$.

\smallskip

A priori, the temperature of the asymptotic bath and the Hawking
temperature determined by $M$ appear to be independent. The Euclidean
geometry, however, is smooth only when these two temperatures agree. To see
this, consider the near-horizon region, where
$
    f(r)
    \simeq
    2\kappa(r-r_s).
$
 Introducing a radial coordinate $\rho$ the metric becomes 
$\rho$ through
\begin{equation}
    r-r_s=\frac{\kappa}{2}\rho^2,
\qquad\Rightarrow\qquad
\mathrm{d}s_E^2
    \simeq
    \mathrm{d}\rho^2
    +
    \kappa^2\rho^2\,\mathrm{d}\tau^2
    +
    r_s^2\mathrm{d}\Omega^2 .
\end{equation}
The $(\rho,\tau)$ submanifold is locally a plane written in polar
coordinates. Notice the parallel with Rindler coordinates, making it clear that stationary observers at fixed radial distance are analogous to the accelerating observer. This geometry is smooth at $\rho=0$ only if $\kappa\tau$ is an angular
coordinate with period $2\pi$. Combining this regularity condition with the
asymptotic identification of Euclidean time gives
\begin{equation}
    \tau\sim\tau+\hbar\beta,
    \qquad
    \hbar\beta
    =
    \frac{2\pi}{\kappa}
    =
    8\pi G_NM .
\end{equation}
Any other period produces a conical singularity at the horizon. This
regularity condition is precisely the
condition for thermal equilibrium between the black hole and the surrounding
bath. Consequently, for a prescribed asymptotic inverse temperature $\beta$,
a smooth black hole saddle exists only when
\begin{equation}
    M=\frac{\hbar\beta}{8\pi G_N}.
\end{equation}
According to the Gibbons-Hawking prescription, it is this smooth geometry
that contributes as the black hole saddle of the gravitational path integral.

\begin{figure}[t!]
    \centering
    \begin{tikzpicture}[
        x=1cm,
        y=1cm,
        line cap=round,
        line join=round,
        every node/.style={font=\small}
    ]
        \begin{scope}[shift={(-3.55,0)},scale=0.78]
            \path[fill=blue!5] (0,0) circle (2.25);
            \foreach \ang in {0,45,...,315} {
                \draw[gray!22] (0,0) -- (\ang:2.25);
            }
            \foreach \rad in {1.65} {
                \draw[blue!50, line width=0.65pt] (0,0) circle (\rad);
            }
            \draw[black, line width=1pt] (0,0) circle (2.25);
            \draw[->, red!75!black, line width=1.1pt]
                (55:1.65) arc[start angle=55,end angle=300,radius=1.65];
            \node[red!75!black, fill=white, inner sep=1.2pt]
                at (1.62,0.95) {$S^1_\tau$};
            \draw[->, line width=0.85pt] (0,0) -- (25:2.08)
                node[pos=0.56, above, sloped] {$\rho$};
            \fill[black] (0,0) circle (1.5pt);
            \node[anchor=north, align=center] at (0,-0.12)
                {horizon\\$r=r_s$};
        \end{scope}
        \node[align=center] at (-3.55,-2.90)
            {(a) $\hbar\beta=2\pi/\kappa$: smooth tip};

        \begin{scope}[
            shift={(3.45,0)},
            scale=0.78,
            every node/.append style={font=\scriptsize}
        ]
            \coordinate (B)  at (0,0);
            \coordinate (L)  at (-3.6,0);
            \coordinate (R)  at ( 3.6,0);
            \coordinate (FL) at (-2.2,2.35);
            \coordinate (FR) at ( 2.2,2.35);

            \draw[line width=1pt] (L)--(FL);
            \draw[line width=1pt] (R)--(FR);
            \draw[line width=1pt] (B)--(FL);
            \draw[line width=1pt] (B)--(FR);
            \draw[line width=1.05pt,
                decoration={zigzag,segment length=1.5mm,amplitude=0.28mm},
                decorate] (FL)--(FR);

            \begin{scope}[cm={1,0,-0.30,0.50,(0,0)}]
                \path[fill=black!3]
                    (-3.6,0) arc[start angle=180,end angle=360,radius=3.6]
                    -- cycle;
                \foreach \a in {220,270,320}
                    \draw[black!18,line width=0.35pt] (0,0)--(\a:3.6);
                \draw[->,black!50,line width=0.6pt]
                    (-2.55,0) arc[start angle=180,end angle=342,radius=2.55];
                \draw[line width=1pt]
                    (-3.6,0) arc[start angle=180,end angle=360,radius=3.6];
            \end{scope}
            \draw[line width=1.25pt] (L)--(B)--(R);
            \fill (B) circle (1.5pt);

            \node[font=\small] at (-1.93,0.78) {$L$};
            \node[font=\small] at ( 1.93,0.78) {$R$};

            \node[below=3pt] at (-2.10,0) {$\tau=\hbar\beta/2$};
            \node[below=3pt] at ( 2.10,0) {$\tau=0$};
        \end{scope}
        \node[align=center] at (3.45,-2.90)
            {(b) Hartle-Hawking state preparation};
    \end{tikzpicture}
    \caption{Euclidean constructions for the black hole. (a) In the smooth
    Euclidean black-hole geometry, the Euclidean-time circle contracts at the
    horizon when $\hbar\beta=2\pi/\kappa$, giving
    $\mathbb{R}^2_{(\rho,\tau)}\times S^2$; the complete disk computes the
    thermal partition function. (b) Cutting the disk in half prepares the
    Hartle-Hawking state: the lower half-disk is glued along its diameter to
    the $t=0$ Einstein-Rosen bridge of the two-sided Lorentzian black hole.
    Evolution through half the thermal circle produces the factors
    $e^{-\beta E_i/2}$ in equation~\eqref{eq:hartle-hawking-state}.}
    \label{fig:euclidean-black-hole}
\end{figure}
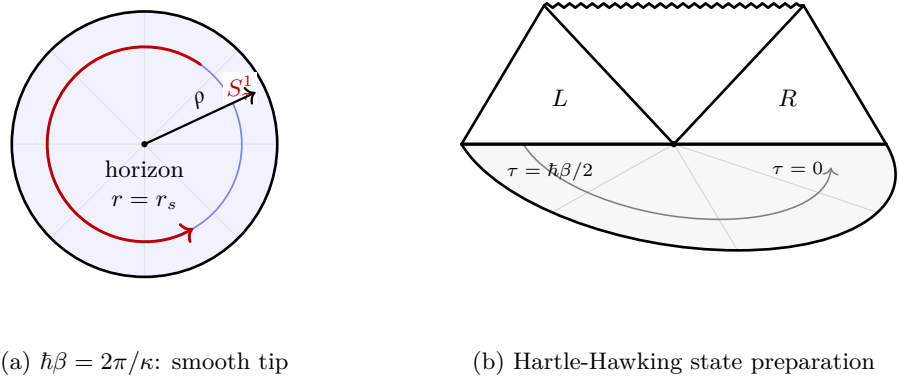

\smallskip

The complete Euclidean geometry shown in panel~(a) of
figure~\ref{fig:euclidean-black-hole} computes the thermal partition function
$
    Z(\beta)=\operatorname{Tr}e^{-\beta H}.
$
The trace arises because the Euclidean-time circle is closed, so the initial
and final field configurations are identified. To prepare a state instead, we cut the $(\rho,\tau)$ disk along the diameter
formed by the rays at $\tau=0$ and $\tau=\hbar\beta/2$. Performing the path
integral over the lower half of the disk, while holding the fields fixed along
this cut, produces a wavefunctional of those field configurations. The two
radial halves of the cut extend to two distinct asymptotic ends which, after
analytic continuation to Lorentzian signature, become the left and right
exterior regions of the maximally extended black hole. The half-disk path
integral therefore prepares a state in
$\mathcal{H}_L\otimes\mathcal{H}_R$, as illustrated in panel~(b) of
figure~\ref{fig:euclidean-black-hole}. The upper half of the disk
prepares the corresponding bra, and gluing the two halves together recovers
the complete Euclidean geometry and hence the thermal trace.

\smallskip 

This construction is the black-hole analogue of the Rindler construction
discussed in Lecture~1. There, a path integral over a Euclidean half-plane
prepares the Minkowski vacuum on a slice containing the left and right Rindler
wedges. Here, the half-disk prepares the Hartle-Hawking state on a slice
connecting the two exterior regions of the eternal black hole. Euclidean evolution through half of the thermal circle weights an energy
eigenstate by $e^{-\beta E_i/2}$. Consequently, the Hartle-Hawking state
takes the thermofield-double form
\begin{equation}
    \label{eq:hartle-hawking-state}
    |\Psi_{\mathrm{HH}}\rangle
    =
    \frac{1}{\sqrt{Z(\beta)}}
    \sum_n
    e^{-\beta E_n/2}
    |n^\ast\rangle_L |n\rangle_R .
\end{equation}
Here, $|n\rangle_R$ is an energy eigenstate in the right exterior and
$|n^\ast\rangle_L$ is its CPT conjugate in the left exterior. Tracing over the
left exterior gives
\begin{align}
    \rho_R
    &=
    \operatorname{Tr}_{\mathcal{H}_L}
    \left(
        |\Psi_{\mathrm{HH}}\rangle
        \langle\Psi_{\mathrm{HH}}|
    \right)
    \nonumber\\
    &=
    \frac{1}{Z(\beta)}
    \sum_n e^{-\beta E_n}
    |n\rangle_R\langle n|
    =
    \frac{e^{-\beta H_R}}{Z(\beta)} .
\end{align}
Thus, although the Hartle-Hawking state on the two-sided geometry is pure,
its restriction to either exterior is thermal. The complete Euclidean
section computes the canonical partition function, while a half-section
prepares its thermofield-double purification. The smoothness condition at the
tip fixes the temperature to the Hawking temperature and ensures that the
resulting state is regular across the horizon.

\smallskip

Let us return to the gravitational partition function $Z(\beta)$. In the
canonical ensemble, it is defined by a path integral over metrics
and matter fields that induce prescribed data at the asymptotic boundary
\begin{equation}
    Z(\beta)
    =
    \int_{\substack{
        \left.g\right|_{\partial M}=h
    }}
    \mathcal{D}g\,\mathcal{D}\varphi\,
    e^{-I_E[g,\varphi]/\hbar}.
\end{equation}
where $h$ is the boundary condition at infinity with Euclidean time compactified to period $\hbar \beta$. The same boundary data admit
two classical fillings. The first is thermal flat space, with topology
$\mathbb{R}^3\times S^1$, in which the Euclidean-time circle is
noncontractible. The second is Euclidean Schwarzschild, with topology
$\mathbb{R}^2\times S^2$, in which the Euclidean-time circle is contractible
and closes off smoothly at the horizon. In the semiclassical approximation
\begin{equation}
    Z(\beta)
    \simeq
    \sum_{\text{saddles}} Z_{\text{one-loop}}
    e^{
        -I_E/\hbar
    },
\end{equation}
the partition function is obtained by
summing over these classical saddles. In this section we will focus on the physics of the black hole saddle. Its leading contribution to the partition function is given by its on-shell action.

\smallskip

Before evaluating this action, however, we encounter an important technical
subtlety. The Einstein-Hilbert term alone does not define a well-posed
variational principle with fixed boundary metric. Its variation produces boundary terms
containing normal derivatives of $\delta g_{\mu\nu}$, which do not vanish
even when the induced metric $h_{ij}$ is held fixed. These terms are canceled
by adding the Gibbons-Hawking-York boundary term
\cite{York:1972sj,Gibbons:1976ue}:
\begin{equation}
    I_E
    =
    -\frac{1}{16\pi G_N}
    \int_M \mathrm{d}^4x\,\sqrt{g}\,R
    -
    \frac{1}{8\pi G_N}
    \int_{\partial M} \mathrm{d}^3x\,\sqrt{h}\,K
    +
    (\text{matter})\end{equation}
Here, $h$ is the determinant of the induced metric $h_{ij}$, and
$K=h^{ij}K_{ij}$ is the trace of the extrinsic curvature of the boundary,
defined using the outward-pointing unit normal. With the
Gibbons-Hawking-York term included, the remaining boundary variation is
proportional to $\delta h_{ij}$ and therefore vanishes when Dirichlet boundary
conditions are imposed. 

\smallskip

A second subtlety that appears is that the action has an infrared divergence. Consider a geometry defined with a radial cutoff at $r=r_c$. For asymptotically flat geometries, we can construct a finite on-shell action by taking the black hole result with a cutoff $r_c$ action and subtracting the action of a flat space with the same boundary metric. This difference is finite as $r_c$ goes to infinity. 

\smallskip

The details of this calculation can be found in many references, so
we will simply quote the result. We strongly suggest students to try to derive it themselves since the details are instructive. The difference between the black hole action $I_{\text{BH}}$ and the flat space action $I_{\text{flat}}$, after taking the large $r_c$ limit, is given by
\begin{equation}
    \begin{aligned}
        I_{\rm BH}-I_{\text{flat}}
        =4\pi G_NM^2
          =\frac{\hbar^2\beta^2}{16\pi G_N}.
    \end{aligned}
\end{equation}
Therefore the saddle point approximation of the partition function given by 
\begin{equation*}
   \log Z(\beta) \approx - \frac{\hbar \beta^2}{16\pi G_N} .
\end{equation*}
Notice that the Hawking temperature of a black hole is $T \sim \hbar$ such that $\hbar \beta$ is of order one. Therefore the on-shell action is of order $1/\hbar$ as expected. This would be more transparent if we wrote the answer in terms of the surface gravity instead of inverse temeprature. 

\smallskip

The energy and entropy are then obtained from well-known thermodynamic relations
\begin{equation}
    \begin{aligned}
        E
        &=-\frac{\partial}{\partial\beta}\log Z= M, \\
        S
        &=\left(1-\beta\frac{\partial}{\partial\beta}\right)\log Z
          = \frac{A}{4\hbar G_N} .
    \end{aligned}
\end{equation}
Taking the path integral seriously, this is a more principled derivation of the Bekenstein-Hawking entropy than the original heuristic arguments and analogies with the laws of thermodynamics. Of course, the drawback is that the nature of the black hole microstates is not evident. This also relies on the assumption that a UV-complete theory of quantum gravity exists, such that the path integral is well-defined. 

\smallskip

\subsubsection*{Comments on generalized entropy}
\label{subsec:smooth-replicas}

The disk calculation above has produced a finite, renormalized black-hole
entropy from a smooth gravitational saddle.  This statement deserves some
comments.  Once the period of Euclidean time is chosen correctly, the horizon is
the smooth center of the disk, not a boundary or a singular point.  There is
therefore no new UV divergence localized at the horizon.  The disk
path integral still has the ordinary short-distance divergences of quantum
field theory at every bulk point.  In gravitational effective field theory
these are removed, order by order, by renormalizing the cosmological constant,
Newton's constant, and the higher-curvature couplings.  Thus ``finite'' here
means finite after this usual bulk renormalization; smoothness tells us that no
independent horizon counterterm is required
\cite{Susskind:1994sm}.

For the stationary black hole, the thermodynamic derivative of $Z(\beta)$ was
enough to obtain the entropy.  For more general problems it is more useful to
compute R\'enyi entropies.  This is the method that, when applied to a boundary
subregion in AdS/CFT, leads to the Ryu-Takayanagi formula
\cite{Lewkowycz:2013nqa}.  For a normalized density matrix $\rho$, define
\begin{equation}
    S_n
    =
    \frac{1}{1-n}\log\operatorname{Tr}\rho^n,
    \qquad
    S
    =
    \lim_{n\to1}S_n
    =
    -\left.\partial_n\log\operatorname{Tr}\rho^n\right|_{n=1}.
    \label{eq:renyi-von-neumann}
\end{equation}
In a path integral, $\operatorname{Tr}\rho^n$ is obtained by cyclically gluing
$n$ copies along the cut that defines the density matrix.

In QFT, we take copies of the spacetime without changing its geometry. Gravity adds one essential ingredient: the geometry is part of the path
integral.  For integer $n$, the asymptotic time circle of the thermal
replica has circumference $n\hbar\beta$, and the dominant smooth bulk filling
is a new saddle $g_n$.  For the formal Schwarzschild saddle its mass is
$M_n=nM$.  Keeping the original metric $g_1$ while merely lengthening the
identification would instead create a conical singularity at the horizon.
More generally, the full replica cover $\mathcal M_n$ is smooth, while its
quotient by the replica symmetry has a wedge of opening angle $2\pi/n$.

Strictly speaking, the exact boundary density matrix $\rho$ does not depend on
$n$; it is the semiclassical saddle representing $\operatorname{Tr}\rho^n$
that changes.  It is nevertheless useful to encode this fact in an effective
wedge operator $\rho_n$ prepared by the $n$-dependent geometry.
The $n$-derivative necessary to compute the entropy now acts in two places.  Writing
$\dot\rho_1=\left.\partial_n\rho_n\right|_{n=1}$ gives
\begin{equation}
    \left.
    \partial_n\operatorname{Tr}\rho_n^n
    \right|_{n=1}
    =
    \operatorname{Tr}(\rho_1\log\rho_1)
    +
    \operatorname{Tr}\dot\rho_1.
    \label{eq:replica-two-derivatives}
\end{equation}
When the derivative acts on the power $n$, it gives the entropy of the exterior
quantum fields.  When it acts on $\rho_n$, it measures the response of the
geometry preparing the state.  The replica entropy therefore separates as
\begin{equation}
    S
    =
    S_{\rm out}+S_{\rm geom},
    \qquad
    S_{\rm out}\equiv-\operatorname{Tr}(\rho_1\log\rho_1),
    \qquad
    S_{\rm geom}\equiv-\operatorname{Tr}\dot\rho_1.
    \label{eq:replica-entropy-split}
\end{equation}

The geometric term is fixed locally near the tip.  
Although we will not reproduce the argument here, one can evaluate it by asking how the Einstein action varies as we introduce a conical singularity at the horizon. The result is precisely the area term
\begin{equation}
 S_{\rm geom}
    =
    \frac{A}{4\hbar G_{N,0}}.
    \label{eq:replica-area-response}
\end{equation}
Here $G_{N,0}$ is the bare Newton constant.   The split in equation~\eqref{eq:replica-entropy-split} makes both terms look
singular.  The matter entropy has the familiar area-law divergence
\cite{Bombelli:1986rw,Srednicki:1993im}, while the bare coefficient of the area
term carries the cutoff dependence required to renormalize Newton's constant.
In four dimensions, schematically,
\begin{equation}
    S_{\rm out}^{(\epsilon)}
    =
    c_{\rm UV}\frac{A}{\epsilon^2}
    +
    S_{\rm out}^{\rm ren}
    +\cdots,
    \qquad
    \frac{1}{\hbar G_{N}}
    =
    \frac{1}{\hbar G_{N,0}}
    +
    \frac{4c_{\rm UV}}{\epsilon^2}
    +\cdots .
    \label{eq:newton-renormalization-entropy}
\end{equation}
Consequently the renormalized replica entropy takes the form
\begin{equation}
    S_{\rm gen}
    =
    \frac{A}{4\hbar G_{N}}
    +
    S_{\rm out, ren},
    \label{eq:renormalized-generalized-entropy}
\end{equation}
and is finite order by order in the gravitational effective theory. This is the Susskind-Uglum mechanism. A more detailed explanation and generalization of this results are reviewed in sections 8 and 9 of \cite{Witten:2024bht}.

\subsubsection*{Charged black holes}
\label{subsec:charged-thermodynamics}

The same Euclidean method also computes the thermodynamics of charged black
holes~\cite{Braden:1990hwCharged}.  The only new subtlety is the choice of
electromagnetic ensemble.  The Euclidean action with fixed boundary value of
the gauge potential is
\begin{equation}
    I_E^\Phi
    =
    -\frac{1}{16\pi G_N}
    \int_M \mathrm{d}^4x\,\sqrt{g}\,R
    +
    \frac{1}{4}
    \int_M \mathrm{d}^4x\,\sqrt{g}\,
    F_{\mu\nu}F^{\mu\nu}
    -
    \frac{1}{8\pi G_N}
    \int_{\partial M} \mathrm{d}^3x\,\sqrt{h}\,(K-K_0).
\end{equation}
Here we sketch how to obtain the partition function at fixed temperature and chemical potential, using the gravitational path integral. From it we derive the Bekenstein-Hawking entropy in the same fashion as for neutral black holes.  

\smallskip

Most of the analysis done for neutral black holes carries over to the charged case.  The Euclidean Reissner-Nordstr\"om solution takes the same structure as Schwarszchild, with a different redshift factor $f(r)$. Regardless of the new charge-dependent terms, the near-horizon analysis when imposing smoothness of the geometry is unchanged. The reason is that, by definition of surface gravity, it is always true that $f(r) \approx 2\kappa(r-r_+)$ near the outer horizon $r_+$, and the same argument as in the neutral case applies. The smoothness condition therefore imposes thermal equilibrium between the black hole radiation and the bath. 

\smallskip

New ingredients appear when we look at the gauge field.  In the
grand-canonical ensemble, the boundary data include the gauge potential, or
equivalently its holonomy around the thermal circle.  Denoting the prescribed
chemical potential by $\Phi_{\rm bath}$, we impose at large radius
\begin{equation}
    \left.A\right|_{r\to\infty}
    =
    \Phi_{\rm bath}\,\mathrm{d}t .
\end{equation}
After Wick rotation, $t=-\mathrm{i}\tau$, this becomes
$A_E|_{r\to\infty}=-\mathrm{i}\Phi_{\rm bath}\,\mathrm{d}\tau$.
Charged bosons and fermions obey their usual periodic and antiperiodic thermal
boundary conditions in this background, understood gauge covariantly.  A
gauge transformation that removes the asymptotic value of $A_E$ instead makes
these boundary conditions twisted, with the twist determined by the Wilson
line $\exp(\mathrm{i}q\oint A_E/\hbar)$ for a field of charge $q$.  These
boundary data compute
$Z(\beta,\Phi_{\rm bath})=\operatorname{Tr}
e^{-\beta(H-\Phi_{\rm bath}Q)}$.  It is important that $\Phi_{\rm bath}$ is
the chemical potential of the radiation bath at infinity, not an independently
prescribed chemical potential of the black hole.

For the Reissner-Nordstr\"om saddle, the corresponding connection is
\begin{equation}
    A
    =
    \left(\Phi_{\rm bath}-\frac{Q}{4\pi r}\right)\mathrm{d}t ,
\end{equation}
or its Euclidean continuation.  We can check again that smoothness implies that the chemical potential of the bath matches the chemical potential of the black hole radiation which we call $\Phi = Q/4\pi r_+$ from now on. We can again obtain the free energy by evaluating the on-shell action and reproduce the thermodynamics of the charged black hole.

\smallskip

There is also a useful path-integral way to see electric-charge quantization.
Let $q_0$ be the smallest electric charge.  Because the Euclidean boundary
holonomy is compact, a large gauge transformation shifts the chemical
potential by an imaginary period,
\begin{equation*}
    \Phi
    \rightarrow
    \Phi_k
    =
    \Phi+\frac{2\pi\mathrm{i}k}{\beta q_0},
    \qquad k\in\mathbb{Z}.
\end{equation*}
These shifts describe the same boundary holonomy, but the corresponding large
gauge transformations do not extend smoothly across the contractible thermal
circle in the bulk.  They therefore label distinct smooth saddle
representatives that must be included in the semiclassical sum.

\subsection{Modern applications}
\label{subsec:modern-applications}

The Gibbons-Hawking evaluation of the free energy is the simplest possible application of the gravitational path integral. This approach has been extended in multiple directions to compute more subtle observables of quantum black holes. It also was applied in cosmology in an approach pioneered by Hartle and Hawking \cite{Hartle:1983ai} although we will not discuss this in these lectures. Below is a brief summary of some fascinating applications.

\paragraph{Holographic Entanglement Entropy:} Let $R$ be a spatial subregion
of a holographic quantum field theory. The Ryu-Takayanagi conjecture identifies
its entanglement entropy with the area of a codimension-two minimal surface
$\gamma_R$ in the bulk, anchored on $\partial R$ and homologous to $R$
\cite{Ryu:2006bv,Ryu:2006ef}. Lewkowycz and Maldacena derived this prescription
from the gravitational path integral using the replica trick
\cite{Lewkowycz:2013nqa}. For time-dependent states, the relevant surface is
the covariant extremal surface of Hubeny, Rangamani, and Takayanagi
\cite{Hubeny:2007xt}. Including bulk quantum corrections leads instead to
quantum extremal surfaces, which extremize the generalized entropy
\cite{Engelhardt:2014gca}.

\paragraph{The Page Curve from Replica Wormholes:} The gravitational replica
trick can also be applied to Hawking radiation collected in an approximately
nongravitating bath. The disconnected replica saddle reproduces Hawking's
result that the radiation entropy grows indefinitely. The path integral also
admits connected saddles, known as replica wormholes, which are subleading at
early times but dominate after the Page time
\cite{Almheiri:2019qdq,Penington:2019kki}. Their contribution gives the island
rule: one extremizes the generalized entropy over possible islands and selects
the smallest extremum. The no-island saddle dominates initially, while a
nonempty island dominates at late times and produces the unitary Page curve dictating that the entropy of Hawking radiation decreases at late times.
The island lies in the entanglement wedge of the radiation, indicating that
part of the black hole interior is encoded in the radiation instead of in the black hole quantum system.

\paragraph{Wormholes and Quantum Chaos:} The Bekenstein-Hawking entropy
suggests that a black hole has an exponentially large number of discrete
states with extremely small level spacing. If the black hole quantum system is strongly interacting and chaotic, this predicts a pattern of correlations between nearby energy levels. These correlations are probed by the
spectral form factor, whose linear ramp in time is a universal signature of level repulsion \cite{Cotler:2016fpe}. In gravity, the ramp arises
from the double-cone, a connected two-boundary saddle obtained by periodically
identifying a two-sided black hole under a Killing-time translation. Its
relative-time zero mode produces the linear growth
\cite{Saad:2018bqo}. In two-dimensional gravity, summing over additional
wormhole topologies is equivalent to a matrix integral and reproduces the
expected random-matrix correlations
\cite{Saad:2019lba,Stanford:2019vob,Maxfield:2020ale}. 

\paragraph{Quantum Effects Near Extremality:} Near extremality, the
conventional semiclassical description of black hole thermodynamics becomes
unreliable because quantum fluctuations in the long near-horizon throat are
strongly enhanced. Recent gravitational path-integral analyses have shown how
these large quantum-gravity corrections modify the spectrum and statistical
mechanics of near-extremal black holes
\cite{Iliesiu:2020qvm,Turiaci:2023wrh}. This will be the subject of
Lecture~4.

\section{Extremal black holes in quantum gravity}

The material in this lecture is more advanced than that covered in the first three lectures and draws heavily on current research. Rather than attempting a comprehensive treatment, we present a brief overview of recent results on near-extremal black holes, intended to serve as an introduction for students interested in the subject.  

\subsection{Near-extremal puzzles}
\label{subsec:extremal-thermo}
\label{subsec:third-law}

In Lectures 2 and 3, we treated a black hole as a thermodynamic object and
used a smooth Euclidean saddle to compute its entropy.  According to the
central dogma, the same black hole should also be an ordinary quantum system
with a spectrum of states.  At sufficiently high temperature, the individual
levels are too closely spaced to resolve and a smooth equation of state is a
good approximation.  As the temperature is lowered toward the bottom of the
spectrum, however, this approximation must eventually fail. This is, in a sense, embodied in the third law of thermodynamics which is the topic of this section.

\smallskip

Extremal black holes provide the cleanest setting for this test.  At fixed
electric charge $Q$ and angular momentum $J$, the Kerr-Newman family has a
minimum mass $M_{\rm ext}(Q,J)$ \cite{Newman:1965my,Page:2000dk}.  We approach
this bound without changing the conserved charges and use
$E\equiv M-M_{\rm ext}$ as the excitation energy.  When $M=M_{\rm ext}$ or $E=0$, the inner and outer horizons coincide and the surface
gravity, and hence the Hawking temperature, vanishes.  The horizon area
nevertheless remains finite. For large enough charge the extremal horizon can display low curvature compared to Planck scale.  Thus any
failure of semiclassical physics near extremality is not caused by a UV effect.  

\smallskip

Let us first review the classical thermodynamics of near-extremal black holes. The Kerr-Newman entropy at extremality is given by 
\begin{equation*}
    S_{\rm ext}
    =
    \frac{1}{\hbar}
    \sqrt{\frac{Q^4}{16}+4\pi^2J^2},
\end{equation*}
which is nonzero and depends on the conserved charges. When either $Q=0$ or $J=0$ we reproduced the formulas we derived in Lecture 1. Taking the expressions for the area and mass as a function of temperature and expanding at low temperature we can find
\begin{eqnarray}
    S(T)
   & =&
    S_{\rm ext}+\frac{4 \pi r_+^{\rm ext} S_{\rm ext} }{\hbar} T+O(T^2),\\
    E(T)
    &=&
    M(T)-M_{\rm ext}
    =
   \frac{2 \pi r_+^{\rm ext} S_{\rm ext}}{\hbar} T^2+O(T^3).
\end{eqnarray}
The range of validity of these approximation is simpler to state in term of geometric quantities, namely when the surface gravity is smaller than $\kappa \ll 1/r_+$. This means that the black hole temperature is small than that of a neutral black hole of the same area.

\smallskip

The third law refers to two related but logically
distinct statements.  One concerns the entropy of equilibrium states as
$T\to0$; the other concerns whether a dynamical process can reach $T=0$ in
finite time.  Extremal black holes described by classical gravity seem inconsistent with both.

\smallskip

For an ordinary system, cooling removes thermally accessible excitations until
only its ground states remain.  The Nernst formulation says that the resulting
zero-temperature entropy should approach a constant independent of continuous
thermodynamic parameters; the stronger Planck formulation sets this constant
to zero when the ground state is unique.  Extremal black holes instead have
\begin{equation*}
    T=0,
    \qquad
    S_{\rm ext}\neq0,
\end{equation*}
with $S_{\rm ext}$ depending on charges such as $Q$ and $J$.  This
violates the Nernst formulation across the family of extremal solutions.\footnote{The large extremal entropy is especially striking in
Einstein-Maxwell-scalar theories with moduli-dependent gauge couplings.
When a stable attractor exists, the near-horizon values of the massless
scalars are fixed by the black hole charges and are independent of their
values at infinity. Consequently, the classical horizon area and entropy
remain unchanged under continuous deformations of the asymptotic moduli.
The attractor mechanism was originally discovered for supersymmetric
extremal black holes in $\mathcal{N}=2$ supergravity
\cite{Ferrara:1995ih}, but it is not essential. The same
mechanism follows directly from the bosonic equations of motion in suitable
non-supersymmetric theories
\cite{Goldstein:2005hq,Sen:2005wa}.}  This is not surprising, quantum effects are expected to be important when we reach small temperatures. The problem is in identifying the origin of such quantum effects in gravity. 

\smallskip

The dynamical version asks a different question. It states that a
nonextremal black hole cannot be driven to zero surface gravity by any physical
process completed in finite time.  Israel formulated and proved an influential
version of this unattainability principle under assumptions on the evolving
horizon and matter stress tensor \cite{Israel:1986ThirdLaw}.  It was long
regarded as the black hole analogue of the impossibility of cooling an
ordinary system to absolute zero in finitely many operations. Kehle and Unger recently disproved the conjecture and constructed smooth initial data in the spherically symmetric
Einstein-Maxwell-charged-scalar system whose collapse becomes exactly
extremal Reissner-Nordstr\"om after a finite advanced time along the event
horizon \cite{Kehle:2022ThirdLaw}.  

\smallskip

There is a second problem, identified by Preskill, Schwarz, Shapere, Trivedi,
and Wilczek, that arises even before the black hole reaches exactly zero
temperature \cite{Preskill:1991tb}. At fixed charge and angular momentum, the
near-extremal equation of state implies that the heat capacity vanishes
linearly with temperature:
\begin{equation}
  C_{Q,J}
        =
        T\left(\frac{\partial S}{\partial T}\right)_{Q,J}
        =
        \frac{4\pi r_+^{\rm ext} S_{\rm ext}}{\hbar} T+O(T^2).
\end{equation}
Consequently, the root-mean-square energy fluctuations vanish as $T\to0$.
The mean excitation energy above extremality, however, scales as $T^2$ and
therefore vanishes even faster. Another to phrase this issue is that when $T$ is too small the average energy of a Hawking quanta, $\sim T$, can be larger than the energy available by the black hole $\sim r_+^{\rm ext} S_{\rm ext}/\hbar T^2$. The fluctuations thus become increasingly
large relative to the mean excitation energy, becoming comparable to it at
the breakdown scale
\begin{equation}
    T
    \lesssim
    E_{\rm brk},
    \qquad
    E_{\rm brk}
    \equiv
    \frac{\hbar \pi}{r_+^{\rm ext}S_{\rm ext}}.
\end{equation}
The precise order-one coefficient defining the onset of this regime is not
meaningful at this level and its just convention. This is the puzzle identified by
Preskill et al. It does not contradict the central dogma; rather, it marks
the scale below which the semiclassical black hole saddle no longer provides
an adequate thermodynamic description. At this scale, the detailed quantum
spectrum and the dynamics of the near-horizon region become essential.

\smallskip

Notice that $E_{\text{brk}}$ is an extremely small energy scale. The Hawking temperature of a Schwarszchild black hole of radius $r_+$ would be of order $\hbar/r_+$. This is known to be extremely small, for example for a black hole with $r_+=3\,\text{km}$ the temperature is $T\sim 10^{-8}\,\text{K}$. The breakdown scale is even smaller by a factor of $S_{\text{ext}}$ which is of order $10^{77}$. For a Kerr black hole with $J$ such that the extremal radius is $3\,\text{km}$, the parameter $a_\star = J/G_N M^2$ needs to be $1-a_\star \lesssim 10^{-152}$ for the temperature to be of order of the gap scale which is absurdly small. For rotating black holes we also have to worry about superradiance, which we mention later.

\smallskip 

The
remaining task is to determine the origin of these enhanced quantum-gravity corrections
and understand how they modify both the black hole spectrum and the spectrum
of Hawking radiation.

\subsection{Leading quantum gravity corrections}
\label{subsec:one-loop-gravity}

The origin of the quantum corrections required to fix these issues has a interesting history. It was discovered through an indirect route involving the Sachdev-Ye-Kitaev (SYK) model and Jackiw-Teitelboim (JT) gravity. Here we will try to explain an alternative history such that the gravitational path integral was taken seriously and the relevant effect was discovered without this detour.

\smallskip

Using the example of the particle moving on a loop from Lecture 3 the approach should be clear. We need to expand the gravitational action to quadratic order in 
\begin{equation}
    g_{\mu\nu}
    \to
    g_{\mu\nu}
    +
    h_{\mu\nu},
\end{equation}
where $h_{\mu\nu}$ represents the metric fluctuations, and setup the eigenvalue problem for the resulting differential operator. The eigenvalues will tell us which fluctuations are suppressed and which are not. The unsuppressed fluctuations will be the source of the enhanced quantum effects that we need to understand.

\smallskip

Trying to do this faces two issues that we will not try to address in these lectures. The first is the gauge fixing of diffeomorphisms. This can be done by a standard Faddeev-Popov procedure. Nothing exciting arises in this sector. The second is the conformal mode problem. For example in pure gravity, modes of the form $h_{\mu\nu}= h g_{\mu\nu}$ have a wrong-sign kinetic term and lead to an unbounded action. One can rotate the integral over $h$ to its imaginary steepest-descent contour. It was verified in \cite{Shi:2025gpi} that this procedure does not introduce spurious phases in the gravitational path integral.\footnote{This is famously not the case in spacetimes with positive cosmological constant \cite{Polchinski:1988ua}. Maldacena proposed that including an observer mostly cancels the de Sitter phase, leaving an overall minus sign \cite{Maldacena:2024spf,Chen:2025dSPhase}.}

\smallskip 

In pure gravity, we can focus on traceless modes $\phi_{\mu\nu}$ in the transverse gauge $\nabla_\mu \phi^{\mu\nu}=0$. The quadratic action for these modes is given by the Lichnerowicz operator
\begin{equation*}
    (\Delta_L h)_{\mu\nu}
    =
    -\nabla^2 h_{\mu\nu}
    -
    2R_{\mu\rho\nu\sigma}h^{\rho\sigma}.
\end{equation*}
The quadratic action of these modes is given by 
\begin{equation*}
    I_E
    =
    \frac{1}{64\pi G_N}
    \int \mathrm{d}^4x\,\sqrt{g}
        \phi^{\mu\nu}(\Delta_L\phi)_{\mu\nu}.
\end{equation*}
Short-distance modes are clearly positive since they are dominated by the Laplacian term in $\Delta_L$ and not the Riemann tensor term. Nevertheless, this sector can still have zero or negative modes. This problem was originally studied by Gross, Perry, and Yaffe \cite{Gross:1982cv}. They found a single negative mode, which is a normalizable transverse-traceless $s$-wave. As opposed to the pure trace, this modes reflects the thermodynamic instability of an asymptotically flat Schwarzschild black hole in the fixed-temperature ensemble. The specific heat of a Schwarszchild black hole is negative since as the mass increases the temperature decreases. Zero-modes are also present in the $p$-wave sector. In this case they do not represent large quantum corrections, they are simply collective coordinates for the geometry. In flat space we can translate the center of the black hole without any energy cost.

\subsubsection*{Which modes become light?}

We can repeat this analysis for the Reissner-Nordstrom black hole. We refer to \cite{Kolanowski:2024zrq} for the details and here we will quote the important aspects of the resulting spectrum. In this case we expand both the metric and vector potential
$$
g_{\mu\nu} \to g_{\mu\nu}+h_{\mu\nu},~~~~A_\mu \to A_\mu + a_\mu
$$
so $h_{\mu\nu}$ and $a_\mu$ represent quantum corrections. The fluctuation spectrum can be expanded in the transverse sphere partial waves labeled by the angular momentum $\ell$. The time dependence can also be diagonalized in terms of Fourier modes $n$ assuming that all components oscillate with frequency $2\pi n/\hbar \beta$.

\begin{figure}[t!]
    \centering
    \begin{tikzpicture}[
        scale=1.2,x=1.12cm,
        y=1.04cm,
        line cap=round,
        line join=round,
        every node/.style={font=\small}
    ]
        \def\xmax{4.35}
        \def\ymax{3.05}

        \begin{scope}
            \path[fill=blue!12]
                (0,1.92)
                .. controls (1.5,1.94) and (3.0,2.05) .. (\xmax,2.18)
                -- (\xmax,\ymax) -- (0,\ymax) -- cycle;
            \draw[blue!65!black, very thick]
                (0,1.92)
                .. controls (1.5,1.94) and (3.0,2.05) .. (\xmax,2.18);

            \draw[orange!85!black, very thick]
                (0,0.92) .. controls (1.5,1.02) and (3.0,1.20) .. (\xmax,1.38);

            \draw[thick,->] (0,0) -- (\xmax+0.28,0)
                node[below right] {$\kappa$};
            \draw[thick,->] (0,0) -- (0,\ymax+0.20)
                node[above] {$\lambda$};
            \node[blue!65!black, anchor=east, font=\footnotesize]
                at (4.18,2.36) {continuum};
            \node[orange!85!black, anchor=west]
                at (0.18,0.20) {$\lambda_j(0)>0$};
            \node[anchor=north] at (2.18,-0.62)
                {\textbf{(a)} Generic sector};
        \end{scope}

        \begin{scope}[shift={(6.05,0)}]
            \path[fill=blue!12]
                (0,1.92)
                .. controls (1.5,1.94) and (3.0,2.05) .. (\xmax,2.18)
                -- (\xmax,\ymax) -- (0,\ymax) -- cycle;
            \draw[blue!65!black, very thick]
                (0,1.92)
                .. controls (1.5,1.94) and (3.0,2.05) .. (\xmax,2.18);

            \draw[orange!85!black, very thick]
                (0,0) -- (\xmax,1.48);

            \draw[thick,->] (0,0) -- (\xmax+0.28,0)
                node[below right] {$\kappa$};
            \draw[thick,->] (0,0) -- (0,\ymax+0.20)
                node[above] {$\lambda$};
            \fill[orange!85!black] (0,0) circle (1.35pt);
            \node[blue!65!black, anchor=east, font=\footnotesize]
                at (4.18,2.34) {continuum};
            \node[orange!85!black, rotate=19, anchor=south]
                at (2.62,0.90) {$\lambda_n\propto\kappa$};
            \node[anchor=north] at (2.18,-0.62)
                {\textbf{(b)} $s$-wave, $|n|>1$};
        \end{scope}
    \end{tikzpicture}
    \caption{Schematic spectra of the coupled Einstein-Maxwell quadratic
    operator.  In a generic sector (a), the continuum has a positive
    threshold and all discrete eigenvalues approach nonzero limits as
    $\kappa\to0$.  In the exceptional $s$-wave sector (b), a representative
    mode with $|n|>1$ vanishes linearly.  The discrete branches are kept
    separate from the continuum throughout the range shown.}
    \label{fig:rn-light-spectrum}
\end{figure}
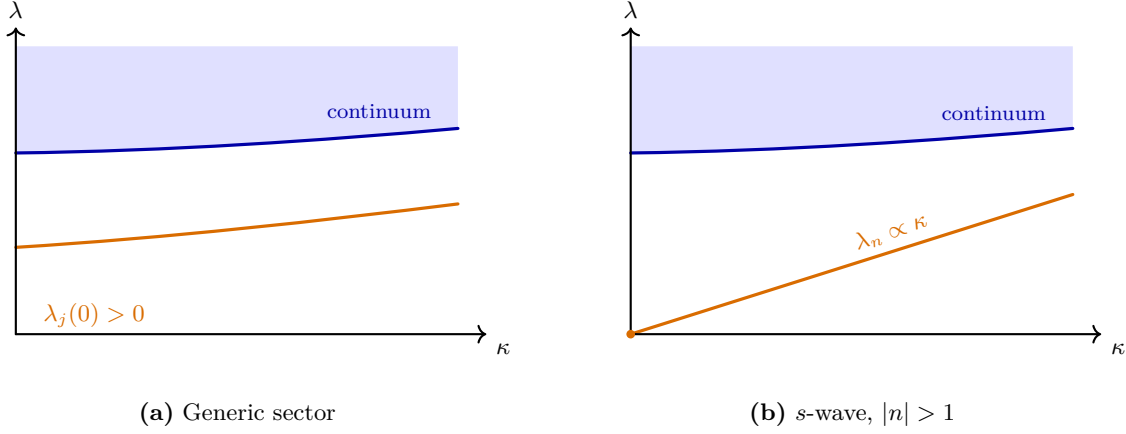

The qualitative outcome of the numerical calculation in \cite{Kolanowski:2024zrq} is schematically shown in figure~\ref{fig:rn-light-spectrum}. Generic sectors have a continuum component of the spectrum and possibly positive discrete modes that always stay above zero. For $s$-wave modes with $\ell=0$ but non-trivial frequency $|n|>1$ a discrete mode is found that becomes arbitrarily light as we decrease the surface gravity
\begin{equation}
    \lambda
    \propto
    |n|\,\kappa,
    \qquad
    |n|\geq 2,
    \qquad
    \kappa\to0 ,
\end{equation}
where $\lambda$ is the eigenvalue of the relevant differential operator for quadratic action of physical fluctuations in Einstein-Maxwell. The constant of proportionality can be found numerically but it also depends on conventions used when defining the quadratic operator. This implies the presence of modes, labeled by $|n|>1$, that become more strongly quantum mechanical as we lower the temperature.

In fact, these modes were found originally by studying SYK and AdS$_2$ gravity although their full implication for quantum gravity in 4d was not initially appreciated until \cite{Iliesiu:2020qvm}. By looking at the numerical eigenmodes of the quadratic action, we can check that as we lower $\kappa$ they become more and more concentrated near the horizon. As we saw in Lecture 1 the geometry close to the horizon is $AdS_2 \times S^2$. AdS has large diffeomorphisms which can be interpreted as fluctuations of the precise location of its boundary, the surface being glued to the exterior of the throat. The lack of $n=-1,0,1$ modes can be interpreted as being due to the $\mathrm{SL}(2,\mathbb R)$ isometries of AdS$_2$. Jackiw-Teitelboim (JT) gravity provides an exact quantization of these modes. A review of these results in 2d gravity can be found in \cite{Mertens:2022irh}. 

\smallskip 

The fact that the modes are pure diffeomorphisms in the throat might be confusing since there are no large diffeomorphisms in asymtotically flat space. The analysis of \cite{Kolanowski:2024zrq} shows that the modes in the full geometry are instead physical, and only seem to be pure gauge close to the horizon.

\smallskip

The implication of $\lambda \sim \kappa$ is the presence of modes whose normalized action is of order $T$. In fact one can check the action is precisely of order $T/E_{\text{brk}}$. Using the mnemotecnic rule introduced in Lecture 3 for zeta-function regularization we can infer the presence of a power of $(T/E_{\text{brk}})^{3/2}$ in the one-loop determinant. All other modes, either discrete or continuous, give a one-loop determinant which is finite in the $\kappa \to 0$. Up to a temperature-independent
normalization and the contribution of modes that remain heavy, the result is
\begin{equation}
    Z_Q(\beta)
    \sim
    e^{S_{\rm ext}-\beta M_{\rm ext}}\,
    \left(\frac{T}{E_{\rm brk}}\right)^{3/2}
    \exp\!\left(\frac{2\pi^2 T}{E_{\rm brk}}\right).
    \label{eq:rn-schwarzian-partition}
\end{equation}
The exponential factor reproduces the classical near-extremal entropy and energy. 

\smallskip

The derivation outlined here only guarantees that this result is valid when $T\gg E_{\text{brk}}$. For the light modes, if $T \sim E_{\text{brk}}$, higher loops could potentially become important and make the one-loop approximation invalid. One can overcome this issue by organizing the gravitaitonal path integral in a smart way: separate the integral in the throat from that in the exterior. The exterior is classical, while the throat can be rewritten as a theory of gravity in AdS$_2$ coupled to matter. Surprisingly this 2d gravity theory can be exactly quantized and used to argue that this result can be extended to $T \sim E_{\text{brk}}$ and fails at much lower temperatures (at least exponentially suppressed in the entropy). This was in fact the approach in \cite{Iliesiu:2020qvm}, that used the decomposition of the geometry that was previously worked out mostly classically, e.g. in \cite{Nayak:2018qej} or \cite{Sachdev:2019bjn}, and combined it with the quantizaiton of JT gravity in \cite{Stanford:2017thb} and \cite{Mertens:2017mtv}. In particular, \cite{Stanford:2017thb} showed that the partition function of JT is one-loop exact, while \cite{Mertens:2017mtv} derived all correlation functions which are not.

\smallskip

These results have been generalized to Kerr-Newman black holes \cite{Rakic:2023vhv}. For rotating neutral black holes this could have astrophysical implications, although $E_{\text{brk}}$ is such a small scale that it would be almost impossible to find a rotating black hole close enough to extremality. Another issue is superradiance which competes with the quantum corrections. In the rest of the section we will explain the implications to the spectrum of black hole microstates and Hawking radiation. 

\subsubsection*{Spectrum around the breakdown scale}
\label{subsec:near-extremal-spectrum}

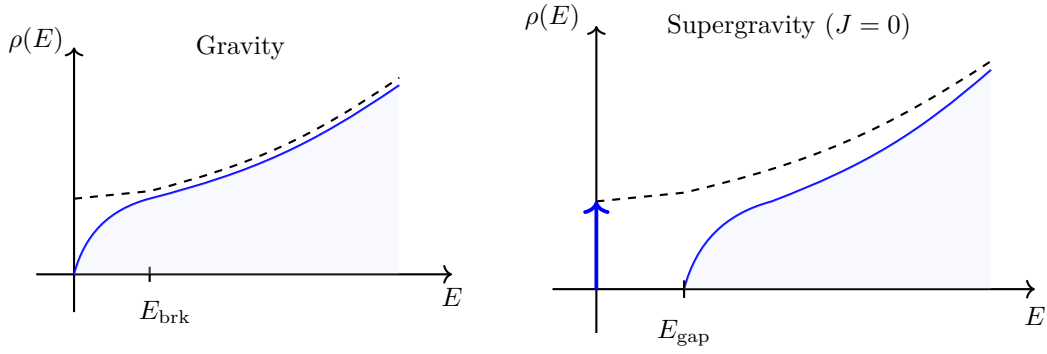
\begin{figure}[t!]
     \centering
\begin{tikzpicture}[scale=1, baseline={([yshift=0cm]current bounding box.center)}]
\draw[thick,->] (0,-0.5) -- (0,3);
\draw[thick,->] (-0.5,0) -- (5,0);
\draw[thick, blue ] (0,0) to[bend left =30] (1,1) to[bend right = 10] (4.3,2.5) ;
\draw[thick, dashed] (0,1) to[bend left =0] (1,1.1) to[bend right = 10] (4.3,2.6) ;
\node at (5,-0.3) {$E$};
\node at (2.2,3) {Gravity};
\node at (-0.5,3.1) {$\rho(E)$};
\draw[thick] (1,-0.1) -- (1,+0.1);
\node at (1.2,-0.5) {$E_{\rm brk}$};
\fill[blue!10,nearly transparent]  (0,0) to[bend left =30] (1,1) to[bend right = 10] (4.3,2.5)-- (4.3,0) -- (0,0);
\end{tikzpicture}
~~~
\begin{tikzpicture}[scale=1.16, baseline={([yshift=0cm]current bounding box.center)}]
\draw[thick,->] (0,-0.5) -- (0,3);
\draw[thick,->] (-0.5,0) -- (5,0);
\draw[thick, blue ] (1,0) to[bend left =30] (2,1) to[bend right = 10] (4.5,2.5) ;
\draw[thick, dashed] (0,1) to[bend left =0] (1,1.1) to[bend right = 10] (4.5,2.6) ;
\node at (5,-0.3) {$E$};
\node at (2.2,3) {Supergravity $(J=0)$};
\node at (-0.5,3.1) {$\rho(E)$};
\draw[thick] (1,-0.1) -- (1,+0.1);
\node at (1,-0.5) {$E_{\rm gap}$};
\fill[blue!10,nearly transparent]  (1,0) to[bend left =30] (2,1) to[bend right = 10] (4.5,2.5)-- (4.5,0) -- (0,0);
\draw[line width = .5mm, blue, ->] (0,0) -- (0,1) ;
\end{tikzpicture}
    \caption{Quantum-corrected near-extremal spectra, following the
    schematic presentation of \cite{Turiaci:2023wrh}.  Without supersymmetry
    (a), the continuum reaches the extremal edge but its density vanishes
    there; the dashed curve is the classical extrapolation.  When the
    extremal limit is supersymmetric (b), protected BPS states remain at
    $E=0$ and are separated from the near-BPS continuum by a gap.}
    \label{fig:schwarzian-density}
\end{figure}

The partition function describes canonical thermodynamics. We can recast the result in equation~\eqref{eq:rn-schwarzian-partition} in terms of the density of states as a function of energy. As a function of $E\equiv M-M_{\rm ext}$, this is given the Laplace transform
\begin{equation}
    Z_{\rm exc}(\beta)
    =
    \int_0^\infty \mathrm{d}E\,
    \rho(E)e^{-\beta E},~~~\Rightarrow~~~\rho(E)
    \propto
    \frac{e^{S_{\rm ext}}}{E_{\rm brk}}
    \sinh\!\left(2\pi\sqrt{\frac{2E}{E_{\rm brk}}}\right)
    \Theta(E).
    \label{eq:near-extremal-density}
\end{equation}
The omitted coefficient is the same temperature-independent normalization
left unspecified in the partition function, but can be absorbed in a subleading shift of $S_{\text{ext}}$.

\smallskip

As shown in figure~\ref{fig:schwarzian-density}, the quantum-corrected spectrum resolves the third law, in the sense that it makes the density of states in quantum gravity deviate from the classical analysis and actually vanish at extremality. Therefore strictly extremal black holes do not exist. 
$$
\text{Classical extremal entropy $S_{\text{ext}}$} \neq \text{Extremal degeneracy.}
$$
At very low temperatures when the $T^{3/2}$ factor overcomes the enhancement in $S_{\text{ext}}$ the gravitational path integral will get contributions from other competing configurations. This is expected since at some point we should observe the discreteness of the spectrum. In this lectures we call the classical quantity $S_{\text{ext}}$ extremal area to follow common practice, even though it is not technically correct.

\smallskip

This also resolves the isue with classical violations of the unattainability principle since there are no extremal states in quantum gravity to end up at. This is clear although it is yet an interesting problem to study the Kehle and Unger process in quantum gravity and understand how it is affected by these quantum corrections \cite{Turiaci:2026gtl}.

\smallskip

An important exception is also shown in figure~\ref{fig:schwarzian-density}. If we study black holes in supergravity that become supersymmetric at extremality, new fermionic strongly quantum-mechanical modes emerge from the gravitini. In this case the analysis is modified. The spectrum of near-supersymmetric black holes in quantum gravity was derived in \cite{Heydeman:2020hhw}. The extremal degeneracy now survives in quantum gravity, as expected from String Theory microstate counting of supersymmetric black holes. We also find a gap that can be evaluated to leading order in $G_N$ as $E_{\text{gap}}=E_{\text{brk}}/8$. 

\smallskip

The fact that the extremal degeneracy survives in supergravity is not necessarily a violation of the third law. Supersymmetry implies that the Witten index  $ \text{Tr} (-1)^{\sf F} e^{-\beta H}$ is protected and independent of the coupling and solely get contributions from supersymmetric states. Unitarity implies that
$$
\Big|\text{Tr}_{\text{susy}} (-1)^{\sf F} \Big| \leq \text{Tr}_{\text{susy}}\,( 1).
$$
If the inequality is not saturated, it means there are states that could combine into non-supersymmetric multiplets and get lifted but they do not. For black holes we can use the supergravity path integral to evaluate $\text{Tr}_{\text{susy}} (-1)^{\sf F}$, which for Reissner-Nordstr\"om was done in \cite{Iliesiu:2021}. We find that to leading order in $G_N$ the inequality is saturated. In fact this happens in all examples we know, the index always matches the entropy. That means that the extremal degeneracy is the minimal possible dictated by supersymmetry. (We could suggest a version of the third law for supergravity: the extremal degeneracy should not be larger than the index in the semiclassical approximation. If it was, it would imply the presence of a large number of states that could potentially combine and be lifted but are not.) The large degeneracy is a consequence of quantum systems with large Witten indices which can be evaluated at weak coupling. This large degeneracy might nevertheless make us worry that collapse to extremality in finite time might be physical even in quantum gravity. Fortunately, Reall recently showed that extremal black holes cannot form by collapse in supergravity \cite{Reall:2024ThirdLaw}.

\subsection{Quantum effects on Hawking radiation}
\label{subsec:quantum-hawking-radiation}

Using a similar approach we can also study how Hawking radiation is modified in quantum gravity close to extremality. In this case the quantum corrections are not one-loop exact anymore and one is forced to use the connection with JT gravity and AdS$_2$.

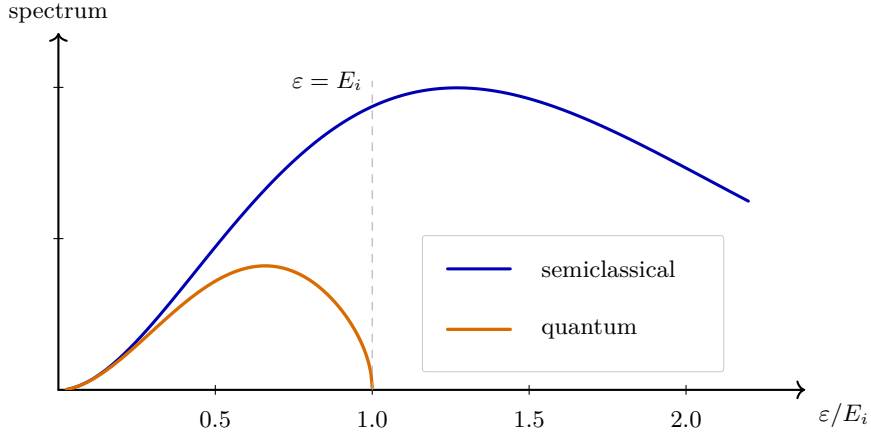
\begin{figure}[t!]
    \centering
    \begin{tikzpicture}[
        x=4.15cm,
        y=4.0cm,
        line cap=round,
        line join=round,
        every node/.style={font=\small}
    ]
        \def\xmax{2.25}
        \def\ymax{1.08}
        \def\A{4.442882938}
        \def\a{2.221441469}
        \def\scale{7.7}

        \draw[->, line width=0.8pt] (0,0) -- (2.38,0)
            node[below right=2pt] {$\varepsilon/E_i$};
        \draw[->, line width=0.8pt] (0,0) -- (0,1.18)
            node[above] {spectrum};

        \foreach \xt in {0.5,1.0,1.5,2.0} {
            \draw (\xt,0.015) -- (\xt,-0.015)
                node[below=3pt] {$\xt$};
        }
        \foreach \yt in {0.5,1.0} {
            \draw (0.015,\yt) -- (-0.015,\yt);
        }

        \begin{scope}
            \clip (0,0) rectangle (\xmax,\ymax);
            \draw[blue!70!black, line width=1.15pt]
                plot[domain=0.025:2.20, samples=260, smooth]
                ({\x},
                 {\scale*(\x*\x*\x)/(exp(\a*\x)-1)});
            \draw[orange!85!black, line width=1.25pt]
                plot[domain=0.025:1.0, samples=280, smooth]
                ({\x},
                 {\scale*(\x*\x*\x)*
                 sinh(\A*sqrt(max(0,1-\x)))/
                 (cosh(\A)-cosh(\A*sqrt(max(0,1-\x))))});
        \end{scope}

        \draw[gray!65, dashed] (1,0) -- (1,1.02);
        \node[anchor=south east, fill=white, inner sep=1pt]
            at (0.98,0.98) {$\varepsilon=E_i$};

        \begin{scope}[shift={(1.24,0.18)}]
            \draw[fill=white, draw=gray!40, rounded corners=1pt]
                (-0.08,-0.12) rectangle (0.88,0.33);
            \draw[blue!70!black, line width=1.15pt]
                (0,0.22) -- (0.20,0.22);
            \node[anchor=west] at (0.27,0.22) {semiclassical};
            \draw[orange!85!black, line width=1.25pt]
                (0,0.02) -- (0.20,0.02);
            \node[anchor=west] at (0.27,0.02) {quantum};
        \end{scope}
    \end{tikzpicture}
    \caption{Semiclassical and quantum-corrected scalar spectra for the
    illustrative value $E_i/E_{\rm brk}=\pi^2$, following
    \cite{Brown:2024ajk}.  The vertical normalization is arbitrary.  The
    Schwarzian-dressed result is nonthermal and has no support for
    $\varepsilon>E_i$, while the canonical Hawking approximation has an
    unphysical tail beyond the energy available above extremality.}
    \label{fig:quantum-hawking-spectrum}
\end{figure}

It has been known for a while how to address this calculation in classical gravity. We can separate the throat from the exterior. Then the Hawking spectrum is a specific Fourier transform of a matter two-point function in AdS$_2$ \cite{Maldacena:1997ih}. Using the classical gravity two-point function reproduces the graybody factor of a near-extremal black hole. In quantum gravity, we need to use the two-point functions computed by quantizing JT gravity for arbitrary temperatures, but still using the leading topology. This was done for all correlators in \cite{Mertens:2017mtv}, see \cite{Mertens:2022irh} for a review. Recently, \cite{Brown:2024ajk} developed the relation between Hawking spectrum and two-point function of \cite{Maldacena:1997ih} more precisely, and used this technique to find explicitly the spectrum of Hawking radiation from near-extremal black holes. They obtained
\begin{equation}
    \frac{\mathrm{d}^2E}
         {\mathrm{d}t\,\mathrm{d}\varepsilon}
    =
    \frac{2\varepsilon}{\pi\hbar}\,
    \left(\frac{r_+\varepsilon}{\hbar}\right)^2
    \frac{
        \sinh\!\left(
            2\pi\sqrt{2(E_i-\varepsilon)/E_{\rm brk}}
        \right)
    }{
        \cosh\!\left(2\pi\sqrt{2E_i/E_{\rm brk}}\right)
        -
        \cosh\!\left(
           2\pi \sqrt{2(E_i-\varepsilon)/E_{\rm brk}}
        \right)
    }
    \Theta(E_i-\varepsilon),
    \label{eq:quantum-hawking-spectrum}
\end{equation}
where $\varepsilon = \hbar \omega$ is the energy of the Hawking quanta of frequency $\omega$. The expression in equation~\eqref{eq:quantum-hawking-spectrum} is the quantum Schwarzian
replacement of the fixed-background Hawking spectrum. When the initial energy of the black hole is large, this formula reproduces the black body spectrum derived in Lecture 2. When the initial energy of the black hole is comparable or lower to the breakdown scale $E_{\text{bkr}}$ we obtain the orange curve in figure~\ref{fig:quantum-hawking-spectrum}, which shows that the average energy of Hawking radiation is much smaller than its semiclassical expectation. Moreover no Hawking quantum is emitted with energy large enough to drive the black hole below extremality.

A low-energy effective description leads to the
same microcanonical picture and can follow the evolving state during
evaporation \cite{Biggs:2025evap}. One can also translate this result to a prediction for the black hole absorption cross section in the quantum gravity regime \cite{Emparan:2025sao}. Besides the emission of Hawking radiation, \cite{Brown:2024ajk} also revisited the Schwinger discharge process in quantum gravity. In this case, the discharge rate is not strongly affected by quantum gravity corrections.


\allowdisplaybreaks
\minitoc


\newpage
\thispagestyle{plain}
\addcontentsline{toc}{chapter}{\nameref{sec:Acknowledgments}}
\section*{Acknowledgments} \label{sec:Acknowledgments}

%
For this first School on Gravity, we are thankful to many people, but especially to all the participants who embraced the ``Copenhagen spirit'' of fusing science, life and creativity~\cite{Joas2022CopenhagenSpirit} and, through innumerable questions, engaging conversations and multiple ice creams, made this a unique event.  
The Center of Gravity is a Center of Excellence funded by the Danish National Research Foundation under grant No. DNRF184.
We acknowledge support by VILLUM Foundation (grant no.\ VIL37766 and no.~VIL53101).
V.C. acknowledges financial support provided under the European Union’s H2020 ERC Advanced Grant “Black holes: gravitational engines of discovery” grant agreement no.\ Gravitas–101052587.
Views and opinions expressed are however those of the author only and do not necessarily reflect those of the European Union or the European Research Council. Neither the European Union nor the granting authority can be held responsible for them.
J.~M.~E is also supported by the Marie Sklodowska-Curie Grant Agreement No.~847523 INTERACTIONS.
The work of J.P.-M. was supported by the European Research
Council (ERC) under the European Union’s Horizon 2020 research and innovation programme, grant agreement ERC-StG-101221094 GravitaS.  J.P-.M acknowledges the use of AI assistants (Claude) for figure generation and proofreading.
L.S. is supported by a University of Nottingham Anne McLaren Fellowship. L.S. acknowledges the use of AI assistant Claude for tutorial set up, figure generation and proofreading. 
GJT is supported by the DOE Early Career Award DE-SC0026287. He used the OpenAI academic subscription, mostly to prepare the figures. 
%
LvS is supported by VI.Veni.242.115, Grant ID \url{https://doi.org/10.61686/XVIAV86753}. 
This project has received funding from the European Union's Horizon 2020 research and innovation programme under the Marie Sk{\l}odowska-Curie grant agreement No 101007855 and No 101131233.
The Tycho supercomputer hosted at the SCIENCE HPC center at the University of Copenhagen was used for supporting this work.
This work is supported by Simons Foundation International \cite{sfi} and the Simons Foundation \cite{sf} through Simons Foundation grant SFI-MPS-BH-00012593-11.
%


\let\i\dotlessitext 
\bibliography{bh_biblio,lieke_references,julio_refs}

\providecommand{\href}[2]{#2}\begingroup\raggedright\begin{thebibliography}{100}

\bibitem{Stanway2016}
E.R.~{Stanway}, J.J.~{Eldridge} and G.D.~{Becker}, \emph{{Stellar population
  effects on the inferred photon density at reionization}},
  \href{https://doi.org/10.1093/mnras/stv2661}{\emph{\mnras} {\bfseries 456}
  (2016) 485} [\href{https://arxiv.org/abs/1511.03268}{{\ttfamily
  1511.03268}}].

\bibitem{Gotberg2020}
Y.~{G{\"o}tberg}, S.E.~{de Mink}, M.~{McQuinn}, E.~{Zapartas}, J.H.~{Groh} and
  C.~{Norman}, \emph{{Contribution from stars stripped in binaries to cosmic
  reionization of hydrogen and helium}},
  \href{https://doi.org/10.1051/0004-6361/201936669}{\emph{\aap} {\bfseries
  634} (2020) A134} [\href{https://arxiv.org/abs/1911.00543}{{\ttfamily
  1911.00543}}].

\bibitem{Hopkins2012}
P.F.~{Hopkins}, E.~{Quataert} and N.~{Murray}, \emph{{Stellar feedback in
  galaxies and the origin of galaxy-scale winds}},
  \href{https://doi.org/10.1111/j.1365-2966.2012.20593.x}{\emph{\mnras}
  {\bfseries 421} (2012) 3522}
  [\href{https://arxiv.org/abs/1110.4638}{{\ttfamily 1110.4638}}].

\bibitem{Agertz2013}
O.~{Agertz}, A.V.~{Kravtsov}, S.N.~{Leitner} and N.Y.~{Gnedin}, \emph{{Toward a
  Complete Accounting of Energy and Momentum from Stellar Feedback in Galaxy
  Formation Simulations}},
  \href{https://doi.org/10.1088/0004-637X/770/1/25}{\emph{\apj} {\bfseries 770}
  (2013) 25} [\href{https://arxiv.org/abs/1210.4957}{{\ttfamily 1210.4957}}].

\bibitem{Sarbadhicary2022}
S.K.~{Sarbadhicary}, D.~{Martizzi}, E.~{Ramirez-Ruiz}, E.~{Koch},
  K.~{Auchettl}, C.~{Badenes} et~al., \emph{{Testing the Momentum-driven
  Supernova Feedback Paradigm in M31}},
  \href{https://doi.org/10.3847/1538-4357/ac3094}{\emph{\apj} {\bfseries 928}
  (2022) 54} [\href{https://arxiv.org/abs/2208.11055}{{\ttfamily 2208.11055}}].

\bibitem{Maiolino2019}
R.~{Maiolino} and F.~{Mannucci}, \emph{{De re metallica: the cosmic chemical
  evolution of galaxies}},
  \href{https://doi.org/10.1007/s00159-018-0112-2}{\emph{\aapr} {\bfseries 27}
  (2019) 3} [\href{https://arxiv.org/abs/1811.09642}{{\ttfamily 1811.09642}}].

\bibitem{Kitadai2017}
N.~{Kitadai} and S.~{Maruyama}, \emph{{Origins of building blocks of life: A
  review}}, \href{https://doi.org/10.1016/j.gsf.2017.07.007}{\emph{Geoscience
  Frontiers} {\bfseries 9} (2018) 1117}.

\bibitem{Vink2022}
J.S.~{Vink}, \emph{{Theory and Diagnostics of Hot Star Mass Loss}},
  \href{https://doi.org/10.1146/annurev-astro-052920-094949}{\emph{\araa}
  {\bfseries 60} (2022) 203}
  [\href{https://arxiv.org/abs/2109.08164}{{\ttfamily 2109.08164}}].

\bibitem{LucySolomon1970}
L.B.~{Lucy} and P.M.~{Solomon}, \emph{{Mass Loss by Hot Stars}},
  \href{https://doi.org/10.1086/150365}{\emph{\apj} {\bfseries 159} (1970)
  879}.

\bibitem{CAK1975}
J.I.~{Castor}, D.C.~{Abbott} and R.I.~{Klein}, \emph{{Radiation-driven winds in
  Of stars.}}, \href{https://doi.org/10.1086/153315}{\emph{\apj} {\bfseries
  195} (1975) 157}.

\bibitem{Vink2001}
J.S.~{Vink}, A.~{de Koter} and H.J.G.L.M.~{Lamers}, \emph{{Mass-loss
  predictions for O and B stars as a function of metallicity}},
  \href{https://doi.org/10.1051/0004-6361:20010127}{\emph{\aap} {\bfseries 369}
  (2001) 574} [\href{https://arxiv.org/abs/astro-ph/0101509}{{\ttfamily
  astro-ph/0101509}}].

\bibitem{2026enap....2..569S}
T.~{Shenar}, \emph{{Wolf-Rayet stars}},  in \emph{Encyclopedia of Astrophysics,
  Volume 2}, vol.~2, pp.~569--583, Jan., 2026,
  \href{https://doi.org/10.1016/B978-0-443-21439-4.00031-6}{DOI}
  [\href{https://arxiv.org/abs/2410.04436}{{\ttfamily 2410.04436}}].

\bibitem{Vink2017}
J.S.~{Vink}, \emph{{Winds from stripped low-mass helium stars and Wolf-Rayet
  stars}}, \href{https://doi.org/10.1051/0004-6361/201731902}{\emph{\aap}
  {\bfseries 607} (2017) L8}
  [\href{https://arxiv.org/abs/1710.02010}{{\ttfamily 1710.02010}}].

\bibitem{SanderVink2020}
A.A.C.~{Sander} and J.S.~{Vink}, \emph{{On the nature of massive helium star
  winds and Wolf-Rayet-type mass-loss}},
  \href{https://doi.org/10.1093/mnras/staa2712}{\emph{\mnras} {\bfseries 499}
  (2020) 873} [\href{https://arxiv.org/abs/2009.01849}{{\ttfamily
  2009.01849}}].

\bibitem{2021ARA&A..59..337D}
L.~{Decin}, \emph{{Evolution and Mass Loss of Cool Ageing Stars: a Daedalean
  Story}},
  \href{https://doi.org/10.1146/annurev-astro-090120-033712}{\emph{\araa}
  {\bfseries 59} (2021) 337}
  [\href{https://arxiv.org/abs/2011.13472}{{\ttfamily 2011.13472}}].

\bibitem{2018MNRAS.475...55B}
E.R.~{Beasor} and B.~{Davies}, \emph{{The evolution of red supergiant mass-loss
  rates}}, \href{https://doi.org/10.1093/mnras/stx3174}{\emph{\mnras}
  {\bfseries 475} (2018) 55}
  [\href{https://arxiv.org/abs/1712.01852}{{\ttfamily 1712.01852}}].

\bibitem{2020MNRAS.492.5994B}
E.R.~{Beasor}, B.~{Davies}, N.~{Smith}, J.T.~{van Loon}, R.D.~{Gehrz} and
  D.F.~{Figer}, \emph{{A new mass-loss rate prescription for red supergiants}},
  \href{https://doi.org/10.1093/mnras/staa255}{\emph{\mnras} {\bfseries 492}
  (2020) 5994} [\href{https://arxiv.org/abs/2001.07222}{{\ttfamily
  2001.07222}}].

\bibitem{2022ApJ...933...41B}
E.R.~{Beasor} and N.~{Smith}, \emph{{The Extreme Scarcity of Dust-enshrouded
  Red Supergiants: Consequences for Producing Stripped Stars via Winds}},
  \href{https://doi.org/10.3847/1538-4357/ac6dcf}{\emph{\apj} {\bfseries 933}
  (2022) 41} [\href{https://arxiv.org/abs/2205.02207}{{\ttfamily 2205.02207}}].

\bibitem{HumphreysDavidson1979}
R.M.~{Humphreys} and K.~{Davidson}, \emph{{Studies of luminous stars in nearby
  galaxies. III. Comments on the evolution of the most massive stars in the
  Milky Way and the Large Magellanic Cloud.}},
  \href{https://doi.org/10.1086/157301}{\emph{\apj} {\bfseries 232} (1979)
  409}.

\bibitem{Smith2025LBVreview}
N.~{Smith}, \emph{{Luminous blue variables}},  in \emph{Encyclopedia of
  Astrophysics, Volume 2}, vol.~2, pp.~508--532, Jan., 2026,
  \href{https://doi.org/10.1016/B978-0-443-21439-4.00147-4}{DOI}
  [\href{https://arxiv.org/abs/2509.22990}{{\ttfamily 2509.22990}}].

\bibitem{Spera2015}
M.~{Spera}, M.~{Mapelli} and A.~{Bressan}, \emph{{The mass spectrum of compact
  remnants from the PARSEC stellar evolution tracks}},
  \href{https://doi.org/10.1093/mnras/stv1161}{\emph{\mnras} {\bfseries 451}
  (2015) 4086} [\href{https://arxiv.org/abs/1505.05201}{{\ttfamily
  1505.05201}}].

\bibitem{SperaMapelli2017}
M.~{Spera} and M.~{Mapelli}, \emph{{Very massive stars, pair-instability
  supernovae and intermediate-mass black holes with the sevn code}},
  \href{https://doi.org/10.1093/mnras/stx1576}{\emph{\mnras} {\bfseries 470}
  (2017) 4739} [\href{https://arxiv.org/abs/1706.06109}{{\ttfamily
  1706.06109}}].

\bibitem{Vink2021}
J.S.~{Vink}, E.R.~{Higgins}, A.A.C.~{Sander} and G.N.~{Sabhahit},
  \emph{{Maximum black hole mass across cosmic time}},
  \href{https://doi.org/10.1093/mnras/stab842}{\emph{\mnras} {\bfseries 504}
  (2021) 146} [\href{https://arxiv.org/abs/2010.11730}{{\ttfamily
  2010.11730}}].

\bibitem{Merritt2026}
J.~{Merritt}, S.~{Stevenson}, A.~{Sander}, I.~{Mandel}, J.~{Riley}, B.~{Farr}
  et~al., \emph{{Implications of Modern Mass-loss Rates for Massive Stars}},
  \href{https://doi.org/10.3847/1538-4357/ae75fd}{\emph{\apj} {\bfseries 1005}
  (2026) 79}.

\bibitem{FishbachvanSon2023}
M.~{Fishbach} and L.~{van Son}, \emph{{LIGO-Virgo-KAGRA's Oldest Black Holes:
  Probing Star Formation at Cosmic Noon With GWTC-3}},
  \href{https://doi.org/10.3847/2041-8213/ad0560}{\emph{\apjl} {\bfseries 957}
  (2023) L31} [\href{https://arxiv.org/abs/2307.15824}{{\ttfamily
  2307.15824}}].

\bibitem{Zahn1975}
J.-P.~{Zahn}, \emph{{Rotational Instabilities and Stellar Evolution}},  in
  \emph{Stellar Instability and Evolution}, P.~{Ledoux}, A.~{Noels} and
  A.W.~{Rodgers}, eds., vol.~59 of \emph{IAU Symposium}, p.~185, Jan., 1974.

\bibitem{ChaboyerZahn1992}
B.~{Chaboyer} and J.-P.~{Zahn}, \emph{{Effect of horizontal turbulent diffusion
  on transport by meridional circulation.}}, {\emph{\aap} {\bfseries 253}
  (1992) 173}.

\bibitem{MeynetMaeder1997}
G.~{Meynet} and A.~{Maeder}, \emph{{Stellar evolution with rotation. I. The
  computational method and the inhibiting effect of the
  {\ensuremath{\mu}}-gradient.}}, {\emph{\aap} {\bfseries 321} (1997) 465}.

\bibitem{Rieutord2025}
M.~{Rieutord}, E.~{Brossier-S{\'e}cher} and J.S.G.~{Mombarg}, \emph{{Angular
  momentum relaxation in models of rotating early-type stars}},
  \href{https://doi.org/10.1051/0004-6361/202555003}{\emph{\aap} {\bfseries
  699} (2025) A146} [\href{https://arxiv.org/abs/2505.10357}{{\ttfamily
  2505.10357}}].

\bibitem{1925Obs....48...73E}
A.S.~{Eddington}, \emph{{Circulating currents in rotating stars}}, {\emph{The
  Observatory} {\bfseries 48} (1925) 73}.

\bibitem{1950MNRAS.110..548S}
P.A.~{Sweet}, \emph{{The importance of rotation in stellar evolution}},
  \href{https://doi.org/10.1093/mnras/110.6.548}{\emph{\mnras} {\bfseries 110}
  (1950) 548}.

\bibitem{2006A&A...451.1025R}
M.~{Rieutord}, \emph{{The dynamics of the radiative envelope of rapidly
  rotating stars. I. A spherical Boussinesq model}},
  \href{https://doi.org/10.1051/0004-6361:20054433}{\emph{\aap} {\bfseries 451}
  (2006) 1025} [\href{https://arxiv.org/abs/astro-ph/0602040}{{\ttfamily
  astro-ph/0602040}}].

\bibitem{2009A&A...495..271D}
T.~{Decressin}, S.~{Mathis}, A.~{Palacios}, L.~{Siess}, S.~{Talon},
  C.~{Charbonnel} et~al., \emph{{Diagnoses to unravel secular hydrodynamical
  processes in rotating main sequence stars}},
  \href{https://doi.org/10.1051/0004-6361:200810665}{\emph{\aap} {\bfseries
  495} (2009) 271} [\href{https://arxiv.org/abs/0812.0363}{{\ttfamily
  0812.0363}}].

\bibitem{Brott2011}
I.~{Brott}, S.E.~{de Mink}, M.~{Cantiello}, N.~{Langer}, A.~{de Koter},
  C.J.~{Evans} et~al., \emph{{Rotating massive main-sequence stars. I. Grids of
  evolutionary models and isochrones}},
  \href{https://doi.org/10.1051/0004-6361/201016113}{\emph{\aap} {\bfseries
  530} (2011) A115} [\href{https://arxiv.org/abs/1102.0530}{{\ttfamily
  1102.0530}}].

\bibitem{deMinkMandel2016}
S.E.~{de Mink} and I.~{Mandel}, \emph{{The chemically homogeneous evolutionary
  channel for binary black hole mergers: rates and properties of
  gravitational-wave events detectable by advanced LIGO}},
  \href{https://doi.org/10.1093/mnras/stw1219}{\emph{\mnras} {\bfseries 460}
  (2016) 3545} [\href{https://arxiv.org/abs/1603.02291}{{\ttfamily
  1603.02291}}].

\bibitem{MandeldeMink2016}
I.~{Mandel} and S.E.~{de Mink}, \emph{{Merging binary black holes formed
  through chemically homogeneous evolution in short-period stellar binaries}},
  \href{https://doi.org/10.1093/mnras/stw379}{\emph{\mnras} {\bfseries 458}
  (2016) 2634} [\href{https://arxiv.org/abs/1601.00007}{{\ttfamily
  1601.00007}}].

\bibitem{Marchant2016}
P.~{Marchant}, N.~{Langer}, P.~{Podsiadlowski}, T.M.~{Tauris} and
  T.J.~{Moriya}, \emph{{A new route towards merging massive black holes}},
  \href{https://doi.org/10.1051/0004-6361/201628133}{\emph{\aap} {\bfseries
  588} (2016) A50} [\href{https://arxiv.org/abs/1601.03718}{{\ttfamily
  1601.03718}}].

\bibitem{2005A&A...443..643Y}
S.-C.~{Yoon} and N.~{Langer}, \emph{{Evolution of rapidly rotating metal-poor
  massive stars towards gamma-ray bursts}},
  \href{https://doi.org/10.1051/0004-6361:20054030}{\emph{\aap} {\bfseries 443}
  (2005) 643} [\href{https://arxiv.org/abs/astro-ph/0508242}{{\ttfamily
  astro-ph/0508242}}].

\bibitem{2006ApJ...637..914W}
S.E.~{Woosley} and A.~{Heger}, \emph{{The Progenitor Stars of Gamma-Ray
  Bursts}}, \href{https://doi.org/10.1086/498500}{\emph{\apj} {\bfseries 637}
  (2006) 914} [\href{https://arxiv.org/abs/astro-ph/0508175}{{\ttfamily
  astro-ph/0508175}}].

\bibitem{2006A&A...460..199Y}
S.-C.~{Yoon}, N.~{Langer} and C.~{Norman}, \emph{{Single star progenitors of
  long gamma-ray bursts. I. Model grids and redshift dependent GRB rate}},
  \href{https://doi.org/10.1051/0004-6361:20065912}{\emph{\aap} {\bfseries 460}
  (2006) 199} [\href{https://arxiv.org/abs/astro-ph/0606637}{{\ttfamily
  astro-ph/0606637}}].

\bibitem{2019A&A...623A...8K}
B.~{Kub{\'a}tov{\'a}}, D.~{Sz{\'e}csi}, A.A.C.~{Sander}, J.~{Kub{\'a}t},
  F.~{Tramper}, J.~{Krti{\v{c}}ka} et~al., \emph{{Low-metallicity massive
  single stars with rotation. II. Predicting spectra and spectral classes of
  chemically homogeneously evolving stars}},
  \href{https://doi.org/10.1051/0004-6361/201834360}{\emph{\aap} {\bfseries
  623} (2019) A8} [\href{https://arxiv.org/abs/1810.01267}{{\ttfamily
  1810.01267}}].

\bibitem{Spruit2002}
H.C.~{Spruit}, \emph{Dynamo action by differential rotation in a stably
  stratified stellar interior}, {\emph{\aap} {\bfseries 381} (2002) 923}.

\bibitem{Suijs2008}
M.P.L.~{Suijs}, N.~{Langer}, A.-J.~{Poelarends}, S.-C.~{Yoon}, A.~{Heger} and
  F.~{Herwig}, \emph{{White dwarf spins from low-mass stellar evolution
  models}}, \href{https://doi.org/10.1051/0004-6361:200809411}{\emph{\aap}
  {\bfseries 481} (2008) L87}
  [\href{https://arxiv.org/abs/0802.3286}{{\ttfamily 0802.3286}}].

\bibitem{2012Natur.481...55B}
P.G.~{Beck}, J.~{Montalban}, T.~{Kallinger}, J.~{De Ridder}, C.~{Aerts},
  R.A.~{Garc{\'\i}a} et~al., \emph{{Fast core rotation in red-giant stars as
  revealed by gravity-dominated mixed modes}},
  \href{https://doi.org/10.1038/nature10612}{\emph{\nat} {\bfseries 481} (2012)
  55} [\href{https://arxiv.org/abs/1112.2825}{{\ttfamily 1112.2825}}].

\bibitem{2012A&A...548A..10M}
B.~{Mosser}, M.J.~{Goupil}, K.~{Belkacem}, J.P.~{Marques}, P.G.~{Beck},
  S.~{Bloemen} et~al., \emph{{Spin down of the core rotation in red giants}},
  \href{https://doi.org/10.1051/0004-6361/201220106}{\emph{\aap} {\bfseries
  548} (2012) A10} [\href{https://arxiv.org/abs/1209.3336}{{\ttfamily
  1209.3336}}].

\bibitem{2012ApJ...756...19D}
S.~{Deheuvels}, R.A.~{Garc{\'\i}a}, W.J.~{Chaplin}, S.~{Basu}, H.M.~{Antia},
  T.~{Appourchaux} et~al., \emph{{Seismic Evidence for a Rapidly Rotating Core
  in a Lower-giant-branch Star Observed with Kepler}},
  \href{https://doi.org/10.1088/0004-637X/756/1/19}{\emph{\apj} {\bfseries 756}
  (2012) 19} [\href{https://arxiv.org/abs/1206.3312}{{\ttfamily 1206.3312}}].

\bibitem{2014A&A...564A..27D}
S.~{Deheuvels}, G.~{Do{\u{g}}an}, M.J.~{Goupil}, T.~{Appourchaux},
  O.~{Benomar}, H.~{Bruntt} et~al., \emph{{Seismic constraints on the radial
  dependence of the internal rotation profiles of six Kepler subgiants and
  young red giants}},
  \href{https://doi.org/10.1051/0004-6361/201322779}{\emph{\aap} {\bfseries
  564} (2014) A27} [\href{https://arxiv.org/abs/1401.3096}{{\ttfamily
  1401.3096}}].

\bibitem{Fuller2019}
J.~{Fuller}, A.L.~{Piro} and A.S.~{Jermyn}, \emph{{Slowing the spins of stellar
  cores}}, \href{https://doi.org/10.1093/mnras/stz514}{\emph{\mnras} {\bfseries
  485} (2019) 3661} [\href{https://arxiv.org/abs/1902.08227}{{\ttfamily
  1902.08227}}].

\bibitem{FullerMa2019}
J.~{Fuller} and L.~{Ma}, \emph{{Most Black Holes Are Born Very Slowly
  Rotating}}, \href{https://doi.org/10.3847/2041-8213/ab339b}{\emph{\apjl}
  {\bfseries 881} (2019) L1}
  [\href{https://arxiv.org/abs/1907.03714}{{\ttfamily 1907.03714}}].

\bibitem{GWTC5_pop}
{The LIGO Scientific Collaboration}, {the Virgo Collaboration}, {the KAGRA
  Collaboration}, A.G.~{Abac}, A.~{Abe}, I.~{Abouelfettouh} et~al.,
  \emph{{GWTC-5.0: Population Properties of Merging Compact Binaries}},
  \href{https://doi.org/10.48550/arXiv.2605.27226}{\emph{arXiv e-prints} (2026)
  arXiv:2605.27226} [\href{https://arxiv.org/abs/2605.27226}{{\ttfamily
  2605.27226}}].

\bibitem{FishbachKalogera2022}
M.~{Fishbach} and V.~{Kalogera}, \emph{{Apples and Oranges: Comparing Black
  Holes in X-Ray Binaries and Gravitational-wave Sources}},
  \href{https://doi.org/10.3847/2041-8213/ac64a5}{\emph{\apjl} {\bfseries 929}
  (2022) L26} [\href{https://arxiv.org/abs/2111.02935}{{\ttfamily
  2111.02935}}].

\bibitem{2024A&A...690A..21B}
K.~{Belczynski}, C.~{Done}, S.~{Hagen}, J.-P.~{Lasota} and K.~{Sen},
  \emph{{Common origin for black holes in both high mass X-ray binaries and
  gravitational-wave sources}},
  \href{https://doi.org/10.1051/0004-6361/202450229}{\emph{\aap} {\bfseries
  690} (2024) A21} [\href{https://arxiv.org/abs/2111.09401}{{\ttfamily
  2111.09401}}].

\bibitem{2024ApJ...962..101Z}
A.A.~{Zdziarski}, S.~{Banerjee}, S.~{Chand}, G.~{Dewangan}, R.~{Misra},
  M.~{Szanecki} et~al., \emph{{Black Hole Spin Measurements in LMC X-1 and Cyg
  X-1 Are Highly Model Dependent}},
  \href{https://doi.org/10.3847/1538-4357/ad1b60}{\emph{\apj} {\bfseries 962}
  (2024) 101} [\href{https://arxiv.org/abs/2308.06167}{{\ttfamily
  2308.06167}}].

\bibitem{2024ApJ...967L...9Z}
A.A.~{Zdziarski}, S.~{Chand}, S.~{Banerjee}, M.~{Szanecki}, A.~{Janiuk},
  P.~{Lubi{\'n}ski} et~al., \emph{{What Is the Black Hole Spin in Cyg X-1?}},
  \href{https://doi.org/10.3847/2041-8213/ad43ed}{\emph{\apjl} {\bfseries 967}
  (2024) L9} [\href{https://arxiv.org/abs/2402.12325}{{\ttfamily 2402.12325}}].

\bibitem{Sana2012}
H.~{Sana} et~al., \emph{Binary interaction dominates the evolution of massive
  stars}, {\emph{Science} {\bfseries 337} (2012) 444}.

\bibitem{Offner2022}
S.S.R.~{Offner}, M.~{Moe}, K.M.~{Kratter}, S.I.~{Sadavoy}, E.L.N.~{Jensen} and
  J.J.~{Tobin}, \emph{{The Origin and Evolution of Multiple Star Systems}},  in
  \emph{Protostars and Planets VII}, S.~{Inutsuka}, Y.~{Aikawa}, T.~{Muto},
  K.~{Tomida} and M.~{Tamura}, eds., vol.~534 of \emph{Astronomical Society of
  the Pacific Conference Series}, p.~275, ASP, July, 2023,
  \href{https://doi.org/10.48550/arXiv.2203.10066}{DOI}
  [\href{https://arxiv.org/abs/2203.10066}{{\ttfamily 2203.10066}}].

\bibitem{Farag2020}
E.~{Farag}, F.X.~{Timmes}, M.~{Taylor}, K.M.~{Patton} and R.~{Farmer},
  \emph{{On Stellar Evolution in a Neutrino Hertzsprung-Russell Diagram}},
  \href{https://doi.org/10.3847/1538-4357/ab7f2c}{\emph{\apj} {\bfseries 893}
  (2020) 133} [\href{https://arxiv.org/abs/2003.05844}{{\ttfamily
  2003.05844}}].

\bibitem{Laplace2021}
E.~{Laplace}, S.~{Justham}, M.~{Renzo}, Y.~{G{\"o}tberg}, R.~{Farmer},
  D.~{Vartanyan} et~al., \emph{{Different to the core: The pre-supernova
  structures of massive single and binary-stripped stars}},
  \href{https://doi.org/10.1051/0004-6361/202140506}{\emph{\aap} {\bfseries
  656} (2021) A58} [\href{https://arxiv.org/abs/2102.05036}{{\ttfamily
  2102.05036}}].

\bibitem{Janka2012}
H.-T.~{Janka}, \emph{{Explosion Mechanisms of Core-Collapse Supernovae}},
  \href{https://doi.org/10.1146/annurev-nucl-102711-094901}{\emph{Annual Review
  of Nuclear and Particle Science} {\bfseries 62} (2012) 407}
  [\href{https://arxiv.org/abs/1206.2503}{{\ttfamily 1206.2503}}].

\bibitem{Muller2016}
B.~{M{\"u}ller}, \emph{{The Status of Multi-Dimensional Core-Collapse Supernova
  Models}}, \href{https://doi.org/10.1017/pasa.2016.40}{\emph{\pasa} {\bfseries
  33} (2016) e048} [\href{https://arxiv.org/abs/1608.03274}{{\ttfamily
  1608.03274}}].

\bibitem{BurrowsVartanyan2021}
A.~{Burrows} and D.~{Vartanyan}, \emph{{Core-collapse supernova explosion
  theory}}, \href{https://doi.org/10.1038/s41586-020-03059-w}{\emph{\nat}
  {\bfseries 589} (2021) 29}
  [\href{https://arxiv.org/abs/2009.14157}{{\ttfamily 2009.14157}}].

\bibitem{Filippenko1997}
A.V.~{Filippenko}, \emph{{Optical Spectra of Supernovae}},
  \href{https://doi.org/10.1146/annurev.astro.35.1.309}{\emph{\araa} {\bfseries
  35} (1997) 309}.

\bibitem{Boccioli2023}
L.~{Boccioli}, L.~{Roberti}, M.~{Limongi}, G.J.~{Mathews} and A.~{Chieffi},
  \emph{{Explosion Mechanism of Core-collapse Supernovae: Role of the Si/Si-O
  Interface}}, \href{https://doi.org/10.3847/1538-4357/acc06a}{\emph{\apj}
  {\bfseries 949} (2023) 17}
  [\href{https://arxiv.org/abs/2207.08361}{{\ttfamily 2207.08361}}].

\bibitem{Burrows2024}
A.~{Burrows}, T.~{Wang} and D.~{Vartanyan}, \emph{{Physical Correlations and
  Predictions Emerging from Modern Core-collapse Supernova Theory}},
  \href{https://doi.org/10.3847/2041-8213/ad319e}{\emph{\apjl} {\bfseries 964}
  (2024) L16} [\href{https://arxiv.org/abs/2401.06840}{{\ttfamily
  2401.06840}}].

\bibitem{2008ApJ...684.1336K}
C.S.~{Kochanek}, J.F.~{Beacom}, M.D.~{Kistler}, J.L.~{Prieto}, K.Z.~{Stanek},
  T.A.~{Thompson} et~al., \emph{{A Survey About Nothing: Monitoring a Million
  Supergiants for Failed Supernovae}},
  \href{https://doi.org/10.1086/590053}{\emph{\apj} {\bfseries 684} (2008)
  1336} [\href{https://arxiv.org/abs/0802.0456}{{\ttfamily 0802.0456}}].

\bibitem{2015MNRAS.450.3289G}
J.R.~{Gerke}, C.S.~{Kochanek} and K.Z.~{Stanek}, \emph{{The search for failed
  supernovae with the Large Binocular Telescope: first candidates}},
  \href{https://doi.org/10.1093/mnras/stv776}{\emph{\mnras} {\bfseries 450}
  (2015) 3289} [\href{https://arxiv.org/abs/1411.1761}{{\ttfamily 1411.1761}}].

\bibitem{2017MNRAS.468.4968A}
S.M.~{Adams}, C.S.~{Kochanek}, J.R.~{Gerke}, K.Z.~{Stanek} and X.~{Dai},
  \emph{{The search for failed supernovae with the Large Binocular Telescope:
  confirmation of a disappearing star}},
  \href{https://doi.org/10.1093/mnras/stx816}{\emph{\mnras} {\bfseries 468}
  (2017) 4968} [\href{https://arxiv.org/abs/1609.01283}{{\ttfamily
  1609.01283}}].

\bibitem{2021MNRAS.508.1156B}
C.M.~{Basinger}, C.S.~{Kochanek}, S.M.~{Adams}, X.~{Dai} and K.Z.~{Stanek},
  \emph{{The search for failed supernovae with the Large Binocular Telescope:
  N6946-BH1, still no star}},
  \href{https://doi.org/10.1093/mnras/stab2620}{\emph{\mnras} {\bfseries 508}
  (2021) 1156} [\href{https://arxiv.org/abs/2007.15658}{{\ttfamily
  2007.15658}}].

\bibitem{2026Sci...391..689D}
K.~{De}, M.~{MacLeod}, J.E.~{Jencson}, E.~{Lovegrove}, A.~{Antoni}, E.~{Kara}
  et~al., \emph{{Disappearance of a massive star in the Andromeda Galaxy due to
  formation of a black hole}},
  \href{https://doi.org/10.1126/science.adt4853}{\emph{Science} {\bfseries 391}
  (2026) 689} [\href{https://arxiv.org/abs/2410.14778}{{\ttfamily
  2410.14778}}].

\bibitem{2026MNRAS.546ag052B}
E.R.~{Beasor}, N.~{Smith}, J.~{Pearson}, B.~{Subrayan}, E.~{Berger},
  D.J.~{Sand} et~al., \emph{{The fate of the failed supernova candidate
  M31-2014-DS1}}, \href{https://doi.org/10.1093/mnras/stag052}{\emph{\mnras}
  {\bfseries 546} (2026) stag052}
  [\href{https://arxiv.org/abs/2601.05317}{{\ttfamily 2601.05317}}].

\bibitem{GW230529}
A.G.~{Abac}, R.~{Abbott}, I.~{Abouelfettouh}, F.~{Acernese}, K.~{Ackley},
  S.~{Adhicary} et~al., \emph{{Observation of Gravitational Waves from the
  Coalescence of a 2.5--4.5 M $_{\odot}$ Compact Object and a Neutron Star}},
  \href{https://doi.org/10.3847/2041-8213/ad5beb}{\emph{\apjl} {\bfseries 970}
  (2024) L34} [\href{https://arxiv.org/abs/2404.04248}{{\ttfamily
  2404.04248}}].

\bibitem{DisbergMandel2025}
P.~{Disberg} and I.~{Mandel}, \emph{{The Kick Velocity Distribution of Isolated
  Neutron Stars}}, \href{https://doi.org/10.3847/2041-8213/adf286}{\emph{\apjl}
  {\bfseries 989} (2025) L8}
  [\href{https://arxiv.org/abs/2505.22102}{{\ttfamily 2505.22102}}].

\bibitem{2005MNRAS.360..974H}
G.~{Hobbs}, D.R.~{Lorimer}, A.G.~{Lyne} and M.~{Kramer}, \emph{{A statistical
  study of 233 pulsar proper motions}},
  \href{https://doi.org/10.1111/j.1365-2966.2005.09087.x}{\emph{\mnras}
  {\bfseries 360} (2005) 974}
  [\href{https://arxiv.org/abs/astro-ph/0504584}{{\ttfamily
  astro-ph/0504584}}].

\bibitem{2017A&A...608A..57V}
F.~{Verbunt}, A.~{Igoshev} and E.~{Cator}, \emph{{The observed velocity
  distribution of young pulsars}},
  \href{https://doi.org/10.1051/0004-6361/201731518}{\emph{\aap} {\bfseries
  608} (2017) A57} [\href{https://arxiv.org/abs/1708.08281}{{\ttfamily
  1708.08281}}].

\bibitem{2020MNRAS.494.3663I}
A.P.~{Igoshev}, \emph{{The observed velocity distribution of young pulsars -
  II. Analysis of complete PSR{\ensuremath{\pi}}}},
  \href{https://doi.org/10.1093/mnras/staa958}{\emph{\mnras} {\bfseries 494}
  (2020) 3663} [\href{https://arxiv.org/abs/2002.01367}{{\ttfamily
  2002.01367}}].

\bibitem{2021ApJ...920L..37W}
R.~{Willcox}, I.~{Mandel}, E.~{Thrane}, A.~{Deller}, S.~{Stevenson} and
  A.~{Vigna-G{\'o}mez}, \emph{{Constraints on Weak Supernova Kicks from
  Observed Pulsar Velocities}},
  \href{https://doi.org/10.3847/2041-8213/ac2cc8}{\emph{\apjl} {\bfseries 920}
  (2021) L37} [\href{https://arxiv.org/abs/2107.04251}{{\ttfamily
  2107.04251}}].

\bibitem{Janka2013}
H.-T.~{Janka}, \emph{{Natal kicks of stellar mass black holes by asymmetric
  mass ejection in fallback supernovae}},
  \href{https://doi.org/10.1093/mnras/stt1106}{\emph{\mnras} {\bfseries 434}
  (2013) 1355} [\href{https://arxiv.org/abs/1306.0007}{{\ttfamily 1306.0007}}].

\bibitem{2023ApJ...957...68B}
A.~{Burrows}, D.~{Vartanyan} and T.~{Wang}, \emph{{Black Hole Formation
  Accompanied by the Supernova Explosion of a 40 M $_{\odot}$ Progenitor
  Star}}, \href{https://doi.org/10.3847/1538-4357/acfc1c}{\emph{\apj}
  {\bfseries 957} (2023) 68}
  [\href{https://arxiv.org/abs/2308.05798}{{\ttfamily 2308.05798}}].

\bibitem{2025ApJ...987..164B}
A.~{Burrows}, T.~{Wang} and D.~{Vartanyan}, \emph{{Channels of Stellar-mass
  Black Hole Formation}},
  \href{https://doi.org/10.3847/1538-4357/addd04}{\emph{\apj} {\bfseries 987}
  (2025) 164} [\href{https://arxiv.org/abs/2412.07831}{{\ttfamily
  2412.07831}}].

\bibitem{FowlerHoyle1964}
W.A.~{Fowler} and F.~{Hoyle}, \emph{{Neutrino Processes and Pair Formation in
  Massive Stars and Supernovae.}},
  \href{https://doi.org/10.1086/190103}{\emph{\apjs} {\bfseries 9} (1964) 201}.

\bibitem{RakavyShaviv1967}
G.~{Rakavy} and G.~{Shaviv}, \emph{{Instabilities in Highly Evolved Stellar
  Models}}, \href{https://doi.org/10.1086/149204}{\emph{\apj} {\bfseries 148}
  (1967) 803}.

\bibitem{Barkat1967}
Z.~{Barkat}, G.~{Rakavy} and N.~{Sack}, \emph{{Dynamics of Supernova Explosion
  Resulting from Pair Formation}},
  \href{https://doi.org/10.1103/PhysRevLett.18.379}{\emph{\prl} {\bfseries 18}
  (1967) 379}.

\bibitem{Fraley1968}
G.S.~{Fraley}, \emph{{Supernovae Explosions Induced by Pair-Production
  Instability}}, \href{https://doi.org/10.1007/BF00651498}{\emph{\apss}
  {\bfseries 2} (1968) 96}.

\bibitem{Renzo2020a}
M.~{Renzo}, R.~{Farmer}, S.~{Justham}, Y.~{G{\"o}tberg}, S.E.~{de Mink},
  E.~{Zapartas} et~al., \emph{{Predictions for the hydrogen-free ejecta of
  pulsational pair-instability supernovae}},
  \href{https://doi.org/10.1051/0004-6361/202037710}{\emph{\aap} {\bfseries
  640} (2020) A56} [\href{https://arxiv.org/abs/2002.05077}{{\ttfamily
  2002.05077}}].

\bibitem{2017ApJ...836..244W}
S.E.~{Woosley}, \emph{{Pulsational Pair-instability Supernovae}},
  \href{https://doi.org/10.3847/1538-4357/836/2/244}{\emph{\apj} {\bfseries
  836} (2017) 244} [\href{https://arxiv.org/abs/1608.08939}{{\ttfamily
  1608.08939}}].

\bibitem{Farmer2019}
R.~{Farmer}, M.~{Renzo}, S.E.~{de Mink}, P.~{Marchant} and S.~{Justham},
  \emph{{Mind the Gap: The Location of the Lower Edge of the Pair-instability
  Supernova Black Hole Mass Gap}},
  \href{https://doi.org/10.3847/1538-4357/ab518b}{\emph{\apj} {\bfseries 887}
  (2019) 53} [\href{https://arxiv.org/abs/1910.12874}{{\ttfamily 1910.12874}}].

\bibitem{Farmer2020}
R.~{Farmer}, M.~{Renzo}, S.E.~{de Mink}, M.~{Fishbach} and S.~{Justham},
  \emph{{Constraints from Gravitational-wave Detections of Binary Black Hole
  Mergers on the $^{12}$C({\ensuremath{\alpha}}, {\ensuremath{\gamma}})$^{16}$O
  Rate}}, \href{https://doi.org/10.3847/2041-8213/abbadd}{\emph{\apjl}
  {\bfseries 902} (2020) L36}
  [\href{https://arxiv.org/abs/2006.06678}{{\ttfamily 2006.06678}}].

\bibitem{Mehta2022}
A.K.~{Mehta}, A.~{Buonanno}, J.~{Gair}, M.C.~{Miller}, E.~{Farag},
  R.J.~{deBoer} et~al., \emph{{Observing Intermediate-mass Black Holes and the
  Upper Stellar-mass gap with LIGO and Virgo}},
  \href{https://doi.org/10.3847/1538-4357/ac3130}{\emph{\apj} {\bfseries 924}
  (2022) 39} [\href{https://arxiv.org/abs/2105.06366}{{\ttfamily 2105.06366}}].

\bibitem{Farag2022}
E.~{Farag}, M.~{Renzo}, R.~{Farmer}, M.T.~{Chidester} and F.X.~{Timmes},
  \emph{{Resolving the Peak of the Black Hole Mass Spectrum}},
  \href{https://doi.org/10.3847/1538-4357/ac8b83}{\emph{\apj} {\bfseries 937}
  (2022) 112} [\href{https://arxiv.org/abs/2208.09624}{{\ttfamily
  2208.09624}}].

\bibitem{vanSon2025}
L.A.C.~{van Son}, S.K.~{Roy}, I.~{Mandel}, W.M.~{Farr}, A.~{Lam}, J.~{Merritt}
  et~al., \emph{{Not Just Winds: Why Models Find That Binary Black Hole
  Formation Is Metallicity-dependent, while Binary Neutron Star Formation Is
  Not}}, \href{https://doi.org/10.3847/1538-4357/ada14a}{\emph{\apj} {\bfseries
  979} (2025) 209} [\href{https://arxiv.org/abs/2411.02484}{{\ttfamily
  2411.02484}}].

\bibitem{LattimerPrakash2004}
J.M.~Lattimer and M.~Prakash, \emph{The physics of neutron stars},
  \href{https://doi.org/10.1126/science.1090720}{\emph{Science} {\bfseries 304}
  (2004) 536} [\href{https://arxiv.org/abs/astro-ph/0405262}{{\ttfamily
  astro-ph/0405262}}].

\bibitem{OzelFreire2016}
F.~{\"O}zel and P.~Freire, \emph{Masses, radii, and the equation of state of
  neutron stars},
  \href{https://doi.org/10.1146/annurev-astro-081915-023322}{\emph{Annual
  Review of Astronomy and Astrophysics} {\bfseries 54} (2016) 401}.

\bibitem{Lattimer2021}
J.M.~Lattimer, \emph{Neutron stars and the nuclear matter equation of state},
  \href{https://doi.org/10.1146/annurev-nucl-102419-124827}{\emph{Annual Review
  of Nuclear and Particle Science} {\bfseries 71} (2021) 433}.

\bibitem{Potekhin2020}
A.Y.~{Potekhin}, D.A.~{Zyuzin}, D.G.~{Yakovlev}, M.V.~{Beznogov} and
  Y.A.~{Shibanov}, \emph{{Thermal luminosities of cooling neutron stars}},
  \href{https://doi.org/10.1093/mnras/staa1871}{\emph{\mnras} {\bfseries 496}
  (2020) 5052} [\href{https://arxiv.org/abs/2006.15004}{{\ttfamily
  2006.15004}}].

\bibitem{Hewish1968}
A.~{Hewish}, S.J.~{Bell}, J.D.H.~{Pilkington}, P.F.~{Scott} and R.A.~{Collins},
  \emph{Observation of a rapidly pulsating radio source},
  \href{https://doi.org/10.1038/217709a0}{\emph{Nature} {\bfseries 217} (1968)
  709}.

\bibitem{StaelinReifenstein1968}
D.H.~{Staelin} and E.C.~{Reifenstein}, III, \emph{{Pulsating Radio Sources near
  the Crab Nebula}},
  \href{https://doi.org/10.1126/science.162.3861.1481}{\emph{Science}
  {\bfseries 162} (1968) 1481}.

\bibitem{CockeDisneyTaylor1969}
W.J.~Cocke, M.J.~Disney and D.J.~Taylor, \emph{Discovery of optical signals
  from pulsar {NP 0532}}, \href{https://doi.org/10.1038/221525a0}{\emph{Nature}
  {\bfseries 221} (1969) 525}.

\bibitem{Fritz1969}
G.~{Fritz}, R.C.~{Henry}, J.F.~{Meekins}, T.A.~{Chubb} and H.~{Friedman},
  \emph{X-ray pulsar in the crab nebula},
  \href{https://doi.org/10.1126/science.164.3880.709}{\emph{Science} {\bfseries
  164} (1969) 709}.

\bibitem{Bradt1969}
H.~{Bradt}, S.~{Rappaport}, W.~{Mayer}, R.E.~{Nather}, B.~{Warner},
  M.~{Macfarlane} et~al., \emph{X-ray and optical observations of the pulsar
  {NP 0532} in the crab nebula},
  \href{https://doi.org/10.1038/222728a0}{\emph{Nature} {\bfseries 222} (1969)
  728}.

\bibitem{BaadeZwicky1934}
W.~{Baade} and F.~{Zwicky}, \emph{Cosmic rays from super-novae},
  \href{https://doi.org/10.1073/pnas.20.5.259}{\emph{Proceedings of the
  National Academy of Sciences} {\bfseries 20} (1934) 259}.

\bibitem{OppenheimerVolkoff1939}
J.R.~{Oppenheimer} and G.M.~{Volkoff}, \emph{On massive neutron cores},
  \href{https://doi.org/10.1103/PhysRev.55.374}{\emph{Physical Review}
  {\bfseries 55} (1939) 374}.

\bibitem{BhattacharyaVanDenHeuvel1991}
D.~{Bhattacharya} and E.P.J.~{van den Heuvel}, \emph{{Formation and evolution
  of binary and millisecond radio pulsars}},
  \href{https://doi.org/10.1016/0370-1573(91)90064-S}{\emph{\physrep}
  {\bfseries 203} (1991) 1}.

\bibitem{Bowyer1965}
S.~{Bowyer}, E.T.~{Byram}, T.A.~{Chubb} and H.~{Friedman}, \emph{Cosmic x-ray
  sources}, \href{https://doi.org/10.1126/science.147.3656.394}{\emph{Science}
  {\bfseries 147} (1965) 394}.

\bibitem{WebsterMurdin1972}
B.L.~{Webster} and P.~{Murdin}, \emph{Cygnus {X-1} -- a spectroscopic binary
  with a heavy companion?},
  \href{https://doi.org/10.1038/235037a0}{\emph{Nature} {\bfseries 235} (1972)
  37}.

\bibitem{Bolton1972}
C.T.~{Bolton}, \emph{Identification of cygnus {X-1} with {HDE 226868}},
  \href{https://doi.org/10.1038/235271b0}{\emph{Nature} {\bfseries 235} (1972)
  271}.

\bibitem{1998ApJ...493..351K}
V.~{Kalogera} and R.F.~{Webbink}, \emph{{Formation of Low-Mass X-Ray Binaries.
  II. Common Envelope Evolution of Primordial Binaries with Extreme Mass
  Ratios}}, \href{https://doi.org/10.1086/305085}{\emph{\apj} {\bfseries 493}
  (1998) 351} [\href{https://arxiv.org/abs/astro-ph/9708223}{{\ttfamily
  astro-ph/9708223}}].

\bibitem{2003MNRAS.341..385P}
P.~{Podsiadlowski}, S.~{Rappaport} and Z.~{Han}, \emph{{On the formation and
  evolution of black hole binaries}},
  \href{https://doi.org/10.1046/j.1365-8711.2003.06464.x}{\emph{\mnras}
  {\bfseries 341} (2003) 385}
  [\href{https://arxiv.org/abs/astro-ph/0207153}{{\ttfamily
  astro-ph/0207153}}].

\bibitem{2025ApJ...983..115S}
C.~{Shariat}, S.~{Naoz}, K.~{El-Badry}, K.A.~{Rocha}, V.~{Kalogera},
  A.P.~{Stephan} et~al., \emph{{Triple Evolution Pathways to Black Hole
  Low-mass X-Ray Binaries: Insights from V404 Cygni}},
  \href{https://doi.org/10.3847/1538-4357/adbf01}{\emph{\apj} {\bfseries 983}
  (2025) 115} [\href{https://arxiv.org/abs/2411.15644}{{\ttfamily
  2411.15644}}].

\bibitem{Shenar2022}
T.~{Shenar}, H.~{Sana}, L.~{Mahy}, K.~{El-Badry}, P.~{Marchant}, N.~{Langer}
  et~al., \emph{{An X-ray-quiet black hole born with a negligible kick in a
  massive binary within the Large Magellanic Cloud}},
  \href{https://doi.org/10.1038/s41550-022-01730-y}{\emph{Nature Astronomy}
  {\bfseries 6} (2022) 1085}
  [\href{https://arxiv.org/abs/2207.07675}{{\ttfamily 2207.07675}}].

\bibitem{Giesers2019}
B.~{Giesers}, S.~{Kamann}, S.~{Dreizler}, T.-O.~{Husser}, A.~{Askar},
  F.~{G{\"o}ttgens} et~al., \emph{{A stellar census in globular clusters with
  MUSE: Binaries in NGC 3201}},
  \href{https://doi.org/10.1051/0004-6361/201936203}{\emph{\aap} {\bfseries
  632} (2019) A3} [\href{https://arxiv.org/abs/1909.04050}{{\ttfamily
  1909.04050}}].

\bibitem{Mahy2022}
L.~{Mahy}, H.~{Sana}, T.~{Shenar}, K.~{Sen}, N.~{Langer}, P.~{Marchant} et~al.,
  \emph{{Identifying quiescent compact objects in massive Galactic single-lined
  spectroscopic binaries}},
  \href{https://doi.org/10.1051/0004-6361/202243147}{\emph{\aap} {\bfseries
  664} (2022) A159} [\href{https://arxiv.org/abs/2207.07752}{{\ttfamily
  2207.07752}}].

\bibitem{ElBadry2024NS}
K.~{El-Badry}, H.-W.~{Rix}, D.W.~{Latham}, S.~{Shahaf}, T.~{Mazeh},
  A.~{Bieryla} et~al., \emph{{A population of neutron star candidates in wide
  orbits from Gaia astrometry}},
  \href{https://doi.org/10.33232/001c.121261}{\emph{The Open Journal of
  Astrophysics} {\bfseries 7} (2024) 58}
  [\href{https://arxiv.org/abs/2405.00089}{{\ttfamily 2405.00089}}].

\bibitem{ElBadry2023BH1}
K.~{El-Badry}, H.-W.~{Rix}, E.~{Quataert}, A.W.~{Howard}, H.~{Isaacson},
  J.~{Fuller} et~al., \emph{{A Sun-like star orbiting a black hole}},
  \href{https://doi.org/10.1093/mnras/stac3140}{\emph{\mnras} {\bfseries 518}
  (2023) 1057} [\href{https://arxiv.org/abs/2209.06833}{{\ttfamily
  2209.06833}}].

\bibitem{ElBadry2023BH2}
K.~{El-Badry}, H.-W.~{Rix}, Y.~{Cendes}, A.C.~{Rodriguez}, C.~{Conroy},
  E.~{Quataert} et~al., \emph{{A red giant orbiting a black hole}},
  \href{https://doi.org/10.1093/mnras/stad799}{\emph{\mnras} {\bfseries 521}
  (2023) 4323} [\href{https://arxiv.org/abs/2302.07880}{{\ttfamily
  2302.07880}}].

\bibitem{Panuzzo2024}
{Gaia Collaboration}, P.~{Panuzzo}, T.~{Mazeh}, F.~{Arenou}, B.~{Holl},
  E.~{Caffau} et~al., \emph{{Discovery of a dormant 33 solar-mass black hole in
  pre-release Gaia astrometry}},
  \href{https://doi.org/10.1051/0004-6361/202449763}{\emph{\aap} {\bfseries
  686} (2024) L2} [\href{https://arxiv.org/abs/2404.10486}{{\ttfamily
  2404.10486}}].

\bibitem{2024A&A...692A.141K}
M.U.~{Kruckow}, J.J.~{Andrews}, T.~{Fragos}, B.~{Holl}, S.S.~{Bavera},
  M.~{Briel} et~al., \emph{{The formation of black holes in non-interacting
  isolated binaries: Gaia black holes as calibrators of stellar winds from
  massive stars}},
  \href{https://doi.org/10.1051/0004-6361/202452356}{\emph{\aap} {\bfseries
  692} (2024) A141} [\href{https://arxiv.org/abs/2410.18501}{{\ttfamily
  2410.18501}}].

\bibitem{2024MNRAS.535L..44G}
A.~{Gilkis} and T.~{Mazeh}, \emph{{Gaia BH1 and BH2 - evolutionary models with
  overshooting of the black hole progenitors within the present-day binary
  separation}}, \href{https://doi.org/10.1093/mnrasl/slae091}{\emph{\mnras}
  {\bfseries 535} (2024) L44}
  [\href{https://arxiv.org/abs/2409.15899}{{\ttfamily 2409.15899}}].

\bibitem{2024MNRAS.535.3577K}
I.~{Kotko}, S.~{Banerjee} and K.~{Belczynski}, \emph{{The enigmatic origin of
  two dormant BH binaries: Gaia BH1 and Gaia BH2}},
  \href{https://doi.org/10.1093/mnras/stae2591}{\emph{\mnras} {\bfseries 535}
  (2024) 3577} [\href{https://arxiv.org/abs/2403.13579}{{\ttfamily
  2403.13579}}].

\bibitem{2024A&A...690A.144I}
G.~{Iorio}, S.~{Torniamenti}, M.~{Mapelli}, M.~{Dall'Amico}, A.A.~{Trani},
  S.~{Rastello} et~al., \emph{{The boring history of Gaia BH3 from isolated
  binary evolution}},
  \href{https://doi.org/10.1051/0004-6361/202450531}{\emph{\aap} {\bfseries
  690} (2024) A144} [\href{https://arxiv.org/abs/2404.17568}{{\ttfamily
  2404.17568}}].

\bibitem{2025arXiv251110728O}
A.~{Olejak}, J.~{Klencki}, A.~{Vigna-Gomez}, S.E.~{de Mink}, L.~{van Son},
  J.~{Cehula} et~al., \emph{{Non-conservative Mass Transfer as a Formation
  Channel for Gaia Black Hole System}},
  \href{https://doi.org/10.48550/arXiv.2511.10728}{\emph{arXiv e-prints} (2025)
  arXiv:2511.10728} [\href{https://arxiv.org/abs/2511.10728}{{\ttfamily
  2511.10728}}].

\bibitem{2024PASP..136a4202N}
P.~{Nagarajan}, K.~{El-Badry}, A.H.M.J.~{Triaud}, T.A.~{Baycroft}, D.~{Latham},
  A.~{Bieryla} et~al., \emph{{ESPRESSO Observations of Gaia BH1: High-precision
  Orbital Constraints and no Evidence for an Inner Binary}},
  \href{https://doi.org/10.1088/1538-3873/ad1ba7}{\emph{\pasp} {\bfseries 136}
  (2024) 014202} [\href{https://arxiv.org/abs/2312.05313}{{\ttfamily
  2312.05313}}].

\bibitem{2024ApJ...964...83G}
A.~{Generozov} and H.B.~{Perets}, \emph{{A Triple Scenario for the Formation of
  Wide Black Hole Binaries Such as Gaia BH1}},
  \href{https://doi.org/10.3847/1538-4357/ad2356}{\emph{\apj} {\bfseries 964}
  (2024) 83} [\href{https://arxiv.org/abs/2312.03066}{{\ttfamily 2312.03066}}].

\bibitem{2024ApJ...975L...8L}
Z.~{Li}, C.~{Zhu}, X.~{Lu}, G.~{L{\"u}}, L.~{Li}, H.~{Liu} et~al., \emph{{A
  Possible Formation Scenario of the Gaia BH1: Inner Binary Merger in Triple
  Systems}}, \href{https://doi.org/10.3847/2041-8213/ad8653}{\emph{\apjl}
  {\bfseries 975} (2024) L8}
  [\href{https://arxiv.org/abs/2410.10581}{{\ttfamily 2410.10581}}].

\bibitem{2025OJAp....8E..79T}
A.~{Tanikawa}, L.~{Wang}, M.S.~{Fujii}, A.A.~{Trani}, T.~{Hayashi} and
  Y.~{Suto}, \emph{{Compact Binary Formation in Open Star Clusters III:
  Probability of Binary Black Holes Hidden Inside of Gaia Black Hole Binary}},
  \href{https://doi.org/10.33232/001c.141294}{\emph{The Open Journal of
  Astrophysics} {\bfseries 8} (2025) 79}
  [\href{https://arxiv.org/abs/2407.03662}{{\ttfamily 2407.03662}}].

\bibitem{2023MNRAS.526..740R}
S.~{Rastello}, G.~{Iorio}, M.~{Mapelli}, M.~{Arca-Sedda}, U.N.~{Di Carlo},
  G.J.~{Escobar} et~al., \emph{{Dynamical formation of Gaia BH1 in a young star
  cluster}}, \href{https://doi.org/10.1093/mnras/stad2757}{\emph{\mnras}
  {\bfseries 526} (2023) 740}
  [\href{https://arxiv.org/abs/2306.14679}{{\ttfamily 2306.14679}}].

\bibitem{2024MNRAS.527.4031T}
A.~{Tanikawa}, S.~{Cary}, M.~{Shikauchi}, L.~{Wang} and M.S.~{Fujii},
  \emph{{Compact binary formation in open star clusters - I. High formation
  efficiency of Gaia BHs and their multiplicities}},
  \href{https://doi.org/10.1093/mnras/stad3294}{\emph{\mnras} {\bfseries 527}
  (2024) 4031} [\href{https://arxiv.org/abs/2303.05743}{{\ttfamily
  2303.05743}}].

\bibitem{2024ApJ...965...22D}
U.N.~{Di Carlo}, P.~{Agrawal}, C.L.~{Rodriguez} and K.~{Breivik}, \emph{{Young
  Star Clusters Dominate the Production of Detached Black Hole--Star
  Binaries}}, \href{https://doi.org/10.3847/1538-4357/ad2f2c}{\emph{\apj}
  {\bfseries 965} (2024) 22}
  [\href{https://arxiv.org/abs/2306.13121}{{\ttfamily 2306.13121}}].

\bibitem{2024A&A...688L...2M}
D.~{Mar{\'\i}n Pina}, S.~{Rastello}, M.~{Gieles}, K.~{Kremer}, L.~{Fitzgerald}
  and B.~{Rando Forastier}, \emph{{Dynamical formation of Gaia BH3 in the
  progenitor globular cluster of the ED-2 stream}},
  \href{https://doi.org/10.1051/0004-6361/202450460}{\emph{\aap} {\bfseries
  688} (2024) L2} [\href{https://arxiv.org/abs/2404.13036}{{\ttfamily
  2404.13036}}].

\bibitem{Balbinot2024}
E.~{Balbinot}, E.~{Dodd}, T.~{Matsuno}, C.~{Lardo}, A.~{Helmi}, P.~{Panuzzo}
  et~al., \emph{{The 33 M$_{\odot}$ black hole Gaia BH3 is part of the
  disrupted ED-2 star cluster}},
  \href{https://doi.org/10.1051/0004-6361/202450425}{\emph{\aap} {\bfseries
  687} (2024) L3} [\href{https://arxiv.org/abs/2404.11604}{{\ttfamily
  2404.11604}}].

\bibitem{vanson2026_post_mt}
L.A.C.~{van Son}, N.~{Yamaguchi}, P.~{Nagarajan}, T.~{Shenar}, K.~{Sen},
  A.~{Laroche} et~al., \emph{{Ongoing and Post-mass-transfer Binaries: A Living
  Catalog and Unified Review of Binary Mass-transfer Products}},
  \href{https://doi.org/10.3847/1538-4365/ae8d04}{\emph{\apjs} {\bfseries 286}
  (2026) 13} [\href{https://arxiv.org/abs/2605.31290}{{\ttfamily 2605.31290}}].

\bibitem{MashianLoeb2017}
N.~{Mashian} and A.~{Loeb}, \emph{{Hunting black holes with Gaia}},
  \href{https://doi.org/10.1093/mnras/stx1410}{\emph{\mnras} {\bfseries 470}
  (2017) 2611} [\href{https://arxiv.org/abs/1704.03455}{{\ttfamily
  1704.03455}}].

\bibitem{Breivik+2017ApJ...850L..13B}
K.~{Breivik}, S.~{Chatterjee} and S.L.~{Larson}, \emph{{Revealing Black Holes
  with Gaia}}, \href{https://doi.org/10.3847/2041-8213/aa97d5}{\emph{\apjl}
  {\bfseries 850} (2017) L13}
  [\href{https://arxiv.org/abs/1710.04657}{{\ttfamily 1710.04657}}].

\bibitem{Yalinewich+2018MNRAS.481..930Y}
A.~{Yalinewich}, P.~{Beniamini}, K.~{Hotokezaka} and W.~{Zhu}, \emph{{Dark
  passengers in stellar surveys}},
  \href{https://doi.org/10.1093/mnras/sty2327}{\emph{\mnras} {\bfseries 481}
  (2018) 930} [\href{https://arxiv.org/abs/1807.00835}{{\ttfamily
  1807.00835}}].

\bibitem{Yamaguchi+2018ApJ...861...21Y}
M.S.~{Yamaguchi}, N.~{Kawanaka}, T.~{Bulik} and T.~{Piran}, \emph{{Detecting
  Black Hole Binaries by Gaia}},
  \href{https://doi.org/10.3847/1538-4357/aac5ec}{\emph{\apj} {\bfseries 861}
  (2018) 21} [\href{https://arxiv.org/abs/1710.09839}{{\ttfamily 1710.09839}}].

\bibitem{ShaoLi2019ApJ...885..151S}
Y.~{Shao} and X.-D.~{Li}, \emph{{Population Synthesis of Black Hole Binaries
  with Normal-star Companions. I. Detached Systems}},
  \href{https://doi.org/10.3847/1538-4357/ab4816}{\emph{\apj} {\bfseries 885}
  (2019) 151} [\href{https://arxiv.org/abs/1909.11328}{{\ttfamily
  1909.11328}}].

\bibitem{Shikauchi+2020PASJ...72...45S}
M.~{Shikauchi}, J.~{Kumamoto}, A.~{Tanikawa} and M.S.~{Fujii}, \emph{{Gaia's
  detectability of black hole-main sequence star binaries formed in open
  clusters}}, \href{https://doi.org/10.1093/pasj/psaa030}{\emph{\pasj}
  {\bfseries 72} (2020) 45} [\href{https://arxiv.org/abs/2001.11199}{{\ttfamily
  2001.11199}}].

\bibitem{Chawla+2022ApJ...931..107C}
C.~{Chawla}, S.~{Chatterjee}, K.~{Breivik}, C.K.~{Moorthy}, J.J.~{Andrews} and
  R.E.~{Sanderson}, \emph{{Gaia May Detect Hundreds of Well-characterized
  Stellar Black Holes}},
  \href{https://doi.org/10.3847/1538-4357/ac60a5}{\emph{\apj} {\bfseries 931}
  (2022) 107} [\href{https://arxiv.org/abs/2110.05979}{{\ttfamily
  2110.05979}}].

\bibitem{Andrews+2022arXiv220700680A}
J.J.~{Andrews}, K.~{Taggart} and R.~{Foley}, \emph{{A Sample of Neutron Star
  and Black Hole Binaries Detected through Gaia DR3 Astrometry}},
  \href{https://doi.org/10.48550/arXiv.2207.00680}{\emph{arXiv e-prints} (2022)
  arXiv:2207.00680} [\href{https://arxiv.org/abs/2207.00680}{{\ttfamily
  2207.00680}}].

\bibitem{Nagarajan+2025PASP..137d4202N}
P.~{Nagarajan}, K.~{El-Badry}, C.~{Chawla}, U.N.~{Di Carlo}, K.~{Breivik},
  C.L.~{Rodriguez} et~al., \emph{{Realistic Predictions for Gaia Black Hole
  Discoveries: Comparison of Isolated Binary and Dynamical Formation Models}},
  \href{https://doi.org/10.1088/1538-3873/adc839}{\emph{\pasp} {\bfseries 137}
  (2025) 044202} [\href{https://arxiv.org/abs/2502.03527}{{\ttfamily
  2502.03527}}].

\bibitem{Chawla+2026ApJ..1007..170C}
C.~{Chawla}, S.~{Chatterjee} and K.~{Breivik}, \emph{{Gaia's Promise to Detect
  Compact-object Binaries: Where We Stand with the Third Data Release}},
  \href{https://doi.org/10.3847/1538-4357/ae8c3c}{\emph{\apj} {\bfseries 1007}
  (2026) 170} [\href{https://arxiv.org/abs/2508.21805}{{\ttfamily
  2508.21805}}].

\bibitem{Whitaker2026}
M.~{Whitaker}, E.~{Kerr}, A.~{Seth}, M.~{H{\"a}berle}, J.~{Strader},
  J.~{Anderson} et~al., \emph{{A Long Period Stellar-mass Black Hole Binary in
  {\ensuremath{\omega}} Centauri}},
  \href{https://doi.org/10.3847/2041-8213/ae7a5c}{\emph{\apjl} {\bfseries 1006}
  (2026) L1} [\href{https://arxiv.org/abs/2606.18350}{{\ttfamily 2606.18350}}].

\bibitem{El-Badry2022_unicorns}
K.~{El-Badry}, R.~{Seeburger}, T.~{Jayasinghe}, H.-W.~{Rix}, S.~{Almada},
  C.~{Conroy} et~al., \emph{{Unicorns and giraffes in the binary zoo: stripped
  giants with subgiant companions}},
  \href{https://doi.org/10.1093/mnras/stac815}{\emph{\mnras} {\bfseries 512}
  (2022) 5620} [\href{https://arxiv.org/abs/2203.06348}{{\ttfamily
  2203.06348}}].

\bibitem{Green2025}
M.J.~{Green}, Y.~{Ziv}, H.-W.~{Rix}, D.~{Maoz}, I.~{Hamoudy}, T.~{Mazeh}
  et~al., \emph{{An upper limit on the frequency of short-period black hole
  companions to Sun-like stars}},
  \href{https://doi.org/10.1051/0004-6361/202453271}{\emph{\aap} {\bfseries
  695} (2025) A210} [\href{https://arxiv.org/abs/2412.02082}{{\ttfamily
  2412.02082}}].

\bibitem{Sahu2022}
K.C.~{Sahu}, J.~{Anderson}, S.~{Casertano}, H.E.~{Bond}, A.~{Udalski},
  M.~{Dominik} et~al., \emph{{An Isolated Stellar-mass Black Hole Detected
  through Astrometric Microlensing}},
  \href{https://doi.org/10.3847/1538-4357/ac739e}{\emph{\apj} {\bfseries 933}
  (2022) 83} [\href{https://arxiv.org/abs/2201.13296}{{\ttfamily 2201.13296}}].

\bibitem{Lam2022}
C.Y.~{Lam}, J.R.~{Lu}, A.~{Udalski}, I.~{Bond}, D.P.~{Bennett}, J.~{Skowron}
  et~al., \emph{{An Isolated Mass-gap Black Hole or Neutron Star Detected with
  Astrometric Microlensing}},
  \href{https://doi.org/10.3847/2041-8213/ac7442}{\emph{\apjl} {\bfseries 933}
  (2022) L23} [\href{https://arxiv.org/abs/2202.01903}{{\ttfamily
  2202.01903}}].

\bibitem{Mroz2022}
P.~{Mr{\'o}z}, A.~{Udalski} and A.~{Gould}, \emph{{Systematic Errors as a
  Source of Mass Discrepancy in Black Hole Microlensing Event
  OGLE-2011-BLG-0462}},
  \href{https://doi.org/10.3847/2041-8213/ac90bb}{\emph{\apjl} {\bfseries 937}
  (2022) L24} [\href{https://arxiv.org/abs/2207.10729}{{\ttfamily
  2207.10729}}].

\bibitem{LamLu2023}
C.Y.~{Lam} and J.R.~{Lu}, \emph{{A Reanalysis of the Isolated Black Hole
  Candidate OGLE-2011-BLG-0462/MOA-2011-BLG-191}},
  \href{https://doi.org/10.3847/1538-4357/aced4a}{\emph{\apj} {\bfseries 955}
  (2023) 116} [\href{https://arxiv.org/abs/2308.03302}{{\ttfamily
  2308.03302}}].

\bibitem{Olejak2020_synthetic_catalog}
A.~{Olejak}, K.~{Belczynski}, T.~{Bulik} and M.~{Sobolewska}, \emph{{Synthetic
  catalog of black holes in the Milky Way}},
  \href{https://doi.org/10.1051/0004-6361/201936557}{\emph{\aap} {\bfseries
  638} (2020) A94} [\href{https://arxiv.org/abs/1908.08775}{{\ttfamily
  1908.08775}}].

\bibitem{HulseTaylor1975}
R.A.~{Hulse} and J.H.~{Taylor}, \emph{{Discovery of a pulsar in a binary
  system.}}, \href{https://doi.org/10.1086/181708}{\emph{\apjl} {\bfseries 195}
  (1975) L51}.

\bibitem{TaylorWeisberg1982}
J.H.~{Taylor} and J.M.~{Weisberg}, \emph{{A new test of general relativity -
  Gravitational radiation and the binary pulsar PSR 1913+16}},
  \href{https://doi.org/10.1086/159690}{\emph{\apj} {\bfseries 253} (1982)
  908}.

\bibitem{GW150914}
B.P.~{Abbott}, R.~{Abbott}, T.D.~{Abbott}, M.R.~{Abernathy}, F.~{Acernese},
  K.~{Ackley} et~al., \emph{{Observing gravitational-wave transient GW150914
  with minimal assumptions}},
  \href{https://doi.org/10.1103/PhysRevD.93.122004}{\emph{\prd} {\bfseries 93}
  (2016) 122004} [\href{https://arxiv.org/abs/1602.03843}{{\ttfamily
  1602.03843}}].

\bibitem{GWTC5}
{The LIGO Scientific Collaboration}, {the Virgo Collaboration}, {the KAGRA
  Collaboration}, A.G.~{Abac}, A.~{Abe}, I.~{Abouelfettouh} et~al.,
  \emph{{GWTC-5.0: Observations from the Second Part of the Fourth
  LIGO-Virgo-KAGRA Observing Run and Updates to the Gravitational-Wave
  Transient Catalog}},
  \href{https://doi.org/10.48550/arXiv.2605.27225}{\emph{arXiv e-prints} (2026)
  arXiv:2605.27225} [\href{https://arxiv.org/abs/2605.27225}{{\ttfamily
  2605.27225}}].

\bibitem{ChristensenMeyer2022}
N.~{Christensen} and R.~{Meyer}, \emph{{Parameter estimation with gravitational
  waves}}, \href{https://doi.org/10.1103/RevModPhys.94.025001}{\emph{Reviews of
  Modern Physics} {\bfseries 94} (2022) 025001}
  [\href{https://arxiv.org/abs/2204.04449}{{\ttfamily 2204.04449}}].

\bibitem{GW190412}
R.~{Abbott} et~al., \emph{{GW190412: Observation of a binary-black-hole
  coalescence with asymmetric masses}},
  \href{https://doi.org/10.1103/PhysRevD.102.043015}{\emph{\prd} {\bfseries
  102} (2020) 043015} [\href{https://arxiv.org/abs/2004.08342}{{\ttfamily
  2004.08342}}].

\bibitem{GW241011_GW241110}
A.G.~{Abac} et~al., \emph{{GW241011 and GW241110: Exploring Binary Formation
  and Fundamental Physics with Asymmetric, High-Spin Black Hole Coalescences}},
  \href{https://doi.org/10.3847/2041-8213/ae0d54}{\emph{\apjl} (2025) }
  [\href{https://arxiv.org/abs/2510.26931}{{\ttfamily 2510.26931}}].

\bibitem{Purrer2016}
M.~{P{\"u}rrer}, M.~{Hannam} and F.~{Ohme}, \emph{{Can we measure individual
  black-hole spins from gravitational-wave observations?}},
  \href{https://doi.org/10.1103/PhysRevD.93.084042}{\emph{\prd} {\bfseries 93}
  (2016) 084042} [\href{https://arxiv.org/abs/1512.04955}{{\ttfamily
  1512.04955}}].

\bibitem{ETBlueBook}
A.~{Abac}, R.~{Abramo}, S.~{Albanesi}, A.~{Albertini}, A.~{Agapito},
  M.~{Agathos} et~al., \emph{{The Science of the Einstein Telescope}},
  \href{https://doi.org/10.1088/1475-7516/2026/03/081}{\emph{\jcap} {\bfseries
  2026} (2026) 081} [\href{https://arxiv.org/abs/2503.12263}{{\ttfamily
  2503.12263}}].

\bibitem{CosmicExplorer2021}
M.~{Evans}, R.X.~{Adhikari}, C.~{Afle}, S.W.~{Ballmer}, S.~{Biscoveanu},
  S.~{Borhanian} et~al., \emph{{A Horizon Study for Cosmic Explorer: Science,
  Observatories, and Community}},
  \href{https://doi.org/10.48550/arXiv.2109.09882}{\emph{arXiv e-prints} (2021)
  arXiv:2109.09882} [\href{https://arxiv.org/abs/2109.09882}{{\ttfamily
  2109.09882}}].

\bibitem{LISA2024}
M.~{Colpi}, K.~{Danzmann}, M.~{Hewitson}, K.~{Holley-Bockelmann}, P.~{Jetzer},
  G.~{Nelemans} et~al., \emph{{LISA Definition Study Report}},
  \href{https://doi.org/10.48550/arXiv.2402.07571}{\emph{arXiv e-prints} (2024)
  arXiv:2402.07571} [\href{https://arxiv.org/abs/2402.07571}{{\ttfamily
  2402.07571}}].

\bibitem{HellingsDowns1983}
R.W.~{Hellings} and G.S.~{Downs}, \emph{{Upper limits on the isotropic
  gravitational radiation background from pulsar timing analysis.}},
  \href{https://doi.org/10.1086/183954}{\emph{\apjl} {\bfseries 265} (1983)
  L39}.

\bibitem{NANOGrav2023}
G.~{Agazie}, A.~{Anumarlapudi}, A.M.~{Archibald}, Z.~{Arzoumanian},
  P.T.~{Baker}, B.~{B{\'e}csy} et~al., \emph{{The NANOGrav 15 yr Data Set:
  Evidence for a Gravitational-wave Background}},
  \href{https://doi.org/10.3847/2041-8213/acdac6}{\emph{\apjl} {\bfseries 951}
  (2023) L8} [\href{https://arxiv.org/abs/2306.16213}{{\ttfamily 2306.16213}}].

\bibitem{BorhanianSathyaprakash2024}
S.~{Borhanian} and B.S.~{Sathyaprakash}, \emph{{Listening to the Universe with
  next generation ground-based gravitational-wave detectors}},
  \href{https://doi.org/10.1103/PhysRevD.110.083040}{\emph{\prd} {\bfseries
  110} (2024) 083040} [\href{https://arxiv.org/abs/2202.11048}{{\ttfamily
  2202.11048}}].

\bibitem{Gupta2024}
I.~{Gupta}, C.~{Afle}, K.G.~{Arun}, A.~{Bandopadhyay}, M.~{Baryakhtar},
  S.~{Biscoveanu} et~al., \emph{{Characterizing gravitational wave detector
  networks: from A to cosmic explorer}},
  \href{https://doi.org/10.1088/1361-6382/ad7b99}{\emph{Classical and Quantum
  Gravity} {\bfseries 41} (2024) 245001}
  [\href{https://arxiv.org/abs/2307.10421}{{\ttfamily 2307.10421}}].

\bibitem{Peters1964}
P.C.~{Peters}, \emph{{Gravitational Radiation and the Motion of Two Point
  Masses}}, \href{https://doi.org/10.1103/PhysRev.136.B1224}{\emph{Physical
  Review} {\bfseries 136} (1964) 1224}.

\bibitem{Zevin2021}
M.~{Zevin}, S.S.~{Bavera}, C.P.L.~{Berry}, V.~{Kalogera}, T.~{Fragos},
  P.~{Marchant} et~al., \emph{{One Channel to Rule Them All? Constraining the
  Origins of Binary Black Holes Using Multiple Formation Pathways}},
  \href{https://doi.org/10.3847/1538-4357/abe40e}{\emph{\apj} {\bfseries 910}
  (2021) 152} [\href{https://arxiv.org/abs/2011.10057}{{\ttfamily
  2011.10057}}].

\bibitem{MandelBroekgaarden2022}
I.~{Mandel} and F.S.~{Broekgaarden}, \emph{Rates of compact object
  coalescences}, {\emph{Living Reviews in Relativity} {\bfseries 25} (2022) }.

\bibitem{2022ApJ...940..171R}
I.~{Romero-Shaw}, P.D.~{Lasky} and E.~{Thrane}, \emph{{Four Eccentric Mergers
  Increase the Evidence that LIGO-Virgo-KAGRA's Binary Black Holes Form
  Dynamically}}, \href{https://doi.org/10.3847/1538-4357/ac9798}{\emph{\apj}
  {\bfseries 940} (2022) 171}
  [\href{https://arxiv.org/abs/2206.14695}{{\ttfamily 2206.14695}}].

\bibitem{1977A&A....57..383Z}
J.-P.~{Zahn}, \emph{{Tidal friction in close binary systems.}}, {\emph{\aap}
  {\bfseries 57} (1977) 383}.

\bibitem{1981A&A....99..126H}
P.~{Hut}, \emph{{Tidal evolution in close binary systems.}}, {\emph{\aap}
  {\bfseries 99} (1981) 126}.

\bibitem{Eggleton1983}
P.P.~{Eggleton}, \emph{{Aproximations to the radii of Roche lobes.}},
  \href{https://doi.org/10.1086/160960}{\emph{\apj} {\bfseries 268} (1983)
  368}.

\bibitem{Soberman1997}
G.E.~{Soberman}, E.S.~{Phinney} and E.P.J.~{van den Heuvel}, \emph{{Stability
  criteria for mass transfer in binary stellar evolution.}},
  \href{https://doi.org/10.48550/arXiv.astro-ph/9703016}{\emph{\aap} {\bfseries
  327} (1997) 620} [\href{https://arxiv.org/abs/astro-ph/9703016}{{\ttfamily
  astro-ph/9703016}}].

\bibitem{HjellmingWebbink1987}
M.S.~{Hjellming} and R.F.~{Webbink}, \emph{{Thresholds for Rapid Mass Transfer
  in Binary Systems. I. Polytropic Models}},
  \href{https://doi.org/10.1086/165412}{\emph{\apj} {\bfseries 318} (1987)
  794}.

\bibitem{Ge2015}
H.~{Ge}, R.F.~{Webbink}, X.~{Chen} and Z.~{Han}, \emph{{Adiabatic Mass Loss in
  Binary Stars. II. From Zero-age Main Sequence to the Base of the Giant
  Branch}}, \href{https://doi.org/10.1088/0004-637X/812/1/40}{\emph{\apj}
  {\bfseries 812} (2015) 40}.

\bibitem{Ge2020}
H.~{Ge}, R.F.~{Webbink}, X.~{Chen} and Z.~{Han}, \emph{{Adiabatic Mass Loss in
  Binary Stars. III. From the Base of the Red Giant Branch to the Tip of the
  Asymptotic Giant Branch}},
  \href{https://doi.org/10.3847/1538-4357/aba7b7}{\emph{\apj} {\bfseries 899}
  (2020) 132} [\href{https://arxiv.org/abs/2007.09848}{{\ttfamily
  2007.09848}}].

\bibitem{Temmink2023}
K.D.~{Temmink}, O.R.~{Pols}, S.~{Justham}, A.G.~{Istrate} and S.~{Toonen},
  \emph{{Coping with loss. Stability of mass transfer from post-main-sequence
  donor stars}}, \href{https://doi.org/10.1051/0004-6361/202244137}{\emph{\aap}
  {\bfseries 669} (2023) A45}
  [\href{https://arxiv.org/abs/2209.12707}{{\ttfamily 2209.12707}}].

\bibitem{Darwin1879RSPS...29..168D}
G.H.~{Darwin}, \emph{{The Determination of the Secular Effects of Tidal
  Friction by a Graphical Method}}, {\emph{Proceedings of the Royal Society of
  London Series I} {\bfseries 29} (1879) 168}.

\bibitem{Hut1980A&A....92..167H}
P.~{Hut}, \emph{{Stability of tidal equilibrium}}, {\emph{\aap} {\bfseries 92}
  (1980) 167}.

\bibitem{2025A&A...703A.121T}
K.D.~{Temmink}, O.R.~{Pols}, S.~{Justham} and N.~{Blagorodnova}, \emph{{Good
  things come to those who wait: Watching donor stars evolve towards a
  mass-transfer instability}},
  \href{https://doi.org/10.1051/0004-6361/202556333}{\emph{\aap} {\bfseries
  703} (2025) A121} [\href{https://arxiv.org/abs/2509.09492}{{\ttfamily
  2509.09492}}].

\bibitem{2026A&A...709L...6T}
K.D.~{Temmink}, S.~{Justham} and O.R.~{Pols}, \emph{{One's loss is (not)
  another's gain: Isotropic re-emission destabilises mass transfer from
  radiative donor stars}},
  \href{https://doi.org/10.1051/0004-6361/202659252}{\emph{\aap} {\bfseries
  709} (2026) L6} [\href{https://arxiv.org/abs/2604.21459}{{\ttfamily
  2604.21459}}].

\bibitem{2021ApJ...923..277R}
M.~{Renzo} and Y.~{G{\"o}tberg}, \emph{{Evolution of Accretor Stars in Massive
  Binaries: Broader Implications from Modeling {\ensuremath{\zeta}} Ophiuchi}},
  \href{https://doi.org/10.3847/1538-4357/ac29c5}{\emph{\apj} {\bfseries 923}
  (2021) 277} [\href{https://arxiv.org/abs/2107.10933}{{\ttfamily
  2107.10933}}].

\bibitem{Inayoshi2017}
K.~{Inayoshi}, R.~{Hirai}, T.~{Kinugawa} and K.~{Hotokezaka}, \emph{{Formation
  pathway of Population III coalescing binary black holes through stable mass
  transfer}}, \href{https://doi.org/10.1093/mnras/stx757}{\emph{\mnras}
  {\bfseries 468} (2017) 5020}
  [\href{https://arxiv.org/abs/1701.04823}{{\ttfamily 1701.04823}}].

\bibitem{vandenHeuvel2017}
E.P.J.~{van den Heuvel}, S.F.~{Portegies Zwart} and S.E.~{de Mink},
  \emph{{Forming short-period Wolf-Rayet X-ray binaries and double black holes
  through stable mass transfer}},
  \href{https://doi.org/10.1093/mnras/stx1430}{\emph{\mnras} {\bfseries 471}
  (2017) 4256} [\href{https://arxiv.org/abs/1701.02355}{{\ttfamily
  1701.02355}}].

\bibitem{Bavera2021}
S.S.~{Bavera}, T.~{Fragos}, M.~{Zevin}, C.P.L.~{Berry}, P.~{Marchant},
  J.J.~{Andrews} et~al., \emph{{The impact of mass-transfer physics on the
  observable properties of field binary black hole populations}},
  \href{https://doi.org/10.1051/0004-6361/202039804}{\emph{\aap} {\bfseries
  647} (2021) A153} [\href{https://arxiv.org/abs/2010.16333}{{\ttfamily
  2010.16333}}].

\bibitem{GallegosGarcia2021}
M.~{Gallegos-Garcia}, C.P.L.~{Berry}, P.~{Marchant} and V.~{Kalogera},
  \emph{{Binary Black Hole Formation with Detailed Modeling: Stable Mass
  Transfer Leads to Lower Merger Rates}},
  \href{https://doi.org/10.3847/1538-4357/ac2610}{\emph{\apj} {\bfseries 922}
  (2021) 110} [\href{https://arxiv.org/abs/2107.05702}{{\ttfamily
  2107.05702}}].

\bibitem{Marchant2021}
P.~{Marchant}, K.M.W.~{Pappas}, M.~{Gallegos-Garcia}, C.P.L.~{Berry},
  R.E.~{Taam}, V.~{Kalogera} et~al., \emph{{The role of mass transfer and
  common envelope evolution in the formation of merging binary black holes}},
  \href{https://doi.org/10.1051/0004-6361/202039992}{\emph{\aap} {\bfseries
  650} (2021) A107} [\href{https://arxiv.org/abs/2103.09243}{{\ttfamily
  2103.09243}}].

\bibitem{vanSon2022_no_peaks}
L.A.C.~{van Son}, S.E.~{de Mink}, M.~{Renzo}, S.~{Justham}, E.~{Zapartas},
  K.~{Breivik} et~al., \emph{{No Peaks without Valleys: The Stable Mass
  Transfer Channel for Gravitational-wave Sources in Light of the Neutron
  Star-Black Hole Mass Gap}},
  \href{https://doi.org/10.3847/1538-4357/ac9b0a}{\emph{\apj} {\bfseries 940}
  (2022) 184} [\href{https://arxiv.org/abs/2209.13609}{{\ttfamily
  2209.13609}}].

\bibitem{Picco2024}
A.~{Picco}, P.~{Marchant}, H.~{Sana} and G.~{Nelemans}, \emph{{Forming merging
  double compact objects with stable mass transfer}},
  \href{https://doi.org/10.1051/0004-6361/202347090}{\emph{\aap} {\bfseries
  681} (2024) A31} [\href{https://arxiv.org/abs/2309.05736}{{\ttfamily
  2309.05736}}].

\bibitem{ChenBreivikvanSon2026}
G.~{Chen}, K.~{Breivik} and L.A.C.~{van Son}, \emph{{A Stellar Role Reversal:
  Multiple Features in the Mass and Mass Ratio Distributions of Merging Binary
  Black Holes from Stable Mass Transfer}},
  \href{https://doi.org/10.48550/arXiv.2606.26262}{\emph{arXiv e-prints} (2026)
  arXiv:2606.26262} [\href{https://arxiv.org/abs/2606.26262}{{\ttfamily
  2606.26262}}].

\bibitem{2026arXiv260523083G}
J.~{Godfrey}, L.~{van Son} and B.~{Farr}, \emph{{A Strongly Parametrized Mass
  Ratio Model for the Stable Mass Transfer Channel: a Case Study of the $10 \,
  \rm{M}_{\odot}$ Peak}},
  \href{https://doi.org/10.48550/arXiv.2605.23083}{\emph{arXiv e-prints} (2026)
  arXiv:2605.23083} [\href{https://arxiv.org/abs/2605.23083}{{\ttfamily
  2605.23083}}].

\bibitem{Klencki2026}
J.~{Klencki}, P.~{Podsiadlowski}, N.~{Langer}, A.~{Olejak}, S.~{Justham},
  A.~{Vigna-G{\'o}mez} et~al., \emph{{Comfort zones of stars: A limit on
  orbital tightening via stable mass transfer shapes the properties of binary
  black hole mergers}},
  \href{https://doi.org/10.1051/0004-6361/202555500}{\emph{\aap} {\bfseries
  706} (2026) A296} [\href{https://arxiv.org/abs/2505.08860}{{\ttfamily
  2505.08860}}].

\bibitem{VignaGomez2018}
A.~{Vigna-G{\'o}mez}, C.J.~{Neijssel}, S.~{Stevenson}, J.W.~{Barrett},
  K.~{Belczynski}, S.~{Justham} et~al., \emph{{On the formation history of
  Galactic double neutron stars}},
  \href{https://doi.org/10.1093/mnras/sty2463}{\emph{\mnras} {\bfseries 481}
  (2018) 4009} [\href{https://arxiv.org/abs/1805.07974}{{\ttfamily
  1805.07974}}].

\bibitem{Broekgaarden2026}
F.S.~{Broekgaarden}, A.~{Lam}, S.~{Levina}, J.~{Klencki}, K.A.~{Rocha}, L.~{van
  Son} et~al., \emph{{How Common Are Common Envelopes? Quantifying Their Role
  in Forming Gravitational-Wave Sources}},
  \href{https://doi.org/10.48550/arXiv.2606.05322}{\emph{arXiv e-prints} (2026)
  arXiv:2606.05322} [\href{https://arxiv.org/abs/2606.05322}{{\ttfamily
  2606.05322}}].

\bibitem{Belczynski2007}
K.~{Belczynski}, R.E.~{Taam}, V.~{Kalogera}, F.A.~{Rasio} and T.~{Bulik},
  \emph{{On the Rarity of Double Black Hole Binaries: Consequences for
  Gravitational Wave Detection}},
  \href{https://doi.org/10.1086/513562}{\emph{\apj} {\bfseries 662} (2007) 504}
  [\href{https://arxiv.org/abs/astro-ph/0612032}{{\ttfamily
  astro-ph/0612032}}].

\bibitem{PostnovYungelson2014}
K.A.~{Postnov} and L.R.~{Yungelson}, \emph{{The Evolution of Compact Binary
  Star Systems}}, \href{https://doi.org/10.12942/lrr-2014-3}{\emph{Living
  Reviews in Relativity} {\bfseries 17} (2014) 3}
  [\href{https://arxiv.org/abs/1403.4754}{{\ttfamily 1403.4754}}].

\bibitem{Belczynski2016}
K.~{Belczynski}, D.E.~{Holz}, T.~{Bulik} and R.~{O'Shaughnessy}, \emph{{The
  first gravitational-wave source from the isolated evolution of two stars in
  the 40-100 solar mass range}},
  \href{https://doi.org/10.1038/nature18322}{\emph{\nat} {\bfseries 534} (2016)
  512} [\href{https://arxiv.org/abs/1602.04531}{{\ttfamily 1602.04531}}].

\bibitem{IvanovaJusthamRicker2020}
N.~{Ivanova}, S.~{Justham} and P.~{Ricker}, \emph{{Common Envelope Evolution}}
  (2020),
  \href{https://doi.org/10.1088/2514-3433/abb6f0}{10.1088/2514-3433/abb6f0}.

\bibitem{ThorneZytkow1975}
K.S.~{Thorne} and A.N.~{Zytkow}, \emph{{Red giants and supergiants with
  degenerate neutron cores.}},
  \href{https://doi.org/10.1086/181839}{\emph{\apjl} {\bfseries 199} (1975)
  L19}.

\bibitem{ThorneZytkow1977}
K.S.~{Thorne} and A.N.~{Zytkow}, \emph{{Stars with degenerate neutron cores. I.
  Structure of equilibrium models.}},
  \href{https://doi.org/10.1086/155109}{\emph{\apj} {\bfseries 212} (1977)
  832}.

\bibitem{Ohlmann2016}
S.T.~{Ohlmann}, F.K.~{R{\"o}pke}, R.~{Pakmor} and V.~{Springel},
  \emph{{Hydrodynamic Moving-mesh Simulations of the Common Envelope Phase in
  Binary Stellar Systems}},
  \href{https://doi.org/10.3847/2041-8205/816/1/L9}{\emph{\apjl} {\bfseries
  816} (2016) L9} [\href{https://arxiv.org/abs/1512.04529}{{\ttfamily
  1512.04529}}].

\bibitem{RoepkeDeMarco2023}
F.K.~{R{\"o}pke} and O.~{De Marco}, \emph{{Simulations of common-envelope
  evolution in binary stellar systems: physical models and numerical
  techniques}}, \href{https://doi.org/10.1007/s41115-023-00017-x}{\emph{Living
  Reviews in Computational Astrophysics} {\bfseries 9} (2023) 2}
  [\href{https://arxiv.org/abs/2212.07308}{{\ttfamily 2212.07308}}].

\bibitem{vandenHeuvel1976}
E.P.J.~{van den Heuvel}, \emph{{Late Stages of Close Binary Systems}},  in
  \emph{Structure and Evolution of Close Binary Systems}, P.~{Eggleton},
  S.~{Mitton} and J.~{Whelan}, eds., vol.~73 of \emph{IAU Symposium}, p.~35,
  Jan., 1976.

\bibitem{TutukovYungelson1979}
A.~{Tutukov} and L.~{Yungelson}, \emph{{Evolution of massive common envelope
  binaries and mass loss.}},  in \emph{Mass Loss and Evolution of O-Type
  Stars}, P.S.~{Conti} and C.W.H.~{De Loore}, eds., vol.~83 of \emph{IAU
  Symposium}, pp.~401--406, Jan., 1979.

\bibitem{IbenTutukov1984}
I.~{Iben}, Jr. and A.V.~{Tutukov}, \emph{{Supernovae of type I as end products
  of the evolution of binaries with components of moderate initial mass.}},
  \href{https://doi.org/10.1086/190932}{\emph{\apjs} {\bfseries 54} (1984)
  335}.

\bibitem{Webbink1984}
R.F.~{Webbink}, \emph{Double white dwarfs as progenitors of r coronae borealis
  stars and type i supernovae}, {\emph{\apj} {\bfseries 277} (1984) 355}.

\bibitem{LivioSoker1988}
M.~{Livio} and N.~{Soker}, \emph{{The Common Envelope Phase in the Evolution of
  Binary Stars}}, \href{https://doi.org/10.1086/166419}{\emph{\apj} {\bfseries
  329} (1988) 764}.

\bibitem{vanSon2022_rate_redshift}
L.A.C.~{van Son}, S.E.~{de Mink}, T.~{Callister}, S.~{Justham}, M.~{Renzo},
  T.~{Wagg} et~al., \emph{{The Redshift Evolution of the Binary Black Hole
  Merger Rate: A Weighty Matter}},
  \href{https://doi.org/10.3847/1538-4357/ac64a3}{\emph{\apj} {\bfseries 931}
  (2022) 17} [\href{https://arxiv.org/abs/2110.01634}{{\ttfamily 2110.01634}}].

\bibitem{Olejak2021}
A.~{Olejak}, K.~{Belczynski} and N.~{Ivanova}, \emph{{Impact of common envelope
  development criteria on the formation of LIGO/Virgo sources}},
  \href{https://doi.org/10.1051/0004-6361/202140520}{\emph{\aap} {\bfseries
  651} (2021) A100} [\href{https://arxiv.org/abs/2102.05649}{{\ttfamily
  2102.05649}}].

\bibitem{FullerLu2022}
J.~{Fuller} and W.~{Lu}, \emph{{The spins of compact objects born from helium
  stars in binary systems}},
  \href{https://doi.org/10.1093/mnras/stac317}{\emph{\mnras} {\bfseries 511}
  (2022) 3951} [\href{https://arxiv.org/abs/2201.08407}{{\ttfamily
  2201.08407}}].

\bibitem{Bavera2020}
S.S.~{Bavera}, T.~{Fragos}, Y.~{Qin}, E.~{Zapartas}, C.J.~{Neijssel},
  I.~{Mandel} et~al., \emph{{The origin of spin in binary black holes.
  Predicting the distributions of the main observables of Advanced LIGO}},
  \href{https://doi.org/10.1051/0004-6361/201936204}{\emph{\aap} {\bfseries
  635} (2020) A97} [\href{https://arxiv.org/abs/1906.12257}{{\ttfamily
  1906.12257}}].

\bibitem{Kippenhahn1974}
R.~{Kippenhahn}, \emph{{Circulation and Mixing}},  in \emph{Late Stages of
  Stellar Evolution}, R.J.~{Tayler} and J.E.~{Hesser}, eds., vol.~66 of
  \emph{IAU Symposium}, p.~20, Jan., 1974.

\bibitem{Zahn1983}
J.-P.~{Zahn}, \emph{{Instability and Mixing Processes in Upper Main Sequence
  Stars}},  in \emph{Saas-Fee Advanced Course 13: Astrophysical Processes in
  Upper Main Sequence Stars}, A.N.~{Cox}, S.~{Vauclair} and J.P.~{Zahn}, eds.,
  p.~253, Jan., 1983.

\bibitem{Maeder1987}
A.~{Maeder}, \emph{{Evidences for a bifurcation in massive star evolution. The
  ON-blue stragglers.}}, {\emph{\aap} {\bfseries 178} (1987) 159}.

\bibitem{Hastings2020}
B.~{Hastings}, N.~{Langer} and G.~{Koenigsberger}, \emph{{Internal circulation
  in tidally locked massive binary stars: Consequences for double black hole
  formation}}, \href{https://doi.org/10.1051/0004-6361/202038499}{\emph{\aap}
  {\bfseries 641} (2020) A86}
  [\href{https://arxiv.org/abs/2007.11299}{{\ttfamily 2007.11299}}].

\bibitem{Sharpe2024}
K.~{Sharpe}, L.A.C.~{van Son}, S.E.~{de Mink}, R.~{Farmer}, P.~{Marchant} and
  G.~{Koenigsberger}, \emph{{Investigating the Chemically Homogeneous Evolution
  Channel and Its Role in the Formation of the Enigmatic Binary Black Hole
  Progenitor Candidate HD 5980}},
  \href{https://doi.org/10.3847/1538-4357/ad2f3e}{\emph{\apj} {\bfseries 966}
  (2024) 9} [\href{https://arxiv.org/abs/2402.12438}{{\ttfamily 2402.12438}}].

\bibitem{Riley2021}
J.~{Riley}, I.~{Mandel}, P.~{Marchant}, E.~{Butler}, K.~{Nathaniel},
  C.~{Neijssel} et~al., \emph{{Chemically homogeneous evolution: a rapid
  population synthesis approach}},
  \href{https://doi.org/10.1093/mnras/stab1291}{\emph{\mnras} {\bfseries 505}
  (2021) 663} [\href{https://arxiv.org/abs/2010.00002}{{\ttfamily
  2010.00002}}].

\bibitem{Naoz2016}
S.~{Naoz}, \emph{{The Eccentric Kozai-Lidov Effect and Its Applications}},
  \href{https://doi.org/10.1146/annurev-astro-081915-023315}{\emph{\araa}
  {\bfseries 54} (2016) 441}
  [\href{https://arxiv.org/abs/1601.07175}{{\ttfamily 1601.07175}}].

\bibitem{vonZeipel1910}
H.~{von Zeipel}, \emph{{Sur l'application des s{\'e}ries de M. Lindstedt {\`a}
  l'{\'e}tude du mouvement des com{\`e}tes p{\'e}riodiques}},
  \href{https://doi.org/10.1002/asna.19091832202}{\emph{Astronomische
  Nachrichten} {\bfseries 183} (1910) 345}.

\bibitem{Lidov1962}
M.L.~{Lidov}, \emph{The evolution of orbits of artificial satellites of planets
  under the action of gravitational perturbations of external bodies},
  {\emph{Planetary and Space Science} {\bfseries 9} (1962) 719}.

\bibitem{Kozai1962}
Y.~{Kozai}, \emph{Secular perturbations of asteroids with high inclination and
  eccentricity}, {\emph{\aj} {\bfseries 67} (1962) 591}.

\bibitem{Su2021b}
Y.~{Su}, D.~{Lai} and B.~{Liu}, \emph{{Spin-orbit misalignments in
  tertiary-induced binary black-hole mergers: Theoretical analysis}},
  \href{https://doi.org/10.1103/PhysRevD.103.063040}{\emph{\prd} {\bfseries
  103} (2021) 063040} [\href{https://arxiv.org/abs/2010.11951}{{\ttfamily
  2010.11951}}].

\bibitem{Toonen2020A&A...640A..16T}
S.~{Toonen}, S.~{Portegies Zwart}, A.S.~{Hamers} and D.~{Bandopadhyay},
  \emph{{The evolution of stellar triples. The most common evolutionary
  pathways}}, \href{https://doi.org/10.1051/0004-6361/201936835}{\emph{\aap}
  {\bfseries 640} (2020) A16}
  [\href{https://arxiv.org/abs/2004.07848}{{\ttfamily 2004.07848}}].

\bibitem{FragioneLoeb2019a}
G.~{Fragione} and A.~{Loeb}, \emph{{Black hole-neutron star mergers from
  triples}}, \href{https://doi.org/10.1093/mnras/stz1131}{\emph{\mnras}
  {\bfseries 486} (2019) 4443}
  [\href{https://arxiv.org/abs/1903.10511}{{\ttfamily 1903.10511}}].

\bibitem{Antonini2017}
F.~{Antonini}, S.~{Toonen} and A.S.~{Hamers}, \emph{{Binary Black Hole Mergers
  from Field Triples: Properties, Rates, and the Impact of Stellar Evolution}},
  \href{https://doi.org/10.3847/1538-4357/aa6f5e}{\emph{\apj} {\bfseries 841}
  (2017) 77} [\href{https://arxiv.org/abs/1703.06614}{{\ttfamily 1703.06614}}].

\bibitem{Liu2019}
B.~{Liu}, D.~{Lai} and Y.-H.~{Wang}, \emph{{Black Hole and Neutron Star Binary
  Mergers in Triple Systems. II. Merger Eccentricity and Spin-Orbit
  Misalignment}}, \href{https://doi.org/10.3847/1538-4357/ab2dfb}{\emph{\apj}
  {\bfseries 881} (2019) 41}
  [\href{https://arxiv.org/abs/1905.00427}{{\ttfamily 1905.00427}}].

\bibitem{Martinez2022}
M.A.S.~{Martinez}, C.L.~{Rodriguez} and G.~{Fragione}, \emph{{On the Mass Ratio
  Distribution of Black Hole Mergers in Triple Systems}},
  \href{https://doi.org/10.3847/1538-4357/ac8d55}{\emph{\apj} {\bfseries 937}
  (2022) 78} [\href{https://arxiv.org/abs/2105.01671}{{\ttfamily 2105.01671}}].

\bibitem{Stegmann2022}
J.~{Stegmann}, F.~{Antonini} and M.~{Moe}, \emph{{Evolution of massive stellar
  triples and implications for compact object binary formation}},
  \href{https://doi.org/10.1093/mnras/stac2192}{\emph{\mnras} {\bfseries 516}
  (2022) 1406} [\href{https://arxiv.org/abs/2112.10786}{{\ttfamily
  2112.10786}}].

\bibitem{SuLai2021}
Y.~{Su}, B.~{Liu} and D.~{Lai}, \emph{{The mass-ratio distribution of
  tertiary-induced binary black hole mergers}},
  \href{https://doi.org/10.1093/mnras/stab1617}{\emph{\mnras} {\bfseries 505}
  (2021) 3681} [\href{https://arxiv.org/abs/2103.01963}{{\ttfamily
  2103.01963}}].

\bibitem{Dorozsmai2024}
A.~{Dorozsmai}, S.~{Toonen}, A.~{Vigna-G{\'o}mez}, S.E.~{de Mink} and
  F.~{Kummer}, \emph{{Stellar triples with chemically homogeneously evolving
  inner binaries}}, \href{https://doi.org/10.1093/mnras/stad3819}{\emph{\mnras}
  {\bfseries 527} (2024) 9782}
  [\href{https://arxiv.org/abs/2307.04793}{{\ttfamily 2307.04793}}].

\bibitem{PortegiesZwart2010}
S.F.~{Portegies Zwart}, S.L.W.~{McMillan} and M.~{Gieles}, \emph{{Young Massive
  Star Clusters}},
  \href{https://doi.org/10.1146/annurev-astro-081309-130834}{\emph{\araa}
  {\bfseries 48} (2010) 431} [\href{https://arxiv.org/abs/1002.1961}{{\ttfamily
  1002.1961}}].

\bibitem{Harris2010}
W.E.~{Harris}, \emph{{A New Catalog of Globular Clusters in the Milky Way}},
  \href{https://doi.org/10.48550/arXiv.1012.3224}{\emph{arXiv e-prints} (2010)
  arXiv:1012.3224} [\href{https://arxiv.org/abs/1012.3224}{{\ttfamily
  1012.3224}}].

\bibitem{Neumayer2020}
N.~{Neumayer}, A.~{Seth} and T.~{B{\"o}ker}, \emph{{Nuclear star clusters}},
  \href{https://doi.org/10.1007/s00159-020-00125-0}{\emph{\aapr} {\bfseries 28}
  (2020) 4} [\href{https://arxiv.org/abs/2001.03626}{{\ttfamily 2001.03626}}].

\bibitem{SigurdssonHernquist1993}
S.~{Sigurdsson} and L.~{Hernquist}, \emph{{Primordial black holes in globular
  clusters}}, \href{https://doi.org/10.1038/364423a0}{\emph{\nat} {\bfseries
  364} (1993) 423}.

\bibitem{PortegiesZwartMcMillan2000}
S.F.~{Portegies Zwart} and S.L.W.~{McMillan}, \emph{{Black Hole Mergers in the
  Universe}}, \href{https://doi.org/10.1086/312422}{\emph{\apjl} {\bfseries
  528} (2000) L17} [\href{https://arxiv.org/abs/astro-ph/9910061}{{\ttfamily
  astro-ph/9910061}}].

\bibitem{MillerLauburg2009}
M.C.~{Miller} and V.M.~{Lauburg}, \emph{{Mergers of Stellar-Mass Black Holes in
  Nuclear Star Clusters}},
  \href{https://doi.org/10.1088/0004-637X/692/1/917}{\emph{\apj} {\bfseries
  692} (2009) 917} [\href{https://arxiv.org/abs/0804.2783}{{\ttfamily
  0804.2783}}].

\bibitem{Rodriguez2015}
C.L.~{Rodriguez}, M.~{Morscher}, B.~{Pattabiraman}, S.~{Chatterjee},
  C.-J.~{Haster} and F.A.~{Rasio}, \emph{{Binary Black Hole Mergers from
  Globular Clusters: Implications for Advanced LIGO}},
  \href{https://doi.org/10.1103/PhysRevLett.115.051101}{\emph{\prl} {\bfseries
  115} (2015) 051101} [\href{https://arxiv.org/abs/1505.00792}{{\ttfamily
  1505.00792}}].

\bibitem{Rodriguez2016a}
C.L.~{Rodriguez}, S.~{Chatterjee} and F.A.~{Rasio}, \emph{{Binary black hole
  mergers from globular clusters: Masses, merger rates, and the impact of
  stellar evolution}},
  \href{https://doi.org/10.1103/PhysRevD.93.084029}{\emph{\prd} {\bfseries 93}
  (2016) 084029} [\href{https://arxiv.org/abs/1602.02444}{{\ttfamily
  1602.02444}}].

\bibitem{Antonini2016}
F.~{Antonini} and F.A.~{Rasio}, \emph{{Merging Black Hole Binaries in Galactic
  Nuclei: Implications for Advanced-LIGO Detections}},
  \href{https://doi.org/10.3847/0004-637X/831/2/187}{\emph{\apj} {\bfseries
  831} (2016) 187} [\href{https://arxiv.org/abs/1606.04889}{{\ttfamily
  1606.04889}}].

\bibitem{Rodriguez2016b}
C.L.~{Rodriguez}, M.~{Zevin}, C.~{Pankow}, V.~{Kalogera} and F.A.~{Rasio},
  \emph{{Illuminating Black Hole Binary Formation Channels with Spins in
  Advanced LIGO}},
  \href{https://doi.org/10.3847/2041-8205/832/1/L2}{\emph{\apjl} {\bfseries
  832} (2016) L2} [\href{https://arxiv.org/abs/1609.05916}{{\ttfamily
  1609.05916}}].

\bibitem{Vitale2017}
S.~{Vitale}, R.~{Lynch}, R.~{Sturani} and P.~{Graff}, \emph{{Use of
  gravitational waves to probe the formation channels of compact binaries}},
  \href{https://doi.org/10.1088/1361-6382/aa552e}{\emph{Classical and Quantum
  Gravity} {\bfseries 34} (2017) 03LT01}
  [\href{https://arxiv.org/abs/1503.04307}{{\ttfamily 1503.04307}}].

\bibitem{Farr2017}
W.M.~{Farr}, S.~{Stevenson}, M.C.~{Miller}, I.~{Mandel}, B.~{Farr} and
  A.~{Vecchio}, \emph{{Distinguishing spin-aligned and isotropic black hole
  populations with gravitational waves}},
  \href{https://doi.org/10.1038/nature23453}{\emph{\nat} {\bfseries 548} (2017)
  426} [\href{https://arxiv.org/abs/1706.01385}{{\ttfamily 1706.01385}}].

\bibitem{Bouffanais2021}
Y.~{Bouffanais}, M.~{Mapelli}, F.~{Santoliquido}, N.~{Giacobbo}, U.N.~{Di
  Carlo}, S.~{Rastello} et~al., \emph{{New insights on binary black hole
  formation channels after GWTC-2: young star clusters versus isolated
  binaries}}, \href{https://doi.org/10.1093/mnras/stab2438}{\emph{\mnras}
  {\bfseries 507} (2021) 5224}
  [\href{https://arxiv.org/abs/2102.12495}{{\ttfamily 2102.12495}}].

\bibitem{Baibhav2023}
V.~{Baibhav}, Z.~{Doctor} and V.~{Kalogera}, \emph{{Dropping Anchor:
  Understanding the Populations of Binary Black Holes with Random and
  Aligned-spin Orientations}},
  \href{https://doi.org/10.3847/1538-4357/acbf4c}{\emph{\apj} {\bfseries 946}
  (2023) 50} [\href{https://arxiv.org/abs/2212.12113}{{\ttfamily 2212.12113}}].

\bibitem{Breivik2016}
K.~{Breivik}, C.L.~{Rodriguez}, S.L.~{Larson}, V.~{Kalogera} and F.A.~{Rasio},
  \emph{{Distinguishing between Formation Channels for Binary Black Holes with
  LISA}}, \href{https://doi.org/10.3847/2041-8205/830/1/L18}{\emph{\apjl}
  {\bfseries 830} (2016) L18}
  [\href{https://arxiv.org/abs/1606.09558}{{\ttfamily 1606.09558}}].

\bibitem{2007PhRvL..98w1101G}
J.A.~{Gonz{\'a}lez}, M.~{Hannam}, U.~{Sperhake}, B.~{Br{\"u}gmann} and
  S.~{Husa}, \emph{{Supermassive Recoil Velocities for Binary Black-Hole
  Mergers with Antialigned Spins}},
  \href{https://doi.org/10.1103/PhysRevLett.98.231101}{\emph{\prl} {\bfseries
  98} (2007) 231101} [\href{https://arxiv.org/abs/gr-qc/0702052}{{\ttfamily
  gr-qc/0702052}}].

\bibitem{2007PhRvL..98w1102C}
M.~{Campanelli}, C.O.~{Lousto}, Y.~{Zlochower} and D.~{Merritt}, \emph{{Maximum
  Gravitational Recoil}},
  \href{https://doi.org/10.1103/PhysRevLett.98.231102}{\emph{\prl} {\bfseries
  98} (2007) 231102} [\href{https://arxiv.org/abs/gr-qc/0702133}{{\ttfamily
  gr-qc/0702133}}].

\bibitem{2011PhRvL.107w1102L}
C.O.~{Lousto} and Y.~{Zlochower}, \emph{{Hangup Kicks: Still Larger Recoils by
  Partial Spin-Orbit Alignment of Black-Hole Binaries}},
  \href{https://doi.org/10.1103/PhysRevLett.107.231102}{\emph{\prl} {\bfseries
  107} (2011) 231102} [\href{https://arxiv.org/abs/1108.2009}{{\ttfamily
  1108.2009}}].

\bibitem{Rodriguez2019}
C.L.~{Rodriguez}, M.~{Zevin}, P.~{Amaro-Seoane}, S.~{Chatterjee}, K.~{Kremer},
  F.A.~{Rasio} et~al., \emph{{Black holes: The next
  generation{\textemdash}repeated mergers in dense star clusters and their
  gravitational-wave properties}},
  \href{https://doi.org/10.1103/PhysRevD.100.043027}{\emph{\prd} {\bfseries
  100} (2019) 043027} [\href{https://arxiv.org/abs/1906.10260}{{\ttfamily
  1906.10260}}].

\bibitem{Pretorius2005}
F.~{Pretorius}, \emph{{Evolution of Binary Black-Hole Spacetimes}},
  \href{https://doi.org/10.1103/PhysRevLett.95.121101}{\emph{\prl} {\bfseries
  95} (2005) 121101} [\href{https://arxiv.org/abs/gr-qc/0507014}{{\ttfamily
  gr-qc/0507014}}].

\bibitem{Buonanno2008}
A.~{Buonanno}, L.E.~{Kidder} and L.~{Lehner}, \emph{{Estimating the final spin
  of a binary black hole coalescence}},
  \href{https://doi.org/10.1103/PhysRevD.77.026004}{\emph{\prd} {\bfseries 77}
  (2008) 026004} [\href{https://arxiv.org/abs/0709.3839}{{\ttfamily
  0709.3839}}].

\bibitem{Hofmann2016}
F.~{Hofmann}, E.~{Barausse} and L.~{Rezzolla}, \emph{{The Final Spin from
  Binary Black Holes in Quasi-circular Orbits}},
  \href{https://doi.org/10.3847/2041-8205/825/2/L19}{\emph{\apjl} {\bfseries
  825} (2016) L19} [\href{https://arxiv.org/abs/1605.01938}{{\ttfamily
  1605.01938}}].

\bibitem{FishbachHolzFarr2017}
M.~{Fishbach}, D.E.~{Holz} and B.~{Farr}, \emph{{Are LIGO's Black Holes Made
  from Smaller Black Holes?}},
  \href{https://doi.org/10.3847/2041-8213/aa7045}{\emph{\apjl} {\bfseries 840}
  (2017) L24} [\href{https://arxiv.org/abs/1703.06869}{{\ttfamily
  1703.06869}}].

\bibitem{Antonini2025a}
F.~{Antonini}, I.M.~{Romero-Shaw} and T.~{Callister}, \emph{{Star Cluster
  Population of High Mass Black Hole Mergers in Gravitational Wave Data}},
  \href{https://doi.org/10.1103/PhysRevLett.134.011401}{\emph{\prl} {\bfseries
  134} (2025) 011401} [\href{https://arxiv.org/abs/2406.19044}{{\ttfamily
  2406.19044}}].

\bibitem{Spitzer1967}
L.~{Spitzer}, Jr., \emph{{Equipartition and the Formation of Compact Nuclei in
  Spherical Stellar Systems}},
  \href{https://doi.org/10.1086/180451}{\emph{\apjl} {\bfseries 158} (1969)
  L139}.

\bibitem{BreenHeggie2013}
P.G.~{Breen} and D.C.~{Heggie}, \emph{{Dynamical evolution of black hole
  subsystems in idealized star clusters}},
  \href{https://doi.org/10.1093/mnras/stt628}{\emph{\mnras} {\bfseries 432}
  (2013) 2779} [\href{https://arxiv.org/abs/1304.3401}{{\ttfamily 1304.3401}}].

\bibitem{Morscher2015}
M.~{Morscher}, B.~{Pattabiraman}, C.~{Rodriguez}, F.A.~{Rasio} and
  S.~{Umbreit}, \emph{{The Dynamical Evolution of Stellar Black Holes in
  Globular Clusters}},
  \href{https://doi.org/10.1088/0004-637X/800/1/9}{\emph{ApJ} {\bfseries 800}
  (2015) 9} [\href{https://arxiv.org/abs/1409.0866}{{\ttfamily 1409.0866}}].

\bibitem{Ye2020}
C.S.~{Ye}, W.-f.~{Fong}, K.~{Kremer}, C.L.~{Rodriguez}, S.~{Chatterjee},
  G.~{Fragione} et~al., \emph{{On the Rate of Neutron Star Binary Mergers from
  Globular Clusters}},
  \href{https://doi.org/10.3847/2041-8213/ab5dc5}{\emph{\apjl} {\bfseries 888}
  (2020) L10} [\href{https://arxiv.org/abs/1910.10740}{{\ttfamily
  1910.10740}}].

\bibitem{Clausen2013}
D.~{Clausen}, S.~{Sigurdsson} and D.F.~{Chernoff}, \emph{{Black hole-neutron
  star mergers in globular clusters}},
  \href{https://doi.org/10.1093/mnras/sts295}{\emph{\mnras} {\bfseries 428}
  (2013) 3618} [\href{https://arxiv.org/abs/1210.8153}{{\ttfamily 1210.8153}}].

\bibitem{Rastello2020}
S.~{Rastello}, M.~{Mapelli}, U.N.~{Di Carlo}, N.~{Giacobbo}, F.~{Santoliquido},
  M.~{Spera} et~al., \emph{{Dynamics of black hole-neutron star binaries in
  young star clusters}},
  \href{https://doi.org/10.1093/mnras/staa2018}{\emph{\mnras} {\bfseries 497}
  (2020) 1563} [\href{https://arxiv.org/abs/2003.02277}{{\ttfamily
  2003.02277}}].

\bibitem{ArcaSedda2020}
M.~{Arca Sedda}, \emph{{Dissecting the properties of neutron star-black hole
  mergers originating in dense star clusters}},
  \href{https://doi.org/10.1038/s42005-020-0310-x}{\emph{Communications
  Physics} {\bfseries 3} (2020) 43}
  [\href{https://arxiv.org/abs/2003.02279}{{\ttfamily 2003.02279}}].

\bibitem{GWTC4pop}
{The LIGO-Virgo-KAGRA Collaboration}, \emph{Gwtc-4.0 population paper}, .

\bibitem{HamersThompson2019}
A.S.~{Hamers} and T.A.~{Thompson}, \emph{{Double Neutron Star Mergers from
  Hierarchical Triple-star Systems}},
  \href{https://doi.org/10.3847/1538-4357/ab3b06}{\emph{\apj} {\bfseries 883}
  (2019) 23} [\href{https://arxiv.org/abs/1907.08297}{{\ttfamily 1907.08297}}].

\bibitem{SamsingHotokezaka2021}
J.~{Samsing} and K.~{Hotokezaka}, \emph{{Populating the Black Hole Mass Gaps in
  Stellar Clusters: General Relations and Upper Limits}},
  \href{https://doi.org/10.3847/1538-4357/ac2b27}{\emph{\apj} {\bfseries 923}
  (2021) 126} [\href{https://arxiv.org/abs/2006.09744}{{\ttfamily
  2006.09744}}].

\bibitem{Lu2021}
W.~{Lu}, P.~{Beniamini} and C.~{Bonnerot}, \emph{{On the formation of
  GW190814}}, \href{https://doi.org/10.1093/mnras/staa3372}{\emph{\mnras}
  {\bfseries 500} (2021) 1817}
  [\href{https://arxiv.org/abs/2009.10082}{{\ttfamily 2009.10082}}].

\bibitem{Hamers2021}
A.S.~{Hamers}, G.~{Fragione}, P.~{Neunteufel} and B.~{Kocsis}, \emph{{First-
  and second-generation black hole and neutron star mergers in 2+2 quadruples:
  population statistics}},
  \href{https://doi.org/10.1093/mnras/stab2136}{\emph{\mnras} {\bfseries 506}
  (2021) 5345} [\href{https://arxiv.org/abs/2103.03782}{{\ttfamily
  2103.03782}}].

\bibitem{deMink2009}
S.E.~{de Mink}, M.~{Cantiello}, N.~{Langer}, O.R.~{Pols}, I.~{Brott} and
  S.-C.~{Yoon}, \emph{{Rotational mixing in massive binaries. Detached
  short-period systems}},
  \href{https://doi.org/10.1051/0004-6361/200811439}{\emph{\aap} {\bfseries
  497} (2009) 243} [\href{https://arxiv.org/abs/0902.1751}{{\ttfamily
  0902.1751}}].

\bibitem{Song2013}
H.F.~{Song}, A.~{Maeder}, G.~{Meynet}, R.Q.~{Huang}, S.~{Ekstr{\"o}m} and
  A.~{Granada}, \emph{{Close-binary evolution. I. Tidally induced shear mixing
  in rotating binaries}},
  \href{https://doi.org/10.1051/0004-6361/201321870}{\emph{\aap} {\bfseries
  556} (2013) A100} [\href{https://arxiv.org/abs/1306.6731}{{\ttfamily
  1306.6731}}].

\bibitem{Marigo2001}
P.~{Marigo}, L.~{Girardi}, C.~{Chiosi} and P.R.~{Wood}, \emph{{Zero-metallicity
  stars. I. Evolution at constant mass}},
  \href{https://doi.org/10.1051/0004-6361:20010309}{\emph{\aap} {\bfseries 371}
  (2001) 152} [\href{https://arxiv.org/abs/astro-ph/0102253}{{\ttfamily
  astro-ph/0102253}}].

\bibitem{Belczynski2004}
K.~{Belczynski}, T.~{Bulik} and B.~{Rudak}, \emph{{The First Stellar Binary
  Black Holes: The Strongest Gravitational Wave Burst Sources}},
  \href{https://doi.org/10.1086/422172}{\emph{\apjl} {\bfseries 608} (2004)
  L45} [\href{https://arxiv.org/abs/astro-ph/0403361}{{\ttfamily
  astro-ph/0403361}}].

\bibitem{Kinugawa2014}
T.~{Kinugawa}, K.~{Inayoshi}, K.~{Hotokezaka}, D.~{Nakauchi} and T.~{Nakamura},
  \emph{{Possible indirect confirmation of the existence of Pop III massive
  stars by gravitational wave}},
  \href{https://doi.org/10.1093/mnras/stu1022}{\emph{\mnras} {\bfseries 442}
  (2014) 2963} [\href{https://arxiv.org/abs/1402.6672}{{\ttfamily 1402.6672}}].

\bibitem{Baruteau2011}
C.~{Baruteau}, J.~{Cuadra} and D.N.C.~{Lin}, \emph{{Binaries Migrating in a
  Gaseous Disk: Where are the Galactic Center Binaries?}},
  \href{https://doi.org/10.1088/0004-637X/726/1/28}{\emph{\apj} {\bfseries 726}
  (2011) 28} [\href{https://arxiv.org/abs/1011.0360}{{\ttfamily 1011.0360}}].

\bibitem{Bellovary2016}
J.M.~{Bellovary}, M.-M.~{Mac Low}, B.~{McKernan} and K.E.S.~{Ford},
  \emph{{Migration Traps in Disks around Supermassive Black Holes}},
  \href{https://doi.org/10.3847/2041-8205/819/2/L17}{\emph{\apjl} {\bfseries
  819} (2016) L17} [\href{https://arxiv.org/abs/1511.00005}{{\ttfamily
  1511.00005}}].

\bibitem{Leigh2018}
N.W.C.~{Leigh}, A.M.~{Geller}, B.~{McKernan}, K.E.S.~{Ford}, M.-M.~{Mac Low},
  J.~{Bellovary} et~al., \emph{{On the rate of black hole binary mergers in
  galactic nuclei due to dynamical hardening}},
  \href{https://doi.org/10.1093/mnras/stx3134}{\emph{\mnras} {\bfseries 474}
  (2018) 5672} [\href{https://arxiv.org/abs/1711.10494}{{\ttfamily
  1711.10494}}].

\bibitem{Yang2019}
Y.~{Yang}, I.~{Bartos}, Z.~{Haiman}, B.~{Kocsis}, Z.~{M{\'a}rka}, N.C.~{Stone}
  et~al., \emph{{AGN Disks Harden the Mass Distribution of Stellar-mass Binary
  Black Hole Mergers}},
  \href{https://doi.org/10.3847/1538-4357/ab16e3}{\emph{\apj} {\bfseries 876}
  (2019) 122} [\href{https://arxiv.org/abs/1903.01405}{{\ttfamily
  1903.01405}}].

\bibitem{Secunda2019}
A.~{Secunda}, J.~{Bellovary}, M.-M.~{Mac Low}, K.E.S.~{Ford}, B.~{McKernan},
  N.W.C.~{Leigh} et~al., \emph{{Orbital Migration of Interacting Stellar Mass
  Black Holes in Disks around Supermassive Black Holes}},
  \href{https://doi.org/10.3847/1538-4357/ab20ca}{\emph{\apj} {\bfseries 878}
  (2019) 85} [\href{https://arxiv.org/abs/1807.02859}{{\ttfamily 1807.02859}}].

\bibitem{McKernan2020}
B.~{McKernan}, K.E.S.~{Ford} and R.~{O'Shaughnessy}, \emph{{Black hole, neutron
  star, and white dwarf merger rates in AGN discs}},
  \href{https://doi.org/10.1093/mnras/staa2681}{\emph{\mnras} {\bfseries 498}
  (2020) 4088} [\href{https://arxiv.org/abs/2002.00046}{{\ttfamily
  2002.00046}}].

\bibitem{Cantiello2021}
M.~{Cantiello}, A.S.~{Jermyn} and D.N.C.~{Lin}, \emph{{Stellar Evolution in AGN
  Disks}}, \href{https://doi.org/10.3847/1538-4357/abdf4f}{\emph{\apj}
  {\bfseries 910} (2021) 94}
  [\href{https://arxiv.org/abs/2009.03936}{{\ttfamily 2009.03936}}].

\bibitem{2023MNRAS.524..426I}
G.~{Iorio}, M.~{Mapelli}, G.~{Costa}, M.~{Spera}, G.J.~{Escobar},
  C.~{Sgalletta} et~al., \emph{{Compact object mergers: exploring uncertainties
  from stellar and binary evolution with SEVN}},
  \href{https://doi.org/10.1093/mnras/stad1630}{\emph{\mnras} {\bfseries 524}
  (2023) 426} [\href{https://arxiv.org/abs/2211.11774}{{\ttfamily
  2211.11774}}].

\bibitem{2022ApJ...937..118W}
T.~{Wagg}, F.S.~{Broekgaarden}, S.E.~{de Mink}, N.~{Frankel}, L.A.C.~{van Son}
  and S.~{Justham}, \emph{{Gravitational Wave Sources in Our Galactic Backyard:
  Predictions for BHBH, BHNS, and NSNS Binaries Detectable with LISA}},
  \href{https://doi.org/10.3847/1538-4357/ac8675}{\emph{\apj} {\bfseries 937}
  (2022) 118} [\href{https://arxiv.org/abs/2111.13704}{{\ttfamily
  2111.13704}}].

\bibitem{gwtc-3pop}
{\scshape LIGO Scientific Collaboration, Virgo Collaboration, and KAGRA
  Collaboration} collaboration, \emph{The population of merging compact
  binaries inferred using gravitational waves through gwtc-3},
  \href{https://doi.org/10.1103/PhysRevX.13.011048}{\emph{Phys. Rev. X}
  {\bfseries 13} (2023) 011048}.

\bibitem{Toubiana2023}
A.~{Toubiana}, M.L.~{Katz} and J.R.~{Gair}, \emph{{Is there an excess of black
  holes around 20 M{\ensuremath{\odot}}? Optimizing the complexity of
  population models with the use of reversible jump MCMC.}},
  \href{https://doi.org/10.1093/mnras/stad2215}{\emph{\mnras} {\bfseries 524}
  (2023) 5844} [\href{https://arxiv.org/abs/2305.08909}{{\ttfamily
  2305.08909}}].

\bibitem{MaganaHernandez2024}
I.~{Maga{\~n}a Hernandez} and A.~{Palmese}, \emph{{A new bump in the night:
  evidence of a new feature in the binary black hole mass distribution at
  $70~M_{\odot}$ from gravitational-wave observations}},
  \href{https://doi.org/10.48550/arXiv.2407.02460}{\emph{arXiv e-prints} (2024)
  arXiv:2407.02460} [\href{https://arxiv.org/abs/2407.02460}{{\ttfamily
  2407.02460}}].

\bibitem{Gennari2025}
V.~{Gennari}, S.~{Mastrogiovanni}, N.~{Tamanini}, S.~{Marsat} and G.~{Pierra},
  \emph{{Searching for additional structure and redshift evolution in the
  observed binary black hole population with a parametric time-dependent mass
  distribution}}, \href{https://doi.org/10.1103/ftw9-7xd5}{\emph{\prd}
  {\bfseries 111} (2025) 123046}
  [\href{https://arxiv.org/abs/2502.20445}{{\ttfamily 2502.20445}}].

\bibitem{Fishbach_2020}
M.~Fishbach, W.M.~Farr and D.E.~Holz, \emph{The most massive binary black hole
  detections and the identification of population outliers},
  \href{https://doi.org/10.3847/2041-8213/ab77c9}{\emph{The Astrophysical
  Journal Letters} {\bfseries 891} (2020) L31}.

\bibitem{Veske2021}
D.~{Veske}, I.~{Bartos}, Z.~{M{\'a}rka} and S.~{M{\'a}rka},
  \emph{{Characterizing the Observation Bias in Gravitational-wave Detections
  and Finding Structured Population Properties}},
  \href{https://doi.org/10.3847/1538-4357/ac27ac}{\emph{\apj} {\bfseries 922}
  (2021) 258} [\href{https://arxiv.org/abs/2105.13983}{{\ttfamily
  2105.13983}}].

\bibitem{TiwariFairhurst2021}
V.~{Tiwari} and S.~{Fairhurst}, \emph{{The Emergence of Structure in the Binary
  Black Hole Mass Distribution}},
  \href{https://doi.org/10.3847/2041-8213/abfbe7}{\emph{\apjl} {\bfseries 913}
  (2021) L19} [\href{https://arxiv.org/abs/2011.04502}{{\ttfamily
  2011.04502}}].

\bibitem{Ray_2023}
A.~Ray, I.M.~Hernandez, S.~Mohite, J.~Creighton and S.~Kapadia,
  \emph{Nonparametric inference of the population of compact binaries from
  gravitational-wave observations using binned gaussian processes},
  \href{https://doi.org/10.3847/1538-4357/acf452}{\emph{The Astrophysical
  Journal} {\bfseries 957} (2023) 37}.

\bibitem{PhysRevD.111.L061305}
J.~Heinzel, M.~Mould and S.~Vitale, \emph{Nonparametric analysis of
  correlations in the binary black hole population with ligo-virgo-kagra data},
  \href{https://doi.org/10.1103/PhysRevD.111.L061305}{\emph{Phys. Rev. D}
  {\bfseries 111} (2025) L061305}.

\bibitem{Sadiq_2023}
J.~Sadiq, T.~Dent and M.~Gieles, \emph{Binary vision: The mass distribution of
  merging binary black holes via iterative density estimation},
  \href{https://doi.org/10.3847/1538-4357/ad0ce6}{\emph{The Astrophysical
  Journal} {\bfseries 960} (2023) 65}.

\bibitem{Edelman2023}
B.~{Edelman}, B.~{Farr} and Z.~{Doctor}, \emph{{Cover Your Basis: Comprehensive
  Data-driven Characterization of the Binary Black Hole Population}},
  \href{https://doi.org/10.3847/1538-4357/acb5ed}{\emph{\apj} {\bfseries 946}
  (2023) 16} [\href{https://arxiv.org/abs/2210.12834}{{\ttfamily 2210.12834}}].

\bibitem{Rinaldi_2024}
S.~Rinaldi, W.~Del~Pozzo, M.~Mapelli, A.~Lorenzo-Medina and T.~Dent,
  \emph{Evidence of evolution of the black hole mass function with redshift},
  \href{https://doi.org/10.1051/0004-6361/202348161}{\emph{Astronomy \&amp;
  Astrophysics} {\bfseries 684} (2024) A204}.

\bibitem{CallisterFarr2024}
T.A.~{Callister} and W.M.~{Farr}, \emph{{Parameter-Free Tour of the Binary
  Black Hole Population}},
  \href{https://doi.org/10.1103/PhysRevX.14.021005}{\emph{Physical Review X}
  {\bfseries 14} (2024) 021005}
  [\href{https://arxiv.org/abs/2302.07289}{{\ttfamily 2302.07289}}].

\bibitem{2023MNRAS.526.4130H}
D.D.~{Hendriks}, L.A.C.~{van Son}, M.~{Renzo}, R.G.~{Izzard} and R.~{Farmer},
  \emph{{Pulsational pair-instability supernovae in gravitational-wave and
  electromagnetic transients}},
  \href{https://doi.org/10.1093/mnras/stad2857}{\emph{\mnras} {\bfseries 526}
  (2023) 4130} [\href{https://arxiv.org/abs/2309.09339}{{\ttfamily
  2309.09339}}].

\bibitem{Antonini_2023}
F.~{Antonini}, M.~{Gieles}, F.~{Dosopoulou} and D.~{Chattopadhyay},
  \emph{{Coalescing black hole binaries from globular clusters: mass
  distributions and comparison to gravitational wave data from GWTC-3}},
  \href{https://doi.org/10.1093/mnras/stad972}{\emph{\mnras} {\bfseries 522}
  (2023) 466} [\href{https://arxiv.org/abs/2208.01081}{{\ttfamily
  2208.01081}}].

\bibitem{2025A&A...695A..71L}
E.~{Laplace}, F.R.N.~{Schneider} and P.~{Podsiadlowski}, \emph{{It's written in
  the massive stars: The role of stellar physics in the formation of black
  holes}}, \href{https://doi.org/10.1051/0004-6361/202451077}{\emph{\aap}
  {\bfseries 695} (2025) A71}
  [\href{https://arxiv.org/abs/2409.02058}{{\ttfamily 2409.02058}}].

\bibitem{2025arXiv251007573W}
R.~{Willcox}, F.R.N.~{Schneider}, E.~{Laplace}, P.~{Podsiadlowski},
  K.~{Maltsev}, I.~{Mandel} et~al., \emph{{Good things always come in 3s:
  trimodality in the binary black-hole chirp-mass distribution supports bimodal
  black-hole formation}},
  \href{https://doi.org/10.48550/arXiv.2510.07573}{\emph{arXiv e-prints} (2025)
  arXiv:2510.07573} [\href{https://arxiv.org/abs/2510.07573}{{\ttfamily
  2510.07573}}].

\bibitem{Roy2025}
S.~{Kishore Roy}, L.A.C.~{van Son} and W.M.~{Farr}, \emph{{A mid-thirties
  crisis: dissecting the properties of gravitational wave sources near the 35
  solar mass peak}},
  \href{https://doi.org/10.1088/1361-6382/ae1921}{\emph{Classical and Quantum
  Gravity} {\bfseries 42} (2025) 225008}
  [\href{https://arxiv.org/abs/2507.01086}{{\ttfamily 2507.01086}}].

\bibitem{2025arXiv250915646B}
S.~{Banagiri}, E.~{Thrane} and P.D.~{Lasky}, \emph{{Evidence for Three
  Subpopulations of Merging Binary Black Holes at Different Primary Masses}},
  \href{https://doi.org/10.48550/arXiv.2509.15646}{\emph{arXiv e-prints} (2025)
  arXiv:2509.15646} [\href{https://arxiv.org/abs/2509.15646}{{\ttfamily
  2509.15646}}].

\bibitem{2025arXiv251022698W}
Y.-Z.~{Wang}, Y.-J.~{Li}, S.-J.~{Gao}, S.-P.~{Tang} and Y.-Z.~{Fan}, \emph{{A
  new group of low-spin $50-70M_\odot$ Black Holes and the high
  pair-instability mass cutoff}},
  \href{https://doi.org/10.48550/arXiv.2510.22698}{\emph{arXiv e-prints} (2025)
  arXiv:2510.22698} [\href{https://arxiv.org/abs/2510.22698}{{\ttfamily
  2510.22698}}].

\bibitem{2025arXiv251018867R}
A.~{Ray} and V.~{Kalogera}, \emph{{Reexamining Evidence of a Pair-Instability
  Mass Gap in the Binary Black Hole Population}},
  \href{https://doi.org/10.48550/arXiv.2510.18867}{\emph{arXiv e-prints} (2025)
  arXiv:2510.18867} [\href{https://arxiv.org/abs/2510.18867}{{\ttfamily
  2510.18867}}].

\bibitem{2025arXiv250904637A}
F.~{Antonini}, I.~{Romero-Shaw}, T.~{Callister}, F.~{Dosopoulou},
  D.~{Chattopadhyay}, M.~{Gieles} et~al., \emph{{Gravitational waves reveal the
  pair-instability mass gap and constrain nuclear burning in massive stars}},
  \href{https://doi.org/10.48550/arXiv.2509.04637}{\emph{arXiv e-prints} (2025)
  arXiv:2509.04637} [\href{https://arxiv.org/abs/2509.04637}{{\ttfamily
  2509.04637}}].

\bibitem{2025arXiv251105316T}
H.~{Tong}, T.A.~{Callister}, M.~{Fishbach}, E.~{Thrane}, F.~{Antonini},
  S.~{Stevenson} et~al., \emph{{A subpopulation of low-mass, spinning black
  holes: signatures of dynamical assembly}},
  \href{https://doi.org/10.48550/arXiv.2511.05316}{\emph{arXiv e-prints} (2025)
  arXiv:2511.05316} [\href{https://arxiv.org/abs/2511.05316}{{\ttfamily
  2511.05316}}].

\bibitem{2025ApJ...994L..47S}
J.~{Stegmann}, D.~{Gerosa}, I.~{Romero-Shaw}, G.~{Fumagalli}, H.~{Tagawa} and
  L.~{Zwick}, \emph{{Distinguishing the Origin of Eccentric Black Hole Mergers
  with Gravitational-wave Spin Measurements}},
  \href{https://doi.org/10.3847/2041-8213/ae1d66}{\emph{\apjl} {\bfseries 994}
  (2025) L47} [\href{https://arxiv.org/abs/2505.13589}{{\ttfamily
  2505.13589}}].

\bibitem{Bailyn1998}
C.D.~{Bailyn}, R.K.~{Jain}, P.~{Coppi} and J.A.~{Orosz}, \emph{{The Mass
  Distribution of Stellar Black Holes}},
  \href{https://doi.org/10.1086/305614}{\emph{\apj} {\bfseries 499} (1998) 367}
  [\href{https://arxiv.org/abs/astro-ph/9708032}{{\ttfamily
  astro-ph/9708032}}].

\bibitem{Ozel2010}
F.~{{\"O}zel}, D.~{Psaltis}, R.~{Narayan} and J.E.~{McClintock}, \emph{{The
  Black Hole Mass Distribution in the Galaxy}},
  \href{https://doi.org/10.1088/0004-637X/725/2/1918}{\emph{\apj} {\bfseries
  725} (2010) 1918} [\href{https://arxiv.org/abs/1006.2834}{{\ttfamily
  1006.2834}}].

\bibitem{Kreidberg2012}
L.~{Kreidberg}, C.D.~{Bailyn}, W.M.~{Farr} and V.~{Kalogera}, \emph{{Mass
  Measurements of Black Holes in X-Ray Transients: Is There a Mass Gap?}},
  \href{https://doi.org/10.1088/0004-637X/757/1/36}{\emph{\apj} {\bfseries 757}
  (2012) 36} [\href{https://arxiv.org/abs/1205.1805}{{\ttfamily 1205.1805}}].

\bibitem{WyrzykowskiMandel2020}
L.~Wyrzykowski and I.~Mandel, \emph{{Constraining the masses of microlensing
  black holes and the mass gap with Gaia DR2}},
  \href{https://doi.org/10.1051/0004-6361/201935842}{\emph{Astronomy \&
  Astrophysics} {\bfseries 636} (2020) A20}
  [\href{https://arxiv.org/abs/1904.07789}{{\ttfamily 1904.07789}}].

\bibitem{Jonker2021}
P.G.~{Jonker}, K.~{Kaur}, N.~{Stone} and M.A.P.~{Torres}, \emph{{The Observed
  Mass Distribution of Galactic Black Hole LMXBs Is Biased against Massive
  Black Holes}}, \href{https://doi.org/10.3847/1538-4357/ac2839}{\emph{\apj}
  {\bfseries 921} (2021) 131}
  [\href{https://arxiv.org/abs/2104.03596}{{\ttfamily 2104.03596}}].

\bibitem{Siegel2023}
J.C.~{Siegel}, I.~{Kiato}, V.~{Kalogera}, C.P.L.~{Berry}, T.J.~{Maccarone},
  K.~{Breivik} et~al., \emph{{Investigating the Lower Mass Gap with Low-mass
  X-Ray Binary Population Synthesis}},
  \href{https://doi.org/10.3847/1538-4357/ace9d9}{\emph{\apj} {\bfseries 954}
  (2023) 212} [\href{https://arxiv.org/abs/2209.06844}{{\ttfamily
  2209.06844}}].

\bibitem{Fishbach2017}
M.~{Fishbach} and D.E.~{Holz}, \emph{{Where Are LIGO{\textquoteright}s Big
  Black Holes?}}, \href{https://doi.org/10.3847/2041-8213/aa9bf6}{\emph{\apjl}
  {\bfseries 851} (2017) L25}
  [\href{https://arxiv.org/abs/1709.08584}{{\ttfamily 1709.08584}}].

\bibitem{SchneiderPodsiadlowskiLaplace2023}
F.R.N.~{Schneider}, P.~{Podsiadlowski} and E.~{Laplace}, \emph{{Bimodal Black
  Hole Mass Distribution and Chirp Masses of Binary Black Hole Mergers}},
  \href{https://doi.org/10.3847/2041-8213/acd77a}{\emph{\apjl} {\bfseries 950}
  (2023) L9} [\href{https://arxiv.org/abs/2305.02380}{{\ttfamily 2305.02380}}].

\bibitem{Kochanek2014}
C.S.~{Kochanek}, \emph{{Failed Supernovae Explain the Compact Remnant Mass
  Function}}, \href{https://doi.org/10.1088/0004-637X/785/1/28}{\emph{\apj}
  {\bfseries 785} (2014) 28} [\href{https://arxiv.org/abs/1308.0013}{{\ttfamily
  1308.0013}}].

\bibitem{Fryer2012}
C.L.~{Fryer}, K.~{Belczynski}, G.~{Wiktorowicz}, M.~{Dominik}, V.~{Kalogera}
  and D.E.~{Holz}, \emph{{Compact Remnant Mass Function: Dependence on the
  Explosion Mechanism and Metallicity}},
  \href{https://doi.org/10.1088/0004-637X/749/1/91}{\emph{\apj} {\bfseries 749}
  (2012) 91} [\href{https://arxiv.org/abs/1110.1726}{{\ttfamily 1110.1726}}].

\bibitem{2020A&A...639L...6S}
T.~{Shenar}, J.~{Bodensteiner}, M.~{Abdul-Masih}, M.~{Fabry}, L.~{Mahy},
  P.~{Marchant} et~al., \emph{{The ``hidden'' companion in LB-1 unveiled by
  spectral disentangling}},
  \href{https://doi.org/10.1051/0004-6361/202038275}{\emph{\aap} {\bfseries
  639} (2020) L6} [\href{https://arxiv.org/abs/2004.12882}{{\ttfamily
  2004.12882}}].

\bibitem{2022MNRAS.511L..24E}
K.~{El-Badry} and K.B.~{Burdge}, \emph{{NGC 1850 BH1 is another stripped-star
  binary masquerading as a black hole}},
  \href{https://doi.org/10.1093/mnrasl/slab135}{\emph{\mnras} {\bfseries 511}
  (2022) 24} [\href{https://arxiv.org/abs/2111.07925}{{\ttfamily 2111.07925}}].

\bibitem{Barr2024}
E.D.~{Barr}, A.~{Dutta}, P.C.C.~{Freire}, M.~{Cadelano}, T.~{Gautam},
  M.~{Kramer} et~al., \emph{{A pulsar in a binary with a compact object in the
  mass gap between neutron stars and black holes}},
  \href{https://doi.org/10.1126/science.adg3005}{\emph{Science} {\bfseries 383}
  (2024) 275} [\href{https://arxiv.org/abs/2401.09872}{{\ttfamily
  2401.09872}}].

\bibitem{ElBadry2024_GaiaNSs}
K.~{El-Badry}, H.-W.~{Rix}, D.W.~{Latham}, S.~{Shahaf}, T.~{Mazeh},
  A.~{Bieryla} et~al., \emph{{A population of neutron star candidates in wide
  orbits from Gaia astrometry}},
  \href{https://doi.org/10.33232/001c.121261}{\emph{The Open Journal of
  Astrophysics} {\bfseries 7} (2024) 58}
  [\href{https://arxiv.org/abs/2405.00089}{{\ttfamily 2405.00089}}].

\bibitem{2020MNRAS.499.3214M}
I.~{Mandel} and B.~{M{\"u}ller}, \emph{{Simple recipes for compact remnant
  masses and natal kicks}},
  \href{https://doi.org/10.1093/mnras/staa3043}{\emph{\mnras} {\bfseries 499}
  (2020) 3214} [\href{https://arxiv.org/abs/2006.08360}{{\ttfamily
  2006.08360}}].

\bibitem{ClaireYe2026}
C.S.~{Ye}, M.~{Fishbach}, K.~{Kremer} and M.~{Reina-Campos}, \emph{{Mass
  Distribution of Binary Black Hole Mergers from Young and Old Dense Star
  Clusters}}, \href{https://doi.org/10.3847/1538-4357/ae317f}{\emph{\apj}
  {\bfseries 997} (2026) 267}
  [\href{https://arxiv.org/abs/2507.07183}{{\ttfamily 2507.07183}}].

\bibitem{2026arXiv260614846O}
C.E.~{O'Connor}, K.~{Kremer}, S.~{Agrawal}, E.~{Gonz{\'a}lez Prieto},
  F.~{Kiro{\u{g}}lu}, M.A.S.~{Martinez} et~al., \emph{{Black hole mergers from
  dense star clusters with realistic binary populations}},
  \href{https://doi.org/10.48550/arXiv.2606.14846}{\emph{arXiv e-prints} (2026)
  arXiv:2606.14846} [\href{https://arxiv.org/abs/2606.14846}{{\ttfamily
  2606.14846}}].

\bibitem{2011ApJ...730...70O}
E.~{O'Connor} and C.D.~{Ott}, \emph{{Black Hole Formation in Failing
  Core-Collapse Supernovae}},
  \href{https://doi.org/10.1088/0004-637X/730/2/70}{\emph{\apj} {\bfseries 730}
  (2011) 70} [\href{https://arxiv.org/abs/1010.5550}{{\ttfamily 1010.5550}}].

\bibitem{2012ApJ...757...69U}
M.~{Ugliano}, H.-T.~{Janka}, A.~{Marek} and A.~{Arcones},
  \emph{{Progenitor-explosion Connection and Remnant Birth Masses for
  Neutrino-driven Supernovae of Iron-core Progenitors}},
  \href{https://doi.org/10.1088/0004-637X/757/1/69}{\emph{\apj} {\bfseries 757}
  (2012) 69} [\href{https://arxiv.org/abs/1205.3657}{{\ttfamily 1205.3657}}].

\bibitem{2014ApJ...783...10S}
T.~{Sukhbold} and S.E.~{Woosley}, \emph{{The Compactness of Presupernova
  Stellar Cores}},
  \href{https://doi.org/10.1088/0004-637X/783/1/10}{\emph{\apj} {\bfseries 783}
  (2014) 10} [\href{https://arxiv.org/abs/1311.6546}{{\ttfamily 1311.6546}}].

\bibitem{2018ApJ...860...93S}
T.~{Sukhbold}, S.E.~{Woosley} and A.~{Heger}, \emph{{A High-resolution Study of
  Presupernova Core Structure}},
  \href{https://doi.org/10.3847/1538-4357/aac2da}{\emph{\apj} {\bfseries 860}
  (2018) 93} [\href{https://arxiv.org/abs/1710.03243}{{\ttfamily 1710.03243}}].

\bibitem{2020ApJ...890...43C}
A.~{Chieffi} and M.~{Limongi}, \emph{{The Presupernova Core Mass-Radius
  Relation of Massive Stars: Understanding Its Formation and Evolution}},
  \href{https://doi.org/10.3847/1538-4357/ab6739}{\emph{\apj} {\bfseries 890}
  (2020) 43} [\href{https://arxiv.org/abs/1911.08988}{{\ttfamily 1911.08988}}].

\bibitem{2021A&A...645A...5S}
F.R.N.~{Schneider}, P.~{Podsiadlowski} and B.~{M{\"u}ller},
  \emph{{Pre-supernova evolution, compact-object masses, and explosion
  properties of stripped binary stars}},
  \href{https://doi.org/10.1051/0004-6361/202039219}{\emph{\aap} {\bfseries
  645} (2021) A5} [\href{https://arxiv.org/abs/2008.08599}{{\ttfamily
  2008.08599}}].

\bibitem{2024A&A...686A..45S}
F.R.N.~{Schneider}, P.~{Podsiadlowski} and E.~{Laplace}, \emph{{Pre-supernova
  evolution and final fate of stellar mergers and accretors of binary mass
  transfer}}, \href{https://doi.org/10.1051/0004-6361/202347854}{\emph{\aap}
  {\bfseries 686} (2024) A45}
  [\href{https://arxiv.org/abs/2403.03984}{{\ttfamily 2403.03984}}].

\bibitem{2016ApJS..227...22F}
R.~{Farmer}, C.E.~{Fields}, I.~{Petermann}, L.~{Dessart}, M.~{Cantiello},
  B.~{Paxton} et~al., \emph{{On Variations Of Pre-supernova Model Properties}},
  \href{https://doi.org/10.3847/1538-4365/227/2/22}{\emph{\apjs} {\bfseries
  227} (2016) 22} [\href{https://arxiv.org/abs/1611.01207}{{\ttfamily
  1611.01207}}].

\bibitem{2024A&A...682A.123T}
D.~{Temaj}, F.R.N.~{Schneider}, E.~{Laplace}, D.~{Wei} and P.~{Podsiadlowski},
  \emph{{Convective-core overshooting and the final fate of massive stars}},
  \href{https://doi.org/10.1051/0004-6361/202347434}{\emph{\aap} {\bfseries
  682} (2024) A123} [\href{https://arxiv.org/abs/2311.05701}{{\ttfamily
  2311.05701}}].

\bibitem{2024RNAAS...8..152R}
M.~{Renzo}, J.A.~{Goldberg}, A.~{Grichener}, O.~{Gottlieb} and M.~{Cantiello},
  \emph{{Progenitor Stars Calculated with Small Reaction Networks should not be
  Used as Initial Conditions for Core Collapse}},
  \href{https://doi.org/10.3847/2515-5172/ad530e}{\emph{Research Notes of the
  American Astronomical Society} {\bfseries 8} (2024) 152}
  [\href{https://arxiv.org/abs/2406.02590}{{\ttfamily 2406.02590}}].

\bibitem{2016MNRAS.460..742M}
B.~{M{\"u}ller}, A.~{Heger}, D.~{Liptai} and J.B.~{Cameron}, \emph{{A simple
  approach to the supernova progenitor-explosion connection}},
  \href{https://doi.org/10.1093/mnras/stw1083}{\emph{\mnras} {\bfseries 460}
  (2016) 742} [\href{https://arxiv.org/abs/1602.05956}{{\ttfamily
  1602.05956}}].

\bibitem{2016ApJ...818..124E}
T.~{Ertl}, H.-T.~{Janka}, S.E.~{Woosley}, T.~{Sukhbold} and M.~{Ugliano},
  \emph{{A Two-parameter Criterion for Classifying the Explodability of Massive
  Stars by the Neutrino-driven Mechanism}},
  \href{https://doi.org/10.3847/0004-637X/818/2/124}{\emph{\apj} {\bfseries
  818} (2016) 124} [\href{https://arxiv.org/abs/1503.07522}{{\ttfamily
  1503.07522}}].

\bibitem{2025A&A...700A..20M}
K.~{Maltsev}, F.R.N.~{Schneider}, I.~{Mandel}, B.~{M{\"u}ller}, A.~{Heger},
  F.K.~{R{\"o}pke} et~al., \emph{{Explodability criteria for the
  neutrino-driven supernova mechanism}},
  \href{https://doi.org/10.1051/0004-6361/202554931}{\emph{\aap} {\bfseries
  700} (2025) A20} [\href{https://arxiv.org/abs/2503.23856}{{\ttfamily
  2503.23856}}].

\bibitem{2020MNRAS.499.2803P}
R.A.~{Patton} and T.~{Sukhbold}, \emph{{Towards a realistic explosion landscape
  for binary population synthesis}},
  \href{https://doi.org/10.1093/mnras/staa3029}{\emph{\mnras} {\bfseries 499}
  (2020) 2803} [\href{https://arxiv.org/abs/2005.03055}{{\ttfamily
  2005.03055}}].

\bibitem{Disberg2023}
P.~{Disberg} and G.~{Nelemans}, \emph{{Failed supernovae as a natural
  explanation for the binary black hole mass distribution}},
  \href{https://doi.org/10.1051/0004-6361/202245693}{\emph{\aap} {\bfseries
  676} (2023) A31} [\href{https://arxiv.org/abs/2306.14332}{{\ttfamily
  2306.14332}}].

\bibitem{2025A&A...694A.186G}
S.~{Galaudage} and A.~{Lamberts}, \emph{{Compactness peaks: An astrophysical
  interpretation of the mass distribution of merging binary black holes}},
  \href{https://doi.org/10.1051/0004-6361/202451654}{\emph{\aap} {\bfseries
  694} (2025) A186} [\href{https://arxiv.org/abs/2407.17561}{{\ttfamily
  2407.17561}}].

\bibitem{2026arXiv260525994G}
S.~{Galaudage}, \emph{{Compactness Peaks and Subpopulations: Probing Stellar
  Physics and Formation Channels of Merging Binary Black Holes}},
  \href{https://doi.org/10.48550/arXiv.2605.25994}{\emph{arXiv e-prints} (2026)
  arXiv:2605.25994} [\href{https://arxiv.org/abs/2605.25994}{{\ttfamily
  2605.25994}}].

\bibitem{2025arXiv250820787W}
R.~{Willcox}, F.R.N.~{Schneider}, E.~{Laplace}, P.~{Podsiadlowski},
  K.~{Maltsev}, I.~{Mandel} et~al., \emph{{New gravitational-wave data support
  a bimodal black-hole mass distribution}},
  \href{https://doi.org/10.48550/arXiv.2508.20787}{\emph{arXiv e-prints} (2025)
  arXiv:2508.20787} [\href{https://arxiv.org/abs/2508.20787}{{\ttfamily
  2508.20787}}].

\bibitem{2026arXiv260401420L}
I.~{Legred}, J.~{Golomb} and K.~{Chatziioannou}, \emph{{Low-mass failed
  supernovae and the $10\,M_{\odot}$ peak in the merging black hole mass
  distribution}}, \href{https://doi.org/10.48550/arXiv.2604.01420}{\emph{arXiv
  e-prints} (2026) arXiv:2604.01420}
  [\href{https://arxiv.org/abs/2604.01420}{{\ttfamily 2604.01420}}].

\bibitem{Farah2021}
A.~{Farah}, M.~{Fishbach}, R.~{Essick}, D.E.~{Holz} and S.~{Galaudage},
  \emph{{Bridging the Gap: Categorizing Gravitational-wave Events at the
  Transition between Neutron Stars and Black Holes}},
  \href{https://doi.org/10.3847/1538-4357/ac5f03}{\emph{\apj} {\bfseries 931}
  (2022) 108} [\href{https://arxiv.org/abs/2111.03498}{{\ttfamily
  2111.03498}}].

\bibitem{2022ApJ...938...45B}
F.S.~{Broekgaarden}, S.~{Stevenson} and E.~{Thrane}, \emph{{Signatures of Mass
  Ratio Reversal in Gravitational Waves from Merging Binary Black Holes}},
  \href{https://doi.org/10.3847/1538-4357/ac8879}{\emph{\apj} {\bfseries 938}
  (2022) 45} [\href{https://arxiv.org/abs/2205.01693}{{\ttfamily 2205.01693}}].

\bibitem{2022MNRAS.517.2738M}
M.~{Mould}, D.~{Gerosa}, F.S.~{Broekgaarden} and N.~{Steinle}, \emph{{Which
  black hole formed first? Mass-ratio reversal in massive binary stars from
  gravitational-wave data}},
  \href{https://doi.org/10.1093/mnras/stac2859}{\emph{\mnras} {\bfseries 517}
  (2022) 2738} [\href{https://arxiv.org/abs/2205.12329}{{\ttfamily
  2205.12329}}].

\bibitem{Maclean2026arXiv260628506M}
D.B.~{Maclean}, P.~{Agrawal}, K.~{Breivik}, A.G.~{Guerrero}, M.~{Zevin},
  M.~{Renzo} et~al., \emph{{Twin Peaks: Resolving Features in the Binary Black
  Hole Mass Function with COSMIC-METISSE}},
  \href{https://doi.org/10.48550/arXiv.2606.28506}{\emph{arXiv e-prints} (2026)
  arXiv:2606.28506} [\href{https://arxiv.org/abs/2606.28506}{{\ttfamily
  2606.28506}}].

\bibitem{2026arXiv260901784G}
M.~{Gallegos-Garcia}, A.~{Ray}, V.~{Kalogera}, M.~{Briel}, M.~{Zevin},
  A.~{Chattaraj} et~al., \emph{{The Mass-Ratio Distribution of the Low-Mass
  Binary Black Hole Subpopulation: A Natural Outcome of Isolated Binary
  Evolution}}, \href{https://doi.org/10.48550/arXiv.2609.01784}{\emph{arXiv
  e-prints} (2026) arXiv:2609.01784}
  [\href{https://arxiv.org/abs/2609.01784}{{\ttfamily 2609.01784}}].

\bibitem{2026arXiv260521580S}
T.B.~{Smith}, F.~{Broekgaarden}, S.~{Levina}, A.~{Romagnolo}, M.~{Komandur},
  M.~{Santiago} et~al., \emph{{Massquerade: Impacts of Mass Ratio Reversals on
  Binary Black Hole Merger Rates and Mass Distributions}},
  \href{https://doi.org/10.48550/arXiv.2605.21580}{\emph{arXiv e-prints} (2026)
  arXiv:2605.21580} [\href{https://arxiv.org/abs/2605.21580}{{\ttfamily
  2605.21580}}].

\bibitem{2024A&A...689A.305O}
A.~{Olejak}, J.~{Klencki}, X.-T.~{Xu}, C.~{Wang}, K.~{Belczynski} and
  J.-P.~{Lasota}, \emph{{Unequal-mass highly spinning binary black hole mergers
  in the stable mass transfer formation channel}},
  \href{https://doi.org/10.1051/0004-6361/202450480}{\emph{\aap} {\bfseries
  689} (2024) A305} [\href{https://arxiv.org/abs/2404.12426}{{\ttfamily
  2404.12426}}].

\bibitem{2023ApJ...958...13A}
C.~{Adamcewicz}, P.D.~{Lasky} and E.~{Thrane}, \emph{{Evidence for a
  Correlation between Binary Black Hole Mass Ratio and Black Hole Spins}},
  \href{https://doi.org/10.3847/1538-4357/acf763}{\emph{\apj} {\bfseries 958}
  (2023) 13} [\href{https://arxiv.org/abs/2307.15278}{{\ttfamily 2307.15278}}].

\bibitem{2024ApJ...964L...6A}
C.~{Adamcewicz}, S.~{Galaudage}, P.D.~{Lasky} and E.~{Thrane}, \emph{{Which
  Black Hole Is Spinning? Probing the Origin of Black Hole Spin with
  Gravitational Waves}},
  \href{https://doi.org/10.3847/2041-8213/ad2df2}{\emph{\apjl} {\bfseries 964}
  (2024) L6} [\href{https://arxiv.org/abs/2311.05182}{{\ttfamily 2311.05182}}].

\bibitem{2026arXiv260700565F}
E.~{Flanagan}, J.~{Stegmann}, I.~{Romero-Shaw}, T.~{Callister}, A.~{Olejak} and
  F.~{Antonini}, \emph{{Distinct spin properties and astrophysical origin of
  low mass binary black holes in gravitational wave data}},
  \href{https://doi.org/10.48550/arXiv.2607.00565}{\emph{arXiv e-prints} (2026)
  arXiv:2607.00565} [\href{https://arxiv.org/abs/2607.00565}{{\ttfamily
  2607.00565}}].

\bibitem{2026ApJ..1007L..45W}
N.E.~{Wolfe}, S.~{Vitale}, M.~{Zevin} and {Spins (Society Of Physicists
  Interested in Non-Aligned Spins)}, \emph{{No Model-independent Evidence for a
  Peak in Binary Black Hole Spin (Mis)alignments}},
  \href{https://doi.org/10.3847/2041-8213/ae90ae}{\emph{\apjl} {\bfseries 1007}
  (2026) L45} [\href{https://arxiv.org/abs/2605.05300}{{\ttfamily
  2605.05300}}].

\bibitem{Stevenson2019}
S.~{Stevenson}, M.~{Sampson}, J.~{Powell}, A.~{Vigna-G{\'o}mez},
  C.J.~{Neijssel}, D.~{Sz{\'e}csi} et~al., \emph{{The Impact of
  Pair-instability Mass Loss on the Binary Black Hole Mass Distribution}},
  \href{https://doi.org/10.3847/1538-4357/ab3981}{\emph{\apj} {\bfseries 882}
  (2019) 121} [\href{https://arxiv.org/abs/1904.02821}{{\ttfamily
  1904.02821}}].

\bibitem{Woosley2020}
S.E.~{Woosley} and A.~{Heger}, \emph{{The Pair-instability Mass Gap for Black
  Holes}}, \href{https://doi.org/10.3847/2041-8213/abf2c4}{\emph{\apjl}
  {\bfseries 912} (2021) L31}
  [\href{https://arxiv.org/abs/2103.07933}{{\ttfamily 2103.07933}}].

\bibitem{deBoer2017}
R.J.~{deBoer}, J.~{G{\"o}rres}, M.~{Wiescher}, R.E.~{Azuma}, A.~{Best},
  C.R.~{Brune} et~al., \emph{{The $^{12}$C({\ensuremath{\alpha}}
  ,{\ensuremath{\gamma}} )$^{16}$O reaction and its implications for stellar
  helium burning}},
  \href{https://doi.org/10.1103/RevModPhys.89.035007}{\emph{Reviews of Modern
  Physics} {\bfseries 89} (2017) 035007}
  [\href{https://arxiv.org/abs/1709.03144}{{\ttfamily 1709.03144}}].

\bibitem{Shen2023}
Y.~{Shen}, B.~{Guo}, R.J.~{deBoer}, E.~{Li}, Z.~{Li}, Y.~{Li} et~al.,
  \emph{{New Determination of the $^{12}$C({\ensuremath{\alpha}},
  {\ensuremath{\gamma}})$^{16}$O Reaction Rate and Its Impact on the Black-hole
  Mass Gap}}, \href{https://doi.org/10.3847/1538-4357/acb7de}{\emph{\apj}
  {\bfseries 945} (2023) 41}.

\bibitem{Winch2024}
E.R.J.~{Winch}, J.S.~{Vink}, E.R.~{Higgins} and G.N.~{Sabhahitf},
  \emph{{Predicting the heaviest black holes below the pair instability gap}},
  \href{https://doi.org/10.1093/mnras/stae393}{\emph{\mnras} {\bfseries 529}
  (2024) 2980} [\href{https://arxiv.org/abs/2401.17327}{{\ttfamily
  2401.17327}}].

\bibitem{2024ApJ...976..121G}
J.~{Golomb}, M.~{Isi} and W.M.~{Farr}, \emph{{Physical Models for the
  Astrophysical Population of Black Holes: Application to the Bump in the Mass
  Distribution of Gravitational-wave Sources}},
  \href{https://doi.org/10.3847/1538-4357/ad8572}{\emph{\apj} {\bfseries 976}
  (2024) 121} [\href{https://arxiv.org/abs/2312.03973}{{\ttfamily
  2312.03973}}].

\bibitem{Hendriks2023}
D.D.~{Hendriks}, L.A.C.~{van Son} et~al., \emph{Pulsational pair-instability
  supernovae in gravitational-wave and electromagnetic transients}, .

\bibitem{2023MNRAS.520.5724B}
M.M.~{Briel}, H.F.~{Stevance} and J.J.~{Eldridge}, \emph{{Understanding the
  high-mass binary black hole population from stable mass transfer and
  super-Eddington accretion in BPASS}},
  \href{https://doi.org/10.1093/mnras/stad399}{\emph{\mnras} {\bfseries 520}
  (2023) 5724} [\href{https://arxiv.org/abs/2206.13842}{{\ttfamily
  2206.13842}}].

\bibitem{Mahapatra2025}
P.~{Mahapatra}, D.~{Chattopadhyay}, A.~{Gupta}, M.~{Favata},
  B.S.~{Sathyaprakash} and K.G.~{Arun}, \emph{{Predictions of a simple
  parametric model of hierarchical black hole mergers}},
  \href{https://doi.org/10.1103/PhysRevD.111.023013}{\emph{\prd} {\bfseries
  111} (2025) 023013} [\href{https://arxiv.org/abs/2209.05766}{{\ttfamily
  2209.05766}}].

\bibitem{AntoniniGieles2020}
F.~{Antonini} and M.~{Gieles}, \emph{{Population synthesis of black hole binary
  mergers from star clusters}},
  \href{https://doi.org/10.1093/mnras/stz3584}{\emph{\mnras} {\bfseries 492}
  (2020) 2936} [\href{https://arxiv.org/abs/1906.11855}{{\ttfamily
  1906.11855}}].

\bibitem{Antonini2023}
F.~{Antonini}, M.~{Gieles}, F.~{Dosopoulou} and D.~{Chattopadhyay},
  \emph{{Coalescing black hole binaries from globular clusters: mass
  distributions and comparison to gravitational wave data from GWTC-3}},
  \href{https://doi.org/10.1093/mnras/stad972}{\emph{\mnras} {\bfseries 522}
  (2023) 466} [\href{https://arxiv.org/abs/2208.01081}{{\ttfamily
  2208.01081}}].

\bibitem{2026ApJ..1005L..54S}
O.~{Sridhar}, A.~{Ray} and V.~{Kalogera}, \emph{{Characterizing Binary Black
  Hole Subpopulations in GWTC-4 with Binned Gaussian Processes: On the Origins
  of the 35 M$_{{\ensuremath{\odot}}}$ Peak}},
  \href{https://doi.org/10.3847/2041-8213/ae8011}{\emph{\apjl} {\bfseries 1005}
  (2026) L54} [\href{https://arxiv.org/abs/2511.22093}{{\ttfamily
  2511.22093}}].

\bibitem{2026ApJ..1005L..55R}
A.~{Ray}, S.~{Mukherjee}, M.~{Zevin} and V.~{Kalogera}, \emph{{On the
  Astrophysical Origin of Binary Black Hole Subpopulations: A Tale of Three
  Channels?}}, \href{https://doi.org/10.3847/2041-8213/ae80cb}{\emph{\apjl}
  {\bfseries 1005} (2026) L55}
  [\href{https://arxiv.org/abs/2603.17987}{{\ttfamily 2603.17987}}].

\bibitem{Banagiri2026}
S.~{Banagiri}, E.~{Thrane} and P.D.~{Lasky}, \emph{{Evidence for Three
  Subpopulations of Merging Binary Black Holes at Different Primary Masses}},
  \href{https://doi.org/10.1103/blyb-lqv6}{\emph{\prl} {\bfseries 137} (2026)
  021403} [\href{https://arxiv.org/abs/2509.15646}{{\ttfamily 2509.15646}}].

\bibitem{2026arXiv260722011G}
N.~{Guttman}, P.D.~{Lasky} and E.~{Thrane}, \emph{{Revealing Four
  Subpopulations of Binary Black-Hole Mergers with the Fifth Gravitational-Wave
  Transient Catalog}},
  \href{https://doi.org/10.48550/arXiv.2607.22011}{\emph{arXiv e-prints} (2026)
  arXiv:2607.22011} [\href{https://arxiv.org/abs/2607.22011}{{\ttfamily
  2607.22011}}].

\bibitem{2026arXiv260516510D}
U.~{Das} and S.~{Mukherjee}, \emph{{Is the Binary Black Hole Population
  Inference from Gravitational-Wave Data Robust?}},
  \href{https://doi.org/10.48550/arXiv.2605.16510}{\emph{arXiv e-prints} (2026)
  arXiv:2605.16510} [\href{https://arxiv.org/abs/2605.16510}{{\ttfamily
  2605.16510}}].

\bibitem{2025arXiv250819208M}
I.~{Maga{\~n}a Hernandez} and A.~{Palmese}, \emph{{Astrophysics informed
  Gaussian processes for gravitational-wave populations: Evidence for the onset
  of the pair-instability supernova mass gap}},
  \href{https://doi.org/10.48550/arXiv.2508.19208}{\emph{arXiv e-prints} (2025)
  arXiv:2508.19208} [\href{https://arxiv.org/abs/2508.19208}{{\ttfamily
  2508.19208}}].

\bibitem{2026arXiv260306566A}
A.J.~{Amsellem}, I.~{Maga{\~n}a Hernandez}, A.~{Palmese} and J.~{Gassert},
  \emph{{Data-Driven Trends and Subpopulations in the Gravitational Wave Binary
  Black Hole Merger Population with UMAP}},
  \href{https://doi.org/10.48550/arXiv.2603.06566}{\emph{arXiv e-prints} (2026)
  arXiv:2603.06566} [\href{https://arxiv.org/abs/2603.06566}{{\ttfamily
  2603.06566}}].

\bibitem{2026MNRAS.545f1938D}
A.~{Dorozsmai}, I.M.~{Romero-Shaw}, A.~{Vijaykumar}, S.~{Toonen},
  F.~{Antonini}, K.~{Kremer} et~al., \emph{{Hierarchical triples versus
  globular clusters: binary black hole merger eccentricity distributions
  compete and evolve with redshift}},
  \href{https://doi.org/10.1093/mnras/staf1938}{\emph{\mnras} {\bfseries 545}
  (2026) staf1938} [\href{https://arxiv.org/abs/2507.23212}{{\ttfamily
  2507.23212}}].

\bibitem{RenzoSmith2024}
M.~{Renzo} and N.~{Smith}, \emph{{Pair-instability evolution and explosions in
  massive stars}}, \href{https://doi.org/10.48550/arXiv.2407.16113}{\emph{arXiv
  e-prints} (2024) arXiv:2407.16113}
  [\href{https://arxiv.org/abs/2407.16113}{{\ttfamily 2407.16113}}].

\bibitem{vanSon2020}
L.A.C.~{van Son}, S.E.~{De Mink}, F.S.~{Broekgaarden}, M.~{Renzo},
  S.~{Justham}, E.~{Laplace} et~al., \emph{{Polluting the Pair-instability Mass
  Gap for Binary Black Holes through Super-Eddington Accretion in Isolated
  Binaries}}, \href{https://doi.org/10.3847/1538-4357/ab9809}{\emph{\apj}
  {\bfseries 897} (2020) 100}
  [\href{https://arxiv.org/abs/2004.05187}{{\ttfamily 2004.05187}}].

\bibitem{DiCarlo2019}
U.N.~{Di Carlo}, N.~{Giacobbo}, M.~{Mapelli}, M.~{Pasquato}, M.~{Spera},
  L.~{Wang} et~al., \emph{{Merging black holes in young star clusters}},
  \href{https://doi.org/10.1093/mnras/stz1453}{\emph{\mnras} {\bfseries 487}
  (2019) 2947} [\href{https://arxiv.org/abs/1901.00863}{{\ttfamily
  1901.00863}}].

\bibitem{DiCarlo2020}
U.N.~{Di Carlo}, M.~{Mapelli}, Y.~{Bouffanais}, N.~{Giacobbo},
  F.~{Santoliquido}, A.~{Bressan} et~al., \emph{{Binary black holes in the pair
  instability mass gap}},
  \href{https://doi.org/10.1093/mnras/staa1997}{\emph{\mnras} {\bfseries 497}
  (2020) 1043} [\href{https://arxiv.org/abs/1911.01434}{{\ttfamily
  1911.01434}}].

\bibitem{Renzo2020}
M.~{Renzo}, M.~{Cantiello}, B.D.~{Metzger} and Y.-F.~{Jiang}, \emph{{The
  Stellar Merger Scenario for Black Holes in the Pair-instability Gap}},
  \href{https://doi.org/10.3847/2041-8213/abc6a6}{\emph{\apjl} {\bfseries 904}
  (2020) L13} [\href{https://arxiv.org/abs/2010.00705}{{\ttfamily
  2010.00705}}].

\bibitem{Tong2026}
H.~{Tong}, M.~{Fishbach}, E.~{Thrane}, M.~{Mould}, T.A.~{Callister},
  A.M.~{Farah} et~al., \emph{{Evidence of the pair-instability gap from
  black-hole masses}},
  \href{https://doi.org/10.1038/s41586-026-10359-0}{\emph{\nat} {\bfseries 652}
  (2026) 874} [\href{https://arxiv.org/abs/2509.04151}{{\ttfamily
  2509.04151}}].

\bibitem{RayKalogera2026}
A.~{Ray} and V.~{Kalogera}, \emph{{Reexamining Evidence of a Pair-instability
  Mass Gap in the Binary Black Hole Population}},
  \href{https://doi.org/10.3847/2041-8213/ae374d}{\emph{\apjl} {\bfseries 998}
  (2026) L20} [\href{https://arxiv.org/abs/2510.18867}{{\ttfamily
  2510.18867}}].

\bibitem{2025PhRvD.112f3040A}
F.~{Antonini}, T.~{Callister}, F.~{Dosopoulou}, I.M.~{Romero-Shaw} and
  D.~{Chattopadhyay}, \emph{{Inferring the pair-instability mass gap from
  gravitational wave data}},
  \href{https://doi.org/10.1103/nxnr-pdyx}{\emph{\prd} {\bfseries 112} (2025)
  063040} [\href{https://arxiv.org/abs/2506.09154}{{\ttfamily 2506.09154}}].

\bibitem{AfrozMukherjee2025}
S.~{Afroz} and S.~{Mukherjee}, \emph{{Binary Black Hole Phase Space Discovers
  the Signature of Pair Instability Supernovae Mass Gap}},
  \href{https://doi.org/10.48550/arXiv.2509.09123}{\emph{arXiv e-prints} (2025)
  arXiv:2509.09123} [\href{https://arxiv.org/abs/2509.09123}{{\ttfamily
  2509.09123}}].

\bibitem{Edelman2021}
B.~{Edelman}, Z.~{Doctor} and B.~{Farr}, \emph{{Poking Holes: Looking for Gaps
  in LIGO/Virgo's Black Hole Population}},
  \href{https://doi.org/10.3847/2041-8213/abfdb3}{\emph{\apjl} {\bfseries 913}
  (2021) L23} [\href{https://arxiv.org/abs/2104.07783}{{\ttfamily
  2104.07783}}].

\bibitem{Ulrich2024}
Y.~{Ulrich}, D.~{Croon}, J.~{Sakstein} and S.~{McDermott},
  \emph{{Multi-Generational Black Hole Population Analysis with an
  Astrophysically Informed Mass Function}},
  \href{https://doi.org/10.48550/arXiv.2406.06109}{\emph{arXiv e-prints} (2024)
  arXiv:2406.06109} [\href{https://arxiv.org/abs/2406.06109}{{\ttfamily
  2406.06109}}].

\bibitem{2026PhRvD.113j3021M}
M.~{Mould}, J.~{Heinzel}, S.~{{\'A}lvarez-L{\'o}pez}, C.~{Plunkett},
  N.E.~{Wolfe} and S.~{Vitale}, \emph{{Measurement prospects for the
  pair-instability mass cutoff with gravitational waves}},
  \href{https://doi.org/10.1103/sqy9-nnqq}{\emph{\prd} {\bfseries 113} (2026)
  103021} [\href{https://arxiv.org/abs/2602.11282}{{\ttfamily 2602.11282}}].

\bibitem{GW190521}
R.~{Abbott}, T.D.~{Abbott}, S.~{Abraham}, F.~{Acernese}, K.~{Ackley},
  C.~{Adams} et~al., \emph{{GW190521: A Binary Black Hole Merger with a Total
  Mass of 150 M{\ensuremath{\odot}}}},
  \href{https://doi.org/10.1103/PhysRevLett.125.101102}{\emph{\prl} {\bfseries
  125} (2020) 101102} [\href{https://arxiv.org/abs/2009.01075}{{\ttfamily
  2009.01075}}].

\bibitem{GW231123_2025}
A.G.~{Abac}, I.~{Abouelfettouh}, F.~{Acernese}, K.~{Ackley}, C.~{Adamcewicz},
  S.~{Adhicary} et~al., \emph{{GW231123: A Binary Black Hole Merger with Total
  Mass 190--265 M$_{{\ensuremath{\odot}}}$}},
  \href{https://doi.org/10.3847/2041-8213/ae0c9c}{\emph{\apjl} {\bfseries 993}
  (2025) L25} [\href{https://arxiv.org/abs/2507.08219}{{\ttfamily
  2507.08219}}].

\bibitem{LIGOScientific:2016aoc}
{\scshape LIGO Scientific, Virgo} collaboration, \emph{{Observation of
  Gravitational Waves from a Binary Black Hole Merger}},
  \href{https://doi.org/10.1103/PhysRevLett.116.061102}{\emph{Phys. Rev. Lett.}
  {\bfseries 116} (2016) 061102}
  [\href{https://arxiv.org/abs/1602.03837}{{\ttfamily 1602.03837}}].

\bibitem{Einstein:1918btx}
A.~Einstein, \emph{{{\"U}ber Gravitationswellen}}, {\emph{Sitzungsber. Preuss.
  Akad. Wiss. Berlin (Math. Phys. )} {\bfseries 1918} (1918) 154}.

\bibitem{Peters:1963ux}
P.C.~Peters and J.~Mathews, \emph{{Gravitational radiation from point masses in
  a Keplerian orbit}},
  \href{https://doi.org/10.1103/PhysRev.131.435}{\emph{Phys. Rev.} {\bfseries
  131} (1963) 435}.

\bibitem{Peters:1964zz}
P.C.~Peters, \emph{{Gravitational Radiation and the Motion of Two Point
  Masses}}, \href{https://doi.org/10.1103/PhysRev.136.B1224}{\emph{Phys. Rev.}
  {\bfseries 136} (1964) B1224}.

\bibitem{Blanchet:2013haa}
L.~Blanchet, \emph{{Post-Newtonian Theory for Gravitational Waves}},
  \href{https://doi.org/10.12942/lrr-2014-2}{\emph{Living Rev. Rel.} {\bfseries
  17} (2014) 2} [\href{https://arxiv.org/abs/1310.1528}{{\ttfamily
  1310.1528}}].

\bibitem{Mino:1996nk}
Y.~Mino, M.~Sasaki and T.~Tanaka, \emph{{Gravitational radiation reaction to a
  particle motion}},
  \href{https://doi.org/10.1103/PhysRevD.55.3457}{\emph{Phys. Rev. D}
  {\bfseries 55} (1997) 3457}
  [\href{https://arxiv.org/abs/gr-qc/9606018}{{\ttfamily gr-qc/9606018}}].

\bibitem{Quinn:1996am}
T.C.~Quinn and R.M.~Wald, \emph{{An Axiomatic approach to electromagnetic and
  gravitational radiation reaction of particles in curved space-time}},
  \href{https://doi.org/10.1103/PhysRevD.56.3381}{\emph{Phys. Rev. D}
  {\bfseries 56} (1997) 3381}
  [\href{https://arxiv.org/abs/gr-qc/9610053}{{\ttfamily gr-qc/9610053}}].

\bibitem{Detweiler:2002mi}
S.L.~Detweiler and B.F.~Whiting, \emph{{Selfforce via a Green's function
  decomposition}},
  \href{https://doi.org/10.1103/PhysRevD.67.024025}{\emph{Phys. Rev. D}
  {\bfseries 67} (2003) 024025}
  [\href{https://arxiv.org/abs/gr-qc/0202086}{{\ttfamily gr-qc/0202086}}].

\bibitem{Poisson:2011nh}
E.~Poisson, A.~Pound and I.~Vega, \emph{{The Motion of point particles in
  curved spacetime}}, \href{https://doi.org/10.12942/lrr-2011-7}{\emph{Living
  Rev. Rel.} {\bfseries 14} (2011) 7}
  [\href{https://arxiv.org/abs/1102.0529}{{\ttfamily 1102.0529}}].

\bibitem{Barack:2018yvs}
L.~Barack and A.~Pound, \emph{{Self-force and radiation reaction in general
  relativity}}, \href{https://doi.org/10.1088/1361-6633/aae552}{\emph{Rept.
  Prog. Phys.} {\bfseries 82} (2019) 016904}
  [\href{https://arxiv.org/abs/1805.10385}{{\ttfamily 1805.10385}}].

\bibitem{LISA:2017pwj}
{\scshape LISA} collaboration, \emph{{Laser Interferometer Space Antenna}},
  \href{https://arxiv.org/abs/1702.00786}{{\ttfamily 1702.00786}}.

\bibitem{Pretorius:2005gq}
F.~Pretorius, \emph{{Evolution of binary black hole spacetimes}},
  \href{https://doi.org/10.1103/PhysRevLett.95.121101}{\emph{Phys. Rev. Lett.}
  {\bfseries 95} (2005) 121101}
  [\href{https://arxiv.org/abs/gr-qc/0507014}{{\ttfamily gr-qc/0507014}}].

\bibitem{Campanelli:2005dd}
M.~Campanelli, C.O.~Lousto, P.~Marronetti and Y.~Zlochower, \emph{{Accurate
  evolutions of orbiting black-hole binaries without excision}},
  \href{https://doi.org/10.1103/PhysRevLett.96.111101}{\emph{Phys. Rev. Lett.}
  {\bfseries 96} (2006) 111101}
  [\href{https://arxiv.org/abs/gr-qc/0511048}{{\ttfamily gr-qc/0511048}}].

\bibitem{Baker:2005vv}
J.G.~Baker, J.~Centrella, D.-I.~Choi, M.~Koppitz and J.~van Meter,
  \emph{{Gravitational wave extraction from an inspiraling configuration of
  merging black holes}},
  \href{https://doi.org/10.1103/PhysRevLett.96.111102}{\emph{Phys. Rev. Lett.}
  {\bfseries 96} (2006) 111102}
  [\href{https://arxiv.org/abs/gr-qc/0511103}{{\ttfamily gr-qc/0511103}}].

\bibitem{Damour:2014afa}
T.~Damour, F.~Guercilena, I.~Hinder, S.~Hopper, A.~Nagar and L.~Rezzolla,
  \emph{{Strong-Field Scattering of Two Black Holes: Numerics Versus
  Analytics}}, \href{https://doi.org/10.1103/PhysRevD.89.081503}{\emph{Phys.
  Rev. D} {\bfseries 89} (2014) 081503}
  [\href{https://arxiv.org/abs/1402.7307}{{\ttfamily 1402.7307}}].

\bibitem{Damour:2022ybd}
T.~Damour and P.~Rettegno, \emph{{Strong-field scattering of two black holes:
  Numerical relativity meets post-Minkowskian gravity}},
  \href{https://doi.org/10.1103/PhysRevD.107.064051}{\emph{Phys. Rev. D}
  {\bfseries 107} (2023) 064051}
  [\href{https://arxiv.org/abs/2211.01399}{{\ttfamily 2211.01399}}].

\bibitem{Buonanno:1998gg}
A.~Buonanno and T.~Damour, \emph{{Effective one-body approach to general
  relativistic two-body dynamics}},
  \href{https://doi.org/10.1103/PhysRevD.59.084006}{\emph{Phys. Rev. D}
  {\bfseries 59} (1999) 084006}
  [\href{https://arxiv.org/abs/gr-qc/9811091}{{\ttfamily gr-qc/9811091}}].

\bibitem{Buonanno:2000ef}
A.~Buonanno and T.~Damour, \emph{{Transition from inspiral to plunge in binary
  black hole coalescences}},
  \href{https://doi.org/10.1103/PhysRevD.62.064015}{\emph{Phys. Rev. D}
  {\bfseries 62} (2000) 064015}
  [\href{https://arxiv.org/abs/gr-qc/0001013}{{\ttfamily gr-qc/0001013}}].

\bibitem{Regge:1957td}
T.~Regge and J.A.~Wheeler, \emph{{Stability of a Schwarzschild singularity}},
  \href{https://doi.org/10.1103/PhysRev.108.1063}{\emph{Phys. Rev.} {\bfseries
  108} (1957) 1063}.

\bibitem{Zerilli:1970se}
F.J.~Zerilli, \emph{{Effective potential for even parity Regge-Wheeler
  gravitational perturbation equations}},
  \href{https://doi.org/10.1103/PhysRevLett.24.737}{\emph{Phys. Rev. Lett.}
  {\bfseries 24} (1970) 737}.

\bibitem{Vishveshwara:1970zz}
C.V.~Vishveshwara, \emph{{Scattering of Gravitational Radiation by a
  Schwarzschild Black-hole}},
  \href{https://doi.org/10.1038/227936a0}{\emph{Nature} {\bfseries 227} (1970)
  936}.

\bibitem{Teukolsky:1973ha}
S.A.~Teukolsky, \emph{{Perturbations of a rotating black hole. 1. Fundamental
  equations for gravitational electromagnetic and neutrino field
  perturbations}}, \href{https://doi.org/10.1086/152444}{\emph{Astrophys. J.}
  {\bfseries 185} (1973) 635}.

\bibitem{Berti:2009kk}
E.~Berti, V.~Cardoso and A.O.~Starinets, \emph{{Quasinormal modes of black
  holes and black branes}},
  \href{https://doi.org/10.1088/0264-9381/26/16/163001}{\emph{Class. Quant.
  Grav.} {\bfseries 26} (2009) 163001}
  [\href{https://arxiv.org/abs/0905.2975}{{\ttfamily 0905.2975}}].

\bibitem{Wald:1984rg}
R.M.~Wald, \emph{{General Relativity}}, Chicago Univ. Pr., Chicago, USA (1984),
  \href{https://doi.org/10.7208/chicago/9780226870373.001.0001}{10.7208/chicago/9780226870373.001.0001}.

\bibitem{Flanagan:2007ix}
E.E.~Flanagan and T.~Hinderer, \emph{{Constraining neutron star tidal Love
  numbers with gravitational wave detectors}},
  \href{https://doi.org/10.1103/PhysRevD.77.021502}{\emph{Phys. Rev. D}
  {\bfseries 77} (2008) 021502}
  [\href{https://arxiv.org/abs/0709.1915}{{\ttfamily 0709.1915}}].

\bibitem{Arkani-Hamed:2017jhn}
N.~Arkani-Hamed, T.-C.~Huang and Y.-t.~Huang, \emph{{Scattering amplitudes for
  all masses and spins}},
  \href{https://doi.org/10.1007/JHEP11(2021)070}{\emph{JHEP} {\bfseries 11}
  (2021) 070} [\href{https://arxiv.org/abs/1709.04891}{{\ttfamily
  1709.04891}}].

\bibitem{Weinberg:1964ew}
S.~Weinberg, \emph{{Photons and Gravitons in $S$-Matrix Theory: Derivation of
  Charge Conservation and Equality of Gravitational and Inertial Mass}},
  \href{https://doi.org/10.1103/PhysRev.135.B1049}{\emph{Phys. Rev.} {\bfseries
  135} (1964) B1049}.

\bibitem{Bautista:2021wfy}
Y.F.~Bautista, A.~Guevara, C.~Kavanagh and J.~Vines, \emph{{Scattering in black
  hole backgrounds and higher-spin amplitudes. Part I}},
  \href{https://doi.org/10.1007/JHEP03(2023)136}{\emph{JHEP} {\bfseries 03}
  (2023) 136} [\href{https://arxiv.org/abs/2107.10179}{{\ttfamily
  2107.10179}}].

\bibitem{Bautista:2022wjf}
Y.F.~Bautista, A.~Guevara, C.~Kavanagh and J.~Vines, \emph{{Scattering in black
  hole backgrounds and higher-spin amplitudes. Part II}},
  \href{https://doi.org/10.1007/JHEP05(2023)211}{\emph{JHEP} {\bfseries 05}
  (2023) 211} [\href{https://arxiv.org/abs/2212.07965}{{\ttfamily
  2212.07965}}].

\bibitem{Bautista:2023sdf}
Y.F.~Bautista, G.~Bonelli, C.~Iossa, A.~Tanzini and Z.~Zhou, \emph{{Black hole
  perturbation theory meets CFT$_2$: Kerr-Compton amplitudes from
  Nekrasov-Shatashvili functions}},
  \href{https://doi.org/10.1103/PhysRevD.109.084071}{\emph{Phys. Rev. D}
  {\bfseries 109} (2024) 084071}
  [\href{https://arxiv.org/abs/2312.05965}{{\ttfamily 2312.05965}}].

\bibitem{Ivanov:2024sds}
M.M.~Ivanov, Y.-Z.~Li, J.~Parra-Martinez and Z.~Zhou, \emph{{Gravitational
  Raman Scattering in Effective Field Theory: A Scalar Tidal Matching at
  $\mathcal{O}(G^3)$}},
  \href{https://doi.org/10.1103/PhysRevLett.132.131401}{\emph{Phys. Rev. Lett.}
  {\bfseries 132} (2024) 131401}
  [\href{https://arxiv.org/abs/2401.08752}{{\ttfamily 2401.08752}}].

\bibitem{Saketh:2024juq}
M.V.S.~Saketh, Z.~Zhou, S.~Ghosh, J.~Steinhoff and D.~Chatterjee,
  \emph{{Investigating tidal heating in neutron stars via gravitational Raman
  scattering}}, \href{https://doi.org/10.1103/PhysRevD.110.103001}{\emph{Phys.
  Rev. D} {\bfseries 110} (2024) 103001}
  [\href{https://arxiv.org/abs/2407.08327}{{\ttfamily 2407.08327}}].

\bibitem{Caron-Huot:2025tlq}
S.~Caron-Huot, M.~Correia, G.~Isabella and M.~Solon, \emph{{Gravitational Wave
  Scattering via the Born Series: Scalar Tidal Matching to $\mathcal{O}(G^7)$
  and Beyond}}, \href{https://doi.org/10.1103/qd3c-nfz6}{\emph{Phys. Rev.
  Lett.} {\bfseries 135} (2025) 191601}
  [\href{https://arxiv.org/abs/2503.13593}{{\ttfamily 2503.13593}}].

\bibitem{Bjerrum-Bohr:2025bqg}
N.E.J.~Bjerrum-Bohr, G.~Chen, C.J.~Eriksen and N.~Shah, \emph{{The
  gravitational Compton amplitude from flat and curved spacetimes at second
  post-Minkowskian order}},
  \href{https://doi.org/10.1007/JHEP10(2025)235}{\emph{JHEP} {\bfseries 10}
  (2025) 235} [\href{https://arxiv.org/abs/2506.19705}{{\ttfamily
  2506.19705}}].

\bibitem{Bautista:2026qse}
Y.F.~Bautista, M.~Driesse, K.~Haddad and G.U.~Jakobsen, \emph{{Gravitational
  wave scattering in spinless WQFT}},
  \href{https://doi.org/10.1007/JHEP05(2026)252}{\emph{JHEP} {\bfseries 05}
  (2026) 252} [\href{https://arxiv.org/abs/2602.06125}{{\ttfamily
  2602.06125}}].

\bibitem{Bjerrum-Bohr:2026fhs}
N.E.J.~Bjerrum-Bohr, G.~Chen, C.~Jordan~Eriksen and N.~Shah, \emph{{The
  gravitational Compton amplitude at third post-Minkowskian order}},
  \href{https://arxiv.org/abs/2602.06947}{{\ttfamily 2602.06947}}.

\bibitem{Ivanov:2026icp}
M.M.~Ivanov, Y.-Z.~Li, J.~Parra-Martinez and Z.~Zhou, \emph{{Gravitational
  Raman Scattering: a Systematic Toolkit for Tidal Effects in General
  Relativity}},  \href{https://arxiv.org/abs/2602.06951}{{\ttfamily
  2602.06951}}.

\bibitem{Bautista:2026fcp}
Y.F.~Bautista, M.~Driesse, K.~Haddad and G.U.~Jakobsen, \emph{{Gravitational
  wave scattering at $\mathcal{O}(G^4)$: Murua construction and elliptics}},
  \href{https://arxiv.org/abs/2606.27544}{{\ttfamily 2606.27544}}.

\bibitem{Brunello:2026rdk}
G.~Brunello, M.~Meo and S.~Smith, \emph{{Gravitational Compton scattering at
  the fourth post-Minkowskian order}},
  \href{https://arxiv.org/abs/2606.28239}{{\ttfamily 2606.28239}}.

\bibitem{Brunello:2026lzf}
G.~Brunello, M.~Meo and S.~Smith, \emph{{Gravitational Compton scattering at
  the fifth post-Minkowskian order}},
  \href{https://arxiv.org/abs/2608.17946}{{\ttfamily 2608.17946}}.

\bibitem{Correia:2026utp}
M.~Correia, G.~Isabella and A.M.~Wolz, \emph{{Gravitational Compton Amplitude
  to All Orders in Perturbation Theory}},
  \href{https://arxiv.org/abs/2608.26284}{{\ttfamily 2608.26284}}.

\bibitem{Kawai:1985xq}
H.~Kawai, D.C.~Lewellen and S.H.H.~Tye, \emph{{A Relation Between Tree
  Amplitudes of Closed and Open Strings}},
  \href{https://doi.org/10.1016/0550-3213(86)90362-7}{\emph{Nucl. Phys. B}
  {\bfseries 269} (1986) 1}.

\bibitem{Bern:2008qj}
Z.~Bern, J.J.M.~Carrasco and H.~Johansson, \emph{{New Relations for
  Gauge-Theory Amplitudes}},
  \href{https://doi.org/10.1103/PhysRevD.78.085011}{\emph{Phys. Rev. D}
  {\bfseries 78} (2008) 085011}
  [\href{https://arxiv.org/abs/0805.3993}{{\ttfamily 0805.3993}}].

\bibitem{Bern:2022wqg}
Z.~Bern, J.J.~Carrasco, M.~Chiodaroli, H.~Johansson and R.~Roiban, \emph{{The
  SAGEX review on scattering amplitudes Chapter 2: An invitation to
  color-kinematics duality and the double copy}},
  \href{https://doi.org/10.1088/1751-8121/ac93cf}{\emph{J. Phys. A} {\bfseries
  55} (2022) 443003} [\href{https://arxiv.org/abs/2203.13013}{{\ttfamily
  2203.13013}}].

\bibitem{Goldberger:2004jt}
W.D.~Goldberger and I.Z.~Rothstein, \emph{{An Effective field theory of gravity
  for extended objects}},
  \href{https://doi.org/10.1103/PhysRevD.73.104029}{\emph{Phys. Rev. D}
  {\bfseries 73} (2006) 104029}
  [\href{https://arxiv.org/abs/hep-th/0409156}{{\ttfamily hep-th/0409156}}].

\bibitem{Goldberger:2007hy}
W.D.~Goldberger, \emph{{Les Houches lectures on effective field theories and
  gravitational radiation}},  in \emph{{Les Houches Summer School - Session 86:
  Particle Physics and Cosmology: The Fabric of Spacetime}}, 1, 2007
  [\href{https://arxiv.org/abs/hep-ph/0701129}{{\ttfamily hep-ph/0701129}}].

\bibitem{Porto:2016pyg}
R.A.~Porto, \emph{{The effective field theorist{\textquoteright}s approach to
  gravitational dynamics}},
  \href{https://doi.org/10.1016/j.physrep.2016.04.003}{\emph{Phys. Rept.}
  {\bfseries 633} (2016) 1} [\href{https://arxiv.org/abs/1601.04914}{{\ttfamily
  1601.04914}}].

\bibitem{Damour:1992we}
T.~Damour and G.~Esposito-Farese, \emph{{Tensor multiscalar theories of
  gravitation}}, \href{https://doi.org/10.1088/0264-9381/9/9/015}{\emph{Class.
  Quant. Grav.} {\bfseries 9} (1992) 2093}.

\bibitem{Bini:2012gu}
D.~Bini, T.~Damour and G.~Faye, \emph{{Effective action approach to
  higher-order relativistic tidal interactions in binary systems and their
  effective one body description}},
  \href{https://doi.org/10.1103/PhysRevD.85.124034}{\emph{Phys. Rev. D}
  {\bfseries 85} (2012) 124034}
  [\href{https://arxiv.org/abs/1202.3565}{{\ttfamily 1202.3565}}].

\bibitem{Binnington:2009bb}
T.~Binnington and E.~Poisson, \emph{{Relativistic theory of tidal Love
  numbers}}, \href{https://doi.org/10.1103/PhysRevD.80.084018}{\emph{Phys. Rev.
  D} {\bfseries 80} (2009) 084018}
  [\href{https://arxiv.org/abs/0906.1366}{{\ttfamily 0906.1366}}].

\bibitem{Damour:2009wj}
T.~Damour and A.~Nagar, \emph{{Effective One Body description of tidal effects
  in inspiralling compact binaries}},
  \href{https://doi.org/10.1103/PhysRevD.81.084016}{\emph{Phys. Rev. D}
  {\bfseries 81} (2010) 084016}
  [\href{https://arxiv.org/abs/0911.5041}{{\ttfamily 0911.5041}}].

\bibitem{Kol:2011vg}
B.~Kol and M.~Smolkin, \emph{{Black hole stereotyping: Induced gravito-static
  polarization}}, \href{https://doi.org/10.1007/JHEP02(2012)010}{\emph{JHEP}
  {\bfseries 02} (2012) 010} [\href{https://arxiv.org/abs/1110.3764}{{\ttfamily
  1110.3764}}].

\bibitem{Hui:2020xxx}
L.~Hui, A.~Joyce, R.~Penco, L.~Santoni and A.R.~Solomon, \emph{{Static response
  and Love numbers of Schwarzschild black holes}},
  \href{https://doi.org/10.1088/1475-7516/2021/04/052}{\emph{JCAP} {\bfseries
  04} (2021) 052} [\href{https://arxiv.org/abs/2010.00593}{{\ttfamily
  2010.00593}}].

\bibitem{Hui:2021vcv}
L.~Hui, A.~Joyce, R.~Penco, L.~Santoni and A.R.~Solomon, \emph{{Ladder
  symmetries of black holes. Implications for love numbers and no-hair
  theorems}}, \href{https://doi.org/10.1088/1475-7516/2022/01/032}{\emph{JCAP}
  {\bfseries 01} (2022) 032}
  [\href{https://arxiv.org/abs/2105.01069}{{\ttfamily 2105.01069}}].

\bibitem{Charalambous:2021mea}
P.~Charalambous, S.~Dubovsky and M.M.~Ivanov, \emph{{On the Vanishing of Love
  Numbers for Kerr Black Holes}},
  \href{https://doi.org/10.1007/JHEP05(2021)038}{\emph{JHEP} {\bfseries 05}
  (2021) 038} [\href{https://arxiv.org/abs/2102.08917}{{\ttfamily
  2102.08917}}].

\bibitem{Charalambous:2022rre}
P.~Charalambous, S.~Dubovsky and M.M.~Ivanov, \emph{{Love symmetry}},
  \href{https://doi.org/10.1007/JHEP10(2022)175}{\emph{JHEP} {\bfseries 10}
  (2022) 175} [\href{https://arxiv.org/abs/2209.02091}{{\ttfamily
  2209.02091}}].

\bibitem{Parra-Martinez:2025bcu}
J.~Parra-Martinez and A.~Podo, \emph{{Naturalness of vanishing black-hole
  tides}},  \href{https://arxiv.org/abs/2510.20694}{{\ttfamily 2510.20694}}.

\bibitem{Riva:2023rcm}
M.M.~Riva, L.~Santoni, N.~Savi{\'c} and F.~Vernizzi, \emph{{Vanishing of
  nonlinear tidal Love numbers of Schwarzschild black holes}},
  \href{https://doi.org/10.1016/j.physletb.2024.138710}{\emph{Phys. Lett. B}
  {\bfseries 854} (2024) 138710}
  [\href{https://arxiv.org/abs/2312.05065}{{\ttfamily 2312.05065}}].

\bibitem{Iteanu:2024dvx}
S.~Iteanu, M.M.~Riva, L.~Santoni, N.~Savi{\'c} and F.~Vernizzi,
  \emph{{Vanishing of quadratic Love numbers of Schwarzschild black holes}},
  \href{https://doi.org/10.1007/JHEP02(2025)174}{\emph{JHEP} {\bfseries 02}
  (2025) 174} [\href{https://arxiv.org/abs/2410.03542}{{\ttfamily
  2410.03542}}].

\bibitem{Hawking:1972hy}
S.W.~Hawking and J.B.~Hartle, \emph{{Energy and angular momentum flow into a
  black hole}}, \href{https://doi.org/10.1007/BF01645515}{\emph{Commun. Math.
  Phys.} {\bfseries 27} (1972) 283}.

\bibitem{Hartle:1973zz}
J.B.~Hartle, \emph{{Tidal Friction in Slowly Rotating Black Holes}},
  \href{https://doi.org/10.1103/PhysRevD.8.1010}{\emph{Phys. Rev. D} {\bfseries
  8} (1973) 1010}.

\bibitem{Goldberger:2005cd}
W.D.~Goldberger and I.Z.~Rothstein, \emph{{Dissipative effects in the worldline
  approach to black hole dynamics}},
  \href{https://doi.org/10.1103/PhysRevD.73.104030}{\emph{Phys. Rev. D}
  {\bfseries 73} (2006) 104030}
  [\href{https://arxiv.org/abs/hep-th/0511133}{{\ttfamily hep-th/0511133}}].

\bibitem{Wardell:2021fyy}
B.~Wardell, A.~Pound, N.~Warburton, J.~Miller, L.~Durkan and A.~Le~Tiec,
  \emph{{Gravitational Waveforms for Compact Binaries from Second-Order
  Self-Force Theory}},
  \href{https://doi.org/10.1103/PhysRevLett.130.241402}{\emph{Phys. Rev. Lett.}
  {\bfseries 130} (2023) 241402}
  [\href{https://arxiv.org/abs/2112.12265}{{\ttfamily 2112.12265}}].

\bibitem{Kalin:2019rwq}
G.~K{\"a}lin and R.A.~Porto, \emph{{From Boundary Data to Bound States}},
  \href{https://doi.org/10.1007/JHEP01(2020)072}{\emph{JHEP} {\bfseries 01}
  (2020) 072} [\href{https://arxiv.org/abs/1910.03008}{{\ttfamily
  1910.03008}}].

\bibitem{Kalin:2019inp}
G.~K{\"a}lin and R.A.~Porto, \emph{{From boundary data to bound states. Part
  II. Scattering angle to dynamical invariants (with twist)}},
  \href{https://doi.org/10.1007/JHEP02(2020)120}{\emph{JHEP} {\bfseries 02}
  (2020) 120} [\href{https://arxiv.org/abs/1911.09130}{{\ttfamily
  1911.09130}}].

\bibitem{Cho:2021arx}
G.~Cho, G.~K{\"a}lin and R.A.~Porto, \emph{{From boundary data to bound states.
  Part III. Radiative effects}},
  \href{https://doi.org/10.1007/JHEP04(2022)154}{\emph{JHEP} {\bfseries 04}
  (2022) 154} [\href{https://arxiv.org/abs/2112.03976}{{\ttfamily
  2112.03976}}].

\bibitem{Westpfahl:1979gu}
K.~Westpfahl and M.~Goller, \emph{{GRAVITATIONAL SCATTERING OF TWO RELATIVISTIC
  PARTICLES IN POSTLINEAR APPROXIMATION}},
  \href{https://doi.org/10.1007/BF02817047}{\emph{Lett. Nuovo Cim.} {\bfseries
  26} (1979) 573}.

\bibitem{Westpfahl:1985tsl}
K.~Westpfahl, \emph{{High-Speed Scattering of Charged and Uncharged Particles
  in General Relativity}},
  \href{https://doi.org/10.1002/prop.2190330802}{\emph{Fortsch. Phys.}
  {\bfseries 33} (1985) 417}.

\bibitem{Shapiro:1964uw}
I.I.~Shapiro, \emph{{Fourth Test of General Relativity}},
  \href{https://doi.org/10.1103/PhysRevLett.13.789}{\emph{Phys. Rev. Lett.}
  {\bfseries 13} (1964) 789}.

\bibitem{Detweiler:2008ft}
S.L.~Detweiler, \emph{{A Consequence of the gravitational self-force for
  circular orbits of the Schwarzschild geometry}},
  \href{https://doi.org/10.1103/PhysRevD.77.124026}{\emph{Phys. Rev. D}
  {\bfseries 77} (2008) 124026}
  [\href{https://arxiv.org/abs/0804.3529}{{\ttfamily 0804.3529}}].

\bibitem{LeTiec:2011ab}
A.~Le~Tiec, L.~Blanchet and B.F.~Whiting, \emph{{The First Law of Binary Black
  Hole Mechanics in General Relativity and Post-Newtonian Theory}},
  \href{https://doi.org/10.1103/PhysRevD.85.064039}{\emph{Phys. Rev. D}
  {\bfseries 85} (2012) 064039}
  [\href{https://arxiv.org/abs/1111.5378}{{\ttfamily 1111.5378}}].

\bibitem{Barack:2009ey}
L.~Barack and N.~Sago, \emph{{Gravitational self-force correction to the
  innermost stable circular orbit of a Schwarzschild black hole}},
  \href{https://doi.org/10.1103/PhysRevLett.102.191101}{\emph{Phys. Rev. Lett.}
  {\bfseries 102} (2009) 191101}
  [\href{https://arxiv.org/abs/0902.0573}{{\ttfamily 0902.0573}}].

\bibitem{Pound:1960zz}
R.V.~Pound and G.A.~Rebka, Jr., \emph{{Apparent Weight of Photons}},
  \href{https://doi.org/10.1103/PhysRevLett.4.337}{\emph{Phys. Rev. Lett.}
  {\bfseries 4} (1960) 337}.

\bibitem{Cheung:2018wkq}
C.~Cheung, I.Z.~Rothstein and M.P.~Solon, \emph{{From Scattering Amplitudes to
  Classical Potentials in the Post-Minkowskian Expansion}},
  \href{https://doi.org/10.1103/PhysRevLett.121.251101}{\emph{Phys. Rev. Lett.}
  {\bfseries 121} (2018) 251101}
  [\href{https://arxiv.org/abs/1808.02489}{{\ttfamily 1808.02489}}].

\bibitem{Bern:2019nnu}
Z.~Bern, C.~Cheung, R.~Roiban, C.-H.~Shen, M.P.~Solon and M.~Zeng,
  \emph{{Scattering Amplitudes and the Conservative Hamiltonian for Binary
  Systems at Third Post-Minkowskian Order}},
  \href{https://doi.org/10.1103/PhysRevLett.122.201603}{\emph{Phys. Rev. Lett.}
  {\bfseries 122} (2019) 201603}
  [\href{https://arxiv.org/abs/1901.04424}{{\ttfamily 1901.04424}}].

\bibitem{Kalin:2020mvi}
G.~K{\"a}lin and R.A.~Porto, \emph{{Post-Minkowskian Effective Field Theory for
  Conservative Binary Dynamics}},
  \href{https://doi.org/10.1007/JHEP11(2020)106}{\emph{JHEP} {\bfseries 11}
  (2020) 106} [\href{https://arxiv.org/abs/2006.01184}{{\ttfamily
  2006.01184}}].

\bibitem{Kalin:2020fhe}
G.~K{\"a}lin, Z.~Liu and R.A.~Porto, \emph{{Conservative Dynamics of Binary
  Systems to Third Post-Minkowskian Order from the Effective Field Theory
  Approach}}, \href{https://doi.org/10.1103/PhysRevLett.125.261103}{\emph{Phys.
  Rev. Lett.} {\bfseries 125} (2020) 261103}
  [\href{https://arxiv.org/abs/2007.04977}{{\ttfamily 2007.04977}}].

\bibitem{Bern:2021dqo}
Z.~Bern, J.~Parra-Martinez, R.~Roiban, M.S.~Ruf, C.-H.~Shen, M.P.~Solon et~al.,
  \emph{{Scattering Amplitudes and Conservative Binary Dynamics at ${\cal
  O}(G^4)$}}, \href{https://doi.org/10.1103/PhysRevLett.126.171601}{\emph{Phys.
  Rev. Lett.} {\bfseries 126} (2021) 171601}
  [\href{https://arxiv.org/abs/2101.07254}{{\ttfamily 2101.07254}}].

\bibitem{Bern:2021yeh}
Z.~Bern, J.~Parra-Martinez, R.~Roiban, M.S.~Ruf, C.-H.~Shen, M.P.~Solon et~al.,
  \emph{{Scattering Amplitudes, the Tail Effect, and Conservative Binary
  Dynamics at $\mathcal{O}(G^4)$}},
  \href{https://doi.org/10.1103/PhysRevLett.128.161103}{\emph{Phys. Rev. Lett.}
  {\bfseries 128} (2022) 161103}
  [\href{https://arxiv.org/abs/2112.10750}{{\ttfamily 2112.10750}}].

\bibitem{Dlapa:2021npj}
C.~Dlapa, G.~K{\"a}lin, Z.~Liu and R.A.~Porto, \emph{{Dynamics of binary
  systems to fourth Post-Minkowskian order from the effective field theory
  approach}}, \href{https://doi.org/10.1016/j.physletb.2022.137203}{\emph{Phys.
  Lett. B} {\bfseries 831} (2022) 137203}
  [\href{https://arxiv.org/abs/2106.08276}{{\ttfamily 2106.08276}}].

\bibitem{Driesse:2024xad}
M.~Driesse, G.U.~Jakobsen, G.~Mogull, J.~Plefka, B.~Sauer and J.~Usovitsch,
  \emph{{Conservative Black Hole Scattering at Fifth Post-Minkowskian and First
  Self-Force Order}},
  \href{https://doi.org/10.1103/PhysRevLett.132.241402}{\emph{Phys. Rev. Lett.}
  {\bfseries 132} (2024) 241402}
  [\href{https://arxiv.org/abs/2403.07781}{{\ttfamily 2403.07781}}].

\bibitem{Bern:2024adl}
Z.~Bern, E.~Herrmann, R.~Roiban, M.S.~Ruf, A.V.~Smirnov, V.A.~Smirnov et~al.,
  \emph{{Amplitudes, supersymmetric black hole scattering at $
  \mathcal{O}\left({G}^5\right) $, and loop integration}},
  \href{https://doi.org/10.1007/JHEP10(2024)023}{\emph{JHEP} {\bfseries 10}
  (2024) 023} [\href{https://arxiv.org/abs/2406.01554}{{\ttfamily
  2406.01554}}].

\bibitem{Dlapa:2025biy}
C.~Dlapa, G.~K{\"a}lin, Z.~Liu and R.A.~Porto, \emph{{Local-in-Time
  Conservative Binary Dynamics at Fifth Post-Minkowskian and First Self-Force
  Orders}}, \href{https://doi.org/10.1103/215k-27sj}{\emph{Phys. Rev. Lett.}
  {\bfseries 135} (2025) 251401}
  [\href{https://arxiv.org/abs/2506.20665}{{\ttfamily 2506.20665}}].

\bibitem{Dlapa:2026oyq}
C.~Dlapa, G.~K{\"a}lin, Z.~Liu and R.A.~Porto, \emph{{Nonlocal-in-time tail
  effects in gravitational scattering to fifth post-Minkowskian and tenth
  self-force orders}}, \href{https://doi.org/10.1103/wkp4-vy6g}{\emph{Phys.
  Rev. D} {\bfseries 114} (2026) 024029}
  [\href{https://arxiv.org/abs/2604.25916}{{\ttfamily 2604.25916}}].

\bibitem{Bern:2025zno}
Z.~Bern, E.~Herrmann, R.~Roiban, M.S.~Ruf, A.V.~Smirnov, V.A.~Smirnov et~al.,
  \emph{{Second-Order Self-Force Potential-Region Binary Dynamics at
  $\mathcal{O}(G^5)$ in Supergravity}},
  \href{https://doi.org/10.1103/jmby-htz9}{\emph{Phys. Rev. Lett.} {\bfseries
  136} (2026) 081401} [\href{https://arxiv.org/abs/2509.17412}{{\ttfamily
  2509.17412}}].

\bibitem{Driesse:2026qiz}
M.~Driesse, G.U.~Jakobsen, G.~Mogull, C.~Nega, J.~Plefka, B.~Sauer et~al.,
  \emph{{Conservative Black Hole Scattering at Fifth Post-Minkowskian and
  Second Self-Force Order}},
  \href{https://doi.org/10.1103/4dvk-nglx}{\emph{Phys. Rev. Lett.} {\bfseries
  137} (2026) 081402} [\href{https://arxiv.org/abs/2601.16256}{{\ttfamily
  2601.16256}}].

\bibitem{Bern:2025wyd}
Z.~Bern, E.~Herrmann, R.~Roiban, M.S.~Ruf, A.V.~Smirnov, S.~Smith et~al.,
  \emph{{Scattering Amplitudes and Conservative Binary Dynamics at
  $\mathcal{O}(G^5)$ without Self-Force Truncation}},
  \href{https://doi.org/10.1103/zjcp-ssrw}{\emph{Phys. Rev. Lett.} {\bfseries
  137} (2026) 021402} [\href{https://arxiv.org/abs/2512.23654}{{\ttfamily
  2512.23654}}].

\bibitem{Weinberg:1965nx}
S.~Weinberg, \emph{{Infrared photons and gravitons}},
  \href{https://doi.org/10.1103/PhysRev.140.B516}{\emph{Phys. Rev.} {\bfseries
  140} (1965) B516}.

\bibitem{Herrmann:2021lqe}
E.~Herrmann, J.~Parra-Martinez, M.S.~Ruf and M.~Zeng, \emph{{Gravitational
  Bremsstrahlung from Reverse Unitarity}},
  \href{https://doi.org/10.1103/PhysRevLett.126.201602}{\emph{Phys. Rev. Lett.}
  {\bfseries 126} (2021) 201602}
  [\href{https://arxiv.org/abs/2101.07255}{{\ttfamily 2101.07255}}].

\bibitem{Jakobsen:2021smu}
G.U.~Jakobsen, G.~Mogull, J.~Plefka and J.~Steinhoff, \emph{{Classical
  Gravitational Bremsstrahlung from a Worldline Quantum Field Theory}},
  \href{https://doi.org/10.1103/PhysRevLett.126.201103}{\emph{Phys. Rev. Lett.}
  {\bfseries 126} (2021) 201103}
  [\href{https://arxiv.org/abs/2101.12688}{{\ttfamily 2101.12688}}].

\bibitem{Damour:2020tta}
T.~Damour, \emph{{Radiative contribution to classical gravitational scattering
  at the third order in $G$}},
  \href{https://doi.org/10.1103/PhysRevD.102.124008}{\emph{Phys. Rev. D}
  {\bfseries 102} (2020) 124008}
  [\href{https://arxiv.org/abs/2010.01641}{{\ttfamily 2010.01641}}].

\bibitem{Dlapa:2022lmu}
C.~Dlapa, G.~K{\"a}lin, Z.~Liu, J.~Neef and R.A.~Porto, \emph{{Radiation
  Reaction and Gravitational Waves at Fourth Post-Minkowskian Order}},
  \href{https://doi.org/10.1103/PhysRevLett.130.101401}{\emph{Phys. Rev. Lett.}
  {\bfseries 130} (2023) 101401}
  [\href{https://arxiv.org/abs/2210.05541}{{\ttfamily 2210.05541}}].

\bibitem{Zeldovich:1974gvh}
Y.B.~Zel'dovich and A.G.~Polnarev, \emph{{Radiation of gravitational waves by a
  cluster of superdense stars}}, {\emph{Sov. Astron.} {\bfseries 18} (1974)
  17}.

\bibitem{Braginsky:1987kwo}
V.B.~Braginsky and K.S.~Thorne, \emph{{Gravitational-wave bursts with memory
  and experimental prospects}},
  \href{https://doi.org/10.1038/327123a0}{\emph{Nature} {\bfseries 327} (1987)
  123}.

\bibitem{Christodoulou:1991cr}
D.~Christodoulou, \emph{{Nonlinear nature of gravitation and gravitational wave
  experiments}}, \href{https://doi.org/10.1103/PhysRevLett.67.1486}{\emph{Phys.
  Rev. Lett.} {\bfseries 67} (1991) 1486}.

\bibitem{Wiseman:1991ss}
A.G.~Wiseman and C.M.~Will, \emph{{Christodoulou's nonlinear gravitational wave
  memory: Evaluation in the quadrupole approximation}},
  \href{https://doi.org/10.1103/PhysRevD.44.R2945}{\emph{Phys. Rev. D}
  {\bfseries 44} (1991) R2945}.

\bibitem{Blanchet:1992br}
L.~Blanchet and T.~Damour, \emph{{Hereditary effects in gravitational
  radiation}}, \href{https://doi.org/10.1103/PhysRevD.46.4304}{\emph{Phys. Rev.
  D} {\bfseries 46} (1992) 4304}.

\bibitem{Favata:2010zu}
M.~Favata, \emph{{The gravitational-wave memory effect}},
  \href{https://doi.org/10.1088/0264-9381/27/8/084036}{\emph{Class. Quant.
  Grav.} {\bfseries 27} (2010) 084036}
  [\href{https://arxiv.org/abs/1003.3486}{{\ttfamily 1003.3486}}].

\bibitem{Georgoudis:2025vkk}
A.~Georgoudis, V.~Goncalves, C.~Heissenberg and J.~Parra-Martinez,
  \emph{{Nonlinear Gravitational Memory in the Post-Minkowskian Expansion}},
  \href{https://doi.org/10.1103/8m17-s2y8}{\emph{Phys. Rev. Lett.} {\bfseries
  136} (2026) 121401} [\href{https://arxiv.org/abs/2506.20733}{{\ttfamily
  2506.20733}}].

\bibitem{Lasky:2016knh}
P.D.~Lasky, E.~Thrane, Y.~Levin, J.~Blackman and Y.~Chen, \emph{{Detecting
  gravitational-wave memory with LIGO: implications of GW150914}},
  \href{https://doi.org/10.1103/PhysRevLett.117.061102}{\emph{Phys. Rev. Lett.}
  {\bfseries 117} (2016) 061102}
  [\href{https://arxiv.org/abs/1605.01415}{{\ttfamily 1605.01415}}].

\bibitem{Blanchet:1987wq}
L.~Blanchet and T.~Damour, \emph{{Tail Transported Temporal Correlations in the
  Dynamics of a Gravitating System}},
  \href{https://doi.org/10.1103/PhysRevD.37.1410}{\emph{Phys. Rev. D}
  {\bfseries 37} (1988) 1410}.

\bibitem{Sahoo:2018lxl}
B.~Sahoo and A.~Sen, \emph{{Classical and Quantum Results on Logarithmic Terms
  in the Soft Theorem in Four Dimensions}},
  \href{https://doi.org/10.1007/JHEP02(2019)086}{\emph{JHEP} {\bfseries 02}
  (2019) 086} [\href{https://arxiv.org/abs/1808.03288}{{\ttfamily
  1808.03288}}].

\bibitem{Saha:2019tub}
A.P.~Saha, B.~Sahoo and A.~Sen, \emph{{Proof of the classical soft graviton
  theorem in $D$ = 4}},
  \href{https://doi.org/10.1007/JHEP06(2020)153}{\emph{JHEP} {\bfseries 06}
  (2020) 153} [\href{https://arxiv.org/abs/1912.06413}{{\ttfamily
  1912.06413}}].

\bibitem{Thorne:1980ru}
K.S.~Thorne, \emph{{Multipole Expansions of Gravitational Radiation}},
  \href{https://doi.org/10.1103/RevModPhys.52.299}{\emph{Rev. Mod. Phys.}
  {\bfseries 52} (1980) 299}.

\bibitem{Hulse:1974eb}
R.A.~Hulse and J.H.~Taylor, \emph{{Discovery of a pulsar in a binary system}},
  \href{https://doi.org/10.1086/181708}{\emph{Astrophys. J. Lett.} {\bfseries
  195} (1975) L51}.

\bibitem{Taylor:1982zz}
J.H.~Taylor and J.M.~Weisberg, \emph{{A new test of general relativity:
  Gravitational radiation and the binary pulsar PS R 1913+16}},
  \href{https://doi.org/10.1086/159690}{\emph{Astrophys. J.} {\bfseries 253}
  (1982) 908}.

\bibitem{Goldberger:2009qd}
W.D.~Goldberger and A.~Ross, \emph{{Gravitational radiative corrections from
  effective field theory}},
  \href{https://doi.org/10.1103/PhysRevD.81.124015}{\emph{Phys. Rev. D}
  {\bfseries 81} (2010) 124015}
  [\href{https://arxiv.org/abs/0912.4254}{{\ttfamily 0912.4254}}].

\bibitem{Brandhuber:2023hhy}
A.~Brandhuber, G.R.~Brown, G.~Chen, S.~De~Angelis, J.~Gowdy and G.~Travaglini,
  \emph{{One-loop gravitational bremsstrahlung and waveforms from a heavy-mass
  effective field theory}},
  \href{https://doi.org/10.1007/JHEP06(2023)048}{\emph{JHEP} {\bfseries 06}
  (2023) 048} [\href{https://arxiv.org/abs/2303.06111}{{\ttfamily
  2303.06111}}].

\bibitem{Herderschee:2023fxh}
A.~Herderschee, R.~Roiban and F.~Teng, \emph{{The sub-leading scattering
  waveform from amplitudes}},
  \href{https://doi.org/10.1007/JHEP06(2023)004}{\emph{JHEP} {\bfseries 06}
  (2023) 004} [\href{https://arxiv.org/abs/2303.06112}{{\ttfamily
  2303.06112}}].

\bibitem{Georgoudis:2023lgf}
A.~Georgoudis, C.~Heissenberg and I.~Vazquez-Holm, \emph{{Inelastic
  exponentiation and classical gravitational scattering at one loop}},
  \href{https://doi.org/10.1007/JHEP06(2023)126}{\emph{JHEP} {\bfseries 2023}
  (2023) 126} [\href{https://arxiv.org/abs/2303.07006}{{\ttfamily
  2303.07006}}].

\bibitem{Blanchet:2023bwj}
L.~Blanchet, G.~Faye, Q.~Henry, F.~Larrouturou and D.~Trestini,
  \emph{{Gravitational-Wave Phasing of Quasicircular Compact Binary Systems to
  the Fourth-and-a-Half Post-Newtonian Order}},
  \href{https://doi.org/10.1103/PhysRevLett.131.121402}{\emph{Phys. Rev. Lett.}
  {\bfseries 131} (2023) 121402}
  [\href{https://arxiv.org/abs/2304.11185}{{\ttfamily 2304.11185}}].

\bibitem{Blanchet:2023sbv}
L.~Blanchet, G.~Faye, Q.~Henry, F.~Larrouturou and D.~Trestini,
  \emph{{Gravitational-wave flux and quadrupole modes from quasicircular
  nonspinning compact binaries to the fourth post-Newtonian order}},
  \href{https://doi.org/10.1103/PhysRevD.108.064041}{\emph{Phys. Rev. D}
  {\bfseries 108} (2023) 064041}
  [\href{https://arxiv.org/abs/2304.11186}{{\ttfamily 2304.11186}}].

\bibitem{Levi:2018nxp}
M.~Levi, \emph{{Effective Field Theories of Post-Newtonian Gravity: A
  comprehensive review}},
  \href{https://doi.org/10.1088/1361-6633/ab12bc}{\emph{Rept. Prog. Phys.}
  {\bfseries 83} (2020) 075901}
  [\href{https://arxiv.org/abs/1807.01699}{{\ttfamily 1807.01699}}].

\bibitem{ET:2019dnz}
{\scshape ET} collaboration, \emph{{Science Case for the Einstein Telescope}},
  \href{https://doi.org/10.1088/1475-7516/2020/03/050}{\emph{JCAP} {\bfseries
  03} (2020) 050} [\href{https://arxiv.org/abs/1912.02622}{{\ttfamily
  1912.02622}}].

\bibitem{Reitze:2019iox}
D.~Reitze et~al., \emph{{Cosmic Explorer: The U.S. Contribution to
  Gravitational-Wave Astronomy beyond LIGO}}, {\emph{Bull. Am. Astron. Soc.}
  {\bfseries 51} (2019) 035}
  [\href{https://arxiv.org/abs/1907.04833}{{\ttfamily 1907.04833}}].

\bibitem{Read:2014qva}
J.I.~Read, \emph{{The Local Dark Matter Density}},
  \href{https://doi.org/10.1088/0954-3899/41/6/063101}{\emph{J. Phys. G}
  {\bfseries 41} (2014) 063101}
  [\href{https://arxiv.org/abs/1404.1938}{{\ttfamily 1404.1938}}].

\bibitem{Garani:2023esk}
R.~Garani, N.~Raj and J.~Reynoso-Cordova, \emph{{Could compact stars in
  globular clusters constrain dark matter?}},
  \href{https://doi.org/10.1088/1475-7516/2023/07/038}{\emph{JCAP} {\bfseries
  07} (2023) 038} [\href{https://arxiv.org/abs/2303.18009}{{\ttfamily
  2303.18009}}].

\bibitem{Benacquista:2011kv}
M.J.~Benacquista and J.M.B.~Downing, \emph{{Relativistic Binaries in Globular
  Clusters}}, \href{https://doi.org/10.12942/lrr-2013-4}{\emph{Living Rev.
  Rel.} {\bfseries 16} (2013) 4}
  [\href{https://arxiv.org/abs/1110.4423}{{\ttfamily 1110.4423}}].

\bibitem{Mallick:2022lfr}
L.~Mallick et~al., \emph{{High-density disc reflection spectroscopy of low-mass
  active galactic nuclei}},
  \href{https://doi.org/10.1093/mnras/stac990}{\emph{Mon. Not. Roy. Astron.
  Soc.} {\bfseries 513} (2022) 4361}
  [\href{https://arxiv.org/abs/2203.04522}{{\ttfamily 2203.04522}}].

\bibitem{Gondolo:1999ef}
P.~Gondolo and J.~Silk, \emph{{Dark matter annihilation at the galactic
  center}}, \href{https://doi.org/10.1103/PhysRevLett.83.1719}{\emph{Phys. Rev.
  Lett.} {\bfseries 83} (1999) 1719}
  [\href{https://arxiv.org/abs/astro-ph/9906391}{{\ttfamily
  astro-ph/9906391}}].

\bibitem{CanevaSantoro:2023aol}
G.~Caneva~Santoro, S.~Roy, R.~Vicente, M.~Haney, O.J.~Piccinni, W.~Del~Pozzo
  et~al., \emph{{First Constraints on Compact Binary Environments from
  LIGO-Virgo Data}},
  \href{https://doi.org/10.1103/PhysRevLett.132.251401}{\emph{Phys. Rev. Lett.}
  {\bfseries 132} (2024) 251401}
  [\href{https://arxiv.org/abs/2309.05061}{{\ttfamily 2309.05061}}].

\bibitem{Caputo:2020irr}
A.~Caputo, L.~Sberna, A.~Toubiana, S.~Babak, E.~Barausse, S.~Marsat et~al.,
  \emph{{Gravitational-wave detection and parameter estimation for accreting
  black-hole binaries and their electromagnetic counterpart}},
  \href{https://doi.org/10.3847/1538-4357/ab7b66}{\emph{Astrophys. J.}
  {\bfseries 892} (2020) 90}
  [\href{https://arxiv.org/abs/2001.03620}{{\ttfamily 2001.03620}}].

\bibitem{Binney1987}
J.~{Binney} and S.~{Tremaine}, \emph{{Galactic dynamics}} (1987).

\bibitem{Katz:2019rgf}
A.~Katz, A.~Kurkela and A.~Soloviev, \emph{{Dynamical Friction in Interacting
  Relativistic Systems}},
  \href{https://doi.org/10.1088/1475-7516/2019/08/017}{\emph{JCAP} {\bfseries
  08} (2019) 017} [\href{https://arxiv.org/abs/1906.05341}{{\ttfamily
  1906.05341}}].

\bibitem{Shakura:1972te}
N.I.~Shakura and R.A.~Sunyaev, \emph{{Black holes in binary systems.
  Observational appearance}}, {\emph{Astron. Astrophys.} {\bfseries 24} (1973)
  337}.

\bibitem{Abramowicz:2011xu}
M.A.~Abramowicz and P.C.~Fragile, \emph{{Foundations of Black Hole Accretion
  Disk Theory}}, \href{https://doi.org/10.12942/lrr-2013-1}{\emph{Living Rev.
  Rel.} {\bfseries 16} (2013) 1}
  [\href{https://arxiv.org/abs/1104.5499}{{\ttfamily 1104.5499}}].

\bibitem{Ostriker:1998fa}
E.C.~Ostriker, \emph{{Dynamical friction in a gaseous medium}},
  \href{https://doi.org/10.1086/306858}{\emph{Astrophys. J.} {\bfseries 513}
  (1999) 252} [\href{https://arxiv.org/abs/astro-ph/9810324}{{\ttfamily
  astro-ph/9810324}}].

\bibitem{Goldreich:1980wa}
P.~Goldreich and S.~Tremaine, \emph{{Disk - satellite interactions}},
  \href{https://doi.org/10.1086/158356}{\emph{Astrophys. J.} {\bfseries 241}
  (1980) 425}.

\bibitem{GoldreichTremaine1}
P.~{Goldreich} and S.~{Tremaine}, \emph{{The excitation of density waves at the
  Lindblad and corotation resonances by an external potential.}},
  \href{https://doi.org/10.1086/157448}{\emph{\apj} {\bfseries 233} (1979)
  857}.

\bibitem{Drasco:2005kz}
S.~Drasco and S.A.~Hughes, \emph{{Gravitational wave snapshots of generic
  extreme mass ratio inspirals}},
  \href{https://doi.org/10.1103/PhysRevD.73.024027}{\emph{Phys. Rev. D}
  {\bfseries 73} (2006) 024027}
  [\href{https://arxiv.org/abs/gr-qc/0509101}{{\ttfamily gr-qc/0509101}}].

\bibitem{Hughes:2021exa}
S.A.~Hughes, N.~Warburton, G.~Khanna, A.J.K.~Chua and M.L.~Katz,
  \emph{{Adiabatic waveforms for extreme mass-ratio inspirals via multivoice
  decomposition in time and frequency}},
  \href{https://doi.org/10.1103/PhysRevD.103.104014}{\emph{Phys. Rev. D}
  {\bfseries 103} (2021) 104014}
  [\href{https://arxiv.org/abs/2102.02713}{{\ttfamily 2102.02713}}].

\bibitem{Brito:2015oca}
R.~Brito, V.~Cardoso and P.~Pani, \emph{{Superradiance}: {New Frontiers in
  Black Hole Physics}},
  \href{https://doi.org/10.1007/978-3-319-19000-6}{\emph{Lect. Notes Phys.}
  {\bfseries 906} (2015) pp.1}
  [\href{https://arxiv.org/abs/1501.06570}{{\ttfamily 1501.06570}}].

\bibitem{Vicente:2022ivh}
R.~Vicente and V.~Cardoso, \emph{{Dynamical friction of black holes in
  ultralight dark matter}},
  \href{https://doi.org/10.1103/PhysRevD.105.083008}{\emph{Phys. Rev. D}
  {\bfseries 105} (2022) 083008}
  [\href{https://arxiv.org/abs/2201.08854}{{\ttfamily 2201.08854}}].

\bibitem{Baumann:2021fkf}
D.~Baumann, G.~Bertone, J.~Stout and G.M.~Tomaselli, \emph{{Ionization of
  gravitational atoms}},
  \href{https://doi.org/10.1103/PhysRevD.105.115036}{\emph{Phys. Rev. D}
  {\bfseries 105} (2022) 115036}
  [\href{https://arxiv.org/abs/2112.14777}{{\ttfamily 2112.14777}}].

\bibitem{Bartos:2016dgn}
I.~Bartos, B.~Kocsis, Z.~Haiman and S.~M{\'a}rka, \emph{{Rapid and Bright
  Stellar-mass Binary Black Hole Mergers in Active Galactic Nuclei}},
  \href{https://doi.org/10.3847/1538-4357/835/2/165}{\emph{Astrophys. J.}
  {\bfseries 835} (2017) 165}
  [\href{https://arxiv.org/abs/1602.03831}{{\ttfamily 1602.03831}}].

\bibitem{Kuntz:2021ohi}
A.~Kuntz, F.~Serra and E.~Trincherini, \emph{{Effective two-body approach to
  the hierarchical three-body problem}},
  \href{https://doi.org/10.1103/PhysRevD.104.024016}{\emph{Phys. Rev. D}
  {\bfseries 104} (2021) 024016}
  [\href{https://arxiv.org/abs/2104.13387}{{\ttfamily 2104.13387}}].

\bibitem{Cusin:2024git}
G.~Cusin, C.~Pitrou, C.~Bonvin, A.~Barrau and K.~Martineau, \emph{{Boosting
  gravitational waves: a review of kinematic effects on amplitude,
  polarization, frequency and energy density}},
  \href{https://doi.org/10.1088/1361-6382/ad7ad0}{\emph{Class. Quant. Grav.}
  {\bfseries 41} (2024) 225006}
  [\href{https://arxiv.org/abs/2405.01297}{{\ttfamily 2405.01297}}].

\bibitem{Vijaykumar:2023tjg}
A.~Vijaykumar, A.~Tiwari, S.J.~Kapadia, K.G.~Arun and P.~Ajith, \emph{{Waltzing
  Binaries: Probing the Line-of-sight Acceleration of Merging Compact Objects
  with Gravitational Waves}},
  \href{https://doi.org/10.3847/1538-4357/acd77d}{\emph{Astrophys. J.}
  {\bfseries 954} (2023) 105}
  [\href{https://arxiv.org/abs/2302.09651}{{\ttfamily 2302.09651}}].

\bibitem{Hawking:1972BlackHoles}
S.W.~Hawking, \emph{{Black Holes in General Relativity}},
  \href{https://doi.org/10.1007/BF01877517}{\emph{Commun. Math. Phys.}
  {\bfseries 25} (1972) 152}.

\bibitem{wald1984general}
R.M.~Wald, \emph{General Relativity}, University of Chicago Press, Chicago
  (1984).

\bibitem{Kunduri:2007vf}
H.K.~Kunduri, J.~Lucietti and H.S.~Reall, \emph{{Near-horizon symmetries of
  extremal black holes}},
  \href{https://doi.org/10.1088/0264-9381/24/16/012}{\emph{Class. Quant. Grav.}
  {\bfseries 24} (2007) 4169}
  [\href{https://arxiv.org/abs/0705.4214}{{\ttfamily 0705.4214}}].

\bibitem{Kerr:1963ud}
R.P.~Kerr, \emph{{Gravitational Field of a Spinning Mass as an Example of
  Algebraically Special Metrics}},
  \href{https://doi.org/10.1103/PhysRevLett.11.237}{\emph{Phys. Rev. Lett.}
  {\bfseries 11} (1963) 237}.

\bibitem{Bardeen:1999px}
J.M.~Bardeen and G.T.~Horowitz, \emph{{The Extreme Kerr Throat Geometry: A
  Vacuum Analog of AdS(2) x S(2)}},
  \href{https://doi.org/10.1103/PhysRevD.60.104030}{\emph{Phys. Rev. D}
  {\bfseries 60} (1999) 104030}
  [\href{https://arxiv.org/abs/hep-th/9905099}{{\ttfamily hep-th/9905099}}].

\bibitem{Miller-Jones:2021cyg}
J.C.A.~Miller-Jones et~al., \emph{{Cygnus X-1 contains a 21-solar mass black
  hole---implications for massive star winds}},
  \href{https://doi.org/10.1126/science.abb3363}{\emph{Science} {\bfseries 371}
  (2021) 1046} [\href{https://arxiv.org/abs/2102.09091}{{\ttfamily
  2102.09091}}].

\bibitem{Zhao:2021cyg}
X.~Zhao, L.~Gou, Y.~Dong, X.~Zheng, J.F.~Steiner, J.C.A.~Miller-Jones et~al.,
  \emph{{Re-estimating the Spin Parameter of the Black Hole in Cygnus X-1}},
  \href{https://doi.org/10.3847/1538-4357/ac0955}{\emph{Astrophys. J.}
  {\bfseries 916} (2021) 108}
  [\href{https://arxiv.org/abs/2102.09093}{{\ttfamily 2102.09093}}].

\bibitem{Bardeen:1973gs}
J.M.~Bardeen, B.~Carter and S.W.~Hawking, \emph{{The Four Laws of Black Hole
  Mechanics}}, \href{https://doi.org/10.1007/BF01645742}{\emph{Commun. Math.
  Phys.} {\bfseries 31} (1973) 161}.

\bibitem{Hawking:1975vcx}
S.W.~Hawking, \emph{{Particle Creation by Black Holes}},
  \href{https://doi.org/10.1007/BF02345020}{\emph{Commun. Math. Phys.}
  {\bfseries 43} (1975) 199}.

\bibitem{Bisognano:1975ih}
J.J.~Bisognano and E.H.~Wichmann, \emph{{On the Duality Condition for a
  Hermitian Scalar Field}}, \href{https://doi.org/10.1063/1.522605}{\emph{J.
  Math. Phys.} {\bfseries 16} (1975) 985}.

\bibitem{Unruh:1976db}
W.G.~Unruh, \emph{{Notes on black-hole evaporation}},
  \href{https://doi.org/10.1103/PhysRevD.14.870}{\emph{Phys. Rev. D} {\bfseries
  14} (1976) 870}.

\bibitem{Witten:2024bht}
E.~Witten, \emph{{Introduction to Black Hole Thermodynamics}},
  \href{https://doi.org/10.1140/epjp/s13360-025-06288-y}{\emph{Eur. Phys. J.
  Plus} {\bfseries 140} (2025) 430}
  [\href{https://arxiv.org/abs/2412.16795}{{\ttfamily 2412.16795}}].

\bibitem{Harlow:2014yka}
D.~Harlow, \emph{{Jerusalem Lectures on Black Holes and Quantum Information}},
  \href{https://doi.org/10.1103/RevModPhys.88.015002}{\emph{Rev. Mod. Phys.}
  {\bfseries 88} (2016) 015002}
  [\href{https://arxiv.org/abs/1409.1231}{{\ttfamily 1409.1231}}].

\bibitem{Davies:1976ei}
P.C.W.~Davies, S.A.~Fulling and W.G.~Unruh, \emph{{Energy Momentum Tensor Near
  an Evaporating Black Hole}},
  \href{https://doi.org/10.1103/PhysRevD.13.2720}{\emph{Phys. Rev. D}
  {\bfseries 13} (1976) 2720}.

\bibitem{Christensen:1977jc}
S.M.~Christensen and S.A.~Fulling, \emph{{Trace Anomalies and the Hawking
  Effect}}, \href{https://doi.org/10.1103/PhysRevD.15.2088}{\emph{Phys. Rev. D}
  {\bfseries 15} (1977) 2088}.

\bibitem{Page:1976df}
D.N.~Page, \emph{{Particle Emission Rates from a Black Hole: Massless Particles
  from an Uncharged, Nonrotating Hole}},
  \href{https://doi.org/10.1103/PhysRevD.13.198}{\emph{Phys. Rev. D} {\bfseries
  13} (1976) 198}.

\bibitem{Page:1976ki}
D.N.~Page, \emph{{Particle Emission Rates from a Black Hole. II. Massless
  Particles from a Rotating Hole}},
  \href{https://doi.org/10.1103/PhysRevD.14.3260}{\emph{Phys. Rev. D}
  {\bfseries 14} (1976) 3260}.

\bibitem{Zeldovich:1971mw}
Y.B.~Zel'dovich, \emph{{Generation of Waves by a Rotating Body}}, {\emph{JETP
  Lett.} {\bfseries 14} (1971) 180}.

\bibitem{Starobinsky:1973aij}
A.A.~Starobinsky and S.M.~Churilov, \emph{{Amplification of Electromagnetic and
  Gravitational Waves Scattered by a Rotating Black Hole}}, {\emph{Sov. Phys.
  JETP} {\bfseries 38} (1974) 1}.

\bibitem{Schwinger:1951nm}
J.S.~Schwinger, \emph{{On Gauge Invariance and Vacuum Polarization}},
  \href{https://doi.org/10.1103/PhysRev.82.664}{\emph{Phys. Rev.} {\bfseries
  82} (1951) 664}.

\bibitem{Israel:1967wqNoHair}
W.~Israel, \emph{{Event Horizons in Static Vacuum Space-Times}},
  \href{https://doi.org/10.1103/PhysRev.164.1776}{\emph{Phys. Rev.} {\bfseries
  164} (1967) 1776}.

\bibitem{Carter:1971zcNoHair}
B.~Carter, \emph{{Axisymmetric Black Hole Has Only Two Degrees of Freedom}},
  \href{https://doi.org/10.1103/PhysRevLett.26.331}{\emph{Phys. Rev. Lett.}
  {\bfseries 26} (1971) 331}.

\bibitem{Robinson:1975bv}
D.C.~Robinson, \emph{{Uniqueness of the Kerr black hole}},
  \href{https://doi.org/10.1103/PhysRevLett.34.905}{\emph{Phys. Rev. Lett.}
  {\bfseries 34} (1975) 905}.

\bibitem{Bekenstein:1973ur}
J.D.~Bekenstein, \emph{{Black Holes and Entropy}},
  \href{https://doi.org/10.1103/PhysRevD.7.2333}{\emph{Phys. Rev. D} {\bfseries
  7} (1973) 2333}.

\bibitem{Hawking:1971tu}
S.W.~Hawking, \emph{{Gravitational Radiation from Colliding Black Holes}},
  \href{https://doi.org/10.1103/PhysRevLett.26.1344}{\emph{Phys. Rev. Lett.}
  {\bfseries 26} (1971) 1344}.

\bibitem{Wald:1993nt}
R.M.~Wald, \emph{{Black Hole Entropy is Noether Charge}},
  \href{https://doi.org/10.1103/PhysRevD.48.R3427}{\emph{Phys. Rev. D}
  {\bfseries 48} (1993) R3427}
  [\href{https://arxiv.org/abs/gr-qc/9307038}{{\ttfamily gr-qc/9307038}}].

\bibitem{Wall:2011hj}
A.C.~Wall, \emph{A proof of the generalized second law for rapidly changing
  fields and arbitrary horizon slices},
  \href{https://doi.org/10.1103/PhysRevD.85.104049}{\emph{Phys. Rev. D}
  {\bfseries 85} (2012) 104049}
  [\href{https://arxiv.org/abs/1105.3445}{{\ttfamily 1105.3445}}].

\bibitem{Susskind:1994sm}
L.~Susskind and J.~Uglum, \emph{{Black hole entropy in canonical quantum
  gravity and superstring theory}},
  \href{https://doi.org/10.1103/PhysRevD.50.2700}{\emph{Phys. Rev. D}
  {\bfseries 50} (1994) 2700}
  [\href{https://arxiv.org/abs/hep-th/9401070}{{\ttfamily hep-th/9401070}}].

\bibitem{Chandrasekaran:2022eqq}
V.~Chandrasekaran, G.~Penington and E.~Witten, \emph{Large $n$ algebras and
  generalized entropy},
  \href{https://doi.org/10.1007/JHEP04(2023)009}{\emph{JHEP} {\bfseries 04}
  (2023) 009} [\href{https://arxiv.org/abs/2209.10454}{{\ttfamily
  2209.10454}}].

\bibitem{tHooft:1993dmi}
G.~'t~Hooft, \emph{{Dimensional Reduction in Quantum Gravity}}, {\emph{Conf.
  Proc. C} {\bfseries 930308} (1993) 284}
  [\href{https://arxiv.org/abs/gr-qc/9310026}{{\ttfamily gr-qc/9310026}}].

\bibitem{Susskind:1994vu}
L.~Susskind, \emph{{The World as a Hologram}},
  \href{https://doi.org/10.1063/1.531249}{\emph{J. Math. Phys.} {\bfseries 36}
  (1995) 6377} [\href{https://arxiv.org/abs/hep-th/9409089}{{\ttfamily
  hep-th/9409089}}].

\bibitem{Maldacena:1997re}
J.M.~Maldacena, \emph{{The Large N limit of superconformal field theories and
  supergravity}}, \href{https://doi.org/10.1023/A:1026654312961,
  10.4310/ATMP.1998.v2.n2.a1}{\emph{Int. J. Theor. Phys.} {\bfseries 38} (1999)
  1113} [\href{https://arxiv.org/abs/hep-th/9711200}{{\ttfamily
  hep-th/9711200}}].

\bibitem{Gubser:1998bc}
S.S.~Gubser, I.R.~Klebanov and A.M.~Polyakov, \emph{{Gauge theory correlators
  from noncritical string theory}},
  \href{https://doi.org/10.1016/S0370-2693(98)00377-3}{\emph{Phys. Lett. B}
  {\bfseries 428} (1998) 105}
  [\href{https://arxiv.org/abs/hep-th/9802109}{{\ttfamily hep-th/9802109}}].

\bibitem{Witten:1998qj}
E.~Witten, \emph{{Anti-de Sitter space and holography}},
  \href{https://doi.org/10.4310/ATMP.1998.v2.n2.a2}{\emph{Adv. Theor. Math.
  Phys.} {\bfseries 2} (1998) 253}
  [\href{https://arxiv.org/abs/hep-th/9802150}{{\ttfamily hep-th/9802150}}].

\bibitem{Gibbons:1976ue}
G.W.~Gibbons and S.W.~Hawking, \emph{{Action Integrals and Partition Functions
  in Quantum Gravity}},
  \href{https://doi.org/10.1103/PhysRevD.15.2752}{\emph{Phys. Rev. D}
  {\bfseries 15} (1977) 2752}.

\bibitem{Ryu:2006ef}
S.~Ryu and T.~Takayanagi, \emph{{Aspects of Holographic Entanglement Entropy}},
  \href{https://doi.org/10.1088/1126-6708/2006/08/045}{\emph{JHEP} {\bfseries
  08} (2006) 045} [\href{https://arxiv.org/abs/hep-th/0605073}{{\ttfamily
  hep-th/0605073}}].

\bibitem{Lewkowycz:2013nqa}
A.~Lewkowycz and J.~Maldacena, \emph{{Generalized gravitational entropy}},
  \href{https://doi.org/10.1007/JHEP08(2013)090}{\emph{JHEP} {\bfseries 08}
  (2013) 090} [\href{https://arxiv.org/abs/1304.4926}{{\ttfamily 1304.4926}}].

\bibitem{Almheiri:2019qdq}
A.~Almheiri, T.~Hartman, J.~Maldacena, E.~Shaghoulian and A.~Tajdini,
  \emph{{Replica Wormholes and the Entropy of Hawking Radiation}},
  \href{https://doi.org/10.1007/JHEP05(2020)013}{\emph{JHEP} {\bfseries 05}
  (2020) 013} [\href{https://arxiv.org/abs/1911.12333}{{\ttfamily
  1911.12333}}].

\bibitem{Penington:2019kki}
G.~Penington, S.H.~Shenker, D.~Stanford and Z.~Yang, \emph{{Replica wormholes
  and the black hole interior}},
  \href{https://doi.org/10.1007/JHEP03(2022)205}{\emph{JHEP} {\bfseries 03}
  (2022) 205} [\href{https://arxiv.org/abs/1911.11977}{{\ttfamily
  1911.11977}}].

\bibitem{Iliesiu:2020qvm}
L.V.~Iliesiu and G.J.~Turiaci, \emph{{The statistical mechanics of
  near-extremal black holes}},
  \href{https://doi.org/10.1007/JHEP05(2021)145}{\emph{JHEP} {\bfseries 05}
  (2021) 145} [\href{https://arxiv.org/abs/2003.02860}{{\ttfamily
  2003.02860}}].

\bibitem{Cabo-Bizet:2018ehj}
A.~Cabo-Bizet, D.~Cassani, D.~Martelli and S.~Murthy, \emph{{Microscopic origin
  of the Bekenstein-Hawking entropy of supersymmetric AdS$_{5}$ black holes}},
  \href{https://doi.org/10.1007/JHEP10(2019)062}{\emph{JHEP} {\bfseries 10}
  (2019) 062} [\href{https://arxiv.org/abs/1810.11442}{{\ttfamily
  1810.11442}}].

\bibitem{Iliesiu:2021}
L.V.~Iliesiu, M.~Kologlu and G.J.~Turiaci, \emph{{Supersymmetric indices
  factorize}}, \href{https://doi.org/10.1007/JHEP05(2023)032}{\emph{JHEP}
  {\bfseries 05} (2023) 032}
  [\href{https://arxiv.org/abs/2107.09062}{{\ttfamily 2107.09062}}].

\bibitem{York:1972sj}
J.W.~York, Jr., \emph{{Role of conformal three geometry in the dynamics of
  gravitation}}, \href{https://doi.org/10.1103/PhysRevLett.28.1082}{\emph{Phys.
  Rev. Lett.} {\bfseries 28} (1972) 1082}.

\bibitem{Bombelli:1986rw}
L.~Bombelli, R.K.~Koul, J.~Lee and R.D.~Sorkin, \emph{{A Quantum Source of
  Entropy for Black Holes}},
  \href{https://doi.org/10.1103/PhysRevD.34.373}{\emph{Phys. Rev. D} {\bfseries
  34} (1986) 373}.

\bibitem{Srednicki:1993im}
M.~Srednicki, \emph{{Entropy and Area}},
  \href{https://doi.org/10.1103/PhysRevLett.71.666}{\emph{Phys. Rev. Lett.}
  {\bfseries 71} (1993) 666}
  [\href{https://arxiv.org/abs/hep-th/9303048}{{\ttfamily hep-th/9303048}}].

\bibitem{Braden:1990hwCharged}
H.W.~Braden, J.D.~Brown, B.F.~Whiting and J.W.~York, Jr., \emph{{Charged Black
  Hole in a Grand Canonical Ensemble}},
  \href{https://doi.org/10.1103/PhysRevD.42.3376}{\emph{Phys. Rev. D}
  {\bfseries 42} (1990) 3376}.

\bibitem{Hartle:1983ai}
J.B.~Hartle and S.W.~Hawking, \emph{{Wave Function of the Universe}},
  \href{https://doi.org/10.1103/PhysRevD.28.2960}{\emph{Phys. Rev. D}
  {\bfseries 28} (1983) 2960}.

\bibitem{Ryu:2006bv}
S.~Ryu and T.~Takayanagi, \emph{{Holographic Derivation of Entanglement Entropy
  from AdS/CFT}},
  \href{https://doi.org/10.1103/PhysRevLett.96.181602}{\emph{Phys. Rev. Lett.}
  {\bfseries 96} (2006) 181602}
  [\href{https://arxiv.org/abs/hep-th/0603001}{{\ttfamily hep-th/0603001}}].

\bibitem{Hubeny:2007xt}
V.E.~Hubeny, M.~Rangamani and T.~Takayanagi, \emph{{A Covariant Holographic
  Entanglement Entropy Proposal}},
  \href{https://doi.org/10.1088/1126-6708/2007/07/062}{\emph{JHEP} {\bfseries
  07} (2007) 062} [\href{https://arxiv.org/abs/0705.0016}{{\ttfamily
  0705.0016}}].

\bibitem{Engelhardt:2014gca}
N.~Engelhardt and A.C.~Wall, \emph{Quantum extremal surfaces: Holographic
  entanglement entropy beyond the classical regime},
  \href{https://doi.org/10.1007/JHEP01(2015)073}{\emph{JHEP} {\bfseries 01}
  (2015) 073} [\href{https://arxiv.org/abs/1408.3203}{{\ttfamily 1408.3203}}].

\bibitem{Cotler:2016fpe}
J.S.~Cotler, G.~Gur-Ari, M.~Hanada, J.~Polchinski, P.~Saad, S.H.~Shenker
  et~al., \emph{{Black Holes and Random Matrices}},
  \href{https://doi.org/10.1007/JHEP05(2017)118}{\emph{JHEP} {\bfseries 05}
  (2017) 118} [\href{https://arxiv.org/abs/1611.04650}{{\ttfamily
  1611.04650}}].

\bibitem{Saad:2018bqo}
P.~Saad, S.H.~Shenker and D.~Stanford, \emph{{A semiclassical ramp in SYK and
  in gravity}},  \href{https://arxiv.org/abs/1806.06840}{{\ttfamily
  1806.06840}}.

\bibitem{Saad:2019lba}
P.~Saad, S.H.~Shenker and D.~Stanford, \emph{{JT gravity as a matrix
  integral}},  \href{https://arxiv.org/abs/1903.11115}{{\ttfamily 1903.11115}}.

\bibitem{Stanford:2019vob}
D.~Stanford and E.~Witten, \emph{{JT gravity and the ensembles of random matrix
  theory}}, \href{https://doi.org/10.4310/ATMP.2020.v24.n6.a4}{\emph{Adv.
  Theor. Math. Phys.} {\bfseries 24} (2020) 1475}
  [\href{https://arxiv.org/abs/1907.03363}{{\ttfamily 1907.03363}}].

\bibitem{Maxfield:2020ale}
H.~Maxfield and G.J.~Turiaci, \emph{{The path integral of 3D gravity near
  extremality; or, JT gravity with defects as a matrix integral}},
  \href{https://doi.org/10.1007/JHEP01(2021)118}{\emph{JHEP} {\bfseries 01}
  (2021) 118} [\href{https://arxiv.org/abs/2006.11317}{{\ttfamily
  2006.11317}}].

\bibitem{Turiaci:2023wrh}
G.J.~Turiaci, \emph{{New insights on near-extremal black holes}}, {\emph{ICTS
  News} {\bfseries IX} (2023) }
  [\href{https://arxiv.org/abs/2307.10423}{{\ttfamily 2307.10423}}].

\bibitem{Newman:1965my}
E.T.~Newman, E.~Couch, K.~Chinnapared, A.~Exton, A.~Prakash and R.~Torrence,
  \emph{{Metric of a Rotating, Charged Mass}},
  \href{https://doi.org/10.1063/1.1704351}{\emph{J. Math. Phys.} {\bfseries 6}
  (1965) 918}.

\bibitem{Page:2000dk}
D.N.~Page, \emph{{Thermodynamics of near extreme black holes}},
  \href{https://arxiv.org/abs/hep-th/0012020}{{\ttfamily hep-th/0012020}}.

\bibitem{Ferrara:1995ih}
S.~Ferrara, R.~Kallosh and A.~Strominger, \emph{$n=2$ extremal black holes},
  \href{https://doi.org/10.1103/PhysRevD.52.R5412}{\emph{Phys. Rev. D}
  {\bfseries 52} (1995) R5412}
  [\href{https://arxiv.org/abs/hep-th/9508072}{{\ttfamily hep-th/9508072}}].

\bibitem{Goldstein:2005hq}
K.~Goldstein, N.~Iizuka, R.P.~Jena and S.P.~Trivedi, \emph{Non-supersymmetric
  attractors}, \href{https://doi.org/10.1103/PhysRevD.72.124021}{\emph{Phys.
  Rev. D} {\bfseries 72} (2005) 124021}
  [\href{https://arxiv.org/abs/hep-th/0507096}{{\ttfamily hep-th/0507096}}].

\bibitem{Sen:2005wa}
A.~Sen, \emph{{Black hole entropy function and the attractor mechanism in
  higher derivative gravity}},
  \href{https://doi.org/10.1088/1126-6708/2005/09/038}{\emph{JHEP} {\bfseries
  09} (2005) 038} [\href{https://arxiv.org/abs/hep-th/0506177}{{\ttfamily
  hep-th/0506177}}].

\bibitem{Israel:1986ThirdLaw}
W.~Israel, \emph{{Third Law of Black-Hole Dynamics: A Formulation and Proof}},
  \href{https://doi.org/10.1103/PhysRevLett.57.397}{\emph{Phys. Rev. Lett.}
  {\bfseries 57} (1986) 397}.

\bibitem{Kehle:2022ThirdLaw}
C.~Kehle and R.~Unger, \emph{{Gravitational Collapse to Extremal Black Holes
  and the Third Law of Black Hole Thermodynamics}},
  \href{https://doi.org/10.4171/JEMS/1591}{\emph{J. Eur. Math. Soc.} (2025) }
  [\href{https://arxiv.org/abs/2211.15742}{{\ttfamily 2211.15742}}].

\bibitem{Preskill:1991tb}
J.~Preskill, P.~Schwarz, A.D.~Shapere, S.~Trivedi and F.~Wilczek,
  \emph{{Limitations on the statistical description of black holes}},
  \href{https://doi.org/10.1142/S0217732391002773}{\emph{Mod. Phys. Lett. A}
  {\bfseries 6} (1991) 2353}.

\bibitem{Shi:2025gpi}
X.~Shi and G.J.~Turiaci, \emph{{The phase of the gravitational path integral}},
  \href{https://doi.org/10.1007/JHEP07(2025)047}{\emph{JHEP} {\bfseries 07}
  (2025) 047} [\href{https://arxiv.org/abs/2504.00900}{{\ttfamily
  2504.00900}}].

\bibitem{Polchinski:1988ua}
J.~Polchinski, \emph{{The phase of the sum over spheres}},
  \href{https://doi.org/10.1016/0370-2693(89)90387-0}{\emph{Phys. Lett. B}
  {\bfseries 219} (1989) 251}.

\bibitem{Maldacena:2024spf}
J.~Maldacena, \emph{{Real observers solving imaginary problems}},
  \href{https://arxiv.org/abs/2412.14014}{{\ttfamily 2412.14014}}.

\bibitem{Chen:2025dSPhase}
Y.~Chen, D.~Stanford, H.~Tang and Z.~Yang, \emph{{On the phase of the de Sitter
  density of states}},
  \href{https://doi.org/10.1007/JHEP05(2026)068}{\emph{JHEP} {\bfseries 05}
  (2026) 068} [\href{https://arxiv.org/abs/2511.01400}{{\ttfamily
  2511.01400}}].

\bibitem{Gross:1982cv}
D.J.~Gross, M.J.~Perry and L.G.~Yaffe, \emph{{Instability of Flat Space at
  Finite Temperature}},
  \href{https://doi.org/10.1103/PhysRevD.25.330}{\emph{Phys. Rev. D} {\bfseries
  25} (1982) 330}.

\bibitem{Kolanowski:2024zrq}
M.~Kolanowski, D.~Marolf, I.~Rakic, M.~Rangamani and G.J.~Turiaci,
  \emph{{Looking at extremal black holes from very far away}},
  \href{https://doi.org/10.1007/JHEP04(2025)020}{\emph{JHEP} {\bfseries 04}
  (2025) 020} [\href{https://arxiv.org/abs/2409.16248}{{\ttfamily
  2409.16248}}].

\bibitem{Mertens:2022irh}
T.G.~Mertens and G.J.~Turiaci, \emph{{Solvable models of quantum black holes: a
  review on Jackiw\textendash{}Teitelboim gravity}},
  \href{https://doi.org/10.1007/s41114-023-00046-1}{\emph{Living Rev. Rel.}
  {\bfseries 26} (2023) 4} [\href{https://arxiv.org/abs/2210.10846}{{\ttfamily
  2210.10846}}].

\bibitem{Nayak:2018qej}
P.~Nayak, A.~Shukla, R.M.~Soni, S.P.~Trivedi and V.~Vishal, \emph{{On the
  Dynamics of Near-Extremal Black Holes}},
  \href{https://doi.org/10.1007/JHEP09(2018)048}{\emph{JHEP} {\bfseries 09}
  (2018) 048} [\href{https://arxiv.org/abs/1802.09547}{{\ttfamily
  1802.09547}}].

\bibitem{Sachdev:2019bjn}
S.~Sachdev, \emph{{Universal low temperature theory of charged black holes with
  AdS$_2$ horizons}}, \href{https://doi.org/10.1063/1.5092726}{\emph{J. Math.
  Phys.} {\bfseries 60} (2019) 052303}
  [\href{https://arxiv.org/abs/1902.04078}{{\ttfamily 1902.04078}}].

\bibitem{Stanford:2017thb}
D.~Stanford and E.~Witten, \emph{{Fermionic Localization of the Schwarzian
  Theory}}, \href{https://doi.org/10.1007/JHEP10(2017)008}{\emph{JHEP}
  {\bfseries 10} (2017) 008}
  [\href{https://arxiv.org/abs/1703.04612}{{\ttfamily 1703.04612}}].

\bibitem{Mertens:2017mtv}
T.G.~Mertens, G.J.~Turiaci and H.L.~Verlinde, \emph{{Solving the Schwarzian via
  the Conformal Bootstrap}},
  \href{https://doi.org/10.1007/JHEP08(2017)136}{\emph{JHEP} {\bfseries 08}
  (2017) 136} [\href{https://arxiv.org/abs/1705.08408}{{\ttfamily
  1705.08408}}].

\bibitem{Rakic:2023vhv}
I.~Rakic, M.~Rangamani and G.J.~Turiaci, \emph{{Thermodynamics of the
  near-extremal Kerr spacetime}},
  \href{https://doi.org/10.1007/JHEP06(2024)011}{\emph{JHEP} {\bfseries 06}
  (2024) 011} [\href{https://arxiv.org/abs/2310.04532}{{\ttfamily
  2310.04532}}].

\bibitem{Turiaci:2026gtl}
G.J.~Turiaci and C.-H.~Wu, \emph{{Collapse to extremality: The third law of
  black hole thermodynamics in quantum gravity}},
  \href{https://arxiv.org/abs/2609.26875}{{\ttfamily 2609.26875}}.

\bibitem{Heydeman:2020hhw}
M.~Heydeman, L.V.~Iliesiu, G.J.~Turiaci and W.~Zhao, \emph{{The statistical
  mechanics of near-BPS black holes}},
  \href{https://doi.org/10.1088/1751-8121/ac3be9}{\emph{J. Phys. A} {\bfseries
  55} (2022) 014004} [\href{https://arxiv.org/abs/2011.01953}{{\ttfamily
  2011.01953}}].

\bibitem{Reall:2024ThirdLaw}
H.S.~Reall, \emph{{Third Law of Black Hole Mechanics for Supersymmetric Black
  Holes and a Quasilocal Mass-Charge Inequality}},
  \href{https://doi.org/10.1103/PhysRevD.110.124059}{\emph{Phys. Rev. D}
  {\bfseries 110} (2024) 124059}
  [\href{https://arxiv.org/abs/2410.11956}{{\ttfamily 2410.11956}}].

\bibitem{Maldacena:1997ih}
J.M.~Maldacena and A.~Strominger, \emph{{Universal low-energy dynamics for
  rotating black holes}},
  \href{https://doi.org/10.1103/PhysRevD.56.4975}{\emph{Phys. Rev. D}
  {\bfseries 56} (1997) 4975}
  [\href{https://arxiv.org/abs/hep-th/9702015}{{\ttfamily hep-th/9702015}}].

\bibitem{Brown:2024ajk}
A.R.~Brown, L.V.~Iliesiu, G.~Penington and M.~Usatyuk, \emph{{The evaporation
  of charged black holes}},
  \href{https://doi.org/10.1007/JHEP01(2026)109}{\emph{JHEP} {\bfseries 01}
  (2026) 109} [\href{https://arxiv.org/abs/2411.03447}{{\ttfamily
  2411.03447}}].

\bibitem{Biggs:2025evap}
A.~Biggs, \emph{{Following the state of an evaporating charged black hole into
  the quantum gravity regime}},
  \href{https://arxiv.org/abs/2503.02051}{{\ttfamily 2503.02051}}.

\bibitem{Emparan:2025sao}
R.~Emparan, \emph{{Quantum cross-section of near-extremal black holes}},
  \href{https://doi.org/10.1007/JHEP04(2025)122}{\emph{JHEP} {\bfseries 04}
  (2025) 122} [\href{https://arxiv.org/abs/2501.17470}{{\ttfamily
  2501.17470}}].

\bibitem{Joas2022CopenhagenSpirit}
C.~Joas, ``The {Copenhagen} spirit: A history of the {Niels Bohr Institute}.''

\bibitem{sfi}
{Simons Foundation International (SFI)}. \url{https://www.sfi.org.bm/}.

\bibitem{sf}
{Simons Foundation}. \url{https://www.simonsfoundation.org/}.

\end{thebibliography}\endgroup
\bibliographystyle{JHEP}

\end{document}